\documentclass[11pt]{article}
\newcommand{\DocumentTitle}{The strengthened quantum Hamming bound of Yu, Lai, and Oh}
\usepackage[a4paper,margin=26.5mm,includefoot]{geometry}
\usepackage{iftex}
\ifPDFTeX
  \input{glyphtounicode}
\fi
\usepackage[T1]{fontenc}
\usepackage{lmodern}
\usepackage[protrusion=false]{microtype}
\usepackage{amsmath,amssymb,amsthm,mathtools}
\usepackage{booktabs,array,longtable}
\usepackage[hidelinks]{hyperref}
\hypersetup{pdftitle={\DocumentTitle},
 pdfauthor={Valery (Binyamin) Grishin; Aryeh Lev Zabokritskiy (Yohananov)},pdflang={en}}
\newtheorem{theorem}{Theorem}[section]
\newtheorem{lemma}[theorem]{Lemma}
\newtheorem{proposition}[theorem]{Proposition}

\newcommand{\Tr}{\operatorname{Tr}}
\newcommand{\E}{\mathbb E}
\newcommand{\Bin}{\operatorname{Bin}}
\newcommand{\Kraw}[2]{P_{#1}^{(#2,Q)}}

\author{
  \normalsize Valery (Binyamin) Grishin\\[0.5em]
  \small Department of Computer Science, Tel-Hai University of Kiryat Shmona\\
  \small and the Galilee, Kiryat Shmona, Israel\\[0.25em]
  \small\href{mailto:valerig.tech@gmail.com}{\texttt{valerig.tech@gmail.com}}\\[2em]
  \normalsize Aryeh Lev Zabokritskiy (Yohananov)\\[0.5em]
  \small Department of Computer Science, MIGAL--Galilee Research Institute,\\
  \small Kiryat Shmona, Israel; and Department of Computer Science, Tel-Hai\\
  \small University of Kiryat Shmona and the Galilee, Kiryat Shmona, Israel\\[0.25em]
  \small\href{mailto:yuhanalev@telhai.ac.il}{\texttt{yuhanalev@telhai.ac.il}}\\[0.5em]
  \small Corresponding author. ORCID: \href{https://orcid.org/0000-0003-3151-6192}{0000-0003-3151-6192}.
}

\title{\DocumentTitle}
\date{}
\begin{document}
\maketitle
\begin{abstract}
We prove a conjecture of Yu, Lai, and Oh: their strengthened quantum
Hamming bounds, a family that interpolates between Hamming-type bounds
and the Singleton bound, hold for all quantum codes of dimension
$K>1$, including impure and nonadditive codes. The result covers
every integer local dimension $p\ge2$, not necessarily a prime power,
every integer $t\ge1$, both distances $2t+1$ and $2t+2$, and every
interpolation parameter $e\in\{0,\ldots,t\}$. The case $e=0$ implies the ordinary quantum Hamming bound for general
codes, recently proved by Zhang and Chen. Yu, Lai, and Oh proved their bounds for pure codes and for
general codes of distance three or four. More generally, for every integer $p\ge2$ and every quantum code of
length $n$, dimension $K>1$ and distance $d\ge3$, we prove
$Kf(0)\le p^n\E f(X)$, where $X\sim\Bin(n,1-p^{-2})$, for every real
polynomial $f$ of degree less than $d$ that is nonnegative on
$\{0,\ldots,n\}$ and positive at zero. For pure codes this inequality is implicit
in the argument of Yu, Lai, and Oh. The proof uses the linear-programming
method with quantum weight enumerators.
\end{abstract}

\clearpage
\setcounter{tocdepth}{2}
\tableofcontents
\clearpage
\section{Introduction}\label{sec:introduction}

The quantum Singleton and Hamming bounds give basic restrictions on
the dimension of a quantum code at a prescribed length and
distance~\cite{RainsNB,YLO}. Throughout, a code is exact: a subspace
of $(\mathbb C^p)^{\otimes n}$ that satisfies the error-detection
condition exactly, with no assumption of purity or additivity. For
such a code of length $n$, dimension $K>1$, distance $d$ and integer
local dimension $p\ge2$, the Hamming bound states that $K$ is at most
$p^n$ divided by the volume of the Hamming ball of radius
$\lfloor(d-1)/2\rfloor$ over an alphabet of size $p^2$. The Singleton
bound, proved for qubits by Knill and Laflamme~\cite{KnillLaflamme},
is known for arbitrary quantum subspaces~\cite[Theorem~2]{RainsNB}.
The Hamming bound was first established for nondegenerate
codes~\cite{EkertMacchiavello,Gottesman96}, and its validity for
degenerate codes became a separate question~\cite[Section~7.3]{GottesmanThesis}.
Partial answers covered degenerate stabilizer codes correcting one or
two errors~\cite{GottesmanThesis,KKKS,Aly}, the asymptotic regime and
all sufficiently large lengths~\cite{AshikhminLitsyn,LiXing}, regions
of rate and distance~\cite{SarvepalliKlappenecker}, and all binary
codes of distance below $127$~\cite{DallasAndreadakisLidar}; Nemec and
Tansuwannont proved a Hamming-like bound for degenerate stabilizer
codes~\cite{NemecTansuwannont}. Below we use the related distinction
between pure and impure codes of Calderbank, Rains, Shor, and
Sloane~\cite{CRSS}: every pure code is nondegenerate, and for
stabilizer codes the two distinctions coincide.

Yu, Lai, and Oh~\cite{YLO} introduced a family of bounds, which they
call the quantum Hamming--Singleton bounds, that interpolates between
Hamming-type bounds and the Singleton bound. For distance
$2t+1+\sigma$, where $t\ge1$ and $\sigma\in\{0,1\}$, the family is
indexed by $e\in\{0,\ldots,t\}$, and member $e$ bounds $K$ by $p^n$ divided by the product of $p^{4e+2\sigma}$ and the
volume of the Hamming ball of radius $t-e$ at length $n-2e-\sigma$. In their strengthened quantum Hamming bounds~\cite[Theorem~2]{YLO},
which we call the rounded bounds, the denominator is $g(0)/\E g(X)$,
where $X\sim\Bin(n,1-p^{-2})$ is the weight of a uniformly random error
operator and
$g(x)\coloneqq\binom{n-x}{2e+\sigma}\prod_\xi(x-\lfloor\xi\rfloor)(x-\lfloor\xi\rfloor-1)$,
a polynomial of degree $2t+\sigma$ in which $\xi$ runs over the roots of
the Lloyd polynomial of degree $t-e$ at length $n-2e-\sigma$, a shifted
Krawtchouk polynomial for the alphabet of size $p^2$. This rounding of
each root to a pair of adjacent integers refines the bounds further.
They proved the rounded inequalities for pure codes and also
established the distance-three and distance-four cases for impure
codes. The $e=t$ member is the Singleton bound, which is known for
general codes.

More recently, Zhang and Chen proved the ordinary quantum Hamming
bound for arbitrary exact subspaces of dimension $K>1$, first for
binary codes~\cite{ZhangChenBinary} and then for every integer local
dimension $p\ge3$~\cite{ZhangChenNonbinary}. Their results establish
\[
 K\sum_{j=0}^{\lfloor(d-1)/2\rfloor}\binom nj(p^2-1)^j\le p^n
\]
for codes of distance $d$, without assumptions of purity or
additivity. Thus the ordinary Hamming bound is available for
general exact codes in every integer local dimension $p\ge2$.

The family of Yu, Lai, and Oh~\cite{YLO} contains bounds stronger
than this ordinary bound. Their $e=0$ member is refined by the rounding;
at odd distance the corresponding unrounded member is the ordinary
Hamming bound. At even distance, the unrounded $e=0$ denominator is
$p^2$ times the volume of the Hamming ball of radius $t$ at length
$n-1$, which exceeds the ordinary denominator at length $n$. These
gains can be strict: the rounding increases the denominator whenever a
zero of the Lloyd polynomial is not an integer
(Section~\ref{sec:extremal-interfaces}), and for $p\ge3$ and
$t(p^2-2)>n-d>p^2-2$ an intermediate unrounded member is strictly
stronger than the Singleton bound and the unrounded $e=0$
member~\cite{YLO}. For impure codes, even
the unrounded members were established, as far as we know, only at
$e=t$, at distances three and four, and, after Zhang and Chen, at
$e=0$ for odd distance; for qubit codes of length at most $50$ they
would also follow from the computer verification reported
in~\cite{CRSS} and discussed below. The standard dimension-increasing length
reduction, which would give the unrounded members from the ordinary
Hamming bound, requires purity~\cite[Theorem~20]{RainsWE}; its
partial-trace construction can fail for impure codes, even for
stabilizer codes~\cite[Section~5]{HuberGrassl}.
At the end of their paper, Yu, Lai, and Oh conjectured that the full
rounded family, including all intermediate values of $e$, holds for
arbitrary exact quantum subspaces~\cite{YLO}.

Our first result proves this conjecture.
Write $S_{t,\sigma}^{n,e}$ for the rounded Yu--Lai--Oh denominator.
It depends also on $p$; it is at least the Hamming-ball volume
$\sum_{j=0}^{t}\binom nj(p^2-1)^j$ when $e=0$, and it equals
$p^{2(d-1)}$, with $d=2t+1+\sigma$, when $e=t$. We prove that every
exact quantum code of dimension $K>1$ and distance at least
$2t+1+\sigma$ satisfies
\[
 K S_{t,\sigma}^{n,e}\le p^n
 \qquad
 \bigl(p\ge2,\ t\ge1,\ 0\le e\le t,\ \sigma\in\{0,1\}\bigr),
\]
where $p,t,e$ are integers. The same bounds therefore
apply to impure and nonadditive codes throughout the full parameter
range, beyond pure codes and the known impure cases at distances
three and four. In particular, the case $e=0$ implies the ordinary
Hamming bound of Zhang and Chen; our proof does not use their result.
Since each rounded denominator is at least the corresponding unrounded
one, the unrounded family follows as well.

Our second result removes the purity assumption from a polynomial
inequality that, for pure codes, follows from the constraints of the
quantum linear program and is implicit in the argument of Yu, Lai, and
Oh~\cite[Appendix]{YLO}.
Let an exact code of length $n$, dimension $K>1$ and distance at
least $d\ge3$ be given. For every integer $p\ge2$ we prove that every
real polynomial $f$ of degree less than $d$ with $f(0)>0$ and
$f(j)\ge0$ for $j=0,\ldots,n$ satisfies
\[
 Kf(0)\le p^n\E f(X).
\]
Equivalently, $K$ is at most $p^n$ times the grid minimum, the least
value of $\E f(X)$ over these polynomials normalized by $f(0)=1$. Its
reciprocal is the polynomial form, in degree at most $d-1$, of
Delsarte's linear-programming lower bound for orthogonal arrays of
strength $d-1$ over an alphabet of size $p^2$~\cite{Delsarte}. The
cases $d\le2$ are elementary.
Applied to $g$, which has degree $2t+\sigma<d$, is positive at zero
and is nonnegative on $\{0,\ldots,n\}$, the inequality gives the first result in every integer local dimension
$p\ge2$. For binary codes we also give a second proof of the first result,
which treats the polynomial prescribed by Yu, Lai, and Oh directly,
without the second result.

For qubits, the removal of the purity assumption was anticipated by
Calderbank, Rains, Shor, and Sloane~\cite{CRSS}. In all cases of
their table, up to length $30$, the linear-programming bound for pure
codes coincided with the one for impure codes. They conjectured
that, for the largest real number $K_{\max}>1$ for which the quantum
linear program has a solution, every solution is pure, that is, its
weight enumerator has no terms of weights $1,\ldots,d-1$; they reported a computer verification of this purity conjecture for
$n\le50$. Every pure solution
satisfies the inequality above, and the weight enumerators of a code of
dimension $K$ form a solution at their dimension $K$, so
$K\le K_{\max}$. Hence their conjecture implies our second result for
$p=2$. We prove this consequence for every length, together with its
analogue for every integer $p\ge3$; the conjecture itself is not
proved here.

Section~\ref{sec:framework} states both results precisely, introduces
the tools they share and outlines the proof.
Sections~\ref{sec:grid-minima}--\ref{sec:extremal-interfaces} study
the polynomials that attain the grid minimum, for every $p$;
Sections~\ref{sec:rank-windows}--\ref{sec:qutrit} complete the proof
for $p\ge3$, Section~\ref{sec:binary} treats binary codes, and
Section~\ref{sec:completion} assembles the cases.

\section{Polynomial bounds for quantum codes}\label{sec:framework}

We state both main results and give a sufficient condition for them,
the coefficient criterion, valid for arbitrary exact codes: inequalities
between the values of a polynomial and its Krawtchouk coefficients.
Most of the paper verifies this condition for a polynomial attaining the
grid minimum; for $p\ge3$ and large distances relative to the length,
where it can fail, the code dimension is bounded directly by $p^n$ times
the grid minimum.
Section~\ref{sec:proof-outline} outlines the proof.

Let $p\ge2$ be an integer. A quantum code of length $n$ and dimension
$K$ is a $K$-dimensional subspace of $(\mathbb C^p)^{\otimes n}$; we call
it exact to stress that the detection condition below holds exactly,
not approximately. No purity or additivity is assumed.
Writing $P$ for its orthogonal projector, the code has distance at
least $d$ if
\begin{equation}\label{eq:detection}
 PEP=\frac{\Tr(PE)}K P
 \qquad\text{whenever }|\operatorname{supp}E|<d.
\end{equation}
Here the support is the set of tensor factors on which the operator
acts nontrivially. We assume $K>1$; for $K=1$ the condition holds for
every $E$. The quantum Singleton bound gives
\begin{equation}\label{eq:singleton}
 K\le p^{n-2(d-1)},\qquad\text{hence }n\ge2(d-1)+1\text{ since }K>1;
\end{equation}
see~\cite[Theorem~2]{RainsNB}, whose proof does not require $p$ to be a
prime power.

Put $Q\coloneqq p^2$. For a nonnegative integer $L$, use the normalization
\begin{equation}\label{eq:kraw}
 \Kraw{j}{L}(x)\coloneqq
 \sum_{a=0}^j(-1)^a(Q-1)^{j-a}\binom xa\binom{L-x}{j-a}.
\end{equation}
For $x\in\{0,\ldots,L\}$, their generating function is
\begin{equation}\label{eq:kraw-generating}
 \sum_{j=0}^L\Kraw{j}{L}(x)z^j
       =(1+(Q-1)z)^{L-x}(1-z)^x.
\end{equation}
For arbitrary real $x$, the coefficient of $z^j$ in the formal power
series on the right is $\Kraw{j}{L}(x)$ for every $j\ge0$.
We write $\mu_L$ for the binomial probability measure
$\Bin(L,(Q-1)/Q)$ and set
$v_j(L)\coloneqq\binom Lj(Q-1)^j$. The Krawtchouk polynomials are orthogonal
for $\mu_L$:
\begin{equation}\label{eq:orthogonality}
 \E_{\mu_L}\bigl[\Kraw{i}{L}(X)\Kraw{j}{L}(X)\bigr]
       =v_i(L)\delta_{ij}.
\end{equation}
Indeed, taking the expectation of the product of the two generating
functions gives $(1+(Q-1)zw)^L$. In particular, the coefficient of
$\Kraw{0}{L}=1$ in a polynomial is its $\mu_L$-expectation.
We use $(x)_j\coloneqq x(x-1)\cdots(x-j+1)$, with $(x)_0\coloneqq1$.
For a variable upper argument, $\binom xj\coloneqq(x)_j/j!$ if $j\ge0$.
For nonnegative integer upper arguments, binomial coefficients outside
their usual lower-index range are zero.

\subsection{The rounded quantity}

Fix integers $t\ge1$, $0\le e\le t$, and $\sigma\in\{0,1\}$, and put
\begin{equation}\label{eq:original-parameters}
 d\coloneqq2t+1+\sigma,\quad D\coloneqq d-1,\quad
 m\coloneqq t-e,\quad r\coloneqq2e+\sigma,\quad N\coloneqq n-r.
\end{equation}
Thus $D=2m+r$. When $m>0$, let $\xi_1<\cdots<\xi_m$ be the roots of
$\Kraw{m}{N-1}(x-1)$ and set $a_j\coloneqq\lfloor\xi_j\rfloor$. Define
\begin{equation}\label{eq:rounded-polynomial}
 h(x)\coloneqq\prod_{j=1}^m(x-a_j)(x-a_j-1),\qquad
 g(x)\coloneqq\binom{n-x}{r}h(x)=\sum_{i=0}^D c_i\Kraw{i}{n}(x).
\end{equation}
The empty product is one. The $\xi_j$ are the zeros of the Lloyd
polynomial $\sum_{j=0}^m\Kraw{j}{N}=\Kraw{m}{N-1}(x-1)$ (by
\eqref{eq:kraw-generating}, these partial sums have generating function
$(1+(Q-1)z)^{N-x}(1-z)^{x-1}$); each paired factor
$(x-a_j)(x-a_j-1)$ replaces the square $(x-\xi_j)^2$ by a factor that
is nonnegative at every integer, and this is the rounding. The rounded
quantity of Yu, Lai, and Oh~\cite{YLO}, the denominator of their
strengthened quantum Hamming--Singleton bound, is
\begin{equation}\label{eq:rounded-quantity}
 S_{t,\sigma}^{n,e}\coloneqq\frac{g(0)}{c_0}.
\end{equation}
By~\cite[Appendix, Eqs.~(21) and~(25)]{YLO}, this quotient equals
their closed form~\cite[Eq.~(11)]{YLO}.
Their normalized root polynomial is $h(x)/h(0)$, so the monic
normalization in~\eqref{eq:rounded-polynomial} leaves this quotient
unchanged. These quantities are well defined for every code with
$K>1$, since~\eqref{eq:singleton} gives $n\ge2D+1$; the required
positivity is checked after Theorem~\ref{thm:grid}. For $e=0$,
Lemma~\ref{lem:radau-rounded} (in Section~\ref{sec:extremal-interfaces};
for the rounded Lloyd polynomial it gives
$h(0)/\E_{\mu_N}h\ge V_m(N)\coloneqq\sum_{j\le m}\binom Nj(Q-1)^j$)
and the identity $S_{t,\sigma}^{n,e}=Q^rh(0)/\E_{\mu_N}h(X)$
of~\eqref{eq:terminal-expectation} below give
\[
 S_{t,\sigma}^{n,0}\ge Q^\sigma\sum_{j=0}^{t}\binom{n-\sigma}{j}(Q-1)^j
 \ge\sum_{j=0}^{t}\binom nj(Q-1)^j.
\]
The middle term is the unrounded $e=0$ denominator of Yu, Lai, and Oh
(the unrounded family $Q^rV_m(N)$ is identified after
Lemma~\ref{lem:radau-rounded}), and for $\sigma=1$ the second
inequality follows from Pascal's rule,
$QV_t(n-1)-V_t(n)=\binom{n-1}{t}(Q-1)^{t+1}$. So the rounded $e=0$
bound contains the ordinary Hamming bound.

Yu, Lai, and Oh proved the rounded bound for pure codes and for
the impure distance-three and distance-four cases~\cite[Theorem~2]{YLO}.
We prove the following extension to arbitrary exact subspaces.

\begin{theorem}\label{thm:rounded}
For every integer $p\ge2$, every exact quantum code of dimension
$K>1$ and distance at least $2t+1+\sigma$ satisfies
\begin{equation}\label{eq:main-bound}
 K S_{t,\sigma}^{n,e}\le p^n
 \qquad(0\le e\le t).
\end{equation}
Here $t\ge1$, $\sigma\in\{0,1\}$, and $S_{t,\sigma}^{n,e}$ is the
quantity~\eqref{eq:rounded-quantity}.
\end{theorem}

We prove a more general statement about every polynomial nonnegative
on the weight grid $\{0,\ldots,n\}$, in every integer local dimension
$p\ge2$. Here $D$ is any degree bound below the distance, not
necessarily $d-1$. For binary codes we also prove the coefficient
inequalities of Lemma~\ref{lem:coefficient-criterion} for each
prescribed rounded polynomial itself (Theorem~\ref{thm:binary-rounded});
Theorem~\ref{thm:rounded} does not need this.

\begin{theorem}\label{thm:grid}
Let $p\ge2$ be an integer, let $D\ge2$, and let an exact quantum
code of dimension $K>1$ have distance at least $D+1$. Every real
polynomial $f$ of degree at most $D$ satisfying
$f(0)>0$ and $f(j)\ge0$ for $j=0,\ldots,n$ obeys
\begin{equation}\label{eq:grid-bound}
 Kf(0)\le p^n\E_{\mu_n}f(X).
\end{equation}
\end{theorem}

The same inequality in degrees zero and one is elementary.
For a constant polynomial it follows from $K\le p^n$.
For a normalized linear polynomial $f(x)=1+ax$, grid nonnegativity
gives $a\ge-1/n$, so $\E_{\mu_n}f\ge p^{-2}$;
Singleton gives $K\le p^{n-2}$ whenever the distance is at least two.
If $f(0)=0$, grid nonnegativity makes the inequality immediate.

We prove Theorem~\ref{thm:grid} by comparing $K$ with the smallest
value of $\E_{\mu_n}f/f(0)$ over all admissible $f$, that is, all
polynomials of degree at most $D$ that are nonnegative on the grid and
positive at the origin: once the
inequality holds for a polynomial attaining this minimum, it holds for
every admissible polynomial. We certify such a minimizer rather than
each prescribed polynomial $g$ of Theorem~\ref{thm:rounded}: one
minimizer gives Theorem~\ref{thm:grid} itself and, since this minimum,
$\beta_{n,D,Q}$ in~\eqref{eq:beta}, is at most $1/S^{n,e}_{t,\sigma}$
for every $e$ (end of Section~\ref{sec:grid-minima}),
Theorem~\ref{thm:rounded} for all $e$
at once; and minimality supplies the structure that the estimates use,
namely the cap, the location of the roots and the positive quadrature
(Sections~\ref{sec:grid-minima} and~\ref{sec:paired-forms}); a
prescribed polynomial need not satisfy the cap. For $p=2$
the prescribed polynomials are also checked directly
(Section~\ref{sec:binary-prescribed}). Section~\ref{sec:grid-minima}
shows that the minimum is attained. The next calculation shows why
Theorem~\ref{thm:grid} implies Theorem~\ref{thm:rounded} for every $p\ge2$.

By~\eqref{eq:singleton}, $n\ge2D+1=4m+2r+1$, so $N=n-r\ge4m+r+1$.
If $m>0$, the positive binomial orthogonality measure implies that
the roots of $\Kraw{m}{N-1}$ are simple and lie in $(0,N-1)$.
Consequently $1<\xi_j<N$, so $1\le a_j\le N-1$ and $h(0)>0$.
Each paired factor in~\eqref{eq:rounded-polynomial} is nonnegative
at every integer. The binomial factor is nonnegative on
$0,\ldots,n$, including its terminal zeros. Hence $g(0)>0$,
$g\ge0$ on this grid, and $c_0>0$ by its origin contribution
to the expectation. These facts do not use that the floors are
distinct; distinctness is needed only in Lemma~\ref{lem:radau-rounded}
and is shown in its proof.

The identity
\begin{equation}\label{eq:terminal-binomial}
 \binom nx\binom{n-x}{r}=\binom nr\binom{N}{x}
\end{equation}
gives
\begin{equation}\label{eq:terminal-expectation}
 c_0=Q^{-r}\binom nr\E_{\mu_N}h(X),\qquad
 S_{t,\sigma}^{n,e}
   =Q^r\frac{h(0)}{\E_{\mu_N}h(X)}.
\end{equation}
Thus Theorem~\ref{thm:grid}, applied to $g$, yields
\eqref{eq:main-bound}. At $e=t$ we have $m=0$, $h=1$ and
$r=D$, and $S_{t,\sigma}^{n,t}=Q^D$; this is exactly
the Singleton endpoint~\eqref{eq:singleton}.

\subsection{The coefficient criterion}

Choose a tensor-product orthogonal unitary error basis, containing
the identity in each local factor and normalized by
$\Tr(E^\dagger F)=p^n\delta_{E,F}$. Such a basis exists for every
integer $p$, using the shift and phase operators on $\mathbb C^p$.
Let $\mathrm{wt}(E)$ count the nonidentity tensor factors, and define
the weight enumerators of Shor and Laflamme~\cite{ShorLaflamme}, in
the form used by Rains~\cite{RainsWE},
\begin{equation}\label{eq:enumerators}
 A_i\coloneqq\frac1{K^2}\sum_{\mathrm{wt}(E)=i}|\Tr(PE)|^2,\qquad
 B_i\coloneqq\frac1K\sum_{\mathrm{wt}(E)=i}\Tr(PEPE^\dagger).
\end{equation}
The normalized enumerator identities are
\begin{equation}\label{eq:enumerator-identities}
 A_0=1,\quad A_i\ge0,\quad \sum_iA_i=\frac{p^n}K,\quad
 B_i=\frac K{p^n}\sum_{x=0}^n A_x\Kraw{i}{n}(x).
\end{equation}
Detection implies $B_i=A_i$ for $i<d$. The last identity is the quantum
MacWilliams identity, and the resulting linear program is the quantum
analogue of Delsarte's linear-programming method~\cite{Delsarte};
Ashikhmin and Litsyn~\cite{AshikhminLitsyn} derived polynomial bounds
from it for arbitrary qubit codes.
The enumerator identities and detection equality follow, for every
integer $p$, from the nonbinary enumerator
identities in \cite[Theorems~17--19]{RainsWE}, after dividing the two
enumerators by $K^2$ and $K$, respectively.
They use the normalization of Yu, Lai, and Oh~\cite{YLO}.
The probability measure $\nu(x)\coloneqq KA_x/p^n$ therefore satisfies
\begin{equation}\label{eq:enumerator-moments}
 \E_\nu\Kraw{i}{n}(X)=A_i\qquad(0\le i<d).
\end{equation}

The next lemma isolates the enumerator argument used in the one-error
proof of~\cite[Eqs.~(14)--(16)]{YLO}. It is also a special case,
under the Krawtchouk transform, of the linear-programming bound
of Li and Xing for general quantum codes~\cite[Theorem~1]{LiXing}.
The lemma itself is therefore not new. Most of the paper verifies its
hypothesis~\eqref{eq:coefficient-criterion} for suitable polynomials:
an extremal polynomial for every $p\ge2$, except, for $p\ge3$, at large
distances relative to the length (outside the root region described in
Section~\ref{sec:proof-outline}), where the code dimension is
bounded directly by $p^n$ times the minimal value of
$\E_{\mu_n}f/f(0)$ (Theorem~\ref{thm:strict-complement}); for $p=4$
and $p=5$ the same direct bounds also settle some parameters inside the
root region (Section~\ref{sec:distance-strips}).
For $p=2$ we additionally verify it for the prescribed rounded polynomials. In the lemma, the Krawtchouk coefficients $c_i$ of
$f$ may have either sign.

\begin{lemma}\label{lem:coefficient-criterion}
Suppose $f=\sum_{i=0}^D c_i\Kraw{i}{n}$ is nonnegative on
$0,\ldots,n$, $f(0)>0$, and $D<d$.
If
\begin{equation}\label{eq:coefficient-criterion}
 c_0f(i)\ge f(0)c_i\qquad(1\le i\le D),
\end{equation}
then $Kf(0)\le p^nc_0$.
\end{lemma}
\begin{proof}
Orthogonality and the origin summand give $c_0>0$.
By~\eqref{eq:enumerator-moments} and $D<d$,
$\E_\nu f=\sum_{i=0}^D c_iA_i$, and the positive mass $\nu(0)=K/p^n$
makes $\E_\nu f>0$. Grid nonnegativity lets us drop the weights
$x>D$; then~\eqref{eq:coefficient-criterion} and $A_i\ge0$ give
\[
 \E_\nu f\ge\frac K{p^n}\sum_{i=0}^D f(i)A_i
   \ge\frac{Kf(0)}{p^nc_0}\sum_{i=0}^D c_iA_i
   =\frac{Kf(0)}{p^nc_0}\E_\nu f.
\]
Cancel $\E_\nu f>0$.
\end{proof}

For the connection with~\cite[Theorem~1]{LiXing}, recall that
theorem: if $F=\sum_jF_j\Kraw{j}{n}$ has $F_j\ge0$ for all $j$,
$F_j>0$ for $j$ in a nonempty set $I\subseteq\{0,\ldots,d-1\}$, and
$F(x)\le0$ for $x\in\{0,\ldots,n\}\setminus I$, then
$K\le p^{-n}\max_{i\in I}F(i)/F_i$. Put $c_i\coloneqq0$ for $i>D$ and take
$F_j\coloneqq f(j)$, that is, $F(x)\coloneqq\sum_{j=0}^n f(j)\Kraw{j}{n}(x)$.
Since this transform is an involution up to the factor $Q^n$,
$F(i)=Q^n c_i$,
and~\eqref{eq:coefficient-criterion} gives $F(i)\le0$ outside the
nonempty set $I\coloneqq\{i\in\{0,\ldots,D\}:f(i)>0\}$.
The theorem therefore yields
$K\le p^{-n}\max_{i\in I}F(i)/f(i)\le p^n c_0/f(0)$.

For a pure code, $A_i=0$ for $1\le i<d$.
Then $\E_\nu f=c_0$, and grid nonnegativity alone proves
\eqref{eq:grid-bound}, as in~\cite{YLO}.
This argument uses only~\eqref{eq:enumerator-moments}, the
nonnegativity of $\nu$ and purity. It therefore applies to every pure
solution of the quantum linear program of Calderbank, Rains, Shor, and
Sloane~\cite{CRSS}. Their purity conjecture states that, for qubits,
every solution at the largest real number $K_{\max}>1$ for which this
program is solvable is pure. The enumerators of a code of dimension
$K$ give a solution at their dimension $K$, so $K\le K_{\max}$ and the
conjecture would
imply $Kf(0)\le K_{\max}f(0)\le2^n\E_{\mu_n}f$, that is,
Theorem~\ref{thm:grid} for $p=2$.
For arbitrary subspaces the low-weight terms remain.
Besides~\eqref{eq:enumerator-identities}, our proof uses $B_i\ge A_i$
(Section~\ref{sec:rank-windows}), the Singleton
bound~\eqref{eq:singleton}, Rains's shadow length bound for $p=2$ and
the integrality of $K$; we do not claim Theorem~\ref{thm:grid} for
arbitrary solutions of the linear program.
Lemma~\ref{lem:coefficient-criterion} controls them without
requiring the individual coefficients $c_i$ to be nonpositive.

\subsection{Coefficient identities}

We record two exact formulas, \eqref{eq:normalized-lift}
and~\eqref{eq:normalized-value} of Lemma~\ref{lem:binomial-lift}; for
a polynomial with a terminal binomial factor they express the two
ratios $c_i/c_0$ and $g(i)/g(0)$ compared
in~\eqref{eq:coefficient-criterion}, and they are used in the analytic
estimates and in the finite calculations.
Lemma~\ref{lem:finite-difference}, which expresses the coefficients
through finite differences and through an expansion at the terminal
end $x=N$, is used only in Appendix~\ref{app:binary-small}. In
Lemmas~\ref{lem:binomial-lift} and~\ref{lem:finite-difference}, $N$,
$r$, $h$ and $g$ denote general data, not the rounded construction
above. The binomial lift in the next lemma is the identity
in~\cite[Appendix, Eq.~(29)]{YLO}, written in the present coefficient
normalization.

\begin{lemma}\label{lem:binomial-lift}
Let $N,r$ be nonnegative integers, $n\coloneqq N+r$, $0\le s\le N$,
$h=\sum_{j=0}^{s}b_j\Kraw{j}{N}$, and write
$g(x)\coloneqq\binom{n-x}{r}h(x)=\sum_i c_i\Kraw{i}{n}(x)$.
Then
\begin{equation}\label{eq:binomial-lift}
 c_i=Q^{-r}\sum_{j=0}^r
       \binom{n-i}{r-j}\binom ij b_{i-j},\qquad
 c_0=Q^{-r}\binom nr b_0,
\end{equation}
where $b_j\coloneqq0$ outside $0,\ldots,s$.
If $b_0h(0)\ne0$, the equivalent normalized formulas, after the change
of index $j\mapsto i-j$, are
\begin{align}
 \frac{c_i}{c_0}
   &=\frac1{(n)_i}\sum_{j=0}^{\min(i,s)}
          \binom ij(N)_j(r)_{i-j}\frac{b_j}{b_0},
       \label{eq:normalized-lift}\\
 \frac{g(i)}{g(0)}
   &=\frac{(N)_i}{(n)_i}\frac{h(i)}{h(0)}
       \qquad(0\le i\le N).
       \label{eq:normalized-value}
\end{align}
\end{lemma}
\begin{proof}
By~\eqref{eq:kraw-generating} and~\eqref{eq:orthogonality},
$\sum_iv_i(n)c_iz^i=\E_{\mu_n}[g(X)(1+(Q-1)z)^{n-X}(1-z)^X]$. Using
\eqref{eq:terminal-binomial} in this expectation gives
\[
 \sum_i v_i(n)c_i z^i
   =Q^{-r}\binom nr(1+(Q-1)z)^r
        \sum_j v_j(N)b_jz^j.
\]
Coefficient extraction and cancellation of the binomial factors
give~\eqref{eq:binomial-lift} and~\eqref{eq:normalized-lift}.
The ratio of the two terminal binomials gives
\eqref{eq:normalized-value}.
\end{proof}

\begin{lemma}\label{lem:finite-difference}
For $0\le s\le N$, $h=\sum_{j=0}^s b_j\Kraw{j}{N}$, and
$\Delta h(x)\coloneqq h(x+1)-h(x)$,
\begin{equation}\label{eq:finite-difference}
 b_i=(-1)^iQ^{-i}\E_{\mu_{N-i}}\Delta^i h(X)
 \qquad(0\le i\le N).
\end{equation}
Let $n\coloneqq N+r$. If $h(N-y)=\sum_{j=0}^{s}\widetilde h_j\,(y)_j$ and
$g(x)\coloneqq\binom{n-x}{r}h(x)=\sum_ic_i\Kraw{i}{n}(x)$, then
\begin{equation}\label{eq:factorial-extraction}
 c_i=\frac1{r!}\sum_{j=0}^{s}
       \widetilde h_j(r+j)!Q^{-r-j}\binom{n-i}{r+j-i}.
\end{equation}
\end{lemma}
\begin{proof}
Write $X=\sum_k\varepsilon_k$ with independent Bernoulli variables
$\varepsilon_k$ of mean $(Q-1)/Q$, and put $w(0)\coloneqq Q-1$, $w(1)\coloneqq-1$.
By~\eqref{eq:kraw-generating},
$\Kraw{i}{N}(X)=\sum_{|J|=i}\prod_{k\in J}w(\varepsilon_k)$, and for
such a variable $\varepsilon$ independent of $Y$,
$\E[w(\varepsilon)\varphi(Y+\varepsilon)]=-\frac{Q-1}{Q}\E\Delta\varphi(Y)$.
Applying this identity once for each coordinate of a subset $J$ with
$|J|=i$, and summing over the $\binom Ni$ subsets, yields
\[
 \E_{\mu_N}\bigl[h(X)\Kraw{i}{N}(X)\bigr]
  =\binom Ni\left(-\frac{Q-1}{Q}\right)^i
       \E_{\mu_{N-i}}\Delta^i h(X).
\]
Divide by $v_i(N)$ to obtain~\eqref{eq:finite-difference}.
For the second formula use
\[
 \binom{n-x}{u}
     =Q^{-u}\sum_i\binom{n-i}{u-i}\Kraw{i}{n}(x),
 \qquad
 (n-x)_r(N-x)_j=(n-x)_{r+j}.
\]
Substitution of the falling-factorial expansion of $h(N-y)$
gives~\eqref{eq:factorial-extraction}.
\end{proof}

\subsection{Outline of the proof}\label{sec:proof-outline}

By Lemma~\ref{lem:coefficient-criterion}, a polynomial that satisfies
\eqref{eq:coefficient-criterion} gives the inequality of
Theorem~\ref{thm:grid} for itself. Theorem~\ref{thm:grid} is
equivalent to $K\le p^n\beta_{n,D,Q}$, where the grid minimum
$\beta_{n,D,Q}$ of Section~\ref{sec:grid-minima} is the least value
of $\E_{\mu_n}f$ over the polynomials of Theorem~\ref{thm:grid} with
$f(0)=1$. Section~\ref{sec:grid-minima} shows that it suffices to
verify the criterion for one minimizer, and determines where its
zeros can lie.
For $p\ge3$, Section~\ref{sec:paired-forms} shows that all zeros of
the minimizer exceed $D$ in the root region $(2Q-2)D\le(Q-2)(n+1)$,
where the degree is at most about $(Q-2)/(2Q-2)$ times the length.
There the minimizer is positive at the indices $1,\ldots,D$ that the
criterion compares. Section~\ref{sec:extremal-interfaces} treats the
minimizers with at most one pair of adjacent zeros below the block of
consecutive zeros ending at $n$. The root-location argument in Section~\ref{sec:paired-forms}
uses only $Q>2$, so its conclusion also holds for binary minimizers
(Lemma~\ref{lem:bg-root}); for $p=2$ the root-region condition
reads $n\ge3D-1$, and Rains's length bound for binary
codes~\cite[Theorem~15]{Rains} shows that every binary code satisfies
it; several tools of Sections~\ref{sec:paired-forms}
and~\ref{sec:extremal-interfaces} serve the binary proofs as well.

Outside the root region a minimizer can vanish at an index $i\le D$;
the criterion then requires a nonpositive Krawtchouk coefficient, and
this can fail even for the unique minimizer
(Section~\ref{sec:paired-forms}). There, again for $p\ge3$,
Sections~\ref{sec:rank-windows}--\ref{sec:complement} bound the code
dimension directly from linear relations between the weight
enumerators, without the extremal polynomial; the bound obtained is
at most $p^n\beta_{n,D,Q}$, as Theorem~\ref{thm:grid} requires.
Section~\ref{sec:rank-windows} derives these bounds and an upper bound
on the distance at each length, Section~\ref{sec:distance-strips}
applies them at this distance ceiling and just below it, and
Section~\ref{sec:complement} covers all parameters outside the root
region; the two theorems of Section~\ref{sec:distance-strips} also
serve $p=4$ and $p=5$ inside it. Inside the root region,
Section~\ref{sec:coefficient-methods} estimates the Krawtchouk
coefficients of the minimizer, and Sections~\ref{sec:higher-dimensions}
and~\ref{sec:qutrit} complete the proof for $p\ge4$ and $p=3$,
respectively; for $p=3$ and $p=4$ an exact finite computation covers
the finitely many remaining parameter values. Finite computations also
enter the distance strips, the qutrit complement and the binary case
(Sections~\ref{sec:distance-strips}--\ref{sec:binary}).
Section~\ref{sec:binary} proves both binary statements: the
grid-minimum bound, which gives both theorems for $p=2$, and the
coefficient criterion for each prescribed rounded polynomial, which
gives a second proof of Theorem~\ref{thm:rounded} for $p=2$ and uses
none of the computations specific to the grid-minimum proof. Both
proofs use Rains's length bound; the grid-minimum proof also compares
the minimizer with a rounded competitor of the same degree.
Section~\ref{sec:completion} combines these cases.

\section{Minimizing a nonnegative grid polynomial}\label{sec:grid-minima}

For every $p\ge2$ the proof uses one extremal polynomial to handle every
polynomial of Theorem~\ref{thm:grid} with the same length and degree.
Fix a real number $Q>1$ (for a code, $Q=p^2$) and integers
$0\le D\le n$, and define the grid minimum
\begin{equation}\label{eq:beta}
 \beta_{n,D,Q}
 \coloneqq\min\{\E_{\mu_n}f(X):\deg f\le D,\ f(0)=1,\
                       f(j)\ge0\ (1\le j\le n)\}.
\end{equation}
If an exact code satisfies $K\le p^n\beta_{n,D,Q}$, then
every polynomial in Theorem~\ref{thm:grid} satisfies its conclusion
after division by its value at zero. It therefore suffices to prove
$K\le p^n\beta_{n,D,Q}$.
By linear-programming duality, $\beta_{n,D,Q}$ is also the largest
mass at $0$ of a probability measure on $\{0,\ldots,n\}$ whose moments
through degree $D$ are those of $\mu_n$. For integer $Q\ge2$,
$1/\beta_{n,D,Q}$ is the degree-at-most-$D$ polynomial form of
Delsarte's linear-programming lower bound for orthogonal arrays of
strength $D$ over an alphabet of size $Q$~\cite{Delsarte}. Lemmas~\ref{lem:grid-vertex}
and~\ref{lem:positive-quadrature} below are the standard extreme-point
and complementary-slackness description of this optimum; we prove them
in the form needed here.

The feasible set is a polyhedron in the $D$ free polynomial
coefficients. We will select an extreme point attaining the minimum;
such a point will be called a minimizing vertex. It need not be
unique; the vertex and quadrature statements in this section apply
to any choice, which we fix. Since
Lemma~\ref{lem:coefficient-criterion}, applied to any minimizing
vertex that satisfies~\eqref{eq:coefficient-criterion}, gives
$K\le p^n\beta_{n,D,Q}$, it suffices to verify the criterion for one
minimizing vertex; later proofs may replace the selected vertex by
another one, for instance a certified one
(Propositions~\ref{prop:four-finite} and~\ref{prop:qutrit-finite}).
From here on, $r$, $m$, $N$ and $a_j$
(and $H\coloneqq n-D=N-2m$) describe the selected minimizing vertex unless a
statement fixes them otherwise. They need not equal the parameters of
the rounded construction
in~\eqref{eq:original-parameters}--\eqref{eq:rounded-polynomial}, which
keep that meaning in Section~\ref{sec:binary-prescribed} and
Appendices~\ref{app:binary-endpoints}--\ref{app:binary-small}. Results
stated for general $m$, $r$ and $N$ are applied with the parameters
named where they are used.

\begin{lemma}\label{lem:grid-vertex}
The minimum in~\eqref{eq:beta} is attained and positive, and
minimizing vertices exist. Every minimizing vertex has degree $D$,
has exactly $D$ distinct zeros in $\{1,\ldots,n\}$, and can be written
\begin{equation}\label{eq:vertex-factorization}
 f_*(x)=\frac{\binom{n-x}{r}}{\binom nr}
       \prod_{j=1}^m
       \frac{(x-a_j)(x-a_j-1)}{a_j(a_j+1)},
 \qquad D=r+2m,\quad N\coloneqq n-r.
\end{equation}
Here $r\ge0$ is the length of its maximal block of consecutive zeros
ending at $n$ (the terminal block), and the remaining zeros form $m$
pairwise disjoint pairs of adjacent integers $a_j,a_j+1$ (the residual
pairs) in $\{1,\ldots,N-1\}$.
\end{lemma}
\begin{proof}
The constant polynomial one is feasible. On a fixed objective
sublevel, every value $f(j)$ is bounded, because every binomial
weight $\mu_n(j)$ is positive. Evaluation at $0,\ldots,D$ is an invertible
linear map on polynomials of degree at most $D$, so this sublevel
is compact. The minimum is attained, and it is at least
$\mu_n(0)>0$.

Choose an extreme point of the compact optimal face. It is
extreme in the full feasible set: a feasible decomposition
would have both endpoint objectives equal to the minimum.
If it had fewer than $D$ zeros in $\{1,\ldots,n\}$, there would be a
nonzero polynomial $v$ of degree at most $D$ vanishing at
zero and at all those zeros. For sufficiently small
$\varepsilon>0$, both $f_*+\varepsilon v$ and
$f_*-\varepsilon v$ would be feasible, contradicting extremality.
Thus there are exactly $D$ distinct zeros and the degree is $D$.

Starting at the positive value $f_*(0)$, the sign changes at
every simple zero. Each interval with negative sign must contain
no integer interior point. Therefore consecutive zeros are
paired as adjacent integers; if $D$ is odd, the sign stays negative
after the last zero, which must therefore be $n$. Removing the maximal consecutive terminal block leaves
adjacent residual pairs, with $r\equiv D\pmod2$.
Maximality says that $N$ is not a residual zero.
Normalization at zero gives~\eqref{eq:vertex-factorization}.
When $D=0$, the assertion is $f_*=1$.
\end{proof}

The following quadrature identity, with nonnegative weights,
distinguishes minimizing polynomials from arbitrary paired products;
it supplies the constraint used in Lemma~\ref{lem:optimizer-cap}.

\begin{lemma}\label{lem:positive-quadrature}
Let $J$ consist of zero and the $D$ zeros of a minimizing
vertex. There are weights $\lambda_j\ge0$, with
$\lambda_0=\beta_{n,D,Q}>0$, such that
\begin{equation}\label{eq:vertex-quadrature}
 \E_{\mu_n}q(X)=\sum_{j\in J}\lambda_jq(j)
 \qquad(\deg q\le D).
\end{equation}
Conversely, a feasible degree-$D$ polynomial with $D$ distinct
zeros in $\{1,\ldots,n\}$ is minimizing if the weights
$\lambda_j\coloneqq\E_{\mu_n}\ell_j$ of its cardinal polynomials $\ell_j$,
defined in the proof, are nonnegative. The finite certificates use
this converse to certify minimizers
(Propositions~\ref{prop:four-finite} and~\ref{prop:qutrit-finite};
Appendices~\ref{app:rank-windows}, \ref{app:higher-details}
and~\ref{app:qutrit-finite}).
\end{lemma}
\begin{proof}
For $j\in J$ let
\[
 \ell_j(x)\coloneqq\prod_{\substack{k\in J\\k\ne j}}\frac{x-k}{j-k},
 \qquad \lambda_j\coloneqq\E_{\mu_n}\ell_j(X).
\]
Polynomial interpolation proves~\eqref{eq:vertex-quadrature}.
Since $\ell_0=f_*$, its weight is the minimum.
For $j\ne0$, if $\lambda_j<0$, the polynomial
$f_*+\varepsilon\ell_j$ would be feasible for sufficiently
small positive $\varepsilon$ and have a smaller objective.
This is impossible. Conversely, every normalized feasible
$q$ has expectation at least $\lambda_0$ under a nonnegative
quadrature, while the polynomial vanishing at all nonzero
nodes has expectation exactly $\lambda_0$.
\end{proof}

The next bound on the number $m$ of residual pairs, which we call the
cap, is used throughout.

\begin{lemma}\label{lem:optimizer-cap}
If $m>0$ in~\eqref{eq:vertex-factorization}, then
\begin{equation}\label{eq:optimizer-cap}
 Qm\le N-1.
\end{equation}
\end{lemma}
\begin{proof}
Multiply any polynomial of degree at most $2m$ by
$\binom{n-x}{r}$ in~\eqref{eq:vertex-quadrature} and use
\eqref{eq:terminal-binomial}.
The terminal nodes disappear and all remaining factors are
positive. After normalization, this gives a nonnegative
quadrature for $\mu_N$, exact through degree $2m$, supported
on zero and the residual paired zeros in $\{1,\ldots,N-1\}$.
Consequently, for every real $q$ of degree at most $m-1$,
\begin{equation}\label{eq:quadrature-rayleigh}
 \E_{\mu_N}\bigl[X(N-1-X)q(X)^2\bigr]\ge0.
\end{equation}

Put $\pi(x)\coloneqq\Kraw{m}{N-1}(x-1)$. By binomial size bias,
$x\mu_N(x)=N\frac{Q-1}{Q}\mu_{N-1}(x-1)$,
and~\eqref{eq:orthogonality} with $L=N-1$, $\pi$ is orthogonal to
every polynomial of degree less than $m$ for the measure $x\mu_N(x)$.
Its endpoint values are
\[
 \pi(N)=(-1)^m\binom{N-1}m,\qquad
 \pi(N-1)=(-1)^m\binom{N-1}m
                  \left(1-\frac{Qm}{N-1}\right).
\]
If $Qm>N-1$, these have opposite signs, so $\pi$ has a real zero
$\xi\in(N-1,N)$. For $q(x)\coloneqq\pi(x)/(x-\xi)$, orthogonality gives
$\E_{\mu_N}[X(X-\xi)q(X)^2]=0$. Hence
\[
 \E_{\mu_N}[X(N-1-X)q(X)^2]
   =(N-1-\xi)\E_{\mu_N}[Xq(X)^2]<0,
\]
contradicting~\eqref{eq:quadrature-rayleigh}.
The final expectation is positive because $q$ is nonzero of degree
$m-1<N$, and $x\mu_N(x)>0$ on $\{1,\ldots,N\}$.
\end{proof}

For $m>0$, recall $H=n-D=N-2m$. The cap is equivalent to
$(Q-2)m\le H-1$, and $2m\le D$ holds since $D=r+2m$; together,
\begin{equation}\label{eq:optimizer-cap-H}
 m\le\min\left\{\left\lfloor\frac D2\right\rfloor,
                \left\lfloor\frac{H-1}{Q-2}\right\rfloor\right\}
 \qquad(Q>2).
\end{equation}
The cap is used in Section~\ref{sec:paired-forms} to apply the
paired-moment estimates to the minimizer, and in
Section~\ref{sec:extremal-interfaces} to restrict its residual degree
$2m$, the degree of the product of its residual pairs.
It restricts only the auxiliary minimizer. Every rounded polynomial
$g$ of Section~\ref{sec:framework} with the same $n$ and $D=d-1$
remains feasible after division by $g(0)$, so
\[
 \beta_{n,D,Q}\le\frac{\E_{\mu_n}g(X)}{g(0)}=\frac{c_0}{g(0)}
 =\frac1{S_{t,\sigma}^{n,e}}.
\]
Hence one proof of $K\le p^n\beta_{n,D,Q}$ gives
$KS_{t,\sigma}^{n,e}\le p^n$ for every $e$ at once.
The inequality can be strict for every $e$: at $(p,n,D)=(2,18,6)$,
exact computation gives $\beta_{18,6,4}=233/6336512$, while the least
value of $1/S^{18,e}_{3,0}$, attained at $e=0$, is $2617/69701632$,
about $2\%$ larger (Section~\ref{sec:bg-minimum}).

\section{Adjacent shifts and the roots of a minimizer}\label{sec:paired-forms}

The residual factors of a minimizing vertex occur in adjacent pairs.
Writing their product as $q(x)q(x-1)$ makes their binomial moments
quadratic forms in $q$. We first give the exact matrices of these
forms. A lower bound for the expectation of a paired product
(Lemma~\ref{lem:paired-constant}) then gives a lower bound for the
minimum in~\eqref{eq:beta}, and a lower bound for its first moment,
the expectation of $X$ times the product (Lemma~\ref{lem:paired-mean}),
locates every residual zero of the minimizer
(Proposition~\ref{prop:optimizer-roots}).

\subsection{Two adjacent-shift forms}

Let $Q>2$, and let $Y$ have distribution
$\Bin(L-1,(Q-1)/Q)$. For a real polynomial $q$ of degree at most $k$,
write
\[
 \operatorname{Forw}(q)\coloneqq\E[q(Y+1)q(Y)],\qquad
 \operatorname{Back}(q)\coloneqq\E[q(Y)q(Y-1)],\qquad W(q)\coloneqq\E[q(Y)^2].
\]
We use these forms when $L\ge Qk$; this condition guarantees the
bounds $\theta_j\le1$ and $\rho_j\le1/(Q-1)$ below. For $k=0$ all three
forms
equal the same positive constant when $q\ne0$.
For real symmetric matrices of the same size, $U\succeq V$
(or $V\preceq U$) means $x^\mathsf T(U-V)x\ge0$ for every
real vector $x$; $U>0$ means $x^\mathsf TUx>0$ for every $x\ne0$.

\begin{lemma}\label{lem:paired-grams}
Assume $k\ge1$ and $L\ge Qk$. In the orthonormal basis
$(-1)^j\Kraw{j}{L-1}/\sqrt{v_j(L-1)}$, $0\le j\le k$, the Gram matrix
of the symmetrized form $\operatorname{Forw}$ is
\begin{equation}\label{eq:forward-gram}
 M_+=(1-\omega)I+\omega\operatorname{Cor}_\theta,\qquad \omega\coloneqq\frac{Q}{2(Q-1)},\qquad
 \theta_j\coloneqq\sqrt{\frac{(Q-1)j}{L-j}}.
\end{equation}
Here $(\operatorname{Cor}_\theta)_{ii}\coloneqq1$ and, for $i<j$,
$(\operatorname{Cor}_\theta)_{ij}\coloneqq\prod_{\ell=i+1}^j\theta_\ell$.
In the basis $\Kraw{j}{L-1}/\sqrt{v_j(L-1)}$, without the alternating
signs, the Gram matrix of the symmetrized form $\operatorname{Back}$ is
\begin{equation}\label{eq:backward-gram}
 M_-=\frac Q2 \operatorname{Cor}_\rho-\frac{Q-2}{2}I,\qquad
 \rho_j\coloneqq\sqrt{\frac{j}{(Q-1)(L-j)}}.
\end{equation}
In particular,
\begin{equation}\label{eq:coarse-paired-forms}
 \operatorname{Forw}(q)\ge\frac{Q-2}{2(Q-1)}W(q)>0,\qquad \operatorname{Back}(q)\ge0
 \quad(q\ne0).
\end{equation}
\end{lemma}
\begin{proof}
Using the formal power-series form of~\eqref{eq:kraw-generating} for
real arguments, which is needed because $Y+1$ can equal $L$, we obtain
\[
 \E\left[
  \sum_{i\ge0}\Kraw{i}{L-1}(Y+1)z^i
  \sum_{j\ge0}\Kraw{j}{L-1}(Y)w^j\right]
  =\frac{1-z}{1+(Q-1)z}\bigl(1+(Q-1)zw\bigr)^{L-1}.
\]
For $\ell\ge1$, the coefficient of $z^\ell$ in the prefactor
is $\pi_\ell=(-1)^\ell Q(Q-1)^{\ell-1}$; put $\pi_0\coloneqq1$.
Hence $\E[\Kraw{i}{L-1}(Y+1)\Kraw{j}{L-1}(Y)]=\pi_{i-j}v_j(L-1)$ for
$i\ge j$, and this expectation vanishes for $i<j$.
Symmetrizing and dividing by the orthogonal norms proves
\eqref{eq:forward-gram} in the basis
$(-1)^j\Kraw{j}{L-1}/\sqrt{v_j(L-1)}$, since
$\frac Q2(Q-1)^{j-i-1}\sqrt{v_i/v_j}=\omega\prod_{\ell=i+1}^j\theta_\ell$
for $i<j$.
The backward shift identity is
\begin{equation}\label{eq:backward-shift}
 \Kraw{i}{L-1}(x-1)=\Kraw{i}{L-1}(x)
                 +Q\sum_{j<i}\Kraw{j}{L-1}(x).
\end{equation}
It proves~\eqref{eq:backward-gram} in the basis without the
alternating signs.

Every correlation matrix $\operatorname{Cor}_\eta$ with $0\le\eta_j\le1$ is
positive semidefinite: take independent centered variables
$\varepsilon_j$ of variance one and set
$Z_0\coloneqq\varepsilon_0$ and
$Z_j\coloneqq\eta_jZ_{j-1}+\sqrt{1-\eta_j^2}\varepsilon_j$.
Their covariance is $\operatorname{Cor}_\eta$.
Since $\theta_j\le1$, this proves the lower bound for $\operatorname{Forw}$.

For $0\le\eta_j<1$, direct multiplication gives
\begin{equation}\label{eq:correlation-inverse}
 y^\mathsf T\operatorname{Cor}_\eta^{-1}y
    =y_0^2+\sum_{j=1}^k
              \frac{(y_j-\eta_jy_{j-1})^2}{1-\eta_j^2}.
\end{equation}
If all $\eta_j\le\eta<1$, each absolute row sum of the
inverse is at most $(1+\eta)/(1-\eta)$.
For an interior row its sum is
$1/(1-\eta_j)+\eta_{j+1}/(1-\eta_{j+1})$;
the endpoint rows are no larger.
Hence
\begin{equation}\label{eq:correlation-lower}
 \operatorname{Cor}_\eta\succeq\frac{1-\eta}{1+\eta}I.
\end{equation}
In~\eqref{eq:backward-gram} every $\rho_j$ is at most
$\rho_k=\sqrt{k/((Q-1)(L-k))}\le1/(Q-1)<1$.
Therefore
\[
 M_-\succeq\frac{1-(Q-1)\rho_k}{1+\rho_k}I\succeq0.
\]
This includes $L=Qk$.
\end{proof}

The same forward matrix also gives an exact recursion, used with $Q=4$
in the binary proof of Section~\ref{sec:binary}
(Lemma~\ref{lem:paired-grams} needs only $Q>2$). This recursion bounds
$\operatorname{Forw}(q)$ for monic $q$ through the last coefficient
of $q$ in the orthonormal basis, which is fixed for monic $q$, rather
than through the smallest eigenvalue of the whole form.
For each leading block of $M_+$ on $0,\ldots,j$, written as
$\bigl(\begin{smallmatrix}M_0&v\\v^\mathsf T&1\end{smallmatrix}\bigr)$,
its Schur complement is $1-v^\mathsf TM_0^{-1}v$; these scalars
for $j=1,\ldots,k$, preceded by the initial entry $1$, are its
successive Schur complements.

\begin{lemma}\label{lem:forward-schur}
Under the hypotheses of Lemma~\ref{lem:paired-grams}, let $\delta_j$ be
the successive Schur complements of $M_+$, with $\delta_0=1$. Then
every $\delta_j$ is positive and
\begin{equation}\label{eq:forward-schur}
 \delta_j=1-\theta_j^2
      \left(2\omega-1+\frac{(1-\omega)^2}{\delta_{j-1}}\right).
\end{equation}
For every monic polynomial $q$ of degree $k$,
\begin{equation}\label{eq:monic-forward}
 \operatorname{Forw}(q)\ge\delta_k\,k!(L-1)_k
                   \left(\frac{Q-1}{Q^2}\right)^k.
\end{equation}
\end{lemma}
\begin{proof}
Since $M_+\succeq(1-\omega)I$ with $1-\omega=(Q-2)/(2(Q-1))>0$, every
leading block of $M_+$ is positive definite, so every $\delta_j$ is
positive. Let $M_0$ and $\operatorname{Cor}_0$ be the preceding principal
blocks and
let $u$ be their last coordinate vector. The new off-diagonal
block is $\omega\theta_j\operatorname{Cor}_0u$, so its Schur complement is
$1-\omega^2\theta_j^2u^\mathsf T\operatorname{Cor}_0M_0^{-1}\operatorname{Cor}_0u$.
Substitute $\omega\operatorname{Cor}_0=M_0-(1-\omega)I$, use
$u^\mathsf TM_0u=1$ and
$u^\mathsf TM_0^{-1}u=1/\delta_{j-1}$, and expand.
This gives~\eqref{eq:forward-schur}.
Completing the square in the final coordinate bounds the
quadratic form by $\delta_k$ times the square of that
coordinate. For a monic $q$, this square is the squared norm
of the monic degree-$k$ binomial orthogonal polynomial,
namely the factor following $\delta_k$ in
\eqref{eq:monic-forward}.
\end{proof}

\subsection{A lower bound for the paired moment}

Set
\[
 V_k(L)\coloneqq\sum_{j=0}^k\binom Lj(Q-1)^j,\qquad
 \operatorname{pref}_Q\coloneqq\frac{(Q-2)^2}{2Q(Q-1)};
\]
$V_k(L)=\sum_{j\le k}v_j(L)$ is the volume of the Hamming ball of
radius $k$ in length $L$ over an alphabet of size $Q$. The next
estimate holds with a constant depending only on $Q$, for every
degree $k$ allowed by $L-1\ge Qk$; for a minimizer this condition is
the cap~\eqref{eq:optimizer-cap}.

\begin{lemma}\label{lem:paired-constant}
Let $Q\ge9$, $k\ge1$, and $L-1\ge Qk$.
For every real polynomial $q$ of degree at most $k$ with
$q(0)q(-1)>0$,
\begin{equation}\label{eq:paired-constant}
 \frac{\E_{\mu_L}[q(X)q(X-1)]}{q(0)q(-1)}
       \ge\frac{\operatorname{pref}_Q}{V_k(L-1)}
       \ge\frac{49}{144V_k(L)}.
\end{equation}
\end{lemma}
\begin{proof}
Write $X=Y+Z$, where $Y\sim\mu_{L-1}$ and $Z$ is an
independent Bernoulli variable of mean $(Q-1)/Q$.
Lemma~\ref{lem:paired-grams} gives
\begin{equation}\label{eq:paired-conditioning}
 \E_{\mu_L}[q(X)q(X-1)]
     =\frac{Q-1}{Q}\operatorname{Forw}(q)+\frac1Q \operatorname{Back}(q)
     \ge\frac{Q-2}{2Q}W(q).
\end{equation}
At length $L-1$, the orthogonal norms satisfy
$v_j/v_{j-1}=(Q-1)(L-j)/j\ge(Q-1)^2$.
By~\eqref{eq:backward-shift},
\[
 \Kraw{j}{L-1}(-1)
    =v_j+Q\sum_{\ell<j}v_\ell
    \le\frac{Q-1}{Q-2}v_j.
\]
Cauchy--Schwarz in that basis yields
\[
 |q(0)|\le\sqrt{V_k(L-1)W(q)},\qquad
 |q(-1)|\le\frac{Q-1}{Q-2}\sqrt{V_k(L-1)W(q)}.
\]
Combining these bounds with~\eqref{eq:paired-conditioning}
proves the first inequality.
The function $\operatorname{pref}_Q$ is increasing for $Q>2$ and
$\operatorname{pref}_9=49/144$, while $V_k(L-1)\le V_k(L)$.
\end{proof}

The first inequality in~\eqref{eq:paired-constant} holds for every
$Q>2$; only the second uses $Q\ge9$. For $k=0$ the normalized paired
moment is exactly one, so the bound $49/144$ still holds.
This lemma applies to the paired factors of a minimizing
vertex because~\eqref{eq:optimizer-cap} supplies its degree
hypothesis. It gives a lower bound for $p^n\beta_{n,D,Q}$
(Proposition~\ref{prop:grid-budget}); outside the root
region~\eqref{eq:root-region}, this lower bound is compared with the upper bounds on $K$ of
Sections~\ref{sec:rank-windows}--\ref{sec:complement}.

\subsection{A degree-sensitive first moment}

The next inequality does not require a sign assumption on
$q(0)q(-1)$. This is essential when one paired factor is
removed from a minimizing polynomial.

\begin{lemma}\label{lem:paired-mean}
Let $Q\ge9$, $L\ge1$, and $k\ge0$ with $L\ge Qk$.
For every nonzero real polynomial $q$ of degree at most $k$, the
paired product $h(x)\coloneqq q(x)q(x-1)$ (a general product, not the rounded
polynomial) satisfies
$\E_{\mu_L}h(X)>0$ and
\begin{equation}\label{eq:paired-mean}
 \frac{\E_{\mu_L}[Xh(X)]}{\E_{\mu_L}h(X)}
       \ge\frac{\bigl(\sqrt{(Q-1)(L-k)}-\sqrt{k}\bigr)^2}{Q}.
\end{equation}
\end{lemma}
\begin{proof}
The proof uses only $Q>2$; the binary proof uses the case $Q=4$
(Lemma~\ref{lem:bg-root}). The assertion is equality for $k=0$.
Otherwise put
$\rho\coloneqq\sqrt{k/((Q-1)(L-k))}$ and $\alpha\coloneqq(Q-1)\rho\le1$.
In the backward correlation matrix, the two off-diagonal
tails in each row are bounded by geometric series. Thus
$\operatorname{Cor}_\rho\preceq(1+\rho)/(1-\rho)I$, and
\eqref{eq:backward-gram} gives
\[
 0\le \operatorname{Back}(q)\le\frac{1+(Q-1)\rho}{1-\rho}W(q).
\]
If $\alpha<1$, apply~\eqref{eq:correlation-lower} to the
forward correlation matrix, whose parameters satisfy
$\theta_j\le\theta_k=\alpha$. At $\alpha=1$ its
positive semidefiniteness gives the same weak bound.
In either case,
\[
 \operatorname{Forw}(q)\ge\frac{1-\rho}{1+(Q-1)\rho}W(q)>0.
\]
Let $a_\rho\coloneqq(1-\rho)/(1+(Q-1)\rho)$.
Conditioning and binomial size bias give
\[
 \E_{\mu_L}h=\frac{Q-1}{Q}\operatorname{Forw}(q)+\frac1Q \operatorname{Back}(q),\qquad
 \E_{\mu_L}[Xh]=\frac{L(Q-1)}Q \operatorname{Forw}(q).
\]
Therefore the ratio is at least
$L(Q-1)/((Q-1)+a_\rho^{-2})$.
The identities
\[
 (Q-1)(1-\rho)^2+(1+(Q-1)\rho)^2
     =Q(1+(Q-1)\rho^2),\qquad
 1+(Q-1)\rho^2=\frac L{L-k}
\]
reduce this expression to~\eqref{eq:paired-mean}.
\end{proof}

\begin{proposition}\label{prop:optimizer-roots}
Let $Q\ge9$ and let a minimizing vertex have the
maximal-terminal representation~\eqref{eq:vertex-factorization}.
If $m>0$, the lower zero $a_j$ of every residual pair satisfies
\begin{equation}\label{eq:optimizer-root-sharp}
 a_j\ge
 \frac{\bigl(\sqrt{(Q-1)(N-m)}-\sqrt{m-1}\bigr)^2}{Q}
       >\frac{(Q-2)(n-D+1)}Q.
\end{equation}
In particular, if $n\ge2D+1$ (for a code of distance at least $D+1$
this follows from~\eqref{eq:singleton}) and
\begin{equation}\label{eq:root-region}
 (2Q-2)D\le(Q-2)(n+1),
\end{equation}
then every positive root of the minimizing vertex exceeds $D$.
\end{proposition}

We call~\eqref{eq:root-region} the root-region condition and the
parameters satisfying it the root region. The proof of
Proposition~\ref{prop:optimizer-roots} uses only $Q>2$, through
Lemma~\ref{lem:paired-mean} and the cap; at $Q=4$, where
\eqref{eq:root-region} reads $n\ge3D-1$, it gives
Lemma~\ref{lem:bg-root}. There the minimizer is
positive at $1,\ldots,D$, so the coefficient
criterion~\eqref{eq:coefficient-criterion} compares $c_i/c_0$ with the
positive values $f_*(i)/f_*(0)$; this is the situation treated in
Sections~\ref{sec:coefficient-methods}--\ref{sec:qutrit}. At an index
$i\le D$ where $f_*$ vanished, the criterion would instead demand the
sign condition $c_i\le0$, which the estimates of those sections, all
of which assume that every root exceeds $D$, do not address, and which
can fail even for the unique minimizer. For example, at
$(p,n,D)=(3,33,16)$, outside the root region, the unique minimizing
vertex is the normalization of
$\binom{33-x}{12}(x-16)(x-17)(x-19)(x-20)$ (the prescribed rounded polynomial with $t=8$,
$\sigma=0$, $e=6$); it vanishes at $16=D$, while for the displayed
unnormalized polynomial the coefficient is
$c_{16}=16!/(12!\,9^{16})>0$ by comparison of leading coefficients,
and normalization preserves its sign.
Minimality and uniqueness follow from the converse in
Lemma~\ref{lem:positive-quadrature}, since an exact computation shows
that all $17$ weights there are positive; no code with $K>1$ exists at
these parameters (case $L=1$ of Theorem~\ref{thm:critical-grid}). Outside
the root region the dimension is therefore bounded directly
(Sections~\ref{sec:rank-windows}--\ref{sec:complement}).

\begin{proof}[Proof of Proposition~\ref{prop:optimizer-roots}]
Fix a residual pair with lower zero $a$ and let $w$ be the product
of the other residual pairs. Replace $(x-a)(x-a-1)$ by
$(x-a-1)(x-a-2)$ and normalize again at zero.
This is a feasible competitor: adjacent pairs are nonnegative
on integers, the new pair may coincide with other zeros, and
$a\le N-2$.
The new polynomial need not retain the same maximal terminal
block.

Let $E_j\coloneqq\E_{\mu_N}[X^jw(X)]$. By~\eqref{eq:terminal-expectation}
with this $r$ and $N$, the objective $\E_{\mu_n}f$ of a polynomial
$\binom{n-x}{r}h(x)$ normalized at zero is $Q^{-r}\E_{\mu_N}h/h(0)$, so
it suffices to compare these residual objectives. As a function of
the lower zero $a$ of the moved pair, the residual objective is
\[
 F(a)\coloneqq\frac{E_2-(2a+1)E_1+a(a+1)E_0}{a(a+1)w(0)}.
\]
Its exact difference is
\[
 F(a+1)-F(a)
     =\frac{2((a+1)E_1-E_2)}{a(a+1)(a+2)w(0)}.
\]
Here $w(0)>0$ and $E_1>0$; for the latter, the contribution
at $X=N$ is positive, because all residual zeros lie below $N$. Minimality therefore implies
$E_2/E_1\le a+1$.
Write $w(x)=q_0(x)q_0(x-1)$, where $\deg q_0=m-1$.
Size bias gives, for $Y\sim\mu_{N-1}$,
\[
 \frac{E_2}{E_1}
      =1+\frac{\E[Yw(Y+1)]}{\E[w(Y+1)]}.
\]
Apply Lemma~\ref{lem:paired-mean} to $q_0(y+1)$,
with length $N-1$ and degree $m-1$.
The cap~\eqref{eq:optimizer-cap} ensures its hypotheses.
This proves the first inequality in
\eqref{eq:optimizer-root-sharp}.

Put $H\coloneqq n-D$ and $k\coloneqq m-1$. Then $N-m=H+1+k$.
Since $H+1>0$, expanding the square reduces
\[
 \bigl(\sqrt{(Q-1)(H+1+k)}-\sqrt{k}\bigr)^2
       \ge(Q-2)(H+1)
\]
to $H+1+Qk\ge2\sqrt{(Q-1)k(H+1+k)}$, which holds because
$(H+1+Qk)^2-4(Q-1)k(H+1+k)=(H+1-(Q-2)k)^2$.
Equality requires $(Q-2)k=H+1$.
But~\eqref{eq:optimizer-cap-H} implies
$(Q-2)k\le H-Q+1<H+1$, so the inequality is strict.
Finally, \eqref{eq:root-region} is equivalent to
$QD\le(Q-2)(H+1)$.
All residual roots consequently exceed $D$.
The terminal roots also exceed $D$, since
$N+1=H+2m+1>D$ when $n\ge2D+1$.
\end{proof}

\section{Comparison polynomials and small residual degree}
\label{sec:extremal-interfaces}

Two comparisons connect the grid minimum with explicit polynomials.
The first is a Gauss--Radau formula at the zeros $\xi_j$ of the Lloyd
polynomial of Section~\ref{sec:framework}. It bounds the objective of
the rounded polynomial by $1/(Q^rV_m(N))$; by minimality, a minimizer
with the same terminal degree $r$ and residual length $N$ then has
residual objective $\E_{\mu_N}h/h(0)\le1/V_m(N)$;
Lemma~\ref{lem:optimizer-product} turns this into a lower bound for
the product of the residual roots of a minimizer, and
Section~\ref{sec:bg-minimum} uses it as the competitor inequality for
$p=2$. The second bounds
that objective from below using the paired moment estimate. We then
dispose of minimizers with zero or one residual pair, and reduce the
grid minimum at fixed $H=n-D$ to one cap minimum
(Lemma~\ref{lem:cap-minimum-identity}), which the finite certificates
for $p=3$ and $p=4$ use.

\subsection{The rounded Gauss--Radau comparison}

We recall that consecutive zeros of a Krawtchouk polynomial are more
than one apart; this makes the rounded floors distinct in
Lemma~\ref{lem:radau-rounded}. For $m<L$,
see~\cite[Section~3.11.4 and Theorem~3.27]{KozhanVaktnas}; the
endpoint Christoffel argument below also covers $m=L$.

\begin{lemma}\label{lem:strict-mesh}
Let $0<s<1$, let $L\ge1$ be an integer, and let $\pi_m^{(L)}$
be the monic orthogonal polynomial of degree $1\le m\le L$
for $\Bin(L,s)$. Its consecutive zeros have distance greater
than one.
\end{lemma}
\begin{proof}
The positive measure on $L+1$ points gives simple roots in
$(0,L)$ and strict interlacing of consecutive degrees.
For completeness, the monic three-term recurrence has positive
subdiagonal coefficient $\|\pi_j^{(L)}\|^2/\|\pi_{j-1}^{(L)}\|^2$.
Evaluation at the preceding roots proves interlacing inductively.
Root location follows also by multiplying $\pi_m^{(L)}$ by its
sign-change factors in $(0,L)$: fewer than $m$ such factors
would contradict orthogonality and positivity.

For $b=0,L$, put
\[
 \operatorname{Chr}_b(x)\coloneqq\frac{\pi_m^{(L)}(x)-\alpha_b\pi_{m-1}^{(L)}(x)}{x-b},
 \qquad \alpha_b\coloneqq\frac{\pi_m^{(L)}(b)}{\pi_{m-1}^{(L)}(b)}.
\]
The endpoint values are nonzero. This is a monic polynomial
of degree $m-1$, orthogonal for the positive measure
$x\Bin(L,s)$ if $b=0$, and for $(L-x)\Bin(L,s)$ if $b=L$.
At each zero $x$ of $\pi_m^{(L)}$,
$\operatorname{Chr}_b(x)=-\alpha_b\pi_{m-1}^{(L)}(x)/(x-b)$, and
$\alpha_b/(x-b)$ has the same nonzero sign at all these zeros. Since
$\pi_{m-1}^{(L)}$ alternates in sign at consecutive zeros of
$\pi_m^{(L)}$, so does $\operatorname{Chr}_b$. Thus $\operatorname{Chr}_b$ has one zero in each
interval between consecutive zeros of $\pi_m^{(L)}$.

If $z_1<\cdots<z_{m-1}$ are the zeros of $\pi_{m-1}^{(L-1)}$,
the binomial size-bias identities show that the zeros of
$\operatorname{Chr}_0$ and $\operatorname{Chr}_L$ are respectively $z_j+1$ and $z_j$.
Writing the zeros of $\pi_m^{(L)}$ as $x_j$, both interlacings give
$x_j<z_j<z_j+1<x_{j+1}$. This also covers $m=L$,
since the transformed measure supports degree $L-1$.
\end{proof}

We use the rounded-polynomial comparison from
\cite[Appendix, Eqs.~(22)--(30)]{YLO} in a Gauss--Radau form valid
for real $Q>1$: a quadrature with nodes at zero and at the Lloyd zeros,
positive weights, and origin weight $1/V_m(N)$. In the lemma, $h_L$ is
the rounded polynomial $h$ of~\eqref{eq:rounded-polynomial} for
arbitrary $m$ and $N$; the subscript refers to Lloyd. Its floors $a_j$
in the lemma are defined from the Lloyd zeros.

\begin{lemma}\label{lem:radau-rounded}
Let $Q>1$, $m\ge1$, and $2m\le N$.
Let $\xi_j$ be the zeros of $\Kraw{m}{N-1}(x-1)$,
$a_j\coloneqq\lfloor\xi_j\rfloor$, and
$h_L(x)\coloneqq\prod_j(x-a_j)(x-a_j-1)$.
Then
\begin{equation}\label{eq:radau-rounded}
 0<\frac{\E_{\mu_N}h_L(X)}{h_L(0)}\le\frac1{V_m(N)}.
\end{equation}
\end{lemma}
\begin{proof}
Let $\pi$ be the monic multiple of $\Kraw{m}{N-1}(x-1)$.
The identity
\[
 x\mu_N(x)=\frac{N(Q-1)}Q\mu_{N-1}(x-1)
\]
shows that $\pi$ is orthogonal for the measure $x\mu_N$.
There are unique weights at $0,\xi_1,\ldots,\xi_m$ such that
\begin{equation}\label{eq:radau-formula}
 \E_{\mu_N}F(X)=\lambda_0F(0)
        +\sum_{j=1}^m\lambda_jF(\xi_j),\qquad \deg F\le2m.
\end{equation}
Indeed, division gives $F=x\pi u+v$ with $\deg u\le m-1$
and $\deg v\le m$. The first term has zero expectation,
and interpolation of $v$ gives the formula. Apply it to
the square of each cardinal polynomial $\ell_j$ to obtain
$\lambda_j=\E\ell_j^2>0$.

The generating function gives the Lloyd identity
\[
 \Kraw{m}{N-1}(x-1)=\sum_{j=0}^m\Kraw{j}{N}(x).
\]
Together with orthogonality this is the reproducing polynomial
at zero through degree $m$. Consequently
$\lambda_0=\E[\pi^2]/\pi(0)^2=1/V_m(N)$.
By Lemma~\ref{lem:strict-mesh}, the floors $a_j$ are distinct.
At $\xi_j$, its own paired factor is nonpositive and every
other paired factor is nonnegative. Thus $h_L(\xi_j)\le0$.
Equation~\eqref{eq:radau-formula} proves the upper bound.
Grid nonnegativity and the positive origin summand prove
the strict lower bound.
\end{proof}

The lemma is used through the following comparison. For every split
$D=r+2m$ with $m\ge1$ and $2m\le N=n-r$, the polynomial
$\binom{n-x}{r}h_L(x)$, divided by its value at zero, is feasible
in~\eqref{eq:beta}. By~\eqref{eq:terminal-expectation}
and~\eqref{eq:radau-rounded},
\[
 \beta_{n,D,Q}\le\frac{\E_{\mu_N}h_L(X)}{Q^rh_L(0)}\le\frac1{Q^rV_m(N)}.
\]
For the rounded polynomial of Section~\ref{sec:framework}, the same
identity gives $S_{t,\sigma}^{n,e}\ge Q^rV_m(N)$, with equality $Q^D$ at
$e=t$. The right side is the denominator of the quantum
Hamming--Singleton bound of~\cite[Theorem~1]{YLO}, so the rounded bound
is at least as strong as this unrounded one. It is strictly stronger
exactly when some $\xi_j$ is not an integer: then $h_L(\xi_j)<0$, since
every other paired factor is positive at $\xi_j$, and $\lambda_j>0$
in~\eqref{eq:radau-formula}.

\subsection{A lower bound for the grid minimum}

\begin{proposition}\label{prop:grid-budget}
Let $p\ge3$, $Q=p^2$, $n\ge2D+1$, $H\coloneqq n-D$, and
\begin{equation}\label{eq:cap-index}
 a\coloneqq\min\left(\left\lfloor\frac D2\right\rfloor,
          \left\lfloor\frac{H-1}{Q-2}\right\rfloor\right),
 \qquad U_j\coloneqq\frac{p^{n-2D+4j}}{V_j(H+2j)}.
\end{equation}
Then
\begin{equation}\label{eq:grid-budget}
 p^n\beta_{n,D,Q}\ge \operatorname{pref}_QU_a,\qquad
 \operatorname{pref}_Q=\frac{(Q-2)^2}{2Q(Q-1)}\ge\frac{49}{144}.
\end{equation}
If $a=0$, the exact value is $p^n\beta_{n,D,Q}=U_0$.
\end{proposition}
\begin{proof}
A minimizing vertex has a residual degree $2j$ with $0\le j\le a$
by~\eqref{eq:optimizer-cap-H}. Its terminal degree is $D-2j$
and residual length is $H+2j$.
For $j>0$, the cap gives $H+2j-1\ge Qj$.
Lemma~\ref{lem:paired-constant} and the terminal expectation
identity therefore give $p^n\beta_{n,D,Q}\ge \operatorname{pref}_QU_j$.
At $j=0$ the value is $U_0$.

It remains to show that $U_j$ decreases up to $a$.
Writing $b\coloneqq Q-1$ and $v_j(L)=\binom Lj b^j$, the exact
binomial recurrence is
\begin{equation}\label{eq:volume-recurrence}
 V_{j+1}(L+2)=Q^2V_j(L)
       +\frac{b\,v_j(L)}{j+1}\bigl(L+1-Q(j+1)\bigr).
\end{equation}
For $L=H+2j$ and $j+1\le a$, the last bracket is
$H-(Q-2)(j+1)-1\ge0$. Hence $U_{j+1}\le U_j$,
which proves the assertion, including the cap endpoint.
\end{proof}

Here $U_j=p^nQ^{-(D-2j)}/V_j(H+2j)$. For $j\ge1$ this is $p^n$ times
the bound given after Lemma~\ref{lem:radau-rounded} by the rounded
polynomial with $r=D-2j$ and $m=j$, at residual length $H+2j$. Hence
$p^n\beta_{n,D,Q}\le U_a$ also when $a\ge1$,
and~\eqref{eq:grid-budget} determines $p^n\beta_{n,D,Q}$ up to the
factor $\operatorname{pref}_Q$.

We also record the elementary estimates at the cap endpoint $j=a$:
\begin{equation}\label{eq:volume-endpoint}
 v_a(H+2a)\le V_a(H+2a)
 \le\frac{v_a(H+2a)}{1-(Q-1)^{-2}}.
\end{equation}
They follow by comparing consecutive terms: through degree $a$,
$v_{j-1}/v_j\le(Q-1)^{-2}$.

\subsection{The exact one-pair inequality}

The following calculation is used for both local-dimension branches:
in Proposition~\ref{prop:one-pair-minimum} for $p\ge3$, and for $p=2$
in the proofs of Theorem~\ref{thm:binary-rounded} and
Proposition~\ref{prop:bg-grid-minimum}. For $r\le1$ it is the
impure-code computation of Yu, Lai, and Oh at distances three and
four~\cite[Eqs.~(14)--(16)]{YLO}. In the theorem,
the convention $1\le\nu\le Q$ makes $a=\lfloor1+(Q-1)(N-1)/Q\rfloor$,
the floor of the
zero of $\Kraw{1}{N-1}(x-1)$. Thus $g$ is the rounded
polynomial~\eqref{eq:rounded-polynomial} with $m=1$, and
$Z=Q^2\E_{\mu_N}[(X-a)(X-a-1)]$. The proof uses this value of $a$, not
only $a>D$. The single exception is binary, with $n=7$ and $D=3$,
that is, $p=2$, $n=7$, $d=4$, the case also set aside in~\cite{YLO}.
Both binary proofs exclude it by a length bound.

\begin{theorem}\label{thm:one-root}
Let $Q=p^2$ with integer $p\ge2$, $r\ge0$, $D\coloneqq r+2$,
$n\ge2D+1$, and $N\coloneqq n-r$. Write
\[
 N-1=Qk+\nu,\quad 1\le\nu\le Q,\quad
 a\coloneqq(Q-1)k+\nu,\quad Z\coloneqq Q(Q-1)k+\nu(\nu+1).
\]
Suppose $a>D$. For
$g(x)\coloneqq\binom{n-x}{r}(x-a)(x-a-1)$, the coefficient
criterion~\eqref{eq:coefficient-criterion} holds unless
$(Q,N,r)=(4,6,1)$.
\end{theorem}
\begin{proof}
Expanding the quadratic in the binomial moments, or using
Lemma~\ref{lem:binomial-lift} with $h=\sum_{j\le2}b_j\Kraw{j}{N}$,
$b_2=2/Q^2$, $b_1=2\nu/Q^2$ and $b_0=Z/Q^2$, gives
\[
 c_i=\frac{\binom nr}{Q^{r+2}\binom ni}
   \left\{Z\binom ri+2N\nu\binom r{i-1}
                         +N(N-1)\binom r{i-2}\right\}.
\]
Thus the criterion is exactly
\begin{align}\label{eq:one-pair-deficit}
 Z\binom Ni(a-i)(a+1-i)\ge a(a+1)
 \left\{Z\binom ri+2N\nu\binom r{i-1}
                         +N(N-1)\binom r{i-2}\right\}.
\end{align}
If $k=0$, then $a=N-1$ and
$g=(r+1)(r+2)\binom{n-x}{r+2}$.
Its coefficient and value ratios are respectively
$(D)_i/(n)_i$ and $(n-D)_i/(n)_i$.
The inequality follows from $n-D\ge D$.

Let $k\ge1$. With $N$, and hence $a$, $Z$, $k$ and $\nu$, held fixed,
the left side of~\eqref{eq:one-pair-deficit} does not depend on $r$,
and its right side is nondecreasing in $r$. Since $a>D=r+2$ gives
$r\le a-3$, it suffices to take $r\coloneqq a-3$ and $1\le i\le a-1$.
Put $\operatorname{lin}\coloneqq2N\nu/Z$, $\operatorname{quad}\coloneqq N(N-1)/Z$ and
$\gamma_i\coloneqq(N)_i/(a+1)_i$.
Divide~\eqref{eq:one-pair-deficit} at $r=a-3$ by
$Za(a+1)\binom{a-1}{i}$ and use
$\binom Ni(a-i)(a+1-i)=\gamma_ia(a+1)\binom{a-1}{i}$; the required
inequality becomes
\begin{equation}\label{eq:one-pair-gamma}
 \gamma_i\ge1+\frac{(\operatorname{lin}-2)i}{a-1}
            +\frac{(\operatorname{quad}-\operatorname{lin}+1)i(i-1)}{(a-1)(a-2)},
\end{equation}
where
$\operatorname{lin}-2=2Qk(\nu-Q+1)/Z$ and $\operatorname{quad}-\operatorname{lin}+1=Q^2k(k+1)/Z$.
Since $N-1-a=k$, the forward differences in $j$ are
\[
 \Delta\gamma_j=\frac{k\gamma_j}{a+1-j},\qquad
 \Delta^2\gamma_j=\frac{k(k+1)\gamma_j}{(a+1-j)(a-j)}.
\]
Thus $\gamma_j$ and $\Delta^2\gamma_j$ increase with $j$. Since
$\gamma_0=1$, summing the second differences from $j=0$ gives
\[
 \gamma_i\ge1+\frac{ki}{a+1}
                   +\frac{k(k+1)i(i-1)}{2a(a+1)}.
\]
The difference from the right side of~\eqref{eq:one-pair-gamma}
is $i(\alpha+(i-1)\beta)$, with
\[
 \alpha\coloneqq\frac{k}{a+1}-\frac{2Qk(\nu-Q+1)}{Z(a-1)},\qquad
 \beta\coloneqq\frac{k(k+1)}{2a(a+1)}-
                    \frac{Q^2k(k+1)}{Z(a-1)(a-2)}.
\]
Always $\alpha\ge0$. This is immediate if $\nu<Q$;
if $\nu=Q$, use $Z=Q(a+1)$ to obtain
$\alpha=k(a-3)/((a+1)(a-1))$.
For $k\ge4$, $Z\ge3Q^2$ and $a\ge13$, so
\[
 Z(a-1)(a-2)-2Q^2a(a+1)
       \ge Q^2(a^2-11a+6)>0.
\]
This proves $\beta>0$ in that range.

For the remaining $k$, it is enough that
$\alpha+(a-2)\beta\ge0$. Its numerator after multiplication
by $2Za(a+1)(a-1)/k$ is
\begin{equation}\label{eq:one-pair-polynomial}
 G=Z(a-1)((k+3)a-2(k+1))
              -2Qa(a+1)(Q(k-1)+2\nu+2).
\end{equation}
For $k=2,3$, write $u\coloneqq Q-4$, $v\coloneqq\nu-1$.
The rows below give the coefficients of $v^0,v^1,\ldots$
in $[u^j]G$; they are all positive.
\begin{center}\small
\begin{tabular}{ccl}\toprule
$k$&$j$&Coefficients\\\midrule
2&0&$940,200,177,58,5$\\
2&1&$2240,380,100,16$\\
2&2&$1432,148,12$\\
2&3&$360,16$\\
2&4&$32$\\
3&0&$7224,1656,582,108,6$\\
3&1&$11464,1684,332,32$\\
3&2&$6310,516,44$\\
3&3&$1476,48$\\
3&4&$126$\\\bottomrule
\end{tabular}
\end{center}
These are polynomial identities obtained by substituting
$a=(Q-1)k+\nu$ and $Z=Q(Q-1)k+\nu(\nu+1)$
in~\eqref{eq:one-pair-polynomial}.
For $k=1$ and $Q\ge9$, use $u\coloneqq Q-9$, $v\coloneqq\nu-1$ instead.
The five rows for $j=0,\ldots,4$ are
\[
\begin{array}{l}
(12464,896,-12,40,4),\quad(7000,532,0,4),\\
(1416,84),\quad(124,4),\quad(4).
\end{array}
\]
The only negative term is absorbed by
$40v^3-12v^2\ge0$ for every nonnegative integer $v$.
Hence $G>0$ here too.

It remains to take $Q=4$, $k=1$ and $\nu=1,2,3,4$.
Now $\gamma_i=(a+2)/(a+2-i)$, $\nu=a-3$,
$\operatorname{lin}-2=8(\nu-3)/Z$ and
$\operatorname{quad}-\operatorname{lin}+1=32/Z$. With $s\coloneqq a+2-i$,
multiplication of~\eqref{eq:one-pair-gamma} by $sZ(a-1)(a-2)/i$ shows
that it is equivalent to
\[
 Z(a-1)(a-2)\ge2Qs(\Theta-Qs),\qquad
 \Theta\coloneqq a^2-Qa+5Q-4,\quad3\le s\le a+1.
\]
For $\nu=2$, the left side is $216$ and the right side is
at most $8\cdot27=216$. For $\nu=3,4$, the left sides
are $480,960$, whereas the right sides are at most
$\Theta^2/2=392,1369/2$, respectively.
For $\nu=1$, $N=6$, $a=4$, $Z=14$ and $r\le1$.
The left side minus the right side of~\eqref{eq:one-pair-deficit}
is $768,660$ for $r=0$ and $i=1,2$, and $488,420,-40$ for $r=1$ and
$i=1,2,3$.
This proves the assertion and identifies its sole exception.
\end{proof}

\begin{proposition}\label{prop:one-pair-minimum}
Let $p\ge3$, $n\ge2D+1$, and let an exact code of dimension $K>1$
have distance at least $D+1$. Suppose
\eqref{eq:root-region} holds. If a minimizing vertex has
at most one residual pair, then $K\le p^n\beta_{n,D,Q}$.
Moreover, if $n-D\le2Q-4$, then every minimizing vertex has at most
one residual pair, and $K\le p^n\beta_{n,D,Q}$ holds
without~\eqref{eq:root-region}.
\end{proposition}
\begin{proof}
With no pair the optimum is $Q^{-D}$, and Singleton suffices.
With one pair, put $r\coloneqq D-2$, $N\coloneqq n-r$, and let its lower zero be $a$.
For $X\sim\mu_N$, set $\kappa_1\coloneqq\E X=N(Q-1)/Q$ and
$\kappa_2\coloneqq\E[X(X-1)]=N(N-1)((Q-1)/Q)^2$.
The normalized residual objective
$F(a)\coloneqq\E_{\mu_N}[(X-a)(X-a-1)]/(a(a+1))$ of a pair, which is the
function $F$ of the proof of Proposition~\ref{prop:optimizer-roots}
with $w=1$, satisfies
\[
 F(a)=1-\frac{2\kappa_1}{a+1}+\frac{\kappa_2}{a(a+1)},\qquad
 F(a+1)-F(a)=\frac{2(\kappa_1a-\kappa_2)}{a(a+1)(a+2)}.
\]
Its minimum over positive integer pairs in the grid is attained
at $a_*\coloneqq\lfloor1+(Q-1)(N-1)/Q\rfloor$, the pair of the rounded
polynomial with $m=1$.
If $1+(Q-1)(N-1)/Q$ is an integer, the pair at $a_*-1$ has the same
objective. Both pairs are feasible;
the pair at $a_*$ may extend the terminal block, which
Theorem~\ref{thm:one-root} allows.

In the root region, Proposition~\ref{prop:optimizer-roots}
gives $a>D$. Both adjacent moves of this pair are feasible,
so its optimality implies $(Q-1)(N-1)/Q\le a
\le1+(Q-1)(N-1)/Q$. Consequently $F(a_*)=F(a)$
and $a_*\ge a>D$.
Theorem~\ref{thm:one-root} applies to the polynomial $g$ with this
equal-objective pair, since $a_*=(Q-1)k+\nu$ in its notation, with no
binary exception. Lemma~\ref{lem:coefficient-criterion} gives
$K\le p^nc_0/g(0)$, and by~\eqref{eq:terminal-expectation}
\[
 \frac{c_0}{g(0)}=Q^{-r}F(a_*)=Q^{-r}F(a)=\beta_{n,D,Q}.
\]

For the last assertion, write $H\coloneqq n-D$. The cap and
$H\le2Q-4$ force at most one residual pair.
If one pair occurs, $H\ge Q-1$ and $N=H+2$.
Thus $1+(Q-1)(N-1)/Q=H+2-(H+1)/Q>H$,
so $a_*\ge H\ge D+1$, now without
Proposition~\ref{prop:optimizer-roots}. The polynomial with pair $a_*$
is feasible and $F(a_*)$ is the least value of $F$, so again
$F(a_*)=F(a)$, and the argument of the preceding paragraph applies.
\end{proof}

\subsection{Reduction to a cap minimum}

For fixed $H=n-D$, the next identity reduces the grid minimum at every
degree $D\ge2a$ to one cap minimum, at length $H+2a$ and degree $2a$;
when $D\ge2a$, the integer $a$ of the lemma equals the cap
index~\eqref{eq:cap-index}. The finite certificates of
Propositions~\ref{prop:four-finite} and~\ref{prop:qutrit-finite} use it
to certify one optimal polynomial for each $H$.

\begin{lemma}\label{lem:cap-minimum-identity}
Let $Q\ge9$, $n\ge2D+1$, $H\coloneqq n-D$, and
$a\coloneqq\lfloor(H-1)/(Q-2)\rfloor$. If $D\ge2a$, then
\begin{equation}\label{eq:cap-minimum-identity}
 \beta_{n,D,Q}=Q^{-(D-2a)}\beta_{H+2a,2a,Q}.
\end{equation}
\end{lemma}
\begin{proof}
The proof uses only $Q>2$, through the cap.
For $0\le j\le a$, put $L_j\coloneqq H+2j$.
Lifting any feasible polynomial at $(L_j,2j)$ by the
normalized terminal factor of degree $2(a-j)$ gives
a feasible polynomial at $(L_a,2a)$, with expectation
multiplied by $Q^{-2(a-j)}$.
A minimizing vertex at $(n,D)$ has residual degree $2j$
with $j\le a$ by the cap. Its expectation is therefore
at least
\[
 Q^{-(D-2j)}\beta_{L_j,2j,Q}
       \ge Q^{-(D-2a)}\beta_{L_a,2a,Q}.
\]
Conversely lift a cap minimizer by a terminal factor of
degree $D-2a\ge0$. The terminal expectation identity gives
the reverse inequality. Values outside the smaller grid
are multiplied by zero, so this construction preserves
grid nonnegativity even for polynomials of smaller degree.
\end{proof}

\section{Rank bounds from invariant windows}\label{sec:rank-windows}

The coefficient comparison will be used in the root
region~\eqref{eq:root-region}, where the roots of a minimizer lie
beyond its degree.  Outside the root region, and for the largest
distances (Section~\ref{sec:distance-strips}), we instead bound the
dimension $K$ of the code, which we also call its rank, the rank of the
projector $P$.  The following construction gives such bounds
from the weight enumerators $A$ and $B$ alone.  The detailed window estimates
for this section and Sections~\ref{sec:distance-strips}--\ref{sec:complement}
are proved in Appendix~\ref{app:rank-windows}.

Define the homogeneous weight enumerators by
\[
 A(x,y)\coloneqq\sum_{i=0}^{n}A_i x^{n-i}y^i,\qquad
 B(x,y)\coloneqq\sum_{i=0}^{n}B_i x^{n-i}y^i.
\]
We use the same coefficient convention for the homogeneous polynomials
$C,{C^{-}},E$ below.  Write $T_j$ for substitution of
\[
 (x,y)\longmapsto ((x+(Q-1)y)/p,(x-y)/p)
\]
on homogeneous polynomials of degree $j$; we write $T$ when the degree
is clear.  For the enumerators of
Section~\ref{sec:framework}, $T_nA=B/K$ and $T_nB=KA$.
Moreover $B_i\ge A_i$: for each element $U$ of the error basis, the
Hilbert--Schmidt inequality on the range of the code projector gives
$\Tr(PUPU^\dagger)\ge |\Tr(PU)|^2/K$.
Consequently
\begin{equation}\label{eq:invariant-enumerators}
 C\coloneqq\frac{KA+B}{K+1},\qquad {C^{-}}\coloneqq\frac{KA-B}{K-1},\qquad
 E\coloneqq C-{C^{-}}=\frac{2K(B-A)}{K^2-1}
\end{equation}
satisfy $TC=C$, $T{C^{-}}=-{C^{-}}$, $C_0=1$, $C_i,E_i\ge0$, and
$E_i=0$ for $i<d$.  Also $C=A+\rho_KE$, where
$\rho_K\coloneqq(K-1)/(2K)$.
This invariant and anti-invariant formulation is the one used by
Rains~\cite[Lemma~5.4 and proof of Theorem~5.5]{RainsAsymptotic}.
We derive the finite coefficient relations needed here explicitly.
They are used as certificates of linear-programming type: if every
term of such a relation is nonnegative and the term of weight zero is
positive, no code with these parameters exists.

Lemma~\ref{lem:window-identity} below needs an odd working length.
An even length can be replaced, for these rank bounds, by $n'\coloneqq n+1$:
tensor the code with a fixed single-qudit state, for example
$\lvert 0\rangle$.  Its rank and detection
distance are unchanged.  We use $n'\coloneqq n$ at odd lengths.  The resulting
bounds on $K$ are compared with $p^n\beta_{n,D,Q}$ at the original
length $n$; the working length never replaces $n$ there.

The name of the next lemma refers to the following mechanism, made
explicit in its proof.  The invariant polynomial $C$ and the
anti-invariant polynomial ${C^{-}}$ have finite expansions in invariant
forms, and their coefficients with the $s$ indices just beyond the end
of these expansions, the window, vanish.  These vanishing coefficients
are linear relations among the $C_j$ and ${C^{-}}_j$; the lemma
combines them through one homogeneous polynomial $W$ of degree
$J=2s-1$.

\begin{lemma}[Invariant-window identity]\label{lem:window-identity}
Let the working length be $n'=2{s_0}+1$.  Choose $1\le s\le {s_0}+1$, set
$M\coloneqq{s_0}+s$ and $J\coloneqq2s-1$, and let $W(x,y)$ be any real homogeneous
polynomial of degree $J$.  Put $V\coloneqq-(W+T_JW)/2$, and define
\[
 a_{M-l}\coloneqq[z^l]\frac{W(1,z)}{(1-z)^{M+1}},\qquad
 b_{M-l}\coloneqq[z^l]\frac{V(1,z)}{(1-z)^{M+1}}.
\]
Coefficients with $l<0$ are zero; thus $l=M-j$ counts the weight $j$
down from $M$.  Then
\begin{equation}\label{eq:window-identity}
 0=\sum_{j=0}^{n'}A_ja_j+
       \sum_{j=d}^{n'}E_j(b_j+\rho_Ka_j).
\end{equation}
In particular, for any real $K_0>1$, the conditions
\begin{equation}\label{eq:window-signs}
 a_{M-l}\ge0\ (0\le l\le M),\quad a_0>0,\qquad
 -a_{M-l}/K_0-a^{\rm ref}_{M-l}\ge0\ (0\le l\le M-d),
\end{equation}
where $a^{\rm ref}$ is extracted from $T_JW$, exclude every $K\ge K_0$.
\end{lemma}

\begin{proof}
Set ${u_+}\coloneqq x+(p-1)y$, ${u_-}\coloneqq x-(p+1)y$, and $v\coloneqq y(x-y)$.
Then $T{u_+}={u_+}$, $T{u_-}=-{u_-}$, $Tv=v$, and ${u_-}^2={u_+}^2-4pv$.
Diagonalizing this reflection, or replacing every even power of ${u_-}$
by a polynomial in ${u_+}^2,v$, gives
\[
 C=\sum_{i=0}^{{s_0}}\alpha_i {u_+}^{2{s_0}+1-2i}v^i,\qquad
 {C^{-}}={u_-}\sum_{i=0}^{{s_0}}\beta_i {u_+}^{2{s_0}-2i}v^i.
\]
Since these expansions stop at $i={s_0}$, their coefficients with indices
${s_0}+1,\ldots,{s_0}+s$ vanish.  To express this, put $x=1,y=z$, and $z_0\coloneqq z(1-z)/{u_+}^2$.  Since $dz_0/dz={u_-}/{u_+}^3$, formal change of variable in a
coefficient residue (Lagrange inversion) gives
\[
 \alpha_i=[z^i]C(1,z){u_-} {u_+}^{2i-2{s_0}-2}(1-z)^{-i-1},\quad
 \beta_i=[z^i]{C^{-}}(1,z){u_+}^{2i-2{s_0}-1}(1-z)^{-i-1}.
\]
These formulas hold for every $i\ge0$, so their right sides vanish
for ${s_0}<i\le {s_0}+s$.  Since $[z^{{s_0}+j}]F=[z^M]z^{s-j}F$ for a formal power
series $F$ and $z^{s-j}(1-z)^{-{s_0}-j-1}=v(1,z)^{s-j}(1-z)^{-M-1}$, with
common denominator $(1-z)^{M+1}$ these vanishing conditions
have numerators ${u_-}{u_+}^{2j-2}v^{s-j}$ and ${u_+}^{2j-1}v^{s-j}$,
$1\le j\le s$.  They form a basis of the degree-$J$ homogeneous
polynomials: the first family spans the anti-invariant subspace and
the second the invariant subspace, each of dimension $s$.
Write $W=W_++W_-$ with $W_\pm\coloneqq(W\pm T_JW)/2$, and let $a_j(\cdot)$
denote the extraction in the statement.  The anti-invariant relations
give $\sum_jC_ja_j(W_-)=0$, and the invariant relations give
$\sum_j{C^{-}}_ja_j(W_+)=0$.  Adding them and substituting ${C^{-}}=C-E$ gives
$0=\sum C_ja_j+\sum E_jb_j$, since $b$ is extracted from $V=-W_+$.
Substitution of $C=A+\rho_KE$ proves \eqref{eq:window-identity}.
Finally $2(b+\rho_Ka)=-a/K-a^{\rm ref}$, which is nondecreasing in $K$
when $a\ge0$.  Under \eqref{eq:window-signs} all summands are
nonnegative and the $j=0$ summand is positive, a contradiction.
\end{proof}

There is a useful polynomial form of this criterion.  To distinguish
the two factorial conventions, write
$(\eta)^{\overline a}\coloneqq\eta(\eta+1)\cdots(\eta+a-1)$.

\begin{lemma}[Factorial reflection]\label{lem:factorial-reflection}
Let $J$ be odd, $M\ge J$, and $\eta\coloneqq M-J+1$.  On polynomials of
degree at most $J$ define
\begin{equation}\label{eq:factorial-reflection}
 \mathcal R(x)_a
   \coloneqq p^{J-2a}\frac{(\eta)^{\overline a}}
                    {(\eta)^{\overline{J-a}}}(x)_{J-a}.
\end{equation}
Then $\mathcal R^2=I$.  For the working length $n'=2M-J$, distance
at least $d$, and a real $K_0>1$, it suffices to find a polynomial $F$
of degree at most $J$ such that
\begin{equation}\label{eq:profile-signs}
 F(l)\ge0\ (0\le l\le M-d),\qquad
 F(l)-K_0\mathcal RF(l)\ge0\ (0\le l\le M),
 \qquad F(M)-K_0\mathcal RF(M)>0.
\end{equation}
All grid variables here are integers.  These conditions exclude
$K\ge K_0$.
\end{lemma}

\begin{proof}
For $w(x)=\sum w_a(x)_a$ associate
$W_w(X,Y)\coloneqq\sum w_a(\eta)^{\overline a}Y^a(X-Y)^{J-a}$.
The generating function for binomial factorial moments gives
\begin{equation}\label{eq:profile-extraction}
 [z^l]\frac{W_w(1,z)}{(1-z)^{M+1}}
       =\binom{\eta+l-1}{l}w(l).
\end{equation}
Since $T(Y)=(X-Y)/p$ and $T(X-Y)=pY$, conjugating $T_J$ by this
association gives \eqref{eq:factorial-reflection}.  Put
$w\coloneqq F-K_0\mathcal RF$.  Then
$-w/K_0-\mathcal Rw=(K_0-K_0^{-1})F$.
The factors in \eqref{eq:profile-extraction} are positive, so
Lemma~\ref{lem:window-identity} applies.  Notice also the exact
identity at the endpoint $l=M$ of the strict condition
in~\eqref{eq:profile-signs},
\[
 \mathcal Rw(M)=p^J\sum_{l=0}^M\binom Ml\frac{(Q-1)^{M-l}}{Q^M}\,w(l),
\]
a binomial average of $w$; it does not imply positivity of the
reflection at the other grid points $l<M$.
\end{proof}

The following two small windows will cover the distances nearest the
distance ceiling of Theorem~\ref{thm:distance-ceiling} below.

\begin{proposition}[Linear and cubic rank bounds]\label{prop:strip-rank}
For every $p\ge2$ and code rank $K>1$ the following bounds hold.
If $M\coloneqq\lfloor n/2\rfloor+1$, $L\coloneqq M-d$, and $0\le L<M/Q$, then
\begin{equation}\label{eq:linear-rank}
 K\le\frac{p(M-L)}{M-QL}.
\end{equation}
If $M\coloneqq\lfloor n/2\rfloor+2$, $M\ge Q$, and
$1\le L\coloneqq M-d\le\lfloor(M-1)/Q\rfloor$, then
\begin{equation}\label{eq:cubic-rank}
 K\le \operatorname{Rank}_3\coloneqq\frac{p^3L(L-1)(M-L)}{(M-1)(M-2)(M-QL)}
       <\frac{pL}{M-QL}.
\end{equation}
Only the last inequality in~\eqref{eq:cubic-rank} is strict.
\end{proposition}

\begin{proof}
We prove a common bound for every odd $J\le n'$, with
$M\coloneqq(n'+J)/2$, $\eta\coloneqq M-J+1$ and $0\le L\coloneqq M-d<M/Q$:
\[
 K\le\frac{p^J(L)_{J-1}(M-L)}{(M-1)_{J-1}(M-QL)}.
\]
For $\varepsilon\ge0$ put
$w_\varepsilon(x)\coloneqq(x)_{J-1}(M+\varepsilon-x)$ and
$A\coloneqq(M-1)_{J-1}>0$. Since
$w_\varepsilon=(\eta+\varepsilon)(x)_{J-1}-(x)_J$,
the factorial reflection of Lemma~\ref{lem:factorial-reflection} gives
\[
 \mathcal Rw_\varepsilon(x)
 =\frac{A}{p^J}\left(Q\left(1+\frac{\varepsilon}{\eta}\right)x-M\right).
\]
At $\varepsilon=0$ the reflection is negative on $0\le l\le L$, and
\[
 \frac{w_0(l)}{-\mathcal Rw_0(l)}
 =\frac{p^J(l)_{J-1}(M-l)}{A(M-Ql)}
 \le\frac{p^J(L)_{J-1}(M-L)}{A(M-QL)}.
\]
Indeed, $(l)_{J-1}$ is nonnegative and nondecreasing on the integer
grid, while $(M-l)/(M-Ql)$ is positive and increasing for $l<M/Q$.
This includes $L<J-1$, when both factorial profiles vanish on the tail.
If $K>1$ exceeded this bound, then on this
finite tail $-\mathcal Rw_\varepsilon-w_\varepsilon/K>0$ would persist
for every sufficiently small $\varepsilon>0$. Put
$F\coloneqq-\mathcal Rw_\varepsilon-w_\varepsilon/K$.
The identity $\mathcal R^2=I$ gives
\[
 F-K\mathcal RF=(K-K^{-1})w_\varepsilon.
\]
This is nonnegative on the whole integer grid $0\le l\le M$ and
positive at $l=M$. Lemma~\ref{lem:factorial-reflection}, with $K_0=K$,
is a contradiction, proving the bound.
Its cases $J=1$ and $J=3$ are the asserted rank bounds, since
$(n'+J)/2=\lfloor n/2\rfloor+(J+1)/2$.
For the final strict comparison in~\eqref{eq:cubic-rank}, put
$u\coloneqq M-QL\ge1$ and use
\[
 (M-1)(M-2)-Q(L-1)(M-L)
 =M(u+Q-3)+QL(L-1)+2>0.
\]
\end{proof}

Here $L$ measures how far $d$ lies below $M$, which is about $n/2$.
At $L=1$ the cubic bound reads $K\le0$: no code with $K>1$ has
$d=M-1=\lfloor n/2\rfloor+1$ when $M\ge Q+1$.
Theorem~\ref{thm:critical-grid} uses this. For $K=p$, that is, for
quantum MDS codes of length $2d-1$, this is the known length bound
$n\le p^2+d-2$~\cite{HuberGrassl} (see~\cite{KKKS} for stabilizer
codes); the cubic bound extends it to every $K>1$.

\begin{theorem}[Finite distance ceiling]\label{thm:distance-ceiling}
For $p\ge3$ and $K>1$, put $M\coloneqq\lfloor n/2\rfloor+2$.  Then
\begin{equation}\label{eq:distance-ceiling}
 d\le M-\lceil M/Q\rceil+1.
\end{equation}
\end{theorem}

Asymptotically~\eqref{eq:distance-ceiling} gives
$\limsup d/n\le(1-p^{-2})/2$, the bound of
Rains~\cite[Theorem~5.5]{RainsAsymptotic}; the theorem is a
finite-length form of that bound for $p\ge3$. Rains's asymptotic
argument uses the same two vanishing indices in the invariant and
anti-invariant expansions; the cubic separator gives the finite-length
bound above.

\begin{proof}
Negating the integer inequality $Qd\ge(Q-1)M+Q+1$ gives
\eqref{eq:distance-ceiling}. Suppose, for contradiction, that
$Qd\ge(Q-1)M+Q+1$.  Singleton gives
$L\coloneqq M-d\ge1$, and hence $M\ge Q(L+1)+1\ge2Q+1$.  Take the window of
Lemma~\ref{lem:window-identity} with $s\coloneqq2$, so $J=3$ and
$M={s_0}+2=\lfloor n/2\rfloor+2$.  Its proof gives, for every cubic $W$
and before $C$ is replaced by $A+\rho_KE$, the relation
\[
 0=\sum_jC_ja_j+\sum_jE_jb_j,
\]
whose coefficients $a_j,b_j$ do not involve $K$.
Lemma~\ref{lem:distance-cubic-certificate} constructs a cubic $W$, the
cubic separator, for which $a_{M-l}$ is $(M+1-l)(M-Ql)^2$ times a
positive factor.  These coefficients are nonnegative for $0\le l\le M$
(those with $l<0$ vanish) and positive at $l=M$, the weight zero.  The
lemma proves that $b_{M-l}>0$ for $0\le l\le L$, that is, on the tail
of weights $j\ge d$ where $E_j$ can be nonzero.  With $C_0=1$,
$C_j\ge0$, $E_j\ge0$ and $E_j=0$ for $j<d$, every term of the relation
is nonnegative and the term of weight zero is positive, a
contradiction.  The lemma is stated for $p\ge3$; this is where
the hypothesis $p\ge3$ enters. It is the only case needed: for $p=2$,
Rains's shadow length bound is stronger
(Section~\ref{sec:proof-outline}).
\end{proof}

\section{The critical and adjacent distance strips}\label{sec:distance-strips}

We now compare the rank bounds with the grid minimum near the largest
distance allowed by Theorem~\ref{thm:distance-ceiling}.  As $n$ varies,
the parameters $(n,d)$ treated here form two strips at the top of the
distance range: the critical strip, where $d$ equals the
ceiling~\eqref{eq:distance-ceiling} (Theorem~\ref{thm:critical-grid}),
and the adjacent strip, where $d$ is one less
(Theorem~\ref{thm:adjacent-grid}, when $M=\lfloor n/2\rfloor+1$
exceeds $Q$ and is not divisible by $Q$).  Throughout this
section $p\ge3$, $b\coloneqq Q-1$, and the length in
$R_{n,D}\coloneqq p^n\beta_{n,D,Q}$ is the original length.  In both theorems
$d$ is the distance assumed of the code, and $K$ is bounded by
$R_{n,d-1}$, the target for $D=d-1$.

\begin{lemma}\label{lem:grid-budget-coarse}
Under the hypotheses of Proposition~\ref{prop:grid-budget},
\begin{equation}\label{eq:grid-budget-coarse}
 R_{n,D}\ge \operatorname{pref}_Q\frac{b^{n-D}}{p^n}.
\end{equation}
For an exact code of dimension $K>1$ and distance at least $D+1$, this
and Theorem~\ref{thm:distance-ceiling} imply
\begin{equation}\label{eq:distance-target-budget}
 R_{n,D}\ge\operatorname{Budget}_Q(n)\coloneqq
 \operatorname{pref}_Q(b/Q)^{n/2}b^{n/(2Q)-2b/Q}.
\end{equation}
The function $\operatorname{Budget}_Q$ is increasing in $n$.
\end{lemma}
\begin{proof}
For $N=H+2a$, evaluation of the binomial generating function at
$b^{-2}$ gives
$V_a(N)\le b^{2a}(1+b^{-1})^N=Q^Nb^{2a-N}$.
Substitute this in \eqref{eq:grid-budget}; its powers of $p$ cancel
to give \eqref{eq:grid-budget-coarse}, also when $a=0$.
The distance ceiling yields $D\le bn/(2Q)+2b/Q$, proving
\eqref{eq:distance-target-budget}.  Finally
$\operatorname{Budget}_Q(n)$ is a positive constant times $c^n$, where
$c\coloneqq[(b/Q)b^{1/Q}]^{1/2}$, and $c^{2Q}=b(b/Q)^Q>b/3>1$, since
$(1-1/Q)^Q>1/3$ for $Q\ge9$: the logarithmic derivative is
$\log(1-1/Q)+1/(Q-1)>0$, and $3\cdot8^9>9^9$.
\end{proof}

Thus, for a code as in the lemma, $\operatorname{Budget}_Q(n)$ bounds
$R_{n,D}$ from below independently of $D$;
Section~\ref{sec:complement} compares rank thresholds with it.

\begin{theorem}[Critical strip]\label{thm:critical-grid}
Let $M\coloneqq\lfloor n/2\rfloor+2$ and
$d\coloneqq M-\lceil M/Q\rceil+1$.  Every code of rank $K>1$ and distance
at least $d$ satisfies $K\le R_{n,d-1}$.
\end{theorem}
\begin{proof}
Put $L\coloneqq M-d=\lceil M/Q\rceil-1$.  The case $L=0$ contradicts
Singleton, and $L=1$ contradicts \eqref{eq:cubic-rank}.
For $L\ge2$ write
\begin{equation}\label{eq:critical-parameters}
 M=QL+u,\quad 1\le u\le Q,\quad
 n=2M-4+\delta,\quad\delta\in\{0,1\},\quad D=M-L-1.
\end{equation}
Proposition~\ref{prop:strip-rank} gives
$K\le\operatorname{Rank}_3<pL/u$, since
$M-QL=u$.  If $u\ge pL/2$, then $K<2$; thus we
may assume $u<pL/2$.

For $L\ge8$, \eqref{eq:grid-budget-coarse} gives
\[
 R_{n,D}>\frac{49}{1296}(b/3)^{L-1}>pL\ge K.
\]
Here the exact base comparison is
$49\cdot8^7>24\cdot1296\cdot3^7$ at $Q=9,L=8$.
The ratio of the middle expression to $\sqrt Q L$ increases in $Q$,
since its logarithmic derivative is
$(L-1)/(Q-1)-1/(2Q)>0$, and increases when $L$ is replaced by $L+1$,
by the factor $(b/3)L/(L+1)>1$.
The first inequality follows explicitly from
\[
 \frac{b^{n-D}}{p^n}
 =b^{L-1}(b/Q)^{M-2}(b/p)^\delta
 >b^{L-1}3^{-L-1}.
\]

For $L=2$, Lemma~\ref{lem:critical-two} proves $K\le R_{n,D}$.  For
$3\le L\le7$, Lemma~\ref{lem:critical-finite} proves it for every
$p\ge3$: the paired-moment bound of Lemma~\ref{lem:paired-constant},
compared at a suitably larger length through the volume
recurrence~\eqref{eq:volume-recurrence}, exceeds the cubic rank bound.
All these comparisons concern the full grid minimum at the original
length, so together they prove the assertion.
\end{proof}

\begin{theorem}[Adjacent strip]\label{thm:adjacent-grid}
Let $L\ge1$, $1\le u\le Q-1$,
$M\coloneqq QL+u$, $n=2M-2+\delta$, $\delta\in\{0,1\}$, and $d\coloneqq M-L$.
Every code of rank $K>1$ and distance at least $d$ satisfies
$K\le R_{n,d-1}$.
\end{theorem}
\begin{proof}
Since $M-QL=u$ and $M-L=u+bL$, the linear bound gives
\begin{equation}\label{eq:adjacent-rank}
 K\le\lfloor \operatorname{Rank}_1\rfloor,\qquad
 \operatorname{Rank}_1\coloneqq p\{1+bL/u\}\le p(bL+1).
\end{equation}
Since $n-D=M+L-1+\delta$, \eqref{eq:grid-budget-coarse} gives
\begin{equation}\label{eq:adjacent-coarse}
 R_{n,D}>\operatorname{Strip}_{p,L}\coloneqq\frac{49}{432}(b/3)^L.
\end{equation}
Indeed $b^{n-D}/p^n=(b/Q)^{u-1}[b(b/Q)^Q]^L(b/p)^\delta$,
and both $(b/Q)^{u-1}$ and $(b/Q)^Q$ exceed $1/3$.
For $L\ge8$, $\operatorname{Strip}_{p,L}>p(bL+1)$: the base comparison is
$49\cdot8^8>432\cdot3^8\cdot195$, and the ratio increases in $Q$
because
\[
 \frac{L}{Q-1}-\frac1{2Q}-\frac{L}{(Q-1)L+1}>0.
\]
It increases with $L$ by at least $(8/3)L/(L+1)>1$.

Here is an explicit reduction for $1\le L\le7$.  If
$2\le L\le7$ and $p\ge200$, then $b\ge9p^2/10$ and
\[
 \operatorname{Strip}_{p,2}\ge\frac{3969p^4}{388800}>2p^3>p(2b+1).
\]
The same monotonicity in $L$ applies.  For $L=1$ first remove
$n-D\le2Q-4$ by Proposition~\ref{prop:one-pair-minimum}.
The remaining parameters have $u\ge Q-3-\delta\ge Q-4$, so
$\operatorname{Rank}_1\le13p/5$.  For $p\ge80$,
$\operatorname{Strip}_{p,1}=49b/1296\ge49p^2/1440>13p/5$.

It remains to treat $3\le p\le79$ for $L=1$, and
$3\le p\le199$ for $2\le L\le7$.  When $\operatorname{Strip}_{p,L}>p$, retain only
\begin{equation}\label{eq:adjacent-finite-domain}
 1\le u\le\min\left(Q-1,
       \left\lceil\frac{pbL}{\operatorname{Strip}_{p,L}-p}\right\rceil-1\right);
\end{equation}
when $\operatorname{Strip}_{p,L}\le p$, retain all $1\le u<Q$.
The discarded values satisfy $\operatorname{Rank}_1\le \operatorname{Strip}_{p,L}$.
For $L=1$, retain only $u\ge Q-3-\delta$ among these; the other values
of $u$ were removed above by Proposition~\ref{prop:one-pair-minimum}.
Lemma~\ref{lem:adjacent-finite} gives exact certificates for every
tuple in this remaining domain: most tuples by rational comparison
with Proposition~\ref{prop:grid-budget}, the others by exactly
certified grid minima and, for eight tuples, explicit rank windows of
higher degree.  The domain includes the seven tuples with
$p=3$, $L=1$ and $u\ge6-\delta$, all of degree at most $15$.  These
certificates bound $R_{n,D}=p^n\beta_{n,D,Q}$ itself.  An assertion
only about rounded Lloyd polynomials would give only
$K\le p^n/S_{t,\sigma}^{n,e}$, which is weaker, because
$R_{n,D}\le p^n/S_{t,\sigma}^{n,e}$ by the end of
Section~\ref{sec:grid-minima}.
\end{proof}

The next lemma gives a degree bound outside the range settled by the
preceding comparisons. In the notation $D=r+2m$ and $N=n-r$,
it bounds $N-r-4m=n-2D$ from below, hence the terminal degree $r$
from above. The tail estimates for $p=4,5$ use this in
Section~\ref{sec:higher-dimensions}; the lemma also applies outside
the root region in Section~\ref{sec:complement}.

\begin{lemma}[Strips or a degree bound]\label{lem:physical-tail-cap}
Let $p\ge3$ and let an exact code of dimension $K>1$ have length $n$
and distance at least $D+1$, where $D\ge2$. Put $\delta\coloneqq n\bmod2$,
$H\coloneqq n-D$, $k\coloneqq n-2D$, and
\[
 M_0\coloneqq\lfloor n/2\rfloor+1,\qquad L_0\coloneqq M_0-(D+1),\qquad
 u_0\coloneqq M_0-QL_0.
\]
If $u_0>0$, then $K\le R_{n,D}$. If $u_0\le0$, then
\begin{equation}\label{eq:physical-tail-cap}
 k\ge2\left\lceil\frac{H+1-\delta}{Q+1}\right\rceil+\delta
       >\frac{2H}{Q+1},\qquad
 D=H-k\le H-\left\lfloor\frac{2H}{Q+1}\right\rfloor-1.
\end{equation}
The latter implication is purely arithmetic and requires only
$u_0\le0$. In the root region~\eqref{eq:root-region}, $n\ge2Q-2$ also
implies $L_0\ge1$, independently of code existence.
\end{lemma}
\begin{proof}
Singleton gives $n\ge2D+1$, hence $L_0\ge0$, with $L_0=0$ possible
only at odd $n$. The distance ceiling gives $u_0\le2Q-1$.
If $Q\le u_0\le2Q-1$, then
$\lceil(M_0+1)/Q\rceil=L_0+2$, so the distance of
Theorem~\ref{thm:critical-grid} is $M_0-L_0=D+1$ and that theorem
applies. If $1\le u_0<Q$ and $L_0\ge1$, apply
Theorem~\ref{thm:adjacent-grid} with $(M,L,u)=(M_0,L_0,u_0)$.
If instead $L_0=0$ and $u_0<Q$, then $n=2M_0-1$ and
$H=M_0\le Q-1\le2Q-4$; the unconditional last assertion of
Proposition~\ref{prop:one-pair-minimum} gives $K\le R_{n,D}$.

For $u_0\le0$, use
$k=2L_0+\delta$ and $u_0=H+1-\delta-(Q+1)L_0$.
These give the first inequality in~\eqref{eq:physical-tail-cap}.
For $\delta=0$, $\lceil(H+1)/(Q+1)\rceil>H/(Q+1)$; for $\delta=1$,
$2\lceil H/(Q+1)\rceil+1>2H/(Q+1)$, including divisibility endpoints.
The integer bound on $D=H-k$ follows.
Finally the root-region inequality gives
\[
 L_0\ge\frac{n-(Q-1)\delta-Q+2}{2(Q-1)}>0\qquad(n\ge2Q-2),
\]
which proves the last assertion.
\end{proof}

\section{The complement of the root region}\label{sec:complement}

In this section the distance is large.  At the lengths treated by
the product windows constructed below, one invariant-window threshold
$K_0$, common to all
degrees $D$ of~\eqref{eq:strict-complement} at the given length,
satisfies $K<K_0<\operatorname{Budget}_Q(n)\le R_{n,D}$
(Lemma~\ref{lem:grid-budget-coarse}).  These windows treat every length
$n\ge Q^2-2$ for $p\ge4$ and $n\ge1300$ for $p=3$; shorter lengths
reduce to the strips of Section~\ref{sec:distance-strips}, except for a
bounded qutrit remainder, which uses $578$ integer windows of degree five.

\begin{theorem}\label{thm:strict-complement}
Let $p\ge3$ and let a code of rank $K>1$, length $n$, and distance
at least $D+1$ satisfy
\begin{equation}\label{eq:strict-complement}
 2(Q-1)D>(Q-2)(n+1).
\end{equation}
Then $K\le p^n\beta_{n,D,Q}$.
\end{theorem}

Singleton and~\eqref{eq:strict-complement} imply $D>Q-2$; in
particular $D\ge2$, as Lemma~\ref{lem:physical-tail-cap} requires.
We first describe the common certificate.  Put $b\coloneqq Q-1$ and let
$\epsilon\coloneqq0$ for odd original $n$, $\epsilon\coloneqq1$ for even $n$ (so
$\epsilon=1-\delta$ for the $\delta=n\bmod2$ of
Section~\ref{sec:distance-strips});
$n'=n+\epsilon$ is the odd working length.  For an integer
$k\ge2$ choose
\begin{equation}\label{eq:complement-profile-parameters}
 \begin{gathered}
 J\coloneqq2k+1,\quad {j_0}\coloneqq2k-2,\quad M\coloneqq(n'+J)/2,\quad\eta=M-2k,\\
 D_0\coloneqq\left\lfloor\frac{(Q-2)(n+1)}{2b}\right\rfloor+1,
 \quad {\tau}\coloneqq M-(D_0+1),\\
 W_k(x)\coloneqq\prod_{a=2k}^{3k-2}(x-a)(x-a-1),\qquad
 F(x)\coloneqq({\tau}-x)W_k(x).
 \end{gathered}
\end{equation}
Here $D_0$ is the least degree satisfying~\eqref{eq:strict-complement},
and ${\tau}=M-(D_0+1)$ is the end of the tail $0\le l\le M-d$ of
Lemma~\ref{lem:factorial-reflection} for $d=D_0+1$.  The polynomial
$W_k=(x-2k)(x-3k+1)\prod_{c=2k+1}^{3k-2}(x-c)^2$ has degree
$2k-2={j_0}$; it vanishes at the $k$ integers $2k,\ldots,3k-1$ and is
positive at all other integers.  The reflection $\mathcal R$ is that in
\eqref{eq:factorial-reflection}.  Since $\deg F=J-2$, it maps $F$ into
combinations of $(x)_{J-a}$ with $J-a\ge2$, so $\mathcal RF$ is
divisible by $(x)_2$; Lemma~\ref{lem:product-window-test} uses this
form.  In the ranges below $M\ge J$
and ${\tau}$ exceeds all roots of $W_k$. Indeed, every product window used
below has $n\ge4b(k+1)-1$, as the table and the choices of $k$ in the
proofs of Lemmas~\ref{lem:large-alphabet-complement}
and~\ref{lem:qutrit-complement} show; the definition gives
${\tau}\ge k-2+(n+1)/(2b)\ge3k$ and $M\ge J$. The paired factors give
$F(l)\ge0$ at every integer $0\le l\le {\tau}$.

\begin{lemma}\label{lem:product-window-test}
Assume the parameters in \eqref{eq:complement-profile-parameters} satisfy
$M\ge J$ and ${\tau}>3k-1$, and suppose
\[
 -\mathcal RF(l)=(l)_2\sum_{j=0}^{{j_0}+1}v_j(M-l)_j,
 \qquad v_j>0.
\]
If a real number $K_0>1$ satisfies
\begin{equation}\label{eq:product-window-budget}
 K_0{\tau}^2v_0>M^{{j_0}+1},
\end{equation}
then every code at this original length with distance at least
$D_0+1$ has $K<K_0$.  The weaker sufficient inequality
$K_0v_0>M^{{j_0}-1}$, which \eqref{eq:product-window-budget} implies
because ${\tau}<M$, may replace \eqref{eq:product-window-budget}.
\end{lemma}
\begin{proof}
At $l=0,1$ the reflection is zero and $F(l)>0$.
For $2\le l\le {\tau}$, $F(l)\ge0$ and $-\mathcal RF(l)>0$.
For $l>{\tau}$, $|F(l)|\le M^{{j_0}+1}$ and $(l)_2\ge {\tau}^2$, so
\eqref{eq:product-window-budget} gives
$F(l)-K_0\mathcal RF(l)>0$.
Alternatively, two positive-root factors of $W_k(l)$ are at most
$l(l-1)$, giving $|F(l)|/(l)_2\le M^{{j_0}-1}$.
Lemma~\ref{lem:factorial-reflection} applies.  A larger distance
only shortens the required tail $0\le l\le M-d$.
\end{proof}

For the windows below, the positivity of the $v_j$ and the threshold
inequality are verified in
Lemmas~\ref{lem:complement-pfive}--\ref{lem:qutrit-growing}.  If a
threshold $K_0$ satisfies Lemma~\ref{lem:product-window-test} and
$K_0<\operatorname{Budget}_Q(n)$, then every code of distance at least
$D+1$ with $D\ge D_0$ has $K<K_0<\operatorname{Budget}_Q(n)\le R_{n,D}$,
by Lemma~\ref{lem:grid-budget-coarse}.

\begin{lemma}\label{lem:large-alphabet-complement}
For $p\ge5$ and every $n\ge Q^2-2$, the parameters
\eqref{eq:complement-profile-parameters} admit a threshold
$K_0<\operatorname{Budget}_Q(n)$ satisfying Lemma~\ref{lem:product-window-test}.
For $p=4$ the same assertion holds for every $n\ge254$.
\end{lemma}
\begin{proof}
The complete coefficient estimates, threshold comparisons, and
overlap checks are in Lemmas~\ref{lem:complement-pfive} and
\ref{lem:complement-pfour}.  There $-\mathcal RF$ is computed by
adjoining the roots of $F$ one at a time: multiplication by $x-\zeta$
acts on the coefficients of the reflection by an explicit tridiagonal
map (Lemma~\ref{lem:reflection-production}), and bounds for the ratios
of consecutive coefficients that this map preserves keep all
coefficients positive (Lemma~\ref{lem:uniform-production-cone}); the
final factor ${\tau}-x$ gives the $v_j$.  The lemmas use the following
fixed windows, where the length intervals are for the original length
$n$:
\begin{center}\small
\begin{tabular}{ccl}\toprule
Local dimension&$k$&Length interval\\\midrule
$p\ge5$&2&$[Q^2-2,39Q)$\\
$p\ge5$&3&$[39Q,3Qb-2)$\\
$p\ge5$&4&$[64Q,4Qb+2)$\\
$p\ge5$&6&$[96Q,8Qb-2)$\\\midrule
$p=4$&2&$[254,544)$\\
$p=4$&3&$[544,768)$\\
$p=4$&4&$[768,1024)$\\
$p=4$&5&$[1024,1536)$\\
$p=4$&6&$[1536,1918)$\\\bottomrule
\end{tabular}
\end{center}
Consecutive intervals meet or overlap, since $3Qb-2\ge64Q$ and
$4Qb+2\ge96Q$ for $Q\ge25$; for $p\ge7$ the first interval is empty
and $Q^2-2$ lies in the second.  For the growing ranges write
$n'=2Qb\ell-1+2s$, $0\le s<Qb$, $\ell\ge4$.
Choose $k\coloneqq2\ell$ for $p\ge5$ and $k\coloneqq\lceil3\ell/2\rceil+1$
for $p=4$.  These cover $n\ge8Qb-2$.
All target comparisons use \eqref{eq:distance-target-budget}
at the original $n$.
\end{proof}

\begin{lemma}\label{lem:qutrit-complement}
For $p=3$, every parameter in \eqref{eq:strict-complement}
satisfies the conclusion of Theorem~\ref{thm:strict-complement}.
\end{lemma}
\begin{proof}
Put $d\coloneqq D+1$. For $n\ge1300$, take
$k\coloneqq\lceil(n'+1)/130\rceil$ in
\eqref{eq:complement-profile-parameters} and $K_0\coloneqq9^{2k+1}$.
Lemma~\ref{lem:qutrit-growing} proves all reflected coefficient
signs, the threshold inequality~\eqref{eq:product-window-budget},
and $K_0<\operatorname{Budget}_9(n)$.
For $n<1300$, put $M_0\coloneqq\lfloor n/2\rfloor+1$ and
$u_0\coloneqq M_0-9(M_0-d)$. Lemma~\ref{lem:physical-tail-cap} covers
$u_0>0$. The remaining parameters have
$u_0\le0$.  Lemma~\ref{lem:qutrit-bounded} covers exactly this finite
domain, including both original parities and every allowed distance,
with $578$ integer windows of degree five checked in exact integer
arithmetic.
\end{proof}

\begin{proof}[Proof of Theorem~\ref{thm:strict-complement}]
The qutrit case is Lemma~\ref{lem:qutrit-complement}.
For $p\ge4$, Lemma~\ref{lem:large-alphabet-complement} and
\eqref{eq:distance-target-budget} cover $n\ge Q^2-2$.
For shorter lengths use the parameters of Lemma~\ref{lem:physical-tail-cap}.
Since $u_0=QD-b\lfloor n/2\rfloor+1$,
\eqref{eq:strict-complement} gives, for both parities,
\[
 u_0>\frac{Q^2-n-2+b^2\delta}{2b}>0\qquad(n<Q^2-2).
\]
Lemma~\ref{lem:physical-tail-cap} gives $K\le R_{n,D}$.
This exhausts the complement of the root region.
\end{proof}

\section{Coefficient estimates in the root region}\label{sec:coefficient-methods}

Inside the root region, minimizing vertices with at most one residual
pair are settled by Proposition~\ref{prop:one-pair-minimum}. For
$m\ge2$, Proposition~\ref{prop:optimizer-roots} places every root of
the minimizer above $D$, so the coefficient
criterion compares each coefficient with a positive product of root
factors. This section develops the estimates for these comparisons;
Sections~\ref{sec:higher-dimensions} and~\ref{sec:qutrit} apply them.
Throughout this section put
\[
 b\coloneqq Q-1,\quad \lambda\coloneqq N/Q,\quad d_0\coloneqq2m,\quad D\coloneqq r+d_0,
 \quad {\widetilde h}(y)\coloneqq h(N-y),
\]
where $h$ is a monic product of $m$ adjacent integer pairs, its $d_0$
zeros are distinct and belong to $\{1,\ldots,N-1\}$, and $N\ge Qm$.
Unless stated otherwise, $h$ is any such product; the residual product
of a minimizing vertex is such a product, with $N\ge Qm+1$ by the
cap~\eqref{eq:optimizer-cap}.
Write ${a_{\min}}$ for its least zero. Except where a stronger quantitative
assumption is specified, we assume
\begin{equation}\label{eq:coefficient-root-domain}
 D\le\frac{(Q-2)(N-d_0+1)}Q<{a_{\min}}.
\end{equation}
We call its second inequality the root barrier.
These are precisely the root inequalities supplied by a minimizing
vertex in the root region: the first is~\eqref{eq:root-region}, since
$N-d_0=n-D$, and the second is~\eqref{eq:optimizer-root-sharp}. Define
\begin{equation}\label{eq:residual-generating}
 G(u)\coloneqq\frac{\E_{\Bin(N,(1+u)/Q)}{\widetilde h}(Y)}{\E_{\Bin(N,1/Q)}{\widetilde h}(Y)}
     =\sum_{\ell=0}^{d_0}G_\ell u^\ell,
 \qquad
 \mathcal L_i\coloneqq\frac{[u^i](1+u)^rG(u)}{\binom Ni}.
\end{equation}
The expectation in the numerator is a polynomial in $u$; no probability
interpretation outside $-1\le u\le Q-1$ is needed. If
$h=\sum b_\ell\Kraw{\ell}{N}$, orthogonality gives
$G_\ell=\binom N\ell b_\ell/b_0$. Indeed, for $Y\sim\Bin(N,(1+u)/Q)$
and $X\coloneqq N-Y$, \eqref{eq:kraw-generating} and the binomial theorem give
$\sum_\ell\E[\Kraw{\ell}{N}(X)]z^\ell=\E[(1+(Q-1)z)^{N-X}(1-z)^X]
=(1+uz)^N$, so $\E\Kraw{\ell}{N}(X)=\binom N\ell u^\ell$, and the
numerator of $G$ is $\sum_\ell b_\ell\binom N\ell u^\ell$. For $g(x)\coloneqq\binom{n-x}{r}h(x)$,
\eqref{eq:normalized-lift} and~\eqref{eq:normalized-value} then give
\[
 \frac{c_i}{c_0}=\frac{\binom Ni}{\binom ni}\mathcal L_i,\qquad
 \frac{g(i)}{g(0)}=\frac{\binom Ni}{\binom ni}\frac{h(i)}{h(0)}.
\]
Hence the criterion~\eqref{eq:coefficient-criterion} for $g$ follows
from the strict inequalities
\begin{equation}\label{eq:residual-comparison}
 \mathcal L_i<\frac{h(i)}{h(0)},\qquad 1\le i\le D,
\end{equation}
which the estimates below prove.

Two complementary representations of $G$ will be useful, for two
ranges of indices. On an initial range of indices $i$, the positive
mixture of Lemma~\ref{lem:positive-mixture} writes $G$ as an average
of products $(1+u)^j(1-au)^{N-j}$;
Lemma~\ref{lem:component-coefficients} controls the initial
coefficients of each product through a Jacobi-matrix test and a
recurrence, and Lemma~\ref{lem:mixture-laplace} averages over the
mixture. The later indices, including $D$, are controlled by the
quadratic-form representation of the second subsection:
Lemma~\ref{lem:weighted-rows} bounds the absolute coefficients
$|G_\ell|$, and Lemma~\ref{lem:positive-tail} converts such a bound
into~\eqref{eq:residual-comparison}. Sections~\ref{sec:higher-dimensions}
and~\ref{sec:qutrit} apply these tools on explicit index ranges. The
exact cutoff $G_\ell=0$ for $\ell>d_0$ is retained in both methods.

\subsection{A positive mixture of signed polynomials}

Put ${J_*}\coloneqq N-{a_{\min}}+1$ and $v\coloneqq{J_*}-d_0$. The root assumptions imply
\begin{equation}\label{eq:mixture-geometry}
 1\le v<2\lambda,\quad {J_*}<2\lambda+d_0\le4\lambda,
 \quad {J_*}\le N-r-d_0,\quad r+{J_*}\le N-d_0.
\end{equation}
Indeed ${J_*}-1$ is the largest reflected root, while there are $d_0$
distinct positive reflected roots. The strict upper bound follows by
substituting~\eqref{eq:coefficient-root-domain}.

\begin{lemma}\label{lem:positive-mixture}
There is a probability measure $\mu$ such that the polynomial identity
\begin{equation}\label{eq:positive-mixture}
 G(u)=\int(1+u)^j(1-au)^{N-j}\,d\mu(j,a),
 \qquad 0\le j\le {J_*},\quad 0\le a\le1/b
\end{equation}
holds. Its atoms with $j<{J_*}$ have $a=1/b$. Its remaining part has
$j={J_*}$, $a=(1-{\Pi})/(b+{\Pi})$, where, for the missing integers
$Z\coloneqq\{0,\ldots,{J_*}-1\}\setminus\{\text{zeros of }{\widetilde h}\}$,
\begin{equation}\label{eq:mixture-tail-density}
 {\Pi}\coloneqq\prod_{z\in Z}t_z,\qquad
 d\mu_{\rm tail}\ \propto\
 \prod_{z\in Z}t_z^{{J_*}-z-1}(b+{\Pi})^{N-{J_*}}\,dt_z.
\end{equation}
Here $|Z|=v$ and $0\in Z$.
\end{lemma}
\begin{proof}
For every integer $y\ge {J_*}$,
\[
 \frac{{\widetilde h}(y)}{(y)_{J_*}}=\prod_{z\in Z}\frac1{y-z}
 =\int_{[0,1]^Z}{\Pi}^{y-{J_*}}\prod_{z\in Z}t_z^{{J_*}-z-1}\,dt_z.
\]
Let $\nu$ be the image of the last positive measure under the product
map. Its exponents are nonnegative. Splitting a binomial expectation
below ${J_*}$ and using factorial size bias above ${J_*}$ gives, for $0<\rho<1$,
\begin{align*}
 \E_{\Bin(N,\rho)}{\widetilde h}
 &=\sum_{j=0}^{{J_*}-1}{\widetilde h}(j)\binom Nj\rho^j(1-\rho)^{N-j}\\
 &\quad +(N)_{J_*}\rho^{J_*}\int(1-\rho+\rho {\Pi})^{N-{J_*}}\,d\nu({\Pi}).
\end{align*}
All terms are nonnegative and the $j=0$ term is positive. Divide by
the value at $\rho=1/Q$, and substitute $\rho=(1+u)/Q$.
The normalized summands and integrands are the stated polynomials;
their baseline masses sum to one. Polynomial equality follows from
equality on an interval. This also proves the conditional density.
\end{proof}

The measure in~\eqref{eq:positive-mixture} is positive, although its
component polynomials can have negative coefficients. The next lemma
gives a sufficient condition for their initial coefficients to be
positive, together with an upper bound for them.

\begin{lemma}\label{lem:component-coefficients}
Let ${k_+},{k_-}$ be nonnegative integers, $0\le a<1$, ${n_{\rm mix}}\coloneqq{k_+}+{k_-}$, and
${F_{\rm mix}}(u)\coloneqq(1+u)^{k_+}(1-au)^{k_-}=\sum {q_i} u^i$. Suppose $I<\min({k_+},{k_-})$,
$I-1<{n_{\rm mix}}/2$, and, for $1\le i\le I$,
\begin{equation}\label{eq:component-jacobi-test}
 {k_+}-a{k_-}-(1-a)(i-1)>2\sqrt{a(i-1)({n_{\rm mix}}-i+2)}.
\end{equation}
Then ${q_i}>0$ for $0\le i\le I$ and
\begin{equation}\label{eq:component-suppression}
 \frac{{q_i}}{\binom {k_+}i}
 \le\exp\left(-\frac{a{k_-}i}{{k_+}-i+1}\right).
\end{equation}
\end{lemma}
\begin{proof}
For $a>0$, ${q_i}=a^i P_i^{({n_{\rm mix}},1+1/a)}({k_-})$, where
$P_i^{(n,R)}$ is the polynomial~\eqref{eq:kraw} with $Q$ replaced by
the real number $R\coloneqq1+1/a$; it is orthogonal for $\Bin(n,1/(1+a))$.
The degree-$i$ Jacobi matrix of this measure has diagonal
$[{n_{\rm mix}}-(1-a)j]/(1+a)$ and off-diagonal
$\sqrt{a(j+1)({n_{\rm mix}}-j)}/(1+a)$, $0\le j<i$.
The diagonal decreases and the off-diagonal increases in the range
under consideration. By Gershgorin's theorem, with the least diagonal
entry and twice the largest off-diagonal entry, its least eigenvalue
is at least
\[
 \frac{{n_{\rm mix}}-(1-a)(i-1)-2\sqrt{a(i-1)({n_{\rm mix}}-i+2)}}{1+a}.
\]
The zeros of $P_i^{({n_{\rm mix}},1+1/a)}$ are the eigenvalues of this
matrix, so condition~\eqref{eq:component-jacobi-test} places ${k_-}$
below every zero. Since the polynomial has positive value at zero,
${q_i}>0$.
For $a=0$ this is immediate.

Set ${\widehat q_i}\coloneqq{q_i}/\binom {k_+}i$. Differentiating ${F_{\rm mix}}$ gives
\begin{equation}\label{eq:component-recurrence}
 {\widehat q_{j+1}}=\left(1-\frac{a({k_-}-j)}{{k_+}-j}\right){\widehat q_j}
 -\frac{aj({n_{\rm mix}}-j+1)}{({k_+}-j)({k_+}-j+1)}{\widehat q_{j-1}}.
\end{equation}
Positivity and $j<{k_-}$ first imply ${\widehat q_j}\le {\widehat q_{j-1}}$. Substituting
this lower bound for ${\widehat q_{j-1}}$ in the negative term, and using
\[
 ({k_-}-j)({k_+}-j+1)+j({n_{\rm mix}}-j+1)={k_-}({k_+}+1),
\]
give ${\widehat q_{j+1}}\le[1-a{k_-}({k_+}+1)/(({k_+}-j)({k_+}-j+1))]{\widehat q_j}$.
The bracket is positive because ${\widehat q_{j+1}}>0$. Bound it by its
exponential and sum the telescoping reciprocal differences to obtain
\eqref{eq:component-suppression}.
\end{proof}

For real random variables $U,V$, write $U\succeq_{\rm st}V$ if
$\Pr(U>t)\ge\Pr(V>t)$ for every real $t$; for $\alpha,\beta,{\operatorname{rate}}>0$,
$\operatorname{Gamma}(\alpha,{\operatorname{rate}})$ and $\operatorname{Beta}(\alpha,\beta)$
denote the laws with densities proportional to $x^{\alpha-1}e^{-{\operatorname{rate}}x}$
on $(0,\infty)$ and $x^{\alpha-1}(1-x)^{\beta-1}$ on $(0,1)$,
respectively.

\begin{lemma}\label{lem:mixture-laplace}
In the conditional tail law~\eqref{eq:mixture-tail-density},
\[
 -\log {\Pi}\ \succeq_{\rm st}\ \operatorname{Gamma}(v,{J_*}+\lambda),
 \qquad
 1-{\Pi}\ \succeq_{\rm st}\ \operatorname{Beta}(v,d_0+\lambda+1),
\]
where the gamma parameter is its rate. Consequently, for $\kappa>0$,
\begin{equation}\label{eq:beta-laplace}
 \E_{\rm tail}e^{-\kappa Na}
 <\left(1+\frac{\kappa\lambda}{{J_*}+\lambda+1}\right)^{-v}.
\end{equation}
If~\eqref{eq:mixture-geometry} holds and $0\le\kappa\le1$, one also has
\begin{equation}\label{eq:gamma-laplace}
 \E_{\rm tail}e^{-\kappa Na}\le e^{-\kappa v/8}.
\end{equation}
\end{lemma}
\begin{proof}
Put $x_z\coloneqq-\log t_z$, $X\coloneqq\sum_zx_z$, and ${\operatorname{rate}}_z\coloneqq{J_*}-z$.
Let $T$ be the sum of independent exponentials of rates ${\operatorname{rate}}_z+\lambda$.
The law of $X$ is that of $T$ reweighted by $R(T)$, where
$R(t)\coloneqq e^{\lambda t}(b+e^{-t})^{N-{J_*}}$ satisfies
$(\log R)'(t)\ge\lambda-(N-{J_*})/Q={J_*}/Q>0$.
An increasing likelihood ratio increases every bounded increasing
expectation: for bounded increasing $g$ and an independent copy $T'$
of $T$, the two-copy covariance identity gives
$2\operatorname{Cov}(g(T),R(T))
=\E[(g(T)-g(T'))(R(T)-R(T'))]\ge0$.
Truncation covers an unbounded integrable $R$. Thus $X\succeq_{\rm st}T$.

Order the missing integers as $z_1<\cdots<z_v$, so $z_j\ge j-1$.
Coupling by common rate-one exponentials shows that $T$ dominates
the sum $S$ of independent exponentials with rates ${J_*}-j+1+\lambda$,
$1\le j\le v$, which in turn dominates
$\operatorname{Gamma}(v,{J_*}+\lambda)$.
For every integer $s\ge0$,
$\E e^{-sS}=\prod_{j=1}^v(d_0+\lambda+j)/(d_0+\lambda+j+s)$.
These are the moments of $\operatorname{Beta}(d_0+\lambda+1,v)$;
moments determine a law on $[0,1]$. Hence
$1-e^{-S}\sim\operatorname{Beta}(v,d_0+\lambda+1)$, proving both
stochastic comparisons.

For $Y\sim\operatorname{Beta}(v,d_0+\lambda+1)$ put
$f(t)\coloneqq\E e^{-tY}$ and $\mu(t)\coloneqq-d\log f/dt$.
Integration of the derivative of
$y^v(1-y)^{d_0+\lambda+1}e^{-ty}$ over $[0,1]$, whose boundary terms
vanish since $v>0$, gives
\[
 v=({J_*}+\lambda+1+t)\mu(t)-t\E_tY^2.
\]
Thus $\mu(t)>v/({J_*}+\lambda+1+t)$ for $t>0$.
Since $Na\ge\lambda(1-{\Pi})$, integration proves~\eqref{eq:beta-laplace}.
Finally~\eqref{eq:mixture-geometry} gives $\lambda>1/2$ and ${J_*}<4\lambda$,
so ${J_*}+\lambda+1<7\lambda$. Consequently, for $\kappa>0$,
\[
 \E_{\rm tail}e^{-\kappa Na}<(1+\kappa/7)^{-v}
 \le\exp\!\left(-\frac{\kappa v}{7+\kappa}\right).
\]
For $\kappa\le1$ this implies~\eqref{eq:gamma-laplace};
at $\kappa=0$ both sides equal one.
\end{proof}

\subsection{Finite quadratic forms and the later coefficients}

For $f(y)f(y+1)={\widetilde h}(y)$ use the coefficients of
$F_z(y)\coloneqq(1+z)^y(1-z/b)^{N+1-y}$ as a basis through degree $m$.
For $Y\sim\Bin(N,(1+u)/Q)$, the symmetric paired expectation
$\frac12\E[F_z(Y)F_w(Y+1)+F_w(Y)F_z(Y+1)]$ is the polynomial kernel
\begin{equation}\label{eq:finite-paired-kernel}
 {\operatorname{Seed}}(z,w)[{\operatorname{Base}}(z,w)+u{\operatorname{Tilt}}(z,w)]^N,
 \quad
 \begin{cases}
 {\operatorname{Seed}}\coloneqq1+\dfrac{b-1}{2b}(z+w)-\dfrac{zw}{b},\\[4pt]
 {\operatorname{Base}}\coloneqq1+\dfrac{zw}{b},\\
 {\operatorname{Tilt}}\coloneqq\dfrac{z+w}{b}+\dfrac{b-1}{b^2}zw.
 \end{cases}
\end{equation}
Normalize its coefficient matrices $M_\ell$ on $0,\ldots,m$ by
dividing the $z^iw^ju^\ell$ coefficient in~\eqref{eq:finite-paired-kernel}
by $\sqrt{\nu_i\nu_j}$, where $\nu_j\coloneqq\binom Nj b^{-j}$.
This diagonal congruence scales row and column $j$ by $\nu_j^{-1/2}$.
Thus, if $x$ is the coefficient vector of $f$ in the basis
$[z^j]F_z(y)/\sqrt{\nu_j}$, then
$G_\ell=x^{\mathsf T}M_\ell x/(x^{\mathsf T}M_0x)$.
The chosen basis need not be orthogonal for the length-$N$ measure.

\begin{lemma}\label{lem:weighted-rows}
Suppose $Q>2$, $N\ge Qm$, and $M_0\succeq {\operatorname{coer}} I>0$. Put
${\operatorname{pref}}\coloneqq2/{\operatorname{coer}}$, ${\omega_1}\coloneqq(b-1)/b$, and $\lambda_+\coloneqq N/b$. Then, for $u\ge0$,
\begin{align}
 \sum_{\ell=0}^{d_0}|G_\ell|u^\ell
 &\le {\operatorname{pref}}\max_{0\le j\le m}{\operatorname{Row}_j}(u),\label{eq:finite-row-bound}\\
 {\operatorname{Row}_j}(u)&\coloneqq\sum_{{j_1}=0}^j\binom j{j_1} u^{j-{j_1}}(1+{\omega_1}u)^{j_1}
           \sum_{{j_2}=0}^{m-{j_1}}\binom{N-j}{j_2}(u/b)^{j_2}.
 \label{eq:finite-row}
\end{align}
For every $a\ge0$ this implies both
\begin{align}
 \sum_\ell|G_\ell|a^{d_0-\ell}
 &\le {\operatorname{pref}}\frac{\lambda_+^m}{m!}(1+a)^{d_0},
 \label{eq:row-factorial-bound}\\
 \sum_\ell|G_\ell|a^{d_0-\ell}
 &\le {\operatorname{pref}}(1+a)^{d_0}\exp\!\left(\frac{\lambda_+}{1+a}\right).
 \label{eq:row-exponential-bound}
\end{align}
One may always take ${\operatorname{coer}}=2(b-1)/(b(m+2)^2)$.
\end{lemma}

The proof, including the retained finite cutoff ${j_1}+{j_2}\le m$, appears
in Appendix~\ref{app:coefficient-details}. The left side
of~\eqref{eq:finite-row-bound} is a sum of absolute coefficients, so it
bounds every $|G_\ell|$, not merely the value $G(u)$.

\begin{lemma}\label{lem:positive-tail}
Assume every root of $h$ exceeds $D$, let $d_0\le i\le D$, and put
$a\coloneqq(D-i)/(i-d_0+1)$, the backward weight. If
$W(a)\ge\sum|G_\ell|a^{d_0-\ell}$, then
\begin{equation}\label{eq:positive-tail-bound}
 \frac{\mathcal L_i}{h(i)/h(0)}\le W(a){\operatorname{TailFac}}(i,D),\qquad
 {\operatorname{TailFac}}(i,D)\coloneqq\frac{\binom{D-d_0}{i-d_0}\binom{D+d_0}{d_0}}
 {\binom Ni\binom{D+d_0-i}{d_0}}.
\end{equation}
If $\bar D\ge D$, $k\coloneqq N-\bar D-d_0\ge0$, and
$(D+1)(i-d_0)>d_0^2$, the positive scalar ${\operatorname{TailFac}}$ may be enlarged to
\begin{equation}\label{eq:tail-factorial-cancellation}
 {\operatorname{TailFac}}(i,\bar D)=\frac{(i)_{d_0}(N-i)_k}{(\bar D)_{d_0}(N)_k}
 \le(i/\bar D)^{d_0}(1-i/N)^k.
\end{equation}
In particular, with ${J_D}\coloneqq\bar D-d_0+1$, $z\coloneqq i-d_0+1$, and
$a_i\coloneqq(\bar D-i)/z$, the positive scalar
\begin{equation}\label{eq:combined-tail-scalar}
 F(i,\bar D)\coloneqq(1+a_i)^{d_0}e^{\lambda_+/(1+a_i)}{\operatorname{TailFac}}(i,\bar D)
\end{equation}
satisfies
$F(i+1,\bar D)/F(i,\bar D)\le
 \exp(\lambda_+/{J_D}-k/(N-i))$.
\end{lemma}
\begin{proof}
Consecutive backward binomial ratios give
$\binom r{i-\ell}/\binom r{i-d_0}\le a^{d_0-\ell}$,
including zero terms and $a=0$ at $i=D$. Ordered integer roots
above $D$ give
$h(i)/h(0)\ge\binom{D+d_0-i}{d_0}/\binom{D+d_0}{d_0}$.
These prove~\eqref{eq:positive-tail-bound}. The consecutive ratio
of ${\operatorname{TailFac}}$ in $D$ exceeds one precisely when
$(D+1)(i-d_0)>d_0^2$. Cancellation of consecutive factorial
blocks gives~\eqref{eq:tail-factorial-cancellation}; all its factors
are positive because $i\le D\le\bar D$ and $k\ge0$.
Finally $1+a_i={J_D}/z$, and, up to a constant independent of $i$,
\[
 F(i,\bar D)=\left[\prod_{j=0}^{d_0-1}(1+j/z)\right]
                    (N-i)_k e^{\lambda_+z/{J_D}}.
\]
The bracket decreases; the next falling-factorial ratio is
$1-k/(N-i)\le e^{-k/(N-i)}$. This proves the last assertion.
\end{proof}

The enlargement to $\bar D$ lets $D$ be replaced by a convenient upper
bound, such as $N-2m-k$ in Section~\ref{sec:higher-dimensions}. The
ratio bound shows where $F(i,\bar D)$ decreases in $i$; on such a tail
range its largest value is at the left end.

\subsection{Uniform coefficient reductions}

The following consequences reduce the number of alphabet-specific
estimates. Their proofs retain all finite offsets and are given in
Appendix~\ref{app:coefficient-details}.

\begin{proposition}\label{prop:terminal-cutoffs}
Under~\eqref{eq:coefficient-root-domain}, all comparisons
\eqref{eq:residual-comparison} hold in either of the regions
\[
 Q\ge16,\quad r\le N/2;
 \qquad Q\ge25,\quad r\le3N/5.
\]
The conclusion holds for every $m\ge1$ and every $h$ as above, not
only for the residual of a minimizer.
\end{proposition}

\begin{proposition}\label{prop:scale-cutoff}
Let $Q\ge9$, $m\ge2$, and let $h$ be the residual of a minimizing
vertex in the root region~\eqref{eq:coefficient-root-domain}.
Set $\xi\coloneqq\sqrt{Qm/N}$. If
\begin{equation}\label{eq:qutrit-scale-cutoff}
 \frac rN\le z(\xi)\coloneqq\frac79-\frac{2\xi}9-\frac{7\xi^2}{18},
\end{equation}
then~\eqref{eq:residual-comparison} holds for every $1\le i\le D$.
In particular $r\le N/6$ is covered. This proposition uses the first
inequality of~\eqref{eq:optimizer-root-sharp}, a sharper lower bound
for the roots of a minimizer, rather than only
\eqref{eq:coefficient-root-domain}.
\end{proposition}

\begin{proposition}\label{prop:original-top}
For $Q\ge9$, every minimizing vertex in the root region satisfies
the strict form of~\eqref{eq:coefficient-criterion} at its top index
$i=D$.
\end{proposition}

Theorem~\ref{thm:qutrit-large-length} uses this proposition for the
index $i=D$, which the estimates of
Proposition~\ref{prop:qutrit-four-ranges} do not cover.

\begin{proposition}\label{prop:large-deficit}
Let $Q\ge16$, $N\ge Qm+1$, and suppose $h$ has distinct paired integer
roots as in the standing assumptions of this section and
${a_{\min}}>(Q-2)(N-2m+1)/Q$. If
\[
 s\coloneqq N-r\ge\lceil20\lambda\rceil+6m
\]
(which implies the first inequality of~\eqref{eq:coefficient-root-domain}),
then~\eqref{eq:residual-comparison} holds for every $1\le i\le D$.
\end{proposition}

For local dimensions $p\ge6$, Proposition~\ref{prop:large-deficit}
leaves a bounded ratio $s/\lambda$, which permits explicit initial
Jacobi ranges and weighted tail bounds. At local dimensions five and
four the distance strips supply, through
Lemma~\ref{lem:physical-tail-cap}, a lower bound on
$N-r-4m=n-2D$. At local dimension three the scale-sensitive cutoff
and the top index leave an interior interval, treated in
Section~\ref{sec:qutrit}.

\section{Local dimensions at least four}\label{sec:higher-dimensions}

We now combine the coefficient estimates at a fixed original length
and degree. For $p\ge6$ the estimates need only the root
inequalities~\eqref{eq:coefficient-root-domain} and $N\ge Qm+1$.
At local dimensions five and four we also use the lower bound on
$N-r-4m=n-2D$ that Lemma~\ref{lem:physical-tail-cap} supplies once
the distance strips have been removed. Local dimension four has a finite
remaining interval; positive quadratures reduce it to 223 minimizing
polynomials. The table in Section~\ref{sec:completion} summarizes the
case division.

\begin{theorem}\label{thm:higher-weak}
Let $p\ge4$, $Q=p^2$, and let an exact code of dimension $K>1$
have distance at least $D+1$, where $D\ge2$. If $(n,D)$ satisfies
the root-region condition
\[
 2(Q-1)D\le(Q-2)(n+1),
\]
then $K\le p^n\beta_{n,D,Q}$.
\end{theorem}

Throughout the proof choose a minimizing vertex, with residual
parameters $m,N,r$ as in~\eqref{eq:vertex-factorization}, and put
$d_0\coloneqq2m$, $s\coloneqq N-r$, $\lambda\coloneqq N/Q$.
The cases $m=0,1$ are supplied by Singleton and
Proposition~\ref{prop:one-pair-minimum}. For $m\ge2$ the cap is
$N\ge Qm+1$, and the root-region condition and
Proposition~\ref{prop:optimizer-roots} give
\eqref{eq:coefficient-root-domain}.

\subsection{Local dimensions at least six}

\begin{proposition}\label{prop:large-alphabet-coefficients}
Let $Q\ge36$ be real, $m\ge2$, $N\ge Qm$, $r\ge0$, and
$s\le\lceil20\lambda\rceil+6m$. Every paired polynomial with distinct
integer roots satisfying~\eqref{eq:coefficient-root-domain}
satisfies~\eqref{eq:residual-comparison} for all $1\le i\le D$.
\end{proposition}

The proof is in
Appendices~\ref{app:higher-initial}--\ref{app:higher-tails-36}.
It treats $Q\ge49$ and $36\le Q<49$ separately, because one tail
estimate for $36\le Q<49$ uses the upper bound on $Q$ to bound the
parameter $a$ of Lemma~\ref{lem:positive-tail} by $12/13$. In each interval
the ratio $s/\lambda$ fixes an initial range $1\le i\le\theta N$ on
which every component of the mixture~\eqref{eq:positive-mixture}
satisfies the hypotheses of Lemma~\ref{lem:component-coefficients};
there that lemma and Lemma~\ref{lem:mixture-laplace} bound
$\mathcal L_i$. Every later index, including $D$, is controlled
by Lemma~\ref{lem:positive-tail} and the row bounds of
Lemma~\ref{lem:weighted-rows}. For $m=2$ some of these tail estimates
use instead exact certificates for $3\times3$ matrices
(Appendix~\ref{app:higher-certificates}): one at $a=1$ for
$Q\ge49$, and one for every $a\ge0$, valid for all $Q\ge25$
and used also at local dimension five.

For $p\ge6$, Proposition~\ref{prop:large-deficit} handles
$s\ge\lceil20\lambda\rceil+6m$, and the preceding proposition
handles the complementary interval. The coefficient criterion applied
to this minimizing vertex proves Theorem~\ref{thm:higher-weak}.

\subsection{Local dimension five}

\begin{proposition}\label{prop:five-cap-coefficients}
Let $Q=25$, $m\ge2$, $N\ge25m+1$, and let $h$ satisfy
\eqref{eq:coefficient-root-domain}. Suppose
\begin{equation}\label{eq:five-physical-cap}
 r>3N/5,\qquad N-r-4m>(N-2m)/13.
\end{equation}
Then~\eqref{eq:residual-comparison} holds for every $1\le i\le D$.
\end{proposition}

The second condition is used only in the tail estimate for
$s\le6\lambda$. There it gives $D\le N-2m-k$ for
$k\coloneqq\lfloor(N-2m)/13\rfloor+1$, so that Lemma~\ref{lem:positive-tail}
applies with $\bar D\coloneqq N-2m-k$ and this $k>(N-2m)/13$. For actual codes
it is the conclusion of Lemma~\ref{lem:physical-tail-cap} at $Q=25$,
after the critical and adjacent distance strips have been removed.
The proposition is proved in Appendix~\ref{app:higher-five}, with the
initial ranges of Appendix~\ref{app:higher-initial} and the
certificates of Appendix~\ref{app:higher-certificates}: after three
initial ranges, the tail uses exact matrix certificates for
$2\le m\le5$, finite rational evaluations for $6\le m\le10$, and one
bound decreasing in $m$ for $m\ge11$.

At $p=5$, Theorems~\ref{thm:critical-grid} and~\ref{thm:adjacent-grid}
already give the target when $u_0>0$ in
Lemma~\ref{lem:physical-tail-cap} (here $L_0\ge1$, since $n\ge N\ge2Q+1$). Otherwise that lemma gives the
second condition of~\eqref{eq:five-physical-cap}. If $r\le3N/5$, use
Proposition~\ref{prop:terminal-cutoffs}; otherwise use
Proposition~\ref{prop:five-cap-coefficients}. This proves
Theorem~\ref{thm:higher-weak} at local dimension five.

\subsection{Local dimension four at unbounded length}

\begin{proposition}\label{prop:four-initial}
Let $Q=16$, $m\ge2$, $N\ge\max(16m,256)$, $r\ge N/2$, and
let $h$ have distinct paired integer roots satisfying
\eqref{eq:coefficient-root-domain}. Then
\eqref{eq:residual-comparison} holds for $1\le i\le\lfloor9N/20\rfloor$.
\end{proposition}

\begin{proposition}\label{prop:four-tail}
Let $Q=16$, $m\ge2$, $N\ge\max(16m,256)$, $r\ge0$, and
$D\coloneqq r+2m<N$. Suppose all $2m$ distinct paired roots exceed $D$ and
\begin{equation}\label{eq:four-physical-cap}
 N-r-4m>2(N-2m)/17.
\end{equation}
Then~\eqref{eq:residual-comparison} holds for
$\lceil9N/20\rceil\le i\le D$.
\end{proposition}

Both proofs, with their finite constants, are in
Appendices~\ref{app:higher-four-initial}
and~\ref{app:higher-four-tail}. The initial estimate bounds each
component coefficient of the mixture~\eqref{eq:positive-mixture}
separately, allowing it to be negative. Where the Jacobi
test~\eqref{eq:component-jacobi-test} is verified,
Lemma~\ref{lem:component-coefficients} gives the bound; otherwise the
coefficient is bounded by a Cauchy integral over a circle about the
origin. The loss in this bound, of order $\sqrt N$, is compensated by
the decay of the integrand away from the positive real axis.
Lemma~\ref{lem:mixture-laplace} then averages over the mixture.
The tail estimate applies Lemma~\ref{lem:positive-tail} with
$\bar D\coloneqq N-2m-k$ and $k\coloneqq\lfloor2(N-2m)/17\rfloor+1$; then $\bar D\ge D$
by~\eqref{eq:four-physical-cap}. It uses both the upper and the lower
bound on $\bar D$ given by this floor, and its final bound is below one
for every $N\ge256$, uniformly in $m$.

When $u_0>0$ in Lemma~\ref{lem:physical-tail-cap}, the distance strips
give the target; otherwise that lemma
supplies~\eqref{eq:four-physical-cap}. For every $N$, the case
$r\le N/2$ is covered by Proposition~\ref{prop:terminal-cutoffs}. For
$r>N/2$ and $N\ge256$, the two preceding propositions cover every
integer index, because their floor and ceiling endpoints overlap or
are consecutive. The case $r>N/2$ and $N\le255$ is treated next.

\subsection{The finite interval at local dimension four}

\begin{proposition}\label{prop:four-finite}
The conclusion of Theorem~\ref{thm:higher-weak} at $p=4$ also holds
if a selected minimizing vertex has $m\ge2$, $N\le255$, and $r>N/2$.
\end{proposition}
\begin{proof}
From the selected vertex we use only that $H\coloneqq n-D=N-2m$ and $D$
satisfy the inequalities below. Put $a\coloneqq\lfloor(H-1)/14\rfloor$, the
index of Lemma~\ref{lem:cap-minimum-identity} at $Q=16$. The
cap~\eqref{eq:optimizer-cap}, $m\ge2$, $N\le255$ and $r>N/2$ give
\[
 29\le H\le251,\qquad D>H/2>2a.
\]
Lemma~\ref{lem:cap-minimum-identity} therefore gives
\[
 \beta_{H+D,D,16}=16^{-(D-2a)}\beta_{H+2a,2a,16}.
\]

For $Q=16$ the exact certificate~\cite{ComputationalCertificates}
provides one cap polynomial for each of the 223 integers
$H=29,\ldots,251$. Its cardinal weights in
Lemma~\ref{lem:positive-quadrature} are positive, and all 4445
quadrature identities for the binomial moments through the respective
degree $2a$ hold exactly; by the converse in
Lemma~\ref{lem:positive-quadrature} it is a cap minimizer.
By Lemmas~\ref{lem:cap-minimum-identity}
and~\ref{lem:grid-vertex}, the terminal lift of this certified cap
minimizer is a global minimizer at the original $(n,D)$; it has $D$
distinct zeros in $\{1,\ldots,n\}$, so it is a minimizing vertex, and
the results stated for a minimizing vertex apply to it. We use it in
place of the selected vertex. To write this lift in the
form~\eqref{eq:vertex-factorization}, merge the terminal factor of
degree $D-2a$ with the maximal terminal root block of the cap
minimizer; the remaining roots give its own residual profile $(m,N,r)$,
which may differ from that of the selected vertex; below, $(m,N,r)$
denotes this profile.
There is one one-pair profile,
at $H=29$, covered by Proposition~\ref{prop:one-pair-minimum}, and
26 profiles with $N\ge256$, at $H=226,\ldots,251$, covered by the
preceding argument for $p=4$ at the unchanged original $n,D$.

For the other 196 profiles, $H=30,\ldots,225$, $m\ge2$ and
$N\le255$. Proposition~\ref{prop:terminal-cutoffs} covers $r\le N/2$.
The remaining integer interval is exactly
\begin{equation}\label{eq:four-finite-lifts}
 \lfloor N/2\rfloor+1\le r
 \le\left\lfloor\frac{7(H+1)}8\right\rfloor-2m.
\end{equation}
Of these lifts, 113 have $u_0>0$ in Lemma~\ref{lem:physical-tail-cap}
and are covered by the adjacent strip,
Theorem~\ref{thm:adjacent-grid}. For the remaining 4578 lifts, the
certificate verifies in exact integer arithmetic all 545042 strict
comparisons~\eqref{eq:residual-comparison}, $1\le i\le D$.
Appendix~\ref{app:four-finite-certificate} specifies the exact sums,
domain generation, positivity tests, and terminal identity. The
coefficient criterion proves the target for this particular global
minimizer, and hence for every admissible polynomial at the same $n,D$.
\end{proof}

This completes the proof of Theorem~\ref{thm:higher-weak} for every
$p\ge4$. Together with Theorem~\ref{thm:strict-complement}, it proves
Theorem~\ref{thm:grid} for $p\ge4$.

\section{Local dimension three}\label{sec:qutrit}

For $Q=9$, the cap leaves less room in the coefficient
estimates. We use an additional consequence of global minimality:
the value $h(0)$ of the residual polynomial, which is the product of
its roots, cannot be too small. This improves the lower bound for the
values $h(i)/h(0)$ with which the coefficients are compared.
The estimates then cover every residual length $N\ge512$.
For the remaining lengths, positive quadrature and integer coefficient
comparisons give a finite certificate for each possible value of $n-D$.

Throughout this section a minimizing vertex has parameters
$D=2m+r$, $n=N+r$, $m\ge2$, and $N\ge9m+1$, and $(n,D)$ lies in the
root region~\eqref{eq:root-region}. Since $n+1=N-2m+1+D$, at $Q=9$
this condition reads $9D\le7(N-2m+1)$. With ${a_{\min}}$ the least
residual root, Proposition~\ref{prop:optimizer-roots} therefore gives
\begin{equation}\label{eq:qutrit-wedge}
 D\le\frac{7(N-2m+1)}9<{a_{\min}}.
\end{equation}
For its monic residual polynomial $h$, set
\begin{equation}\label{eq:qutrit-generating}
 {\widetilde h}(y)\coloneqq h(N-y),\qquad
 G(u)\coloneqq\frac{\E_{\Bin(N,(1+u)/9)}{\widetilde h}(Y)}
             {\E_{\Bin(N,1/9)}{\widetilde h}(Y)},\qquad
 \mathcal L_i\coloneqq\frac{[u^i](1+u)^rG(u)}{\binom Ni}.
\end{equation}
These are the quantities~\eqref{eq:residual-generating} at $Q=9$; the
expectation with parameter $(1+u)/9$ is interpreted as
a polynomial in $u$. By Lemma~\ref{lem:binomial-lift}, the coefficient
criterion for the lifted polynomial is $\mathcal L_i\le h(i)/h(0)$,
$1\le i\le D$; the estimates below prove the strict
comparison~\eqref{eq:residual-comparison}.

\subsection{Minimality constrains the root product}

\begin{lemma}\label{lem:optimizer-product}
Let $Q=p^2\ge9$, and let a minimizing vertex have residual
degree $2m>0$, length $N\ge Qm+1$, and monic reflected
polynomial ${\widetilde h}(y)\coloneqq h(N-y)$. Put $b\coloneqq Q-1$ and
\[
 {\operatorname{coer}_m}\coloneqq\frac{(b-1)(m+2)}{2b(m+1)},\qquad
 \eta_m\coloneqq m!(N)_m\left(\frac b{Q^2}\right)^m.
\]
If $z_1,\ldots,z_{2m}$ are the roots of ${\widetilde h}$, then
\begin{align}
 {\widetilde h}(N)&\ge {\operatorname{coer}_m}\eta_m V_m(N)
      \ge {\operatorname{coer}_m}(N)_m^2(b/Q)^{2m},\label{eq:optimizer-product}\\
 \sum_{j=1}^{2m}-\log(1-z_j/N)&\le {\operatorname{LogBud}},\label{eq:optimizer-log-budget}\\
 {\operatorname{LogBud}}&\coloneqq\log(1/{\operatorname{coer}_m})-2\sum_{j=0}^{m-1}\log(1-j/N)-2m\log(b/Q).
       \label{eq:product-budget}
\end{align}
In particular, if $\max z_j/N\le\gamma<1$ and
$0<w\coloneqq i/N<1-\gamma$, then
\begin{equation}\label{eq:product-endpoint}
 \frac{h(i)}{h(0)}\ge(1-w)^{2m}
       \exp\{-{\operatorname{LogBud}}\kappa_\gamma(w)\},\qquad
 \kappa_\gamma(w)\coloneqq\frac w2
             \left(\frac1{1-w}+\frac1{1-\gamma-w}\right).
\end{equation}
\end{lemma}
\begin{proof}
Write ${\widetilde h}(y)=q(y)q(y+1)$ with $q$ monic of degree $m$.
The lower bound for monic paired expectations in
Lemma~\ref{lem:monic-paired-energy}
(Appendix~\ref{app:coefficient-details}) gives
$\E_{\Bin(N,1/Q)}{\widetilde h}\ge {\operatorname{coer}_m}\eta_m$.
As a competitor take the rounded Lloyd polynomial $h_L$ of
Lemma~\ref{lem:radau-rounded}, at the same length $N$, times the same
terminal factor $\binom{n-x}{r}$. This polynomial of degree $D$ is
nonnegative on the grid and positive at zero, and
Lemma~\ref{lem:radau-rounded} applies because $2m\le N$.
By~\eqref{eq:terminal-expectation}, both objectives are $Q^{-r}$
times the corresponding ratios $\E_{\mu_N}h/h(0)$, and reflection
gives $\E_{\mu_N}h=\E_{\Bin(N,1/Q)}{\widetilde h}$ and
$h(0)={\widetilde h}(N)$. Minimality and Lemma~\ref{lem:radau-rounded}
therefore imply
\[
 \frac{\E_{\Bin(N,1/Q)}{\widetilde h}}{{\widetilde h}(N)}\le\frac1{V_m(N)}.
\]
The first bound in~\eqref{eq:optimizer-product} follows.
Its second bound uses $V_m(N)\ge\binom Nm b^m$.
Divide by $N^{2m}$ and take logarithms to obtain
\eqref{eq:optimizer-log-budget}.

Let ${\ell_j}\coloneqq-\log(1-z_j/N)$ and ${\ell_\gamma}\coloneqq-\log(1-\gamma)$.
Then $0<{\ell_j}\le {\ell_\gamma}$ and $\sum {\ell_j}\le {\operatorname{LogBud}}$.
The concave function $f(t)\coloneqq\log(1-we^t)$ lies above its
endpoint chord on $[0,{\ell_\gamma}]$. Hence
\[
 \log\frac{{\widetilde h}(N-i)}{{\widetilde h}(N)}
 \ge2m\log(1-w)-\frac {\operatorname{LogBud}} {\ell_\gamma}
        \log\frac{(1-w)(1-\gamma)}{1-\gamma-w}.
\]
The function $-f'(t)=we^t/(1-we^t)$ is convex.
Its integral average is at most its endpoint trapezoid average,
which is $\kappa_\gamma(w)$. This proves
\eqref{eq:product-endpoint}. The restriction $w<1-\gamma$
makes every logarithm and denominator positive where required.
\end{proof}

For the qutrit estimates, write $x\coloneqq2m/N$ and
\begin{equation}\label{eq:qutrit-product-functions}
 {\operatorname{LogBud}}_9(x)\coloneqq x+(2-x)\log(1-x/2)-x\log(8/9),\quad
 \kappa(w)\coloneqq\frac w2\left(\frac1{1-w}+\frac1{49/81-w}\right).
\end{equation}
At $Q=9$, ${\operatorname{coer}_m}=7(m+2)/(16(m+1))>7/16$. With this
bound and the left Riemann sum for the increasing function
$-\log(1-t)$, \eqref{eq:product-budget} is at most
$\log(16/7)+N{\operatorname{LogBud}}_9(x)$.
By~\eqref{eq:qutrit-wedge}, $\max z_j/N\le\gamma\coloneqq(2+7x)/9\le32/81$.
Since ${\operatorname{LogBud}}_9$ is convex and vanishes at zero,
\begin{equation}\label{eq:qutrit-product-simplified}
 \frac1N\log\frac{h(i)}{h(0)}
 \ge x\log(1-w)
   -\left(\frac{9x}{2}{\operatorname{LogBud}}_9(2/9)+\frac{\log(16/7)}N\right)
       \kappa(w),\qquad 0<w<49/81.
\end{equation}
This lower bound is specific to a global minimizer; it is used in
the proofs of all four rows of
Proposition~\ref{prop:qutrit-four-ranges}.

\subsection{All coefficients at residual length at least 512}

The next proposition groups the four estimates used at this length.
Rows 1 and 2 of its table use the positive
mixture~\eqref{eq:positive-mixture}: at the first indices its
components have positive coefficients, and at later indices each
component coefficient is bounded by a Cauchy integral over a circle
before the average over the mixture is taken. Rows 3 and 4 bound the
coefficients of $(1+u)^rG(u)$ through the finite quadratic
forms~\eqref{eq:finite-paired-kernel} and a nonnegative matrix
majorant (Lemma~\ref{lem:schur-coefficient}). The proofs, including
the exact scalar certificates, are in
Appendix~\ref{app:qutrit-estimates}; its first three sections,
Appendices~\ref{app:qutrit-one-coefficient}--\ref{app:qutrit-initial},
contain the estimates shared by the rows.

\begin{proposition}\label{prop:qutrit-four-ranges}
Let a minimizing vertex with $m\ge2$ satisfy
\eqref{eq:qutrit-wedge} and $N\ge512$.
The strict inequality $\mathcal L_i<h(i)/h(0)$ holds on
each of the following ranges:
\begin{center}
\begin{tabular}{lll}\toprule
Terminal degree&Indices&Proof\\\midrule
$r\ge N/3$&$1\le i\le\lfloor N/4\rfloor$
 &\ref{app:qutrit-quarter}\\
$N/6<r<N/3$&$1\le i\le\lfloor N/8\rfloor$
 &\ref{app:qutrit-eighth}\\
$r\le N/3$&$\lceil N/8\rceil\le i\le\min(\lfloor N/4\rfloor,D-1)$
 &\ref{app:qutrit-middle}\\
all $r$&$\lceil N/4\rceil\le i\le D-1$
 &\ref{app:qutrit-tail}\\\bottomrule
\end{tabular}
\end{center}
An empty index interval imposes no condition.
The last two rows hold also with $|\mathcal L_i|$ on the left.
\end{proposition}

The top index $i=D$ is supplied by
Proposition~\ref{prop:original-top}.

\begin{theorem}\label{thm:qutrit-large-length}
If a minimizing vertex in~\eqref{eq:qutrit-wedge} has
$m\ge2$ and $N\ge512$, it satisfies every coefficient
comparison~\eqref{eq:coefficient-criterion}.
\end{theorem}
\begin{proof}
If $r\le N/6$, Proposition~\ref{prop:scale-cutoff} applies:
for $\xi=\sqrt{9m/N}<1$, its allowance
\[
 z(\xi)=\frac79-\frac{2\xi}9-\frac{7\xi^2}{18}
\]
is at least $1/6$. If $r>N/6$, use
Proposition~\ref{prop:qutrit-four-ranges}.
For $r\ge N/3$, its first and fourth rows cover all
indices $1,\ldots,D-1$.
For $N/6<r<N/3$, use the second, third and fourth rows.
At each junction, the floor and ceiling coincide or are
consecutive integers. Proposition~\ref{prop:original-top}
supplies the remaining top index $i=D$.
Finally multiply each residual comparison by
$\binom Ni/\binom ni>0$.
\end{proof}

\subsection{Reduction of the finite remainder}

The finite calculation certifies an optimal polynomial, rather
than searching all rounded root patterns at every original degree.
Lemma~\ref{lem:cap-minimum-identity} explains why one cap problem
for each $H=n-D$ suffices.

\begin{proposition}\label{prop:qutrit-finite}
For $15\le H\le507$ and
\begin{equation}\label{eq:qutrit-finite-domain}
 2\left\lfloor\frac{H-1}7\right\rfloor
       \le D\le\left\lfloor\frac{7(H+1)}9\right\rfloor,
 \qquad n=H+D,
\end{equation}
every qutrit code of rank $K>1$ and distance at least $D+1$
satisfies $K\le3^n\beta_{n,D,9}$.
\end{proposition}
\begin{proof}
The exact certificate in Appendix~\ref{app:qutrit-finite}
specifies an optimal cap polynomial for every $H$ in the
stated interval; its nonnegative cardinal quadrature certifies
optimality by the converse in Lemma~\ref{lem:positive-quadrature}.
Lemma~\ref{lem:cap-minimum-identity} supplies an optimal
polynomial for every original degree in
\eqref{eq:qutrit-finite-domain}.
For $H=16,\ldots,507$, all coefficients of every such
terminal lift satisfy the strict residual
comparison~\eqref{eq:residual-comparison}: the certificate checks
each comparison as the positivity of an exact integer, for all 64526
lifts and 11543154 indices, without removing the distance strips.
Lemma~\ref{lem:coefficient-criterion}
gives the bound.
For $H=15$, the cap roots are $15,16,18,19$ at length
$19$. Merging the terminal pair leaves $N=17$ and
$h=(x-15)(x-16)$; this lift is a minimizing vertex with one residual
pair. Its nine original degrees are
$D=4,\ldots,12$. Proposition~\ref{prop:one-pair-minimum}
therefore gives the same optimal target.
\end{proof}

\begin{theorem}\label{thm:qutrit-weak}
For $p=3$, the grid bound of Theorem~\ref{thm:grid}
holds throughout the root region $16D\le7(n+1)$.
\end{theorem}
\begin{proof}
Choose a minimizing vertex. With at most one residual pair,
apply Proposition~\ref{prop:one-pair-minimum}.
For $m\ge2$, Proposition~\ref{prop:scale-cutoff} (the scale cutoff,
$r/N\le z(\xi)$) or Theorem~\ref{thm:qutrit-large-length} handles every
case
except $N\le511$ and $r/N>z(\xi)$, where
$\xi=\sqrt{9m/N}<1$.
Put $H\coloneqq N-2m$ and $a\coloneqq\lfloor(H-1)/7\rfloor$.
The cap gives $15\le H\le507$.
Moreover, since $r/N>z(\xi)$, $2m/N=2\xi^2/9$ and
$H/N=1-2\xi^2/9$,
\[
 \frac DN-\frac{2H}{7N}
 >\frac{31}{63}-\frac{2\xi}9-\frac{13\xi^2}{126}
 \ge\frac16.
\]
The final quadratic decreases on $[0,1]$ and equals
$1/6$ at one. Thus $D>2H/7+N/6>2a$.
The root-region inequality gives $9D\le7(H+1)$, so the original
$(n,D)$ lies in~\eqref{eq:qutrit-finite-domain}.
Proposition~\ref{prop:qutrit-finite} completes the proof.
Every normalized polynomial nonnegative on the grid has expectation
at least $\beta_{n,D,9}$, so the assertion follows for all of them.
\end{proof}

Together with Theorem~\ref{thm:strict-complement}, this proves
Theorem~\ref{thm:grid} for $p=3$.

\section{Binary codes}\label{sec:binary}

We prove for $p=2$ both the coefficient criterion for each prescribed
rounded polynomial and the general grid-minimum bound. The grid-minimum bound,
Proposition~\ref{prop:bg-grid-minimum} in Section~\ref{sec:bg-minimum},
is the binary case of Theorem~\ref{thm:grid}, and therefore also gives
Theorem~\ref{thm:rounded} for $p=2$. Theorem~\ref{thm:binary-rounded}
below is an additional statement: each prescribed rounded polynomial
itself satisfies the coefficient criterion, by an argument that uses
none of the computations of
Appendices~\ref{app:bg-minima}--\ref{app:bg-finite}, so it proves
Theorem~\ref{thm:rounded} for $p=2$ without them. Its proof treats
the prescribed rounded Lloyd roots. The proof in
Section~\ref{sec:bg-minimum} treats a minimizing polynomial and
compares it with a rounded competitor of the same degree. Both use
Rains's length bound for binary codes, stated below, to place the
relevant roots beyond the coefficient indices being tested. For the
grid-minimum route to Theorem~\ref{thm:rounded}, a reader needs from
Section~\ref{sec:binary-prescribed} only this length bound,
Lemma~\ref{lem:binary-root-distance}, the ratios
of~\eqref{eq:binary-endpoint} and the matrices $\operatorname{Mat}_i$
and $\operatorname{Gram}_0$; Proposition~\ref{prop:binary-index-cover}
and the proof of Theorem~\ref{thm:binary-rounded} may be skipped.

\subsection{The prescribed rounded polynomial}\label{sec:binary-prescribed}

In this subsection $m=t-e$, $r=2e+\sigma$, $N=n-r$ and the floors
$a_j$ are those of the rounded
construction~\eqref{eq:original-parameters}--\eqref{eq:rounded-polynomial},
with $Q=4$.

\begin{theorem}\label{thm:binary-rounded}
Let an exact binary quantum code have dimension $K>1$ and distance
at least $2t+1+\sigma$, where $t\ge1$ and $\sigma\in\{0,1\}$.
For every $0\le e\le t$, the polynomial
in~\eqref{eq:rounded-polynomial}, with $Q=4$, satisfies
\eqref{eq:coefficient-criterion}. Consequently
$KS_{t,\sigma}^{n,e}\le2^n$.
\end{theorem}

Rains's length bound for binary codes~\cite[Theorem~15]{Rains},
obtained from shadow enumerators, states that,
if $n=6{\ell_R}-1+\ell$, $0\le\ell\le5$ and $K>1$, then
\[
 d\le2{\ell_R}+1\quad(\ell<5),\qquad d\le2{\ell_R}+2\quad(\ell=5).
\]
The bound is nondecreasing with $n$. At $n=6t-2$ it is $2t$,
and at $n=6t+3$ it is $2t+1$. Thus the parameters of our code obey
\begin{equation}\label{eq:binary-length}
 n\ge6t-1+5\sigma,
 \qquad N\ge6m+2r-1.
\end{equation}
Here Singleton gives $n\ge2D+1\ge5$, so every relevant length has the
above form with $\ell_R\ge1$.
Rains's theorem applies to every binary quantum code with $K>1$, with
no purity or additivity assumption.

\begin{lemma}\label{lem:binary-root-distance}
Let $m\ge1$, $r\ge0$, $D\coloneqq2m+r$ and $N\ge6m+2r-1$.
Every zero $\xi$ of $P_m^{(N-1,4)}(x-1)$ satisfies
$\xi>D+1$. In particular its floor is at least $D+1$.
\end{lemma}
\begin{proof}
Reflect by $y\coloneqq N-x$. By~\eqref{eq:orthogonality}, the reflected
polynomial is the orthogonal polynomial of degree $m$ for
$\Bin(N-1,1/4)$, so its zeros are the eigenvalues of the Jacobi matrix
of this law: the symmetric tridiagonal matrix with diagonal
$(N-1)/4+j/2$ and off-diagonal entries
$\sqrt{3j(N-j)}/4$, for $0\le j<m$ and $1\le j<m$,
respectively. These entries follow by coefficient comparison in the
Krawtchouk generating function. Bounding the off-diagonal row sums
by twice their largest entry bounds the largest reflected zero by
\[
 \frac{N-1}{4}+\frac{m-1}{2}
       +\frac{\sqrt{3(m-1)(N-m+1)}}2.
\]
Write $N=2D+2m-1+z$, $z\ge0$. The positive difference between
$N-D-1$ and the first two terms is
$(D+2m-2)/2+3z/4$. Its square minus
$3(m-1)(N-m+1)/4$ is
\[
 \frac{(D-m+1)^2-3(m-1)}4
 +\frac{3(D+m-1)z}4+\frac{9z^2}{16}>0.
\]
Indeed $D\ge2m$ makes the first numerator at least
$m^2-m+4$. The reflected zero is therefore less than $N-D-1$.
\end{proof}

For the rest of this section, for given $m$ and $N$, define $L$ and
the Lloyd ratios $U_i$ by
\begin{equation}\label{eq:binary-endpoint}
 L(x)\coloneqq\frac{m!}{4^m}P_m^{(N-1,4)}(x-1)
      =\prod_{j=1}^m(\xi_j-x),\qquad
 {U_i}\coloneqq\frac{L(i)L(i+1)}{L(0)L(1)}.
\end{equation}
In Section~\ref{sec:bg-minimum} the same functions are evaluated at
the parameters of a minimizing vertex.
With this sign convention, $L>0$ on $0\le x\le D+1$.
For $1\le i\le D$, the function
$(a-i)(a+1-i)/(a(a+1))$ increases for $a>i$.
Since $a_j\ge\xi_j-1$, Lemma~\ref{lem:binary-root-distance} gives
\begin{equation}\label{eq:binary-floor-comparison}
 \frac{h(i)}{h(0)}\ge {U_i}>0.
\end{equation}
For a real polynomial $q$ of degree at most $m$, the ratio
$(n)_ic_i/((N)_ic_0)$ for the lift $\binom{n-x}{r}q(x)q(x-1)$ is,
by~\eqref{eq:normalized-lift}, a ratio of two quadratic forms in the
coefficients of $q$; Appendix~\ref{app:binary-matrices} writes their
matrices as $\operatorname{Mat}_i$ and $\operatorname{Gram}_0$, where
$\operatorname{Gram}_0$ is positive definite and represents the
constant Krawtchouk coefficient $b_0=\E_{\mu_N}q(X)q(X-1)$. The matrix
comparison~\eqref{eq:binary-matrix-target},
$\operatorname{Mat}_i\preceq U_i\operatorname{Gram}_0$, therefore gives
$c_i/c_0\le(N)_iU_i/(n)_i$ for every such $q$, in particular for
$q(x)=\prod_j(x-a_j)$; \eqref{eq:normalized-value}
and~\eqref{eq:binary-floor-comparison} turn this into
$c_0g(i)\ge c_ig(0)$. At $i=1$ the proof uses the rounded polynomial
itself, through its Gauss--Radau bound (Lemma~\ref{lem:radau-rounded}),
rather than a comparison valid for every $q$.

\begin{proposition}\label{prop:binary-index-cover}
Suppose $m\ge2$, $r\ge0$, $D\coloneqq2m+r\ge16$ and
$N\ge6m+2r-1$. The rounded polynomial satisfies
$c_0g(i)\ge c_i g(0)$ for every $1\le i\le D$.
\end{proposition}
\begin{proof}
In Appendix~\ref{app:binary-endpoints}, Lemma~\ref{lem:binary-first}
proves $i=1$ and Lemma~\ref{lem:binary-top} proves $i=D$; both use
Lemma~\ref{lem:forward-schur} at $Q=4$, through
Lemma~\ref{lem:binary-monic-moment}.
Proposition~\ref{prop:binary-low-band} in Appendix~\ref{app:binary-matrices}
proves the matrix comparison for $2\le i\le\min(54,D-1)$ by exact
polynomial certificates.
Proposition~\ref{prop:binary-high-band} in Appendix~\ref{app:binary-high}
proves it for $55\le i\le D-1$.
These ranges cover every index, including $16\le D\le55$,
when the last range is empty. The matrix comparisons in the two
interior propositions imply the required rounded inequalities by
\eqref{eq:binary-floor-comparison} and
\eqref{eq:normalized-lift}--\eqref{eq:normalized-value}, as just
explained.
\end{proof}

\begin{proof}[Proof of Theorem~\ref{thm:binary-rounded}]
At $m=0$ the coefficient and value ratios of
$g(x)=\binom{n-x}{D}$ are $(D)_i/(n)_i$ and
$(n-D)_i/(n)_i$. Singleton gives $n-D>D$, so the criterion holds.
At $m=1$ the Lloyd zero is $\xi_1=1+3(N-1)/4$, whose floor is the
integer $a$ of Theorem~\ref{thm:one-root}. By
Lemma~\ref{lem:binary-root-distance}, $a>D$, so
Theorem~\ref{thm:one-root} applies. The only exception in its
statement, $(Q,N,r)=(4,6,1)$, has $n=7$ and distance four, whereas
\eqref{eq:binary-length} requires $n\ge10$.

When $m\ge2$ and $2\le t\le7$,
Proposition~\ref{prop:binary-small-radii} in Appendix~\ref{app:binary-small}
applies: uniform bounds cover $N\ge82$ for $t=2$, $N\ge100$ for
$3\le t\le5$, $N\ge128$ for $t=6$ and $N\ge176$ for $t=7$, and an
exact finite certificate covers the shorter
lengths (Lemma~\ref{lem:binary-small-finite}).
When $t\ge8$, $D=2t+\sigma\ge16$, and
Proposition~\ref{prop:binary-index-cover} applies.
The degree is exactly $D$, so all coefficients beyond $D$ vanish.
The positivity facts proved after Theorem~\ref{thm:grid} and
Lemma~\ref{lem:coefficient-criterion} now give the dimension bound.
\end{proof}

\subsection{The binary grid minimum}\label{sec:bg-minimum}

We now prove the binary case of Theorem~\ref{thm:grid}.
The polynomial in this argument minimizes the binomial expectation over
all feasible polynomials in~\eqref{eq:beta}.
Its roots need not be the rounded Lloyd
roots used in Theorem~\ref{thm:binary-rounded}, nor lie to their
right: at $(n,D)=(18,6)$ the unique minimizing vertex has residual
pairs with lower zeros $9$, $13$, $16$, whereas the Lloyd floors are
$10$, $13$, $16$, and $\beta_{18,6,4}$ is about $2\%$ below
$\min_{0\le e\le3}1/S^{18,e}_{3,0}$. Minimality supplies an
additional inequality that replaces the individual-root comparison
\eqref{eq:binary-floor-comparison}.
The proof has four steps. Lemma~\ref{lem:bg-root} places every root of
a minimizing vertex above $D$, and Lemma~\ref{lem:bg-low-indices} gives
the first two coefficients. For the higher indices,
Lemma~\ref{lem:bg-cap-comparison} turns a matched coefficient bound and
endpoint inequality into the criterion;
Propositions~\ref{prop:bg-coefficients} and~\ref{prop:bg-endpoints}
supply them for $D\ge16$, and Proposition~\ref{prop:bg-small-degree}
treats $D\le15$. Proposition~\ref{prop:bg-grid-minimum} assembles the
cases.

Fix an exact binary code of dimension $K>1$ and distance at least $D+1$,
where $D\ge 2$. Take a minimizing vertex $F$ for $\beta_{n,D,4}$, as in
Lemma~\ref{lem:grid-vertex}, normalized by $F(0)=1$. Write its maximal
terminal block and its residual adjacent pairs as
\[
 \begin{gathered}
 F(x)=\frac{\binom{n-x}{r}}{\binom nr}
       \frac{q(x)q(x-1)}{q(0)q(-1)},\qquad
 q(x)\coloneqq\prod_{j=1}^{m}(x-a_j),\\
 D=2 m+r,\quad N\coloneqq n-r,\quad H\coloneqq n-D=N-2 m.
 \end{gathered}
\]
These $m,r,N$ are the parameters of the minimizing vertex. They are not
fixed by a choice of $e$ in~\eqref{eq:original-parameters}. As in
Section~\ref{sec:paired-forms}, $a_j$ is the lower zero of the $j$th
residual pair.
Applying the same Rains bound
as in~\eqref{eq:binary-length}, now with
$\sigma\coloneqq D\bmod 2=r\bmod 2$, gives
\[
 n\ge 3D-1+2\sigma,\qquad
 N\ge 6 m+2 r-1+2\sigma,\qquad H+1\ge 2D.
\]

\begin{lemma}\label{lem:bg-root}
For a minimizing vertex of degree $D=2 m+r$ with maximal terminal
block, if $n\ge 3D-1$ and $m\ge 1$, the lower zero of each residual pair satisfies
\[
 a_j\ge\frac{\bigl(\sqrt{3(N-m)}-\sqrt{m-1}\bigr)^2}{4}
       >\frac{H+1}{2}\ge D.
\]
Every positive root of the vertex, including the terminal roots,
therefore exceeds $D$.
\end{lemma}

Lemma~\ref{lem:bg-root} has the form of
Proposition~\ref{prop:optimizer-roots} at $Q=4$: there the root-region
condition~\eqref{eq:root-region} reads $n\ge3D-1$, which the length
bound above guarantees for every binary code, and the two bounds of
that proposition become those of the lemma. The proof in
Appendix~\ref{sec:bg-roots} is the proof of
Proposition~\ref{prop:optimizer-roots} at $Q=4$:
Lemma~\ref{lem:paired-mean} is stated for $Q\ge9$, but its proof uses
only $Q>2$, and Appendix~\ref{sec:bg-roots} writes that calculation
out at $Q=4$ with its constants before applying it to variations of a
single pair.
It uses minimality.

We first isolate the two low coefficient indices. The first uses a
feasible competitor, and the second uses the positive quadrature of
Lemma~\ref{lem:positive-quadrature}; both arguments use minimality.

\begin{lemma}\label{lem:bg-low-indices}
Let $F=\sum_{i=0}^{D}c_iP_i^{(n,4)}$ be a normalized minimizing
vertex with at least two residual pairs, $D\ge 2$, and
$n\ge 3D-1+2(D\bmod 2)$. Then
$c_i\le c_0F(i)$ for $i=1,2$.
\end{lemma}

\begin{proof}[Proof of Lemma~\ref{lem:bg-low-indices} for $i=1$]
The $D$ distinct roots are integers $z_j\ge D+j$, by
Lemma~\ref{lem:bg-root}. Hence
\[
 u\coloneqq F(1)=\prod_{j=1}^{D}\frac{z_j-1}{z_j}\ge\frac 12.
\]
Put $G(x)\coloneqq(n-x)F(x+1)+xF(x)$.
Since $G(x)=nF(x)+(n-x)(F(x+1)-F(x))$, its degree is at most $D$.
It is nonnegative on the grid (at $x=n$ the first term vanishes) and
$G(0)=nu$, so minimality gives $\E_{\mu_n}G\ge nu c_0$.
Binomial size bias gives $\E_{\mu_n}G=(4/3)\E_{\mu_n}[XF(X)]$.
Thus orthogonality with $P_1^{(n,4)}(x)=3 n-4 x$ yields
\[
 \frac{c_1}{c_0}
 =1-\frac{\E_{\mu_n}[XF(X)]}{(3 n/4)c_0}
 \le 1-u\le u.
\qedhere
\]
\end{proof}
Appendix~\ref{sec:bg-low-indices} proves the second index using a
degree-preserving difference operator, a comparison with the rounded
competitor introduced below, and
Lemma~\ref{lem:bg-second-finite} for a finite remainder; that lemma
combines a scalar test in directed-rounding rational arithmetic with
nine cases settled by exact quadrature.

For the higher indices suppose $m\ge 2$. Define quadratic forms on
real polynomials of degree at most $m$ by
\[
 \mathcal G(q)\coloneqq\E_{\mu_N}[q(X)q(X-1)],\qquad
 \mathcal E_i(q)\coloneqq q(i)q(i-1).
\]
If $q(x)q(x-1)=\sum_{\ell=0}^{2 m}b_\ell(q)P_\ell^{(N,4)}(x)$,
put
\[
 \mathcal C_i(q)\coloneqq\sum_{\ell=0}^{2 m}
 \frac{\binom N\ell\binom r{i-\ell}}{\binom Ni}b_\ell(q)
 \quad(0\le i\le D),\qquad
 V\coloneqq V_m(N)=\sum_{j=0}^{m}\binom Nj 3^j.
\]
In particular $\mathcal C_0=\mathcal G$. These are polynomial
quadratic forms; the code enumerators retain the notation $A_i,B_i$.
The matrices of $\mathcal C_i$ and $\mathcal G$ are exactly
$\operatorname{Mat}_i$ and $\operatorname{Gram}_0$ of
Appendix~\ref{app:binary-matrices}. In our range $N\ge 6 m-1$,
$\mathcal G$ is positive definite. Indeed its normalized matrix
$\operatorname{Gram}_0$ in the basis $P_j^{(N,4)}-P_{j-1}^{(N,4)}$ is
tridiagonal, with diagonal entries $1-j/(N-j+1)\ge4/5$ and adjacent
entries $\sqrt{j/[3(N-j+1)]}\le1/\sqrt{15}$ for $j\le m$
(Appendix~\ref{app:binary-matrices}), so strict diagonal dominance
proves this assertion.

For the actual minimizing vertex, the normalized lift identities
\eqref{eq:normalized-lift}--\eqref{eq:normalized-value} read
\[
 c_0=4^{-r}\frac{\mathcal G(q)}{\mathcal E_0(q)},\qquad
 \frac{c_i}{c_0}=\frac{(N)_i}{(n)_i}
                    \frac{\mathcal C_i(q)}{\mathcal G(q)},\qquad
 F(i)=\frac{(N)_i}{(n)_i}
                    \frac{\mathcal E_i(q)}{\mathcal E_0(q)}.
\]
Here $D<N$, and all common factors are positive. It is enough to show
$\mathcal C_i(q)/\mathcal G(q)\le
\mathcal E_i(q)/\mathcal E_0(q)$.

Residual minimality supplies the needed extra information. Every
normalized feasible polynomial of degree at most $2 m$ on
$\{0,\ldots,N\}$ lifts by the same terminal factor to a competitor
for $F$. The normalized rounded Lloyd polynomial $h_L/h_L(0)$ of
Lemma~\ref{lem:radau-rounded}, at the pair count $m$ and residual
length $N$ of $F$, is such a competitor and has expectation at most
$1/V$; its floors are determined by the Lloyd zeros and need not equal
the $a_j$ of $q$. Consequently the minimizing residual polynomial satisfies
\[
 \frac{\mathcal G(q)}{\mathcal E_0(q)}\le\frac 1V,
 \qquad\text{or equivalently}\qquad
 \mathcal E_0(q)\ge V\mathcal G(q)>0.
\]

\begin{lemma}\label{lem:bg-cap-comparison}
Suppose $\mathcal E_0(q)\ge V\mathcal G(q)>0$. If
$\lambda,\tau\ge 0$ satisfy
\[
 \mathcal C_i(q)\le\lambda\mathcal G(q),\qquad
 \mathcal E_i(q)-(\lambda+\tau)\mathcal E_0(q)
                 +\tau V\mathcal G(q)\ge 0,
\]
then $\mathcal C_i(q)/\mathcal G(q)
       \le\mathcal E_i(q)/\mathcal E_0(q)$.
\end{lemma}
\begin{proof}
The second inequality gives
$\mathcal E_i(q)\ge\lambda\mathcal E_0(q)
+\tau(\mathcal E_0(q)-V\mathcal G(q))
\ge\lambda\mathcal E_0(q)$. Combine this with the first inequality.
\end{proof}

The next two propositions provide matched pairs for this lemma: a
coefficient inequality, which bounds $\mathcal C_i$ by $\lambda\mathcal G$,
and an endpoint inequality with the same $\lambda$, which compares the
products $\mathcal E_i$ and $\mathcal E_0$ of values of $q$ at the
endpoints of the unit intervals $[i-1,i]$ and $[-1,0]$. With the linear
coefficient factor $1-2i/(3D)$ of Proposition~\ref{prop:bg-coefficients}
and the linear endpoint inequality (the second inequality of
Proposition~\ref{prop:bg-endpoints}), the lemma applies with
$\lambda+\tau=U_i$; with the quarter coefficient factor $1/4$ and the
quarter endpoint inequality (the first), it applies with
$\lambda+\tau=5U_i/4$.
The first pair serves $3\le i\le\min(99,D-1)$, since the second
endpoint inequality is proved only for $i\le99$; the second pair
serves $100\le i\le D-1$ and, for every $D\ge16$, the top index $i=D$.
Use $U_i$ from~\eqref{eq:binary-endpoint}, and
Lemma~\ref{lem:binary-root-distance}, now with the parameters $m$, $r$
and $N$ of the minimizing vertex; both depend only on these parameters
(and $U_i$ on $i$).
Equivalently (see the opening of Appendix~\ref{app:bg-endpoints}), if
$p_j(x)\coloneqq P_j^{(N-1,4)}(x)$ and $w_j\coloneqq\binom{N-1}{j}3^j$, then
\[
 U_i=\frac{p_m(i)p_m(i-1)}{w_m V},\qquad
 p_m(0)=w_m,\quad p_m(-1)=V.
\]
In the following ranges $U_i>0$. For $i\le D$ in
Proposition~\ref{prop:bg-coefficients}, every zero of $p_m$ exceeds
$D$ by Lemma~\ref{lem:binary-root-distance}. For
$1\le i\le\lfloor(H+1)/2\rfloor$ in
Proposition~\ref{prop:bg-endpoints}, Lemma~\ref{lem:bg-ratio-hull}
gives positive ratios $p_{j-1}/p_j$ on $[-1,(H+1)/2]$; since
$p_0=1$, also $p_m>0$ there.

\begin{proposition}\label{prop:bg-coefficients}
Let $m\ge 2$, $r\ge 0$, $D\coloneqq 2 m+r\ge 16$, and
$N\ge 6 m+2 r-1$. For every nonzero real polynomial $q$ of degree
at most $m$,
\[
 \mathcal C_i(q)<\left(1-\frac{2 i}{3D}\right)U_i\mathcal G(q)
 \qquad(2\le i\le\min(99,D-1)),
\]
and
\[
 \mathcal C_i(q)<\frac{U_i}{4}\mathcal G(q)
 \qquad(100\le i\le D-1\text{ or }i=D).
\]
\end{proposition}

Appendix~\ref{sec:bg-coefficients} strengthens the matrix
comparisons~\eqref{eq:binary-matrix-target}
of Appendices~\ref{app:binary-matrices} and~\ref{app:binary-high},
which have the factor $1$ in place of $1-2i/(3D)$ or $1/4$, to these
smaller factors. The interior indices $2\le i\le\min(54,D-1)$ use the
exact
polynomial certificates of Lemma~\ref{lem:bg-low-coeff-finite}.
Appendix~\ref{sec:bg-coefficients} treats the larger indices separately
inside and outside the recurrence strip
$i\le N(2N-3m+6)/(3(8N-9m))$, reusing estimates of
Appendix~\ref{app:binary-high}; the indices $55\le i\le 61$ outside
the strip use Lemma~\ref{lem:bg-outside-finite}, and the top index
$i=D$ follows the calculation of Lemma~\ref{lem:binary-top}.
The proof actually gives the linear factor at every interior index,
but only the stated range is needed here.

\begin{proposition}\label{prop:bg-endpoints}
Let $m\ge 2$, $N\ge 6 m-1$, and $H\coloneqq N-2 m$. For every nonzero real
polynomial $q$ of degree at most $m$ and every integer
$1\le i\le\lfloor(H+1)/2\rfloor$,
\[
 \frac{\mathcal E_i(q)}{VU_i}
       -\frac{5\mathcal E_0(q)}{4V}+\mathcal G(q)>0.
\]
For the same range with $i\le 99$, we also have
\[
 \frac{\mathcal E_i(q)}{VU_i}
       -\frac{\mathcal E_0(q)}{V}
       +\frac{4 i}{3(H+1)}\mathcal G(q)>0.
\]
\end{proposition}

The numbers $U_i$, $V$ and the forms $\mathcal G$, $\mathcal E_i$
depend only on
$m$, $N$ and $i$, so this proposition is uniform in $r$. Its proof in
Appendix~\ref{app:bg-endpoints} writes $\mathcal E_i/(VU_i)$ and
$\mathcal E_0/V$ as symmetrized products of normalized evaluation
vectors (Lemma~\ref{lem:bg-endpoint-normalization}) and compares
them in the inner product $\langle a,b\rangle_{B^{-1}}$, where $B$
is the matrix of $\mathcal G$.
The first assertion is proved analytically for every triple $(m,N,i)$,
by comparison with fixed covariance weights and an explicit
three-dimensional operator bound. For the second, analytic estimates
cover $12i\le H+1$, and the remaining triples, about $5.7\times10^6$,
are covered by Lemma~\ref{lem:bg-finite-linear}, whose certified
domain is larger and which uses
outward-rounded rational interval arithmetic.
Both endpoint statements hold for every polynomial $q$; in this
comparison, minimality enters only through the hypothesis of
Lemma~\ref{lem:bg-cap-comparison}.

\begin{proposition}\label{prop:bg-small-degree}
Let $4\le D\le 15$, and let a normalized minimizing vertex $F$ have
at least two residual pairs and
$n\ge 3D-1+2(D\bmod 2)$. Its Krawtchouk coefficients satisfy
$c_i\le c_0F(i)$ for every $1\le i\le D$.
\end{proposition}

Appendix~\ref{sec:bg-small-degree} proves this proposition by the same
comparison principle. The first two indices use
Lemma~\ref{lem:bg-low-indices}. For $N\ge 13D$ the coefficient and
endpoint bounds are analytic. For $N<13D$, the paired matrix
certificates of Lemma~\ref{lem:bg-small-finite} verify, at every
$(m,r,N,i)$ with $3\le i\le D$, both the coefficient inequality and
the endpoint inequality of the comparison. At $(m,r,N,i)=(2,0,11,4)$ they use the factor
$1/2$ in place of $1/4$, and the endpoint inequality with $3/2$ in
place of $5/4$, so that $\lambda=U_i/2$ and $\tau=U_i$ in
Lemma~\ref{lem:bg-cap-comparison}.

\begin{proposition}\label{prop:bg-grid-minimum}
For every exact binary quantum code of dimension $K>1$ and distance
at least $D+1$, where $D\ge 2$,
\[
 K\le 2^n\beta_{n,D,4}.
\]
\end{proposition}
\begin{proof}
Choose the minimizing vertex above. If $m=0$, its expectation is
$4^{-D}$ and the assertion follows from Singleton. If $m=1$, the
one-pair argument in the proof of
Proposition~\ref{prop:one-pair-minimum} applies with $Q=4$, with
Lemma~\ref{lem:bg-root} in place of
Proposition~\ref{prop:optimizer-roots}; apart from its remark that
there is no binary exception, which is treated below, the steps of
that proof hold at $Q=4$. The lower zero $a$ of the vertex's pair
satisfies $a>D$, moving
its pair by one in either direction gives a feasible competitor, and
minimality, through the difference formula for the normalized
residual objective in that proof, gives $3(N-1)/4\le a\le 1+3(N-1)/4$. Hence the
representative whose pair has lower zero
$a_*\coloneqq\lfloor 1+3(N-1)/4\rfloor$ has the
same objective, and $a_*\ge a>D$; the preceding integer ties only
when $3(N-1)/4$ is an integer. Choose this representative, allowing
its pair to merge with the terminal block when necessary.
Theorem~\ref{thm:one-root} supplies the coefficient criterion for it.
Its only binary exception $(N,r)=(6,1)$
would give $n=7,D=3$, which the length bound
$n\ge3D-1+2\sigma=10$ above excludes.

Suppose $m\ge 2$. Proposition~\ref{prop:bg-small-degree} handles
$D\le 15$. For $D\ge 16$, Lemma~\ref{lem:bg-low-indices} gives the
first two coefficients. Every index $i\le D$ lies in the range
$i\le\lfloor(H+1)/2\rfloor$ of Proposition~\ref{prop:bg-endpoints},
because $H+1\ge 2D$. At $3\le i\le\min(99,D-1)$ use
Proposition~\ref{prop:bg-coefficients} and the linear endpoint in
Proposition~\ref{prop:bg-endpoints}. Since $H+1\ge 2D$,
\[
 \frac{4 i}{3(H+1)}\le\frac{2 i}{3D}.
\]
We may therefore enlarge the coefficient of the positive form
$\mathcal G$ in the endpoint inequality to $2 i/(3D)$.
Lemma~\ref{lem:bg-cap-comparison}, whose hypothesis is the competitor
inequality above, applies with
\[
 \lambda\coloneqq\left(1-\frac{2 i}{3D}\right)U_i,
 \qquad\tau\coloneqq\frac{2 i}{3D}U_i.
\]
At $100\le i\le D-1$ and at $i=D$, use the quarter coefficient
and quarter endpoint, with $\lambda\coloneqq U_i/4$ and $\tau\coloneqq U_i$.
The ranges, including empty interior ranges, are thus
\[
 \begin{array}{c|c}
 \text{index}&\text{comparison}\\\hline
 1,2&\text{competitor and positive quadrature}\\
 3,\ldots,\min(99,D-1)&\text{linear coefficient and endpoint}\\
 100,\ldots,D-1&\text{quarter coefficient and endpoint}\\
 D&\text{quarter coefficient and endpoint}.
 \end{array}
\]
The normalized lift now gives $c_i\le c_0F(i)$ for every required
index. Since $F$ is grid-nonnegative, $F(0)=1$, and
$c_0=\beta_{n,D,4}>0$, Lemma~\ref{lem:coefficient-criterion}
proves $K\le 2^nc_0$. Finally, any polynomial $f$ in
Theorem~\ref{thm:grid} satisfies
$\E_{\mu_n}f/f(0)\ge\beta_{n,D,4}$, so this binary branch gives
the stated bound for every admissible $f$.
\end{proof}

\section{Completion of the two bounds}\label{sec:completion}

The preceding arguments cover complementary parameter regions.
We assemble them here at the original length and distance.
Table~\ref{tab:completion} lists the result that covers each case. For Theorem~\ref{thm:grid},
$m$, $N$ and $r$ are the parameters of a minimizing vertex
in~\eqref{eq:vertex-factorization}, for every $p\ge2$.
In the separate direct binary proof of Theorem~\ref{thm:rounded},
they are instead those of~\eqref{eq:original-parameters}.

\begin{table}[!tb]\centering\small
\begin{tabular}{@{}p{9mm}>{\raggedright\arraybackslash}p{36mm}%
>{\raggedright\arraybackslash}p{101mm}@{}}\toprule
$p$&Case&Covered by\\\midrule
\multicolumn{3}{@{}l}{Theorem~\ref{thm:grid} (Singleton gives $n\ge2D+1$)}\\
$2$&all parameters, all in the root region by Rains's bound&Proposition~\ref{prop:bg-grid-minimum}:
 $m=0$ by Singleton; $m=1$ by the one-pair argument of
 Proposition~\ref{prop:one-pair-minimum}, with Lemma~\ref{lem:bg-root}
 and Theorem~\ref{thm:one-root}; $m\ge2$ by
 Proposition~\ref{prop:bg-small-degree} ($D\le15$), or by
 Lemma~\ref{lem:bg-low-indices} ($i=1,2$) and
 Propositions~\ref{prop:bg-coefficients} and~\ref{prop:bg-endpoints}
 with Lemma~\ref{lem:bg-cap-comparison} ($3\le i\le D$, $D\ge16$)\\
$\ge3$&outside the root region&Theorem~\ref{thm:strict-complement},
 from the rank bounds of
 Sections~\ref{sec:rank-windows}--\ref{sec:complement}, including
 Theorems~\ref{thm:critical-grid} and~\ref{thm:adjacent-grid}\\
$\ge3$&root region, $m\le1$&Proposition~\ref{prop:one-pair-minimum}\\
$\ge6$&root region, $m\ge2$&Theorem~\ref{thm:higher-weak}:
 Proposition~\ref{prop:large-deficit} ($N-r\ge\lceil20N/Q\rceil+6m$)
 or Proposition~\ref{prop:large-alphabet-coefficients}\\
$5$&root region, $m\ge2$&Theorem~\ref{thm:higher-weak}:
 Theorems~\ref{thm:critical-grid} and~\ref{thm:adjacent-grid} if
 $u_0>0$ in Lemma~\ref{lem:physical-tail-cap}; otherwise
 Proposition~\ref{prop:terminal-cutoffs} ($r\le3N/5$) or
 Proposition~\ref{prop:five-cap-coefficients}\\
$4$&root region, $m\ge2$&Theorem~\ref{thm:higher-weak}:
 Theorems~\ref{thm:critical-grid} and~\ref{thm:adjacent-grid} if
 $u_0>0$; otherwise Proposition~\ref{prop:terminal-cutoffs}
 ($r\le N/2$), Propositions~\ref{prop:four-initial}
 and~\ref{prop:four-tail} ($r>N/2$, $N\ge256$), or the finite interval
 of Proposition~\ref{prop:four-finite} ($r>N/2$, $N\le255$)\\
$3$&root region, $m\ge2$&Theorem~\ref{thm:qutrit-weak}:
 Proposition~\ref{prop:scale-cutoff} ($r/N\le z(\xi)$),
 Theorem~\ref{thm:qutrit-large-length} ($N\ge512$), or the finite
 remainder of Proposition~\ref{prop:qutrit-finite}\\\midrule
\multicolumn{3}{@{}l}{Theorem~\ref{thm:rounded} for every $p\ge2$:
 Theorem~\ref{thm:grid} applied to $g$
 in~\eqref{eq:rounded-polynomial}}\\\midrule
\multicolumn{3}{@{}l}{Additional direct proof of Theorem~\ref{thm:rounded} for $p=2$:
 Theorem~\ref{thm:binary-rounded}, where $n\ge6t-1+5\sigma$
 by~\eqref{eq:binary-length}}\\
$2$&$m=0$&direct comparison, using Singleton\\
$2$&$m=1$&Theorem~\ref{thm:one-root}, with
 Lemma~\ref{lem:binary-root-distance}\\
$2$&$m\ge2$, $t\le7$&Proposition~\ref{prop:binary-small-radii}: uniform
 bounds and the finite certificate of
 Lemma~\ref{lem:binary-small-finite}\\
$2$&$m\ge2$, $t\ge8$&Proposition~\ref{prop:binary-index-cover}:
 Lemma~\ref{lem:binary-first} ($i=1$),
 Proposition~\ref{prop:binary-low-band} ($2\le i\le\min(54,D-1)$),
 Proposition~\ref{prop:binary-high-band} ($55\le i\le D-1$),
 Lemma~\ref{lem:binary-top} ($i=D$)\\\bottomrule
\end{tabular}
\caption{The result that covers each case in the proofs of
Theorems~\ref{thm:grid} and~\ref{thm:rounded}.}
\label{tab:completion}
\end{table}

\begin{proof}[Proof of Theorem~\ref{thm:grid}]
Singleton gives $n\ge2D+1$.
For $p=2$, where Rains's length bound places every code in the root
region, Proposition~\ref{prop:bg-grid-minimum} gives
$K\le2^n\beta_{n,D,4}$. Suppose now $p\ge3$.
If $2(Q-1)D>(Q-2)(n+1)$, Theorem~\ref{thm:strict-complement}
gives $K\le p^n\beta_{n,D,Q}$.
Otherwise $(n,D)$ lies in the root region, where a minimizing vertex
has every positive root above $D$ by
Proposition~\ref{prop:optimizer-roots}.
With at most one residual pair, use
Proposition~\ref{prop:one-pair-minimum}.
With at least two pairs, apply Theorem~\ref{thm:higher-weak}
for $p\ge4$ or Theorem~\ref{thm:qutrit-weak} for $p=3$.
Thus $K\le p^n\beta_{n,D,Q}$ throughout the parameter range.
For any polynomial in the theorem, $f/f(0)$ is feasible for
the minimum, so
$\E_{\mu_n}f/f(0)\ge\beta_{n,D,Q}$.
This proves the asserted inequality.
\end{proof}

\begin{proof}[Proof of Theorem~\ref{thm:rounded}]
For every $p\ge2$, the original rounded polynomial is grid-nonnegative
and positive at the origin, as shown in Section~\ref{sec:framework}.
Apply Theorem~\ref{thm:grid} to that polynomial, with
$D=2t+\sigma\ge2$, and use $S_{t,\sigma}^{n,e}=g(0)/c_0$.
No restriction on $t$, $e$ or $\sigma$ is imposed in this
application. At $e=t$, the formula is the Singleton bound.
For $p=2$, Theorem~\ref{thm:binary-rounded} additionally supplies the
direct coefficient criterion for each prescribed rounded polynomial.
\end{proof}

\section{Conclusion}\label{sec:conclusion}

We have proved that the rounded Yu--Lai--Oh bounds hold for
arbitrary exact quantum subspaces of dimension greater than one,
throughout all integer local dimensions, interpolation parameters,
and both distance parities. More generally, we prove $Kf(0)\le p^n\E f(X)$, where
$X\sim\Bin(n,1-p^{-2})$, for every integer $p\ge2$ and every real
polynomial of degree below the code distance that is nonnegative on
$\{0,\ldots,n\}$ and positive at zero. This extends the
pure-code polynomial inequality to general exact subspaces.

Minimization reduces the polynomial statement to one minimizing
polynomial. For nonbinary codes, estimates of its roots and Krawtchouk
coefficients combine with direct dimension bounds at large distances.
For binary codes, Rains's length bound~\cite[Theorem~15]{Rains} places
every root of the minimizer above its degree, and a comparison with a
rounded polynomial of the same degree, together with coefficient and
endpoint estimates, yields the coefficient criterion of
Lemma~\ref{lem:coefficient-criterion}. The proof is computer-assisted:
analytic estimates are combined with finite computations and sign
certificates, checked in exact rational arithmetic or in
outward-rounded interval arithmetic, and together they cover every
parameter; the opening paragraph of the appendices lists
the computer-verified steps.
We also give a direct binary proof that each prescribed rounded
polynomial satisfies the coefficient criterion. Thus the prescribed
polynomials themselves certify the rounded bounds for $p=2$, and this proof uses the computations of
Appendices~\ref{app:binary-endpoints}--\ref{app:binary-small} but none
of those of Appendices~\ref{app:bg-minima}--\ref{app:bg-finite}, which
serve only the binary grid-minimum proof.
The purity conjecture of Calderbank, Rains, Shor, and
Sloane~\cite{CRSS}, which would make the linear-programming bound for
pure codes valid for all qubit codes, is not proved
here.

\medskip
\noindent\textbf{Computational support.}
The computational support for Appendices A--I is archived
in~\cite{ComputationalCertificates}, apart from three short computations
specified in their proofs. Version~2.0.0 of the supporting
package~\cite{ComputationalCertificatesV2} also includes those three
computations and the additional certificates for the binary grid-minimum
proof in Appendices J--L.

\appendix
\addtocontents{toc}{\protect\setcounter{tocdepth}{1}}
\addcontentsline{toc}{part}{Appendices}
\section*{Overview of the appendices}
Appendices J--L contain the proofs and certificates of the binary
grid-minimum argument of Section~\ref{sec:bg-minimum}; Appendices
A--I contain the other deferred proofs and certificates.

\medskip\noindent\textbf{Computer-verified steps.}
The following results rest on finite computations. They use exact
rational or integer arithmetic, except where outward-rounded rational
intervals are named below. Finite parameter domains checked case by case:
the remaining domain of the adjacent strip
(Lemma~\ref{lem:adjacent-finite} for Theorem~\ref{thm:adjacent-grid});
the qutrit complement for $n<1300$
(Lemma~\ref{lem:qutrit-bounded}, part of
Lemma~\ref{lem:qutrit-complement}); the interval
$29\le n-D\le251$ at $p=4$ (Proposition~\ref{prop:four-finite},
Appendix~\ref{app:four-finite-certificate}); the qutrit finite
remainder $15\le n-D\le507$ (Proposition~\ref{prop:qutrit-finite},
Appendix~\ref{app:qutrit-finite}); and the binary small radii below
the length cutoffs (Lemma~\ref{lem:binary-small-finite}, part of
Proposition~\ref{prop:binary-small-radii}). Polynomial and matrix sign
certificates, that is, exact expansions whose coefficients, minors or
pivots are checked to be positive on whole parameter regions: the case $L=2$ of the critical
strip (Lemma~\ref{lem:critical-two}); the qutrit complement for
$n\ge1300$ (Lemma~\ref{lem:qutrit-growing}); the scale-sensitive
cutoff (Proposition~\ref{prop:scale-cutoff}); the matrix certificates
of Appendix~\ref{app:higher-certificates} and the rational
reconstructions of Appendix~\ref{app:higher-five}, used in
Propositions~\ref{prop:large-alphabet-coefficients}
and~\ref{prop:five-cap-coefficients}; the scalar certificates of the
four qutrit rows (Proposition~\ref{prop:qutrit-four-ranges},
Appendices~\ref{app:qutrit-quarter}--\ref{app:qutrit-middle}); and, for
$p=2$, the endpoint Lemmas~\ref{lem:binary-first}
and~\ref{lem:binary-top}, the binary low band
(Proposition~\ref{prop:binary-low-band},
Appendix~\ref{app:binary-certificate-spec}), and
Lemmas~\ref{lem:binary-strip} and~\ref{lem:binary-affine-gap} of the
high band.
The binary grid-minimum argument reuses Lemmas~\ref{lem:binary-top},
\ref{lem:binary-strip} and~\ref{lem:binary-affine-gap}, and
Appendix~\ref{app:bg-finite} adds five computations: the second-index
remainder (Lemma~\ref{lem:bg-second-finite}, a scalar test with
directed rounding and nine exact quadrature cases); the strengthened
row certificates (Lemma~\ref{lem:bg-low-coeff-finite}); the
transition outside the recurrence strip
(Lemma~\ref{lem:bg-outside-finite}, with rational bounds for square
roots and exponentials); the linear endpoint certificate
(Lemma~\ref{lem:bg-finite-linear}),
in outward-rounded dyadic interval arithmetic with an independent
replay at lower precision; and the small-degree matrix certificates
(Lemma~\ref{lem:bg-small-finite}, by exact fraction-free
elimination). Short exact evaluations displayed in full within a proof
are not listed. Each computation is specified in the appendix where it
is used, by its defining expression and its complete domain. The
data for Appendices A--I are archived
in~\cite{ComputationalCertificates}, except for the computations of
Lemmas~\ref{lem:binary-first} and~\ref{lem:binary-top} and the
expansions of Lemma~\ref{lem:binary-strip}, which are specified in
their proofs. Version~2.0.0 of the supporting
package~\cite{ComputationalCertificatesV2}
also includes those three computations and the data for
Appendix~\ref{app:bg-finite}.

\section{Detailed rank-window estimates}\label{app:rank-windows}

This appendix proves the parts of the rank bounds of
Sections~\ref{sec:rank-windows}--\ref{sec:complement} that were deferred:
coefficient identities, finite certificates and long estimates.
Appendix~\ref{app:rank-separator} proves
Lemma~\ref{lem:distance-cubic-certificate}, the sign conditions of the
cubic separator in the proof of Theorem~\ref{thm:distance-ceiling}.
Appendix~\ref{app:rank-critical} completes
Theorem~\ref{thm:critical-grid} for $2\le L\le7$: the case $L=2$
(Lemma~\ref{lem:critical-two}) and the cases $3\le L\le7$ for every
$p\ge3$ (Lemma~\ref{lem:critical-finite}).
Appendix~\ref{app:rank-adjacent} proves
Lemma~\ref{lem:adjacent-finite}, the finite domain left in the proof of
Theorem~\ref{thm:adjacent-grid}.  The remaining subsections serve
Theorem~\ref{thm:strict-complement}.  Appendix~\ref{app:rank-production}
gives the tools for the product windows of
Lemma~\ref{lem:product-window-test}; Appendices~\ref{app:rank-pfive}
and~\ref{app:rank-pfour} apply them to prove
Lemma~\ref{lem:large-alphabet-complement} for $p\ge5$
(Lemma~\ref{lem:complement-pfive}) and for $p=4$
(Lemma~\ref{lem:complement-pfour}), and Appendix~\ref{app:rank-qutrit}
proves the two parts of Lemma~\ref{lem:qutrit-complement}, for
$n\ge1300$ (Lemma~\ref{lem:qutrit-growing}) and for the remaining
lengths $n<1300$ (Lemma~\ref{lem:qutrit-bounded}).

Every argument here is of one of two kinds.  Either a window of
Lemma~\ref{lem:window-identity} or~\ref{lem:factorial-reflection} is
exhibited whose sign conditions exclude every rank $K\ge K_0$, or a rank
bound of Proposition~\ref{prop:strip-rank} is compared with a lower bound
for the target $R_{n,D}=p^n\beta_{n,D,Q}$ at the original length $n$
(Proposition~\ref{prop:grid-budget}, Lemma~\ref{lem:grid-budget-coarse},
or an exactly certified grid minimum).  As in
Sections~\ref{sec:distance-strips}--\ref{sec:complement}, $b=Q-1$ and
$\delta=n\bmod2$; symbols introduced inside a proof are local to that
proof.

The supplementary package~\cite{ComputationalCertificates} contains the rational
coefficient arrays and integer windows specified below.
Each certificate is defined by its
mathematical expression and its complete domain; regenerating the
expression and checking its coefficients or rows is sufficient, and the
description of each certificate lists exactly these checks.

\subsection{The cubic separator}\label{app:rank-separator}

Here a cubic invariant relation means the relation
$0=\sum_jC_ja_j+\sum_jE_jb_j$ obtained in the proof of
Lemma~\ref{lem:window-identity} for the window $s=2$ (so that $W$ is a
cubic) and $M={s_0}+2$,
before $C$ is replaced by $A+\rho_KE$; its coefficients of weight
$j=M-l$ are $a_{M-l}$ and $b_{M-l}$.  The polynomial $W$ constructed in
the proof is the cubic separator of
Theorem~\ref{thm:distance-ceiling}, which applies the lemma with
$M=\lfloor n/2\rfloor+2$ and $L=M-d$.

\begin{lemma}[Cubic separator]\label{lem:distance-cubic-certificate}
Let $p\ge3$, $Q=p^2$, and $M\ge Q(L+1)+1$, where $L\ge1$.
There is a cubic invariant relation whose $C$ coefficients are
nonnegative for $0\le l\le M$ and positive at $l=M$, and whose
$E$ coefficients are positive for $0\le l\le L$.
\end{lemma}
\begin{proof}
Put $\Lambda\coloneqq M(M-1)(M-2)$ and write
$f(l)\coloneqq(M+1-l)(M-Ql)^2=\sum_{i=0}^3f_i(l)_i$.
Its falling-factorial coefficients are
\[
 \begin{aligned}
 f_0&=M^2(M+1),&
 f_1&=M\{p^4-(2p^2+1)M\},\\
 f_2&=p^2\{(p^2+2)M-2p^2\},& f_3&=-p^4.
 \end{aligned}
\]
With $J\coloneqq3$ and $\eta\coloneqq M-2$ in Lemma~\ref{lem:factorial-reflection},
choose the homogeneous cubic
\[
 W(x,y)\coloneqq\frac1\Lambda\sum_{i=0}^3
 f_i(M-2)^{\overline i}y^i(x-y)^{3-i}.
\]
By~\eqref{eq:profile-extraction}, its $C$ coefficients are
\[
 a_{M-l}=\binom{M+l-3}{l}\frac{f(l)}\Lambda,
\]
which are nonnegative on the full grid $0\le l\le M$ and positive at
$l=M$.

The $E$ coefficients are extracted from $V=-(W+T_3W)/2$.  The
substitution $T_3$ maps $x-y$ to $py$ and $y$ to $(x-y)/p$, so the same
extraction~\eqref{eq:profile-extraction} gives
\[
 b_{M-l}=\binom{M+l-3}{l}
 \frac{G(l)}{2\Lambda(M-1)(M-2)},
\]
where
\[
 \begin{aligned}
 \Sigma&\coloneqq(p-1)M^2-(3p+1)M+2p,\\
 \Omega&\coloneqq(M-2)\{(p-1)(p^2-p+1)M^2+(p^4-3p^3-2p)M+2p^3\},\\
 G(l)&\coloneqq\Sigma\{\Lambda+p^3(l)_3\}-\Omega l\{M-1+p(l-1)\}.
 \end{aligned}
\]
Equivalently, $G=-(M-1)(M-2)(f+\mathcal Rf)$ with $\mathcal R$ as
in~\eqref{eq:factorial-reflection}, which allows a direct check of
these formulas.  All denominators are positive, so
it suffices to prove $G(l)>0$ for $0\le l\le L$.

Put $l_*\coloneqq M/Q-1$; the hypothesis gives $L<l_*$.  Since $M\ge Q(L+1)+1\ge2Q+1\ge19$,
$(p-1)M-(3p+1)\ge6p-10>0$, so $\Sigma>0$ and $G(0)=\Lambda\Sigma>0$.
Next, $G$ is strictly concave on $[0,M/Q]$.  Indeed, write
$G(l)=\sum_{i=0}^3G_il^i$; then $G_3=p^3\Sigma>0$ and $G_2=-3p^3\Sigma-p\Omega$, and
direct collection of terms gives
\[
 -\frac{G_2+3G_3M/Q}{p}
 =\kappa_3M^3+M(\kappa_2M-\kappa_1)+2p^3,
\]
where
\[
 \kappa_3\coloneqq(p-2)(p^2-1),\quad
 \kappa_2\coloneqq p^2(p-1)^2+3p+5,\quad
 \kappa_1\coloneqq2p^4+p^3+3p^2+2p.
\]
For $p\ge3$ these terms are positive, since $\kappa_3>0$ and
$\kappa_2M-\kappa_1\ge Q\kappa_2-\kappa_1=p(p-1)^2(p^3-2p-2)>0$.
Since $G_3>0$, the function $G''(l)/2=G_2+3G_3l$ increases in $l$, so
it is negative on $[0,M/Q]$.

At the other endpoint, the terms in $M^5$ cancel, and
\[
 p^3G(l_*)=\gamma_4M^4+\gamma_3M^3+\gamma_2M^2+\gamma_1M-4p^6(p-1),
\]
with
\[
 \begin{aligned}
 \gamma_4&\coloneqq(p^2-1)^2(p^2-3),\\
 \gamma_3&\coloneqq-p^2(p-1)(p+1)^2(p^2-2p+2),\\
 \gamma_2&\coloneqq-p^2(p-1)(2p^5+p^4-4p^3-p^2+4p+4),\\
 \gamma_1&\coloneqq2p^4(p-1)(p+2)(2p^2+1).
 \end{aligned}
\]
For $p\ge3$ we have $\gamma_4\ge\tfrac12p^6$, $\gamma_3\ge-(p-1)p^6$, $\gamma_2\ge-2p^8$ and
$\gamma_1>0$.  Indeed, $\gamma_4/p^6=(1-p^{-2})^2(1-3p^{-2})$ increases in $p$ and
equals $(8/9)^2(2/3)>1/2$ at $p=3$, while
\[
 \gamma_3+(p-1)p^6=p^2(p-1)(p^2-2p-2)>0,\qquad
 \gamma_2+2p^8=p^4\{p(p^2-3)+5(p^2-1)\}+4p^2>0.
\]
Since $M\ge Q$ and $(p-1)/p^2$ decreases in $p$, we have
$(p-1)/M\le2/9$ and $2p^2/M^2\le2/9$.  Hence
\[
 p^3G(l_*)\ge p^6\left\{\frac{M^4}{18}-4(p-1)\right\}
 \ge p^6\,\frac{M(M^3-24)}{18}>0,
\]
using $p\le M/3$ and $M\ge9$.  By concavity, $G$ lies above the chord
joining its positive values at $0$ and $l_*$, so $G(l)>0$ on $[0,l_*]$,
which contains every integer $0\le l\le L$.
\end{proof}

\subsection{Moment estimates on the critical strip}\label{app:rank-critical}

The two lemmas of this subsection use the notation
\eqref{eq:critical-parameters} of the proof of
Theorem~\ref{thm:critical-grid}; there the cubic rank bound gives
$K\le\operatorname{Rank}_3<pL/u$ and $u<pL/2$.  For a minimizing
vertex with $m$ residual pairs, $r=D-2m$ and $N=n-r$,
\eqref{eq:terminal-expectation} gives
$R_{n,D}=p^{n-2r}\E_{\mu_N}h/h(0)$, and $n-2r=2L-2+\delta+4m$ on the
critical strip.

For Lemma~\ref{lem:critical-two} define, in ordinary Krawtchouk
coordinates through degree $m$,
\begin{equation}\label{eq:rank-moment-matrices}
 \begin{gathered}
 v_i\coloneqq\binom Ni b^i,\quad w_i\coloneqq v_i+Q\sum_{j<i}v_j,\\
 G_{ii}\coloneqq v_i,\quad G_{ij}\coloneqq\tfrac Q2v_{\min(i,j)}\ (i\ne j),
 \quad O\coloneqq\tfrac12(vw^{\mathsf T}+wv^{\mathsf T}).
 \end{gathered}
\end{equation}
Since $v_i$ and $w_i$ are the values of $\Kraw{i}{N}$ at $0$ and,
by~\eqref{eq:backward-shift}, at $-1$, the form
$q^{\mathsf T}Gq=\E_{\mu_N}q(X)q(X-1)$ is the paired expectation and
$q^{\mathsf T}Oq=q(0)q(-1)$ is its origin product.  This notation for quadratic forms identifies $q$ with
its coefficient vector only in these matrix formulas.

\begin{lemma}\label{lem:critical-two}
Theorem~\ref{thm:critical-grid} holds when $L=2$.
\end{lemma}
\begin{proof}
\smallskip\noindent\emph{Step 1 (Rank, cap and the target).}
Here $\operatorname{Rank}_3=2p^3/[u(2Q+u-1)]\le p/u$, so $u<p$ for a code
of rank greater than one.  Since $H=n-D=2Q+u-1+\delta$, the
cap~\eqref{eq:optimizer-cap-H} gives
$m\le2+\lfloor(u+2+\delta)/(Q-2)\rfloor\le2+\lfloor(u+3)/(Q-2)\rfloor=2$,
as $u+3\le p+2<Q-2$.
The case $m=0$ gives a target $p^{2+\delta}>p$.
For $m=1,2$, let the minimizing vertex have residual factor
$q(x)q(x-1)$, and take $G$ and $O$ from~\eqref{eq:rank-moment-matrices}
at its residual length.  This length and the target are
\[
 N=2Q+2m+u-1+\delta,\qquad
 R_{n,D}=p^{4m+2+\delta}\frac{q^{\mathsf T}Gq}{q^{\mathsf T}Oq}.
\]
Since $K\le p/u\le p$ and $q^{\mathsf T}Oq=q(0)q(-1)>0$, it suffices to
prove $q^{\mathsf T}Gq>p^{-(4m+1)}q^{\mathsf T}Oq$, which gives
$R_{n,D}>p^{1+\delta}\ge K$.

\smallskip\noindent\emph{Step 2 (A Bernstein certificate for $m=1$, and
for $m=2$ when $p\ge4$).}
The two polynomial matrix inequalities
\begin{equation}\label{eq:critical-small-matrix}
 G-p^{-(4m+1)}O>0
\end{equation}
hold on the entire real domains
$m=1,p\ge3,2p^2+2\le N\le2p^2+p+1$ and
$m=2,p\ge4,2p^2+4\le N\le2p^2+p+3$.  For $1\le u<p$ and
$\delta\in\{0,1\}$ these contain every residual length with $m=1$, and
with $m=2$ when $p\ge4$; the case $p=3$, $m=2$ is treated below.
Here is their finite coefficient certificate.  Set
$p=p_0+x$, $N=2p^2+2m+s(p-1)$, $x\ge0$, $0\le s\le1$,
with $(p_0,m)=(3,1)$ or $(4,2)$; thus $x$ covers $p\ge p_0$, and $s$
runs linearly through the range of $N$.  Put $V_i\coloneqq2v_i$, $W_i\coloneqq2w_i$,
and form the polynomial matrix $\operatorname{Cert}\coloneqq8p^{4m+1}(G-p^{-(4m+1)}O)$,
whose positive scale factor clears the denominators:
\[
 \operatorname{Cert}_{ii}=4p^{4m+1}V_i-2V_iW_i,\quad
 \operatorname{Cert}_{ij}=2p^{4m+1}QV_{\min(i,j)}-V_iW_j-W_iV_j\quad(i\ne j).
\]
Every leading principal minor has an exact expansion, of degree $a_x$
in $x$ and Bernstein degree $a_s$ in $s$,
\begin{equation}\label{eq:critical-bernstein}
 \Delta(x,s)=\sum_{i=0}^{a_x}\sum_{j=0}^{a_s}
 \operatorname{bern}_{ij}x^i\binom{a_s}j s^j(1-s)^{a_s-j},
 \qquad \operatorname{bern}_{ij}\ge0,\quad \operatorname{bern}_{0j}>0,
\end{equation}
which is positive for $x\ge0$, $0\le s\le1$.
The pairs of degrees $(a_x,a_s)$ for the five minors, the two leading
principal minors for $m=1$ followed by the three for $m=2$, are
$(5,0),(14,2),(9,0),(22,2),(39,5)$; the respective minima
of $\operatorname{bern}_{0j}$ are
\[
 \begin{gathered}
 1936,\quad104753088,\quad2097144,\\
 2088570990882816,\quad96734090864111971309977600.
 \end{gathered}
\]
All $370$ rational coefficients are in
the supplementary Bernstein certificate for the five principal
minors~\cite{ComputationalCertificates}.
The defining check forms the displayed matrix, expands its
determinants, converts monomials in $s$ to Bernstein coefficients
using
$s^k=\sum_{j=k}^{a_s}\binom jk\binom{a_s}k^{-1}\binom{a_s}j
s^j(1-s)^{a_s-j}$, and checks equality and every sign.
The scale $8p^{4m+1}$ is positive.  Thus
\eqref{eq:critical-bernstein} and Sylvester's criterion prove
\eqref{eq:critical-small-matrix}, giving $R_{n,D}>p\ge K$.

\smallskip\noindent\emph{Step 3 (The case $p=3$, $m=2$, $n=35$).}
For $p=3,m=2$, the rank bound forces $u=1$, leaving the original
lengths $34,35$ and $D=16$.  At $n=35$ the weaker inequality
$G-3^{-10}O>0$ suffices, since the target has the factor $p^\delta=3$;
at $n=34$ the matrix $G-3^{-9}O$ is not positive definite, so there
the grid minimum is computed exactly.  At $n=35$, $N=23$, and
\[
 G-3^{-10}O=\frac1{59049}
 \begin{pmatrix}59048&265532&248696\\
 265532&10829504&45687200\\248696&45687200&666980864\end{pmatrix}.
\]
Its positive $LDL^{\mathsf T}$ pivots are
$59048/59049$, $878014366/5380749$, and
$3418289331072/439007183$.  Hence
$q^{\mathsf T}Gq>3^{-10}q^{\mathsf T}Oq$, and with the factor
$p^{4m+2+\delta}=3^{11}$ of Step~1 the target exceeds $3\ge K$.

\smallskip\noindent\emph{Step 4 (The case $p=3$, $m=2$, $n=34$).}
Here we certify a minimizer.  The polynomial
$\binom{34-x}{12}(x-17)(x-18)(x-20)(x-21)$ is nonnegative on the grid,
positive at zero and of degree $16$.  Take as nodes zero and its $16$
zeros,
\[
 \{0,17,18,20,21,23,24,\ldots,34\},
\]
and their degree-$16$ cardinal polynomials $\ell_j$.
Their exact expectations are all positive; specifically, in node
order they are
\begin{align*}
 &\frac{383}{1984584622260321711},\quad
 \frac{1928750560}{73503134157789693},\quad
 \frac{12373116800}{38913423965888661},\quad
 \frac{19796986880}{4323713773987629},\\
 &\frac{781980981760}{12971141321962887},\quad
 \frac{638824509440}{617673396283947},\quad
 \frac{17494941601280}{12971141321962887},\quad
 \frac{4068259285376}{480412641554181},\\
 &\frac{11212066556416}{617673396283947},\quad
 \frac{5627526608642048}{116740271897665983},\quad
 \frac{14814515651584}{160137547184727},\quad
 \frac{95970767461888}{617673396283947},\\
 &\frac{382063112364032}{1853020188851841},\quad
 \frac{34133064593408}{160137547184727},\quad
 \frac{690895767202432}{4323713773987629},\\
 &
 \frac{143580227822896}{1853020188851841},\quad
 \frac{148893112823744}{8167014906421077}.
\end{align*}
These identities can be checked directly by summing
$9^{-34}\sum_{x=0}^{34}\binom{34}x8^x\ell_j(x)$.
Since these weights are nonnegative, Lagrange interpolation and the
converse in Lemma~\ref{lem:positive-quadrature} show that the
displayed polynomial, normalized at zero, is a minimizing vertex; its
expectation is the first weight, so
$3^{34}\beta_{34,16,9}=383/119>3\ge K$.
This completes both parities $\delta=0,1$.
\end{proof}

\begin{lemma}\label{lem:critical-finite}
Theorem~\ref{thm:critical-grid} holds when $3\le L\le7$, for every
$p\ge3$.
\end{lemma}
\begin{proof}
\smallskip\noindent\emph{Step 1 (Rank bound and a comparison at a
larger length).}
By the reduction in the proof of Theorem~\ref{thm:critical-grid} we may
assume $1\le u\le\min(Q,\lceil pL/2\rceil-1)$, so $u<pL/2$; here
$H=n-D=(Q+1)L+u-3+\delta$.  For $L\ge3$ the cubic rank bound
improves to
\[
 K\le\operatorname{Rank}_3<\frac{p(L-1)}u .
\]
Indeed $(M-1)(M-2)-QL(M-L)$ increases in $M\ge QL+1$, since its
derivative $2M-3-QL$ is positive, and equals $QL(L-2)>0$ at $M=QL+1$.

A minimizing vertex with $m=0$ gives $R_{n,D}=p^{2L-2+\delta}$
exactly.  For $m\ge1$, the cap~\eqref{eq:optimizer-cap-H}, the first
inequality of Lemma~\ref{lem:paired-constant} and the terminal
expectation~\eqref{eq:terminal-expectation} give, as in the proof of
Proposition~\ref{prop:grid-budget},
\begin{equation}\label{eq:critical-paired}
 R_{n,D}\ge\operatorname{pref}_Q\frac{p^{2L-2+\delta+4m}}{V_m(H+2m-1)}.
\end{equation}
Now let integers $\bar H\ge H$ and $a\ge1$ satisfy
$\bar H-1\ge(Q-2)a$, and suppose that every residual degree allowed by
the cap satisfies $m\le a$; this $a$ is chosen case by case and is not
the cap index of~\eqref{eq:cap-index}.  Replace $V_m(H+2m-1)$ by the
larger $V_m(\bar H+2m)$ and drop the factor $p^\delta\ge1$.  At length
$\ell=\bar H+2j$ and for $j+1\le a$, the bracket in~\eqref{eq:volume-recurrence} is
$\bar H-1-(Q-2)(j+1)\ge0$, so $p^{4j}/V_j(\bar H+2j)$ does not
increase for $0\le j\le a$.  Since $\operatorname{pref}_Q<1$, every
$m\le a$, including $m=0$, therefore gives
\begin{equation}\label{eq:critical-uniform}
 R_{n,D}\ge\operatorname{pref}_Q\frac{p^{2L-2+4a}}{V_a(\bar N)},
 \qquad \bar N\coloneqq\bar H+2a.
\end{equation}
This is a numerical comparison of the expression
in~\eqref{eq:critical-paired}; no code or minimizer at the length
$\bar N$ is involved.  If moreover $Qa\le\bar N+1$, then
$v_{j-1}(\bar N)/v_j(\bar N)=j/((\bar N-j+1)b)\le b^{-2}$ for
$j\le a$, so
$V_a(\bar N)\le b^a\binom{\bar N}a/(1-b^{-2})\le b^a\bar N^a/(a!(1-b^{-2}))$.
Finally
$\operatorname{pref}_Q(1-b^{-2})=(Q-2)^3/(2(Q-1)^3)\ge343/1024>1/3$
for $Q\ge9$.

\smallskip\noindent\emph{Step 2 ($p\ge4$).}
Here $Q\ge16$, $b<Q$, and $\operatorname{Rank}_3<p(L-1)$.  In each case
below the two conditions on $\bar H$ and on $m$ are quadratic
inequalities in $p$ whose margins are nonnegative at $p=4$ and increase
with $p$; the conditions $\bar H-1\ge(Q-2)a$ and $Qa\le\bar N+1$ are
immediate.
\begin{itemize}
\item $L=3$: take $a\coloneqq3$ and $\bar H\coloneqq4Q-6$, so $\bar N=4Q$.  Then
 $H<3Q+3p/2+1\le4Q-6$, and $H-1<4(Q-2)$ gives $m\le3$.
 By~\eqref{eq:critical-uniform},
 $R_{n,D}\ge\tfrac13\cdot6p^{16}/(b^3(4Q)^3)\ge p^4/32\ge2p
 >\operatorname{Rank}_3\ge K$.
\item $4\le L\le6$: take $a\coloneqq L+1$ and $\bar H\coloneqq aQ$, so
 $\bar N=a(Q+2)\le\tfrac98aQ$.  The inequalities $L(1+p/2)-2\le Q$ and
 $L(3+p/2)+1<2Q$ give $H<\bar H$ and $m\le a$.  Hence
 \[
  R_{n,D}\ge\frac{a!}{3(9a/8)^a}\,p^{2L-2}.
 \]
 Its ratio to $p(L-1)$ is $268435456/110716875>1$ at $p=4$, $L=4$; it
 increases with $p$, and replacing $L$ by $L+1$ multiplies it by
 $p^2\tfrac89\bigl(\tfrac{L+1}{L+2}\bigr)^{L+1}\tfrac{L-1}L
 >\tfrac{8}{27}p^2\tfrac{L-1}L\ge\tfrac{32}9$, using
 $(1+1/k)^k<3$.
\item $L=7$: take $a\coloneqq9$ and $\bar H\coloneqq9Q$, so
 $\bar N=9(Q+2)\le\tfrac{81}8Q$.  Then $H<7Q+7p/2+5\le9Q$, and
 $H-1<10(Q-2)$ gives $m\le9$.  Hence
 $R_{n,D}\ge 9!\,p^{12}/(3(81/8)^9)$, whose ratio to $6p$ is
 $9!\,8^9\,4^{11}/(18\cdot81^9)>1$ at $p=4$ and grows like $p^{11}$.
\end{itemize}
In every case $R_{n,D}\ge p(L-1)>\operatorname{Rank}_3\ge K$.

\smallskip\noindent\emph{Step 3 ($p=3$, $4\le L\le7$).}
Here $Q=9$, $b=8$ and $\operatorname{pref}_9(1-1/64)=343/1024$.  Take
\begin{center}
\begin{tabular}{c|cccc}
$L$&$4$&$5$&$6$&$7$\\\hline
largest $u$&$5$&$7$&$8$&$9$\\
$\bar H$&$43$&$55$&$66$&$77$\\
$a$&$6$&$7$&$9$&$10$\\
$\bar N$&$55$&$69$&$84$&$97$
\end{tabular}
\end{center}
In each column $H\le\bar H$, the cap gives
$m\le\lfloor(\bar H-1)/7\rfloor=a$, $\bar H-1\ge7a$ and
$9a\le\bar N+1$.  Then~\eqref{eq:critical-uniform} gives
\[
 R_{n,D}\ge B_L\coloneqq\frac{343\cdot3^{2L-2+4a}}{1024\cdot8^a\binom{\bar N}a},
\]
and the four integer inequalities $B_L>3(L-1)$ hold; their values are
approximately $9.08$, $22.2$, $60.1$ and $160.3$ against $9$, $12$,
$15$ and $18$.  Hence $R_{n,D}>3(L-1)>\operatorname{Rank}_3\ge K$.

\smallskip\noindent\emph{Step 4 ($p=3$, $L=3$).}
Now $u\le4$ and $\operatorname{Rank}_3<6/u$, so $u\ge3$ gives $K<2$.
The exact rank bounds are $75/13$ for $u=1$ and $39/14$ for $u=2$, so
$K\le5$ and $K\le2$.  For $(u,\delta)\in\{(1,1),(2,0),(2,1)\}$ we have
$H\le30$ and $m\le4$; with $\bar H\coloneqq30$ and $a\coloneqq4$,
the derivation of~\eqref{eq:critical-uniform}, keeping the factor
$3^\delta$ and using the exact volume, gives
\[
 R_{n,D}\ge3^\delta\frac{49\cdot3^{20}}{144\,V_4(38)}
 =3^\delta\,\frac{6327867987}{1635790768}>3^\delta\cdot3.8,
\]
which exceeds $5$ when $\delta=1$ and exceeds $2$ in all three cases.
There remains $(u,\delta)=(1,0)$, that is $(n,D)=(52,24)$ and $H=28$,
where the cap gives $m\le3$.  Here we keep the denominator
$V_m(H+2m-1)=V_m(27+2m)$ of~\eqref{eq:critical-paired}.  At length
$\ell=27+2j$ the bracket in~\eqref{eq:volume-recurrence} is $19-7j\ge0$
for $j\le2$, so the bound decreases for $1\le m\le3$, and at $m=3$
\[
 R_{n,D}\ge\frac{49\cdot3^{16}}{144\,V_3(33)}
 =\frac{234365481}{45240464}>5\ge K.
\]
The case $m=0$ gives $R_{n,D}=81$.
\end{proof}

\subsection{The finite part of the adjacent strip}\label{app:rank-adjacent}

\begin{lemma}\label{lem:adjacent-finite}
Every tuple $(p,L,u,\delta)$ retained at the end of the proof of
Theorem~\ref{thm:adjacent-grid}, by \eqref{eq:adjacent-finite-domain}
and the rules stated there, satisfies $K\le R_{n,D}$.
\end{lemma}
\begin{proof}
Here $K\le\lfloor\operatorname{Rank}_1\rfloor$ by~\eqref{eq:adjacent-rank}.
Steps 1--4 treat the tuples with $D>15$, and Step 5 the seven tuples
with $D\le15$.

\smallskip\noindent\emph{Step 1 (A rational screen).}
First retain the tuples with $D>15$; the seven tuples with $D\le15$
are treated at the end of the proof.  The exact domain has $1549$
tuples, distributed over $L=1,\ldots,7$ as
$343,1086,72,24,16,6,2$.
For $1514$ of them, direct rational comparison gives
$\lfloor p(1+bL/u)\rfloor\le \operatorname{pref}_QU_a$ in
Proposition~\ref{prop:grid-budget}.  The full domain, the exact
fractions, and the $35$ remaining tuples are recorded in
the supplementary reduction certificate for the finite adjacent
strip~\cite{ComputationalCertificates}.
The defining check uses the finite ranges in the proof of
Theorem~\ref{thm:adjacent-grid}, computes all binomial coefficients
as integers, and compares these rational numbers by cross
multiplication.  It does not truncate any unbounded parameter.

\smallskip\noindent\emph{Step 2 (Reduction to residual grid minima).}
For the $35$ remaining tuples, let $H\coloneqq n-D$ and let $a$ be the cap
index \eqref{eq:cap-index}.  Define $\beta_j$ to be the grid minimum
of degree $2j$ at length $H+2j$.  Then
\begin{equation}\label{eq:residual-grid-minima}
 R_{n,D}=\min_{0\le j\le a}p^{n-2D+4j}\beta_j.
\end{equation}
The forward inequality follows from the vertex representation of
Lemma~\ref{lem:grid-vertex} and its cap~\eqref{eq:optimizer-cap-H}.
Conversely any residual feasible polynomial can be
multiplied by the terminal factor of degree $D-2j$; the values
outside its residual grid vanish.  The terminal expectation
identity~\eqref{eq:terminal-expectation} proves
the reverse inequality.  Thus this is an equality for the full
degree-$D$ problem, independent of any Lloyd-root description.  When
$D\ge2\lfloor(H-1)/(Q-2)\rfloor$, Lemma~\ref{lem:cap-minimum-identity}
shows that the minimum is attained at $j=a$.

\smallskip\noindent\emph{Step 3 (Exact residual minima).}
For each distinct residual problem, the data in
the supplementary quadrature certificate for the residual problems on
the adjacent strip~\cite{ComputationalCertificates} specify a set
$\mathcal Z$ of $2j$
nonzero integer roots and weights $\lambda_x\ge0$ at
$\{0\}\cup\mathcal Z$.
There are $109$ such problems, including the $33$ with $j=0$, and $j\le4$.  The normalized product
$f(x)\coloneqq\prod_{z\in\mathcal Z}(x-z)/\prod_{z\in\mathcal Z}(-z)$ is nonnegative
at every grid point $x=0,\ldots,N$.  The exact certificate conditions are
\begin{equation}\label{eq:finite-quadrature-contract}
 \sum_{x\in\{0\}\cup\mathcal Z}\lambda_x x^i
 =Q^{-N}\sum_{x=0}^N\binom Nx b^x x^i
 \quad(0\le i\le2j),\qquad N\coloneqq H+2j.
\end{equation}
By the converse in Lemma~\ref{lem:positive-quadrature} (interpolation,
and equality for $f$), they imply $\beta_j=\lambda_0$.  The supplement verifies all the grid signs and every moment
in \eqref{eq:finite-quadrature-contract}, with positive rational
denominators.  The resulting minimum in
\eqref{eq:residual-grid-minima} is at least the integer linear
rank bound for $30$ of the $35$ tuples.

\smallskip\noindent\emph{Step 4 (Explicit rank windows).}
The five others, together with the three even small-degree cases
below, have the following explicit rank windows.  In the table,
$s$ is the window index, $K_0$ the first excluded rank, and the last
column is the exact target at the original length.
\begin{center}\small
\begin{tabular}{rrrrrl}\toprule
$p$&$n$&$d$&$s$&$K_0$&$R_{n,d-1}$\\\midrule
4&60&30&3&8&$41472/5797$\\
3&36&17&6&24&$29889/1265$\\
3&38&18&3&23&$513/23$\\
4&64&31&3&87&$1801472/20757$\\
5&100&49&3&238&$5194375/21879$\\
3&28&14&3&5&$24/5$\\
3&30&15&3&5&$1467/323$\\
3&32&16&3&4&$1215/323$\\\bottomrule
\end{tabular}
\end{center}
In this order the integer vectors of coefficients of $W(1,z)$ are
\begin{align*}
 &(1,-34,561,-1924,2528,-1260),\\
 &(152,-3800,45600,-341760,1726800,-5723760,12372948,\\
 &\hspace{13mm}-17777940,17941020,-14655335,11094824,-5800680),\\
 &(3919,-90137,991507,-3682030,5875994,-3672004),\\
 &(0,0,12031317,-59851979,106814871,-70970913),\\
 &(0,0,614389146,-3058794473,5477141208,-3646368381),\\
 &(5,-90,765,-2330,3051,-1644),\\
 &(415,-7885,70965,-218560,285064,-150168),\\
 &(284,-5680,53960,-172400,232975,-126392).
\end{align*}
For each row put $M\coloneqq n/2+s$, $J\coloneqq2s-1$, and form the integer polynomials
and rows
\begin{equation}\label{eq:integer-window-contract}
 \begin{split}
 \operatorname{Ref}_W(z)&\coloneqq\sum_{i=0}^Jw_i(1+bz)^{J-i}(1-z)^i,\\
 \operatorname{Row}_W(l)&\coloneqq\sum_{i=0}^{\min(J,l)}w_i\binom{M+l-i}{M},\\
 Z_W(l)&\coloneqq-p^J\operatorname{Row}_W(l)-K_0[z^l]\operatorname{Ref}_W(z)(1-z)^{-M-1}.
 \end{split}
\end{equation}
Here $\operatorname{Row}_W(l)$ is the coefficient $a_{M-l}$ of
Lemma~\ref{lem:window-identity}, and $\operatorname{Ref}_W=p^JT_JW$
gives $Z_W(l)=K_0p^J(-a_{M-l}/K_0-a^{\rm ref}_{M-l})$; the three sign
conditions below are therefore those of~\eqref{eq:window-signs}.
Literal integer expansion gives $\operatorname{Row}_W(l)\ge0$ for $0\le l\le M$,
$\operatorname{Row}_W(M)>0$, and $Z_W(l)\ge0$ for $0\le l\le M-d$.
For the first five table rows the strict values
$\operatorname{Row}_W(M)$ at the origin $l=M$ are respectively
\[
 19939373770140162520,\quad215112462975900,\quad862052806517760,
\]
\[
 466445517832028382654006,\quad
 1942652768118693055982586732665178216.
\]
The data~\cite{ComputationalCertificates} contain all $172+57$
full-grid margins and $27+12$ tail margins, for the first five and the
last three table rows respectively.  They are independently defined by
\eqref{eq:integer-window-contract}; zeros in the tail are allowed.
Thus Lemma~\ref{lem:window-identity} excludes
$K\ge K_0$.  Integer rank gives $K\le K_0-1\le R_{n,d-1}$ in
every row.  Every table length $n$ is even, so the working length is
$2M-J=n+1$, the length $n'$ of
Section~\ref{sec:rank-windows}; it is used only in the invariant
relation, while the target is at the original length $n$.

\smallskip\noindent\emph{Step 5 (The seven tuples with $D\le15$).}
Finally, after the one-pair region has been removed, the tuples with
$D\le15$ are exactly $p=3,L=1$ with
$(\delta,u)=(0,6)$, $(0,7)$, $(0,8)$, $(1,5)$, $(1,6)$, $(1,7)$, $(1,8)$.
The seven quadratures in
the supplementary certificate for small-degree quadratures on the
adjacent strip~\cite{ComputationalCertificates} satisfy
\eqref{eq:finite-quadrature-contract}; the four odd lengths meet
the linear rank bound.  The other three are the last three rows of
the preceding table.  These certificates are exact on the full grid and complete the finite
domain.
\end{proof}

\subsection{Production of reflected products}\label{app:rank-production}

The product windows of Lemma~\ref{lem:product-window-test} need the
expansion $-\mathcal RF(l)=(l)_2\sum_jv_j(M-l)_j$ of the reflected
product $F=({\tau}-x)W_k$ of~\eqref{eq:complement-profile-parameters},
with every $v_j>0$ and a lower bound for $v_0$ in the threshold
inequality~\eqref{eq:product-window-budget}.  This subsection gives the
tools.  Lemma~\ref{lem:reflection-production} shows that adjoining one
root to $f$ acts on the coefficients of $\mathcal Rf$ by a tridiagonal
map.  The ratio argument below and
Lemma~\ref{lem:uniform-production-cone} keep the ratios of consecutive
coefficients of $\mathcal RW_k$ in a cone, which makes them positive and
bounds the first coefficient from below.  Finally~\eqref{eq:final-S}
and the formulas after it adjoin the last root ${\tau}$.
Appendices~\ref{app:rank-pfive}--\ref{app:rank-qutrit} apply these
tools on explicit length intervals.

\begin{lemma}\label{lem:reflection-production}
Suppose $f$ has degree $a<J$, and write
$\mathcal Rf(l)=(l)_{J-a}\sum_j u_j(M-l)_j$.
Multiplication of $f$ by $x-{\zeta}$ transforms each basis term
$(y)_j$, where $y\coloneqq M-l$, by
\begin{equation}\label{eq:reflection-production}
 (y)_j\longmapsto({\zeta}-a+j)(y)_{j+1}+\operatorname{diag}_j(y)_j
       +\frac{b j({M_a}-j+1)({M_a}-j)}Q(y)_{j-1},
\end{equation}
where ${M_a}\coloneqq\eta+a$ and
\begin{equation}\label{eq:production-diagonal}
 \operatorname{diag}_j\coloneqq\frac{({M_a}-j)[M+ba-Q{\zeta}-(2Q-1)j]}Q.
\end{equation}
The initial constant is $u_0=p^J/(\eta)^{\overline J}$.
\end{lemma}
\begin{proof}
On each falling factorial, \eqref{eq:factorial-reflection} and
$x(x)_a=(x)_{a+1}+a(x)_a$ give, for $U(l)\coloneqq\mathcal Rf(l)$,
\[
 \mathcal R(xf)(l)=\frac{(M-l)(\eta+l)}{Q(l+1)}U(l+1)
  +\left(J-l+\frac{2l-J}Q\right)U(l)
  +\frac{bl}Q U(l-1).
\]
This is a polynomial identity after cancellation, including $l=-1$;
$U(0)=0$ since $a<J$.
Substitute $U(l)=(l)_{J-a}\sum u_j(M-l)_j$ and use
$y(y)_j=(y)_{j+1}+j(y)_j$.  The terms in $(y)_{j+2}$ cancel,
leaving exactly \eqref{eq:reflection-production} after subtracting
${\zeta}U$.  The initial constant is the image of $1$ in
\eqref{eq:factorial-reflection}.
\end{proof}

\smallskip\noindent\emph{The ratio argument.}
We record the ratio argument used in all the product-window estimates
of Appendices~\ref{app:rank-pfive}--\ref{app:rank-qutrit}.
By~\eqref{eq:reflection-production}, one multiplication acts on the
coefficient vector $u=(u_j)$ by the production matrix
$\operatorname{Prod}$, whose nonzero entries are
$\operatorname{Prod}_{j+1,j}={\zeta}-a+j$, $\operatorname{Prod}_{jj}=\operatorname{diag}_j$, and
$\operatorname{Prod}_{j,j+1}=b(j+1)({M_a}-j)({M_a}-j-1)/Q$.
If its entries and its $2$ by $2$ minors are nonnegative (total
positivity of order two), it preserves
likelihood-ratio order: for positive vectors $u$ and $r$ with
$u_j/r_j$ nondecreasing,
$(\operatorname{Prod}u)_j/(\operatorname{Prod}r)_j$ is nondecreasing.  Cross-multiplication at adjacent
output indices proves this directly as a sum of the matrix minors
times $u_kr_j-u_jr_k$ for $j<k$.  The reversed order follows too.
For the reference vector $r_j\coloneqq\binom aj {\rho}^j$, the output divided
by $\binom{a+1}j {\rho}^j$ is $F_{\rho}(j)/(a+1)$, with
\begin{equation}\label{eq:production-reference}
 \begin{split}
 F_{\rho}(j)\coloneqq{}&\frac{j({\zeta}-a+j-1)}{\rho}+(a+1-j)\operatorname{diag}_j\\
 &+\frac{b{\rho}}Q(a-j)(a+1-j)({M_a}-j)({M_a}-j-1).
 \end{split}
\end{equation}
This includes both endpoints, with absent terms zero.  If $F_{{\rho}_-}$
increases and $F_{{\rho}_+}$ decreases, induction gives
\begin{equation}\label{eq:production-ratio-cone}
 \frac{a-j}{j+1}{\rho}_-\le\frac{u_{j+1}}{u_j}
                     \le\frac{a-j}{j+1}{\rho}_+.
\end{equation}
Indeed the input is above the ${\rho}_-$ reference in likelihood-ratio
order, and below the ${\rho}_+$ reference; the two stated monotonicities
make the same comparison with the next-degree references.
Thus, if at every step $a$ the entries and the adjacent $2$ by $2$
minors of $\operatorname{Prod}$ are nonnegative, $F_{{\rho}_-}$ increases
and $F_{{\rho}_+}$ decreases, then~\eqref{eq:production-ratio-cone}
holds at every degree, starting from the single coefficient at degree
zero, and every coefficient is positive.

\begin{lemma}[A uniform production cone]\label{lem:uniform-production-cone}
For real $Q\ge16$, integer $k\ge3$, and $M\ge8Qk$, the polynomial
$W_k$ in \eqref{eq:complement-profile-parameters} satisfies
\begin{equation}\label{eq:uniform-production-cone}
 \mathcal RW_k(l)=(l)_3\sum_{j=0}^{{j_0}}u_j(M-l)_j,
 \qquad u_j>0,
\end{equation}
with \eqref{eq:production-ratio-cone} at $a={j_0}$ and
\begin{equation}\label{eq:uniform-cone-constants}
 {\rho}_-\coloneqq\frac{Qk}{2M^2},\qquad {\rho}_+\coloneqq\frac{14Qk}{M^2},\qquad
 u_0\ge\frac{p^J}{(\eta)^{\overline J}}
                \left(\frac{11M^2}{30Q}\right)^{j_0}.
\end{equation}
Here $p\coloneqq\sqrt Q$ in this algebraic statement.
\end{lemma}
\begin{proof}
\smallskip\noindent\emph{Step 1 (Entries and minors of the production
matrix).}
Order the roots as $2k,2k+1,2k+1,\ldots,3k-2,3k-1$.
At degree $0\le a\le {j_0}-1$, the next root satisfies
${\zeta}-a\ge k+2$, ${\zeta}+a\le5k$, ${\zeta}\le3k$, and ${\zeta}>a$.
Thus all entries in \eqref{eq:reflection-production} are positive.
Since ${M_a}-j\ge M-2k$ and its other diagonal factor is at least
$M-Q({\zeta}+a)\ge M-5Qk$,
\[
 \frac{11M^2}{30Q}<\frac{189M^2}{512Q}\le \operatorname{diag}_j\le\frac{M^2}Q.
\]
The upper bound uses $ba-Q{\zeta}<0$.
Also direct subtraction gives
\[
 Q(\operatorname{diag}_j-\operatorname{diag}_{j+1})=M+ba-Q{\zeta}+(2Q-1)({M_a}-2j-1),
 \qquad0<\operatorname{diag}_j-\operatorname{diag}_{j+1}\le17M/8.
\]
For the last bound the numerator is at most $2QM+2bk$.
The only $2$ by $2$ minors that need comparison subtract an
off-diagonal product at most $6k^2M^2$ from $\operatorname{diag}_j\operatorname{diag}_{j+1}$.
The latter is greater than $121M^4/(900Q^2)$, and
$121M^2/(900Q^2k^2)\ge7744/900>6$.
All other minors are nonnegative from the tridiagonal support.

\smallskip\noindent\emph{Step 2 (The reference-vector test).}
In \eqref{eq:production-reference}, the first-term increment is
$({\zeta}-a+2j)/{\rho}$, at least $2M^2/Q$ for ${\rho}_-$ and at most
$5M^2/(14Q)$ for ${\rho}_+$.
The second-term drop is $\operatorname{diag}_j+(a-j)(\operatorname{diag}_j-\operatorname{diag}_{j+1})$, between
$11M^2/(30Q)$ and $(1+17/32)M^2/Q$.
The quartic drop is exactly
\[
 2(a-j)({M_a}-j-1)({M_a}+a-2j-1)\le5kM^2.
\]
Its contribution at ${\rho}_-$ is at most
$5bk^2/2\le5M^2/(128Q)$, and it is nonnegative for both ratios.
Consequently
\[
 \Delta F_{{\rho}_-}\ge\frac{55M^2}{128Q}>0,
 \qquad \Delta F_{{\rho}_+}\le-\frac{M^2}{105Q}<0.
\]
The ratio induction proves the assertion about every $u_j$.
Keeping only the positive diagonal inflow at index zero proves
the lower bound for $u_0$.
\end{proof}

For the last factor $x-{\tau}$, Lemma~\ref{lem:reflection-production}
with $f=W_k$, $a={j_0}$, ${M_a}=M-2$ and ${\zeta}={\tau}$ gives
$\mathcal R((x-{\tau})W_k)=-\mathcal RF$.  Put ${M_*}\coloneqq M-2$ and
\begin{equation}\label{eq:final-S}
 S\coloneqq M+b{j_0}-Q{\tau},\qquad
 -\mathcal RF(l)=(l)_2\sum_{j=0}^{{j_0}+1}v_j(M-l)_j.
\end{equation}
For $0\le j\le {j_0}$, divide $v_j$ by the positive number
$({M_*}-j)u_j/Q$.  The diagonal contribution is $S-(2Q-1)j$.
If ${\rho}_+=\gamma Q/M^2$, the lower-neighbor contribution for $j\ge1$
is at least
\begin{equation}\label{eq:final-lower-neighbor}
 \frac M\gamma\frac{{\tau}-{j_0}+j-1}{{j_0}-j+1}j,
\end{equation}
and the upper-neighbor contribution is nonnegative.  At $j=0$
the latter is exactly $b(M-3)u_1/u_0$.
The remaining coefficient is $v_{{j_0}+1}={\tau}u_{j_0}>0$.
These contributions determine every final coefficient exactly; where
the simpler bounds below do not suffice, the proofs use them directly,
as for the four coefficients written out in the first interval of the
proof of Lemma~\ref{lem:complement-pfive}.

\subsection{The complement at local dimensions at least five}\label{app:rank-pfive}

\begin{lemma}\label{lem:complement-pfive}
Lemma~\ref{lem:large-alphabet-complement} holds for $p\ge5$.
\end{lemma}
\begin{proof}
Here $Q\ge25$.  On each interval we prove, in this order, the signs of
the $v_j$, the threshold inequality~\eqref{eq:product-window-budget},
which says that the all-row ratio $K_0{\tau}^2v_0/M^{{j_0}+1}$ exceeds
one, and the target comparison $K_0<\operatorname{Budget}_Q(n)$; the
last paragraph shows that the intervals cover every $n\ge Q^2-2$.
The notation is that of
Lemmas~\ref{lem:reflection-production}
and~\ref{lem:uniform-production-cone} and of~\eqref{eq:final-S}: $v_j$
is normalized by $(M-2-j)u_j/Q$, the upper-neighbor term at zero is
$b(M-3)u_1/u_0$, and the effective origin diagonal is
$S+b(M-3)u_1/u_0$, the normalized $v_0$.  The floor in
\eqref{eq:complement-profile-parameters} gives
\begin{equation}\label{eq:complement-T-bounds}
 \frac{n+1}{2b}+k-2+\frac\epsilon2\le {\tau}
 <\frac{n+1}{2b}+k-1+\frac\epsilon2,
\end{equation}
and therefore
$S>b(k-1-\epsilon/2)+1-(n+1)/(2b)$.
The elementary inequalities used for the targets are
$\operatorname{pref}_Q>1/3$, $(b/Q)^Q>1/3$, and $b/4>\sqrt Q$.
The last follows from $Q^2-18Q+1>0$ for $Q\ge25$.

\emph{The first interval.}
For $Q^2-2\le n<39Q$ take $k\coloneqq2$, $J\coloneqq5$, ${j_0}\coloneqq2$, and $K_0\coloneqq Q^5$.
Let $\operatorname{init}\coloneqq p^5/(\eta)^{\overline5}$, the initial constant
of Lemma~\ref{lem:reflection-production}, and
$Z\coloneqq(M-4Q)(M-4Q-1)+4Qb$.
Two applications of \eqref{eq:reflection-production} give
\[
 u_0=\operatorname{init}(M-3)(M-4)Z/Q^2,\quad
 u_1=8\operatorname{init}(M-4)(M-5Q)/Q,\quad u_2=20\operatorname{init}.
\]
The final four coefficients are
\begin{align*}
 v_0&=\frac{\operatorname{init}(M-2)(M-3)(M-4)}{Q^3}
             [SZ+8Qb(M-5Q)],\\
 v_1&=\frac{\operatorname{init}(M-3)(M-4)}{Q^2}
       [({\tau}-2)Z+8(M-5Q)(S-2Q+1)+40bQ],\\
 v_2&=\frac{\operatorname{init}(M-4)}Q
       [8({\tau}-1)(M-5Q)+20(S-4Q+2)],\qquad v_3=20{\tau}\operatorname{init}.
\end{align*}
Put $x\coloneqq M/Q$.  The interval implies $Q/2\le x<20$,
${\tau}\ge(Q+1)/2\ge13$, and $S/Q\ge-1/3$.
The upper-neighbor term at zero, divided by $Q$, is at least
\[
 \frac{192}{25}\frac{x-5}{(x-4)^2+4}\ge\frac{144}{325},
\]
because $(x-5)/((x-4)^2+4)$ decreases for $x\ge25/2$ and
its value at $20$ is $3/52$.
Thus the effective origin diagonal exceeds $107Q/975>Q/10$.
The bracket for $v_1$, divided by $Q^2$, is greater than $424$:
one may use
$({\tau}-2)\ge11$, $Z/Q^2\ge(x-4)^2-(x-4)/Q$, and
$8(M-5Q)(S-2Q+1)/Q^2\ge-(56/3)(x-5)$;
the resulting quadratic increases for $x\ge25/2$ and is already
greater than $424$ there.
For $v_2$, the lower bound
$96Q(x-5)-(260/3)Q+40$ is positive.
All four coefficients are therefore positive.

The same bounds give
$u_0\ge(9/25)\operatorname{init}M^4/Q^2$ and $v_0\ge81\sqrt Q/2500$.
For example $M\ge Q^2/2$ makes
$(M-3)(M-4)>9M^2/10$ and $Z>2M^2/5$;
$(\eta)^{\overline5}=(M)_5\le M^5$ and $M-2\ge9M/10$
then prove these inequalities.
The ratio in \eqref{eq:product-window-budget} exceeds
\[
 \frac{6561}{5000000}Q^{5/2}\ge\frac{6561}{1600}>1.
\]
Here ${\tau}\ge(M-2)/b\ge9M/(10b)$ and $M<20Q$ were used.
At the lower original length the target bound is greater than
$b^{10}/3^{13}>Q^5$, since $b^2/Q\ge23$ and $23^5>3^{13}$.
To see the first inequality directly,
$\operatorname{Budget}_Q(Q^2-2)
 >b^{Q/2-2}/3^{Q/2+1}$;
this exceeds $b^{10}/3^{13}$ for $Q\ge25$, because
$Q/2-12\ge1/2$ and $(b/3)^{Q/2-12}>1$.

\emph{The second interval.}
For $39Q\le n<3Qb-2$ take $k\coloneqq3$, ${j_0}\coloneqq4$, $J\coloneqq7$, $K_0\coloneqq Q^7$.
The ordered roots are $6,7,7,8$, and $M\ge39Q/2$.
Lemma~\ref{lem:uniform-production-cone} does not apply on the whole
interval, since it assumes $M\ge8Qk=24Q$; its proof, with
these four roots, gives
\begin{equation}\label{eq:pfive-special-cone}
 \operatorname{diag}_j\ge\frac{3M^2}{7Q},\quad
 {\rho}_-\coloneqq\frac{3Q}{2M^2},\quad {\rho}_+\coloneqq\frac{27Q}{M^2},\quad
 u_0\ge \operatorname{init}\left(\frac{3M^2}{7Q}\right)^4,
 \quad \operatorname{init}\coloneqq\frac{p^7}{(\eta)^{\overline7}}.
\end{equation}
Here are explicit scalar checks for that application.
${\zeta}+a\le11$, ${\zeta}-a\ge5$, and $a\le3$.
The lower diagonal follows from
$(1-6/M)(1-11Q/M)\ge(1-4/325)(17/39)>3/7$.
The adjacent off-diagonal product is at most $21M^2$, whereas
$\operatorname{diag}_j\operatorname{diag}_{j+1}\ge(9/49)(39/2)^2M^2>54M^2$.
In \eqref{eq:production-reference}, the lower-reference increment
has positive reserve at least $415M^2/(676Q)$, using
$5/{\rho}_-=10M^2/(3Q)$, the second-term bound
$[1+51/(8\cdot39/2)]M^2/Q$, and the quartic bound
$5kM^2$.
For the upper reference, the reserve is at most
$(11/27-3/7)M^2/Q=-4M^2/(189Q)$.
Thus the same ratio induction proves
\eqref{eq:pfive-special-cone}.

Equation~\eqref{eq:complement-T-bounds} gives $S>-1/2$, ${\tau}>41/2$.
The lower-neighbor term \eqref{eq:final-lower-neighbor}
is greater than $(403/180)Qj$ for $j\ge1$.
At zero, the upper neighbor is at least
$6bQ(M-3)/M^2\ge297/80>3$.
For this last bound use
\begin{equation}\label{eq:pfive-closed-endpoint}
 M\le3Qb/2+2<8Qb/5,
\end{equation}
including equality in the first bound, and $M-3\ge99M/100$.
All final coefficients are positive and
$v_0\ge(M-2)u_0/Q$.
Using ${\tau}\ge M/b$ in \eqref{eq:product-window-budget}, its ratio
is greater than
$(9/16)(3/7)^4Q^{3/2}\ge91125/38416>1$.
The original target at $39Q$ exceeds $\operatorname{pref}_QQ^{31/4}>Q^7$:
indeed \eqref{eq:distance-target-budget} is greater than
$\operatorname{pref}_Q(b/4)^{39/2}/b^2>\operatorname{pref}_QQ^{31/4}$,
and $Q^{3/4}>3$.

\emph{Two further fixed intervals.}
Use $(k,\operatorname{width})\coloneqq(4,2)$ on $[64Q,4Qb+2)$ and $(6,4)$ on
$[96Q,8Qb-2)$, in both cases $K_0\coloneqq Q^{2k+1}$; here $\operatorname{width}$
is the upper bound for $M/Q^2$ on the interval.
The uniform cone applies, since $n\ge16Qk$ gives $M\ge8Qk$.  The
endpoint and floor bounds give
$M<\operatorname{width}Q^2$, ${\tau}>9k-2$, and $S\ge Q/3$.
For $k=4$ this follows from
$S>(k-3/2-\operatorname{width})Q-k+5/2-2/b>Q/3$;
for $k=6$ the same inequality without $2/b$ suffices.
The lower-neighbor term is greater than $2Qj$, since
\[
 \frac{M}{14k}\frac{{\tau}-{j_0}-1}{{j_0}+1}
 >\frac{4Q}7\frac{7k-1}{2k-1}>2Q.
\]
Thus $v_j>0$ and $v_0\ge(M-2)Su_0/Q$.
Using $M-2\ge9M/10$, $(\eta)^{\overline J}=(M)_J\le M^J$,
and ${\tau}>8k$, the all-row ratio exceeds
\begin{equation}\label{eq:fixed-window-ratio}
 \frac{96k^2}{5\operatorname{width}^3}3^{-(2k-2)}Q^{k-5/2}.
\end{equation}
At $Q=25$ its two values are $1600/243$ and $31250/2187$,
both greater than one, and the expression increases with $Q$.
At the lower length $16Qk$,
$\operatorname{Budget}_Q(n)>\operatorname{pref}_Q(b/4)^{8k}/b^2
 >\operatorname{pref}_QQ^{4k-2}>Q^{2k+1}$.

\emph{The growing interval.}
For $n\ge8Qb-2$ write $n'=2Qb\ell-1+2s$,
$0\le s<Qb$, $\ell\ge4$, and take $k\coloneqq2\ell$.
Put
\[
 c\coloneqq\left\lceil\frac{2s+\epsilon(b-1)}{2b}\right\rceil,
 \qquad\delta_1\coloneqq Qc-s.
\]
The exact floor identities are
\begin{equation}\label{eq:growing-parity-identities}
 M=Qb\ell+k+s,\quad {\tau}=Q\ell+k+c-2,\quad
 0\le\delta_1\le3Q-2,\quad
 S=bk-Q\ell+2-\delta_1.
\end{equation}
They follow by writing $n+1=2Qb\ell+2s-\epsilon$ in
\eqref{eq:complement-profile-parameters}; in particular both
original parities are retained.  Now $M/(Qk)\ge b/2\ge12$ and
$S\ge Q-4$.
The lower-neighbor term is at least
$Qb(Q-3)j/112>(2Q-1)j$.
For the strict inequality, its cleared difference is
$Q^3-4Q^2-221Q+112$, whose expansion at $Q=25+x$ has
positive coefficients (constant $7712$, followed by $1454,71,1$).
Thus all $v_j$ are positive, with $v_0\ge(M-2)(Q-4)u_0/Q$.
Here ${\tau}\ge Q\ell$ and $M<Q^2(\ell+1)\le5Q^2\ell/4$.
The all-row ratio is at least
\[
 \frac{288}{625}\frac{Q-4}{\sqrt Q\,\ell}
                  \left(\frac{11\sqrt Q}{30}\right)^{4\ell-2}>1.
\]
It increases in $Q\ge25$; replacing $11\sqrt Q/30$ by $3/2$
shows that it increases with integer $\ell\ge4$ and is at least
$6048/3125$.
For the original target, monotonicity permits the even lower endpoint
$n=2Qb\ell-2$, and direct substitution gives
\[
 \operatorname{Budget}_Q(n)>\operatorname{pref}_Q(b/4)^{b\ell}/b^2
       >\operatorname{pref}_QQ^{5\ell-2}>Q^{4\ell+1}.
\]
The middle inequality uses $(b/4)^b>Q^5$, which holds for $Q\ge25$
already from $b/4>\sqrt Q$ and $b\ge24$.

\emph{Coverage.}
Finally the four fixed intervals and the growing interval cover every
$n\ge Q^2-2$.  If $Q^2-2\ge39Q$, the first interval is empty
and the second or a later interval starts earlier; otherwise its
upper endpoint meets the second exactly.
The overlaps $3Qb-2\ge64Q$ and $4Qb+2\ge96Q$ hold for $Q\ge25$.
The last fixed interval ends at the start of the growing one.
This proves the lemma.
\end{proof}

\subsection{The complement at local dimension four}\label{app:rank-pfour}

\begin{lemma}\label{lem:complement-pfour}
Lemma~\ref{lem:large-alphabet-complement} holds for $p=4$.
\end{lemma}
\begin{proof}
Here $Q=16$, $b=15$, the notation is that of the proof of
Lemma~\ref{lem:complement-pfive}, and
\begin{equation}\label{eq:pfour-target}
 \operatorname{Budget}_{16}(n)=\frac{49}{120}(15/16)^{n/2}15^{n/32-15/8}
       >\frac{49}{120}\frac{5^{n/32}}{225}.
\end{equation}
The latter follows from $(15/16)^{16}>1/3$.
For $k\ge3$, \eqref{eq:complement-T-bounds} gives ${\tau}\ge M/15$.

\emph{Lengths $254\le n<544$.}
Use $k\coloneqq2$, so $130\le M\le274$.
In the explicit coefficients $v_0,\ldots,v_3$ of the first interval in
the proof of Lemma~\ref{lem:complement-pfive} put
\[
 Z\coloneqq(M-64)(M-65)+960,\quad S=M+30-16{\tau},\quad
 W\coloneqq SZ+1920(M-80).
\]
Their first coefficient simplifies to $v_0=W/[4M(M-1)]$.
The other three have positive prefactors times
\[
 ({\tau}-2)Z+8(M-80)(S-31)+9600,\quad
 8({\tau}-1)(M-80)+20(S-62),\quad {\tau}.
\]
The exact parity formulas are
${\tau}=\lceil(M+5)/15\rceil$ for even original $n$ and
${\tau}=\lceil(M-2)/15\rceil$ for odd $n$.
The even formula maximizes ${\tau}$ and minimizes $S,W$ at each $M$.
In this range $9\le {\tau}\le19$ and $S\ge-8$.
Putting $x\coloneqq M-80\ge50$, the first two displayed brackets are at
least $7x^2-95x+18000>0$ and $64x-1400>0$.
For fixed ${\tau}$,
\[
 \frac{dW}{dM}=Z+S(2M-129)+1920\ge M^2-145M+8072>0.
\]
In a full even block $10\le {\tau}\le19$ the first $M$ is $15{\tau}-19$.
Its value is
\[
 W_{\rm first}\coloneqq-225{\tau}^3+4980{\tau}^2-6687{\tau}-102828,
\]
and
$W_{\rm first}-24624=(19-{\tau})(225{\tau}^2-705{\tau}-6708)\ge0$.
The initial partial ${\tau}=9$ block starts at $M=130$, with value
$180000$.  Thus $W\ge24624$ throughout; for $M\le154$ the
same calculation with ${\tau}=9,10,11$ gives $W\ge103302$.
All final coefficients are positive.

Take $K_0\coloneqq1024$ for $254\le n<304$ and $K_0\coloneqq4096$ for
$304\le n<544$.  The weaker sufficient condition in
Lemma~\ref{lem:product-window-test} gives the ratios
\[
 \frac{1024\cdot103302}{4\cdot154^2\cdot153}
       =\frac{734592}{100793}>1,\qquad
 \frac{4096\cdot24624}{4\cdot274^2\cdot273}
       =\frac{2101248}{1707979}>1.
\]
For the first target comparison $\operatorname{Budget}_{16}(254)>1024$,
the following exact integer inequalities suffice:
\[
 20\cdot15^{16}>7\cdot16^{16},\quad
 21^{15}>9^{16}2^{14},\quad15\cdot3^8>4^8,\quad
 49\cdot21^7>1024\cdot4500\cdot4^7.
\]
In order, they give $(15/16)^{16}>7/20$,
$(21/4)^{15/16}>9/2$, $15^{1/8}>4/3$, and
$\operatorname{Budget}_{16}(254)>(49/4500)(21/4)^7>1024$.
At $n=304$, \eqref{eq:pfour-target} and $\sqrt5>2$ give
$\operatorname{Budget}_{16}(304)>765625/108>4096$.
Monotonicity covers the two full intervals.

\emph{Three small production cones.}
For $k=3,4,5$, respectively, the following constants are valid:
\begin{center}
\begin{tabular}{c|rrrrrr}
$k$&${j_0}$&Lower $M/Q$&$c_{\rm diag}$&$\gamma$&Maximal ${\zeta}+a$&Maximal $a({\zeta}-1)$\\\hline
3&4&17&$1/3$&34&11&21\\
4&6&24&$8/25$&56&16&50\\
5&8&32&$1/3$&70&21&91
\end{tabular}
\end{center}
They mean $\operatorname{diag}_j\ge c_{\rm diag}M^2/Q$, ${\rho}_-\coloneqq Qk/(2M^2)$,
${\rho}_+\coloneqq\gamma Q/M^2$, and
\[
 u_0\ge p^J(\eta)^{-\overline J}(c_{\rm diag}M^2/Q)^{j_0},
\]
where $(\eta)^{-\overline J}$ means $1/(\eta)^{\overline J}$.  The last
two columns are maxima over the steps $a$ of
Lemma~\ref{lem:reflection-production} and their roots ${\zeta}$; they
enter the diagonal and minor bounds below.
To verify them without a length enumeration, use
${M_a}-j\ge M-2k$ and the other diagonal factor
$M-Q({\zeta}+a)$.  At the three lower endpoints their normalized
products are $1596/4624>1/3$, $47/144>8/25$, and
$2761/8192>1/3$, and increase with $M$.
The upper diagonal bound is $M^2/Q$ and its difference bound is
$17M/8$.  The adjacent minor test is implied by
\[
 289/9>21,\qquad36864/625>50,\qquad1024/9>91.
\]
In the reference-vector test the first increment at ${\rho}_-$ is at
least $2M^2/Q$, the second drop at most $3M^2/(2Q)$, and
the last drop at most $M^2/(10Q)$: these follow from
$17({j_0}-1)/(8(M/Q))<1/2$ and
$5k^2/[2(M/Q)^2]<1/10$ in each row.
At ${\rho}_+$ the increment is bounded above by
$11/34-1/3$, $16/56-8/25$, or $21/70-1/3$, respectively,
times $M^2/Q$, and is negative.
The ratio argument after \eqref{eq:production-reference} proves
all three cones and their lower bounds for $u_0$.

\emph{Lengths $544\le n<768$.}
Take $k\coloneqq3$ and $K_0\coloneqq2Q^7$.  For $n<718$,
$M>17Q$, ${\tau}\ge20$, and $S>-1/2$; the lower-neighbor
term in \eqref{eq:final-lower-neighbor} exceeds $2Qj$.
For $718\le n<768$, $M\ge22Q$, ${\tau}\ge25$, and $S>-32/15$;
the same term is at least $(231/68)Qj$.
Both bounds prove positivity for $j\ge1$.
At zero keep the upper neighbor:
\[
 b(M-3)u_1/u_0\ge6bQ(M-3)/M^2\ge237/65.
\]
Here $M\ge240$, $M-3\ge79M/80$, and $M<13Qb/8$.
Its sum with $S$ exceeds $59/39>3/2$, so
$v_0\ge(3/2)(M-2)u_0/Q$.
With ${\tau}\ge M/b$, $M-2\ge9M/10$, and
$(\eta)^{\overline7}\le M^7$, the all-row ratio exceeds
$(4/195)Q^{3/2}=256/195>1$.  The factor $2$ in $K_0$ is needed for
this bound, which is proportional to $K_0$: for $K_0=Q^7$ it would be
$(2/195)Q^{3/2}=128/195<1$.
The original target exceeds $2Q^7$ because
$49\cdot5^{17}>240\cdot225\cdot16^7$ in
\eqref{eq:pfour-target} at $n=544$.

\emph{Lengths $768\le n<1024$.}
Take $k\coloneqq4$, $K_0\coloneqq Q^9$.  Now ${\tau}\ge28$, $S>4$, and
$M<21Q^2/10$.  For $j\ge2$, the lower-neighbor term divided
by $j$ is at least $(69/35)Q>2Q-1$.
For $j=1,n<832$, use $S>10$ and lower-neighbor bound
$(11/7)Q$; for $n\ge832$, use $M\ge26Q$, ${\tau}\ge30$,
and the bound $(13/7)Q$, with $S>4$.
All final coefficients are positive and $v_0\ge4(M-2)u_0/Q$.
The ratio in \eqref{eq:product-window-budget} is greater than
\[
 \frac{12}{7}(8/25)^6Q^{5/2}
       =\frac{3221225472}{1708984375}>1.
\]
The target comparison follows from
$49\cdot5^{24}>120\cdot225\cdot16^9$.

\emph{Lengths $1024\le n<1536$.}
Take $k\coloneqq5$, $K_0\coloneqq Q^{11}$.  Here ${\tau}\ge38$, $S>2$, and
$M<31Q^2/10$.  For $j\ge2$ the lower neighbor divided by $j$
is at least $(496/245)Q>2Q$.
For $j=1,n<1280$, the bounds $S>10$ and $(12/7)Q$ suffice.
For $n\ge1280$, use $M\ge40Q,{\tau}\ge46$ to obtain
$(19/7)Q>2Q$.
Thus $v_0\ge2(M-2)u_0/Q$ and the all-row ratio is greater than
$(18/31)3^{-8}Q^{7/2}=32768/22599>1$.
The target comparison is $49\cdot5^{32}>120\cdot225\cdot16^{11}$.
For this and the preceding ratio, substitution into
$K_0{\tau}^2v_0/M^{{j_0}+1}$ uses only ${\tau}\ge M/b$,
$M-2\ge9M/10$, $(\eta)^{\overline J}\le M^J$,
and the displayed upper bound on $M$.

\emph{Lengths $1536\le n<1918$.}
Take $k\coloneqq6$, $K_0\coloneqq Q^{13}$.  The uniform cone applies, and its
diagonal constant may be reduced to $1/3$.
Here $S>9/2>4$, $M<4Q^2$, and the lower neighbor exceeds
$2Qj$.  Thus $v_0\ge4(M-2)u_0/Q$, and the ratio is greater
than $(9/10)3^{-10}Q^{9/2}=131072/32805>1$.
The target comparison is $49\cdot5^{48}>120\cdot225\cdot16^{13}$.

\emph{The growing range.}
For $n\ge1918$, write $n'=2Qb\ell-1+2s$ as in
\eqref{eq:growing-parity-identities}, with $\ell\ge4$.
Take $k\coloneqq\lceil3\ell/2\rceil+1$, so $k\le9\ell/5$ and
$M/(Qk)\ge25/3>8$.
The identities \eqref{eq:growing-parity-identities} hold unchanged.
For $\ell\ge5$ they give
$S\ge(Q-3)\ell/2-2Q+3\ge7/2$.
For $\ell=4$, $k=7$ and $S\ge-3$.
In all cases ${\tau}-{j_0}-1\ge(Q-2)\ell$, so the lower neighbor divided
by $j$ is at least
\[
 \frac{25Qb(Q-2)}{2268}>2Q+2;
\]
the cleared margin at $Q=16$ is $6888$.
Thus every final coefficient with $j\ge1$ is positive.
For $\ell=4$ the upper neighbor at zero satisfies
\[
 b(M-3)u_1/u_0\ge42bQ(M-3)/M^2
       \ge189b/(25Q)=2835/400,
\]
using $M\le5Q^2$ and $M-3\ge9M/10$.
With $S\ge-3$ it makes the effective origin diagonal greater than $4$.
For $\ell\ge5$ the bound $S\ge7/2$ already suffices.
Denote these lower bounds by $S_{\rm eff}$, so
$v_0\ge(M-2)S_{\rm eff}u_0/Q$.
Since ${\tau}\ge Q\ell$, $M<Q^2(\ell+1)$, the all-row ratio is at least
\[
 \frac9{10}\frac{S_{\rm eff}}{\sqrt Q}
     \frac{\ell^2}{(\ell+1)^3}
       \left(\frac{11\sqrt Q}{30}\right)^{j_0}.
\]
Here $11\sqrt Q/30=22/15$, ${j_0}\ge3\ell$, and
$(22/15)^3>3$.  For $\ell\ge5$, use
$\ell^2/(\ell+1)^3\ge1/(2\ell)$ to bound this below by
$(63/160)3^\ell/\ell\ge15309/800>1$.
For $\ell=4$ it exceeds $(9/10)(16/125)3^4=5832/625>1$.
The original target comparison is
\[
 R_{n,D}>\operatorname{pref}_Q(b/4)^{b\ell}/b^2
       >\operatorname{pref}_QQ^{5\ell-2}>Q^{2k+1},
\]
because $(15/4)^{15}>16^5$, $k\le2\ell$, and
$\operatorname{pref}_QQ^{\ell-3}>1$.
The first inequality includes the even lower endpoint, by
substitution in \eqref{eq:distance-target-budget}.
All intervals meet at the stated endpoints, proving the lemma.
\end{proof}

\subsection{The qutrit complement}\label{app:rank-qutrit}

\begin{lemma}\label{lem:qutrit-growing}
For $p=3$, $n\ge1300$, and \eqref{eq:strict-complement}, the
choice $k\coloneqq\lceil(n'+1)/130\rceil$ and $K_0\coloneqq9^{2k+1}$ in
\eqref{eq:complement-profile-parameters} satisfies the product-window
test of Lemma~\ref{lem:product-window-test} and
$K_0<\operatorname{Budget}_9(n)$.
\end{lemma}
\begin{proof}
\smallskip\noindent\emph{Step 1 (The window and its domain).}
The even integer $n'+1$ satisfies
$130(k-1)+2\le n'+1\le130k$.  Consequently $k\ge11$ and
\begin{equation}\label{eq:qutrit-window-domain}
 66k-64\le M\le66k,\qquad n\ge130(k-1).
\end{equation}
In particular $n\ge4b(k+1)-1$, so $M\ge J$ and ${\tau}\ge3k$, as
Lemma~\ref{lem:product-window-test} requires.

\smallskip\noindent\emph{Step 2 (Root steps and diagonal entries).}
Steps 2--4 prove the
ratio cone~\eqref{eq:production-ratio-cone} for $W_k$ by the ratio
argument of Appendix~\ref{app:rank-production}, for every integer
$k\ge10$ in this full interval; only $k\ge11$ is needed.
At step $a=0,\ldots,2k-3$, the next root is
${\zeta}=2k+\lfloor(a+1)/2\rfloor$.  Thus
\begin{equation}\label{eq:qutrit-root-step}
 {\zeta}-a\ge k+2,\quad {\zeta}+a\le5k-4,\quad
 11k\le9{\zeta}-8a\le18k+1.
\end{equation}
In \eqref{eq:production-diagonal} one has
$\operatorname{diag}_j=({M_a}-j)[M+8a-9{\zeta}-17j]/9$.  Every existing entry is positive,
and
\begin{equation}\label{eq:qutrit-corrected-diagonal}
 \begin{gathered}
 \operatorname{diag}_j\ge\frac{(M-2k)(M-45k+36)}9\eqqcolon\operatorname{diag}_{\min},\\
 \operatorname{diag}_j\le\frac{M^2-11kM}9,
 \quad0<\operatorname{diag}_j-\operatorname{diag}_{j+1}\le17M/8.
 \end{gathered}
\end{equation}
For the last bound the exact numerator is
$M+8a-9{\zeta}+17({M_a}-2j-1)$, at most $18M+16k\le153M/8$,
since $M\ge59k$ for $k\ge10$.

\smallskip\noindent\emph{Step 3 (Adjacent minors).}
Here it is useful to retain the exact adjacent minor.  With
$\operatorname{step}_j\coloneqq M+8a-9{\zeta}-17j$, it is
\begin{equation}\label{eq:qutrit-exact-minor}
 \frac{({M_a}-j)({M_a}-j-1)}{81}
       \{\operatorname{step}_j(\operatorname{step}_j-17)-72(j+1)({\zeta}-a+j)\}.
\end{equation}
For $0\le j\le a-1$, the bracket is at least
\[
 (M-45k+53)(M-45k+36)-72(2k-3)(3k-2).
\]
It increases with $M$ in \eqref{eq:qutrit-window-domain}.
At $M=66k-64$, $k=10+x$, it is
$1946+297x+9x^2>0$.
All other minors are nonnegative by tridiagonal support.

\smallskip\noindent\emph{Step 4 (The reference-vector test).}
Take
\begin{equation}\label{eq:qutrit-cone-ratios}
 {\rho}_-\coloneqq6k/M^2,\qquad {\rho}_+\coloneqq(146k-48)/M^2.
\end{equation}
At ${\rho}_+$, the first increment in
\eqref{eq:production-reference} is at most
$(5k-4)M^2/(146k-48)$, and the second drop is at least
$\operatorname{diag}_{\min}$.  The third term is nonincreasing.
The sufficient positive difference, after clearing denominators, is
\[
 (146k-48)(M-2k)(M-45k+36)-9(5k-4)M^2.
\]
After division by $M^2$ it increases with $M$; at the lower endpoint
and $k=10+x$ its coefficients are
$963360,277072,20728,204$, all positive.
At ${\rho}_-$, the first increment is at least
$(3/2+3/k)M^2/9$.
The second drop is at most
$M^2/9+(109k/36-51/8)M$.
The exact quartic drop is at most $5kM^2$, so its contribution
is at most $(80/3)k^2$.
After division by $M^2/9$ the total increment is at least
\[
 \frac12+\frac3k-\frac{109k/4-459/8}{M}-\frac{240k^2}{M^2}.
\]
This increases with $M$.  At its lower endpoint, multiplication
by $8kM^2$ gives
$4(k+6)M^2-k(218k-459)M-1920k^3$,
whose coefficients at $k=10+x$ are
$10556664,2419016,148478,1116$, all positive.
Thus $F_{{\rho}_-}$ increases and $F_{{\rho}_+}$ decreases, proving the
ratio cone \eqref{eq:production-ratio-cone} for $W_k$.

\smallskip\noindent\emph{Step 5 (A lower bound for $u_0$).}
At index zero retain the stronger diagonal bound
\[
 \operatorname{diag}_0\ge(M-2k)(M-18k-1)/9\ge2M^2/27.
\]
For the latter inequality, after clearing denominators its
polynomial at $M=66k-64,k=10+x$ has coefficients
$6688,5280,504$; the normalized inequality increases with $M$.
It follows that
\begin{equation}\label{eq:qutrit-origin-cone}
 u_0\ge\frac{3^J}{(\eta)^{\overline J}}(2M^2/27)^{j_0}.
\end{equation}
All these inequalities use the $45k-36$ diagonal and the
$18k+1$ root-step bound in the displayed formulas.

\smallskip\noindent\emph{Step 6 (Signs of the final coefficients).}
We next prove the final reflected signs.  The exact parity formulas
are ${\tau}=k+\lceil(M-k)/8\rceil-2$ for odd original $n$ and
${\tau}=k+\lceil(M-k)/8+7/16\rceil-2$ for even $n$.  Hence
\begin{equation}\label{eq:qutrit-T-bounds}
 (M+7k)/8-2\le {\tau}<(M+7k)/8-9/16.
\end{equation}
Normalize $v_j$ by $(M-2-j)u_j/9$ for $0\le j\le {j_0}$.
The diagonal is $S-17j$, the lower neighbor is at least
\[
 \frac{9M}{146k-48}\frac{{\tau}-{j_0}+j-1}{{j_0}-j+1}j,
\]
and the upper neighbor is at least $c_{\rm up}({j_0}-j)$, where
$c_{\rm up}\coloneqq48k(M-2k-1)/M^2\ge7/10$.
The last expression decreases with $M$ in the domain; at $M=66k$
the comparison reduces to $114k\ge240$.

Put $t\coloneqq66k-M$, $S_{\rm eff}\coloneqq381k/40-M/8-987/80$, and
\begin{equation}\label{eq:qutrit-final-polynomial}
 \Phi(k,t,j)\coloneqq
 (S_{\rm eff}-177j/10)(2k-1-j)
 +\frac{9M}{146k-48}\{(M-9k)/8+j-1\}j,
 \qquad M=66k-t.
\end{equation}
Here $S_{\rm eff}-177j/10$ bounds the diagonal plus the upper neighbor
from below: the upper bound for ${\tau}$ in
\eqref{eq:qutrit-T-bounds} gives $S=M+8{j_0}-9{\tau}>-M/8+65k/8-175/16$,
the upper neighbor is at least $(7/10)({j_0}-j)$, and $S_{\rm eff}$ is
the sum of these two bounds at $j=0$, while $177/10=17+7/10$.  At
$j=0$, $S_{\rm eff}$ is thus a lower bound for the effective origin
diagonal, as in the proof of Lemma~\ref{lem:complement-pfour}.
The normalized final coefficient, multiplied by ${j_0}-j+1=2k-1-j>0$,
is at least $\Phi$.  Indeed the upper bound for ${\tau}$ in
\eqref{eq:qutrit-T-bounds} supplies the lower bound for $S$,
and its lower bound supplies the nonnegative lower-neighbor term;
these are separate valid inequalities.

Here is the exact finite polynomial identity proving positivity
throughout $k\ge11$, $0\le t\le64$, $0\le j\le2k-2$.
Form $\operatorname{Poly}\coloneqq80(146k-48)\Phi$, a polynomial with integer
coefficients (the factor $80(146k-48)$ clears the denominators of
$\Phi$), and substitute
\begin{equation}\label{eq:qutrit-positive-chamber}
 k=11+x,\qquad t=64y/(1+y),\qquad
 j=(2k-2)z/(1+z).
\end{equation}
After multiplication by $(1+y)^2(1+z)^{64}$ the resulting
polynomial has degree $(3,2,64)$ in $(x,y,z)$.  The factor
$(1+y)^2(1+z)^2$ clears the denominators; the further factor
$(1+z)^{62}$ is a P\'olya-type multiplier, without which some
coefficients are negative.
Every one of its $4\cdot3\cdot65=780$ coefficients is a positive
integer, the least being $29784$.
The complete array in
the supplementary coefficient certificate for the qutrit
complement~\cite{ComputationalCertificates} is specified
by \eqref{eq:qutrit-final-polynomial}--\eqref{eq:qutrit-positive-chamber}.
An exact check expands these defining expressions, cancels only
the displayed positive denominators, and compares every
coefficient, including the highest-degree rows.
Positivity holds for $x,y,z\ge0$; the endpoints $t=64$ and $j={j_0}$
follow from the positive leading coefficients in $y$ and $z$.
Thus $\Phi>0$ on the whole closed domain, proving $v_j>0$.
The last coefficient $v_{{j_0}+1}={\tau}u_{j_0}$ is positive separately.
At zero the same bounds give
\[
 v_0\ge(M-2)S_{\rm eff}u_0/9,\qquad
 S_{\rm eff}\ge(102k-987)/80\ge k/10\quad(k\ge11).
\]

\smallskip\noindent\emph{Step 7 (The threshold inequality).}
All roots of $W_k$ lie below ${\tau}$ by
\eqref{eq:qutrit-T-bounds}.  Use ${\tau}\ge M/8$, $M-2\ge9M/10$,
$(\eta)^{\overline J}\le M^J$, and $M\le66k$ in
\eqref{eq:product-window-budget}.  Its ratio is at least
\[
 \frac{19683S_{\rm eff}\,4^{k-1}}{42240k}
 \ge\frac{19683\cdot4^{10}}{422400}>1.
\]
Hence $K<K_0=9^{2k+1}$.

\smallskip\noindent\emph{Step 8 (The target comparison).}
For the original target, \eqref{eq:distance-target-budget} gives
\[
 R_{n,D}\ge\frac{49}{144}(8/9)^{n/2}8^{n/18-16/9}
 >\frac{49}{144\cdot64}(8/3)^{n/18}.
\]
Now $n/18\ge7(k-1)$ and $8^7>3^{13}$, so this is greater than
$(49/(144\cdot64))9^{3k-3}>9^{2k+1}$.
The last comparison already holds at $k=11$ because
$49\cdot9^7>144\cdot64$ and strengthens with $k$.
This proves the lemma.
\end{proof}

\begin{lemma}\label{lem:qutrit-bounded}
Let $p=3$ and $n<1300$, and let $d\coloneqq D+1$ and $M_0\coloneqq\lfloor n/2\rfloor+1$
as in Lemma~\ref{lem:qutrit-complement}.  The remaining part of the
complement of the root region, namely the parameters
of~\eqref{eq:strict-complement} with
$M_0\le9(M_0-d)$ (those with $u_0\le0$ in that lemma), satisfies
$K\le R_{n,d-1}=R_{n,D}$.
\end{lemma}
\begin{proof}
\smallskip\noindent\emph{Step 1 (The target exponent and the domain).}
First retain the parity $\delta=n\bmod2$ in the target estimate.
The condition $M_0\le9(M_0-d)$ reads $Qd\le bM_0$; with $D=d-1$ and
$M_0=(n-\delta)/2+1$ it gives
$n-D\ge n+1-bM_0/Q=((Q+1)n+2+b\delta)/(2Q)$.
From~\eqref{eq:grid-budget-coarse} one therefore obtains
\begin{equation}\label{eq:residual-complement-target}
 R_{n,D}\ge \operatorname{pref}_Qb^{\lceil((Q+1)n+2+b\delta)/(2Q)\rceil}/p^n.
\end{equation}
Indeed the exponent is a lower bound for the integer $n-D$.
In the variables $M_0,L_0\coloneqq M_0-d$, the complement of the root region
is exactly
$M_0>bL_0+[Q+(Q-2)\delta]/2$.
Since $M_0\le QL_0$, it forces
$L_0>[Q+(Q-2)\delta]/2$.
For even $n$, its least possible integer $M_0$ is
$Q(\lfloor Q/2\rfloor+1)$; for odd $n$ it is $Q^2$.
Consequently the remaining qutrit domain starts at $n=88$ for
even lengths and $n=161$ for odd lengths, and
\[
 R_{n,D}\ge\frac{49}{144}\frac{8^{z_{\rm bd}}}{3^n},
 \qquad {z_{\rm bd}}\coloneqq\left\lceil\frac{5n+1+4\delta}9\right\rceil.
\]

\smallskip\noindent\emph{Step 2 (Integer windows of degree five).}
The finite data consist of one six-integer vector
$(w_0,\ldots,w_5)$ for each even $88\le n_0\le1298$ satisfying
\begin{equation}\label{eq:qutrit-bounded-domain}
 D_0+1=\lfloor7(n_0+1)/16\rfloor+2,\qquad
 9(D_0+1)\le8(n_0/2+1).
\end{equation}
Here $D_0$ is evaluated at the representative length $n_0$.
There are $578$ such representatives.  For each define
\[
 z_*\coloneqq\lceil(5n_0+1)/9\rceil,\quad
 \operatorname{low}_0\coloneqq\frac{49\cdot8^{z_*}}{144\cdot3^{n_0}},\quad
 K_0\coloneqq\lfloor \operatorname{low}_0\rfloor+1,\quad J\coloneqq5,\quad M\coloneqq n_0/2+3.
\]
The vectors in the supplementary integer-vector certificate for
bounded qutrit ranks~\cite{ComputationalCertificates}
satisfy the exact integer contract
\eqref{eq:integer-window-contract}, defined in the proof of
Lemma~\ref{lem:adjacent-finite}, with $p=3$, $b=8$:
$\operatorname{Row}_W(l)\ge0$ for every $0\le l\le M$, $\operatorname{Row}_W(M)>0$, and
$Z_W(l)\ge0$ for every $0\le l\le M-(D_0+1)$.  As shown there, these are
the sign conditions~\eqref{eq:window-signs}, here at the working length
$2M-J=n_0+1$.
There are $210471$ full-grid rows, including $578$ strict origins,
and $27218$ tail rows.  The verification consists solely of
enumerating \eqref{eq:qutrit-bounded-domain}, expanding the degree-five
polynomial $\operatorname{Ref}_W$ by integer binomial multiplication, and checking
every displayed inequality.  It also reconstructs the rational
number $\operatorname{low}_0$ and its floor. These checks certify the stored vectors
against the stated inequalities.
Lemma~\ref{lem:window-identity} gives $K<K_0$, and integer rank
then gives $K\le K_0-1=\lfloor \operatorname{low}_0\rfloor\le \operatorname{low}_0$.
It is this last inference that compares the excluded rank with
the target; no assertion $K_0\le \operatorname{low}_0$ is needed.

\smallskip\noindent\emph{Step 3 (Coverage).}
To check coverage, take any remaining $(n,d)$ in this
lemma and set $n_0\coloneqq n-(n\bmod2)$.
The necessary length bounds imply $n_0\ge88$.
The actual distance satisfies $d\ge D_0+1$ and
$9d\le8(n_0/2+1)$, so its representative satisfies
\eqref{eq:qutrit-bounded-domain}.
At original even length the target lower bound is $\operatorname{low}_0$.
At its odd partner $n=n_0+1$, the exponent is $z_*+1$, so
the target lower bound is $(8/3)\operatorname{low}_0>\operatorname{low}_0$.
The same odd working length $n_0+1$ and window therefore prove
both original targets, and larger distance only shortens the
required invariant tail.  This covers every remaining pair $(n,d)$
with $n<1300$, completing the proof.
\end{proof}

\section{Details of the coefficient estimates}\label{app:coefficient-details}

This appendix proves the coefficient estimates of
Section~\ref{sec:coefficient-methods} whose proofs were deferred; all of
them serve the strict comparisons~\eqref{eq:residual-comparison},
$\mathcal L_i<h(i)/h(0)$, in the root region.
Appendix~\ref{app:coef-rows} proves Lemma~\ref{lem:weighted-rows} and
Lemma~\ref{lem:monic-paired-energy}, a lower bound for monic paired
moments and an upper bound for the top Krawtchouk coefficient of a
paired product; they are used in the proofs of
Lemma~\ref{lem:optimizer-product} and
Proposition~\ref{prop:original-top}.  Appendix~\ref{app:coef-recurrence}
proves a recurrence mechanism that reduces the comparisons to the
positivity of two scalar polynomials, and with it
Proposition~\ref{prop:terminal-cutoffs};
Appendix~\ref{app:coef-scale} applies the mechanism to prove
Proposition~\ref{prop:scale-cutoff}.  Appendix~\ref{app:coef-power}
proves Lemma~\ref{lem:approximate-power}, a bound for the normalized
Krawtchouk coefficients of a paired product, and deduces
Proposition~\ref{prop:large-deficit}; its proof also extends
Lemma~\ref{lem:paired-mean} to real alphabets, as used again in
Appendix~\ref{app:qutrit-initial}.  Appendix~\ref{app:coef-top} proves
Proposition~\ref{prop:original-top}.

We use the notation of Section~\ref{sec:coefficient-methods}: $b=Q-1$,
$\lambda=N/Q$, $\lambda_+=N/b$, $d_0=2m$, ${a_{\min}}$, ${J_*}$, $G(u)$,
and $G_\ell=\binom N\ell b_\ell/b_0$, where the $b_\ell$ are the
Krawtchouk coefficients of $h$ at length $N$.  In a formula that
contains these coefficients we write $Q-1$ instead of $b$.

\subsection{The finite kernel and weighted rows}\label{app:coef-rows}

\begin{proof}[Proof of Lemma~\ref{lem:weighted-rows}]
\smallskip\noindent\emph{Step 1 (The kernel and the coercivity
constant).}
For
$Y\sim\Bin(N,(1+u)/Q)$, the binomial generating function gives
$\E[F_z(Y)F_w(Y+1)]=(1+w)(1-z/b)[{\operatorname{Base}}+u{\operatorname{Tilt}}]^N$;
symmetrizing in $z,w$ replaces $(1+w)(1-z/b)$ by ${\operatorname{Seed}}$
and gives~\eqref{eq:finite-paired-kernel}.
The base matrix has diagonal and off-diagonal entries
\begin{equation}\label{eq:base-matrix-entries}
 (M_0)_{jj}=1-\frac j{N-j+1},\qquad
 (M_0)_{j,j+1}=\frac{b-1}{2}\sqrt{\frac{j+1}{b(N-j)}}.
\end{equation}
The hypothesis $N\ge Qm$ makes its diagonal at least
${\omega_1}=(b-1)/b$ and its off-diagonal at most ${\omega_1}/2$. Taking absolute values in the quadratic
form and comparing with the path matrix on $0,\ldots,m$ with diagonal
$1$ and off-diagonal $-1/2$, whose least eigenvalue is
$1-\cos(\pi/(m+2))$, gives
\[
 M_0\succeq {\omega_1}\left(1-\cos\frac\pi{m+2}\right)I
       \succeq\frac{2{\omega_1}}{(m+2)^2}I.
\]
The last inequality follows from $\sin t\ge2t/\pi$ on
$[0,\pi/2]$.  This gives the value of ${\operatorname{coer}}$ stated in
Lemma~\ref{lem:weighted-rows}.

\smallskip\noindent\emph{Step 2 (The row bound~\eqref{eq:finite-row-bound}).}
We use the Schur test: for a symmetric matrix $A$ with nonnegative
entries and a vector $s$ with positive entries,
$y^{\mathsf T}Ay\le\max_j\bigl((As)_j/s_j\bigr)\|y\|^2$ for every real
vector $y$.  Consider first $[{\operatorname{Base}}+u{\operatorname{Tilt}}]^N$. Choose ${j_1}$ factors containing both
$z,w$, $j-{j_1}$ containing $z$ alone, and ${j_2}$ containing $w$ alone.
After division by $\nu_j$, the sum of its coefficients in row $j$
is exactly \eqref{eq:finite-row}; for the normalized matrices this is
the row sum weighted by the Schur vector $(\sqrt{\nu_j})$ used below.
The column cutoff is ${j_1}+{j_2}\le m$.
Replace the negative $zw$ coefficient of ${\operatorname{Seed}}$ by its absolute value.
The corresponding row is at most
\[
 (1+{\omega_1}/2){\operatorname{Row}_j}(u)
 +({\omega_1}/2+1/b)\frac{\nu_{j-1}}{\nu_j}{\operatorname{Row}_{j-1}}(u).
\]
The absent row $j=-1$ contributes zero. Since
$\nu_{j-1}/\nu_j=bj/(N-j+1)\le1$, the two multipliers sum to
at most two. This is a coefficientwise inequality in nonnegative
polynomials. For the normalized symmetric matrices use the positive
Schur vector $(\sqrt{\nu_j})$: its weighted row bound controls
the quadratic form on the absolute coordinates of any vector.
Consequently
\[
 \sum_\ell |x^{\mathsf T}M_\ell x|u^\ell
 \le2\max_j{\operatorname{Row}_j}(u)\|x\|^2.
\]
Division by $x^{\mathsf T}M_0x\ge {\operatorname{coer}}\|x\|^2$ proves
\eqref{eq:finite-row-bound}.

\smallskip\noindent\emph{Step 3 (The factorial bound~\eqref{eq:row-factorial-bound}).}
We now prove the factorial bound~\eqref{eq:row-factorial-bound} for
all $a\ge0$.
At $a>0$, use $\binom{N-j}{j_2} b^{-{j_2}}\le\lambda_+^{j_2}/{j_2}!$ and ${\omega_1}\le1$.
Because $\lambda_+\ge m$, these last coefficients increase for
$0\le {j_2}\le m$. Thus
\[
 a^{2m}{\operatorname{Row}_j}(1/a)\le\frac{\lambda_+^m}{m!}F_{m,j}(a),
 \quad
 F_{m,j}(a)\coloneqq\sum_{{j_1}=0}^j\binom j{j_1}(a+1)^{j_1}
                   \sum_{{j_2}=0}^{m-{j_1}}a^{2m-j-{j_2}}.
\]
To compare its rows, set
${\operatorname{Sum}_{j,l}}\coloneqq\sum_{{j_1}=0}^{\min(j,l)}\binom j{j_1}(a+1)^{j_1}
\sum_{{j_2}=0}^{l-{j_1}}a^{-{j_2}}$. Then
$F_{m,j}=a^{2m-j}{\operatorname{Sum}_{j,m}}$ and
${\operatorname{Sum}_{j+1,m}}={\operatorname{Sum}_{j,m}}+(a+1){\operatorname{Sum}_{j,m-1}}$.
For $j<m$ the difference
${\operatorname{Diff}_j}\coloneqq{\operatorname{Sum}_{j,m}}-{\operatorname{Sum}_{j,m-1}}=a^{-m}(1+a+a^2)^j$ also satisfies
${\operatorname{Sum}_{j,m-1}}-{\operatorname{Sum}_{j,m-2}}=a{\operatorname{Diff}_j}$ (with ${\operatorname{Sum}_{j,-1}}=0$).
Hence
\[
 F_{m,j+1}-F_{m,j}
 =a^{2m-j-1}\{2{\operatorname{Sum}_{j,m-1}}+(1-a){\operatorname{Diff}_j}\}\ge0.
\]
For $a\le1$ this is immediate; for $a\ge1$ use
${\operatorname{Sum}_{j,m-1}}\ge a{\operatorname{Diff}_j}$. At $j=m$, the inequality
$\binom mk\binom k{j_1}\ge\binom m{j_1}$ yields
\[
 F_{m,m}(a)=\sum_{k=0}^m a^k\sum_{{j_1}=0}^k\binom m{j_1}(a+1)^{j_1}
 \le\sum_{k=0}^m\binom mk[a(a+2)]^k=(1+a)^{2m}.
\]
This proves~\eqref{eq:row-factorial-bound}, and continuity gives $a=0$.

\smallskip\noindent\emph{Step 4 (The exponential bound~\eqref{eq:row-exponential-bound}).}
For the exponential bound retain the simultaneous cutoff in the identity
\[
 {\operatorname{Row}_j}(u)=\sum_{k=0}^m[w^k]
       (u+(1+{\omega_1}u)w)^j(1+uw/b)^{N-j};
\]
here $\nu_j(u+(1+{\omega_1}u)w)^j(1+uw/b)^{N-j}$ is the coefficient of
$z^j$ in $[{\operatorname{Base}}+u{\operatorname{Tilt}}]^N$, and $w$ marks
the column as in Step~2.
All coefficients are nonnegative. For a real number $0<w\le1$, the sum
is therefore at most
$w^{-m}(u+(1+{\omega_1}u)w)^j(1+uw/b)^{N-j}$.
Substitute $u=1/a$, multiply by $a^{2m}$, and maximize the resulting
geometric sequence in $j$. This gives
\begin{equation}\label{eq:row-cutoff-two-endpoints}
 \begin{gathered}
 \sum_\ell|G_\ell|a^{2m-\ell}
 \le {\operatorname{pref}}w^{-m}\max\left\{
 a^{2m}{\operatorname{Factor}_w}^N, a^m[1+(a+{\omega_1})w]^m{\operatorname{Factor}_w}^{N-m}\right\},\\
 {\operatorname{Factor}_w}\coloneqq1+\frac w{ab}.
 \end{gathered}
\end{equation}
Choosing $w\coloneqq a/(1+a)$ bounds both algebraic factors by
$(1+a)^{2m}$ and gives~\eqref{eq:row-exponential-bound}.
For $Q\ge3$ the second endpoint is maximal, since its successive-row
ratio exceeds one precisely when $1+a-2a/b\ge0$. Thus one also has
\begin{equation}\label{eq:row-finite-exponential}
 \sum_\ell|G_\ell|a^{2m-\ell}
 \le {\operatorname{pref}}[(1+a)^2-a/b]^m
       \left(1+\frac1{b(1+a)}\right)^{N-m}.
\end{equation}
Continuity supplies all $a=0$ endpoints.
\end{proof}

The base matrix $M_0$ also gives the following two bounds.  The first
is used in the proof of Lemma~\ref{lem:optimizer-product} and the second
in that of Proposition~\ref{prop:original-top}.  With $g(x)\coloneqq f(N-x)$, the
left side of~\eqref{eq:monic-paired-energy} is
$\E_{\mu_N}[g(X)g(X-1)]$, the backward form $\operatorname{Back}(g)$ of
Section~\ref{sec:paired-forms} at length $N+1$; the first bound is thus
the backward analogue of the monic bound~\eqref{eq:monic-forward} for
$\operatorname{Forw}$.

\begin{lemma}\label{lem:monic-paired-energy}
For a monic polynomial $f$ of degree $m\ge1$, $Q>2$, and $N\ge Qm$,
\begin{equation}\label{eq:monic-paired-energy}
 \E_{\Bin(N,1/Q)}[f(Y)f(Y+1)]
 \ge\frac{(Q-2)(m+2)}{2(Q-1)(m+1)}
          m!(N)_m\left(\frac{Q-1}{Q^2}\right)^m.
\end{equation}
For every nonzero real polynomial $f$ of degree at most $m$, the paired
polynomial $h(x)\coloneqq f(N-x)f(N-x+1)$, so that $h(N-y)=f(y)f(y+1)$ as in
Section~\ref{sec:coefficient-methods}, has Krawtchouk coefficients
$h=\sum_\ell b_\ell\Kraw{\ell}{N}$ at length $N$ that obey
\begin{equation}\label{eq:top-residual-bound}
 \frac{b_{2m}}{b_0}
 \le\frac{2(Q-1)}{Q-2}
       \frac{\binom{2m}m}{\binom Nm(Q-1)^m}.
\end{equation}
\end{lemma}
\begin{proof}
\smallskip\noindent\emph{Step 1 (An entrywise bound for $M_0^{-1}$).}
Here $M_0$ is the normalized base matrix of
Section~\ref{sec:coefficient-methods}, with
entries~\eqref{eq:base-matrix-entries}: if $x$ is the coefficient vector
of $f$ in the basis $[z^j]F_z(y)/\sqrt{\nu_j}$, $0\le j\le m$, then
$\E_{\Bin(N,1/Q)}[f(Y)f(Y+1)]=x^{\mathsf T}M_0x$.
Conjugate $M_0$ by $\operatorname{diag}(1,-1,1,\ldots)$ and write
the result as ${J_{\rm pair}}=\mathcal D-\mathcal F$, with $\mathcal F$ nonnegative.
For ${\omega_2}\coloneqq(b-1)/(2b)$ and the adjacency matrix $\operatorname{Path}$
of the path with vertices $0,\ldots,m$, the entry
bounds~\eqref{eq:base-matrix-entries} imply
$0\le\mathcal D^{-1}\mathcal F\le\operatorname{Path}/2$ and
$\mathcal D^{-1}\le I/(2{\omega_2})$ entrywise. Since
$\rho(\operatorname{Path}/2)=\cos(\pi/(m+2))<1$, the nonnegative
Neumann series gives
\begin{equation}\label{eq:entrywise-inverse-path}
 0\le {J_{\rm pair}}^{-1}\le {\omega_2}^{-1}(2I-\operatorname{Path})^{-1}
 \quad\text{entrywise}.
\end{equation}
The path inverse has entry
$(\min(j,k)+1)(m+1-\max(j,k))/(m+2)$.
In particular $(M_0^{-1})_{mm}\le {\omega_2}^{-1}(m+1)/(m+2)$.
This is an entrywise inverse comparison, obtained from a nonnegative
series.

\smallskip\noindent\emph{Step 2 (The monic bound~\eqref{eq:monic-paired-energy}).}
The leading coefficient of $[z^m]F_z(y)$ is $(Q/b)^m/m!$.
In the normalized basis the last coordinate of a monic $f$ therefore
has square $\eta_m\coloneqq m!(N)_m(b/Q^2)^m$. Minimizing a positive
quadratic form with its last coordinate fixed gives
$\eta_m/(M_0^{-1})_{mm}$, proving~\eqref{eq:monic-paired-energy}.

\smallskip\noindent\emph{Step 3 (The top coefficient~\eqref{eq:top-residual-bound}).}
Here $b_{2m}/b_0=G_{2m}/\binom N{2m}$, and the numerator
matrix $M_{2m}/\binom N{2m}$ of this quotient has just one entry, at
$(m,m)$, equal to $\binom{2m}m/(\binom Nm(Q-1)^m)$. Since
$x_m^2\le(M_0^{-1})_{mm}\,x^{\mathsf T}M_0x$, the weaker bound
$(M_0^{-1})_{mm}<{\omega_2}^{-1}$ proves~\eqref{eq:top-residual-bound}.
\end{proof}

\subsection{Positive recurrence and a finite degree cutoff}\label{app:coef-recurrence}

We prove a general mechanism that yields Propositions
\ref{prop:terminal-cutoffs} and~\ref{prop:scale-cutoff}. Besides the
standing assumptions of Section~\ref{sec:coefficient-methods}, it
assumes real parameters with
$d_0/N\le\delta$, $r/N\le z$, $L/N\ge a$ for $L\coloneqq{a_{\min}}-1$, $b\ge b_*$,
${y_{\max}}\coloneqq z+\delta<a<1$ and $\delta<\theta<{y_{\max}}$, and it reduces the comparisons
to the positivity of a cubic ${\operatorname{Init}}$ on $[0,\theta]$ and
of a quadratic ${\operatorname{Tail}}$ on $[\theta,{y_{\max}}]$, both defined below:
if ${\operatorname{Init}}>0$ on $[0,\theta]$ and
${\operatorname{Tail}}\ge0$ on $[\theta,{y_{\max}}]$, then
\eqref{eq:residual-comparison} holds for $1\le i\le D$.

\smallskip\noindent\emph{Step 1 (A majorant and the target).}
Put $L\coloneqq{a_{\min}}-1=N-{J_*}$, $d_0=2m$. The mixture identity of
Lemma~\ref{lem:positive-mixture}, with absolute values taken before
averaging, gives
\[
 |G_\ell|\le {q_\ell^{\rm maj}},\qquad
 \sum {q_\ell^{\rm maj}} u^\ell=(1+u)^{J_*}(1+u/b)^L;
\]
only the support of the mixture is used: $j\le{J_*}$, and mixing
parameters at most $1/b$.
Increasing $j$ to ${J_*}$ in the positive comparison adds
$(1-1/b)u(1+u)^j(1+u/b)^{N-j-1}$, whose coefficients are
nonnegative. Let ${\operatorname{Maj}_{d_0}}\coloneqq\sum_{\ell=0}^{d_0}{q_\ell^{\rm maj}} u^\ell$,
$f_i\coloneqq[u^i](1+u)^r{\operatorname{Maj}_{d_0}}$, and
\begin{equation}\label{eq:cutoff-target}
 t_i\coloneqq\binom Ni\frac{(L)_i}{(L+d_0)_i}.
\end{equation}
Ordered distinct roots imply $h(i)/h(0)\ge(L)_i/(L+d_0)_i$ for
$i\le D$.  Since $(1+u)^r$ has nonnegative coefficients,
$\binom Ni\mathcal L_i\le f_i$, while $t_i\le\binom Nih(i)/h(0)$.
Thus $f_i<t_i$ implies~\eqref{eq:residual-comparison} at $i$.

\smallskip\noindent\emph{Step 2 (Comparison through a recurrence).}
The untruncated coefficients
${\bar f_i}\coloneqq[u^i](1+u)^{r+{J_*}}(1+u/b)^L$, which satisfy
$f_i\le{\bar f_i}$, satisfy, with $\mu\coloneqq r+N-(1-1/b)L$,
\begin{equation}\label{eq:positive-majorant-recurrence}
 i{\bar f_i}=[\mu-(1+1/b)(i-1)]{\bar f_{i-1}}
              +(N+r-i+2){\bar f_{i-2}}/b.
\end{equation}
Both recurrence coefficients are positive through $D$: $L\ge D$
and $r+{J_*}\ge D$. For $x\coloneqq i-1$, the normalized supersolution margin
of $t_i$ is exactly
\begin{align}\label{eq:cutoff-margin}
 \mathfrak M\coloneqq{}&\frac{(N-x)(L-x)}{L+d_0-x}-\mu+(1+1/b)x\notag\\
 &-\frac{x(N+r-x+1)(L+d_0-x+1)}
 {b(N-x+1)(L-x+1)}.
\end{align}
Here the margin is
$(it_i-[\mu-(1+1/b)(i-1)]t_{i-1}-(N+r-i+2)t_{i-2}/b)/t_{i-1}$.
Since ${\bar f_0}=t_0=1$, ${\bar f_{-1}}=0$ and both recurrence coefficients are
positive, $\mathfrak M>0$ for $x=0,\ldots,i-1$ gives ${\bar f_i}<t_i$ by
induction on $i\le D$.
It increases with $L,b$ and decreases with $d_0,r$ in the positive
domain. For $L,d_0$ write the last ratio as
$1+d_0/(L-x+1)$; the first ratio has the same monotonicities.
The terms that depend on $b$ sum to $1/b$ times
$-(L-x)-x(N+r-x+1)(L+d_0-x+1)/((N-x+1)(L-x+1))$, which is negative.
Hence $\mathfrak M$ does not increase when $L$ and $b$ are replaced by
lower bounds and $d_0$ and $r$ by upper bounds, also when these bounds
hold with equality.

\smallskip\noindent\emph{Step 3 (The initial range).}
Suppose $d_0/N\le\delta$, $r/N\le z$, $L/N\ge a$, $b\ge b_*$,
and ${y_{\max}}=z+\delta<a<1$. The finite $+1$ offsets improve the lower
bound, since $(u+v+1)/(u+1)\le(u+v)/u$ for $u>0,v\ge0$.
For $y\coloneqq x/N$, multiplication of the resulting lower bound for
$\mathfrak M/N$ by the positive denominator
$b_*(a+\delta-y)(1-y)(a-y)$ gives
\begin{align}\label{eq:cutoff-cubic}
 {\operatorname{Init}}(y)\coloneqq{}&b_*(1-y)^2(a-y)^2\notag\\
 &-[b_*\mu_*-(b_*+1)y](a+\delta-y)(1-y)(a-y)\notag\\
 &-y(1+z-y)(a+\delta-y)^2,
 \quad \mu_*\coloneqq z+1-(1-1/b_*)a.
\end{align}
The fourth-degree coefficient cancels. Positivity on $[0,\theta]$
therefore proves $f_i\le{\bar f_i}<t_i$ for $1\le i\le\min(I,D)$, where
$I\coloneqq\lceil\theta N\rceil$: there $(i-1)/N<\theta$, and the recurrence
coefficients are positive.  If $I\ge D$, this proves all the
comparisons.

\smallskip\noindent\emph{Step 4 (The tail).}
If $\delta<\theta<{y_{\max}}$, then $I>d_0$. For $I\le i<D$ and
$0\le\ell\le d_0$, a nonzero summand
$U_\ell(i)\coloneqq\binom r{i-\ell}/t_i$ has ratio
\[
 \frac{U_\ell(i+1)}{U_\ell(i)}
 =\frac{(r-i+\ell)(i+1)(L+d_0-i)}
 {(i+1-\ell)(N-i)(L-i)}
 \le\frac{(z+\delta-y)y(a+\delta-y)}
 {(y-\delta)(1-y)(a-y)},\quad y\coloneqq i/N.
\]
The first ratio increases with $\ell$; at $\ell=d_0$ use
$(i+1)/(i+1-d_0)\le i/(i-d_0)$ and the parameter bounds.
The final ratio is at most one when
\begin{equation}\label{eq:cutoff-quadratic}
 {\operatorname{Tail}}(y)\coloneqq-(1-z-\delta)y^2+
       [(a+\delta)(1-z)-\delta^2]y-\delta a\ge0.
\end{equation}
Terms that become zero remain zero.  Since $D=r+d_0\le{y_{\max}}N$, these
indices have $i/N<{y_{\max}}$.  Consequently
$f_i/t_i=\sum_\ell{q_\ell^{\rm maj}}U_\ell(i)$, a
nonnegative combination of these summands, is nonincreasing for
$I\le i\le D$, and $f_i/t_i\le f_I/t_I<1$ by Step~3.
This proves the comparisons from positivity of ${\operatorname{Init}}$ on
$[0,\theta]$ and ${\operatorname{Tail}}$ on $[\theta,{y_{\max}}]$.

\begin{proof}[Proof of Proposition~\ref{prop:terminal-cutoffs}]
For $Q\ge16,r\le N/2$, choose
\[
 b_*\coloneqq15,\quad a\coloneqq3/4,\quad\delta\coloneqq1/8,
 \quad z\coloneqq1/2,\quad\theta\coloneqq1/3.
\]
Here $b_*$, $a$ and $\delta$ are the values at $Q=16$ of $Q-1$,
$1-4/Q$ and $2/Q$, which bound $b$, $L/N$ and $d_0/N=2m/N\le2/Q$ from
the required side for every $Q\ge16$; $z$ is the bound on $r/N$, and
$\theta$ is a choice in $(\delta,{y_{\max}})$, ${y_{\max}}=5/8$.
The strict root barrier, the second inequality
of~\eqref{eq:coefficient-root-domain}, and $N\ge Qm$ give
$L>(Q-2)(N-2m)/Q-2/Q\ge(Q-2)^2N/Q^2-2/Q>(1-4/Q)N\ge aN$.
Positivity on $[0,\theta]$ is certified by the substitution
$y=\theta t/(1+t)$, which maps $t\in[0,\infty)$ onto $[0,\theta)$:
after clearing the denominator, positive coefficients in $t$ give
positivity there, and the leading coefficient is the value at
$y=\theta$; similarly for ${\operatorname{Tail}}$ with
$y=(\theta+{y_{\max}}t)/(1+t)$ on $[\theta,{y_{\max}})$.
The coefficient lists, in increasing order, are
\begin{align*}
 (1+t)^3{\operatorname{Init}}(\theta t/(1+t))&:
       \quad 9/16, 103/128, 25/72, 139/3456,\\
 (1+t)^2{\operatorname{Tail}}((\theta+{y_{\max}}t)/(1+t))&:
       \quad 1/192, 31/512, 3/128.
\end{align*}
They are strictly positive, including the right interval endpoints
by the leading coefficients. This proves the first part of
Proposition~\ref{prop:terminal-cutoffs}. For $Q\ge25,r\le3N/5$ use
$b_*\coloneqq24,a\coloneqq21/25,\delta\coloneqq2/25,z\coloneqq3/5,{y_{\max}}=17/25$, the same bounds at
$Q=25$, and compare the
untruncated recurrence on the whole interval: Step~3 with ${y_{\max}}$ in
place of $\theta$ covers every $i\le D$. The four coefficients
of $(1+t)^3{\operatorname{Init}}({y_{\max}}t/(1+t))$ are
\[
 34209/15625,\quad467977/390625,\quad65264/390625,\quad516/390625.
\]
This proves its second part without a tail argument.
\end{proof}

\subsection{The scale-sensitive cutoff}\label{app:coef-scale}

\begin{proof}[Proof of Proposition~\ref{prop:scale-cutoff}]
\smallskip\noindent\emph{Step 1 (A rational lower bound for $L/N$).}
Here the root bound of a minimizer, the first inequality
of~\eqref{eq:optimizer-root-sharp}, is essential. If
$N\ge Q{s_{\rm cap}}m$, $m\ge2$, ${s_{\rm cap}},v_0\ge1$, its square-root term can be
bounded by
$2\sqrt{b(N-m)(m-1)}\le(N-m)/v_0+bv_0(m-1)$.
Subtracting one from the root bound gives
\[
 L\ge\frac{(b-1/v_0)N-(b-1-1/v_0+bv_0)m+bv_0-Q-1}{Q}.
\]
Use $1/N\le(m/N)/2$ on a negative last coefficient, and otherwise
discard that last coefficient. It follows that
\begin{align}\label{eq:rational-root-family}
 \frac LN\ge\mathfrak a(Q,{s_{\rm cap}},v_0)\coloneqq{}&
 1-\frac{(1+1/v_0)+(1+v_0)/{s_{\rm cap}}}{Q}
 +\frac{2+v_0+1/v_0}{{s_{\rm cap}}Q^2}\notag\\
 &+\frac{\min(0,(Q-1)v_0-Q-1)}{2{s_{\rm cap}}Q^2}.
\end{align}
For fixed ${s_{\rm cap}},v_0$, this increases with $Q\ge9$. On the branch
with nonnegative final numerator it has form $1-U/Q+V/Q^2$,
where $2V/U=2(v_0+1)/({s_{\rm cap}}+v_0)\le2$.
On the other branch,
$U=1+1/v_0+(v_0+3)/(2{s_{\rm cap}})$ and
$V=((v_0+3)/2+1/v_0)/{s_{\rm cap}}$, so
$QU-2V=Q(1+1/v_0)+[(Q-2)(v_0+3)/2-2/v_0]/{s_{\rm cap}}>0$.
The branches agree at their boundary.

\smallskip\noindent\emph{Step 2 (Parameters for the mechanism).}
Set ${s_{\rm cap}}\coloneqq\xi^{-2}$, the value for which $N=Q{s_{\rm cap}}m$
(recall $\xi^2=Qm/N$), and $v_0\coloneqq5/(4\xi)$, and apply this monotonicity
at $Q=9$: for every $Q\ge9$,
$L/N\ge\mathfrak a(Q,{s_{\rm cap}},v_0)\ge\mathfrak a(9,{s_{\rm cap}},v_0)$.
At $Q=9$ the last term of~\eqref{eq:rational-root-family} vanishes,
since $(Q-1)v_0-Q-1=10/\xi-10\ge0$, and
\begin{equation}\label{eq:scale-root-lower}
 L/N\ge a(\xi)\coloneqq8/9-86\xi/405-7\xi^2/81+4\xi^3/405.
\end{equation}
Use in~\eqref{eq:cutoff-cubic}--\eqref{eq:cutoff-quadratic}
\[
 b_*\coloneqq8,\quad \delta\coloneqq2\xi^2/9,\quad
 z\coloneqq7/9-2\xi/9-7\xi^2/18,\quad
 {J_\xi}\coloneqq3+7\xi^2,\quad \theta\coloneqq3\xi^2/{J_\xi}.
\]
These satisfy the hypotheses of the mechanism for every $Q\ge9$:
$d_0/N=2m/N=2\xi^2/Q\le\delta$, $r/N\le z=z(\xi)$ is the hypothesis
of Proposition~\ref{prop:scale-cutoff}, $L/N\ge a=a(\xi)$
by~\eqref{eq:scale-root-lower}, and $b\ge8=b_*$.
For $0<\xi\le1$,
$0<\delta<\theta<{y_{\max}}=z+\delta<a<1$.
Indeed $\theta/\delta\ge27/20$, $\theta\le3/10$,
${y_{\max}}\ge7/18$, and
$a-{y_{\max}}=1/9+4\xi/405+13\xi^2/162+4\xi^3/405>0$.
Thus the cutoff mechanism applies once the two scalar polynomials
are positive.

\smallskip\noindent\emph{Step 3 (The polynomial certificate).}
Here is a complete specification of the finite polynomial certificate.
Form
\begin{align*}
 {\operatorname{Init}}_*(\xi,u)&\coloneqq{J_\xi}^3(1+u)^3{\operatorname{Init}}(\theta u/(1+u)),\\
 {\operatorname{Tail}}_*(\xi,u)&\coloneqq{J_\xi}^2(1+u)^2{\operatorname{Tail}}((\theta+{y_{\max}}u)/(1+u));
\end{align*}
the factors $J_\xi^3(1+u)^3$ and $J_\xi^2(1+u)^2$ clear the
denominators, since $\theta=3\xi^2/J_\xi$.
Their degrees in $(\xi,u)$ are $(15,3)$ and $(11,2)$.
Substitute $\xi=t/(1+t)$ and multiply respectively by $(1+t)^{15}$
and $(1+t)^{12}$.  The factor $(1+t)^{12}$ contains, beyond the
$(1+t)^{11}$ that clears the denominators, one P\'olya-type factor
$1+t$: with $(1+t)^{11}$ alone, the coefficient of $t^{10}u^0$ is
$-11/162$.  The first resulting polynomial has exactly the
60 positive coefficients with exponents $1\le j\le15$, $0\le k\le3$;
their least value is $7552/1215$. The second has the 36 positive
coefficients with $0\le j\le12$, $0\le k\le2$, excluding
$(0,0),(1,0),(2,0)$; their least value is $14/81$.
The exact coefficient tables and rational expansion algorithm are
included in the supplementary package~\cite{ComputationalCertificates}. Every clearing factor is positive for
$t>0,u\ge0$. Leading rows and columns give the endpoints $\xi=1$
and $y=\theta,{y_{\max}}$; $\xi=0$ is not an actual finite parameter.
This proves Proposition~\ref{prop:scale-cutoff}.
\end{proof}

\subsection{A finite approximate-power estimate}\label{app:coef-power}

The large-deficit reduction, Proposition~\ref{prop:large-deficit},
uses the following bound, derived from a
positive distribution and a polynomial recurrence.

\begin{lemma}\label{lem:approximate-power}
Let $Q\ge9$, $N\ge Qm$, and $h(x)\coloneqq q(x)q(x-1)$ be nonzero and
nonnegative on $0,\ldots,N$, with $\deg q\le m$ and Krawtchouk
coefficients $h=\sum_\ell b_\ell\Kraw{\ell}{N}$. Put $b\coloneqq Q-1$ and
\[
 \alpha\coloneqq m/N,\quad
 \beta\coloneqq(b-1)\alpha/b+2\sqrt{\alpha(1-\alpha)/b},\quad
 \gamma\coloneqq\max(\beta,1/b).
\]
Then for $1\le i\le2m$,
\begin{equation}\label{eq:approximate-power}
 |b_i/b_0|\le
 \left(\frac{\gamma+4\sqrt{i/((Q-1)N)}+4i/N}{1-(i-1)/N}\right)^i.
\end{equation}
\end{lemma}
\begin{proof}
\smallskip\noindent\emph{Step 1 (The paired mean for real alphabets).}
The paired mean of Lemma~\ref{lem:paired-mean} is valid for every real
alphabet $R>2$ in place of $Q$; this extension is used in Step~2 and
in Appendix~\ref{app:qutrit-initial}.
To check this extension explicitly, let $L\ge Rk$, $b_R\coloneqq R-1$,
$\rho\coloneqq\sqrt{k/(b_R(L-k))}\le1/b_R$. The forward and backward
correlation matrices of Section~\ref{sec:paired-forms} give, for
$W\coloneqq\E q(Y_R)^2$, $Y_R\sim\Bin(L-1,b_R/R)$,
\[
 {\operatorname{Forw}(q)}\ge\frac{1-\rho}{1+b_R\rho}W>0,
 \qquad0\le {\operatorname{Back}(q)}\le\frac{1+b_R\rho}{1-\rho}W.
\]
The lower forward inequality at correlation ratio one uses positive
semidefiniteness instead of an inverse. Conditioning gives
$\E h=(b_R{\operatorname{Forw}(q)}+{\operatorname{Back}(q)})/R$ and $\E Xh=Lb_R{\operatorname{Forw}(q)}/R$; hence
\[
 \frac{\E Xh}{\E h}
 \ge\frac{[\sqrt{b_R(L-k)}-\sqrt k]^2}{R}.
\]
The identities reducing the ratio are
$b_R(1-\rho)^2+(1+b_R\rho)^2=R(1+b_R\rho^2)$ and
$1+b_R\rho^2=L/(L-k)$. Thus the extension holds also at the cap.

\smallskip\noindent\emph{Step 2 (An exponential bound for $G$).}
Use the probability distribution proportional to
$\binom Nx b^xh(x)$, and let $Y\coloneqq1-QX/(bN)\in[-1/b,1]$.
The exact generating function is
\[
 G(u)=\E[(1+u)^{N-X}(1-u/b)^X],
\]
the function~\eqref{eq:residual-generating} for this $h$.
For $0\le u\le1$ the effective alphabet is $R\coloneqq Q/(1+u)>2$,
and $N\ge Rm$: differentiation gives
$(\log G)'(u)=N/(1+u)-Q\,\E_uX/((1+u)(b-u))$, where $\E_u$ is taken
for the law proportional to $\binom Nx(b-u)^x(1+u)^{N-x}h(x)$, that
is, for $\mu_N$ at the alphabet $R$ weighted by $h$.  Step~1, with
$L=N$ and $k=m$, bounds $\E_uX$ from below and gives
\[
 (\log G)'(u)\le N\left[
 \frac{\alpha(b-1-2u)}{(1+u)(b-u)}
 +2\sqrt{\frac{\alpha(1-\alpha)}{(1+u)(b-u)}}\right]
 \le N\beta;
\]
the middle expression decreases in $u$ and equals $N\beta$ at $u=0$.
Thus $G(u)\le e^{N\beta u}$. We next convert this real inequality
into a moment bound for $Y$.

\smallskip\noindent\emph{Step 3 (A moment bound for $Y$).}
Write $(1+u)^{N-X}(1-u/b)^X=\prod_{c=1}^N(1+y_cu)$, where $y_c\coloneqq1$ for
$N-X$ coordinates and $y_c\coloneqq-1/b$ for the other $X$, so that the average
of the $y_c$ is $Y$.  For these coordinate values
$y_c\in\{1,-1/b\}$,
$\log(1+y_cu)\ge y_cu-y_c^2u^2/(2(1-u/b))$.
Their average square is $[1+(b-1)Y]/b$. Put
\[
 v(u)\coloneqq u-\frac{(b-1)u^2}{2(b-u)},\qquad
 \mu\coloneqq\frac{1+(b-1)\beta}{b},\qquad {\operatorname{scale}}\coloneqq2+1/b.
\]
It follows that
$\E e^{Nv(u)(Y-\beta)}\le
\exp(N\mu u^2/(2(1-u/b)))$.
For every $0\le v<1/{\operatorname{scale}}$, a $u\in[0,1]$ with $v(u)=v$ exists
by continuity, since $v(0)=0,v(1)=1/2$.
Moreover $v(u)\ge u-u^2/2\ge u/(1+u)$, and so
\[
 \frac{u^2}{1-u/b}\le
 \frac{v^2}{(1-v)(1-Qv/b)}\le\frac{v^2}{1-{\operatorname{scale}}v}.
\]
All denominators are positive. With $t\coloneqq Nv$ this proves
\[
 \E e^{t(Y-\beta)}\le
 \exp\left(\frac{\mu t^2}{2N(1-{\operatorname{scale}}t/N)}\right),\qquad0\le t<N/{\operatorname{scale}}.
\]
Chernoff's inequality, with
$t\coloneqq s/(1+{\operatorname{scale}}s/N)$ and $s\coloneqq\sqrt{2Nx/\mu}$, yields
$\Pr\{Y-\beta\ge\sqrt{2\mu x/N}+{\operatorname{scale}}x/N\}\le e^{-x}$.
Consequently, by the $L^i$ triangle inequality and $i!\le i^i$,
\[
 \|(Y-\beta)_+\|_i\le\sqrt{2\mu i/N}+{\operatorname{scale}}i/N.
\]
The cap gives $\beta\le(3Q-4)/(Q(Q-1))$, $\mu<4/b$, and
${\operatorname{scale}}\le17/8$. Since $|Y|\le\gamma+(Y-\beta)_+$,
\begin{equation}\label{eq:tilted-moment-bound}
 \|Y\|_i\le\gamma+\sqrt{8i/(bN)}+17i/(8N).
\end{equation}

\smallskip\noindent\emph{Step 4 (The Krawtchouk recurrence).}
For $f_j(Y)\coloneqq\Kraw{j}{N}(X)/(\binom Nj b^j)$ the exact recurrence is
\[
 (N-j)f_{j+1}=[NY-(b-1)j/b]f_j-(j/b)f_{j-1},
 \quad f_0=1, f_1=Y.
\]
Fix $i$ and put $k\coloneqq i-1$. Its absolute recurrence is dominated by
$|f_i(Y)|\le S_i(Y)^i$, where
\[
 S_i(Y)\coloneqq\frac{|Y|+k/N+\sqrt{k/(bN)}}{1-k/N}.
\]
Indeed $S_i\ge\kappa_1+\sqrt{\kappa_2}$ for the recurrence constants
$\kappa_1\coloneqq(|Y|+k/N)/(1-k/N)$ and $\kappa_2\coloneqq k/(bN(1-k/N))$, so
$S_i^2\ge\kappa_1S_i+\kappa_2$ and induction applies.
Finally $b_i/b_0=\E f_i(Y)$. Combining this identity with
\eqref{eq:tilted-moment-bound}, $\sqrt8+1<4$, and $17/8+1<4$
proves~\eqref{eq:approximate-power}.
\end{proof}

\begin{proof}[Proof of Proposition~\ref{prop:large-deficit}]
Lemma~\ref{lem:approximate-power} applies to $h$, since $Q\ge16$,
$N\ge Qm+1$, and $h$, a monic product of $m$ adjacent integer pairs,
has the form $q(x)q(x-1)$ with $q$ monic of degree $m$.
In~\eqref{eq:approximate-power}, for $i\le2m$ the numerator
$\gamma+4\sqrt{i/((Q-1)N)}+4i/N$ of the base is at most
$[11+4\sqrt{2Q/(Q-1)}]/Q$, by $\gamma\le3/Q$ and $i/N\le2/Q$, and the
denominator $1-(i-1)/N$ is at least $(Q-2)/Q$. Thus
$|b_i/b_0|\le(20/Q)^i$ for $Q\ge16$.
Let ${J_{\rm cap}}\coloneqq\lceil20\lambda\rceil+2m$. For $1\le\ell\le2m$,
\[
 G_\ell\le(20\lambda)^\ell/\ell!\le\binom{{J_{\rm cap}}}\ell,
\]
because every falling factor of $({J_{\rm cap}})_\ell$ exceeds $20\lambda$.
At zero equality holds, and beyond $2m$ the actual coefficients vanish.
Since $(1+u)^r$ has nonnegative coefficients, multiplication by it
therefore gives
\[
 \mathcal L_i\le\frac{\binom{r+{J_{\rm cap}}}i}{\binom Ni}
 =\prod_{j=0}^{i-1}\left(1-\frac{s-{J_{\rm cap}}}{N-j}\right).
\]
Here $r+{J_{\rm cap}}\le N-4m$ and $i\le D\le r+{J_{\rm cap}}$, so all factors
are positive. The root barrier gives ${a_{\min}}+2m-1>N-2\lambda$.
Every root exceeds $i$, since
$i\le D\le N-\lceil20\lambda\rceil-4m<N-2m+1-2\lambda<{a_{\min}}$.
Replacing the roots by the $2m$ consecutive integers from ${a_{\min}}$
(packing them) can therefore only decrease $h(i)/h(0)$, and gives
\[
 \frac{h(i)}{h(0)}>
 \prod_{j=0}^{i-1}\left(1-\frac{2m}{N-2\lambda-j}\right).
\]
For $0\le j<i$, $N-j\ge s-2m+1>4\lambda$, and one has
$s-{J_{\rm cap}}\ge4m>2m(N-j)/(N-j-2\lambda)$. This compares each pair of positive
factors strictly, including $i=D$.
\end{proof}

\subsection{The top index}\label{app:coef-top}

\begin{proof}[Proof of Proposition~\ref{prop:original-top}]
Suppose first $m\ge1$. At $i=D$ only the term $j=2m$ of the
lift~\eqref{eq:normalized-lift} survives, because $(r)_{D-j}=0$ for
$j<2m$; with~\eqref{eq:normalized-value} the assertion becomes
\[
 \frac{\binom N{2m}}{\binom ND}\frac{b_{2m}}{b_0}
 <\frac{h(D)}{h(0)}.
\]
By~\eqref{eq:top-residual-bound} and
$\binom N{2m}\binom{2m}m=\binom Nm\binom{N-m}m$, the left side is at most
${\operatorname{pref}_{\rm top}}\binom{N-m}m/(\binom ND b^m)$, ${\operatorname{pref}_{\rm top}}\coloneqq2b/(b-1)$.
The $2m$ distinct roots above $D$ give
$h(D)/h(0)\ge1/\binom{D+2m}{2m}$.
The root-region inequality and the cap $N\ge Qm+1$ imply
$2m\le D\le N-3m$: the difference of the root-region upper bound and
$N-3m$, multiplied by $Q$, is
$-2N+(Q+4)m+Q-2\le-(Q-4)(m-1)\le0$.

Set $F({j})\coloneqq\binom N{j}/\binom{{j}+2m}{2m}$.
Its ratio $F({j}+1)/F({j})=(N-{j})/({j}+2m+1)$ decreases, so its minimum
on $[2m,N-3m]$ is at an endpoint. At ${j}=N-3m$,
$F({j})=\binom Nm/\binom{3m}m$; it suffices that
$b^m>{\operatorname{pref}_{\rm top}}\binom{3m}m$. The case $m=1$ is $b>3{\operatorname{pref}_{\rm top}}$, true for $b\ge8$,
and the binomial consecutive ratio is
$3(3m+1)(3m+2)/(2(2m+1)(m+1))<27/4<8$.
At ${j}=2m$ the required inequality becomes
$b^m\binom Nm>{\operatorname{pref}_{\rm top}} {\operatorname{Comb}_m}$, ${\operatorname{Comb}_m}\coloneqq\binom{2m}m\binom{4m}{2m}$.
Here ${\operatorname{Comb}_1}=12$, ${\operatorname{Comb}_{m+1}}/{\operatorname{Comb}_m}=4(4m+1)(4m+3)/(m+1)^2<64$,
${\operatorname{pref}_{\rm top}}\le16/7$, and $\binom Nm\ge(N/m)^m>Q^m$.
Thus ${\operatorname{pref}_{\rm top}}{\operatorname{Comb}_m}\le(192/7)64^{m-1}<72^m\le(Qb)^m$.
Both endpoints are strict, proving the result for $m\ge1$.

If $m=0$, then $r=D$ and the normalized terminal polynomial has
$c_D/c_0=1/\binom nD$, whereas its value at $D$ is
$\binom{n-D}D/\binom nD>1/\binom nD$ since $n\ge2D+1$.
\end{proof}

\section{Alphabet-specific coefficient estimates}\label{app:higher-details}

This appendix proves the coefficient estimates behind
Theorem~\ref{thm:higher-weak}: Propositions~\ref{prop:large-alphabet-coefficients}
($Q\ge36$), \ref{prop:five-cap-coefficients} ($Q=25$),
\ref{prop:four-initial} and~\ref{prop:four-tail} ($Q=16$); it also
specifies the finite certificate of Proposition~\ref{prop:four-finite}.
The four propositions assert the strict
comparisons~\eqref{eq:residual-comparison}, $\mathcal L_i<h(i)/h(0)$, on
ranges of indices, and their proofs follow the route of
Section~\ref{sec:coefficient-methods} in two parts. On an initial range
$1\le i\le\lfloor\theta N\rfloor$, the coefficient of every component of the
positive mixture~\eqref{eq:positive-mixture} is bounded separately (by
Lemma~\ref{lem:component-coefficients}, and at $Q=16$ also by a Cauchy
integral), these bounds are averaged by Lemma~\ref{lem:mixture-laplace},
and the average is compared with a lower bound for $h(i)/h(0)$ obtained by
packing the roots. On the remaining indices $\lceil\theta N\rceil\le i\le D$,
called a tail, Lemma~\ref{lem:positive-tail} is combined with the row
bounds of Lemma~\ref{lem:weighted-rows} or with exact matrix certificates.

We use the notation of Sections~\ref{sec:coefficient-methods}
and~\ref{sec:higher-dimensions}: $b=Q-1$, $\lambda=N/Q$, $\lambda_+=N/b$,
$d_0=2m$, ${a_{\min}}$ the least root of $h$, ${J_*}=N-{a_{\min}}+1$,
$v={J_*}-d_0$, and $s=N-r$, which we call the deficit.
The letter $a$ denotes the mixture parameter of
Lemma~\ref{lem:positive-mixture} in Appendices~\ref{app:higher-initial}
and~\ref{app:higher-four-initial}, and elsewhere the weight $a$ of
Lemma~\ref{lem:positive-tail}, which we call the backward weight. We write
${\operatorname{pref}}_m$ for a value of the constant ${\operatorname{pref}}$ of
Lemma~\ref{lem:weighted-rows} for $m$ pairs (not the constant
$\operatorname{pref}_Q$ of Section~\ref{sec:paired-forms}).
In a tail, Lemma~\ref{lem:positive-tail} is applied with an enlarged degree
$\bar D\ge D$, called the comparison degree. The enlargement changes only
the positive scalar ${\operatorname{TailFac}}$ and upper bounds for the
backward weight, and it is made after the coefficients of $(1+u)^rG(u)$ have
been bounded through absolute values; the residual parameters $N,r,D$
themselves remain fixed.
Appendices~\ref{app:higher-initial} and~\ref{app:higher-certificates}
prove the initial ranges and the matrix bounds used later.
Appendices~\ref{app:higher-tails-49} and~\ref{app:higher-tails-36}
complete Proposition~\ref{prop:large-alphabet-coefficients};
Appendices~\ref{app:higher-five}, \ref{app:higher-four-initial}
and~\ref{app:higher-four-tail} prove
Propositions~\ref{prop:five-cap-coefficients}, \ref{prop:four-initial}
and~\ref{prop:four-tail}; and Appendix~\ref{app:four-finite-certificate}
specifies the certificate of Proposition~\ref{prop:four-finite}.

\subsection{Initial ranges at \texorpdfstring{$Q\ge25$}{Q at least 25}}
\label{app:higher-initial}

This subsection proves~\eqref{eq:residual-comparison} on the initial range
$1\le i\le I\coloneqq\lfloor\theta N\rfloor$ of every row of the following table.
The rows with $Q\ge36$ serve
Proposition~\ref{prop:large-alphabet-coefficients}, and those with $Q=25$
serve Proposition~\ref{prop:five-cap-coefficients}, where $r>3N/5$; real
alphabets $25<Q<36$ are not needed.

\smallskip\noindent\emph{Step 1 (Cutoffs and the hypotheses of
Lemma~\ref{lem:component-coefficients}).}
For each alphabet the rows divide the possible deficits into consecutive
intervals; their lower endpoints are used only in the tails, which start at
$\lceil\theta N\rceil$ with the same $\theta$
(Appendices~\ref{app:higher-tails-49}--\ref{app:higher-five}).
The column $\bar S$ is the upper bound $s/N\le\bar S/Q$, that is
$s\le\bar S\lambda$, used for that row.
At the last row of the first two blocks this bound is strict: the
hypothesis $s\le\lceil20\lambda\rceil+6m$ of
Proposition~\ref{prop:large-alphabet-coefficients}, with $m\le\lambda$
and $\lambda\ge2$, gives $s<26\lambda+1\le53\lambda/2\le\bar S\lambda$.
The choice of $\theta$ balances the two parts of the proof: the Jacobi test
below needs $\theta$ small, while the tail estimates improve as $\theta$
grows.
\begin{center}
\begin{tabular}{c|c|c|c}
Alphabet&Deficit interval&$\theta$&$\bar S$\\\hline
$Q\ge49$&$s\le10\lambda$&$1/2$&$10$\\
$Q\ge49$&$10\lambda<s\le20\lambda$&$11/32$&$20$\\
$Q\ge49$&$20\lambda<s$&$3/16$&$27$\\\hline
$36\le Q<49$&$s\le15\lambda/2$&$1/2$&$15/2$\\
$36\le Q<49$&$15\lambda/2<s\le16\lambda$&$5/16$&$16$\\
$36\le Q<49$&$16\lambda<s\le23\lambda$&$11/64$&$23$\\
$36\le Q<49$&$23\lambda<s$&$7/64$&$53/2$\\\hline
$Q=25$&$s\le6\lambda$&$27/64$&$6$\\
$Q=25$&$6\lambda<s\le7\lambda$&$25/64$&$7$\\
$Q=25$&$7\lambda<s<10\lambda$&$19/64$&$10$
\end{tabular}
\end{center}
In the last block $s<10\lambda$ follows from $r>3N/5$.
The coefficient $[u^i](1+u)^rG(u)$ is the average over the
mixture~\eqref{eq:positive-mixture} of the coefficients of
$(1+u)^{r+j}(1-au)^{N-j}$. For each such component put
${k_+}\coloneqq r+j$, ${k_-}\coloneqq N-j$, ${n_{\rm mix}}\coloneqq N+r$. Every table row satisfies
$I<{k_+},{k_-}$ and $I-1<{n_{\rm mix}}/2$, using ${k_+}\ge r=N-s$ and
${k_-}\ge N-{J_*}>N(1-4/Q)$ (by~\eqref{eq:mixture-geometry}). The bound
$N\ge Qm\ge2Q$ is used below in the form $1/N\le1/(2Q)$.

\smallskip\noindent\emph{Step 2 (The Jacobi test for $Q\ge49$).}
We verify~\eqref{eq:component-jacobi-test} for $1\le i\le I$. Its left
side decreases with the mixture parameter $a$ and increases with $j$,
while its right side increases with $a$ and does not depend on $j$; the
worst component therefore has $a=1/b$ and $j=0$. As $i$ increases through
$1\le i\le I$, the left side decreases and the right side increases (since
$I-1<{n_{\rm mix}}/2$), so it suffices to take $i=I$, where $(i-1)/N<\theta$;
this applies to every block of the table. We call the resulting lower bound
for the left side, divided by $N$, the main term, and the square of the
main term minus the upper bound for the squared right side, divided by
$N^2$, the squared margin. The test holds when both are positive.

For $Q\ge49$ the main term is $1-\theta-\bar S/Q-1/(Q-1)$.
The respective upper bounds for the squared right side are
$49/(16b)$, $37/(16b)$, and $11/(8b)$, each at least
$4[\theta(2-\theta)+\theta/98]/b$: use ${n_{\rm mix}}\le2N$, $1/N\le1/98$ and
$(i-1)({n_{\rm mix}}+2-i)/N^2\le\theta(2-\theta)+\theta/98$.
At $Q=49$ the main terms and their squared margins are
\[
 (647/2352,32831/2765952),\quad
 (1069/4704,76717/22127616),\quad
 (283/1176,80945/2765952).
\]
The main terms increase and the bounds for the squared right side
decrease with $Q$.

\smallskip\noindent\emph{Step 3 (The Jacobi test for $36\le Q<49$ and
$Q=25$).}
For $36\le Q<49$ keep the exact value ${n_{\rm mix}}=2N-s$ instead of the
bound ${n_{\rm mix}}\le2N$, and the factor $1-a=1-1/b$ of $i-1$. With $t\coloneqq s/N$
it suffices that
\begin{equation}\label{eq:higher-jacobi-scalar}
 [1-t-1/b-(1-1/b)\theta]^2
 >4\theta(2-t-\theta+1/(2Q))/b,
\end{equation}
in which $1/(2Q)$ bounds $1/N$. The difference of the two sides decreases
with $t$ when the positive main term exceeds $2\theta/b$, as it does in all
four rows; so $t=\bar S/Q$ is the worst case. At $Q=36,t=\bar S/Q$ the
main terms and squared margins are
\begin{center}
\begin{tabular}{c|c}
Main term&Squared margin\\\hline
$233/840$&$1649/705600$\\
$563/2520$&$15947/3175200$\\
$1669/10080$&$96079/25401600$\\
$1301/10080$&$52159/25401600$
\end{tabular}
\end{center}
As functions of $Q$ at $t=\bar S/Q$, the main terms increase, and the
right side of~\eqref{eq:higher-jacobi-scalar} decreases: its derivative in
$Q$ has the sign of $(2\bar S-1)(Q-1/2)-(2-\theta)Q^2$, which is negative
because $(2-\theta)Q\ge2\bar S-1$. The same argument at $Q=25$, using
$1/N\le1/50$, gives respectively the main terms and squared margins
\[
 \begin{gathered}
 (12059/38400,4609081/1474560000),\\
 (3891/12800,2239643/491520000),\\
 (701/2560,187067/19660800).
 \end{gathered}
\]
Their main terms exceed $\theta/12$, which is $2\theta/b$ at $b=24$.
This verifies every hypothesis of Lemma~\ref{lem:component-coefficients}
on the initial ranges.

\smallskip\noindent\emph{Step 4 (Packed endpoint and suppression; $Q\ge49$).}
It remains to prove~\eqref{eq:residual-comparison} on these ranges. Write
$h(i)/h(0)=\prod_z(1-i/z)$ over the $d_0$ roots $z$ of $h$. Each factor
increases with $z$, and the roots are distinct integers at least
${a_{\min}}=N-v-d_0+1$; so $h(i)/h(0)$ is at least the value obtained by
packing the roots at ${a_{\min}},\ldots,{a_{\min}}+d_0-1$. We call this value,
$(N-v-d_0)_i/(N-v)_i$, the packed endpoint. Set
$U_0(i)\coloneqq\binom{N-d_0}i/\binom Ni$, the same product with $v=0$. Comparing
the two products factor by factor gives
\begin{equation}\label{eq:higher-packed-loss}
 \frac{h(i)}{h(0)U_0(i)}\ge
 \prod_{j=0}^{i-1}(1-{\varepsilon_j^{\rm pack}}),\qquad
 {\varepsilon_j^{\rm pack}}\coloneqq\frac{d_0v}{(N-v-j)(N-d_0-j)}.
\end{equation}
We call $-\log\prod_j(1-{\varepsilon_j^{\rm pack}})$ the loss. On the other
side, suppression is the exponential factor
in~\eqref{eq:component-suppression}: it bounds each component coefficient
by $\binom{k_+}i$ times a negative exponential, and
Lemma~\ref{lem:mixture-laplace} averages these factors over the mixture.
This bounds $\mathcal L_i$ by $U_0(i)$ times a negative exponential; we
call the absolute value of its exponent the saving. Once the loss is
smaller than the saving,
$\mathcal L_i<U_0(i)\prod_j(1-{\varepsilon_j^{\rm pack}})\le h(i)/h(0)$.
We first give the argument in its simplest form, for $Q\ge49$, and then a
refinement valid for every $Q\ge36$.

Put ${\operatorname{gap}}\coloneqq1-4/Q$, so that ${k_-}\ge N-{J_*}>{\operatorname{gap}}N$.
For $Q\ge49$, since also ${k_+}-i+1\le r+{J_*}\le N-d_0<N$
by~\eqref{eq:mixture-geometry}, the suppression factor is at most
$\exp(-a{\operatorname{gap}}i)$, and
$\binom{r+j}i/\binom{r+{J_*}}i\le\exp(-({J_*}-j)i/N)$. The second ratio
refers every component to the common binomial
$\binom{r+{J_*}}i\le\binom{N-d_0}i$, which after division by $\binom Ni$
gives the factor $U_0(i)$.
Lemma~\ref{lem:mixture-laplace}, with $\kappa={\operatorname{gap}}i/N<1$, gives
$\mathcal L_i\le U_0(i)e^{-{\operatorname{gap}}vi/(8N)}$.
The atoms with $j<{J_*}$ in~\eqref{eq:positive-mixture} (the heads),
for which $a=1/b$, have $aN=N/b>\lambda>v/2$ and satisfy the same bound.
For $i\le N/2$, each denominator in~\eqref{eq:higher-packed-loss}
is at least $(45/49)(N-j)$ and ${\varepsilon_j^{\rm pack}}<1/125$.
Using $-\log(1-\varepsilon)\le125\varepsilon/124$ for
$0\le\varepsilon<1/125$, $d_0\le2N/Q$ and
\begin{equation}\label{eq:reciprocal-integral-sum}
 \sum_{j=0}^{i-1}(N-j)^{-2}\le\frac{i}{N(N-i)},
\end{equation}
the loss is at most $2(49/45)^2(125/124)vi/(Q(N-i))$. Since
$2(49/45)^2(125/124)=12005/5022<5/2$ and $N-i\ge N/2$, it is at most
$5vi/(QN)$. The strict inequality $5/Q<{\operatorname{gap}}/8$ holds for $Q>44$,
so~\eqref{eq:residual-comparison} holds on the initial ranges of the three
rows with $Q\ge49$.

\smallskip\noindent\emph{Step 5 (The same comparison for $Q\ge36$).}
Since $5/Q<{\operatorname{gap}}/8$ fails for $Q\le44$, the alphabets $36\le Q<49$
need a refinement, which is valid for every $Q\ge36$. Keep instead the
finer bound ${k_+}-i+1\le N-d_0-i+1$ in the suppression factor, so that
Lemma~\ref{lem:mixture-laplace} applies with
$\kappa={\operatorname{gap}}i/(N-d_0-i+1)$; here
${\operatorname{gap}}i/(N-i)<\kappa\le1$, the upper bound because $i\le N/2$
and $d_0\le2N/Q$. Thus
$\mathcal L_i<U_0(i)e^{-{\operatorname{gap}}vi/(8(N-i))}$.
The denominator factors in~\eqref{eq:higher-packed-loss} are at
least $(8/9)(N-j)$, and ${\varepsilon_j^{\rm pack}}<1/64$.
Since $2(9/8)^2(64/63)=18/7<3$, the loss is at most
$3vi/(Q(N-i))$, smaller than ${\operatorname{gap}}vi/(8(N-i))$ for $Q>28$.
This proves~\eqref{eq:residual-comparison} on the initial ranges of the
rows with $36\le Q<49$, and again for $Q\ge49$.

\smallskip\noindent\emph{Step 6 (The same comparison for $Q=25$).}
At $Q=25$ put $\alpha\coloneqq m/N\le1/25$,
${\operatorname{gap}}_\alpha\coloneqq1-2/25-2\alpha$, and
$\kappa\coloneqq{\operatorname{gap}}_\alpha i/(N-d_0-i+1)$; now
${k_-}\ge N-{J_*}>N-2\lambda-d_0={\operatorname{gap}}_\alpha N$
by~\eqref{eq:mixture-geometry}. Since $1/N\le\alpha/2$ (as $m\ge2$),
\[
 \frac{\kappa}{i/(N-i)}\ge\frac{{\operatorname{gap}}_\alpha}{1-3\alpha/2}
 \ge42/47,\qquad \kappa\le27/37.
\]
The middle quotient decreases in $\alpha$, so its worst case is
$\alpha=1/25$; and $\kappa\le\theta/(1-\theta)\le27/37$, since
$\theta\le27/64$. By~\eqref{eq:mixture-geometry} ($1\le v<2\lambda$,
${J_*}<4\lambda$), ${J_*}+\lambda+1<7\lambda$,
so~\eqref{eq:beta-laplace} gives
$\E_{\rm tail}e^{-\kappa Na}<(1+\kappa/7)^{-v}$. With
$\log(1+\kappa/7)\ge\kappa/(7+\kappa)\ge37\kappa/286$ and
$777/6721=(42/47)(37/286)$, it follows that
\[
 \mathcal L_i\le U_0(i)
 \exp\left(-\frac{777}{6721}\frac{vi}{N-i}\right).
\]
Here $N-j\ge37N/64$; the two denominators $N-v-j$ and $N-d_0-j$
of~\eqref{eq:higher-packed-loss} are at least $(797/925)(N-j)$ and
${\varepsilon_j^{\rm pack}}<1/38$. The loss is at most
$(70300/635209)vi/(N-i)$. The difference of the two constants is
$21071093/4269239689>0$, which proves~\eqref{eq:residual-comparison} on
the three initial ranges with $Q=25$. This proves all initial intervals in
the table.

\subsection{Finite coercivity and signed matrix certificates}
\label{app:higher-certificates}

This subsection proves two kinds of bounds used in the tails. Step~1 gives
lower bounds $M_0\succeq{\operatorname{coer}}I$ for the base matrix $M_0$ of
Section~\ref{sec:coefficient-methods}, hence admissible constants
${\operatorname{pref}}=2/{\operatorname{coer}}$ in Lemma~\ref{lem:weighted-rows}.
Steps~2--4 give, for $2\le m\le5$, exact certificates that bound
$\sum_\ell|G_\ell|a^{2m-\ell}$ directly.

\smallskip\noindent\emph{Step 1 (Coercivity and admissible values of
${\operatorname{pref}}$).}
Let ${J_m^{\rm path}}$ be the symmetric $(m+1)\times(m+1)$ matrix, indexed by
$0\le j\le m$, with zero diagonal and entries $\sqrt{(j+1)/m}/2$ at
$(j,j+1)$ and $(j+1,j)$ (a weighted path).
By~\eqref{eq:base-matrix-entries} in
Appendix~\ref{app:coefficient-details} and $N\ge Qm$, the diagonal entries
of $M_0$ are at least $(b-1)/b$ and its off-diagonal entries are at most
$(b-1)/b$ times those of ${J_m^{\rm path}}$. Comparing the quadratic forms on
the absolute values of the coordinates gives
\begin{equation}\label{eq:retained-coercivity}
 M_0\succeq\frac{b-1}{b}(1-\|{J_m^{\rm path}}\|)I.
\end{equation}
For a nonnegative symmetric matrix and a positive weight vector, the
norm is at most the largest row ratio, the weighted row sum divided by the
weight of that row. The positive row weights
$(m+1-j)(m+10)^{j/2}/\sqrt{j!}$ give the row bound
\begin{equation}\label{eq:finite-tridiagonal-norm}
 \|{J_m^{\rm path}}\|\le\frac{m+2}{\sqrt{m(m+10)}}.
\end{equation}
Indeed, with $u\coloneqq m+1-j$, the row ratio is at most
$(2m+10-u-9/u)/(2\sqrt{m(m+10)})$, and $u+9/u\ge6$.
For $2\le m\le15$ this is at most $8/9$, as follows from
$17m^2-316m+324\le0$. The sharper estimates
$\|{J_3^{\rm path}}\|=\sqrt{(3+\sqrt6)/12}<11/16$ and
$\|{J_4^{\rm path}}\|\le6/\sqrt{56}<13/16$ give, for $Q\ge36$,
\begin{equation}\label{eq:large-alphabet-P}
 {\operatorname{pref}}=20\ (2\le m\le15),\quad {\operatorname{pref}}_3=20/3,\quad {\operatorname{pref}}_4=12,
 \quad {\operatorname{pref}}=2(m+2)^2\ (m\ge2)
\end{equation}
as valid choices in Lemma~\ref{lem:weighted-rows}. Indeed, with
$(b-1)/b\ge34/35$, \eqref{eq:retained-coercivity} gives
${\operatorname{pref}}=2/{\operatorname{coer}}$ at most $315/17<20$, $112/17<20/3$ and
$560/51<12$, respectively, while ${\operatorname{pref}}=2(m+2)^2$ follows from the
choice of ${\operatorname{coer}}$ allowed for every $m$ in
Lemma~\ref{lem:weighted-rows}. Below, ${\operatorname{pref}}_m$ is the value used for
a given $m$; in~\eqref{eq:Q49-five-scalars} and in
Appendix~\ref{app:higher-tails-36} these are
${\operatorname{pref}}_3=20/3$, ${\operatorname{pref}}_4=12$, and ${\operatorname{pref}}_m=20$ for
$5\le m\le7$.

\smallskip\noindent\emph{Step 2 (The signed matrices $Y$ and $Z(a)$).}
The certificates use the coefficient matrices of the
kernel~\eqref{eq:finite-paired-kernel} before the normalization by
$\sqrt{\nu_j\nu_k}$ of Section~\ref{sec:coefficient-methods}, in
coordinates rescaled by the diagonal matrix $\operatorname{diag}((-1)^jb^j)$,
$0\le j\le m$, whose signs alternate. Put $v_j\coloneqq v_j(N)=b^j\binom Nj$ and
$v_{-1}\coloneqq0$. In these coordinates the base matrix, the coefficient of $u^0$,
becomes the matrix $Y$ with entries
\begin{equation}\label{eq:signed-certificate-base}
 Y_{jj}=v_j-bv_{j-1},\qquad
 Y_{j,j+1}=Y_{j+1,j}=-(b-1)v_j/2;
\end{equation}
equivalently, $Y$ is congruent to $M_0$ by
$\operatorname{diag}((-1)^j\sqrt{v_j})$.
For $0\le j,k\le m$ and $a\ge0$ define
\begin{align}\label{eq:signed-certificate-majorant}
 {\operatorname{Entry}_{jk}}(a)&\coloneqq\sum_{{j_1}=0}^{\min(j,k)}\sum_{t=0}^{j_1}
 \frac{(N)_{j+k-{j_1}}\binom {j_1}t b^{{j_1}-t}(b-1)^t
 a^{2m-j-k+2{j_1}-t}}{(j-{j_1})!(k-{j_1})!{j_1}!},\notag\\
 Z_{jk}(a)&\coloneqq{\operatorname{Entry}_{jk}}(a)+\frac{b-1}{2}
       [{\operatorname{Entry}_{j-1,k}}(a)+{\operatorname{Entry}_{j,k-1}}(a)]+b{\operatorname{Entry}_{j-1,k-1}}(a).
\end{align}
A negative index contributes zero. All exponents are nonnegative.
Here ${\operatorname{Entry}_{jk}}(a)$ is the sum over $\ell$ of $a^{2m-\ell}$ times the
rescaled, nonnegative $z^jw^ku^\ell$ coefficient of
$[{\operatorname{Base}}+u{\operatorname{Tilt}}]^N$, and $Z_{jk}(a)$ is the same weighted sum
for the absolute kernel, in which the coefficient $-1/b$ of $zw$ in
${\operatorname{Seed}}$ is replaced by $+1/b$. These are exact coefficients; no
eigenvalue approximation is involved. Let $q$ be a coefficient vector in
these coordinates. The true denominator $q^{\mathsf T}Yq$ of $G_\ell$, the
quadratic form of $M_0$, is at least $|q|^{\mathsf T}Y|q|$, because the
off-diagonal entries of $Y$ are negative. The absolute values of the
numerator forms of the $G_\ell$, weighted by $a^{2m-\ell}$ and summed, are
at most $|q|^{\mathsf T}Z(a)|q|$. Hence, if $V$ is a positive multiple of $Y$
minus a positive multiple of $Z(a)$ and $V$ is positive definite, then
$\sum_\ell|G_\ell|a^{2m-\ell}$ is less than the ratio of the two multiples.

\smallskip\noindent\emph{Step 3 (The three certificates).}
The following three certificates will be used. Each asserts positivity
of every nonzero coefficient of every leading principal minor of $V$, as
a polynomial in the specified nonnegative variables. The complete rational
coefficient lists of these minors are in
the certificate package~\cite{ComputationalCertificates}. Certificate~1 is
used in Appendix~\ref{app:higher-tails-49} (critical deficit, $m=2$),
certificate~2 for $m=2$ in Appendices~\ref{app:higher-tails-36}
and~\ref{app:higher-five}, and certificate~3 for $m=3,4,5$ in
Appendix~\ref{app:higher-five}.
\begin{enumerate}
\item For $m=2$, $Q\ge49$, at $a=1$, use
$V\coloneqq14N^2Y-Q^2Z(1)$ and variables $x\coloneqq Q-49$, $y\coloneqq N-2Q$.
The three minor supports have sizes $6,27,85$, and their constant
terms are
\[
132055,\quad253264397028895/4,\quad5969157740628909704237129.
\]
This gives $\sum_{\ell=0}^4|G_\ell|<14\lambda^2$.
For direct construction at $a=1$ one may equivalently use
${\operatorname{Entry}_{jk}}(1)=\sum_{j_1}(N)_{j+k-{j_1}}(2b-1)^{j_1}/((j-{j_1})!(k-{j_1})!{j_1}!)$.
\item For $m=2$, $Q\ge25$, put
${\operatorname{Bound}_2}(a)\coloneqq\frac98(1+a)^4-\frac72a^2$ and
$V\coloneqq N^2{\operatorname{Bound}_2}(a)Y-Q^2Z(a)$.
In variables $x\coloneqq Q-25$, $y\coloneqq N-2Q$, $a$, its three minor supports
have sizes $30,243,1105$, with constants
\[
5625/2,\quad132099609375/16,\quad8697328033447265625/2.
\]
It proves
\begin{equation}\label{eq:two-pair-allweight}
 \sum_{\ell=0}^4|G_\ell|a^{4-\ell}<\lambda^2{\operatorname{Bound}_2}(a).
\end{equation}
All coefficients of ${\operatorname{Bound}_2}$ are positive.
\item At $Q=25$, for $m=3,4,5$ put
$\gamma_3\coloneqq1/3$, $\gamma_4\coloneqq1/12$, $\gamma_5\coloneqq1/70$ and
$V\coloneqq\gamma_mN^m(1+a)^{2m}Y-25^mZ(a)$.
Expand in $x\coloneqq N-25m$, $a$. The support sizes by minor order are
\begin{center}
\begin{tabular}{c|l}
$m$&Support sizes by minor order\\\hline
$3$&$28,104,247,475$\\
$4$&$45,170,400,759,1271$\\
$5$&$66,252,589,1107,1836,2806$.
\end{tabular}
\end{center}
All 10155 coefficients (the sum of the support sizes in the table) and
each minor's constant term are positive.
It follows that
\begin{equation}\label{eq:five-low-degree-mass}
 \sum_\ell|G_\ell|a^{2m-\ell}
 <\gamma_m\lambda^m(1+a)^{2m}.
\end{equation}
\end{enumerate}

\smallskip\noindent\emph{Step 4 (From positive coefficients to the signed
bounds).}
Every minor is a polynomial with rational coefficients obtained
from the displayed falling-factorial sums. The factors contributed by
factorials, powers of $Q$ and the diagonal congruences are positive on the
full stated domains, so they do not change the signs of the minors. A
polynomial whose nonzero coefficients are all positive is positive wherever
its variables are nonnegative; the positive constant terms cover the corner
$x=0,y=0,a=0$, where every other monomial vanishes.
Sylvester's criterion therefore proves $V>0$ on each whole domain, and the
absolute-coordinate comparison of Step~2 proves the claimed signed bounds.
The negative off-diagonal signs in $Y$ are essential there: they give the
lower bound $q^{\mathsf T}Yq\ge|q|^{\mathsf T}Y|q|$ for the true denominator.

\subsection{The tails for \texorpdfstring{$Q\ge49$}{Q at least 49}}
\label{app:higher-tails-49}

We prove Proposition~\ref{prop:large-alphabet-coefficients} first
on this interval, by proving~\eqref{eq:residual-comparison} on the tail
$i_0\le i\le D$ of each row of the table, where $i_0\coloneqq\lceil\theta N\rceil$
with the $\theta$ of that row; with Appendix~\ref{app:higher-initial} this
covers every index. If $i_0>D$ the tail is empty. Otherwise
$i\ge i_0\ge4d_0$, so the hypothesis $(D+1)(i-d_0)>d_0^2$ of
Lemma~\ref{lem:positive-tail} holds (as $D\ge i$), and every comparison
degree $\bar D\ge D$ used below is admissible. We treat the middle and
last deficits first, since they share one argument. We call the first
deficit interval of each alphabet critical: there $D$ can be as large
as~\eqref{eq:coefficient-root-domain} allows. (This use of the word is
unrelated to the critical distance strip of Section~\ref{sec:distance-strips}.)

\smallskip\noindent\emph{Step 1 (Middle deficit).}
For the middle deficit $10\lambda<s\le20\lambda$ take
$\bar D\coloneqq N-\lceil10\lambda\rceil+d_0$, which is at least $D=N-s+d_0$
because $s\ge\lceil10\lambda\rceil$.
Then $k=N-\bar D-d_0\ge6m=3d_0$ and
$\bar D/N\ge39/49$. With $\theta=11/32$, $\zeta=21/32$,
the backward weight $a=(D-i)/(i-d_0+1)$ of Lemma~\ref{lem:positive-tail}
is at most $\zeta/\theta=21/11$ for $i\ge i_0$, because $i\ge\theta N$ and
$N-\bar D\ge8\lambda>(\zeta/\theta)(d_0-1)$; this bounds the factor
$(1+a)^{d_0}$ of~\eqref{eq:row-factorial-bound} by $\theta^{-2m}$.
The factorial ratio ${\operatorname{TailFac}}(i,\bar D)$ decreases with $i$, since
$d_0(N-i)-k(i+1)<0$; so it suffices to take $i=i_0$. The ceiling in
$i_0=\lceil\theta N\rceil$ contributes to $(i_0/\bar D)^{2m}$ at most the
factor $\exp(2/(\theta Q))<8/7$.
Equations~\eqref{eq:row-factorial-bound} and
\eqref{eq:tail-factorial-cancellation}, with $\lambda_+\le(49/48)\lambda$,
$(1+a)^{2m}(i_0/\bar D)^{2m}\le(8/7)(N/\bar D)^{2m}$, $N/\bar D\le49/39$
and $k\ge10\lambda-4m$, give
\[
 \frac{\mathcal L_i}{h(i)/h(0)}
 <{\operatorname{pref}}\frac87(49/48)^m\frac{\lambda^m}{m!}
       (49/39)^{2m}\zeta^{10\lambda-4m}.
\]
The factor depending on $\lambda$ decreases for $\lambda\ge m$,
since its logarithmic derivative is at most $1+10\log(21/32)<0$.
Using $m^m/m!<3^m$, the resulting base is
\[
3(49/48)(49/39)^2(21/32)^6
=1121144263281/2903397892096<7/18.
\]
Retain the exact corrections $m^m/(m!3^m)=2/9,1/6$ for $m=2,3$.
With ${\operatorname{pref}}=20$ the bounds at $m=2,3,4$ are
$560/729$, $490/2187$, $3430/6561$.
For $m\ge5$, use ${\operatorname{pref}}=2(m+2)^2$, which is valid for every $m$:
its value at five is $117649/118098<1$, and the next ratio is at most
$32/63$.

\smallskip\noindent\emph{Step 2 (Last deficit).}
For the last deficit $s>20\lambda$ take
$\bar D\coloneqq N-\lceil20\lambda\rceil+d_0$, again at least $D$.
Now $k\ge16m=8d_0$, $\bar D/N\ge29/49$, $\theta=3/16$ and
$\zeta=13/16$, and the backward weight is at most $\zeta/\theta=13/3$.
The argument of Step~1 applies because $N-\bar D\ge18\lambda>
(\zeta/\theta)(d_0-1)$; the ceiling factor is less than $9/7$, and the
factor depending on $\lambda$ decreases since $1+20\log(13/16)<0$.
The base is
\[
3(49/48)(49/29)^2(13/16)^{16}
=78285598653791925113809/248219388255835726544896<8/25.
\]
With ${\operatorname{pref}}=2(m+2)^2$ and the same two exact corrections, the bounds
at $m=2,3,4$ are $4096/4375$, $1536/4375$, and
$2654208/2734375$. From four onward the next ratio is at most
$(7/6)^2(8/25)<1$.

\smallskip\noindent\emph{Step 3 (Critical deficit: the factorial ratio).}
For the critical deficit $s\le10\lambda$ take instead the largest degree
allowed by~\eqref{eq:coefficient-root-domain},
\begin{equation}\label{eq:critical-comparison-degree}
 \bar D\coloneqq\left\lfloor\frac{(Q-2)(N-d_0+1)}Q\right\rfloor,
 \quad k\coloneqq\left\lceil2\lambda-4m/Q-1+2/Q\right\rceil,
\end{equation}
so that $\bar D\ge D$ automatically. Here $\theta=1/2$, so the tail is
$i\ge N/2$; $\bar D/N\ge45/49$, and the backward weight is at most one for
$i\ge N/2$, because $\bar D+d_0-1<N\le2i$.
For $u\coloneqq i/N$, the factorial ratio is at most
$(49/45)^{2m}u^{2m}(1-u)^k$.
If $k\ge2m$, its maximum on $u\ge1/2$ is $2^{-2m-k}$.
If $0<k<2m$, its unconstrained maximum is at $u=2m/(2m+k)$,
and is at most
\begin{equation}\label{eq:beta-maximum}
 2^{-2m-k}\exp\left(\frac{(2m-k)^2}{4k}\right).
\end{equation}
Indeed, with $\delta\coloneqq(2m-k)/(2m+k)$, the logarithmic excess is
$(2m+k)J(\delta)$, where $J'(\delta)=\operatorname{atanh}\delta
\le\delta/(1-\delta)$ and
$J(\delta)\le\delta^2/(2(1-\delta))$.
The bounds $k\ge139m/98$ and $(2m-k)_+\le1+4m/49$, which follow from
$k\ge2m-4m/49-1$ (by $\lambda\ge m$ and $Q\ge49$) and $m\ge2$, give
the exponent in~\eqref{eq:beta-maximum} less than $1/(4m)+m/128$.
In detail, the bound before simplification is
$49/(278m)+4/139+8m/6811$;
use $4/139<1/(16m)+m/256$.
Since $2^{4/49}<17/16$ and $e^{1/128}<128/127$,
\begin{equation}\label{eq:Q49-critical-R}
 {\operatorname{TailFac}(i,\bar D)}<2^{1-2m-2\lambda}(49/45)^{2m}
                  (136/127)^m e^{1/(4m)}.
\end{equation}

\smallskip\noindent\emph{Step 4 (Critical deficit, $m=2$).}
For $m=2$, the backward weight $a\le1$ gives
$\sum_\ell|G_\ell|a^{4-\ell}\le\sum_\ell|G_\ell|<14\lambda^2$ by the first
signed certificate. With~\eqref{eq:Q49-critical-R}, the decrease of
$\lambda^2 2^{-2\lambda}$ for $\lambda\ge2$, and $8/7>e^{1/8}$ for the
factor $e^{1/(4m)}$, the entire tail reduces to
\[
56\,2^{-7}(49/45)^4(136/127)^2(8/7)
=53312879648/66138980625<1.
\]

\smallskip\noindent\emph{Step 5 (Critical deficit, $3\le m\le7$).}
For $3\le m\le7$ retain the positive row polynomial
\[
 {\operatorname{Row}_m}(\lambda)\coloneqq\sum_{{j_1}=0}^m\binom m{j_1}2^{j_1}
                    \sum_{{j_2}=0}^{m-{j_1}}[(49/48)\lambda]^{j_2}/{j_2}!.
\]
For $a\le1$, the absolute mass $\sum_\ell|G_\ell|a^{d_0-\ell}$ is at most
$\sum_\ell|G_\ell|$, which~\eqref{eq:finite-row-bound} at $u=1$ bounds by
${\operatorname{pref}}_m{\operatorname{Row}_m}(\lambda)$, with ${\operatorname{pref}}_m$ from
\eqref{eq:large-alphabet-P}: use $\binom j{j_1}\le\binom m{j_1}$,
$1+\omega_1<2$, and $\binom{N-j}{j_2}b^{-j_2}\le\lambda_+^{j_2}/j_2!$ with
$\lambda_+\le(49/48)\lambda$. Its positive coefficients and degree $m$
imply that ${\operatorname{Row}_m}(\lambda)2^{-2\lambda}$ decreases for $\lambda\ge m$.
For $m=3,\ldots,7$ the five exact quantities
\begin{equation}\label{eq:Q49-five-scalars}
 {\operatorname{pref}}_m{\operatorname{Row}_m}(m)2^{1-4m}(49/45)^{2m}(136/127)^m\frac{4m}{4m-1},
\end{equation}
in which $4m/(4m-1)\ge e^{1/(4m)}$, are respectively less than
$9/10,3/4,3/5,3/10,3/20$.
These comparisons are exact rational evaluations of the displayed finite
sums at these five values of $m$.

\smallskip\noindent\emph{Step 6 (Critical deficit, $m\ge8$).}
For larger degrees,
${\operatorname{Row}_m}(m)=[x^m](1+2x)^me^{49mx/48}/(1-x)$.
Positive coefficient extraction at $x=1/2$ gives
${\operatorname{Row}_m}(m)\le2\cdot4^me^{49m/96}<2\cdot7^m$.
For the last inequality use
$e^x\le1+x+x^2/(2(1-x/3))$ at $x=8/15>49/96$,
which gives $e^{8/15}\le947/555<7/4$.
In~\eqref{eq:Q49-critical-R} the resulting base is
$285719/514350<5/9$, and the prefactor $4e^{1/(4m)}<9/2$ for $m\ge8$.
The ratio is therefore less than $(9/2){\operatorname{pref}}(5/9)^m$.
For $8\le m\le12$ use ${\operatorname{pref}}=20$; at eight the bound is
$3906250/4782969<1$. For $m\ge13$ use ${\operatorname{pref}}=2(m+2)^2$;
at thirteen it is $30517578125/31381059609<1$, and the next
ratio is at most $256/405$.
This completes every $Q\ge49$ tail, and hence that part of the proposition.

\subsection{The tails for \texorpdfstring{$36\le Q<49$}{Q at least 36 and less than 49}}
\label{app:higher-tails-36}

This subsection proves~\eqref{eq:residual-comparison} on the tail
$i_0\le i\le D$, $i_0\coloneqq\lceil\theta N\rceil$, of each row with
$36\le Q<49$, which completes
Proposition~\ref{prop:large-alphabet-coefficients}. Here
$i_0\ge63d_0/32$, so $(D+1)(i-d_0)>i(i-d_0)\ge(63/32)(31/32)d_0^2>d_0^2$
for $D\ge i\ge i_0$, and Lemma~\ref{lem:positive-tail} permits every
comparison degree $\bar D\ge D$ below. The weighted sum
$\sum_\ell|G_\ell|a^{d_0-\ell}$ has nonnegative coefficients in $a$, and the
backward weight is at most $a_i\coloneqq(\bar D-i)/(i-d_0+1)$. Hence
Lemma~\ref{lem:positive-tail}, with~\eqref{eq:row-exponential-bound} at
the weight $a_i$, gives
\[
 \frac{\mathcal L_i}{h(i)/h(0)}\le{\operatorname{pref}}\,F(i,\bar D),
\]
where $F$ is the combined scalar~\eqref{eq:combined-tail-scalar}. This
bound is used also in Appendices~\ref{app:higher-five}
and~\ref{app:higher-four-tail}.

\smallskip\noindent\emph{Step 1 (Comparison degrees and decrease in $i$).}
For the three noncritical deficits, whose intervals are bounded below
by $s>S\lambda$ with respectively $S=15/2,16,23$ (these lower endpoints are
not the column $\bar S$ of the table), choose
$\bar D\coloneqq N-\lceil S\lambda\rceil+d_0$, which is at least $D$ because
$s\ge\lceil S\lambda\rceil$. Then
\[
 k\ge(S-4)\lambda,\qquad {J_D}\coloneqq\bar D-d_0+1\ge(Q-S)\lambda.
\]
The combined scalar decreases in $i$, by the ratio bound of
Lemma~\ref{lem:positive-tail}, because
$\lambda_+/{J_D}\le Q/((Q-1)(Q-S))<(S-4)/Q\le k/(N-i)$, that is,
$(Q-1)(Q-S)(S-4)-Q^2>0$.
In $x\coloneqq Q-36$ the coefficients of this quadratic are
$S-5$, $(S-4)(71-S)-72$, and
$35(36-S)(S-4)-1296$, all positive for these three $S$.
For the critical deficit $s\le15\lambda/2$
use~\eqref{eq:critical-comparison-degree}.
Now $k<2\lambda$, $k\ge(3/2-4/Q)\lambda$, and
${J_D}\ge(Q-6)\lambda$. The tail is $i\ge N/2$, where decrease follows
from $2Q^3-29Q^2+74Q-48>0$; in $x\coloneqq Q-36$ this cubic is
$2x^3+187x^2+5762x+58344$.

\smallskip\noindent\emph{Step 2 (Noncritical deficits, $m\ge3$).}
For the noncritical deficits write $\zeta\coloneqq1-\theta$. At $i_0$, with
$z\coloneqq i_0-d_0+1$ and $1+a_i={J_D}/z$ (from the proof of
Lemma~\ref{lem:positive-tail}),
\[
 (1+a_i)^{d_0}\frac{(i_0)_{d_0}}{(\bar D)_{d_0}}
 =\prod_{j=0}^{d_0-1}\frac{1+j/z}{1+j/{J_D}}
 \le e^{d_0(d_0-1)/(2z)},\qquad
 z\ge(\theta Q-2)\lambda,\quad z\le\theta N.
\]
The bound $z\le\theta N$ enters the exponential factor of $F$.
Consequently
\begin{equation}\label{eq:Q36-noncritical-F}
 F(i_0,\bar D)\le
 e^{2m/(\theta Q-2)}
 e^{\theta Q^2\lambda/((Q-1)(Q-S))}\zeta^{S\lambda-4m}.
\end{equation}
The $\lambda$ factor decreases for $\lambda\ge m$, and the positive
exponent terms decrease with $Q$. At $Q=36$ and $\lambda=m$ the resulting
bases are $e^E\zeta^{S-4}$, where $E\coloneqq2/(\theta Q-2)+\theta Q^2/((Q-1)(Q-S))$
at $Q=36$; the three pairs are
\[
 (E,\zeta)=
 (3062/4921,11/16),\quad
 (149297/187600,53/64),\quad
 (10831/8060,57/64).
\]
Each base is less than $51/100$: use respectively
$e^{5/8}\le569/304<15/8$ and square (the first row has
$\zeta^{S-4}=\zeta^{7/2}$), $e^{4/5}\le123/55<9/4$, and $e^{11/8}<4$.
The first two follow from the Taylor bound $e^x\le1+x+x^2/(2(1-x/3))$ of
Appendix~\ref{app:higher-tails-49}; the third follows from
$\log2>56/81>11/16$, where $56/81=2(1/3+1/81)$ is a partial sum of
$\log2=2\operatorname{atanh}(1/3)$.
For $m\ge3$ the ratio is therefore less than ${\operatorname{pref}}_m(51/100)^m$.
With ${\operatorname{pref}}_3=20/3$, ${\operatorname{pref}}_4=12$ and ${\operatorname{pref}}_m=20$ for
$5\le m\le7$, these bounds are below one. For $m\ge8$,
${\operatorname{pref}}=2(m+2)^2$ starts at $200(51/100)^8<1$ and its next ratio is
at most $(11/10)^2(51/100)<1$.

\smallskip\noindent\emph{Step 3 (Noncritical deficits, $m=2$).}
For $m=2$ use~\eqref{eq:two-pair-allweight} and discard its negative term:
$\sum_\ell|G_\ell|a^{4-\ell}<(9/8)\lambda^2(1+a)^4$. The ratio is then at
most $(9/8)\lambda^2$ times $F$ without its exponential factor, that is,
times the product form of $F$ in the proof of
Lemma~\ref{lem:positive-tail} without $e^{\lambda_+z/{J_D}}$; this product
decreases in $i$. Since $\lambda^2\zeta^{S\lambda}$ decreases for
$\lambda\ge2$, the three bounds reduce, at $\lambda=2$, to
$(9/2)e^{4/(36\theta-2)}\zeta^{2S-8}$.
The exponents $16/37,64/67,64/31$ are smaller than
$\log2,\log3,\log16$, respectively. Thus the rational upper bounds
are $9(11/16)^7$, $(27/2)(53/64)^{24}$,
and $72(57/64)^{38}$, all below one.

\smallskip\noindent\emph{Step 4 (Critical deficit, $m=2$).}
For the critical deficit with $m=2$, $k>2\lambda-5/4$ and
$\bar D/N\ge8/9$, and the backward weight is at most one. At $a=1$,
\eqref{eq:two-pair-allweight} and ${\operatorname{Bound}_2}(1)=29/2$ give absolute
mass $\sum_\ell|G_\ell|$ less than $(29/2)\lambda^2$. For every $i\ge N/2$
the resulting bound for $\mathcal L_i/(h(i)/h(0))$ is at most
\[
 \frac{29}2\lambda^2(9/8)^4
       \max_{0<u<1}u^4(1-u)^{2\lambda-5/4}.
\]
It decreases for $\lambda\ge2$: its logarithmic derivative is at most
$2/\lambda-8/(2\lambda+11/4)<0$.
At $\lambda=2$ the maximizer is $16/27$, and the fourth power of
the bound is
\[
 \frac{211597943060738805334016}{239299329230617529590083}<1.
\]

\smallskip\noindent\emph{Step 5 (Critical deficit, $3\le m\le7$).}
Here we use the upper bound $Q<49$, which gives $4/Q-8/Q^2>1/13$; this is
why Proposition~\ref{prop:large-alphabet-coefficients} treats
$36\le Q<49$ separately. Since $a_i$ decreases in $i$, for $i\ge N/2$,
$1-a_i\ge(k+1)/(N/2-d_0+1)\ge4/Q-8/Q^2>1/13$.
Thus $a_i\le a_{\rm fix}\coloneqq12/13$. By~\eqref{eq:finite-row-bound} at
$u=1/a$, $\sum_\ell|G_\ell|a^{2m-\ell}\le{\operatorname{pref}}\max_ja^{2m}{\operatorname{Row}_j}(1/a)$,
where
\[
 a^{2m}{\operatorname{Row}_j}(1/a)=\sum_{j_1=0}^j\binom j{j_1}(a+\omega_1)^{j_1}
 \sum_{j_2=0}^{m-j_1}\binom{N-j}{j_2}b^{-j_2}a^{2m-j-j_2}.
\]
For $a\le a_{\rm fix}$ this is at most the row polynomial (different from
that of Appendix~\ref{app:higher-tails-49})
\[
 {\operatorname{Row}_m}(\lambda)\coloneqq\sum_{{j_1}=0}^m\binom m{j_1}(1+a_{\rm fix})^{j_1}
       \sum_{{j_2}=0}^{m-{j_1}}[(36/35)\lambda]^{j_2}a_{\rm fix}^{m-{j_2}}/{j_2}!,
\]
because $a_{\rm fix}<1$ permits $a_{\rm fix}^{m-j}\le1$, while
$\binom j{j_1}\le\binom m{j_1}$, $\omega_1\le1$ and $N/b\le(36/35)\lambda$.
Set $E_m\coloneqq(1+m/9)^2/(4((17/9)m-1))<1$. The beta
maximum~\eqref{eq:beta-maximum} with $k\ge2\lambda-m/9-1$ (so that
$(2m-k)_+\le1+m/9$ and $k\ge(17/9)m-1$ for $\lambda\ge m$), the bound
$e^{E_m}\le1/(1-E_m)$, and $2^{1/9}<13/12$ give
\[
 {\operatorname{TailFac}(i,\bar D)}\le
 \frac{2^{1-2m-2\lambda}(9/8)^{2m}(13/12)^m}{1-E_m}.
\]
Since ${\operatorname{Row}_m}(\lambda)2^{-2\lambda}$ decreases, it suffices to evaluate
at $\lambda=m$. For $m=3,\ldots,7$ the endpoint values
${\operatorname{pref}}_m{\operatorname{Row}_m}(m)2^{1-4m}(9/8)^{2m}(13/12)^m/(1-E_m)$ are
respectively
\begin{align*}
&12889431529509897/13726494018764800,\\
&579660491442892280669253/719750881215093145600000,\\
&735397480235698517037519729/1115373902780392356246978560,\\
&2286423785096725190097841457625333/6933102702381190833309220864000000,\\
&1252786663244096956564041490798729023/7502952816660916207205696405504000000.
\end{align*}
They are all below one.

\smallskip\noindent\emph{Step 6 (Critical deficit, $m\ge8$).}
Finally, for $m\ge8$ decrease of $F$ places its maximum at
$i_0=\lceil N/2\rceil$, where $z\coloneqq i_0-d_0+1$ satisfies
$(Q/2-2)\lambda\le z\le N/2$. Retain the denominator in the product:
\[
 \log\prod_{j=0}^{d_0-1}\frac{1+j/z}{1+j/{J_D}}
 \le\frac{d_0(d_0-1)}2(1/z-1/{J_D})
 \le2m\left(\frac1{Q/2-2}-\frac1Q\right).
\]
Thus
\[
 F(i_0,\bar D)\le
 2e^{2m(1/(Q/2-2)-1/Q)}
 e^{Q^2\lambda/(2(Q-1)(Q-6))}2^{-2\lambda+4m/Q}.
\]
Each factor depending on $Q$ decreases in $Q$, so $Q=36$ is the worst
case, and the $\lambda$ factor decreases. At $Q=36,\lambda=m$, the positive
exponent apart from powers of two is $8651/12600<69/100$;
$e^{69/100}\le30787/15400<2$. The resulting base is
$2^{-8/9}<6/11$. The ratio is less than $2{\operatorname{pref}}(6/11)^m$.
For $8\le m\le10$ use ${\operatorname{pref}}=20$, beginning at
$67184640/214358881<1$. For $m\ge11$ use ${\operatorname{pref}}=2(m+2)^2$,
beginning at $245250809856/\allowbreak285311670611<1$, with next ratio
at most $(14/13)^2(6/11)<1$.
The initial ranges $1\le i\le\lfloor\theta N\rfloor$ of
Appendix~\ref{app:higher-initial} and these tails
$\lceil\theta N\rceil\le i\le D$ cover every index $1\le i\le D$.
Together with the $Q\ge49$ proof this completes
Proposition~\ref{prop:large-alphabet-coefficients}.

\subsection{The tails for \texorpdfstring{$p=5$}{p=5}}
\label{app:higher-five}

We prove Proposition~\ref{prop:five-cap-coefficients}. Its three
initial ranges were proved in Appendix~\ref{app:higher-initial}; here we
prove~\eqref{eq:residual-comparison} on the tails $i_0\le i\le D$,
$i_0\coloneqq\lceil\theta N\rceil$. All of them start at $i_0>3d_0$, so
$(D+1)(i-d_0)>3d_0\cdot2d_0>d_0^2$ for $D\ge i\ge i_0$,
Lemma~\ref{lem:positive-tail} applies with every comparison degree below,
and the bound $\mathcal L_i/(h(i)/h(0))\le{\operatorname{pref}}F(i,\bar D)$ of
Appendix~\ref{app:higher-tails-36} holds.

\smallskip\noindent\emph{Step 1 (Values of ${\operatorname{pref}}_m$ at $Q=25$).}
At $Q=25$,~\eqref{eq:retained-coercivity} has factor $23/24$.
The following values of ${\operatorname{pref}}_m$ are valid in
Lemma~\ref{lem:weighted-rows}; they are used for $m\ge6$, while for
$2\le m\le5$ the signed certificates of
Appendix~\ref{app:higher-certificates} replace that lemma:
\begin{equation}\label{eq:five-P-constants}
 \begin{gathered}
 {\operatorname{pref}}_6=9,\quad {\operatorname{pref}}_7=10,\quad {\operatorname{pref}}_8=11,\quad {\operatorname{pref}}_9={\operatorname{pref}}_{10}=12,
 \\
 {\operatorname{pref}}=(3m+25)/4\quad(m\ge11).
 \end{gathered}
\end{equation}
For $6\le m\le10$ set $\beta_m\coloneqq1-48/(23{\operatorname{pref}}_m)$. Positive
definiteness of $\beta_mI-{J_m^{\rm path}}$ gives, through the
absolute-coordinate comparison behind~\eqref{eq:retained-coercivity},
$M_0>(23/24)(1-\beta_m)I=2I/{\operatorname{pref}}_m$. The leading principal minors
of $\beta_mI-{J_m^{\rm path}}$ obey
\[
 \Delta_0=1,\quad\Delta_1=\beta_m,\quad
 \Delta_j=\beta_m\Delta_{j-1}-\frac{j-1}{4m}\Delta_{j-2}
 \quad(2\le j\le m+1).
\]
The $45=7+8+9+10+11$ minors $\Delta_j$, $1\le j\le m+1$, at the five
displayed pairs $(m,{\operatorname{pref}}_m)$ are positive rational numbers; their
values are included in the exact scalar table
of~\cite{ComputationalCertificates}. Sylvester's criterion gives the
positive definiteness, hence $M_0>2I/{\operatorname{pref}}_m$.
For $m\ge11$, it is enough to verify
$(m+2)/\sqrt{m(m+10)}\le(69m+383)/(69m+575)$.
After squaring, the cleared difference at $m=11+x$ is
\[
516920+757920x+76470x^2+2070x^3>0.
\]
This proves~\eqref{eq:five-P-constants} for every $m\ge6$ and $N\ge25m$.

\smallskip\noindent\emph{Step 2 (Comparison degrees and decrease in $i$).}
For the critical deficit $s\le6\lambda$ (the first interval, as in
Appendix~\ref{app:higher-tails-49}) define
\begin{equation}\label{eq:five-cap-degree}
 k\coloneqq\lfloor(N-d_0)/13\rfloor+1,\qquad
 \bar D\coloneqq N-d_0-k,\qquad {J_D}\coloneqq\bar D-d_0+1.
\end{equation}
The second condition of~\eqref{eq:five-physical-cap} gives
$\bar D\ge D$. Moreover $k>(25\lambda-2m)/13$ and
${J_D}\ge(12N-50m)/13\ge(250/13)\lambda$.
The combined scalar decreases in $i$, since
$\lambda_+/{J_D}\le13/240<23/325<k/(N-i)$.
For the other two deficits, whose intervals are bounded below by
$s>S\lambda$ with respectively $S=6,7$, use
\[
 \bar D\coloneqq N-\lceil S\lambda\rceil+d_0,\quad
 k\ge S\lambda-4m,\quad {J_D}\ge(25-S)\lambda.
\]
Decrease follows from $625<24(25-S)(S-4)$.
Lemma~\ref{lem:positive-tail} needs the roots of $h$ to exceed only the
actual degree $D$, as they do by~\eqref{eq:coefficient-root-domain}; no
root bound at $\bar D$ is required, since only the positive scalar and
the nonnegative backward weights are enlarged.

\smallskip\noindent\emph{Step 3 (The product factor; $m\ge11$).}
At $i_0$, with $z\coloneqq i_0-d_0+1\ge\theta N-d_0+1$ and ${J_D}\le N$, the
inequalities $\log(1+x)-\log(1+y)\le x-y$ for $x\ge y\ge0$,
$m(2m-1)\le2m^2$ and $\lambda\ge m$ give
\begin{equation}\label{eq:five-product-loss}
 \prod_{j=0}^{d_0-1}\frac{1+j/z}{1+j/{J_D}}
 \le\exp\left\{2m\left(\frac1{25\theta-2}-\frac1{25}\right)\right\}.
\end{equation}
The bound $z\le\theta N$ enters the exponential factor of $F$.
For the critical case $\theta=27/64$ this yields
\[
 F(i_0,\bar D)\le
 e^{2m(1/(25\theta-2)-1/25)}
 e^{65\theta\lambda/48}(1-\theta)^{(25\lambda-2m)/13}.
\]
The $\lambda$ factor decreases. Indeed the first term of the series
$\log a=2\sum_{j\ge0}y^{2j+1}/(2j+1)$, $y\coloneqq(a-1)/(a+1)$, whose terms are
positive for $a>1$, gives $\log(64/37)>54/101>13/25$; hence
$(25/13)\log(64/37)>1>65\theta/48=1755/3072$. At $\lambda=m$ the base is
$e^{{E_{\rm crit}}}(37/64)^{23/13}$, where
${E_{\rm crit}}\coloneqq128/547-2/25+1755/3072$.
For the other two cases the corresponding bound is
\[
 F(i_0,\bar D)\le
 e^{2m(1/(25\theta-2)-1/25)}
 e^{625\theta\lambda/(24(25-S))}(1-\theta)^{S\lambda-4m}.
\]
The $\lambda$ factor again decreases, since
$\log(1-\theta)\le-\theta$ and $625<24S(25-S)$.
All three bases are less than $98/125$. In this subsection these
exponential comparisons have the following exact rational verification:
\begin{equation}\label{eq:rational-exponential-upper}
 U_{20}(x)\coloneqq\sum_{j=0}^{20}\frac{x^j}{j!}
       +\frac{x^{21}}{21!(1-x/22)}\ge e^x,
       \qquad0\le x<22.
\end{equation}
The inequality follows by bounding the remaining positive series
geometrically. With the exponents $E_S$ of the cases $S=6,7$ (indexed by
$S$, not by $m$), the required three rational inequalities are
\begin{align*}
 U_{20}({E_{\rm crit}})^{13}(37/64)^{23}&<(98/125)^{13},\\
 U_{20}(E_6)(39/64)^2&<98/125,
 &E_6&=\frac2{625/64-2}-\frac2{25}+\frac{(25/64)625}{24\cdot19},\\
 U_{20}(E_7)(45/64)^3&<98/125,
 &E_7&=\frac2{475/64-2}-\frac2{25}+\frac{(19/64)625}{24\cdot18}.
\end{align*}
Their inputs lie in $[0,22)$ and every denominator is positive.
For $m\ge11$ the full ratio is less than ${\operatorname{pref}}(98/125)^m$.
At eleven this is
$116106045858720413103104/116415321826934814453125<1$;
its next ratio is at most $(61/58)(98/125)<1$.

\smallskip\noindent\emph{Step 4 ($2\le m\le5$, by the signed certificates).}
For $2\le m\le5$ use the signed matrix bounds
\eqref{eq:two-pair-allweight} and~\eqref{eq:five-low-degree-mass}; for
$m=2$, discarding the negative term of ${\operatorname{Bound}_2}$ gives the form
of~\eqref{eq:five-low-degree-mass} with $\gamma_2\coloneqq9/8$. The ratio is then
at most $\gamma_m\lambda^m$ times $F$ without its exponential factor, a
positive product that decreases in $i$ (Appendix~\ref{app:higher-tails-36},
Step~3). Equation~\eqref{eq:five-product-loss} therefore
bounds the ratio by
\[
 \gamma_m\lambda^m e^{mE_\theta}(1-\theta)^k,
 \qquad E_\theta\coloneqq2/(25\theta-2)-2/25.
\]
This decreases in $\lambda\ge m$ after using the relevant lower bound
on $k$: in the critical case its logarithmic derivative is at most
$1+(25/13)\log(37/64)<0$, and in the other cases it is at most
$1-S\theta<0$. The twelve endpoint inequalities are
\begin{align}\label{eq:five-low-degree-scalars}
 \gamma_m m^m e^{mE_\theta}(37/64)^{23m/13}&<1
 &&(\theta=27/64),\notag\\
 \gamma_m m^m e^{mE_\theta}(1-\theta)^{(S-4)m}&<1
 &&((S,\theta)=(6,25/64),(7,19/64)).
\end{align}
Each is verified by replacing the exponential by $U_{20}(mE_\theta)$;
in the first case take the thirteenth power. These are exact rational
evaluations for the four specified degrees.

\smallskip\noindent\emph{Step 5 ($6\le m\le10$, critical deficit).}
For $6\le m\le10$ use~\eqref{eq:finite-row-bound} itself at $u=1/a$,
instead of its consequences~\eqref{eq:row-factorial-bound}
and~\eqref{eq:row-exponential-bound}. Define, for $a>0$,
\begin{equation}\label{eq:five-finite-row-polynomial}
 {\operatorname{Row}_{m,j}}(a;\lambda)\coloneqq a^{2m-j}\sum_{{j_1}=0}^j\binom j{j_1}(a+1)^{j_1}
                 \sum_{{j_2}=0}^{m-{j_1}}\frac{(25\lambda/(24a))^{j_2}}{{j_2}!},
 \quad 0\le j\le m.
\end{equation}
It bounds $a^{2m}{\operatorname{Row}_j}(1/a)$ (compare the formula in
Appendix~\ref{app:higher-tails-36}, Step~5, with $\omega_1\le1$ and
$\binom{N-j}{j_2}b^{-j_2}\le(25\lambda/24)^{j_2}/j_2!$) and has positive
coefficients in $\lambda$ of degree at most $m$. Its exponents of $a$ are
nonnegative, so it increases with $a$, and it may be evaluated at any fixed
weight that dominates $a_i$. Put $\lambda_{\min}\coloneqq m+1/25$, the least value
of $\lambda$ allowed by the hypothesis $N\ge25m+1$. For the critical case
the fixed weight
\begin{equation}\label{eq:five-fixed-critical-weight}
 {a_{\rm crit}}(m)\coloneqq\frac{(12/13-27/64)(25m+1)-24m/13}
 {(27/64)(25m+1)-2m+1}
\end{equation}
dominates $a_i$ for all $i\ge i_0$ and $N\ge25m+1$. Indeed $a_i$
decreases in $i$, and $i\ge\theta N$ and $\bar D<(12/13)(N-2m)$ give
$a_i<((12/13-\theta)N-24m/13)/(\theta N-2m+1)$; this ratio equals
${a_{\rm crit}}(m)$ at $N=25m+1$, and its derivative in $N$ has numerator
$(12/13-27/64)(1-2m)+24(27/64)m/13<0$ on these five degrees (it is
negative for every $m\ge3$).
Also $\bar D/N\ge {\operatorname{degfrac}}_m\coloneqq276/325-1/(25m)>0$.
For $k'\coloneqq(25\lambda-2m)/13$, multiply any positive row polynomial
by ${\operatorname{degfrac}}_m^{-2m}\max_{0<u<1}u^{2m}(1-u)^{k'}$.
Its logarithmic derivative is at most
\[
 m/\lambda+(25/13)\log\frac{k'}{k'+2m}
 \le m/\lambda-\frac{50m}{25\lambda+24m}<0.
\]
Thus it suffices to evaluate at $\lambda_{\min}$, where
$k'_0\coloneqq(23m+1)/13$ and the maximum over $u$ is attained at
$u_{\max}\coloneqq26m/(49m+1)$.
With $M_m(a;\lambda)\coloneqq\max_{0\le j\le m}{\operatorname{Row}_{m,j}}(a;\lambda)$,
the five exact bounds
\begin{equation}\label{eq:five-critical-finite-scalars}
 {\operatorname{pref}}_mM_m({a_{\rm crit}}(m);\lambda_{\min}){\operatorname{degfrac}}_m^{-2m}
          u_{\max}^{2m}(1-u_{\max})^{k'_0}
\end{equation}
are less than $19/20,4/5,7/10,3/5,1/2$, respectively.
Taking thirteenth powers makes the comparisons rational. The finite
maximum is evaluated over all $m+1$ rows, rather than inferred from
a general row monotonicity claim.

\smallskip\noindent\emph{Step 6 ($6\le m\le10$, the other two deficits).}
For $S=6,7$, with $\theta=25/64,19/64$, put
\[
 a_S(m)\coloneqq\frac{(25-S-25\theta)\lambda_{\min}+2m}
 {25\theta\lambda_{\min}-2m+1}.
\]
It dominates $a_i$: the derivative of the same ratio at $\lambda$
has numerator $(25-S-25\theta)(1-2m)-50\theta m<0$.
Retain $\bar D/N\ge {\operatorname{degfrac}}(\lambda)\coloneqq1-S/25+(2m-1)/(25\lambda)$.
Multiplying a row polynomial by ${\operatorname{degfrac}}(\lambda)^{-2m}$ and the beta
maximum with $k'\coloneqq S\lambda-4m$ gives logarithmic derivative at most
\[
 [1+4/(25-S)]m/\lambda-2Sm/(S\lambda-2m)<0.
\]
At $\lambda_{\min}$ put ${\operatorname{degfrac}}_0\coloneqq1-S/25+(2m-1)/(25m+1)$,
$k'_0\coloneqq S\lambda_{\min}-4m$, and $u_{\max}\coloneqq2m/(k'_0+2m)$.
The five endpoint quantities
\begin{equation}\label{eq:five-noncritical-finite-scalars}
 {\operatorname{pref}}_mM_m(a_S(m);\lambda_{\min}){\operatorname{degfrac}}_0^{-2m}
              u_{\max}^{2m}(1-u_{\max})^{k'_0}
\end{equation}
are less than $4/5,7/10,3/5,1/2,2/5$ for $S=6$, and
$9/10,4/5,7/10,3/5,1/2$ for $S=7$.
Twenty-fifth powers make every comparison rational.

\smallskip\noindent\emph{Step 7 (Conclusion).}
The certificate package~\cite{ComputationalCertificates} contains a
reconstruction in rational arithmetic of the 45 minors $\Delta_j$ of
Step~1, of
\eqref{eq:rational-exponential-upper}--\eqref{eq:five-low-degree-scalars},
and of every row and maximum in
\eqref{eq:five-critical-finite-scalars}--\eqref{eq:five-noncritical-finite-scalars}.
Their formulas and complete finite domains have been specified here. The
initial ranges of Appendix~\ref{app:higher-initial} and the tails
$\lceil\theta N\rceil\le i\le D$ cover every index $1\le i\le D$, and the
degree ranges exhaust $m\ge2$.
This proves Proposition~\ref{prop:five-cap-coefficients} under exactly
its stated hypotheses.

\subsection{Integrated coefficient extraction at \texorpdfstring{$p=4$}{p=4}}
\label{app:higher-four-initial}

We prove Proposition~\ref{prop:four-initial}, that
is,~\eqref{eq:residual-comparison} for $1\le i\le\lfloor9N/20\rfloor$ when
$Q=16$, $N\ge256$ and $r\ge N/2$. Steps~1--6 bound the coefficient of
each component of the mixture~\eqref{eq:positive-mixture} separately: by
Lemma~\ref{lem:component-coefficients} where its Jacobi test holds, and
otherwise by Cauchy's integral over a circle $|u|=\rho$, which bounds the
absolute value of the coefficient. Step~7 averages these bounds over the
mixture by Lemma~\ref{lem:mixture-laplace}, and Step~8 compares the
average with the packed endpoint of Appendix~\ref{app:higher-initial}.

\smallskip\noindent\emph{Step 1 (Geometry and the target estimate).}
Put $Q=16$, $b=15$, $M\coloneqq N-d_0$, ${k_{-,0}}\coloneqq N-{J_*}$. At $Q=16$ the root
barrier~\eqref{eq:coefficient-root-domain} gives $v<(N-d_0+1)/8$, hence
the integer inequality $8v\le N-d_0$; with $d_0\le N/8$, from
$N\ge16m$, it gives the finite inequalities
\begin{equation}\label{eq:four-mixture-geometry}
 v\ge1,\quad8v\le N-d_0,\quad r+{J_*}\le M,\quad
 d_0/N\le1/8,\quad {J_*}/N\le15/64,\quad {k_{-,0}}/N\ge49/64.
\end{equation}
Also $r\le7N/8-15d_0/8+7/8$, and in particular $N+r<15N/8$.
For a component let ${k_+}\coloneqq r+j$, ${k_-}\coloneqq N-j$, $E\coloneqq M-{k_+}\ge0$, and
\[
 \kappa_i\coloneqq\frac{{k_{-,0}}i}{N(M-i+1)}.
\]
We will prove, for every component of~\eqref{eq:positive-mixture}, that
is, for every $0\le j\le{J_*}$ and $0\le a\le1/15$, the pointwise
coefficient estimate
\begin{equation}\label{eq:four-component-bound}
 \frac{[u^i](1+u)^{k_+}(1-au)^{k_-}}{\binom Mi}
 \le e^{-aN\kappa_i},\qquad1\le i\le9N/20.
\end{equation}
It bounds the coefficient from above, and the coefficient may be
negative; in the Cauchy cases below its absolute value is bounded.

\smallskip\noindent\emph{Step 2 ($i\le2N/5$: the Jacobi case).}
First consider $i\le2N/5$; the estimates for these indices use only
$N\ge128$. Here $i<{k_+},{k_-}$ and $i-1<{n_{\rm mix}}/2$, since
${k_+}\ge r\ge N/2$, ${k_-}\ge{k_{-,0}}\ge49N/64$ and ${n_{\rm mix}}=N+r$.
If $i\le N/8$, the main term of the Jacobi test (as in
Appendix~\ref{app:higher-initial}, at the worst $a=1/15$) is
at least $19/60$. If $E\le N/32$ and $i\le2N/5$, it is at least
$323/800$. With $2/N\le1/64$, the squared margins are
\[
 (19/60)^2-\frac4{15}\frac18(15/8-1/8+1/64)=1193/28800>0,
\]
\[
 (323/800)^2-\frac4{15}\frac25(15/8-2/5+1/64)=2569/640000>0.
\]
Smaller indices have a larger main term and a smaller right side, so the
test holds at every index up to $i$. Hence the
component coefficients are positive through $i$, and
\eqref{eq:component-suppression}, ${k_+}\le M$, ${k_-}\ge {k_{-,0}}$ prove
\eqref{eq:four-component-bound} in these cases.

\smallskip\noindent\emph{Step 3 ($N/8<i\le2N/5$ and $E>N/32$: the circle).}
For $E>N/32$ and $N/8<i\le2N/5$ put
${\theta_0}\coloneqq i/M$, ${\bar\theta_0}\coloneqq1-{\theta_0}$, $\rho\coloneqq{\theta_0}/{\bar\theta_0}$.
Then $1/8<{\theta_0}\le16/35$ and $\rho\le16/19$.
On $u=\rho e^{\mathrm{i}t}$, where $\mathrm{i}$ is the imaginary unit
and $\zeta\coloneqq\sin^2(t/2)$, the exact relative
modulus is
\[
 \frac{|(1+\rho e^{\mathrm{i}t})^{k_+}(1-a\rho e^{\mathrm{i}t})^{k_-}|}
 {(1+\rho)^{k_+}(1-a\rho)^{k_-}}
 =(1-4{\theta_0}{\bar\theta_0}\zeta)^{{k_+}/2}
  \left(1+\frac{4a\rho\zeta}{(1-a\rho)^2}\right)^{{k_-}/2}.
\]
Here $4a\rho/(1-a\rho)^2=4{\theta_0}{\bar\theta_0}\cdot a(1+\rho)^2/(1-a\rho)^2$,
because $\rho/(1+\rho)^2={\theta_0}{\bar\theta_0}$. Since
$a(1+\rho)^2/(1-a\rho)^2\le18375/72361<4/15$, the inequalities
$\log(1-x)\le-x$ and $\log(1+x)\le x$, with ${k_+}=M-E$ and ${k_-}\le N$,
give an upper bound
$\exp(-2\nu {\theta_0}{\bar\theta_0}\sin^2(t/2))$, where
$\nu\coloneqq M-E-4N/15\ge7N/30>0$.

\smallskip\noindent\emph{Step 4 (Angular integral and binomial mode).}
For $V>0$ the angular integral satisfies
\begin{equation}\label{eq:finite-angular-integral}
 J(V)\coloneqq\frac1{2\pi}\int_{-\pi}^{\pi}e^{-2V\sin^2(t/2)}dt
 \le\sqrt{\frac8{13\pi V}}+e^{-117V/128}.
\end{equation}
Indeed on $|t|\le3/2$,
$\sin^2(t/2)\ge(t^2/4)(1-t^2/12)\ge13t^2/64$.
On the remaining arc it is at least $\sin^2(3/4)\ge117/256$.
Integrate the first Gaussian over the real line and bound the other
arc by its maximum.

For the binomial probability at its integer mean,
$\operatorname{Mode}_i\coloneqq\binom Mi {\theta_0}^i {\bar\theta_0}^{M-i}$, the finite
Stirling bounds (those of Robbins~\cite{Robbins}) give
\begin{equation}\label{eq:finite-mode-bound}
 \operatorname{Mode}_i^{-1}\le\sqrt{2\pi M{\theta_0}{\bar\theta_0}}e^\epsilon,
 \qquad\epsilon\coloneqq\frac1{12i}+\frac1{12(M-i)}.
\end{equation}
One elementary derivation of the required factorial inequalities sets
$S_n\coloneqq\log(n!)-(n+1/2)\log n+n-\frac12\log(2\pi)$.
The Wallis product gives $S_n\to0$, while at $t=1/(2n+1)$,
$S_n-S_{n+1}=\sum_{k\ge1}t^{2k}/(2k+1)$ lies strictly between
zero and $1/(12n(n+1))$. Summation gives
$0<S_n<1/(12n)$, which proves~\eqref{eq:finite-mode-bound}.
All its factorial arguments here are positive integers.

\smallskip\noindent\emph{Step 5 (Assembly for $i\le2N/5$).}
On the circle $|u|=\rho$, Cauchy's integral bounds the absolute value of
the coefficient by
$\rho^{-i}(1+\rho)^{k_+}(1-a\rho)^{k_-}J(\nu{\theta_0}{\bar\theta_0})$. Here
$\rho^{-i}(1+\rho)^M=\binom Mi/\operatorname{Mode}_i$; also
$(1+\rho)^{-E}\le e^{-{\theta_0}E}$ since $\log(1+\rho)\ge {\theta_0}$, and
$(1-a\rho)^{k_-}\le e^{-a{k_-}\rho}\le e^{-aN\kappa_i}$ since
${k_-}\rho\ge N\kappa_i$. With~\eqref{eq:finite-angular-integral}
and~\eqref{eq:finite-mode-bound} this gives
\begin{equation}\label{eq:four-cauchy-ratio}
 \frac{|[u^i](1+u)^{k_+}(1-au)^{k_-}|}
 {\binom Mi e^{-aN\kappa_i}}
 \le e^{\epsilon-{\theta_0}E}
 \left\{\sqrt{\frac{16M}{13\nu}}+
 \sqrt{2\pi M{\theta_0}{\bar\theta_0}}\,e^{-117\nu {\theta_0}{\bar\theta_0}/128}\right\}.
\end{equation}
For fixed $N,M,i$ each term decreases with $E$ on $E\ge N/32$, the
present case. The first logarithmic derivative is
$1/(2\nu)-{\theta_0}\le15/(7N)-1/8<0$; the second is
$-{\theta_0}+(117/128){\theta_0}{\bar\theta_0}<0$.
At the worst case $E=N/32$,
\[
 \nu\ge277N/480,\quad M/\nu\le420/277,
 \quad {\theta_0}{\bar\theta_0}\ge7/64,\quad\epsilon\le16/(19N).
\]
Using $M{\theta_0}{\bar\theta_0}\le N/4$ and $\pi<7/2$, the right side of
\eqref{eq:four-cauchy-ratio} is at most
\[
 e^{16/(19N)-N/256}
 \left\{\sqrt{6720/3601}+\sqrt{7N/4}\,
 e^{-226863N/3932160}\right\}.
\]
It decreases for $N\ge128$. At 128 use
$\sqrt{6720/3601}<11/8$, $\sqrt{224}<15$,
$226863/30720>7$, and
$e^7>\sum_{j=0}^{11}7^j/j!=2959911103/2851200>1000$.
At $N=128$ the exponent of the prefactor is $-75/152$, and
$e^{75/152}>1+75/152+(75/152)^2/2>3/2$.
The bound is less than $(2/3)(11/8+15/1000)=139/150<1$.
This proves~\eqref{eq:four-component-bound} through $2N/5$.

\smallskip\noindent\emph{Step 6 ($2N/5<i\le9N/20$).}
For the added interval $2N/5<i\le9N/20$ assume $N\ge256$ and
split at $E=N/192$. If $E\le N/192$, put $x\coloneqq d_0/N$, $w\coloneqq i/N$.
The normalized Jacobi main term is at least
$(14/15)(1-w)-x-1/192$. Retaining the residual degree in the
bound on $r$, that is, using $r\le7N/8-15d_0/8+7/8$, the squared right
side is at most
\[
 \frac4{15}w\left(\frac{15}8(1-x)-w+\frac{23}{8N}\right).
\]
The squared margin decreases with $w$ through $9/20$. At that
endpoint it decreases with $x\le1/8$, since its derivative is
$-2(77/150-x-1/192)+9/40<0$.
The worst value is therefore
\[
 (613/1600)^2-\frac4{15}\frac9{20}
       \left(\frac{15}8\frac78-\frac9{20}+\frac{23}{2048}\right)
 =6559/2560000>0.
\]
Lemma~\ref{lem:component-coefficients}, through its
recurrence~\eqref{eq:component-recurrence}, proves
\eqref{eq:four-component-bound} in this case.

If $E>N/192$, then $2/5<{\theta_0}\le18/35$, $\rho\le18/17$, and
$a\rho<1$. Now
$a(1+\rho)^2/(1-a\rho)^2\le6125/18723<1/3$.
The same circle argument works with $\nu\coloneqq M-E-N/3\ge N/6$.
At a possible zero of $1+\rho e^{\mathrm{i}t}$ (only $\rho=1$, $t=\pi$)
the modulus bound holds by continuity, its left side being zero there.
In~\eqref{eq:four-cauchy-ratio},
$\epsilon\le55/(136N)$; both terms decrease with $E$, because
$1/(2\nu)-{\theta_0}\le3/N-2/5<0$.
At $E=N/192$,
$\nu\ge103N/192$, $M/\nu\le168/103$, and
${\theta_0}{\bar\theta_0}\ge6/25$. The ratio is at most
\[
 e^{55/(136N)-N/480}
 \left\{\sqrt{2688/1339}+\sqrt{7N/4}\,e^{-12051N/102400}\right\}.
\]
It decreases for $N\ge256$. At 256 use
$\sqrt{2688/1339}<10/7$, $\sqrt{448}<22$,
$12051/400>30$, $e^{30}>2^{30}>22000$.
At $N=256$ the exponent of the prefactor is $-a_*$, where
$a_*\coloneqq277703/522240=8/15-55/34816$, and
\[
 1+a_*+a_*^2/2+a_*^3/6-5/3
 =26943598854413207/854597560172544000>0
\]
shows $e^{-a_*}<3/5$. The ratio is strictly less than
$(3/5)(10/7+1/1000)=30021/35000<1$.
Thus~\eqref{eq:four-component-bound} holds on the full required interval.

\smallskip\noindent\emph{Step 7 (Averaging over the mixture).}
Average this pointwise-in-$a$ estimate over the same positive
mixture~\eqref{eq:positive-mixture}. Since $\binom Mi/\binom Ni=U_0(i)$,
Lemma~\ref{lem:mixture-laplace}, in the form~\eqref{eq:beta-laplace} with
$\kappa=\kappa_i$, gives
\begin{equation}\label{eq:four-beta-average}
 \mathcal L_i<U_0(i)(1+t_i)^{-v},\qquad
 t_i\coloneqq\frac{\kappa_i\lambda}{{J_*}+\lambda+1}.
\end{equation}
The heads, the atoms with $j<{J_*}$ and $a=1/15$, obey the stronger
bound $aN=N/15>\lambda$: their factor $e^{-N\kappa_i/15}$ is at most
$e^{-\lambda\kappa_i}\le e^{-vt_i}\le(1+t_i)^{-v}$.

\smallskip\noindent\emph{Step 8 (Comparison with the packed endpoint).}
It remains to compare this expression with the packed endpoint
$(N-v-d_0)_i/(N-v)_i$ of Appendix~\ref{app:higher-initial}.
Write $x\coloneqq d_0/N$, $y\coloneqq v/N$, $\eta\coloneqq1/N$, $w\coloneqq i/N$.
Then
\[
 t_i=\frac{(1-x-y)w}
 {(1-x-w+\eta)(1+16(x+y+\eta))},
 \quad x\le1/8,\quad y\le(1-x)/8,\quad\eta\le1/256.
\]
This decreases in $y,\eta$. After inserting their upper bounds it
also decreases in $x$: its logarithmic derivative is
$-1/(1-x)+1/(1-x-w+1/256)-14/(49/16+14x)<0$.
The positive middle term is less than $5/2$, while the last magnitude
is at least $224/77>5/2$. Hence
\begin{equation}\label{eq:four-beta-rational}
 1+t_i\ge\frac{17325-16576w}{17325-19712w}.
\end{equation}

Move the packed block of roots down one step at a time, from
$N-d_0+1,\ldots,N$ (value $U_0(i)$) to ${a_{\min}},\ldots,{a_{\min}}+d_0-1$
(the packed endpoint). The step from the block ending at $N-\ell$ to the
block ending at $N-\ell-1$ multiplies the value by $e^{-g(\ell/N)}$, so the
negative logarithm of the packed endpoint divided by $U_0(i)$
is $\sum_{\ell=0}^{v-1}g(\ell/N)$, where
\[
 g(z)\coloneqq\log\frac{(1-w-z)(1-x-z)}{(1-z)(1-x-w-z)}
 =\int_0^x\int_0^w\frac{ds\,dt}{(1-z-s-t)^2}.
\]
The integral proves that $g$ is increasing and convex. The left
Riemann sum and the trapezoid inequality bound the sum by
$v(g(0)+g(y))/2$. Since the bound~\eqref{eq:four-beta-average} is
$e^{-v\log(1+t_i)}$ times $U_0(i)$, it suffices that
$(1+t_i)^2>e^{g(0)+g(y)}$. The bound $v(g(0)+g(y))/2$ increases in $y$;
after substituting
$y=(1-x)/8$ it also increases in $x$.
For the last assertion write $1-y=(7+x)/8$ and
$1-y-x=7(1-x)/8$; differentiation of the two logarithmic ratios
is positive. At $x=1/8,y=7/64$,
\begin{equation}\label{eq:four-packed-rational}
 e^{g(0)+g(y)}=
 \frac{7(1-w)}{7-8w}\frac{49(57-64w)}{57(49-64w)}.
\end{equation}
All denominators are positive through $w=9/20$. Subtract
\eqref{eq:four-packed-rational} from the square of
\eqref{eq:four-beta-rational} and clear these positive denominators.
The numerator is $w{\operatorname{Gap}}(w)$, with
\[
 {\operatorname{Gap}}(w)\coloneqq210366161775-901294123968w
                  +1202435325952w^2-511027707904w^3.
\]
Its four Bernstein coefficients on $[0,9/20]$ are
\[
 210366161775,\quad375860215899/5,\quad
 528557727159/25,\quad213694951491/125.
\]
They are positive, so $(1+t_i)^{-v}$ is strictly smaller than the
packed endpoint divided by $U_0(i)$. Equation~\eqref{eq:four-beta-average}
proves Proposition~\ref{prop:four-initial}.

\subsection{The tail for \texorpdfstring{$p=4$}{p=4} at every residual degree}
\label{app:higher-four-tail}

We prove Proposition~\ref{prop:four-tail},
that is,~\eqref{eq:residual-comparison} for $\lceil9N/20\rceil\le i\le D$,
by bounding the combined scalar $F(i,\bar D)$ in the bound
${\operatorname{pref}}\,F(i,\bar D)$ of Appendix~\ref{app:higher-tails-36}.

\smallskip\noindent\emph{Step 1 (The constant ${\operatorname{pref}}$ and the comparison degree).}
The exact identity
\[
 m(7m+55)^2-49(m+10)(m+2)^2=84m^2+869m-1960>0
 \quad(m\ge2)
\]
is the squared form of
$(m+2)/\sqrt{m(m+10)}<(7m+55)/(7(m+10))=1-15/(7(m+10))$. With
\eqref{eq:finite-tridiagonal-norm} and~\eqref{eq:retained-coercivity} at
$b=15$ it gives
$M_0>2I/(m+10)$. Thus ${\operatorname{pref}}=m+10$ is valid in
\eqref{eq:row-exponential-bound}. Set
\[
 k\coloneqq\lfloor2(N-d_0)/17\rfloor+1,\quad
 \bar D\coloneqq N-d_0-k,\quad {J_D}\coloneqq\bar D-d_0+1,
 \quad i_0\coloneqq\lceil9N/20\rceil.
\]
Condition~\eqref{eq:four-physical-cap} gives $\bar D\ge D$. In a
nonempty tail
$i\ge i_0>3d_0$, so Lemma~\ref{lem:positive-tail} applies.
Its combined scalar $F$ decreases for real $i\ge i_0$:
the product $\prod_{j=0}^{d_0-1}(1+j/z)$ is nonincreasing, and
\[
 \lambda_+/{J_D}\le17/165,\qquad
 k/(N-i)>35/187>17/165,
\]
because ${J_D}\ge(15N-32d_0)/17\ge11N/17$, $N-i\le11N/20$ and $d_0\le N/8$.
It suffices to bound $F(i_0,\bar D)$.

\smallskip\noindent\emph{Step 2 (Reduction to a function of $x=m/\lambda$).}
Put $\lambda=N/16\ge16$ and $x\coloneqq m/\lambda\in(0,1]$. We bound
$\log F(i_0,\bar D)$ by $\lambda$ times a function of $x$ that is less
than $-6/25$. The floor gives both bounds
\[
 (240\lambda-64m)/17\le {J_D}<(240\lambda-64m)/17+1.
\]
Define $Z(x)\coloneqq36/5-2x$, $U(x)\coloneqq(240-64x)/17+1/16$.
At $z_0\coloneqq i_0-d_0+1$,
\[
 z_0\ge\lambda Z(x),\qquad
 z_0\le\lambda(293/40-2x),\qquad {J_D}\le\lambda U(x).
\]
The finite offsets use $2/\lambda\le1/8$ and $1/\lambda\le1/16$.
The two factors of $F$ need opposite bounds on ${J_D}$: the product uses
$z_0\ge\lambda Z(x)$ and the upper bound ${J_D}\le\lambda U(x)$, while the
exponential factor uses $z_0\le\lambda(293/40-2x)$ and the lower bound
${J_D}\ge\lambda(240-64x)/17$:
\begin{align*}
 \log\prod_{j=0}^{d_0-1}\frac{1+j/z_0}{1+j/{J_D}}
 &\le\lambda {\operatorname{Loss}}(x),&
 {\operatorname{Loss}}(x)&\coloneqq2x^2(1/Z(x)-1/U(x)),\\
 \frac{N}{15}\frac{z_0}{{J_D}}&\le\lambda E(x),&
 E(x)&\coloneqq\frac{17}{15}\frac{293/40-2x}{15-4x}.
\end{align*}
The remaining quotient $(N-i_0)_k/(N)_k$ is at most
$(1-i_0/N)^k\le(11/20)^k$, with
$k>\lambda(32-4x)/17$. Let
\[
 \ell\coloneqq2\left(\frac9{31}+\frac{(9/31)^3}{3}+\frac{(9/31)^5}{5}\right)
 =85570218/143145755<\log(20/11).
\]
The positive series $\log(20/11)=2\sum_{j\ge0}(9/31)^{2j+1}/(2j+1)$ proves
this strict inequality. Thus
\[
 \log F(i_0,\bar D)<\lambda {\operatorname{Save}}(x),\qquad
 {\operatorname{Save}}(x)\coloneqq{\operatorname{Loss}}(x)+E(x)-(32-4x)\ell/17.
\]

\smallskip\noindent\emph{Step 3 (${\operatorname{Save}}$ increases; evaluation at $x=1$).}
This function increases on $[0,1]$, so its largest value is at $x=1$,
that is, at $m=\lambda$. Indeed $U>3Z/2$,
with difference at one equal to $3557/1360$, and
$(1/Z-1/U)'=2/Z^2-(64/17)/U^2>0$.
Hence ${\operatorname{Loss}}$ is nondecreasing. Also
$E'(x)=-119/(150(15-4x)^2)\ge-119/18150$,
whereas $4\ell/17>2/17>119/18150$.
At one,
\[
 U(1)=2833/272,\quad {\operatorname{Loss}}(1)=7093/36829,\quad E(1)=1207/2200,
\]
\[
 {\operatorname{Save}}(1)=-9596481216935939/39433924281494600<-6/25,
\]
the difference being $26467877875447/7886784856298920>0$.
Therefore the full coefficient ratio ${\operatorname{pref}}F(i_0,\bar D)$ is less than
$(m+10)e^{-6\lambda/25}\le(\lambda+10)e^{-6\lambda/25}<1$.
This decreases for $\lambda\ge16$; at sixteen,
$e^{96/25}>e^{7/2}>\sum_{j=0}^{10}(7/2)^j/j!
=1951237723/58982400>26$.
This proves the whole tail, including the top index $D$.

\subsection{The exact finite certificate for \texorpdfstring{$p=4$}{p=4}}
\label{app:four-finite-certificate}

We specify the finite claim used in Proposition~\ref{prop:four-finite}:
for every $H=n-D$ with $29\le H\le251$, a certified cap minimizer, its
lift to every original degree, and the exact verification of all
comparisons~\eqref{eq:residual-comparison} for the lifts that no other
result covers.

\smallskip\noindent\emph{Step 1 (Cap minimizers certified by positive quadrature).}
For every $H=29,\ldots,251$ let
$a\coloneqq\lfloor(H-1)/14\rfloor$, the cap index of
Lemma~\ref{lem:cap-minimum-identity} at $Q=16$, and $N_a\coloneqq H+2a$.
For each $H$ the certificate package~\cite{ComputationalCertificates} lists
$2a$ distinct roots $Z_H\subseteq\{1,\ldots,N_a\}$.
Put ${h_H^{\rm cap}}(x)\coloneqq\prod_{z\in Z_H}(x-z)$ and
$w_x\coloneqq\binom{N_a}x15^x$. Check ${h_H^{\rm cap}}(0)>0$ and ${h_H^{\rm cap}}(x)\ge0$
at every integer $0\le x\le N_a$.
For the nodes $\mathcal Z_H=\{0\}\cup Z_H$, compute the exact rational
weights
\begin{equation}\label{eq:four-cardinal-weights}
 \lambda_z\coloneqq16^{-N_a}\sum_{x=0}^{N_a}w_x
       \prod_{v\in\mathcal Z_H\setminus\{z\}}\frac{x-v}{z-v}.
\end{equation}
All denominators $z-v$ are nonzero; their signs are retained in rational
arithmetic. All the resulting weights are positive. Verify the
4445 moment identities
\[
 \sum_{z\in\mathcal Z_H}\lambda_z z^k
 =16^{-N_a}\sum_{x=0}^{N_a}w_xx^k,
 \qquad H=29,\ldots,251,\quad0\le k\le2a.
\]
For exact cardinal weights these identities hold by interpolation;
verifying them checks the recorded weights. Since the weights are
positive, the quadrature is nonnegative, and every feasible normalized
polynomial of degree at most $2a$ has expectation at least $\lambda_0$;
with the feasibility check above, the converse in
Lemma~\ref{lem:positive-quadrature} therefore certifies that
${h_H^{\rm cap}}/{h_H^{\rm cap}}(0)$ is a cap minimizer and
$\lambda_0=\beta_{N_a,2a,16}$.

\smallskip\noindent\emph{Step 2 (Terminal merging and the objective identity).}
Remove the maximal terminal root block of ${h_H^{\rm cap}}$, of length
${\ell_{\rm end}}$. The $2a$ roots of the feasible polynomial ${h_H^{\rm cap}}$ pair
into adjacent integers (as in the proof of Lemma~\ref{lem:grid-vertex}),
so this block consists of whole adjacent pairs and ${\ell_{\rm end}}$ is even;
this parity and the residual pair structure are verified from the list.
Set $d_0\coloneqq2a-{\ell_{\rm end}}$, $m\coloneqq d_0/2$, $N\coloneqq N_a-{\ell_{\rm end}}=H+d_0$, the
residual parameters of the lift, and let $h_H$ be the
monic product of the remaining roots. For every $D\ge2a$ the lift
merges exactly to terminal degree $r=D-d_0$, since
\[
 (n-x)_{D-2a}(N_a-x)_{\ell_{\rm end}}=(n-x)_{D-d_0}.
\]
Normalization cancels its positive constants. The exact objective
identity is also checked:
\begin{equation}\label{eq:four-merged-objective}
 16^{d_0}\frac{\E_{\mu_N}h_H}{h_H(0)}
       =16^{2a}\beta_{N_a,2a,16}.
\end{equation}
Each nonzero residual satisfies the cap~\eqref{eq:optimizer-cap}, here
$N\ge16m+1$, and the root barrier $8\min\operatorname{root}(h_H)>7(H+1)$,
which is the second inequality of~\eqref{eq:coefficient-root-domain} at
$Q=16$.

\smallskip\noindent\emph{Step 3 (The result covering each profile and lift).}
There are no zero-pair profiles. The one-pair profile is
$H=29,N=31,m=1$, covered by Proposition~\ref{prop:one-pair-minimum}. The
26 profiles $H=226,\ldots,251$ have $N\ge256$; they are covered, at the
unchanged original $n,D$, by the argument of
Section~\ref{sec:higher-dimensions} for $N\ge256$ (the distance strips,
Proposition~\ref{prop:terminal-cutoffs}, and
Propositions~\ref{prop:four-initial} and~\ref{prop:four-tail}).
For the remaining 196 profiles, enumerate exactly the
lifts~\eqref{eq:four-finite-lifts}. Form the original values $n\coloneqq N+r$,
$D\coloneqq r+d_0$, and the strip parameters $M_0,L_0,u_0$ of
Lemma~\ref{lem:physical-tail-cap}, written here $M_*,L_*,u_*$ because
$M_0$ denotes an integer moment below:
\[
 M_* \coloneqq\lfloor n/2\rfloor+1,\quad
 L_*\coloneqq M_*-(D+1),\quad u_*\coloneqq M_*-16L_*.
\]
Every lift has $L_*\ge1$. The 113 lifts with $u_*>0$ have
$1\le u_*\le14$ and go to the adjacent strip.
Every other lift satisfies~\eqref{eq:four-physical-cap}, in the exact
form $17(N-r-2d_0)>2(N-d_0)$.
There are 4578 such lifts, with $52\le n\le422$ and
$22\le D\le197$; lifts of both parities of $n$ are included, and every
index $1\le i\le D$ of each is checked in Step~4.

\smallskip\noindent\emph{Step 4 (Integer moments and the integers $\Delta_i$).}
For each remaining residual compute its defining integer moments (these
integers $M_\ell$ are not the matrices $M_\ell$ of
Section~\ref{sec:coefficient-methods})
\begin{equation}\label{eq:four-integer-moments}
 M_\ell\coloneqq\sum_{x=0}^N\binom Nx15^xh_H(x)\Kraw{\ell}{N}(x),
 \qquad0\le\ell\le d_0.
\end{equation}
They may be computed from the exact integer recurrence
\[
 \ell\Kraw{\ell}{N}(x)=[15N-16x-14(\ell-1)]\Kraw{\ell-1}{N}(x)
       -15(N-\ell+2)\Kraw{\ell-2}{N}(x),
\]
with $\Kraw{0}{N}=1$ and $\Kraw{-1}{N}=0$, checking each division for
exactness. The integer $M_0$ is positive, and orthogonality gives
$G_\ell=M_\ell/(15^\ell M_0)$, so that
$\mathcal L_i=S_i/(15^{d_0}M_0\binom Ni)$ with $S_i$ as below.
For every enumerated $r$ and every $1\le i\le D$, compute
\begin{align}\label{eq:four-integer-deficit}
 S_i&\coloneqq\sum_{\ell=0}^{d_0}\binom r{i-\ell}M_\ell15^{d_0-\ell},
 \notag\\
 \Delta_i&\coloneqq M_0\,15^{d_0}\binom Ni h_H(i)-h_H(0)S_i.
\end{align}
The binomial coefficient is zero outside $0\le i-\ell\le r$.
The root inequality makes $h_H(i)>0$ for every required index,
and the other factors $M_0$, $15^{d_0}$, $\binom Ni$ and $h_H(0)$ are
positive. Therefore
$\Delta_i>0$ is exactly~\eqref{eq:residual-comparison}.
All 545042 integers are strictly positive.

\smallskip\noindent\emph{Step 5 (Largest ratio and data).}
The largest recorded value of $\mathcal L_i/(h(i)/h(0))$, that is, of
$h_H(0)S_i$ divided by $15^{d_0}M_0\binom Nih_H(i)$, is
\[
 \frac{15021033622792475655438720920517}
 {17423669850982121099386341818000}<1,
\]
at $H=131,N=147,m=8,r=99,D=115,i=1$.
This maximum is obtained after checking every index.
The finite data in~\cite{ComputationalCertificates} contain the root
lists, rational weights, defining moments, all lift parameters, and each
exact maximum.
The check reconstructs the signs directly from
\eqref{eq:four-integer-moments}--\eqref{eq:four-integer-deficit} for
every $H$, root, moment, lift and index, and verifies every terminal
identity and the positivity of every $\Delta_i$. The discovery of the root lists is unnecessary
for this verification. Thus the finite certificate establishes the
full declared domain at the original $n,D$.

\section{Coefficient estimates for the qutrit minimizer}
\label{app:qutrit-estimates}

This appendix proves Proposition~\ref{prop:qutrit-four-ranges}: the strict
comparisons $\mathcal L_i<h(i)/h(0)$ on the four index ranges of its
table, for a qutrit minimizing vertex with $m\ge2$ and $N\ge512$.
Together with Propositions~\ref{prop:scale-cutoff}
and~\ref{prop:original-top} they give Theorem~\ref{thm:qutrit-large-length}.
Appendices~\ref{app:qutrit-one-coefficient}--\ref{app:qutrit-initial}
record the shared estimates, including the functions $H_2$,
$\operatorname{curv}$, $\Psi_x$ and $I$ used later. The four rows are
proved as follows.
\begin{itemize}
\item Row~1 ($r\ge N/3$, $1\le i\le\lfloor N/4\rfloor$) in
Appendix~\ref{app:qutrit-quarter}, and row~2 ($N/6<r<N/3$,
$1\le i\le\lfloor N/8\rfloor$) in Appendix~\ref{app:qutrit-eighth}. Both use
the positive mixture~\eqref{eq:positive-mixture}: the component coefficients
are positive at the first indices
(Appendix~\ref{app:qutrit-initial}), and later each component coefficient
is bounded by a Cauchy integral over a circle before the average over the
mixture is taken (Appendix~\ref{app:qutrit-circle-factors}).
\item Row~4 ($\lceil N/4\rceil\le i\le D-1$, every $r$) in
Appendix~\ref{app:qutrit-tail}, and row~3 ($r\le N/3$,
$\lceil N/8\rceil\le i\le\min(\lfloor N/4\rfloor,D-1)$) in
Appendix~\ref{app:qutrit-middle}. Both use the one-coefficient bound of
Lemma~\ref{lem:schur-coefficient}. Row~4 is proved first because the proof
of row~3 reuses its notation and its bound for $N/4\le i\le N/2$.
\end{itemize}
In each row the coefficients are compared with a lower bound for
$h(i)/h(0)$: the bound~\eqref{eq:qutrit-product-simplified}, specific to a
global minimizer, or distinct-root packing.
All parameters are those of Section~\ref{sec:qutrit}; in particular
$d_0=2m$, $x=d_0/N$, $w=i/N$, and $s=r/N$. Here $s$ is the terminal
ratio, not the deficit $N-r$ of Section~\ref{sec:higher-dimensions} and
Appendix~\ref{app:higher-details}. In Appendices~\ref{app:qutrit-one-coefficient}
and~\ref{app:qutrit-circle-factors} the polynomial variable of $G$ and of
the mixture components is written $v$, and $u>0$ denotes a real radius at
which bounds are evaluated.
The finite range $N<512$ is treated in
Appendix~\ref{app:qutrit-finite}.

\subsection{One coefficient at a time}
\label{app:qutrit-one-coefficient}

Lemma~\ref{lem:weighted-rows} bounds the sum $\sum_\ell|G_\ell|u^\ell$ of
all absolute coefficients. The next lemma bounds a single coefficient of
$(1+v)^rG(v)$, which is what rows~3 and~4 need. Its constant
${\operatorname{pref}_0}$ plays the role of ${\operatorname{pref}}$ in
Lemma~\ref{lem:weighted-rows}.

\begin{lemma}\label{lem:schur-coefficient}
Let $Q>2$, $b\coloneqq Q-1$, ${\omega_1}\coloneqq(b-1)/b$, $m\ge1$, $N\ge Qm$,
and $0\le r\le R$ be integers. In the finite normalized paired
basis $[z^j]F_z(y)/\sqrt{\nu_j}$ of Section~\ref{sec:coefficient-methods},
suppose $M_0\succeq {\operatorname{coer}}I$ with ${\operatorname{coer}}>0$, and put ${\operatorname{pref}_0}\coloneqq2/{\operatorname{coer}}$.
For the function $G$ of~\eqref{eq:residual-generating}, written in the
variable $v$, define
\[
 {\operatorname{Row}_j}(v)\coloneqq\sum_{k=0}^m[z^k]
           (v+(1+{\omega_1}v)z)^j(1+vz/b)^{N-j},\qquad0\le j\le m.
\]
Then
\begin{equation}\label{eq:schur-one-coefficient}
 |[v^i](1+v)^rG(v)|
       \le {\operatorname{pref}_0}\max_{0\le j\le m}[v^i](1+v)^R{\operatorname{Row}_j}(v).
\end{equation}
For every $u>0$, with $\theta\coloneqq u/(1+u)$ and
$V\coloneqq(N-m)(\theta/b)(1-\theta/b)$, this implies
\begin{equation}\label{eq:schur-variance}
 |[v^i](1+v)^rG(v)|
 \le {\operatorname{pref}_0}u^{-i}\frac{(1+u)^{N+R}}{(1+{\omega_1}u)^{N-2m}}
                  \frac{\sqrt\pi}{2\sqrt{2V}}.
\end{equation}
\end{lemma}
The integer $R\ge r$ is used in Appendix~\ref{app:qutrit-tail}, where one
bound with $R\coloneqq{D_{\rm cap}}-2m$ serves every terminal degree $r\le R$.
Inequality~\eqref{eq:schur-variance} is a Cauchy-type bound at the real
radius $u$, valid for every $u>0$.
\begin{proof}
\smallskip\noindent\emph{Step 1 (Absolute majorant and weighted rows).}
We give the coefficient comparison explicitly, since taking coefficients
of a real bound on a signed polynomial would not justify it. The first
part follows the proof of Lemma~\ref{lem:weighted-rows} in
Appendix~\ref{app:coefficient-details}; it is repeated because here a
single coefficient is extracted before the Schur comparison.
In the generating basis
$F_z(y)=(1+z)^y(1-z/b)^{N+1-y}$, the symmetric paired kernel is the
kernel~\eqref{eq:finite-paired-kernel} in the variable $v$ (here $z,w$ are
the kernel variables, not $w=i/N$):
\[
 \begin{gathered}
 {\operatorname{Seed}}(z,w)[{\operatorname{Base}}(z,w)+v{\operatorname{Tilt}}(z,w)]^N,\\
 {\operatorname{Seed}}=1+\frac {\omega_1}2(z+w)-\frac{zw}b,\\
 {\operatorname{Base}}=1+\frac{zw}b,\quad {\operatorname{Tilt}}=\frac{z+w}b+\frac{{\omega_1}zw}b.
 \end{gathered}
\]
Normalize entry $(j,k)$ by $\sqrt{\nu_j\nu_k}$, where
$\nu_j=\binom Nj b^{-j}$, retaining only $0\le j,k\le m$.
Replace the minus sign in ${\operatorname{Seed}}$ by a plus sign to obtain
an entrywise absolute coefficient majorant $M^{\rm abs}(v)$, a matrix
polynomial in $v$. Its coefficients are nonnegative.

For a symmetric matrix $A$ with nonnegative entries, the Schur test with
the positive weights $\sqrt{\nu_j}$ gives
$\|A\|\le\max_j\nu_j^{-1/2}\sum_kA_{jk}\nu_k^{1/2}$; we call
$\nu_j^{-1/2}\sum_kA_{jk}\nu_k^{1/2}$ the weighted row $j$ of $A$. For the
normalized coefficients of
$[{\operatorname{Base}}+v{\operatorname{Tilt}}]^N$ it is the sum of the unnormalized
$(j,k)$ coefficients over $k\le m$, divided by $\nu_j$. Since
${\operatorname{Base}}+v{\operatorname{Tilt}}=(1+vz/b)+(w/b)(v+(1+{\omega_1}v)z)$, the
coefficient of $w^j$ in $[{\operatorname{Base}}+v{\operatorname{Tilt}}]^N$ is
$\binom Nj b^{-j}(v+(1+{\omega_1}v)z)^j(1+vz/b)^{N-j}$, so this weighted row
is exactly ${\operatorname{Row}_j}(v)$. Expanded, this is the row
polynomial~\eqref{eq:finite-row} of Lemma~\ref{lem:weighted-rows} in the
variable $v$:
\[
 {\operatorname{Row}_j}(v)=\sum_{{j_1}=0}^j\binom j{j_1} v^{j-{j_1}}(1+{\omega_1}v)^{j_1}
                \sum_{{j_2}=0}^{m-{j_1}}\binom{N-j}{j_2}(v/b)^{j_2}.
\]
In ${\operatorname{Seed}}$, the terms $({\omega_1}/2)z$ and $zw/b$ contain $z$, which
only shortens this positive row (the column $k=m$ is lost). The terms
$({\omega_1}/2)w$ and $zw/b$ contain $w$, which changes it to row $j-1$ and
contributes $\nu_{j-1}/\nu_j=bj/(N-j+1)\le1$, by $N\ge Qm$.
Thus the weighted row of $M^{\rm abs}(v)$ is bounded coefficientwise by
\[
 (1+{\omega_1}/2){\operatorname{Row}_j}(v)+({\omega_1}/2+1/b)\frac{bj}{N-j+1}{\operatorname{Row}_{j-1}}(v).
\]
The two multipliers sum to at most $1+{\omega_1}+1/b=2$; this factor $2$,
divided by the coercivity, is ${\operatorname{pref}_0}$.

\smallskip\noindent\emph{Step 2 (Schur test for one coefficient).}
Take the single coefficient $[v^i](1+v)^r$ first: it is the signed matrix
$\sum_\ell\binom r{i-\ell}M_\ell$, whose quadratic form at the coefficient
vector $q$ of $f$ (called $x$ in Section~\ref{sec:coefficient-methods}),
divided by $q^{\mathsf T}M_0q$, is $[v^i](1+v)^rG(v)$. Its absolute value
is bounded entrywise by $[v^i](1+v)^RM^{\rm abs}(v)$, because $R-r$ is a
nonnegative integer, so $(1+v)^r$ is coefficientwise at most $(1+v)^R$.
Schur comparison for this nonnegative
symmetric matrix, followed by division by the base quadratic
form at least ${\operatorname{coer}}\|q\|^2$, proves
\eqref{eq:schur-one-coefficient}.

\smallskip\noindent\emph{Step 3 (Domination of the rows at the radius $u$).}
Fix $u>0$ and $z_0\coloneqq(1+{\omega_1}u)^{-1}\le1$.
For a polynomial $P$ with nonnegative coefficients,
$\sum_{k\le m}[z^k]P\le z_0^{-m}P(z_0)$. Applied with coefficients in $v$,
this gives the following inequality coefficient by coefficient in $v$
(written $\preceq$):
\[
 {\operatorname{Row}_j}(v)\preceq
 z_0^{j-m}(1+({\omega_1}+1/z_0)v)^j(1+z_0v/b)^{N-j}.
\]
This $z_0$ makes the evaluation at $u$ collapse:
$1+({\omega_1}+1/z_0)u=(1+u)(1+{\omega_1}u)$ and
$1+z_0u/b=(1+u)/(1+{\omega_1}u)$.
After multiplication by $(1+v)^R$, the value at $u$ is therefore
\[
 (1+u)^{N+R}(1+{\omega_1}u)^{m-N+j}
       \le\frac{(1+u)^{N+R}}{(1+{\omega_1}u)^{N-2m}}.
\]

\smallskip\noindent\emph{Step 4 (Binomial point-mass bound).}
The coefficients of this positive polynomial, times $u^i$ and divided by
its value at $u$, form a convolution of three binomial distributions. The
last factor contributes $\Bin(N-j,\theta/b)$, since
$(z_0u/b)/(1+z_0u/b)=\theta/b$; its variance is at least $V$.
The maximum point mass of a binomial distribution $\Bin(L,q)$ of variance
$W\coloneqq Lq(1-q)>0$ is at most $\sqrt\pi/(2\sqrt{2W})$: Fourier inversion
and $\sin(|t|/2)\ge|t|/\pi$ give
\[
 \frac1{2\pi}\int_{-\pi}^{\pi}
     |1-q+qe^{it}|^L\,dt
 \le\frac1{2\pi}\int_{\mathbb R}e^{-2Wt^2/\pi^2}\,dt.
\]
Convolution cannot increase maximum point mass.
Applying this to each row proves~\eqref{eq:schur-variance}.
\end{proof}

\begin{lemma}\label{lem:qutrit-finite-norm}
For $m\ge2$ and $N\ge9m$, the finite qutrit paired matrix
satisfies $M_0\succ7I/(4(m+10))$.
Thus Lemma~\ref{lem:schur-coefficient} applies with
${\operatorname{pref}_0}=8(m+10)/7$.
\end{lemma}
\begin{proof}
By~\eqref{eq:base-matrix-entries} at $Q=9$, its diagonal at $j$ is
$1-j/(N-j+1)$ and its adjacent entry is $(7/2)\sqrt{(j+1)/(8(N-j))}$.
The cap $N\ge9m$ bounds the diagonal below by $7/8$ and the adjacent
entry above in absolute value by $(7/16)\sqrt{(j+1)/m}$.
Let ${J_m^{\rm path}}$ be the matrix of Appendix~\ref{app:higher-certificates}:
zero diagonal and adjacent entries $\tfrac12\sqrt{(j+1)/m}$. The comparison
of quadratic forms on absolute coordinates, as
in~\eqref{eq:retained-coercivity}, gives
$M_0\succeq(7/8)(1-\|{J_m^{\rm path}}\|)I$.
The finite tridiagonal estimate~\eqref{eq:finite-tridiagonal-norm} gives
\[
 \|{J_m^{\rm path}}\|\le\frac{m+2}{\sqrt{m(m+10)}}.
\]
The identity
\[
 m(m+8)^2-(m+10)(m+2)^2=2(m^2+10m-20)>0
\]
is the squared form of $(m+2)/\sqrt{m(m+10)}<(m+8)/(m+10)$, so the last
bound is less than $1-2/(m+10)$. The conclusion follows.
\end{proof}

\subsection{Finite binomial and circle factors}
\label{app:qutrit-circle-factors}

Let $H_2(w)\coloneqq-w\log w-(1-w)\log(1-w)$.
The finite Stirling bounds of Robbins~\cite{Robbins} imply
\begin{equation}\label{eq:binomial-three}
 \binom Ni\ge\frac{e^{NH_2(i/N)}}{3\sqrt i},\qquad1\le i<N.
\end{equation}
Indeed, by~\eqref{eq:binomial-mode} below, the normalized binomial mode
$\binom Ni w^i(1-w)^{N-i}$, $w=i/N$, is greater than
$e^{-1/6}/\sqrt{2\pi i}$, and $e^{-1/6}/\sqrt{2\pi i}>1/(3\sqrt i)$ means
$2\pi e^{1/3}<9$. For a rational check of this constant, $\pi<22/7$ and
$e<87/32<(7/5)^3$ give $2\pi e^{1/3}<44/5<9$.
The same factorial inequalities give the sharper inverse-mode bound
\begin{equation}\label{eq:binomial-mode}
 \left(\binom Ni w^i(1-w)^{N-i}\right)^{-1}
 \le\sqrt{2\pi Nw(1-w)}e^{\varepsilon_i},\qquad
 \varepsilon_i\coloneqq\frac1{12i}+\frac1{12(N-i)},\quad w=i/N;
\end{equation}
this is~\eqref{eq:finite-mode-bound} of Appendix~\ref{app:higher-four-initial}
with $M=N$ and $\theta_0=w$.

Consider a component of the mixture~\eqref{eq:positive-mixture}, multiplied
by $(1+v)^r$, that is,
${F_{\rm mix}}(v)\coloneqq(1+v)^{k_+}(1-av)^{k_-}$, where ${k_+}\coloneqq r+j$, ${k_-}\coloneqq N-j$, and
$0\le a\le1/8=1/b$. Choose a radius $0<u<8$, so that $au<1$, and set
\[
 q_u\coloneqq\frac u{1+u},\qquad
 {\operatorname{curv}(u)}\coloneqq\frac{(1+u)^2}{8(1-u/8)^2};
\]
$q_u$ is the $\theta$ of Lemma~\ref{lem:schur-coefficient}, and
$\operatorname{curv}(u)$ is the value at the largest parameter $a=1/8$ of
$a(1+u)^2/(1-au)^2$, which increases with $a$.
The exact modulus on the circle is
\[
 \frac{|{F_{\rm mix}}(ue^{it})|}{{F_{\rm mix}}(u)}
 =(1-4q_u(1-q_u)\sin^2(t/2))^{{k_+}/2}
  \left(1+\frac{4au\sin^2(t/2)}{(1-au)^2}\right)^{{k_-}/2}.
\]
Here $au/(1-au)^2=q_u(1-q_u)\,a(1+u)^2/(1-au)^2$, since
$u/(q_u(1-q_u))=(1+u)^2$. Using $\log(1-z)\le-z$ and $\log(1+z)\le z$,
together with ${k_+}\ge r\ge s_0N$ and ${k_-}\le N$, yields
\begin{equation}\label{eq:qutrit-circle-curvature}
 \frac{|{F_{\rm mix}}(ue^{it})|}{{F_{\rm mix}}(u)}
 \le e^{-2V\sin^2(t/2)},\qquad
 V\coloneqq N(s_0-{\operatorname{curv}(u)})q_u(1-q_u),
\end{equation}
whenever $s=r/N\ge s_0>{\operatorname{curv}(u)}$; the threshold $s_0$ is
chosen for each row.
This lower curvature is valid uniformly in $j,a$.
For its angular average $J(V)$, defined
in~\eqref{eq:finite-angular-integral}, we use the two elementary bounds
\begin{equation}\label{eq:angular-two-bounds}
 J(V)\le\frac{\sqrt\pi}{2\sqrt{2V}},\qquad
 J(V)\le\frac1{\sqrt{1+2V}}.
\end{equation}
The first follows from the bound $\sin(|t|/2)\ge|t|/\pi$ and the Gaussian
integral in the proof of Lemma~\ref{lem:schur-coefficient}, with $W$
replaced by $V$. For the second, apply $e^{-z}\le(1+z)^{-1}$ and evaluate
the resulting trigonometric integral by $t=2\arctan y$.
Cauchy's integral over the circle $|v|=u$ thus gives
$|[v^i]{F_{\rm mix}}(v)|\le u^{-i}{F_{\rm mix}}(u)J(V)$ for every component.
Averaging these absolute Cauchy bounds over the positive
mixture~\eqref{eq:positive-mixture} is valid even when the extracted
coefficients have signs; since the components average to $(1+v)^rG(v)$,
\[
 |[v^i](1+v)^rG(v)|\le u^{-i}(1+u)^rG(u)\,J(V),
 \qquad s\ge s_0>\operatorname{curv}(u).
\]

\subsection{Integrated tilt and the positive initial coefficients}
\label{app:qutrit-initial}

This subsection proves two shared estimates: the tilt
bound~\eqref{eq:qutrit-integrated-tilt} for $G$ (Step~1), used in all four
rows, and the conditional initial estimate~\eqref{eq:qutrit-positive-initial}
(Steps~2--6), which holds at every index $i\le N/24$ through which all
mixture components have positive coefficients. The latter is used in
Appendix~\ref{app:qutrit-quarter}, where positivity holds through
$\lfloor N/24\rfloor$, and in Appendix~\ref{app:qutrit-eighth}, where it
holds through $i=3,5,10$ in three bands of $s$.

\smallskip\noindent\emph{Step 1 (The tilt bound for $G$).}
For $0\le u\le1$, the degree-sensitive tilted mean, defined below, gives
\begin{equation}\label{eq:qutrit-integrated-tilt}
 \frac1N\log G(u)\le\Psi_x(u)
 \coloneqq\frac x2\log((1+u)(1-u/8))+
                 \frac{2+7x}{16}I(u),\quad
 I(u)\coloneqq\int_0^u\frac{dt}{\sqrt{(1+t)(1-t/8)}}.
\end{equation}
For the derivation, write $X\coloneqq N-Y$ in~\eqref{eq:qutrit-generating}: the
numerator of $G(u)$ is $\E h(X)$ for $X\sim\Bin(N,(8-u)/9)$, which is the
binomial law $\mu_N$ of the real alphabet $R\coloneqq9/(1+u)$. Differentiation in
$u$ gives
\[
 (\log G)'(u)=\frac N{1+u}-\frac9{(1+u)(8-u)}\,\frac{\E[Xh(X)]}{\E h(X)}.
\]
The degree-sensitive tilted mean is the lower bound for
$\E[Xh(X)]/\E h(X)$ given by Lemma~\ref{lem:paired-mean} at the
alphabet $R$, with $L=N$ and $k=m$. That lemma is stated for $Q\ge9$, but
its proof uses only $Q>2$ and $L\ge Qk$; the proof of
Lemma~\ref{lem:approximate-power} writes out this extension to every real
alphabet $R>2$. Here $R=9/(1+u)\in[9/2,9]$ for $0\le u\le1$, so $R>2$, and
the degree condition $N\ge Rm$ follows from $N\ge9m$. The result is the
derivative bound in the proof of Lemma~\ref{lem:approximate-power} at
$Q=9$ and $m/N=x/2$:
\[
 \frac1N(\log G)'(u)\le\frac{x(7-2u)}{2(1+u)(8-u)}
       +\frac{\sqrt{x(2-x)/8}}{\sqrt{(1+u)(1-u/8)}}.
\]
The tangent-line inequality $\sqrt{x(2-x)/8}\le(2+7x)/16$, with equality at
$x=2/9$, follows by squaring, with difference $(9x-2)^2/256$; it makes the
bound affine in $x$. Integration from $G(0)=1$
proves~\eqref{eq:qutrit-integrated-tilt}, since $(x/2)\log((1+u)(1-u/8))$
has derivative $x(7-2u)/(2(1+u)(8-u))$.
The derivative bound decreases in $u$ on $[0,1]$ and equals $N\beta_x$
at $u=0$, so a simpler consequence is
\begin{equation}\label{eq:qutrit-linear-tilt}
 G(u)\le e^{N\beta_x u},\qquad
 \beta_x\coloneqq\frac{7x}{16}+\sqrt{\frac{x(2-x)}8}
                 \le\frac{1+7x}8.
\end{equation}

\smallskip\noindent\emph{Step 2 (Suppression under component positivity).}
The next argument isolates the role of component positivity.
Put $M\coloneqq N-2m$, ${J_*}\coloneqq N-{a_{\min}}+1$, $v\coloneqq{J_*}-2m$ as in
Section~\ref{sec:coefficient-methods} (not the polynomial variable of
Appendix~\ref{app:qutrit-one-coefficient}), and ${J_0}\coloneqq M-r$.
The root assumptions give
\begin{equation}\label{eq:qutrit-mixture-ranges}
 v\ge1,\qquad {J_*}\le {J_0},\qquad r+{J_*}\le M,
 \qquad9v\le2M+1;
\end{equation}
the last two follow from $D<{a_{\min}}$ and, by~\eqref{eq:qutrit-wedge},
$v<2(M+1)/9$.
Suppose every mixture component, that is, every atom $j<{J_*}$ (with
$a=1/8$) and every $a$ in the tail part $j={J_*}$, has positive coefficients
through index $i$, where $i\le N/24$ and $i<{k_+},{k_-}$; this premise is
verified in Appendices~\ref{app:qutrit-quarter} and~\ref{app:qutrit-eighth}.
For ${q_\ell}\coloneqq[u^\ell](1+u)^{k_+}(1-au)^{k_-}$, let
${\widehat q_\ell}\coloneqq{q_\ell}/\binom {k_+}\ell$.
The exact recurrence is the recurrence~\eqref{eq:component-recurrence} of
Lemma~\ref{lem:component-coefficients}:
\[
 {\widehat q_{\ell+1}}=\left(1-\frac{a({k_-}-\ell)}{{k_+}-\ell}\right){\widehat q_\ell}
 -\frac{a\ell(N+r-\ell+1)}{({k_+}-\ell)({k_+}-\ell+1)}{\widehat q_{\ell-1}}.
\]
Positivity implies successively ${\widehat q_\ell}\le {\widehat q_{\ell-1}}$;
substitution in the negative term and telescoping gives, as in the proof
of~\eqref{eq:component-suppression}, which uses only this positivity,
\[
 {\widehat q_i}\le\exp\left(-\frac{a{k_-}i}{{k_+}-i+1}\right).
\]
Put $\eta\coloneqq\log((M+1)/(M-i+1))$ and $T_{j,a}\coloneqq{J_0}-j+a(N-j)$. The binomial
ratio satisfies $\binom{k_+}i/\binom Mi\le((M+1-i)/(M+1))^{M-{k_+}}
=e^{-\eta(M-{k_+})}$, and $\eta\le i/(M+1-i)\le i/({k_+}-i+1)$; together,
since $M-{k_+}={J_0}-j$ and ${k_-}=N-j$, they yield
\[
 \frac{{q_i}}{\binom Mi}\le e^{-\eta T_{j,a}}.
\]

\smallskip\noindent\emph{Step 3 (Joint tilt comparison).}
Set $u\coloneqq\eta/(1+\eta/8)$. Then $\log(1+u)\le u\le\eta$, and
$-\log(1-au)\le au/(1-au)=a\eta/(1+\eta/8-a\eta)\le a\eta$ for every
$a\le1/8$. Hence $e^{-\eta T_{j,a}}\le(1+u)^{j-{J_0}}(1-au)^{N-j}$ for every
component. Averaging over the mixture~\eqref{eq:positive-mixture}, whose
components evaluated at $u$ average to $G(u)$, the entire mixture satisfies
\begin{equation}\label{eq:joint-tilt-comparison}
 \mathcal L_i\le\frac{\binom Mi}{\binom Ni}
                  \frac{G(u)}{(1+u)^{{J_0}}}.
\end{equation}

\smallskip\noindent\emph{Step 4 (The tilt saving at $N\ge512$).}
Here are finite constants valid at $N\ge512$.
Write $\epsilon\coloneqq1/N$. Then
$0<x\le2/9$, $0<w\le1/24$, and
$u\le72w/53<1$, because $u\le\eta\le i/(M+1-i)\le w/(7/9-w)$ with
$M\ge7N/9$, and $7/9-1/24=53/72$.
The left side below is $(1/N)\log((1+u)^{J_0}/G(u))$ bounded
through~\eqref{eq:qutrit-linear-tilt}. Since $D\le7(N-2m+1)/9$
by~\eqref{eq:qutrit-wedge}, ${J_0}=N-D$ gives
${J_0}/N\ge(2+7x)/9-7\epsilon/9$; with $\log(1+u)\ge u-u^2/2$ and
$(2+7x)/9-(1+7x)/8=7(1-x)/72$ this gives
\[
 \frac{{J_0}}{N}\log(1+u)-\beta_x u
 \ge\frac7{72}(1-x-8\epsilon)u-\frac{2+7x}{18}u^2.
\]
Use $u\ge w/(1-x+\epsilon+w/8)$ in its first term and
$u\le72w/53$ in the second. The ratio of the resulting
lower bound to $w$ decreases in $x$, $\epsilon$ and $w$, so it is at least
its value at $x=2/9$, $\epsilon=1/512$, $w=1/24$:
\[
 \frac7{72}\frac{7/9-8/512}{7/9+1/512+1/192}
       -\frac{16}{81}\frac{(72/53)^2}{24}
 =\frac{7243129}{91441377}>\frac5{64}.
\]
Consequently~\eqref{eq:joint-tilt-comparison} is less than
$\binom Mi/\binom Ni$ times $e^{-5i/64}$.

\smallskip\noindent\emph{Step 5 (Loss at the packed endpoint).}
Distinct root packing gives
$h(i)/h(0)\ge(N-v-2m)_i/(N-v)_i$, the packed endpoint of
Appendix~\ref{app:higher-initial} (roots ${a_{\min}},\ldots,{a_{\min}}+2m-1$).
Relative to $U_0(i)\coloneqq\binom Mi/\binom Ni$, the logarithmic loss is, as in
Appendix~\ref{app:higher-four-initial},
\[
 -\log\left(\frac{h(i)}{h(0)U_0(i)}\right)
 \le\sum_{\ell=0}^{v-1}g(\ell/N),\quad
 g(z)\coloneqq\log\frac{(1-w-z)(1-x-z)}{(1-z)(1-x-w-z)}.
\]
The function $g$ is increasing and convex, as follows from
$g(z)=\int_0^x\int_0^w(1-z-s-t)^{-2}\,dt\,ds$; the same integral gives
$g(z)\le xw/((1-z-w)(1-z-x-w))$. The left Riemann sum, the trapezoid
inequality and this bound show that the loss divided by $Nw$ is at most
\[
 \frac{xy}{2}\left(
  \frac1{(1-w)(1-x-w)}+
  \frac1{(1-y-w)(1-x-y-w)}\right),\quad y\coloneqq v/N.
\]
This increases with $y$, so use the bound $y\le2(1-x)/9+\epsilon/9$
from $9v\le2M+1$. After this substitution it increases with $x,w,\epsilon$
on the rectangle $0<x\le2/9$, $0<w\le1/24$, $0<\epsilon\le1/512$.
For the second summand, its numerator's logarithmic derivative
in $x$ is at least $1/x-1/(1-x)\ge45/14$, while that of
the denominator is at most $(2/9)/(1-y-w)<1/3$.
For the first, use $x(1-x)/(1-x-w)=x+xw/(1-x-w)$.
All denominators are positive, with $1-y-w>2/3$.
At $x=2/9,w=1/24,\epsilon=1/512$ the bound is
\[
 \frac{14165796445273}{200235323964168}<\frac3{40}.
\]

\smallskip\noindent\emph{Step 6 (Conclusion).}
Combining the two strict estimates, $5/64-3/40=1/320$, proves
\begin{equation}\label{eq:qutrit-positive-initial}
 \mathcal L_i<e^{-i/320}\frac{h(i)}{h(0)}
\end{equation}
for $N\ge512$, $m\ge2$ and every index $i\le N/24$ with $i<{k_+},{k_-}$
through which all mixture components have positive coefficients
(the premise of Step~2). This estimate uses only distinct roots
and~\eqref{eq:qutrit-wedge}; the bound~\eqref{eq:qutrit-product-simplified},
specific to a global minimizer, enters in
Appendices~\ref{app:qutrit-quarter}--\ref{app:qutrit-middle}.

\subsection{The first quarter when the terminal degree is large}
\label{app:qutrit-quarter}

We prove the first row of Proposition~\ref{prop:qutrit-four-ranges}: for a
minimizing vertex as in that proposition ($m\ge2$,~\eqref{eq:qutrit-wedge},
$N\ge512$) with $r\ge N/3$, $\mathcal L_i<h(i)/h(0)$ for
$1\le i\le\lfloor N/4\rfloor$. For $i\le N/24$ the component coefficients are
positive and~\eqref{eq:qutrit-positive-initial} applies (Step~1). For the
other indices we use the averaged Cauchy bound of
Appendix~\ref{app:qutrit-circle-factors} at a radius $u$, the tilt
bound~\eqref{eq:qutrit-integrated-tilt} for $G(u)$,~\eqref{eq:binomial-mode}
for $\binom Ni$, and~\eqref{eq:qutrit-product-simplified} for $h(i)/h(0)$
(Step~2). The exponent is $-N$ times a continuous saving, certified by
polynomials (Steps~3--4), plus a finite loss (Step~5); the remaining
prefactor (Step~6) is absorbed in Step~7. Lemma~\ref{lem:schur-coefficient}
is not used here.

\smallskip\noindent\emph{Step 1 (Positive initial coefficients).}
Assume $r\ge N/3$, $N\ge512$, and $1\le i\le N/4$.
For $i\le N/24$, the Jacobi test~\eqref{eq:component-jacobi-test} of
Lemma~\ref{lem:component-coefficients} holds for every component and every
index $\ell\le N/24$; the other hypotheses of that lemma hold because
$\ell\le N/24<N/3\le r\le{k_+}$, ${k_-}=N-j\ge N-{J_*}\ge r+2m>\ell$ and
${n_{\rm mix}}=N+r$.
Its worst values are $a=1/8,j=0$, since the left side
of~\eqref{eq:component-jacobi-test} increases with $j$ and decreases
with $a$, while its right side increases with $a$; with
$t\coloneqq(\ell-1)/N$ its squared margin, as defined in
Appendix~\ref{app:higher-initial}, is
\[
 F(s,t)\coloneqq(s-1/8-7t/8)^2-\frac t2(1+s-t+2/N),
\]
in which $2/N$ replaces the exact term $1/N$.
On $s\ge1/3$, $0\le t\le1/24$, the unsquared term is
positive, $F_s>0$ and $F_t<0$.
At $s=1/3,t=1/24,N=512$, the worst case, it is $47/18432>0$.
As in Appendix~\ref{app:higher-initial}, the left side of the test decreases
and its right side increases through these indices, so the largest index
is the worst. All component coefficients are therefore
positive through $\lfloor N/24\rfloor$, and
\eqref{eq:qutrit-positive-initial} applies.

\smallskip\noindent\emph{Step 2 (Radius and the continuous saving).}
For the remaining indices choose
\begin{equation}\label{eq:quarter-radius}
 u\coloneqq\frac{w}{7/9-w},\qquad 0<w\le\frac14,
 \qquad0<u\le\frac9{19}.
\end{equation}
This is the saddle point of $u^{-i}(1+u)^{7N/9}$, where $7/9$ is the leading
term of the bound $s\le7/9-16x/9+7/(9N)$, which follows from $r=D-2m$
and~\eqref{eq:qutrit-wedge}; it gives $q_u=u/(1+u)=9w/7$.
Set ${\operatorname{LogUp}}\coloneqq7801/200000$ and ${\operatorname{LogLow}}\coloneqq2513144/10^7$, one-sided
rational bounds in
\begin{equation}\label{eq:quarter-log-constants}
 {\operatorname{LogBud}_9}(2/9)<{\operatorname{LogUp}},\qquad-\log(7/9)>{\operatorname{LogLow}},
 \qquad\log(16/7)<5/6,
\end{equation}
which will be certified below by rational series.
For $i>N/24$, the averaged Cauchy bound of
Appendix~\ref{app:qutrit-circle-factors} with $s_0=1/3$,
\eqref{eq:qutrit-integrated-tilt}, \eqref{eq:binomial-mode},
\eqref{eq:qutrit-product-simplified} and the bound on $s$ give, with the
finite loss $\operatorname{Fin}(w)$ of Step~5,
\[
 \frac{|\mathcal L_i|}{h(i)/h(0)}\le
 \exp\{-N{\operatorname{Save}}(x,w)+\operatorname{Fin}(w)\}\,
 \sqrt{2\pi Nw(1-w)}\,e^{\varepsilon_i}J(V),
\]
with the continuous saving
\begin{align}\label{eq:quarter-saving}
 {\operatorname{Save}}(x,w)\coloneqq{}&H_2(w)+w\log u-(7/9-16x/9)\log(1+u)
       -\Psi_x(u)\notag\\
       &+x\log(1-w)-(9x/2){\operatorname{LogUp}}\kappa(w).
\end{align}
Here $H_2(w)$ and $\sqrt{2\pi Nw(1-w)}\,e^{\varepsilon_i}$ come
from~\eqref{eq:binomial-mode}, and ${\operatorname{LogUp}}$ replaces
${\operatorname{LogBud}_9}(2/9)$, which lowers the saving.

\smallskip\noindent\emph{Step 3 (Endpoint expressions and the saving table).}
The saving~\eqref{eq:quarter-saving} is affine in $x$, so it suffices to
bound it at the endpoints $x=0$ and $x=2/9$ of the range of $x$ (the
subscripts $0$ and $*$ below). With $w=7u/(9(1+u))$,
$H_2(w)+w\log u=w\log(9/7)+\log(1+u)-(1-w)\log(1+2u/9)$; substituting
$x=0$ and $x=2/9$, and replacing $-\log(7/9)$ by ${\operatorname{LogLow}}$, which
again lowers the saving, gives the lower bounds
\begin{align}
 {\operatorname{Save}}_0(u)\coloneqq{}&\frac29\log(1+u)
  -\frac{1+2u/9}{1+u}\log(1+2u/9)
        +\frac{7u{\operatorname{LogLow}}}{9(1+u)}-\frac18I(u),\label{eq:quarter-zero}\\
 {\operatorname{Save}}_*(u)\coloneqq{}&\frac{23}{81}\log(1+u)-\frac19\log(1-u/8)
  -\frac7{9(1+u)}\log(1+2u/9)\notag\\
  &+\frac{7u{\operatorname{LogLow}}}{9(1+u)}-\frac29I(u)-{\operatorname{LogUp}}\kappa(w).
       \label{eq:quarter-cap}
\end{align}
Here $\kappa(w)$ of~\eqref{eq:qutrit-product-functions} is written in
terms of $u$:
\[
 \kappa(w)=\frac{7u}{18}
          \left(\frac9{9+2u}+\frac{81}{49-14u}\right).
\]
In each row of the next table, the entry ${\operatorname{save}}$ in the last
column gives ${\operatorname{Save}}_*(u)\ge {\operatorname{save}}\,w$ for all $0<w$ up to the
upper $w$ of that row (equivalently $0<u\le U$, the upper $u$), with $w=i/N$:
\begin{center}
\begin{tabular}{ccc}\toprule
Upper $w$&Upper $u$ ($U$)&${\operatorname{save}}$, a lower bound for ${\operatorname{Save}}_*(u)/w$\\\midrule
$1/16$&$9/103$&$1/15$\\
$1/8$&$9/47$&$1/16$\\
$1/6$&$3/11$&$1/20$\\
$2/9$&$2/5$&$1/24$\\
$1/4$&$9/19$&$1/32$\\\bottomrule
\end{tabular}
\end{center}
Also ${\operatorname{Save}}_0(u)>w/14$ throughout $0<u\le9/19$. Since every
${\operatorname{save}}$ in the table is at most $1/15<1/14$ and the saving is affine
in $x$, the five bounds hold for~\eqref{eq:quarter-saving}
at every $0\le x\le2/9$.

\smallskip\noindent\emph{Step 4 (The polynomial certificate).}
Here is the precise polynomial certificate for these continuous
inequalities. Each replacement below lowers ${\operatorname{Save}}_0$ and
${\operatorname{Save}}_*$, given the signs of their terms. For $\log(1+u)$ use its
alternating lower polynomial of degree $12$; for the subtracted
$\log(1+2u/9)$ use its upper polynomial of degree $11$.
For $-\log(1-u/8)$ retain its first $12$ positive terms.
With $(1+t)(1-t/8)=1+z$, $z\coloneqq7t/8-t^2/8\ge0$, Taylor's theorem gives
\begin{equation}\label{eq:qutrit-integral-polynomial}
 I(u)\le I_{12}(u)\coloneqq\int_0^u\sum_{j=0}^{12}
     \frac{(-1)^j\binom{2j}j}{4^j}(7t/8-t^2/8)^j\,dt.
\end{equation}
Its direction follows because the next derivative of
$(1+z)^{-1/2}$ is negative for $z\ge0$.
In~\eqref{eq:quarter-cap}, multiply the resulting lower
bound minus ${\operatorname{save}}\,w$ by $(1+u)(9+2u)(49-14u)>0$.
The polynomial has degree $28$.
For each pair $(U,{\operatorname{save}})$ in the table, substitute
$u=Ut/(1+t)$, which maps $0\le t<\infty$ onto $0\le u<U$, and clear
$(1+t)^{28}$. Its constant coefficient is zero and all $28$ remaining
coefficients are positive; the leading coefficient is the value at $u=U$
($t\to\infty$), so the endpoint is included.
For~\eqref{eq:quarter-zero}, first subtract $w/14$ from
the resulting lower bound, then multiply by $1+u$ and use
the same substitution with $U=9/19$;
the resulting degree-$26$ polynomial has all $26$
nonconstant coefficients positive.
These $166$ rational coefficients, defined by the formulas
above, are listed in the certificate
package~\cite{ComputationalCertificates}.
For~\eqref{eq:quarter-log-constants}, the rational rule is
\begin{equation}\label{eq:rational-log-series}
 \left|\log c-2\sum_{j=0}^{k-1}\frac{z^{2j+1}}{2j+1}\right|
 \le\frac{2|z|^{2k+1}}{(2k+1)(1-z^2)},
 \qquad z\coloneqq\frac{c-1}{c+1}.
\end{equation}
Use $k=60$ and take the lower or upper end of this interval according
to the direction required in the comparison.

\smallskip\noindent\emph{Step 5 (Finite loss).}
The finite loss is the part of the exponent that is not proportional to
$N$: the terms $\log(16/7)\kappa(w)$ from~\eqref{eq:qutrit-product-simplified}
and $(7/9)\log(1+u)$ from the term $7/(9N)$ in the bound on $s$. By
$\log(16/7)<5/6$ it is at most
\[
 \operatorname{Fin}(w)\coloneqq\frac56\kappa(w)+\frac79\log(1+u).
\]
For $w\le1/16$, $\operatorname{Fin}(w)/w\le1498840/651681<5/2$,
using $\log(1+u)\le u$ and the increase of the resulting quotient in $w$.
At the next four upper endpoints, the alternating cubic
upper bound gives respectively
\[
 \operatorname{Fin}(w)\le\frac{206499809}{678068013}<\frac13,\quad
 \frac{40658}{94501}<\frac12,\quad
 \frac{458807}{732375}<\frac23,\quad
 \frac{1052203}{1419813}<\frac34.
\]
The function $\operatorname{Fin}$ increases, so these bounds hold throughout
the corresponding intervals.

\smallskip\noindent\emph{Step 6 (Prefactor).}
For the angular factor use~\eqref{eq:qutrit-circle-curvature}
with $s_0=1/3$ and $q_u=9w/7$.
For $i>N/24$, $N\ge512$ implies
$\varepsilon_i<19/(9N)<41/9728<1/200$.
The prefactor is $\sqrt{2\pi Nw(1-w)}\,e^{\varepsilon_i}J(V)$. The Gaussian
bound in~\eqref{eq:angular-two-bounds}, $\pi/2<11/7$,
$e^{\varepsilon_i}\le1/(1-\varepsilon_i)\le200/199$ and $1-w\le1$ bound it by
\begin{equation}\label{eq:quarter-prefactor}
 \frac{11}{7}\frac{200}{199}
 \sqrt{\frac{7/9}{(1/3-{\operatorname{curv}(u)})(1-9w/7)}}.
\end{equation}
It increases with $w$.
At the five upper endpoints its squares are respectively
\[
 \frac{51437584000000}{4443993370821},\quad
 \frac{208606323200}{13272552357},\quad
 \frac{6358000000}{299502363},\quad
 \frac{349448000}{7959801},\quad
 \frac{395892640000}{3686100681}.
\]
They are strictly smaller than $(7/2)^2,4^2,5^2,7^2,11^2$.
All curvature denominators in~\eqref{eq:quarter-prefactor}
are positive on this range, since $\operatorname{curv}(9/19)<1/3$.

\smallskip\noindent\emph{Step 7 (Absorption by index interval).}
In each row of the next table, the entry ``Saving'' is a lower bound for
$N{\operatorname{save}}\,w-\operatorname{Fin}(w)$, followed by its value at the first index
$i$ allowed by $N\ge512$, and the prefactor is below the last entry. The
bound of Step~2 is therefore at most the prefactor times $e^{-\rm Saving}$,
which is less than one by the exponential bounds after the table:
\begin{center}\small
\begin{tabular}{llll}\toprule
Index interval&First possible $i$&Saving&Prefactor\\\midrule
$N/24<i\le N/16$&22&$i/15-5i/(2N)>4/3$&$<7/2$\\
$N/16<i\le N/8$&33&$i/16-1/3>3/2$&$<4$\\
$N/8<i\le N/6$&65&$i/20-1/2>2$&$<5$\\
$N/6<i\le2N/9$&86&$i/24-2/3>2$&$<7$\\
$2N/9<i\le N/4$&114&$i/32-3/4>5/2$&$<11$\\\bottomrule
\end{tabular}
\end{center}
For the first row, the lower bound at $N=512,i=22$ is
$10439/7680>4/3$.
The positive exponential series gives
$e^{4/3}>7/2$, $e^{3/2}>4$, $e^2>7$, and $e^{5/2}>11$.
Each saving increases with $i$.
Together with the positive initial interval, this proves the
first row of Proposition~\ref{prop:qutrit-four-ranges}.

\subsection{The first eighth for smaller terminal degree}
\label{app:qutrit-eighth}

We prove the second row of Proposition~\ref{prop:qutrit-four-ranges}: for a
minimizing vertex as in that proposition with $N/6<r<N/3$,
$\mathcal L_i<h(i)/h(0)$ for $1\le i\le\lfloor N/8\rfloor$. The route is that
of Appendix~\ref{app:qutrit-quarter}, with three changes: $s$ enters exactly,
the radius is capped in three bands of $s$, and positivity at the first
indices is proved directly, by residue classes of $N$ (Step~4).

\smallskip\noindent\emph{Step 1 (Radius and the saving table).}
Assume $N/6<r<N/3$ and $1\le i\le N/8$.
Use the same product bound~\eqref{eq:qutrit-product-simplified} and tilt
bound~\eqref{eq:qutrit-integrated-tilt}, but choose
\[
 u\coloneqq\min\left(\frac{w}{3/5-w},U\right),
\]
with the cap $U$ of the band of $s$ in the table below; in the uncapped
part $q_u=u/(1+u)=5w/3$.
The continuous saving is
\begin{equation}\label{eq:eighth-saving}
 {\operatorname{Save}}_s(x,w,u)\coloneqq H_2(w)+w\log u-s\log(1+u)-\Psi_x(u)
          +x\log(1-w)-(9x/2){\operatorname{LogUp}}\kappa(w).
\end{equation}
Since $s$ enters exactly, the finite loss is only the term
$\log(16/7)\kappa(w)$ from~\eqref{eq:qutrit-product-simplified}; it is at
most $(5/6)\kappa(w)<3w/2$.
This last inequality follows by division by $w$ and evaluation
at $w=1/8$, since $\kappa(w)/w$ increases.
The needed continuous bounds are (the weaker bounds of the last column are
used for $1/32<w\le1/8$)
\begin{center}
\begin{tabular}{cccc}\toprule
Range of $s$&$U$&Bound for $w\le1/32$&Bound for $w\le1/8$\\\midrule
$[1/6,3/16]$&$1/8$&${\operatorname{Save}}_s\ge39w/100$&${\operatorname{Save}}_s\ge3w/20$\\
$[3/16,1/4]$&$3/16$&${\operatorname{Save}}_s\ge2w/7$&${\operatorname{Save}}_s\ge3w/20$\\
$[1/4,1/3]$&$1/4$&${\operatorname{Save}}_s\ge2w/15$&${\operatorname{Save}}_s\ge2w/15$\\\bottomrule
\end{tabular}
\end{center}
The shared endpoints may be assigned to either adjacent row.

\smallskip\noindent\emph{Step 2 (Certificates for the uncapped part).}
To certify the table, ${\operatorname{Save}}_s$ decreases with $s$ and is affine
in $x$, so use the upper $s$ endpoint and $x=0,2/9$.
In the uncapped part, $w=3u/(5(1+u))$, and
$H_2(w)+w\log u=w\log(5/3)+\log(1+u)-(1-w)\log(1+2u/5)$, as in
Appendix~\ref{app:qutrit-quarter}.
In this subsection set ${\operatorname{LogLow}}\coloneqq5108256/10^7<-\log(3/5)$.
The lower endpoint expressions of this subsection, with $\kappa(w)$ written
in terms of $u$, are
\begin{align}
 {\operatorname{Save}}_0\coloneqq{}&(1-s)\log(1+u)
 -\frac{1+2u/5}{1+u}\log(1+2u/5)+w{\operatorname{LogLow}}-I(u)/8,
       \label{eq:eighth-zero}\\
 {\operatorname{Save}}_*\coloneqq{}&(2/3-s)\log(1+u)-\tfrac19\log(1-u/8)
 -\frac{7/9+8u/45}{1+u}\log(1+2u/5)\notag\\
 &+w{\operatorname{LogLow}}-2I(u)/9-{\operatorname{LogUp}}\kappa(w),\qquad
 \kappa(w)=\frac{3u}{10}
     \left(\frac5{5+2u}+\frac{405}{245+2u}\right).
       \label{eq:eighth-cap}
\end{align}
Use the same degree-$12$ and $I_{12}$ substitutions as in
Appendix~\ref{app:qutrit-quarter}, and the degree-$11$ upper polynomial now
for $\log(1+2u/5)$.
For ${\operatorname{Save}}_*-{\operatorname{save}}\,w$, multiply by
$(1+u)(5+2u)(245+2u)>0$, substitute $u=Ut/(1+t)$,
and clear $(1+t)^{28}$.
For the three triples
\[
 (s,U,{\operatorname{save}})=(3/16,1/8,39/100),\quad
 (1/4,3/16,2/7),\quad(1/3,1/4,2/15),
\]
all $28$ nonconstant coefficients are positive.
For ${\operatorname{Save}}_0$ use $s=1/3,U=1/4,{\operatorname{save}}=7/10$ and multiplier
$1+u$; all $26$ nonconstant coefficients are positive. This single check
covers the three bands, since ${\operatorname{Save}}_0$ decreases with $s$, $U=1/4$
is the largest cap, and $7/10$ exceeds every constant of the table.
The $110$ rational coefficients, specified by these
formulas, are listed in the certificate
package~\cite{ComputationalCertificates}.

\smallskip\noindent\emph{Step 3 (The capped portion).}
The capped portions start at $w=3U/(5(1+U))>1/32$.
At fixed $u=U$, ${\operatorname{Save}}_s-{\operatorname{save}}\,w$ is concave in $w$:
entropy and $x\log(1-w)$ are concave and $\kappa$ is convex.
The left endpoint was just certified: the uncapped certificate includes
$u=U$, with the larger constant of the column $w\le1/32$. At $w=1/8$, the six
checks at $x=0,2/9$ use ${\operatorname{save}}=3/20,3/20,2/15$ in the
three rows and are all strictly positive.
They use~\eqref{eq:rational-log-series} with $k=60$
and the upper bound $I_{12}$.
Concavity, followed by affine interpolation in $x$, proves
the table on the whole continuous domain.

\smallskip\noindent\emph{Step 4 (Positive first coefficients by residue classes).}
Here the Jacobi test of Lemma~\ref{lem:component-coefficients} is
replaced by an exact argument over residue classes. The unnormalized
coefficients $q_\ell$ of a
component $(1+u)^{k_+}(1-au)^{k_-}$ satisfy the recurrence
\begin{equation}\label{eq:qutrit-residue-recurrence}
 \ell {q_\ell}=[{k_+}-a{k_-}-(1-a)(\ell-1)]{q_{\ell-1}}
                 -a({k_+}+{k_-}-\ell+2){q_{\ell-2}}
\end{equation}
with ${q_0}=1,{q_{-1}}=0$ (compare~\eqref{eq:component-recurrence}).
In the three bands use respectively the worst values
\[
 {k_{+,0}}\coloneqq\lceil N/6\rceil,\ \lceil3N/16\rceil,\ \lceil N/4\rceil,
 \qquad {k_-}=N,\qquad a=1/8;
\]
here ${k_{+,0}}$ is a uniform integer lower bound for $r$ in the band (for example $r>N/6$
gives $r\ge\lceil N/6\rceil$), and ${k_+}=r+j\ge{k_{+,0}}$; why these values
are the worst is shown below.
For each residue class of $N$ modulo $c_1$ write $N=N_0+c_1y$, $y\ge0$
integral, where $N_0$ is its smallest representative at least $512$,
and $(c_1,c_2)\coloneqq(6,1),(16,3),(4,1)$ in the three bands.
Then ${k_{+,0}}={k_{+,0}}(N_0)+c_2y$.
Expansion of~\eqref{eq:qutrit-residue-recurrence} gives the coefficients
$q_\ell$ with $\ell\le3,5,10$, respectively, at these worst values, as
polynomials in $y$ with all coefficients positive, including the
constant term, in all $26=6+16+4$ residue classes; the circle bound
below is used from the next indices on.
Their complete rational coefficient lists are in the same part of the
certificate package~\cite{ComputationalCertificates}.
To see why these are the worst values, induct on the index $i$, with
base $q_0=1$: assume $q_{i-1}\ge0$ on the whole parameter range
$k_+\ge k_{+,0}$, $k_-\le N$, $0\le a\le1/8$. The identities
\[
 \Delta_{k_+}{q_i}={q_{i-1}}\ge0,\qquad
 {q_i}({k_+},{k_-},a)-{q_i}({k_+},{k_-}+1,a)=a{q_{i-1}}({k_+},{k_-},a)\ge0,
\]
and $\partial_a{q_i}({k_+},{k_-},a)=-{k_-}{q_{i-1}}({k_+},{k_-}-1,a)\le0$,
where $\Delta_{k_+}$ is the forward difference in $k_+$, follow by
differentiating or multiplying the generating polynomial. They show that
$q_i$ increases with $k_+$ and decreases with $k_-$ and $a$, so at every
actual component $q_i$ is at least its value at $(k_{+,0},N,1/8)$.
Thus the residue certificates prove positivity for every
actual component. Equation~\eqref{eq:qutrit-positive-initial}
applies through $i=3,5,10$ in the respective bands.
These indices are below $N/24$ and below both ${k_+},{k_-}$.

\smallskip\noindent\emph{Step 5 (Curvature and the ratio bound).}
After that interval, use the averaged Cauchy bound of
Appendix~\ref{app:qutrit-circle-factors},
with~\eqref{eq:qutrit-circle-curvature} and $s_0$ the lower end of the band.
The worst positive curvatures $s_0-\operatorname{curv}(U)$ at the radius caps
are
\[
 \begin{gathered}
 1/6-{\operatorname{curv}(1/8)}=1/294,\qquad
 3/16-{\operatorname{curv}(3/16)}=667/250000,\\
 1/4-{\operatorname{curv}(1/4)}=161/3844.
 \end{gathered}
\]
As in Step~2 of Appendix~\ref{app:qutrit-quarter},~\eqref{eq:binomial-mode}
with $2\pi Nw(1-w)\le2\pi i<44i/7$, the saving $N{\operatorname{Save}}_s$, and the
finite loss bound $3w/2$ give
\begin{equation}\label{eq:eighth-absolute-ratio}
 \frac{|\mathcal L_i|}{h(i)/h(0)}
 \le e^{-N{\operatorname{Save}}_s+3w/2}\sqrt{44i/7}\,e^{\varepsilon_i}J(V).
\end{equation}
For $w\le1/32$, the radius is uncapped, $q_u=5w/3$.
In the three bands respectively,
$V>i/25,i/14,i/6$.
These rational comparisons evaluate
$(s_0-{\operatorname{curv}(5/91)})(5/3)(1-5/96)$ at each lower $s_0$; here
$5/91$ is $u$ and $5/96$ is $q_u$ at $w=1/32$.
For all $w\le1/8$ in the third band, the weaker bound
$V>i/20$ follows from
$(1/4-{\operatorname{curv}(1/4)})(32/25)>1/20$ and
$q_u(1-q_u)/w\ge32/25$ in both capped and uncapped portions.

\smallskip\noindent\emph{Step 6 (Small $w$).}
For small $w$, the first indices after the positive range
are $4,6,11$. Here $e^{\varepsilon_i}<47/46$.
In this table the prefactor is $\sqrt{44i/7}\,e^{\varepsilon_i}J(V)$, and
the bound $J(V)\le(1+2V)^{-1/2}$ yields
\begin{center}
\begin{tabular}{cccc}\toprule
Band&First $i$&Prefactor&Saving lower bound\\\midrule
1&4&$<9/2$&$(39/100-3/1024)4>20/13$\\
2&6&$<5$&$(2/7-3/1024)6>5/3$\\
3&11&$<4$&$(2/15-3/1024)11>10/7$\\\bottomrule
\end{tabular}
\end{center}
The prefactors are checked by squaring
$\sqrt{(44i/7)/(1+2\nu i)}(47/46)$, with
$\nu\coloneqq1/25,1/14,1/6$ the lower bounds for $V/i$ of Step~5.
The exponential series through degree seven gives
$e^{20/13}>9/2$, $e^{5/3}>5$, and $e^{10/7}>4$.
For larger $i$, the exponential divided by
$\sqrt{i/(1+2\nu i)}$ increases: its logarithmic derivative
is at least ${\operatorname{save}}-3/1024-1/(2i)>0$ in each row.

\smallskip\noindent\emph{Step 7 (The range $w>1/32$).}
If $w>1/32$, then $i\ge17$ and
$e^{\varepsilon_i}<196/195$.
For the first two bands, at $i=17$,
\[
 (3/20-3/1024)17>5/2,\qquad
 \sqrt{(44/7)17}\frac{196}{195}<\frac{21}2<e^{5/2}.
\]
The exponential divided by $\sqrt i$ increases thereafter,
so $J(V)\le1$ suffices.
For the third band, $V>i/20$ and $J(V)\le(1+2V)^{-1/2}<(2V)^{-1/2}$ make
the angular prefactor uniformly
less than $\sqrt{440/7}(196/195)<8$, while
$(2/15-3/1024)17>11/5$ and $e^{11/5}>8$.
Equation~\eqref{eq:eighth-absolute-ratio} is therefore
strictly below one at every remaining index.
This proves the second row of
Proposition~\ref{prop:qutrit-four-ranges}.

\subsection{The tail from one quarter of the residual length}
\label{app:qutrit-tail}

We prove the fourth row of Proposition~\ref{prop:qutrit-four-ranges}: for a
minimizing vertex as in that proposition, $|\mathcal L_i|<h(i)/h(0)$ for
$\lceil N/4\rceil\le i\le D-1$, for every $r$. This row is proved before
row~3 (Appendix~\ref{app:qutrit-middle}), whose proof reuses its notation and
its bound for $N/4\le i\le N/2$. Both ranges use the one-coefficient bound
of Lemma~\ref{lem:schur-coefficient} (Step~1). For $N/4\le i\le N/2$ the
coefficient is compared with the product bound~\eqref{eq:qutrit-product-simplified}
(Steps~2--4); for $i\ge N/2$, with distinct-root packing, through a saddle
point estimate certified by a polynomial (Steps~5--7). Step~8 shows that
the resulting bounds are below one for $N\ge512$.

\smallskip\noindent\emph{Step 1 (Set-up and the extraction bound).}
Assume $\lceil N/4\rceil\le i\le D-1$ and put
\[
 {D_{\rm cap}}\coloneqq\left\lfloor\frac{7(N-2m+1)}9\right\rfloor,
 \quad R\coloneqq{D_{\rm cap}}-2m,\quad S\coloneqq N/m-2,\quad a\coloneqq{D_{\rm cap}}/m.
\]
Here $R$ is the integer of Lemma~\ref{lem:schur-coefficient}, chosen so that
one bound serves every terminal degree $r\le R$; $S$ and $a$ normalize by
$m$, as the analysis for $i\ge N/2$ requires ($S$ is unrelated to the deficit
bounds of Appendix~\ref{app:higher-details}). By~\eqref{eq:qutrit-wedge},
$D\le{D_{\rm cap}}$, hence $r\le R$, and ${a_{\min}}>7(N-2m+1)/9\ge{D_{\rm cap}}$,
hence ${a_{\min}}-1\ge {D_{\rm cap}}$. Moreover
\begin{equation}\label{eq:qutrit-floor-offset}
 a=\frac{7S}9+\frac\phi m,\qquad-\frac29<\phi\le\frac79,
\end{equation}
where $\phi\coloneqq7/9-\{7(N-2m+1)/9\}$, with $\{\cdot\}$ the fractional part;
this offset is used to choose the regions of Step~6.
Set ${\operatorname{pref}_0}\coloneqq8(m+10)/7$.
Lemmas~\ref{lem:schur-coefficient} and
\ref{lem:qutrit-finite-norm} give, for every $u>0$,
\begin{equation}\label{eq:qutrit-tail-extraction}
 |[v^i](1+v)^rG(v)|
 \le {\operatorname{pref}_0}u^{-i}\frac{(1+u)^{N+R}}{(1+7u/8)^{N-2m}}
          \frac{\sqrt\pi}{2\sqrt{2V}},\quad
 V\coloneqq(N-m)\frac\theta8\left(1-\frac\theta8\right),
 \quad\theta\coloneqq\frac u{1+u}.
\end{equation}
Here $r$ is replaced by $R\ge r$ only in the nonnegative matrix
majorant of Lemma~\ref{lem:schur-coefficient}.
We call $|\mathcal L_i|/(h(i)/h(0))$ the relative coefficient.

\smallskip\noindent\emph{Step 2 (Indices between $N/4$ and $N/2$: the ratio bound).}
Choose $u\coloneqq w/(2/3-w)$, so $\theta=3w/2$.
Then $1/4\le w\le1/2<49/81$, within the domain $0<w<49/81$
of~\eqref{eq:qutrit-product-simplified}. Put
${\operatorname{LogBud}_{\rm cap}}\coloneqq{\operatorname{LogBud}_9}(2/9)$, the constant bounded by
${\operatorname{LogUp}}$ in Appendix~\ref{app:qutrit-quarter}, and
\begin{align}
 {\operatorname{Num}_x}(w)&\coloneqq(1-x)\left(\frac{16}9\log(1+u)-\log(1+7u/8)\right)
                                     -w\log u,\notag\\
 F_x(w)&\coloneqq H_2(w)+x\log(1-w)-(9x/2){\operatorname{LogBud}_{\rm cap}}\kappa(w)-{\operatorname{Num}_x}(w).
                                      \label{eq:qutrit-tail-gap}
\end{align}
Here $N{\operatorname{Num}_x}(w)+(7/9)\log(1+u)$ bounds the logarithm of the
evaluated bound in~\eqref{eq:qutrit-tail-extraction} after removing
both ${\operatorname{pref}}_0$ and the final angular factor: since ${D_{\rm cap}}\le7(N-2m+1)/9$, $N+R\le(16/9)(N-2m)+7/9$, and
$N-2m=(1-x)N$. The exponent $F_x$ (the gap) must be positive. Dividing~\eqref{eq:qutrit-tail-extraction}
by $\binom Ni$, bounded below by~\eqref{eq:binomial-three}, and by $h(i)/h(0)$,
bounded below by~\eqref{eq:qutrit-product-simplified}, shows that the
relative coefficient is at most
\begin{equation}\label{eq:qutrit-middle-tail-ratio}
 3{\operatorname{pref}_0}\sqrt{\frac{\pi i}{8V}}
 \exp\left(-NF_x(w)+\frac79\log(1+u)
                         +\log(16/7)\kappa(w)\right).
\end{equation}
The cap $m\le N/9$ gives $N-m\ge8N/9$, and with $w\le1/2$ it gives
$V\ge29i/192$; so the square-root factor apart from ${\operatorname{pref}_0}$ is
at most $3\sqrt{24\pi/29}<5$.
The finite loss, the part of the exponent not proportional to $N$,
consists of $(7/9)\log(1+u)\le(14/9)\log2$, since $u\le3$, and
$\log(16/7)\kappa(w)\le\log(16/7)\kappa(1/2)$, with $\kappa(1/2)=49/17$;
using $\log2<7/10$ and $\log(16/7)<5/6$ (certified in Step~8), it is less
than
\[
 \frac{49}{45}+\frac{245}{102}<\frac72.
\]

\smallskip\noindent\emph{Step 3 (Positivity of the gap at $x=0$).}
We now prove $F_x(w)>1/50$ on the rectangle $0<x\le2/9$, $1/4\le w\le1/2$.
The expression is affine in $x$, so it suffices to treat $x=0$ (this step)
and $x=2/9$ (Step~4).
Define
\begin{align*}
 F_0(w)&\coloneqq H_2(w)-\tfrac{16}9\log(1+u)+\log(1+7u/8)+w\log u,\\
 C(w)&\coloneqq\tfrac{16}9\log(1+u)-\log(1+7u/8)+\log(1-w),\\
 f(w)&\coloneqq F_0(w)+\tfrac29C(w)-\tfrac1{25}\kappa(w).
\end{align*}
Since ${\operatorname{LogBud}_{\rm cap}}=2/9-(14/9)\log(9/8)$ and $\log(9/8)\ge2/17$,
one has ${\operatorname{LogBud}_{\rm cap}}\le2/51<1/25$ and hence $F_{2/9}>f$.
Direct differentiation gives
\[
 F_0'(w)=\log\frac{3(1-w)}{2-3w}
              -\frac1{3(2-3w)}-\frac3{16-3w}.
\]
The lower bound $\log c\ge2(c-1)/(c+1)$ for $c\ge1$, at
$c=3(1-w)/(2-3w)$, reduces positivity to
${\operatorname{Poly}_1}(w)\coloneqq22+30w-126w^2>0$.
Its cleared coefficients under $w=(1+2t)/(4(1+t))$, which maps
$0\le t<\infty$ onto $1/4\le w<1/2$, are $(346,560,88)$.
So $F_0$ increases, and a directed logarithm calculation using
\eqref{eq:rational-log-series} with $k=40$ gives
$F_0(1/4)-1/50>1/1000$; this settles $x=0$.

\smallskip\noindent\emph{Step 4 (Positivity at $x=2/9$ by concavity of $f'$).}
For the endpoint $x=2/9$, where $F_{2/9}>f$, set
${\delta_1}\coloneqq1-w$, ${\delta_2}\coloneqq2-3w$, ${\delta_3}\coloneqq16-3w$, ${\delta_4}\coloneqq49-81w$.
All are positive on $[1/4,1/2]$. Direct differentiation of $f$ gives
\[
 -f'''(w)=\frac1{\delta_1^2}+\frac4{9\delta_1^3}+\frac3{25\delta_1^4}
 -\frac9{\delta_2^2}-\frac{10}{3\delta_2^3}+\frac{42}{\delta_3^3}
 +\frac{3\cdot49\cdot81^3}{25\,\delta_4^4},
\]
and, with $\delta_2=3\delta_1-1$ and $1-2w=2\delta_1-1$,
the polynomial
\begin{align}
 {\operatorname{Poly}_3}\coloneqq{}&225{\delta_1}^2{\delta_2}{\delta_3}^3{\delta_4}^4
 -1350(1-2w){\delta_1}^3{\delta_3}^3{\delta_4}^4
 -2100{\delta_1}^4{\delta_3}^3{\delta_4}^4\notag\\
 &+9450{\delta_1}^4{\delta_2}^3{\delta_4}^4+100{\delta_1}{\delta_2}^3{\delta_3}^3{\delta_4}^4
 +27{\delta_2}^3{\delta_3}^3{\delta_4}^4
 +27\cdot49\cdot81^3{\delta_1}^4{\delta_2}^3{\delta_3}^3
                      \label{eq:qutrit-third-polynomial}
\end{align}
equals $225{\delta_1}^4{\delta_2}^3{\delta_3}^3{\delta_4}^4(-f'''(w))$.
Under the same substitution, the coefficients of
$4^{11}(1+t)^{11}{\operatorname{Poly}_3}(w)$ are
\[
\begin{array}{r}
465694333819955250,\quad10264575742405337800,\\
56688383984862941320,\quad155855251140497110656,\\
258920596005162578496,\quad282374438699290851072,\\
209772949617663227136,\quad107169365861600888832,\\
37120525788202363392,\quad8336157859983722496,\\
1095646242853169152,\quad63998126164541440.
\end{array}
\]
Thus $f'''<0$ and $f'$ is strictly concave.
For the derivative calculation use
\[
 C'=\frac7{3(2-3w)}+\frac3{16-3w}-\frac1{1-w},\qquad
 \kappa'=\frac12\left(\frac1{(1-w)^2}
                  +\frac{49/81}{(49/81-w)^2}\right).
\]
The same directed logarithm rule, \eqref{eq:rational-log-series} with
$k=40$, gives
\[
 f'(1/4)>3/20,\qquad
 f(1/4)-1/50>1/125,\qquad
 f(1/2)-1/50>37/1000.
\]
Concavity of $f'$ and its positive initial value imply that
$f$ first increases and then may decrease; its minimum is
at an endpoint. Consequently $f>1/50$, and affine
interpolation between $x=0$ and $x=2/9$ proves $F_x>1/50$.
With the square-root factor below $5$ and the finite loss below $7/2$
(Step~2), equation~\eqref{eq:qutrit-middle-tail-ratio} is therefore
strictly less than $5{\operatorname{pref}_0}e^{7/2-N/50}$.

\smallskip\noindent\emph{Step 5 (Indices at least $N/2$: packed entropy and the saddle).}
Here we use distinct-root packing and retain the exact floor ${D_{\rm cap}}$:
the product bound~\eqref{eq:qutrit-product-simplified} holds only for
$w<49/81$, while $D-1$ can be close to $7N/9>49N/81$.
Since ${a_{\min}}-1\ge {D_{\rm cap}}$, the paired roots give
$h(i)/h(0)\ge({D_{\rm cap}})_i/({D_{\rm cap}}+2m)_i$.
The logarithm of this product is
$\sum_{j=0}^{i-1}\log(({D_{\rm cap}}-j)/({D_{\rm cap}}+2m-j))$.
The summand decreases, so its left Riemann sum is at least
the integral through $i$, which with $z\coloneqq i/m$ equals
$m\int_0^z\log((a-y)/(a+2-y))\,dy
=m[(a+2-z)H_2(2/(a+2-z))-(a+2)H_2(2/(a+2))]$.
Together with~\eqref{eq:binomial-three}, written as
$\binom Ni\ge e^{m(S+2)H_2(z/(S+2))}/(3\sqrt i)$, this proves
\begin{equation}\label{eq:qutrit-packed-entropy}
 \binom Ni\frac{h(i)}{h(0)}
 \ge\frac{e^{mE_a(i/m)}}{3\sqrt i},
\end{equation}
where
\[
 E_a(z)\coloneqq(S+2)H_2\!\left(\frac z{S+2}\right)
 +(a+2-z)H_2\!\left(\frac2{a+2-z}\right)
 -(a+2)H_2\!\left(\frac2{a+2}\right).
\]
For the numerator define
\[
 {\operatorname{Num}_a}(z)\coloneqq\inf_{u>0}
       \left((S+a)\log(1+u)-S\log(1+7u/8)-z\log u\right);
\]
since $N+R=m(S+a)$ and $N-2m=mS$, the bracket is $1/m$ times the
logarithm of the evaluated bound in~\eqref{eq:qutrit-tail-extraction}
after removing both ${\operatorname{pref}}_0$ and the final angular factor.
For $2<a<S$, $7a>S$, and $0<z<a$, the infimum is attained at a unique
positive saddle point $u_K$, which satisfies
\[
 M(u_K)=z,\qquad
 M(u)\coloneqq u\left(\frac{S+a}{1+u}-\frac{7S/8}{1+7u/8}\right).
\]
Indeed $M$ increases: since $(1+u)/(1+7u/8)\le8/7$,
$M'(u)\ge(a-S/7)/(1+u)^2>0$; and $M(0)=0$, $M(u)\to a$ as $u\to\infty$.
By the envelope identity, ${\operatorname{Num}_a}'(z)=-\log u_K$. The packed
entropy $E_a$ satisfies $E_a'(z)=-\log u_E$, which defines
\[
 u_E\coloneqq\frac{z(a+2-z)}{(S+2-z)(a-z)}.
\]

\smallskip\noindent\emph{Step 6 (The saddle-gap claim and its certificate).}
We claim
\begin{equation}\label{eq:qutrit-saddle-gap}
 E_a(z)-{\operatorname{Num}_a}(z)>z\log(103/100)\qquad(0<z<a)
\end{equation}
on each of the following closed regions of the parameters $(S,a)$:
\[
\begin{array}{ll}
 S\ge7,&7S/9-2/(9\cdot16)\le a\le7S/9+7/(9\cdot16),\\
 S\ge16,&7S/9-2/(9\cdot2)\le a\le7S/9+7/(9\cdot2).
\end{array}
\]
They contain every actual parameter, since $a=7S/9+\phi/m$ with
$-2/9<\phi\le7/9$ by~\eqref{eq:qutrit-floor-offset} and $S\ge7$ by
$N\ge9m$: use the first for $m\ge16$, and the second when $2\le m\le15$,
since $N\ge512$ then implies $S>16$.
Both give $2<a<S$ and $7a>S$.

Here is the complete polynomial specification of the claim.
For $(m_0,S_0)=(16,7)$ or $(2,16)$, introduce
$z_0,t,w_0\ge0$ and set
\begin{align*}
 S&\coloneqq S_0+z_0,&q&\coloneqq9m_0(1+t),\\
 \alpha&\coloneqq7m_0S(1+t)+9t-2(1+t),&
 a&\coloneqq\alpha/q,\qquad y\coloneqq\frac{\alpha w_0}{q(1+w_0)},\\
 Y&\coloneqq\alpha w_0,&V_0&\coloneqq q(S+2)(1+w_0)-\alpha,\\
 J_0&\coloneqq Y+2q(1+w_0),&
 {\operatorname{Aux}_0}&\coloneqq7\alpha-Sq-(8\alpha+Sq)w_0.
\end{align*}
These symbols are local to the specification. The parameter $t$ moves the
offset $a-7S/9=(9t/(1+t)-2)/(9m_0)$ over $[-2/(9m_0),7/(9m_0))$; $w_0$
gives $y=aw_0/(1+w_0)\in[0,a)$, so that $z=a-y\in(0,a]$; and
$z_0=S-S_0\ge0$. The polynomial
\begin{equation}\label{eq:qutrit-saddle-certificate}
 \operatorname{Cert}\coloneqq8\cdot100^2YV_0^2+103\cdot100{\operatorname{Aux}_0}J_0V_0
                         -7\cdot103^2\alpha J_0^2
\end{equation}
has $60$ nonzero positive integer coefficients in each region;
their smallest values are $13031424000$ and $25452000$.
The complete $120$ coefficients are listed in the certificate
package~\cite{ComputationalCertificates}.
The degree in $t$ is three, with a positive coefficient
at $z_0^0t^3w_0^0$, so the upper offset endpoint obtained
as $t\to\infty$ is included.

To verify why this polynomial suffices, put
$\Lambda\coloneqq103/100$ and ${J_{\rm den}}\coloneqq S+2-a+y$, which is $S+2-z$, the factor
of the denominator of $u_E$.
The expression
\[
 {\operatorname{Poly}(y)}\coloneqq y{J_{\rm den}}^2+\Lambda\left(\frac{7a-S}{8}-\frac{15y}{8}\right)
                 (y+2){J_{\rm den}}-\frac78\Lambda^2(a-y)(y+2)^2
\]
becomes~\eqref{eq:qutrit-saddle-certificate} after multiplication
by $8\cdot100^2q^3(1+w_0)^3>0$.
With $z=a-y$, direct substitution gives
\[
 z-M(\Lambda u_E)
  =\frac{z{\operatorname{Poly}(y)}}{{J_{\rm den}}^2y(1+\Lambda u_E)(1+7\Lambda u_E/8)}>0.
\]
Since $M$ increases, this gives $u_K>\Lambda u_E$.
Both functions vanish at zero ($E_a(0)=0$, and ${\operatorname{Num}_a}(0)=0$
because the bracket is nonnegative at $z=0$ and tends to $0$ as $u\to0$),
and $E_a'-{\operatorname{Num}_a}'=\log(u_K/u_E)>\log\Lambda$; integration proves
\eqref{eq:qutrit-saddle-gap}.

\smallskip\noindent\emph{Step 7 (The ratio bound for $i\ge N/2$).}
Both regions also give $a\le13S/16$, since $S\ge7$ in the first and
$S\ge16$ in the second.
Take $u\coloneqq u_K$ in~\eqref{eq:qutrit-tail-extraction}. Writing
$\theta=u_K/(1+u_K)$, the saddle equation $M(u_K)=z=i/m$ reads
\[
 i=m\theta\left(a+\frac S8\frac{1-\theta}{1-\theta/8}\right)
       \le\frac{15mS\theta}{16}.
\]
Meanwhile $V\ge7m(S+1)\theta/64$, hence $V/i\ge7/60$, and the square-root
factor is at most $3\sqrt{60\pi/56}<6$.
Equations~\eqref{eq:qutrit-tail-extraction},
\eqref{eq:qutrit-packed-entropy}, and
\eqref{eq:qutrit-saddle-gap} bound the relative coefficient by
\[
 6{\operatorname{pref}_0}(103/100)^{-i}\le6{\operatorname{pref}_0}e^{-3N/206},
\]
using $i\ge N/2$ and $\log(103/100)\ge3/103$.
The bound $V/i\ge7/60$ holds at every interior index $i\le D-1$, so the
prefactor does not grow as $i$ approaches ${D_{\rm cap}}$.

\smallskip\noindent\emph{Step 8 (The finite length threshold).}
Since $m\le N/9$, the bounds of Steps~4 and~7, for $N/4\le i\le N/2$ and
for $i\ge N/2$, are at most
\[
 T_1(N)\coloneqq\frac{40}7(N/9+10)e^{7/2-N/50},\qquad
 T_2(N)\coloneqq\frac{48}7(N/9+10)e^{-3N/206}.
\]
Their logarithmic derivatives are
$1/(N+90)-1/50$ and $1/(N+90)-3/206$, both negative
for $N\ge512$.
At $512$, their prefactors are $24080/63$ and $9632/21$,
and the positive exponents to absorb are $337/50>6$
and $768/103>7$.
The rational Taylor sums
\[
 \sum_{j=0}^{11}\frac{6^j}{j!}=\frac{760997}{1925}>\frac{24080}{63},
 \qquad
 \sum_{j=0}^{7}\frac{7^j}{j!}=\frac{47273}{72}>\frac{9632}{21}
\]
prove both bounds below one.
The constants $\log2<7/10$ and $\log(16/7)<5/6$ of Step~2 follow similarly
from
\[
 \sum_{j=0}^{5}\frac{(7/10)^j}{j!}
       =\frac{24162857}{12000000}>2,\qquad
 \sum_{j=0}^{5}\frac{(5/6)^j}{j!}
       =\frac{429319}{186624}>16/7.
\]
The intervals $N/4\le i\le N/2$ and $i\ge N/2$
cover every integer from $\lceil N/4\rceil$ to $D-1$,
including odd residual lengths. This proves the fourth row, with
$|\mathcal L_i|$ on the left.

\subsection{The middle interval for smaller terminal degree}
\label{app:qutrit-middle}

We prove the third row of Proposition~\ref{prop:qutrit-four-ranges}: for a
minimizing vertex as in that proposition with $r\le N/3$,
$|\mathcal L_i|<h(i)/h(0)$ for
$\lceil N/8\rceil\le i\le\min(\lfloor N/4\rfloor,D-1)$. The argument
follows Step~2 of Appendix~\ref{app:qutrit-tail}, with the exact terminal
degree in place of $R$.

\smallskip\noindent\emph{Step 1 (The ratio bound).}
Assume $r\le N/3$ and
$\lceil N/8\rceil\le i\le\min(\lfloor N/4\rfloor,D-1)$.
Use $u\coloneqq w/(2/3-w)$ again, but keep $R\coloneqq r$ in
Lemma~\ref{lem:schur-coefficient}, so that the hypothesis $r\le N/3$ is
used instead of $r\le{D_{\rm cap}}-2m$. Its real numerator, the evaluated bound
$u^{-i}(1+u)^{N+r}/(1+7u/8)^{N-2m}$, is at most the value with exponent
$4N/3$ on $1+u$. Since $4N/3$ need not be an integer, this compares
positive values at $u$, not coefficients.
The same binomial and product bounds give a relative coefficient
at most
\[
 3{\operatorname{pref}_0}\sqrt{\frac{\pi i}{8V}}
       e^{-N\widetilde F_x(w)+\log(16/7)\kappa(w)},
\]
where now the gap is
\[
 \begin{aligned}
 \widetilde F_x(w)&\coloneqq H_2(w)+x\log(1-w)-(9x/2){\operatorname{LogBud}_{\rm cap}}\kappa(w)\\
       &\quad-\frac43\log(1+u)+(1-x)\log(1+7u/8)+w\log u.
 \end{aligned}
\]
Again $V\ge29i/192$ and the prefactor is less than $5{\operatorname{pref}_0}$.
Since the exponent $4N/3$ has no additive offset, the finite loss is only
$\log(16/7)\kappa(w)$, which is less than $(5/6)\kappa(1/4)=179/414<1/2$.

\smallskip\noindent\emph{Step 2 (Reduction to one function).}
Replace ${\operatorname{LogBud}_{\rm cap}}$ by $1/25$ to lower the gap $\widetilde F_x$.
Its coefficient of $x$ is then
$\log(1-w)-\log(1+7u/8)-(9/50)\kappa(w)<0$, so $x=2/9$ is the worst case.
It therefore suffices to show that
\[
 \tilde f(w)\coloneqq H_2(w)+\frac29\log(1-w)-\frac1{25}\kappa(w)
       -\frac43\log(1+u)+\frac79\log(1+7u/8)+w\log u
\]
is greater than $1/50$ on $[1/8,1/4]$.

\smallskip\noindent\emph{Step 3 (Monotonicity).}
The derivative of $\tilde f$ is
\begin{align*}
 \tilde f'(w)={}&\log\frac{3(1-w)}{2-3w}+\frac1{3(2-3w)}
  -\frac7{3(16-3w)}-\frac2{9(1-w)}\\
 &-\frac1{50}\left(\frac1{(1-w)^2}
                       +\frac{49/81}{(49/81-w)^2}\right).
\end{align*}
Bound the logarithm below by $2/(5-6w)$, which is the bound
$\log c\ge2(c-1)/(c+1)$ of Appendix~\ref{app:qutrit-tail}, Step~3, at
$c=3(1-w)/(2-3w)$.
Using ${\delta_1},{\delta_2},{\delta_3},{\delta_4}$ of
Appendix~\ref{app:qutrit-tail} and ${\delta_5}\coloneqq5-6w$,
the positive denominator $450{\delta_5}{\delta_1}^2{\delta_2}{\delta_3}{\delta_4}^2$
clears the resulting lower bound to
\begin{align*}
 {\operatorname{Poly}_{\rm mid}}\coloneqq{}&900{\delta_1}^2{\delta_2}{\delta_3}{\delta_4}^2
 +150{\delta_5}{\delta_1}^2{\delta_3}{\delta_4}^2-1050{\delta_5}{\delta_1}^2{\delta_2}{\delta_4}^2\\
 &-100{\delta_5}{\delta_1}{\delta_2}{\delta_3}{\delta_4}^2-9{\delta_5}{\delta_2}{\delta_3}{\delta_4}^2
 -9\cdot3969{\delta_5}{\delta_1}^2{\delta_2}{\delta_3}.
\end{align*}
Under $w=(1+2t)/(8(1+t))$, which maps $0\le t<\infty$ onto $1/8\le w<1/4$,
the coefficients of $8^6(1+t)^6{\operatorname{Poly}_{\rm mid}}(w)$ are
\[
\begin{array}{r}
2853982080100,\quad14411376857744,\quad30163349021104,\\
33502905607552,\quad20831128307648,\quad6875444613376,\\
941179096320.
\end{array}
\]
All are positive, including the upper endpoint coefficient,
so $\tilde f'>0$.

\smallskip\noindent\emph{Step 4 (The value at $w=1/8$ and conclusion).}
The value of $\tilde f$ at $w=1/8$ is
\[
 -\frac{47}{72}\log(7/8)-\frac43\log(16/13)
 +\frac79\log(125/104)+\frac18\log(24/13)
 -\frac{439}{2177\cdot25}.
\]
Formula~\eqref{eq:rational-log-series} with $k=40$ proves
$\tilde f(1/8)-1/50>1/600$; the rational interval is listed in the
certificate package~\cite{ComputationalCertificates}.
Thus the relative coefficient is less than
$5{\operatorname{pref}_0}e^{1/2-N/50}<5{\operatorname{pref}_0}e^{7/2-N/50}\le T_1(N)$,
and $T_1(N)<1$ for $N\ge512$ by Step~8 of Appendix~\ref{app:qutrit-tail}.
This proves the third row of
Proposition~\ref{prop:qutrit-four-ranges}, with $|\mathcal L_i|$ on the left.

\section{The finite qutrit certificates}\label{app:qutrit-finite}

This appendix specifies the exact data and checks used in
Proposition~\ref{prop:qutrit-finite}: for every $15\le H\le507$ and every
degree $D$ of~\eqref{eq:qutrit-finite-domain}, an explicit global minimizer
of~\eqref{eq:beta} at $(n,D)=(H+D,D)$, with every coefficient
comparison~\eqref{eq:residual-comparison} verified exactly. The
certificate has two parts. For each $H$, positive cardinal quadrature
certifies an optimal polynomial of the cap problem, of length $H+2a$ and
degree $2a$ (Appendix~\ref{app:qutrit-finite-cap}). Integer deficits, that
is, exact integer differences of the two sides of each comparison
(unrelated to the deficit $s=N-r$ of Appendix~\ref{app:higher-details}),
then verify every required coefficient of each original terminal lift
(Appendix~\ref{app:qutrit-finite-deficits}). Discovery of the
roots is separate from verification of these identities. The construction
parallels the certificate for $p=4$ of
Appendix~\ref{app:four-finite-certificate}, and the same names are used
for the same objects where possible.

\subsection{Cap polynomials and positive quadrature}
\label{app:qutrit-finite-cap}

\smallskip\noindent\emph{Step 1 (The cap minimizer, certified by positive quadrature).}
For each integer $H=15,\ldots,507$, put
\[
 a\coloneqq\left\lfloor\frac{H-1}7\right\rfloor,\qquad N_a\coloneqq H+2a,
\]
the cap index of Lemma~\ref{lem:cap-minimum-identity} at $Q=9$ and the cap
length. For each $H$ the certificate package~\cite{ComputationalCertificates}
lists $2a$ distinct roots
$Z_H\subseteq\{1,\ldots,N_a\}$, which define the monic polynomial
${h_H^{\rm cap}}(x)\coloneqq\prod_{z\in Z_H}(x-z)$.
The verification checks ${h_H^{\rm cap}}(0)>0$ and ${h_H^{\rm cap}}(x)\ge0$ on every
integer $0\le x\le N_a$. At the nodes $\mathcal Z_H\coloneqq\{0\}\cup Z_H$,
the rational cardinal weights are
\begin{equation}\label{eq:qutrit-cardinal-data}
 \lambda_z\coloneqq9^{-N_a}\sum_{x=0}^{N_a}\binom{N_a}x8^x
                   \prod_{v\in\mathcal Z_H\setminus\{z\}}
                        \frac{x-v}{z-v}.
\end{equation}
Every weight is positive. The following identities are checked
with rational arithmetic through the full cap degree:
\begin{equation}\label{eq:qutrit-cardinal-moments}
 \sum_{z\in\mathcal Z_H}\lambda_z(z)_j
                    =(N_a)_j(8/9)^j,\qquad0\le j\le2a.
\end{equation}
For exact cardinal weights these identities hold by interpolation;
verifying them checks the recorded weights.
Falling factorials form a basis for this polynomial space, so the
quadrature with nodes $\mathcal Z_H$ and weights $\lambda_z$ is exact for
$\E_{\mu_{N_a}}$ on all polynomials of degree at most $2a$. Since the
weights are positive, every such polynomial that is nonnegative on the grid
and equals one at zero has expectation at least $\lambda_0$, while
${h_H^{\rm cap}}/{h_H^{\rm cap}}(0)$, feasible by the check above, attains
equality. This proves $\beta_{N_a,2a,9}=\lambda_0$ directly; it is the
converse in Lemma~\ref{lem:positive-quadrature}.

\smallskip\noindent\emph{Step 2 (Terminal merging and global minimality of the lift).}
Remove the maximal consecutive block of roots ending at $N_a$,
and let its length be ${\ell_{\rm end}}$. As in
Appendix~\ref{app:four-finite-certificate}, the pairing of the roots of the
feasible polynomial ${h_H^{\rm cap}}$ makes ${\ell_{\rm end}}$ even and the remaining
roots distinct adjacent pairs; the certificate also verifies both facts. Put
\[
 d_0\coloneqq2a-{\ell_{\rm end}},\quad m\coloneqq d_0/2,\quad N\coloneqq N_a-{\ell_{\rm end}}=H+d_0,
 \quad h_H(x)\coloneqq\prod_{z\text{ remaining}}(x-z).
\]
For each original degree in~\eqref{eq:qutrit-finite-domain},
set $r\coloneqq D-d_0\ge0$ and $n=N+r=H+D$.
Since $n-(D-2a)=N_a$, the identity
\[
 (n-x)_{D-2a}(N_a-x)_{{\ell_{\rm end}}}=(n-x)_{D-d_0}
\]
shows that merging this terminal block preserves the normalized
polynomial at the original $(n,D)$: the terminal lift of
${h_H^{\rm cap}}$ by the factor of degree $D-2a$, normalized at zero, is
$\binom{n-x}r h_H(x)/(\binom nrh_H(0))$, of the
form~\eqref{eq:vertex-factorization} with residual $h_H$.
The corresponding exact objective identity is checked as well:
\begin{equation}\label{eq:qutrit-merge-objective}
 9^{d_0}\frac{\E_{\mu_N}h_H}{h_H(0)}
                 =9^{2a}\beta_{N_a,2a,9}.
\end{equation}
Lemma~\ref{lem:cap-minimum-identity} consequently identifies
every such lift as a global minimum at that $(n,D)$; it applies because
$Q=9$, $D\ge2a$, and $n=H+D\ge2D+1$, as $D\le\lfloor7(H+1)/9\rfloor\le H-1$.
This is a statement about the supplied representative;
no uniqueness of the minimizing face is required, because
Lemma~\ref{lem:coefficient-criterion} needs only one global minimizer
whose comparisons hold, and it then gives $K\le3^n\beta_{n,D,9}$.

\smallskip\noindent\emph{Step 3 (Residual root checks).}
Let $Z_{\mathrm{res}}$ be the set of remaining roots in $h_H$.
For each nonempty residual profile the data also verify
$N\ge9m+1$ and $9\min Z_{\mathrm{res}}>7(H+1)$, the cap~\eqref{eq:optimizer-cap}
and the root barrier (the second inequality of~\eqref{eq:qutrit-wedge}) of
a minimizing vertex. The first is a consistency check of the data. The
second gives $h_H(i)>0$ for every required $1\le i\le D$, so the ratios
reported in Appendix~\ref{app:qutrit-finite-domain} are defined.
The residual $N$ of this representative can exceed $511$: only $H$ and $D$
are taken from the selected vertex in the proof of
Theorem~\ref{thm:qutrit-weak}, and the representative's residual length
$N=H+d_0$ depends on its own root structure.
All such cases are included in the direct checks below; unlike
Appendix~\ref{app:four-finite-certificate}, no profile or lift is routed to
another result, except the one-pair profile $H=15$.

\subsection{Integer coefficient deficits}
\label{app:qutrit-finite-deficits}

For every residual profile of degree $d_0\ge4$, form the integer moments
(integers, not the matrices $M_\ell$ of Section~\ref{sec:coefficient-methods})
\begin{equation}\label{eq:qutrit-integer-moments}
 M_\ell\coloneqq\sum_{x=0}^N\binom Nx8^xh_H(x)\Kraw{\ell}{N}(x),
 \qquad0\le\ell\le d_0.
\end{equation}
These are integers. One exact evaluation uses the recurrence of the
Krawtchouk polynomials $P_\ell\coloneqq\Kraw{\ell}{N}$ at $Q=9$:
\[
 P_0=1,\quad P_{-1}=0,\qquad
 \ell P_\ell(x)=[8N-9x-7(\ell-1)]P_{\ell-1}(x)
                       -8(N-\ell+2)P_{\ell-2}(x).
\]
Every division by $\ell$ is exact and $M_0>0$.
Let
\[
 g_H\coloneqq\gcd_{0\le\ell\le d_0}|M_\ell8^{d_0-\ell}|,
 \qquad f_\ell\coloneqq M_\ell8^{d_0-\ell}/g_H,
 \qquad {a_H^{\rm norm}}\coloneqq f_0>0.
\]
Orthogonality gives $[u^\ell]G(u)=f_\ell/{a_H^{\rm norm}}$, that is,
$G_\ell=M_\ell/(8^\ell M_0)$, as in Appendix~\ref{app:four-finite-certificate}
(where the same quantities appear without the reduction by $g_H$).
For every integer
\begin{equation}\label{eq:qutrit-all-lifts}
 2a\le D\le\left\lfloor\frac{7(H+1)}9\right\rfloor,
 \quad r=D-d_0,\quad1\le i\le D,
\end{equation}
compute
\begin{align}
 {\operatorname{Int}_i}^{(r)}&\coloneqq\sum_{\ell=0}^{d_0}\binom r{i-\ell}f_\ell,
       \label{eq:qutrit-integer-lift}\\
 \Delta_i&\coloneqq{a_H^{\rm norm}}\binom Ni h_H(i)-h_H(0){\operatorname{Int}_i}^{(r)}.
       \label{eq:qutrit-integer-deficit}
\end{align}
All these deficits are strictly positive.
Their sign is precisely the residual coefficient comparison,
because
\[
 \mathcal L_i=\frac{{\operatorname{Int}_i}^{(r)}}{{a_H^{\rm norm}}\binom Ni}.
\]
Multiplying by $\binom Ni/\binom ni>0$ gives the original
coefficient comparison at the same $n,D$.
No positivity of ${\operatorname{Int}_i}^{(r)}$ is needed.
Every index $1\le i\le D$ is tested, so
Proposition~\ref{prop:original-top}, which supplies $i=D$ in
Theorem~\ref{thm:qutrit-large-length}, is not needed here. Every degree
$D$ of~\eqref{eq:qutrit-all-lifts} is checked directly, without removing
the distance strips.
Successive terminal degrees may be evaluated efficiently by
${\operatorname{Int}_i}^{(r+1)}={\operatorname{Int}_i}^{(r)}+{\operatorname{Int}_{i-1}}^{(r)}$; this is exactly
Pascal's identity, with zero coefficients outside the support.

The only residual degree-two profile is $H=15$.
Its cap roots are $15,16,18,19$, and its merged polynomial
is $(x-15)(x-16)$ at $N=17$.
For $D=4,\ldots,12$, Proposition~\ref{prop:one-pair-minimum} gives the
optimal target, as used in Proposition~\ref{prop:qutrit-finite}; it applies
because the lift has one residual pair and the root-region condition
$9D\le7(H+1)=112$ holds (it is needed since $H=15>2Q-4$).
There are no degree-zero profiles.

\subsection{Complete domain}
\label{app:qutrit-finite-domain}

\smallskip\noindent\emph{Step 1 (Domain totals).}
The exact totals provide checks on domain generation:
\begin{center}
\begin{tabular}{lr}\toprule
Quantity&Count\\\midrule
Cap profiles&493\\
Cardinal weights and moments&36,695\\
Profiles checked by integer deficits&492\\
Original lifts checked directly&64,526\\
Strictly positive integer deficits&11,543,154\\
One-pair lifts&9\\
Canonical profiles with residual $N>511$&106\\\bottomrule
\end{tabular}
\end{center}
Here $36{,}695=\sum_H(2a+1)$ is the number of cardinal weights, and also of
moment identities; $492=493-1$ excludes the one-pair profile $H=15$; and the
canonical profiles are the residual profiles of the certified
representatives.
The lift and deficit totals are independently specified by
\begin{align*}
 64526&=\sum_{H=16}^{507}
 \left(\left\lfloor\frac{7(H+1)}9\right\rfloor
                  -2\left\lfloor\frac{H-1}7\right\rfloor+1\right),\\
 11543154&=\sum_{H=16}^{507}
       \sum_{D=2\lfloor(H-1)/7\rfloor}^{\lfloor7(H+1)/9\rfloor}D.
\end{align*}
The maximum cap length is $651$ and maximum cap degree is $144$.
The merged residual length is at most $649$ and $m\le71$.
Among the directly checked indices the largest exact value of the ratio
$\mathcal L_i/(h(i)/h(0))$, that is, of $h_H(0){\operatorname{Int}_i}^{(r)}$ divided
by ${a_H^{\rm norm}}\binom Nih_H(i)$, is
$4041/5225<1$, at $(H,N,m,r,n,D,i)=(21,25,2,13,38,17,1)$.
Its location is a result of checking all indices, rather than
an assumed ordering of the coefficient ratios.

\smallskip\noindent\emph{Step 2 (Data and verification).}
The data are listed in the certificate
package~\cite{ComputationalCertificates}.
Each row specifies $H$, cap roots, rational cardinal weights,
cap objective, the terminal block, merged roots, and the full
original-degree interval.
For a direct row, that is, a profile with $d_0\ge4$ checked by integer
deficits, it also specifies $h_H(0)$, all moments
$M_\ell$, the gcd $g_H$, reduced coefficients $f_\ell$, and
each lift's degree, terminal degree, index count and maximal ratio.
The check reconstructs~\eqref{eq:qutrit-cardinal-data}--
\eqref{eq:qutrit-integer-deficit} with exact arithmetic for every row, index and lift, and verifies every identity and the positivity of every deficit.

\section{The first and last binary indices}\label{app:binary-endpoints}

This appendix proves the two extreme indices $i=1$ and $i=D$ of the
coefficient criterion~\eqref{eq:coefficient-criterion} for the binary
rounded polynomial $g$ of~\eqref{eq:rounded-polynomial}, that is, the
cases $i=1$ and $i=D$ of Proposition~\ref{prop:binary-index-cover}.
Lemma~\ref{lem:binary-first} treats $i=1$ through the rounded
polynomial itself. Lemma~\ref{lem:binary-top} proves, for the lift of
every real polynomial $q$ of degree at most $m$, the bound
$c_D/c_0\le(N)_D{U_D}/(n)_D$, which is the comparison of
Section~\ref{sec:binary} at $i=D$ (there the lift has a single term),
and deduces the case $i=D$. Both lemmas use the lower bound for monic
paired moments in Lemma~\ref{lem:binary-monic-moment}, a consequence of
Lemma~\ref{lem:forward-schur} at $Q=4$.

Throughout this appendix $Q=4$, $m\ge2$, $r\ge0$,
$D=2m+r$, $n=N+r$ and $N\ge6m+2r-1$. The lemmas do not need the
assumption $D\ge16$ of Proposition~\ref{prop:binary-index-cover}.
Besides that proposition, Appendix~\ref{app:bg-minima} uses
Lemma~\ref{lem:binary-monic-moment} and the calculation and
monotonicities of Lemma~\ref{lem:binary-top}. We use the polynomial
$L$ and the ratios ${U_i}$ of~\eqref{eq:binary-endpoint}. By
Lemma~\ref{lem:binary-root-distance}, every zero of $L$ exceeds $D+1$,
so $L(x)>0$ for $x\le D+1$.

\begin{lemma}\label{lem:binary-monic-moment}
If $Y\sim\Bin({L_{\rm aux}}-1,3/4)$, ${L_{\rm aux}}\ge6m-1$, and $q$ is monic
of degree $m$, then
\[
 \E[q(Y+1)q(Y)]\ge\frac23m!({L_{\rm aux}}-1)_m(3/16)^m.
\]
\end{lemma}
\begin{proof}
Lemma~\ref{lem:forward-schur} applies with $Q=4$, $k=m$ and length
$L={L_{\rm aux}}$, since ${L_{\rm aux}}\ge6m-1\ge4m$; we write
${L_{\rm aux}}$ for the length $L$ of Section~\ref{sec:paired-forms}
because $L$ denotes the polynomial~\eqref{eq:binary-endpoint}.
Then ${\omega}=2/3$, and
$\theta_j^2=3j/({L_{\rm aux}}-j)\le3m/(5m-1)\le2/3$ for $j\le m$,
since ${L_{\rm aux}}-m\ge5m-1$ and $m\ge2$.
Starting with $\delta_0=1$, the implication
\[
 \delta_{j-1}\ge2/3
 \ \Longrightarrow\
 \delta_j\ge1-\tfrac23(\tfrac13+\tfrac16)=2/3
\]
follows from~\eqref{eq:forward-schur}: there $2\omega-1=1/3$ and
$(1-\omega)^2/\delta_{j-1}\le(1/9)/(2/3)=1/6$, and the right side
decreases in $\theta_j^2$ and increases in $\delta_{j-1}$.
Hence $\delta_m\ge2/3$, and~\eqref{eq:monic-forward}, with $k=m$,
$L={L_{\rm aux}}$, $(Q-1)/Q^2=3/16$ and
$\operatorname{Forw}(q)=\E[q(Y+1)q(Y)]$, proves the claim.
\end{proof}

\begin{lemma}\label{lem:binary-first}
The rounded polynomial $g$ of~\eqref{eq:rounded-polynomial} satisfies
$c_0g(1)\ge c_1g(0)$.
\end{lemma}
\begin{proof}
We reduce the claim to the inequality~\eqref{eq:binary-negative-endpoint}
between two Krawtchouk values, and prove that inequality by a series
bound and a polynomial certificate.

\smallskip\noindent\emph{Step 1 (Reduction to a first-moment inequality).}
Write $h=\sum_{j=0}^{2m}b_jP_j^{(N,4)}$ and $s\coloneqq3/4$, and let
$X\sim\mu_N$. Since $P_1^{(N,4)}(x)=3N-4x$ and $v_1(N)=3N$,
orthogonality at degree one gives the first identity below,
and~\eqref{eq:normalized-lift} and~\eqref{eq:normalized-value} at $i=1$
give the other two:
\[
 b_1=b_0-\frac{\E[Xh(X)]}{Ns},\qquad
 \frac{c_1}{c_0}=\frac rn+\frac Nn\frac{b_1}{b_0},
 \qquad \frac{g(1)}{g(0)}=\frac Nn\frac{h(1)}{h(0)}.
\]
Since $c_0,g(0)>0$ (Section~\ref{sec:framework}) and
$b_0=\E h(X)>0$ (Lemma~\ref{lem:radau-rounded}), the claim is
equivalent to
\begin{equation}\label{eq:binary-first-moment-target}
 \E[Xh(X)]\ge s b_0\{r+N(1-h(1)/h(0))\}.
\end{equation}

\smallskip\noindent\emph{Step 2 (Bounds through the Lloyd zeros).}
Set $\alpha\coloneqq m!/4^m$, $V\coloneqq V_m(N)$ and
$v\coloneqq P_m^{(N-1,4)}(0)=\binom{N-1}m3^m$, which is $v_m(N-1)$.
By~\eqref{eq:binary-endpoint}, $L(1)=\alpha v$, and the Lloyd identity
in the proof of Lemma~\ref{lem:radau-rounded} gives $L(0)=\alpha V$.
Let $q(x)\coloneqq\prod_j(x-a_j)$, so that $h(x)=q(x)q(x-1)$. Size bias gives
$\E[Xh(X)]=Ns\,\E[q(Y+1)q(Y)]$ with $Y\sim\Bin(N-1,3/4)$, and
Lemma~\ref{lem:binary-monic-moment} with ${L_{\rm aux}}=N$ gives
$\E[Xh(X)]\ge(2/3)Ns\alpha^2v$, since
$\alpha^2v=m!(N-1)_m(3/16)^m$.
Lemma~\ref{lem:radau-rounded}, applied to $h_L=h$, gives
$b_0\le h(0)/V$.

Let $F(x)\coloneqq\prod_j(a_j-x)$, so that $h(x)=F(x)F(x-1)$; in
particular $h(0)=F(0)F(-1)$ and $h(1)=F(0)F(1)$. Increasing each
$a_j$ to $\xi_j$ turns $F$ into $L$, and it does not decrease
$F(0)F(-1)=\prod_ja_j(a_j+1)$ or $F(0)[F(-1)-F(1)]$.
For the second assertion, expand over the subsets $S$ of
$\{1,\ldots,m\}$:
$\prod_j(a_j+1)-\prod_j(a_j-1)=\sum_S\bigl(1-(-1)^{m-|S|}\bigr)\prod_{j\in S}a_j$.
Every coefficient is $0$ or $2$, so this bracket, and its product with
$F(0)=\prod_ja_j$, does not decrease when one of the nonnegative
variables $a_j$ increases. Thus
\[
 h(0)\le L(0)L(-1),\qquad
 h(0)-h(1)\le L(0)[L(-1)-L(1)].
\]

\smallskip\noindent\emph{Step 3 (Reduction to an endpoint ratio).}
The brace in~\eqref{eq:binary-first-moment-target} is positive, since
$h(1)/h(0)=\prod_j(a_j-1)/(a_j+1)<1$. Hence the bounds of Step~2 bound
the right side of~\eqref{eq:binary-first-moment-target} by
\[
 \frac sV\{rh(0)+N[h(0)-h(1)]\}
 \le\frac{sL(0)}V\{(N+r)L(-1)-NL(1)\}
 =s\alpha\{(N+r)L(-1)-N\alpha v\},
\]
while its left side is at least $(2/3)Ns\alpha^2v=(2/3)Ns\alpha L(1)$.
So~\eqref{eq:binary-first-moment-target} holds if
\[
 (1+2/3)NL(1)\ge(N+r)L(-1),
\]
where the $2/3$ comes from Lemma~\ref{lem:binary-monic-moment}. Since
$L(-1)/L(1)=P_m^{(N-1,4)}(-2)/P_m^{(N-1,4)}(0)$, this is
\begin{equation}\label{eq:binary-negative-endpoint}
 (1+r/N)\frac{P_m^{(N-1,4)}(-2)}{P_m^{(N-1,4)}(0)}\le\frac53.
\end{equation}

\smallskip\noindent\emph{Step 4 (A series bound for the endpoint ratio).}
Put ${N_0}\coloneqq N-1$, which occurs throughout the remaining computation;
the standing assumption gives ${N_0}\ge6m-2$. Writing the right side
of~\eqref{eq:kraw-generating} as $(1+3z)^{N_0}(1-4z/(1+3z))^x$ gives
$P_m^{({N_0},4)}(x)=\sum_j(-4)^j3^{m-j}\binom xj\binom{{N_0}-j}{m-j}$.
At $x=-2$, $\binom{-2}j=(-1)^j(j+1)$, and division by
$P_m^{({N_0},4)}(0)=\binom{N_0}m3^m$ gives the exact positive series
\[
 \frac{P_m^{({N_0},4)}(-2)}{P_m^{({N_0},4)}(0)}
 =\sum_{j=0}^m(j+1)(4/3)^j\frac{(m)_j}{({N_0})_j}.
\]
Put $u\coloneqq4m/(3{N_0})$, $v_*\coloneqq4(m-1)/(3({N_0}-1))$ and
$z\coloneqq4(m-2)/(3({N_0}-2))$. Since ${N_0}\ge6m-2$, $0\le z<1$.
The ratios $(m-k)/({N_0}-k)$ decrease with $k$, so the term with
index $j\ge2$ is at most $uv_*(j+1)z^{j-2}$; since
$\sum_{i\ge0}(i+3)z^i=(3-2z)/(1-z)^2$, the series
is at most $1+2u+uv_*(3-2z)/(1-z)^2$.
Also $1+r/N\le(3{N_0}-6m+4)/(2({N_0}+1))$, because
$r\le(N-6m+1)/2$.
With $Z_1\coloneqq3{N_0}-4m+2=3({N_0}-2)(1-z)$ and
$Z_3\coloneqq9{N_0}-8m-2=3({N_0}-2)(3-2z)$, set
\[
 {\operatorname{Num}}\coloneqq({N_0}-1)Z_1^2(3{N_0}+8m)+16m(m-1)({N_0}-2)Z_3.
\]
The preceding series bound equals ${\operatorname{Num}}/[3{N_0}({N_0}-1)Z_1^2]$.

\smallskip\noindent\emph{Step 5 (Polynomial certificate).}
Both bounds of Step~4 are positive, so~\eqref{eq:binary-negative-endpoint}
follows if their product is at most $5/3$. The numerator of $5/3$
minus this product is
\[
 {\operatorname{Cert}}\coloneqq10({N_0}+1){N_0}({N_0}-1)Z_1^2-(3{N_0}-6m+4){\operatorname{Num}},
\]
over the positive denominator $6({N_0}+1){N_0}({N_0}-1)Z_1^2$.
With $w\coloneqq{N_0}-6m+2\ge0$, write
${\operatorname{Cert}}=\sum_{k=0}^5w^k{\operatorname{Cert}}_k$, where each
${\operatorname{Cert}}_k$ is a polynomial in $m-2$.
The coefficients of each ${\operatorname{Cert}}_k$, in increasing powers
of $m-2$, are
\begin{center}
\begin{tabular}{c|l}
$k$&Coefficients of ${\operatorname{Cert}}_k$\\\hline
$0$&$50688,148736,170592,97504,28416,3456$\\
$1$&$39168,106912,107992,48456,8208$\\
$2$&$16056,33936,23652,5472$\\
$3$&$3150,4398,1528$\\
$4$&$279,192$\\
$5$&$9$.
\end{tabular}
\end{center}
All of them are positive, so ${\operatorname{Cert}}>0$ whenever $m\ge2$
and $w\ge0$, that is, $N\ge6m-1$. This
proves~\eqref{eq:binary-negative-endpoint} throughout the standing
domain.
\end{proof}

\begin{lemma}\label{lem:binary-top}
Under the standing assumptions of this appendix, let $q$ be any real
polynomial of degree at most $m$, and write
\[
 q(x)q(x-1)=\sum_{i=0}^{2m} b_iP_i^{(N,4)}(x).
\]
Then
\begin{equation}\label{eq:binary-top-comparison}
 \frac{D!}{(N-2m)_r}\frac{b_{2m}}{(2m)!}
 \le {U_D} b_0.
\end{equation}
In particular the rounded polynomial $g$ of~\eqref{eq:rounded-polynomial}
satisfies $c_0g(D)\ge c_Dg(0)$.
\end{lemma}
\begin{proof}
Step~1 explains~\eqref{eq:binary-top-comparison} and deduces the
rounded case. Steps~2--6 prove~\eqref{eq:binary-top-comparison}: a
reduction to a scalar inequality, monotonicity in $N$, the case $r=0$,
and the passage from $r$ to $r+1$.

\smallskip\noindent\emph{Step 1 (Meaning of the comparison; the rounded case).}
For nonzero $q$ we have $b_0>0$ (Step~2). Write the lift as
$\binom{n-x}{r}q(x)q(x-1)=\sum_ic_iP_i^{(n,4)}(x)$. At $i=D$ only the
term $j=2m$ remains in~\eqref{eq:normalized-lift}, which is the
binomial lift~\eqref{eq:binomial-lift} divided by $c_0$: indeed
$b_j=0$ for $j>2m$ and $(r)_{D-j}=0$ for $j<2m$. As
$(N)_D=(N)_{2m}(N-2m)_r$, this gives
\[
 \frac{(n)_D}{(N)_D}\frac{c_D}{c_0}
 =\frac{D!}{(N-2m)_r}\frac{b_{2m}}{(2m)!\,b_0}.
\]
Thus~\eqref{eq:binary-top-comparison} is the comparison
$c_D/c_0\le(N)_D{U_D}/(n)_D$ of Section~\ref{sec:binary} at $i=D$, for
every such $q$. For $q(x)=\prod_j(x-a_j)$ the product $q(x)q(x-1)$ is
the rounded $h$; \eqref{eq:binary-floor-comparison} at $i=D$ gives
${U_D}\le h(D)/h(0)$, and~\eqref{eq:normalized-value} turns the
comparison into $c_D/c_0\le g(D)/g(0)$. Since $c_0,g(0)>0$, this is
$c_0g(D)\ge c_Dg(0)$.

\smallskip\noindent\emph{Step 2 (Reduction to a scalar inequality).}
Scaling $q$ by a constant multiplies $b_0$ and $b_{2m}$ by its square.
If $q\ne0$ has degree below $m$, then $b_{2m}=0$ while $b_0>0$:
conditioning on one coordinate gives, as in the first equality
of~\eqref{eq:paired-conditioning} with length $L=N$,
$b_0=\E_{\mu_N}[q(X)q(X-1)]=\tfrac34\operatorname{Forw}(q)+\tfrac14\operatorname{Back}(q)$,
which is positive by Lemma~\ref{lem:paired-grams} (with $Q=4$ and
$N\ge4m$; for constant $q$ it equals $q^2$).
It therefore suffices to treat monic $q$ of degree $m$. Then
$q(x)q(x-1)$ is monic of degree $2m$, and by~\eqref{eq:kraw} the
leading coefficient of $P_{2m}^{(N,4)}$ is $4^{2m}/(2m)!$; hence
$b_{2m}=(2m)!/4^{2m}$. Lemma~\ref{lem:binary-monic-moment}, with
length parameter ${L_{\rm aux}}=N+1$ and polynomial $q(x-1)$, gives
\[
 b_0=\E_{\mu_N}[q(X)q(X-1)]\ge\tfrac23m!(N)_m(3/16)^m.
\]
Inserting both values into~\eqref{eq:binary-top-comparison}, the factor
$16^m=4^{2m}$ cancels, and it is enough to show
\begin{equation}\label{eq:binary-top-target}
 {U_D}\ge {\operatorname{Top}}(m,r,N)\coloneqq
 \frac{3\,D!}{2\cdot m!\,(N)_m\,3^m\,(N-2m)_r}.
\end{equation}

\smallskip\noindent\emph{Step 3 (Monotonicity in $N$).}
The proof of Lemma~\ref{lem:binary-root-distance} gives the Jacobi
matrix of the reflected polynomial. Subtracting it from $NI$ and
changing the signs of the off-diagonal entries (a diagonal similarity)
gives the symmetric tridiagonal matrix $J_N$ whose eigenvalues are the
zeros of $L$; we write them $\xi_j(N)$ when the length varies. Thus
$L(x)=\det(J_N-xI)$, the diagonal entries of $J_N$ are $(3N+1)/4-j/2$
for $0\le j<m$, and its off-diagonal entries are
$e_j\coloneqq\sqrt{3j(N-j)}/4$ for $1\le j<m$.
Since $N-j\ge5m$ for $j<m$,
\[
 e_j(N+2)-e_j(N)\le\frac{\sqrt{3j}}{4\sqrt{N-j}}
 \le\tfrac14\sqrt{3(m-1)/(5m)}.
\]
Thus $J_{N+2}-J_N-I$, whose diagonal entries equal $1/2$, has absolute
off-diagonal row sums at most $\tfrac12\sqrt{3/5}<\tfrac12$; it is
strictly diagonally dominant with positive diagonal, and is positive
definite. Similarly $J_{N+1}-J_N$, with diagonal entries $3/4$ and
off-diagonal row sums at most $\tfrac14\sqrt{3/5}$, is positive
definite. By the Courant--Fischer theorem, adding a positive definite
matrix strictly increases every ordered eigenvalue, which yields
\begin{equation}\label{eq:binary-jacobi-shift}
 \xi_j(N+2)>\xi_j(N)+1,
 \qquad \xi_j(N+1)>\xi_j(N).
\end{equation}
The first inequality is used in Step~6. Each factor
$(\xi-D)(\xi-D-1)/(\xi(\xi-1))$ of ${U_D}$ increases for $\xi>D+1$,
and every $\xi_j(N)>D+1$ by Lemma~\ref{lem:binary-root-distance}.
Hence, by the second inequality, ${U_D}$ increases with $N$, whereas
${\operatorname{Top}}$ decreases. It suffices to take $N=6m+2r-1$.

\smallskip\noindent\emph{Step 4 (The case $r=0$, $m\ge15$).}
Here $N=6m-1$ and $D=2m$. The determinant
$L(2m+1)=\det(J_N-(2m+1)I)$ is the product of the Cholesky pivots of
$J_N-(2m+1)I$; each pivot is the diagonal entry minus the squared
off-diagonal entry divided by the preceding pivot. If $m\ge15$, the
pivots are at least $15m/8$ on the first
$\lceil m/2\rceil$ indices, and at least $9m/8$ thereafter.
Indeed the first block has diagonal at least $9m/4-3/2$
and squared off-diagonal at most $33m^2/64$. A preceding
pivot $15m/8$ gives the next lower bound
$79m/40-3/2$, which is at least $15m/8$ exactly when $m\ge15$; the
first pivot, $(5m-3)/2$, is larger.
In the second block, diagonal $2m-1$ and squared
off-diagonal at most $15m(m-1)/16$ give
$7m/6-1/6\ge9m/8$ from a preceding pivot $9m/8$;
the last pivot of the first block meets this premise.
Therefore
\[
 L(2m+1)\ge(15m/8)^{\lceil m/2\rceil}
                  (9m/8)^{\lfloor m/2\rfloor}.
\]
The same lower bound holds for $L(2m)\ge L(2m+1)$, since every
$\xi_j>2m+1$. The matrices $J_N$ and $J_N-I$ are positive definite,
because every $\xi_j>1$, and their diagonal entries are less than
$9m/2$; Hadamard's inequality gives
$L(0)L(1)=\det J_N\det(J_N-I)\le(9m/2)^{2m}$. Since
$(15/8)(9/8)/(81/4)=5/48$, and an odd $m$ only adds a factor $5/3$,
we obtain ${U_{2m}}\ge(5/48)^m$.

Let $R_m\coloneqq{\operatorname{Top}}(m,0,6m-1)/(5/48)^m$. Exact cancellation gives
\[
 \frac{R_{m+1}}{R_m}
 =\frac{32(2m+1)\prod_{k=0}^4(5m+k)}
        {5\prod_{k=0}^5(6m+k)}<1.
\]
Indeed, as a polynomial in $m-1$, the denominator minus the numerator
has the positive coefficients, in increasing powers of $m-1$,
\[
 \begin{gathered}
 211680,\quad970920,\quad1826640,\quad1803800,\\
 986000,\quad282880,\quad33280.
 \end{gathered}
\]
Also
$R_{15}=79551583817872441344/345374486968994140625<1$.
Hence $R_m<1$ for every $m\ge15$, that is,
${\operatorname{Top}}(m,0,6m-1)<(5/48)^m\le{U_{2m}}$.

\smallskip\noindent\emph{Step 5 (The case $r=0$, $2\le m\le14$).}
For these degrees, with $N=6m-1$, the exact ratios
${U_{2m}}/{\operatorname{Top}}(m,0,6m-1)$ are as follows.
\begin{center}\small
\begin{tabular}{cl}\toprule
$m$&Ratio\\\midrule
2&$473/207$\\
3&$606050/103089$\\
4&$651039725/43742883$\\
5&$505190614/13511617$\\
6&$429266667500/4591398001$\\
7&$573495001314326/2460755044495$\\
8&$1280385120922625/2208914666143$\\
9&$327641806939792115/227655629714136$\\
10&$45847427114945971673489/12847423145385551646$\\
11&$24602774964378373420205/2783396911370153301$\\
12&$103649727918724355606273005508/4738425222577544623580091$\\
13&$1374041123184422053960325/25401670569658905198$\\
14&$146243267501698931784996583245/1093976879394279668327807$\\
\bottomrule\end{tabular}
\end{center}
Every numerator exceeds its denominator. The entries are the exact
values of
\[
 {U_{2m}}=\frac{{\operatorname{Rec}_m}(2m){\operatorname{Rec}_m}(2m+1)}
               {{\operatorname{Rec}_m}(0){\operatorname{Rec}_m}(1)},
 \qquad {\operatorname{Rec}_j}(x)\coloneqq j!P_j^{(N-1,4)}(x-1),
\]
divided by ${\operatorname{Top}}(m,0,6m-1)$; note that
$L=4^{-m}{\operatorname{Rec}_m}$. These polynomials follow from
${\operatorname{Rec}_0}(x)=1$, ${\operatorname{Rec}_1}(x)=3(N-1)-4(x-1)$ and the
three-term recurrence of the Krawtchouk polynomials,
\[
 {\operatorname{Rec}_{j+1}}(x)=[3(N-1)-2j-4(x-1)]{\operatorname{Rec}_j}(x)-3j(N-j){\operatorname{Rec}_{j-1}}(x),
\]
which is the expansion of the leading principal minors of
$4(J_N-xI)$ along their last row.

\smallskip\noindent\emph{Step 6 (From $r$ to $r+1$).}
Fix $m$, and write $L_N$ and ${U_D}(N)$ for $L$ and ${U_D}$ at length
$N$. Passing from $r$ to $r+1$ along $N=6m+2r-1$ replaces $(N,D)$ by
$(N+2,D+1)$. By~\eqref{eq:binary-jacobi-shift},
$\xi_j(N+2)-1>\xi_j(N)>D+1$, so $L_{N+2}(x+1)>L_N(x)>0$ for $x=D,D+1$;
hence
$L_{N+2}(D+1)L_{N+2}(D+2)\ge L_N(D)L_N(D+1)$.
The endpoint identities $L_N(0)=\alpha V_m(N)$ and
$L_N(1)=\alpha\binom{N-1}m3^m$, with $\alpha\coloneqq m!/4^m$,
give
\[
 \frac{L_{N+2}(0)L_{N+2}(1)}{L_N(0)L_N(1)}\le {\operatorname{Grow}_r},
 \quad {\operatorname{Grow}_r}\coloneqq
 \frac{(N+2)(N+1)}{(N+2-m)(N+1-m)}
 \frac{N(N+1)}{(N-m)(N+1-m)}.
\]
Here $\binom{N+2}j/\binom Nj$ increases with $j\le m$, so
$V_m(N+2)/V_m(N)$ is at most this ratio at $j=m$, which is the first
factor of ${\operatorname{Grow}_r}$; the second factor equals
$L_{N+2}(1)/L_N(1)$.
Consequently ${U_{D+1}}(N+2)\ge{U_D}(N)/{\operatorname{Grow}_r}$.
At $N=6m+2r-1$, the product of ${\operatorname{Grow}_r}$ and
${\operatorname{Top}}(m,r+1,N+2)/{\operatorname{Top}}(m,r,N)$ is
\[
 \frac{(2m+r+1)(4m+r)(6m+2r-1)(6m+2r)}
 {(4m+2r+1)(4m+2r)(5m+2r-1)(5m+2r)}<1.
\]
Indeed, subtracting the numerator from the denominator gives, by
powers $r^0,\ldots,r^4$ and with coefficients in increasing powers of
$m-1$, the positive coefficient lists
\[
 (40,228,448,372,112),\quad(246,808,874,312),\quad
 (252,526,272),\quad(94,96),\quad(12).
\]
Thus ${\operatorname{Top}}/{U_D}$ decreases along $N=6m+2r-1$ as $r$
increases. Steps~4 and~5 give ${U_D}>{\operatorname{Top}}$ at $r=0$, so
\eqref{eq:binary-top-target} holds at $N=6m+2r-1$ for every $r$, and
Step~3 extends it to every $N\ge6m+2r-1$. By Step~2 this
proves~\eqref{eq:binary-top-comparison}, and Step~1 completes the proof.
\end{proof}

\section{Binary coefficient matrices and finite polynomial certificates}
\label{app:binary-matrices}

This appendix proves Proposition~\ref{prop:binary-low-band}: if
$D\ge16$, the matrix comparison
${\operatorname{Mat}_i}\preceq{U_i}{\operatorname{Gram}_0}$
of~\eqref{eq:binary-matrix-target} holds for every
$2\le i\le\min(54,D-1)$. By Step~5 of Appendix~\ref{app:binary-kernel}
(see also Section~\ref{sec:binary}), this gives $c_0g(i)\ge c_ig(0)$ at
these indices of Proposition~\ref{prop:binary-index-cover}. All cases
are proved in the same way: with the rational lower
bound~\eqref{eq:binary-rational-endpoint} in place of ${U_i}$, exact
polynomial certificates show strict weighted diagonal dominance of
${U_i}{\operatorname{Gram}_0}-{\operatorname{Mat}_i}$.

We assume $m\ge2$, $r\ge0$, $N\ge6m+2r-1$ and $D=2m+r$, and put
$n=N+r$. The matrices below have rows and columns $0,\ldots,m$,
indexed by the orthonormal basis of Step~1 of
Appendix~\ref{app:binary-kernel}; entries with an index above $m$ do
not belong to them. In particular the last row $j=m$ has no upper
neighbours, and the certificates treat it separately.
Appendix~\ref{app:binary-kernel} writes the lifted coefficient ratios
through the matrices $\operatorname{Mat}_i$ and
$\operatorname{Gram}_0$, reduces the coefficient criterion to the
matrix comparison~\eqref{eq:binary-matrix-target}, and states the
weighted row conditions that the certificates verify.
Appendices~\ref{app:binary-second}, \ref{app:binary-signed-rows}
and~\ref{app:binary-scalar} give the certificates for $i=2$, for
$3\le i\le13$ and for $14\le i\le54$, and
Appendix~\ref{app:binary-certificate-spec}
specifies them and proves
Proposition~\ref{prop:binary-low-band}. The indices $55\le i\le D-1$
are treated in Appendix~\ref{app:binary-high}.

\subsection{An exact congruence and its lifted kernel}
\label{app:binary-kernel}

\smallskip\noindent\emph{Step 1 (Coefficient forms and the translation matrix).}
Put ${v_j(N)}\coloneqq\binom Nj3^j$. For a real polynomial $q$ of degree at
most $m$, let $\vec q\in\mathbb R^{m+1}$ be its coordinate vector in the
basis $P_j^{(N,4)}/\sqrt{{v_j(N)}}$, $0\le j\le m$, which is
orthonormal for $\mu_N$ by~\eqref{eq:orthogonality}. Let $M_i$ be the
symmetric matrix with $\vec q^{\,\mathsf T}M_i\vec q=b_i$, the coefficient
in $q(x)q(x-1)=\sum b_iP_i^{(N,4)}(x)$.
By the shift identity~\eqref{eq:backward-shift}, used at length $N$
with $Q=4$, the coordinate vector of $q(x-1)$ is $T\vec q$, where
\[
 T\coloneqq(I+3R)(I-R)^{-1},\qquad
 R_{j-1,j}\coloneqq\sqrt{{v_{j-1}(N)}/{v_j(N)}},
\]
and all other entries of $R$ are zero. Since $R$ is nilpotent, the
inverse is the finite sum $\sum_{k\le m}R^k$.

\smallskip\noindent\emph{Step 2 (Product matrices).}
Let ${\operatorname{Sph}_i}$ have entries
\begin{equation}\label{eq:binary-sphere-matrix}
 ({\operatorname{Sph}_i})_{jk}\coloneqq\frac{[z^jw^k](z+w+2zw)^i(1+3zw)^{N-i}}
                   {\sqrt{{v_j(N)}{v_k(N)}}}.
\end{equation}
This entry is the coefficient of $P_i^{(N,4)}$ in the product of the
$j$th and $k$th basis polynomials of Step~1, that is,
$\E_{\mu_N}[P_jP_kP_i]/(v_i(N)\sqrt{v_j(N)v_k(N)})$ with
$P_j\coloneqq P_j^{(N,4)}$: by~\eqref{eq:kraw-generating}, the expectation of
the product of three Krawtchouk generating functions is
$[1+3(zw+zt+wt)+6zwt]^N$, and extracting $t^i$ and dividing by
$\binom Ni3^i$ gives~\eqref{eq:binary-sphere-matrix}.
Combinatorially, these are the radial matrices of averaging
translation by a uniform vector of Hamming weight $i$ (a Hamming
sphere, hence the name) on an alphabet group of size four,
compressed to weights at most $m$; this interpretation is not used
below. By Step~1, $b_i=\vec q^{\,\mathsf T}{\operatorname{Sph}_i}T\vec q$,
that is, $M_i=\operatorname{sym}({\operatorname{Sph}_i}T)$, where
$\operatorname{sym}(A)\coloneqq(A+A^{\mathsf T})/2$.

\smallskip\noindent\emph{Step 3 (Congruence; positivity of ${\operatorname{Gram}_0}$).}
Since $T(I-R)=I+3R$, the congruence by $I-R$, that is, the change of
coordinates $\vec q=(I-R)\vec y$, removes the inverse:
\begin{equation}\label{eq:binary-congruence}
 {\operatorname{Gram}_i}\coloneqq(I-R^{\mathsf T})M_i(I-R)
     ={\operatorname{Sph}_i}+{\operatorname{Sph}_i}R+R^{\mathsf T}{\operatorname{Sph}_i}-3R^{\mathsf T}{\operatorname{Sph}_i}R,
\end{equation}
and $b_i=\vec y^{\,\mathsf T}{\operatorname{Gram}_i}\vec y$.
Multiplication by $R$ on the right, or by $R^{\mathsf T}$ on the left,
multiplies the generating function of the scaled entries
$\sqrt{v_j(N)v_k(N)}A_{jk}$ by $w$, or by $z$. Hence
$\sqrt{v_j(N)v_k(N)}({\operatorname{Gram}_i})_{jk}$ is the coefficient of
$z^jw^k$ in the kernel
\begin{equation}\label{eq:binary-C-kernel}
 (1+z+w-3zw)(z+w+2zw)^i(1+3zw)^{N-i}.
\end{equation}
In particular ${\operatorname{Gram}_0}$ is tridiagonal, with diagonal
entries $1-j/(N-j+1)$ and entries $\theta_j\coloneqq\sqrt{j/[3(N-j+1)]}$
between $j-1$ and $j$ (this $\theta_j$ is not that of
Section~\ref{sec:paired-forms}); all certificates below use these
entries. ${\operatorname{Gram}_0}$ is the Gram matrix of the form
$b_0=\E_{\mu_N}[q(X)q(X-1)]$ in the coordinates $\vec y$, and it is
positive definite: conditioning on one coordinate gives, as in the
first equality of~\eqref{eq:paired-conditioning} with $L=N$,
$b_0=\tfrac34\operatorname{Forw}(q)+\tfrac14\operatorname{Back}(q)$,
which is positive for $q\ne0$ by Lemma~\ref{lem:paired-grams} (with
$Q=4$ and $N\ge4m$; for constant $q$ it equals $q^2$), and the
congruence is invertible. For $i\ge1$ the matrices
${\operatorname{Gram}_i}$, named after ${\operatorname{Gram}_0}$, need
not be definite.

\smallskip\noindent\emph{Step 4 (Lifted matrices and their kernel).}
Define
\begin{equation}\label{eq:binary-lifted-matrix}
 {\operatorname{Mat}_i}\coloneqq\sum_{\ell=0}^{\min(i,2m)}
 \frac{\binom i\ell(N)_\ell(r)_{i-\ell}}{(N)_i}{\operatorname{Gram}_\ell}.
\end{equation}
Its weights are those of~\eqref{eq:normalized-lift}. Hence, for the
lift $\binom{n-x}{r}q(x)q(x-1)=\sum_ic_iP_i^{(n,4)}(x)$ of any
$q\ne0$,
\[
 \frac{(n)_i}{(N)_i}\frac{c_i}{c_0}
 =\frac{\vec y^{\,\mathsf T}{\operatorname{Mat}_i}\vec y}
       {\vec y^{\,\mathsf T}{\operatorname{Gram}_0}\vec y}.
\]
The normalization by $(N)_i$ matches~\eqref{eq:normalized-value}, so
that the factor $(N)_i/(n)_i$ cancels in Step~5.
Write ${\operatorname{Step}}(z,w)\coloneqq z+w+2zw$ and
${\operatorname{Stay}}(z,w)\coloneqq1+3zw$; these are the parts without and with
$t$ of the factor
$1+3(zw+zt+wt)+6zwt={\operatorname{Stay}}+3t\,{\operatorname{Step}}$ of
Step~2. The exact kernel of ${\operatorname{Mat}_i}$ is
\begin{equation}\label{eq:binary-lifted-kernel}
 \binom Ni({\operatorname{Mat}_i})_{jk}\sqrt{{v_j(N)}{v_k(N)}}
 =[z^jw^ku^i](1+z+w-3zw)({\operatorname{Stay}}(z,w)+u{\operatorname{Step}}(z,w))^N(1+u)^r.
\end{equation}
Indeed, by the binomial theorem,
\[
 [u^i]({\operatorname{Stay}}+u{\operatorname{Step}})^N(1+u)^r
 =\sum_\ell\binom N\ell\binom r{i-\ell}
   {\operatorname{Step}}^\ell{\operatorname{Stay}}^{N-\ell},
\]
and the lift coefficient $\binom i\ell(N)_\ell(r)_{i-\ell}/(N)_i$
equals $\binom N\ell\binom r{i-\ell}/\binom Ni$.
The terms with $\ell>2m$ contain no monomial $z^jw^k$ with
$j,k\le m$, because ${\operatorname{Step}}^\ell$ has total degree at
least $\ell$; they matter only for the extended rows of Step~6.

\smallskip\noindent\emph{Step 5 (The rounded criterion from the matrix comparison).}
Let $1\le i\le D$, so that $i<N$. A proof of
\begin{equation}\label{eq:binary-matrix-target}
 {\operatorname{Mat}_i}\preceq {U_i} {\operatorname{Gram}_0}
\end{equation}
implies the rounded coefficient criterion at the index $i$. Indeed,
take $q(x)\coloneqq\prod_j(x-a_j)$, so that $q(x)q(x-1)$ is the rounded $h$,
with coordinates $\vec y$ as in Step~3. Then
$\vec y^{\,\mathsf T}{\operatorname{Gram}_0}\vec y=b_0>0$, and Step~4,
\eqref{eq:binary-matrix-target}, \eqref{eq:binary-floor-comparison}
and~\eqref{eq:normalized-value} give
\[
 \frac{(n)_i}{(N)_i}\frac{c_i}{c_0}
 =\frac{\vec y^{\,\mathsf T}{\operatorname{Mat}_i}\vec y}
       {\vec y^{\,\mathsf T}{\operatorname{Gram}_0}\vec y}
 \le{U_i}\le\frac{h(i)}{h(0)}=\frac{(n)_i}{(N)_i}\frac{g(i)}{g(0)}.
\]
Thus $c_i/c_0\le g(i)/g(0)$, that is, $c_0g(i)\ge c_ig(0)$, since
$c_0,g(0)>0$. This is the reduction used in Section~\ref{sec:binary}.
The comparison~\eqref{eq:binary-matrix-target} is proved below for
$2\le i\le\min(54,D-1)$ and in Appendix~\ref{app:binary-high} for
$55\le i\le D-1$; Lemma~\ref{lem:binary-top} proves its analogue at
$i=D$, and Lemma~\ref{lem:binary-first} treats $i=1$ through the
rounded polynomial itself.

\smallskip\noindent\emph{Step 6 (Entrywise nonnegativity and extended rows).}
For a row index $j\le m$, the extended row of
${\operatorname{Gram}_\ell}$, or of ${\operatorname{Mat}_i}$, consists
of the kernel coefficients at $z^jw^k$ for all $k\ge0$, normalized as
the entries are; it contains the finite row $k\le m$ and serves as an
upper bound for it. If $N-\ell\ge2m-1$, all coefficients of the
kernel~\eqref{eq:binary-C-kernel} in rows $j\le m$ are nonnegative, so
the entries of ${\operatorname{Gram}_\ell}$ and of its extended rows
are nonnegative. To see this, split the multiplier
in~\eqref{eq:binary-C-kernel} as $z+w$ and $1-3zw$. The coefficient of
$(zw)^c$ in $(1-3zw)(1+3zw)^{N-\ell}$ is
$3^c[\binom{N-\ell}c-\binom{N-\ell}{c-1}]\ge0$ for $2c\le N-\ell+1$.
The remaining factors have nonnegative coefficients, and in row $j$
only $c\le j\le m$ occurs. The extended rows of
${\operatorname{Mat}_i}$ involve ${\operatorname{Gram}_\ell}$ for every
$\ell\le i<D$, and our domain gives
$N-\ell\ge N-D\ge4m+r-1\ge2m-1$ for all $\ell\le D$.

\smallskip\noindent\emph{Step 7 (Signed rows and weighted diagonal dominance).}
Since ${U_i}>0$, ${\operatorname{Gram}_0}$ is tridiagonal and
${\operatorname{Mat}_i}$ is entrywise nonnegative, the entries of
$Z\coloneqq{U_i}{\operatorname{Gram}_0}-{\operatorname{Mat}_i}$ with
$|j-k|\ge2$ are nonpositive; only the diagonal and the two nearest
off-diagonal entries can have either sign.
For any positive weights $w_j$, the weighted diagonal-dominance
condition of row $j$ is therefore equivalent to the four inequalities
\begin{equation}\label{eq:binary-four-signs}
 Z_{jj}-\varepsilon_-Z_{j,j-1}\frac{w_{j-1}}{w_j}
       -\varepsilon_+Z_{j,j+1}\frac{w_{j+1}}{w_j}
       -\sum_{|k-j|\ge2}|Z_{jk}|\frac{w_k}{w_j}>0,
 \qquad \varepsilon_\pm\in\{-1,1\}.
\end{equation}
Absent neighbors are omitted. We call the left side
of~\eqref{eq:binary-four-signs} the signed margin of row $j$, and, for
a symmetric matrix $A$, we call
$A_{jj}-\varepsilon_-A_{j,j-1}w_{j-1}/w_j-\varepsilon_+A_{j,j+1}w_{j+1}/w_j$
its signed row. Strict weighted diagonal dominance
of a symmetric matrix implies positive definiteness: apply ordinary
strict diagonal dominance to the symmetric congruence by
$\operatorname{diag}(w_j)$.

All certificates of
Appendices~\ref{app:binary-second}--\ref{app:binary-scalar} use the
weights $w_j\coloneqq\sqrt{{v_j(N)}}4^{-j}$. With them a weighted entry is
$A_{jk}w_k/w_j=4^{j-k}\sqrt{v_j(N)v_k(N)}A_{jk}/v_j(N)$, so the weighted
extended rows are obtained by evaluating the kernels at $w=1/4$.

\smallskip\noindent\emph{Step 8 (A rational lower bound for ${U_i}$).}
Since ${U_i}$ involves the irrational zeros $\xi_j$, the certificates
use instead the following endpoint lower bound:
\begin{equation}\label{eq:binary-rational-endpoint}
 {U_i}\ge {v_i^{\rm end}}\coloneqq
 \frac{(N-m)(3N-4m+3)}{3N(N-m+1)}\,{u_i^{\rm end}}{u_{i+1}^{\rm end}},
 \qquad
 {u_i^{\rm end}}\coloneqq\frac{P_{i-1}^{(N-1,4)}(m)}{\binom{N-1}{i-1}3^{i-1}}.
\end{equation}
To prove it, write ${U_i}=[L(i)/L(1)]\,[L(i+1)/L(1)]\,[L(1)/L(0)]$.
Krawtchouk duality,
$v_x(N-1)P_j^{(N-1,4)}(x)=v_j(N-1)P_x^{(N-1,4)}(j)$ for integers
$0\le x,j\le N-1$ (in the sum~\eqref{eq:kraw},
$\binom{N-1}x\binom xa\binom{N-1-x}{j-a}$ is symmetric in $x$ and $j$),
gives $L(i)/L(1)=P_m^{(N-1,4)}(i-1)/P_m^{(N-1,4)}(0)={u_i^{\rm end}}$,
and likewise $L(i+1)/L(1)={u_{i+1}^{\rm end}}$.
Moreover $L(1)/L(0)=\binom{N-1}m3^m/V_m(N)$.
The ratio of the $(j-1)$th term of $V_m(N)$ to the $j$th is
$j/[3(N-j+1)]\le m/[3(N-m+1)]$, so a geometric series bounds $V_m(N)$
by $\binom Nm3^m/(1-m/[3(N-m+1)])$; this gives the displayed
lower bound for $L(1)/L(0)$. All factors are positive by
Lemma~\ref{lem:binary-root-distance}.

\subsection{The second index}
\label{app:binary-second}

At $i=2$, ${\operatorname{Mat}_2}={\operatorname{Gram}_2}+\alpha {\operatorname{Gram}_1}+\beta {\operatorname{Gram}_0}$, where
$\alpha\coloneqq2r/(N-1)$ and $\beta\coloneqq r(r-1)/[N(N-1)]$.
The lower bound in~\eqref{eq:binary-rational-endpoint} is $F/G$, with
\begin{align*}
F&\coloneqq(N-m)(3N-4m+3)(3N-4m-3)W,\\
W&\coloneqq9N^2-27N+18-24mN+32m+16m^2,\\
G&\coloneqq81N(N-m+1)(N-1)^2(N-2).
\end{align*}
Here $3N-4m-3=P_1^{(N-1,4)}(m)$ and $W=2P_2^{(N-1,4)}(m)$; in the
notation of Appendix~\ref{app:binary-signed-rows}, $F=E_2$ and
$G=(N)_2\mathcal D_2$.
Since ${U_2}\ge F/G$ and ${\operatorname{Gram}_0}$ is positive definite,
${U_2}{\operatorname{Gram}_0}-{\operatorname{Mat}_2}\succeq
(F/G-\beta){\operatorname{Gram}_0}-{\operatorname{Gram}_2}-\alpha {\operatorname{Gram}_1}$,
so it is enough to check the signed margins~\eqref{eq:binary-four-signs}
of the latter matrix. The following explicit formulas obtain them
directly from~\eqref{eq:binary-C-kernel}; they include every nonzero
entry of the row, since ${\operatorname{Gram}_1}$ and
${\operatorname{Gram}_2}$ vanish at distance more than two and three,
respectively, from the diagonal. For an interior row $j<m$, set
\begin{align*}
{a_{\rm row}}\coloneqq{}&1728(N-2j+1)-432\varepsilon_+(N-j+1)-2304\varepsilon_-j,\\
{d_{\rm row}}\coloneqq{}&1152j(N-1)(N-j+1)-768(j)_3+2304rj(2N-3j+3),\\
{u_{\rm row,+}}\coloneqq{}&48j(21N-29j+8)(N-j+1)+288r(3N-4j)(N-j+1),\\
V_1\coloneqq{}&256(j)_2(21N-29j+37)+1536rj(3N-4j+4),\\
U_f\coloneqq{}&36(N-j)(3N-2j-3)(N-j+1)+216r(N-j)(N-j+1)\\
 &+27(N-j)(N-j-1)(N-j+1),\\
V_f\coloneqq{}&1024(j)_2(3N-2j+1)+6144r(j)_2+4096(j)_3,\\
{\operatorname{rem}_{\rm row}}\coloneqq{}&-{d_{\rm row}}+\varepsilon_+{u_{\rm row,+}}+\varepsilon_-V_1-U_f-V_f.
\end{align*}
Here ${a_{\rm row}}/[1728(N-j+1)]$ is the signed row of
${\operatorname{Gram}_0}$. After division by $1728N(N-1)(N-j+1)$,
${d_{\rm row}}$ is the weighted diagonal entry of
${\operatorname{Gram}_2}+\alpha{\operatorname{Gram}_1}$, the terms
${u_{\rm row,+}}$ and $V_1$ are its weighted upper and lower nearest
entries, and $U_f$ and $V_f$ are the sums of its weighted upper and
lower entries at distances two and three.
Near the last row some of these upper entries lie beyond column $m$
and are absent from the matrix; including them only decreases the
signed margin, since they are nonnegative. The signed margin of row
$j$ is therefore at least
$(F/G-\beta){a_{\rm row}}/[1728(N-j+1)]+{\operatorname{rem}_{\rm row}}/[1728N(N-1)(N-j+1)]$,
with equality when no upper entry is absent. Over a common denominator
this lower bound is precisely
\begin{equation}\label{eq:binary-second-numerator}
 \frac{N(N-1){a_{\rm row}}F+[{\operatorname{rem}_{\rm row}}-r(r-1){a_{\rm row}}]G}
      {1728N(N-1)(N-j+1)G}.
\end{equation}
For $j=m$, remove the upper-neighbour terms, namely
$-432\varepsilon_+(N-j+1)$ from ${a_{\rm row}}$ and both
$\varepsilon_+{u_{\rm row,+}}$ and $-U_f$ from
${\operatorname{rem}_{\rm row}}$; the formula is then the exact last-row
margin. At $j=0$ the lower-neighbour terms, which carry a factor $j$,
vanish.

The numerator has nonnegative coefficients and a positive constant
under every substitution in the table. The polynomial certificates
are obtained by exact expansion of~\eqref{eq:binary-second-numerator}.
\begin{center}\small
\begin{tabular}{clrr}\toprule
Rows&Substitution and signs&Polynomials&Nonzero terms each\\\midrule
$1\le j<m$&$j=a+1,m=a+b+2$; all four signs&4&495\\
$j=0$&$m=a+2$; both upper signs&2&165\\
$j=m$&$m=a+2$; both lower signs&2&165\\\bottomrule
\end{tabular}
\end{center}
In every table row $N=6m+2r-1+s$ and $a,b,r,s\ge0$.
There are $2640$ nonzero coefficients in all eight polynomials;
their minimum coefficients, in the four interior sign orders
$(\varepsilon_+,\varepsilon_-)=(+,+)$, $(+,-)$, $(-,+)$, $(-,-)$, are
$94041,94041,164025,164025$; the two $j=0$ minima are
$94041,164025$ and both last-row minima are $139968$. These counts and
minima allow a recomputation to be compared with the certificate.
The positive denominator and exhaustive row partition prove
\eqref{eq:binary-matrix-target} at $i=2$ on the whole standing domain;
the assumption $D\ge16$ is not needed here.

\subsection{Direct signed-row certificates through index thirteen}
\label{app:binary-signed-rows}

For $3\le i\le13$ each signed margin is bounded below through the
extended row and the two exact nearest entries, and the bound is
cleared to an integer polynomial.

\smallskip\noindent\emph{Step 1 (The extended weighted row).}
Put
\[
 W_j(u)\coloneqq(1+u)^r(1+u/4)^{N-j}(1+2u)^j,
 \qquad
 {\operatorname{RowExt}_{\rm raw}}\coloneqq i![u^i]\{\tfrac54(N-j+1)W_j+\tfrac j3W_{j-1}\}.
\]
The second term is omitted at $j=0$.
At $w=1/4$ the multiplier in~\eqref{eq:binary-lifted-kernel} becomes
$5/4+z/4$ and ${\operatorname{Stay}}+u{\operatorname{Step}}$ becomes
$(1+u/4)+\tfrac34z(1+2u)$. Extracting $z^j$ and using the weights of
Step~7 of Appendix~\ref{app:binary-kernel} shows that
${\operatorname{RowExt}_{\rm raw}}$ is $(N)_i(N-j+1)$ times the extended
weighted row of ${\operatorname{Mat}_i}$ at $j$, that is, of the sum of
the weighted entries $({\operatorname{Mat}_i})_{jk}w_k/w_j$ over all
$k\ge0$.

\smallskip\noindent\emph{Step 2 (Exact nearest entries).}
To retain the two nearest neighbors exactly,
define $V_{\delta_{\rm off}}$, ${\delta_{\rm off}}=-1,1$, by the finite sum
\begin{equation}\label{eq:binary-neighbor-sum}
 V_{\delta_{\rm off}}\coloneqq\sum
 \frac{i!\gamma_\delta\,2^{j_2}3^{-\delta_z-j_0-{j_2}}4^{-{\delta_{\rm off}}}
       (j)_{\delta_z+j_0+{j_2}}(N-j+1)_{{j_1}-\delta_z+1}(r)_{j_3}}
      {j_0!{j_1}!{j_2}!j_3!}.
\end{equation}
Here $(\delta_z,\delta_w,\gamma_\delta)$ ranges over
$(0,0,1),(1,0,1),(0,1,1),(1,1,-3)$, the monomials $1,z,w,zw$ of the
multiplier $1+z+w-3zw$ with their coefficients, and the nonnegative
integers $j_0,{j_1},{j_2},j_3$ obey
$j_0+{j_1}+{j_2}+j_3=i$ and ${j_1}-j_0+\delta_w-\delta_z={\delta_{\rm off}}$.
In~\eqref{eq:binary-lifted-kernel}, $j_0$, $j_1$ and $j_2$ count the
factors $z$, $w$ and $2zw$ selected from ${\operatorname{Step}}^\ell$,
where $\ell=j_0+j_1+j_2$, and $j_3$ counts the factors $u$ selected
from $(1+u)^r$; the remaining $z$-degree $j-\delta_z-j_0-j_2$ comes from
factors $3zw$ of ${\operatorname{Stay}}^{N-\ell}$. Dividing the resulting
multinomial counts by $v_j(N)\binom Ni$ and applying the weights
gives the falling factorials, which proves
\[
 V_{\delta_{\rm off}}=(N)_i(N-j+1)({\operatorname{Mat}_i})_{j,j+{\delta_{\rm off}}}w_{j+{\delta_{\rm off}}}/w_j.
\]
All falling-factorial indices in~\eqref{eq:binary-neighbor-sum}
are nonnegative; $V_-=0$ at $j=0$.

\smallskip\noindent\emph{Step 3 (A lower bound for each signed margin).}
For $j<m$ set
\begin{align*}
s_{\rm raw}&\coloneqq N-2j+1-\tfrac43\varepsilon_-j
                         -\tfrac14\varepsilon_+(N-j+1),\\
{\operatorname{RowCert}_{\rm raw}}&\coloneqq{\operatorname{RowExt}_{\rm raw}}-(1+\varepsilon_-)V_-
                         -(1+\varepsilon_+)V_+.
\end{align*}
By the entries of ${\operatorname{Gram}_0}$ in Step~3 of
Appendix~\ref{app:binary-kernel}, whose weighted nearest entries are
$4j/[3(N-j+1)]$ below and $1/4$ above the diagonal,
$s_{\rm raw}/(N-j+1)$ is the signed row of ${\operatorname{Gram}_0}$;
moreover $s_{\rm raw}\ge(9N-37j+9)/12>0$, since $N\ge6m-1>6j$.
Every signed margin is at least
\[
 \frac{{v_i^{\rm end}}s_{\rm raw}}{N-j+1}
 -\frac{{\operatorname{RowCert}_{\rm raw}}}{(N)_i(N-j+1)}.
\]
Indeed, the signed margin is ${U_i}s_{\rm raw}/(N-j+1)$ minus the
weighted diagonal and far entries of ${\operatorname{Mat}_i}$, plus its
weighted nearest entries multiplied by $\varepsilon_\pm$. These
weighted entries are nonnegative, so their sum is at most the
extended row ${\operatorname{RowExt}_{\rm raw}}/((N)_i(N-j+1))$, and
subtracting $(1+\varepsilon_\pm)V_\pm$ restores the nearest entries
with their signs. Finally ${U_i}\ge{v_i^{\rm end}}$ and
$s_{\rm raw}>0$. At the last row use instead
\[
s_{\rm raw}\coloneqq N-2m+1-\tfrac43\varepsilon_-m,
\qquad {\operatorname{RowCert}_{\rm raw}}\coloneqq{\operatorname{RowExt}_{\rm raw}}-(1+\varepsilon_-)V_- -V_+.
\]
The upper ${\operatorname{Gram}_0}$ neighbor is absent, whereas the
upper ${\operatorname{Mat}_i}$ neighbor, in column $m+1$, belongs to
the extended row but not to the matrix; it is therefore subtracted
once and carries no sign. Here $s_{\rm raw}\ge N-10m/3+1>0$.

\smallskip\noindent\emph{Step 4 (Clearing denominators).}
For exact polynomial numerators define
\begin{align}
F_0&\coloneqq1,\quad F_1\coloneqq3N-3-4m,\quad
F_{k+1}\coloneqq(3N-3-2k-4m)F_k-3k(N-k)F_{k-1},
 \label{eq:binary-endpoint-recurrence}\\
E_i&\coloneqq(N-m)(3N-4m+3)F_{i-1}F_i,
\qquad \mathcal D_i\coloneqq3^{2i}(N-m+1)(N-1)_i.
 \label{eq:binary-endpoint-clearing}
\end{align}
The recurrence~\eqref{eq:binary-endpoint-recurrence} is the
three-term recurrence of ${\operatorname{Rec}_k}$ in Step~5 of the proof
of Lemma~\ref{lem:binary-top}, at $x=m+1$.
Thus $F_k={\operatorname{Rec}_k}(m+1)=k!P_k^{(N-1,4)}(m)$ and
${v_i^{\rm end}}=E_i/[(N)_i\mathcal D_i]$.
With $S_i\coloneqq3^{i+1}4^{i+1}$ the cleared margin is
\begin{equation}\label{eq:binary-four-sign-polynomial}
 {\operatorname{Cert}}\coloneqq12S_i s_{\rm raw}E_i-12S_i {\operatorname{RowCert}_{\rm raw}}\mathcal D_i.
\end{equation}
The multiplier relative to the margin is
$12S_i(N)_i(N-j+1)\mathcal D_i>0$.
The factor $12$ clears the denominators of $s_{\rm raw}$, and $S_i$
clears all powers of three and four in the finite sums; the factorial
quotients are multinomial integers. Thus ${\operatorname{Cert}}$ is a
polynomial with integer coefficients.

\smallskip\noindent\emph{Step 5 (Orthant cover).}
For $D\ge16$ and $3\le i\le13$ the complete coefficient
certificates use the following cover by nonnegative orthants in the
variables $a,b,c,s$.
\begin{center}\small
\begin{tabular}{p{22mm}p{81mm}rr}\toprule
Family&Substitution&Polynomials&Terms\\\midrule
Bulk&$j=a+7,m=a+b+8,r=c$&44&511808\\
Small rows&each $j=0,\ldots,6$; $m=a+8,r=c$&308&532840\\
Last row&$j=m=a+8,r=c$&22&38060\\
Small $m$&each $m=2,\ldots,7$; $r=16-2m+c$;
 every row&1320&277200\\\bottomrule
\end{tabular}
\end{center}
Always $N=6m+2r-1+s$, and $a,b,c,s\ge0$.
There are four sign choices for interior rows (at $j=0$ the choice
of $\varepsilon_-$ does not matter, and both copies are kept) and two
for the last row. The table contains $1694$ polynomials and $1359908$
nonzero coefficients. Every coefficient is nonnegative and every
constant positive. Their minimum nonzero coefficients by family
are $132505956,132505956,177855588,132505956$; as in
Appendix~\ref{app:binary-second}, such counts and minima allow a
recomputation to be compared with the certificate.
For $m\ge8$ the first three families cover every row index:
$j\le6$ (small rows), $7\le j<m$ (bulk) and $j=m$ (last row).
For $2\le m\le7$, $D\ge16$ is exactly $r\ge16-2m$,
which gives the last family. Hence every required signed margin
is strictly positive throughout its full domain: $3\le i\le13$,
$m\ge2$, $r\ge0$, $D\ge16$, $N\ge6m+2r-1$, every row $0\le j\le m$ and
every sign choice.

\subsection{Two scalar margins for indices fourteen through fifty-four}
\label{app:binary-scalar}

For $14\le i\le54$ two scalar inequalities suffice: the extended row
increases with the row index, so only the rows $j=m-1$ and $j=m$ are
checked, with the smallest signed row of ${\operatorname{Gram}_0}$ and
without retaining nearest entries.

\smallskip\noindent\emph{Step 1 (Monotonicity of the extended row in $j$).}
The extended weighted row of Appendix~\ref{app:binary-signed-rows},
which equals ${\operatorname{RowExt}_{\rm raw}}/((N)_i(N-j+1))$, is
\begin{equation}\label{eq:binary-full-row}
 {\operatorname{RowExt}_j}\coloneqq\frac{[u^i]\{\tfrac54W_j+jW_{j-1}/[3(N-j+1)]\}}{\binom Ni}.
\end{equation}
By entrywise nonnegativity it bounds the weighted absolute sum of the
finite row from above. We show that the cleared quantity
$(N-j+1)\binom Ni{\operatorname{RowExt}_j}$ increases with $j$; Step~2
uses this to reduce the rows $j<m$ to $j=m-1$.
Since $W_{j+1}=W_j(1+2u)/(1+u/4)$ and $W_{j-1}=W_j(1+u/4)/(1+2u)$, its
forward difference, before the coefficient of $u^i$ is extracted, is
\[
 u\{\tfrac{35}{16}(N-j)W_j/(1+u/4)
       +\tfrac{7j}{12}W_j/(1+2u)\}-\tfrac{11}{12}W_j.
\]
For $c=1/4,2,1$ let $X_c$ be the coefficient of $u^{i-1}$ in
$W_j/(1+cu)$. When $j=0$ (or $r=0$), $W_j/(1+2u)$ (or $W_j/(1+u)$) is
not a polynomial, but it enters only multiplied by $j$ (or $r$), and
we set $X_2\coloneqq0$ (or $X_1\coloneqq0$).
Then $X_2,X_1\le X_{1/4}$, because $1+2u$ and $1+u$ dominate $1+u/4$
coefficientwise and the remaining factors of $W_j$ have nonnegative
coefficients. Differentiating $W_j$ gives
$i[u^i]W_j=(N-j)X_{1/4}/4+2jX_2+rX_1$, and the coefficient of $u^i$ in
the forward difference is ${\operatorname{Lead}}-\tfrac{11}{12}[u^i]W_j$,
where ${\operatorname{Lead}}\coloneqq(35/16)(N-j)X_{1/4}+(7j/12)X_2$. Now
\[
 {\operatorname{Lead}}-\tfrac{11}{4}i[u^i]W_j
 \ge\frac{18N-77j-33r}{12}X_{1/4}
 \ge\frac{3N+44j-33}{24}X_{1/4}\ge0,
\]
since $r\le(N-6j+1)/2$ (from $N\ge6m+2r-1$ and $j\le m$) and $N\ge11$.
The difference coefficient is consequently at least
$11(3i-1)[u^i]W_j/12>0$.

\smallskip\noindent\emph{Step 2 (Reduction to two scalar inequalities).}
For $j<m$, the smallest signed row of ${\operatorname{Gram}_0}$ over the
four signs is $s_{\rm raw}/(N-j+1)$ with $\varepsilon_\pm=1$, namely
$(9N-37j+9)/[12(N-j+1)]$.
For every sign choice, the signed margin is at least ${U_i}$ times the
signed row of ${\operatorname{Gram}_0}$ minus ${\operatorname{RowExt}_j}$,
because the weighted entries of ${\operatorname{Mat}_i}$ are
nonnegative and their total is at most ${\operatorname{RowExt}_j}$.
After multiplication by $N-j+1$, the resulting lower bound
${v_i^{\rm end}}(9N-37j+9)/12$ decreases with $j$, while
$(N-j+1){\operatorname{RowExt}_j}$ increases by Step~1; so among the
rows $j<m$ it is enough to compare at $j=m-1$. At $j=m$ the upper
${\operatorname{Gram}_0}$ neighbour is absent, and the smallest signed
row is $(3N-10m+3)/[3(N-m+1)]$.
Since ${U_i}\ge{v_i^{\rm end}}$, the following two scalar inequalities
imply the four-sign conditions~\eqref{eq:binary-four-signs} for every
row:
\begin{equation}\label{eq:binary-two-scalars}
 {v_i^{\rm end}}\frac{9N-37j+9}{12(N-j+1)}>{\operatorname{RowExt}_j}\ (j=m-1),
 \qquad
 {v_i^{\rm end}}\frac{3N-10m+3}{3(N-m+1)}>{\operatorname{RowExt}_m}.
\end{equation}

\smallskip\noindent\emph{Step 3 (Cleared numerators).}
Set
\[
 \begin{gathered}
 G_i(j)\coloneqq4^ii![u^i]W_j,\qquad Z_i(j)\coloneqq(N-1)_iG_i(j),\\
 {\operatorname{RowCert}_i}(j)\coloneqq15(N-j+1)Z_i(j)+4jZ_i(j-1).
 \end{gathered}
\]
Put $S\coloneqq12s_{\rm raw}$ with $\varepsilon_\pm=1$, that is,
$S=9N-37j+9$ for $j=m-1$ and $S=12N-40m+12=4(3N-10m+3)$ for $j=m$.
The exact cleared numerator in either case is
\begin{equation}\label{eq:binary-scalar-polynomial}
 {\operatorname{Cert}}_i\coloneqq4^iSE_i-3^{2i}(N-m+1){\operatorname{RowCert}_i}(j).
\end{equation}
Its denominator relative to~\eqref{eq:binary-two-scalars} is
$12\cdot4^i(N)_i(N-j+1)3^{2i}(N-m+1)(N-1)_i>0$.

\smallskip\noindent\emph{Step 4 (Exact recurrences).}
Multiplying $W_j'/W_j=r/(1+u)+(N-j)/(4+u)+2j/(1+2u)$ by
$(1+u)(1+u/4)(1+2u)=1+13u/4+11u^2/4+u^3/2$ and extracting
coefficients gives, for $G_k=G_k(j)$, the integer recurrence
\begin{align}\label{eq:binary-factor32}
G_0&=1,\quad G_1=N+7j+4r,\nonumber\\
G_{k+1}&=(N+7j+4r-13k)G_k
 +4k(3N+7j+9r-11k+11)G_{k-1}\nonumber\\
&\hspace{22mm}+32k(k-1)(N+r-k+2)G_{k-2}.
\end{align}
Terms with negative subscripts are absent.
The scaling $4^kk!$ makes the coefficients integers; for instance the
coefficient $32$ comes from $4^3/2$, where $1/2$ is the coefficient of
$u^3$ in the logarithmic derivative denominator
$1+13u/4+11u^2/4+u^3/2$.
For example $N=11,j=r=0$ gives $G_3=990$ directly from
$(1+u/4)^{11}$.
The two remaining recurrences serve only the generation of the
coefficient arrays (Appendix~\ref{app:binary-certificate-spec}).
Multiplying by $(N-1)_{k+1}$ also gives a recurrence that uses
only multiplications of bounded polynomial degree:
\begin{align*}
Z_{k+1}=(N-k-1)\{&(N+7j+4r-13k)Z_k\\
 &+4k(3N+7j+9r-11k+11)(N-k)Z_{k-1}\\
 &+32k(k-1)(N+r-k+2)(N-k)(N-k+1)Z_{k-2}\}.
\end{align*}
The symmetric-square recurrence for $F_k^2,F_kF_{k-1},F_{k-1}^2$
likewise follows directly from~\eqref{eq:binary-endpoint-recurrence}.
These identities specify the complete coefficient arrays in
\eqref{eq:binary-scalar-polynomial} without interpolation, that is,
without evaluating the polynomials at sample points.

\smallskip\noindent\emph{Step 5 (Families and exhaustive partition).}
For $14\le i\le\min(54,D-1)$ the exact positive expansions are listed
below. A task is a parameter family together with its list of
indices. For $m\ge28$, $D\ge56$, so every index $14\le i\le54$ is
required. For $2\le m\le27$, a tail task fixes $m$ and covers all
$r\ge55-2m$, where again $D-1\ge54$; a fixed-lift task fixes $(m,r)$
with $\max(0,16-2m)\le r\le54-2m$.
\begin{center}\small
\begin{tabular}{p{25mm}p{78mm}rr}\toprule
Family&Substitution&Polynomials&Terms\\\midrule
$m\ge28$&$m=28+a,r=b,N=6m+2r-1+s$;
 $j=m-1,m$, all $i=14,\ldots,54$&82&6753192\\
$2\le m\le27$&26 fixed-$m$ tails $r=55-2m+b$, and
634 fixed $(m,r)$ with $\max(0,16-2m)\le r\le54-2m$;
 $N=6m+2r-1+s$&33700&7920800\\\bottomrule
\end{tabular}
\end{center}
All free variables are nonnegative. Tail tasks use all indices
$14,\ldots,54$; fixed-lift tasks retain precisely
$14\le i\le\min(54,2m+r-1)$. Both choices $j=m-1,m$ are
included. Every nonzero coefficient and every constant is positive.
The first family's minimum coefficient is
$55268479587031452650529$.
For the small-$m$ family, the $m=2,\ldots,14$ group has
478 tasks, 21870 polynomials and 4219284 nonzero coefficients;
the $m=15,\ldots,27$ group has 182 tasks, 11830 polynomials
and 3701516 coefficients. The two groups retain every one of
the 660 tasks. A tail or fixed-lift interval contains every
allowed $r$ once, so the partition is exhaustive. As before, the
counts and the minimum allow a recomputation to be compared with the
certificate.

\subsection{Finite certificate specification and conclusion}
\label{app:binary-certificate-spec}

The finite polynomial calculations above are over the integers.
The certificate package~\cite{ComputationalCertificates} contains the
complete domain lists and coefficient records. Each polynomial is
defined by
\eqref{eq:binary-second-numerator},
\eqref{eq:binary-four-sign-polynomial} or
\eqref{eq:binary-scalar-polynomial}, followed by its stated
substitution. Exact addition, multiplication and the displayed
recurrences produce its full sparse coefficient array.
The check verifies that no coefficient is negative and every constant
is positive, over the complete listed domain. Zero coefficients may be
omitted from the records. All these operations have finite specified ranges.

For indices three through fifty-four there are $35476$
polynomials and $16033900$ nonzero coefficients.
One implementation packs exponent tuples in base $256$;
the total degree in the substitution variables is at most $110$, so
polynomial multiplication cannot produce a carry between exponent
coordinates. An equivalent implementation using exponent tuples gives
the same integer arrays. These are implementation details of the
check; they play no role in the mathematics.
The eight second-index polynomials are additional.
Thus the statements of positivity are finite symbolic certificates
for entire orthants, not tests at finitely many values of $m,r,N$.

\begin{proposition}\label{prop:binary-low-band}
If $D\ge16$, then~\eqref{eq:binary-matrix-target} holds
for every $2\le i\le\min(54,D-1)$.
\end{proposition}
\begin{proof}
The second-index certificate proves $i=2$.
The direct signed-row certificate proves $3\le i\le13$,
all strictly below $D$.
For larger indices the scalar certificate proves
\eqref{eq:binary-two-scalars} in every parameter region, and Step~2 of
Appendix~\ref{app:binary-scalar} turns these two inequalities into the
four-sign conditions. All denominators are positive. The four-sign
conditions and strict weighted diagonal dominance prove the asserted
matrix comparison, even with strict inequality, in each case.
\end{proof}

\section{Uniform control of the larger binary indices}\label{app:binary-high}

This appendix proves Proposition~\ref{prop:binary-high-band}: for every
$55\le i\le D-1$, the strict form
${\operatorname{Mat}_i}\prec{U_i}{\operatorname{Gram}_0}$ of the matrix
comparison~\eqref{eq:binary-matrix-target} holds. By Step~5 of
Appendix~\ref{app:binary-kernel}, this gives $c_0g(i)\ge c_ig(0)$ at
these indices of Proposition~\ref{prop:binary-index-cover}. The indices
are divided at the threshold $X$ of~\eqref{eq:binary-strip-endpoint}.
For $55\le i\le X$ (Lemma~\ref{lem:binary-strip}), the ratio recurrence
of $L$ gives ${U_i}\ge{\operatorname{gap}_{\rm bin}}^{2i}$, but only up to
$X$, and fixed weights bound the rows of ${\operatorname{Mat}_i}$; the
lower limit $55$ is the least index at which the resulting margin is
positive. For $i>X$ the weights are adapted to $i$:
Lemma~\ref{lem:binary-adaptive-row} reduces the comparison to
$\mathcal B_i/{U_i}\le1/8$ for a saddle-point bound $\mathcal B_i$ of
the rows, and Lemma~\ref{lem:binary-affine-gap} compares the exponent
$E_{\nu,\rho}$ of $\mathcal B_i$ with a continuous model $L_\nu$ of
$\log{U_i}$; it needs $\alpha/\nu\ge1/13$, which holds for $i>X$.
Lemmas~\ref{lem:binary-binomial-prefactor} and~\ref{lem:binary-midpoint}
control the finite corrections.

We keep $m\ge2$, $r\ge0$, $D=2m+r$ and $N\ge6m+2r-1$, and call this
the finite domain. The matrices ${\operatorname{Mat}_i}$,
${\operatorname{Gram}_\ell}$ and ${\operatorname{Sph}_\ell}$ and the
four-sign conditions~\eqref{eq:binary-four-signs} are those of
Appendix~\ref{app:binary-matrices}, and ${U_i}$ is the Lloyd ratio
of~\eqref{eq:binary-endpoint}; for symmetric matrices, $A\prec B$
means $B-A>0$. Throughout, $R_x\coloneqq L(x+1)/L(x)$.

\subsection{A finite recurrence strip}

\begin{lemma}\label{lem:binary-strip}
Put
\begin{equation}\label{eq:binary-strip-endpoint}
 X\coloneqq\frac{N(2N-3m+6)}{3(8N-9m)}.
\end{equation}
Then ${\operatorname{Mat}_i}\prec {U_i}{\operatorname{Gram}_0}$ whenever
$55\le i\le\min(D,\lfloor X\rfloor)$.
\end{lemma}
\begin{proof}
\smallskip\noindent\emph{Step 1 (The ratio recurrence and ${U_i}\ge{\operatorname{gap}_{\rm bin}}^{2i}$).}
Let ${\operatorname{gap}_{\rm bin}}\coloneqq1-3m/(2N)>0$. By Step~8 of
Appendix~\ref{app:binary-kernel}, $L(x+1)/L(1)=F_x/((N-1)_x3^x)$, with
$F_x$ as in~\eqref{eq:binary-endpoint-recurrence}; hence
$R_x=F_x/(3(N-x)F_{x-1})$ for $1\le x\le D$, where $L(x)>0$, and the
recurrence~\eqref{eq:binary-endpoint-recurrence} becomes the exact
Krawtchouk recurrence in the argument,
\begin{equation}\label{eq:binary-ratio-recurrence}
 3(N-x)R_x=3N-1-2x-4m-\frac{x-1}{R_{x-1}}.
\end{equation}
At $x=1$ the last term vanishes.
If $R_{x-1}\ge {\operatorname{gap}_{\rm bin}}$, the right side is at least
$3N-1-2x-4m-(x-1)/{\operatorname{gap}_{\rm bin}}$. After multiplication
by ${\operatorname{gap}_{\rm bin}}$, the comparison of this lower bound
with $3(N-x){\operatorname{gap}_{\rm bin}}$ is linear in $x$, and the lower
bound is at least $3(N-x){\operatorname{gap}_{\rm bin}}$ precisely when
\[
 x\le\frac{(m/2){\operatorname{gap}_{\rm bin}}+1-{\operatorname{gap}_{\rm bin}}}{(1-{\operatorname{gap}_{\rm bin}})(3{\operatorname{gap}_{\rm bin}}+1)}=X.
\]
The same computation without the last term gives the base case
$R_1\ge{\operatorname{gap}_{\rm bin}}$, since $1\le X$ whenever the strip
is nonempty. Moreover $X\le N/6$,
since this is equivalent to $3m+12\le4N$.
Thus every denominator ($N-x$, $8N-9m$ and, inductively, $R_{x-1}$) is
positive, and $R_x\ge{\operatorname{gap}_{\rm bin}}$ for $1\le x\le i$
whenever $i\le\min(D,X)$. Step~8 of Appendix~\ref{app:binary-kernel}
also gives
\[
 R_0\ge\frac{(N-m)(3N-4m+3)}{3N(N-m+1)}\ge {\operatorname{gap}_{\rm bin}},
\]
where the final difference is $m(N-m+3)/[6N(N-m+1)]$.
Consequently, for $i\le\min(D,X)$,
\begin{equation}\label{eq:binary-strip-H}
 {U_i}=R_0R_i\prod_{x=1}^{i-1}R_x^2\ge {\operatorname{gap}_{\rm bin}}^{2i}.
\end{equation}
This is where the upper limit $X$ of the strip comes from.

\smallskip\noindent\emph{Step 2 (A row bound with fixed weights).}
Take $w_j\coloneqq\tau_0^j\sqrt{\binom Nj3^j}$ with $\tau_0\coloneqq1/4$, the weights of
Appendix~\ref{app:binary-matrices}, and let
${\operatorname{Row}_{\ell,j}}\coloneqq\sum_{k\le m}({\operatorname{Sph}_\ell})_{jk}w_k/w_j$
be the weighted row of the sphere matrix ${\operatorname{Sph}_\ell}$
of~\eqref{eq:binary-sphere-matrix}.
By~\eqref{eq:binary-congruence}, ${\operatorname{Gram}_\ell}$ is
${\operatorname{Sph}_\ell}$ plus two shifted copies minus a term with
nonnegative entries; they correspond to the terms $1$, $w$, $z$ and
$-3zw$ of the multiplier in~\eqref{eq:binary-C-kernel}. Dropping only
the negative term increases the weighted row. With these weights the
copy for $w$ is the row of ${\operatorname{Sph}_\ell}$ moved one
column to the right, multiplied by $\tau_0$ and without its last
entry; the copy for $z$ is the row $j-1$ of ${\operatorname{Sph}_\ell}$
multiplied by $\theta_j^2/\tau_0$. This gives
\[
 \sum_{k\le m}({\operatorname{Gram}_\ell})_{jk}\frac{w_k}{w_j}
 \le(1+\tau_0){\operatorname{Row}_{\ell,j}}+
       \frac{\theta_j^2}{\tau_0}{\operatorname{Row}_{\ell,j-1}}
 \le\frac{91}{60}\max_{j'\le m}{\operatorname{Row}_{\ell,j'}}.
\]
Here $\theta_j^2=j/[3(N-j+1)]\le1/15$; the second term
is absent at $j=0$. Since the shifted row has a smaller cutoff, its
replacement by ${\operatorname{Row}_{\ell,j}}$ is in the upper-bound
direction. Extending sphere rows to all columns $k\ge0$ gives
\[
 {\operatorname{Row}_{\ell,j}}\le
 \frac{[u^\ell](1+u/4)^{N-j}(1+2u)^j}{\binom N\ell}
 \le\left(\frac14+\frac{7j}{4N}\right)^\ell.
\]
The middle expression is the extended row: it is the kernel of
${\operatorname{Sph}_\ell}$ evaluated at $w=1/4$, as in Step~1 of
Appendix~\ref{app:binary-signed-rows} without the multiplier and the
lift. The last step is Maclaurin's inequality: the $\ell$th elementary
symmetric function of $N$ nonnegative numbers, here $N-j$ copies of
$1/4$ and $j$ copies of $2$, is at most $\binom N\ell$ times the $\ell$th
power of their mean $1/4+7j/(4N)$. A proof is the last paragraph of the
proof of Lemma~\ref{lem:binary-geometric} in
Appendix~\ref{app:binary-small}, which does not depend on this
appendix.
Using~\eqref{eq:binary-lifted-matrix}, including its $\ell=0$ term
(${\operatorname{Sph}_0}$ is the identity), therefore yields
\begin{equation}\label{eq:binary-strip-row}
 \sum_{k\le m}({\operatorname{Mat}_i})_{jk}\frac{w_k}{w_j}
 \le\frac{91}{60}\left(\lambda+\frac r{N-i+1}\right)^i,
 \qquad\lambda\coloneqq\frac14+\frac{7m}{4N}.
\end{equation}
Indeed the lift weight equals $\binom i\ell(r)_{i-\ell}/(N-\ell)_{i-\ell}$,
$(r)_{i-\ell}/(N-\ell)_{i-\ell}\le[r/(N-i+1)]^{i-\ell}$, the row bounds
above hold with $j$ replaced by $m$, and the binomial theorem sums the
lift over $0\le\ell\le i$.

\smallskip\noindent\emph{Step 3 (The base of the row bound).}
We next verify $\lambda+r/(N-i+1)\le(32/33){\operatorname{gap}_{\rm bin}}^2$;
the factor $32/33$ is chosen so that $91(32/33)^i<17$ in Step~4.
The left side increases with $r$ and with $i$, so we may substitute the
upper bounds $r\le(N-6m+1)/2$ and $i\le X$.
For $Z_{\rm strip}\coloneqq22N^2-24Nm+18N-27m$, one has
$N-X+1=Z_{\rm strip}/[3(8N-9m)]>0$.
Clearing positive denominators reduces the claim to $F_{\rm strip}>0$,
where
\begin{align*}
F_{\rm strip}\coloneqq{}&32(2N-3m)^2Z_{\rm strip}
 -33\{N(N+7m)Z_{\rm strip}+6N^2(N-6m+1)(8N-9m)\}\\
={}&506N^4-4524mN^3+126N^3+10404m^2N^2-11853mN^2\\
 &-6912m^3N+21789m^2N-7776m^3.
\end{align*}
At $N=6m-1+s$ it equals
\begin{align*}
506s^4&+(7620m-1898)s^3+(38268m^2-32445m+2658)s^2\\
 &+(66528m^3-183375m^2+42030m-1646)s\\
 &+(18m^2-525m+95)(648m^2-159m+4).
\end{align*}
All coefficients in $s$ are positive for $m\ge29$:
the first constant factor is $8$ at $m=29$ and increases, and the
second is positive for every $m\ge1$.
At $N=10m+s$ the expansion is
\begin{align*}
506s^4&+(15716m+126)s^3+(178284m^2-8073m)s^2\\
 &+(867968m^3-177471m^2)s+m^3(1507280m-849186),
\end{align*}
positive for $m\ge2$.
These regions cover the strip, that is, all parameters with
$55\le i\le\min(D,X)$: $55\le i\le X\le N/6$
forces $N\ge330$, so $m\le28$ implies $N>10m$.

\smallskip\noindent\emph{Step 4 (Signed margins).}
The weighted nearest entries of ${\operatorname{Gram}_0}$ are $\tau_0$
above and $\theta_j^2/\tau_0$ below the diagonal, as in Step~3 of
Appendix~\ref{app:binary-signed-rows}. Hence every signed weighted
${\operatorname{Gram}_0}$ row is at least
\[
 1-\frac j{N-j+1}-\tau_0-\frac{\theta_j^2}{\tau_0}\ge\frac{17}{60}.
\]
Absent neighbors improve this bound. By nonnegative entries of
${\operatorname{Mat}_i}$, equations~\eqref{eq:binary-strip-H} and
\eqref{eq:binary-strip-row} bound every signed margin below by
\[
 \frac{{\operatorname{gap}_{\rm bin}}^{2i}}{60}\{17-91(32/33)^i\}>0.
\]
The last inequality follows from the exact integer comparison
$17\cdot33^{55}>91\cdot32^{55}$. Since $55$ is the least index with
$17\cdot33^i>91\cdot32^i$, the strip, and with it this appendix,
starts at $i=55$.
Weighted diagonal dominance proves the lemma.
\end{proof}

\subsection{An adaptive bound retaining the finite cutoff}
\label{app:binary-adaptive}

For $i>X$ the weights are adapted to the index.
Write $\nu\coloneqq N/m$, $\rho\coloneqq r/m$, $v\coloneqq\nu-1$ and
$\tau\coloneqq1/\sqrt{3v}$; in the finite domain
$\nu\ge6+2\rho-1/m\ge11/2+2\rho$. For $u>0$ let $a=a(u)>0$ solve
\begin{equation}\label{eq:binary-adaptive-root}
 v(3+2u)a^2+(v-1)ua-1=0,
 \qquad b\coloneqq(2+a^{-1})/3.
\end{equation}
Then $0<a<\tau$, since the left side is $-1$ at $a=0$ and
$2u/3+(v-1)u\tau>0$ at $a=\tau$. As shown after
Lemma~\ref{lem:binary-adaptive-row}, $a(u)$ is the weight tilt that
minimizes the bound of that lemma at the point $u$. Define
\begin{equation}\label{eq:binary-adaptive-B}
 \Phi(u)\coloneqq(1+au)^{N-m}(1+bu)^m(1+u)^r,
 \qquad \mathcal B_i\coloneqq\inf_{u>0}\frac{\Phi(u)}{\binom Ni u^i},
\end{equation}
with $a=a(u)$ and $b=b(u)$. By the proof of
Lemma~\ref{lem:binary-adaptive-row}, $\mathcal B_i$ bounds the weighted
rows of ${\operatorname{Mat}_i}$ before the congruence.

\begin{lemma}\label{lem:binary-adaptive-row}
If $\mathcal B_i/{U_i}\le1/8$, then
${\operatorname{Mat}_i}\prec {U_i}{\operatorname{Gram}_0}$.
\end{lemma}
\begin{proof}
\smallskip\noindent\emph{Step 1 (Rows before the congruence).}
Use the weights $w_j\coloneqq\tau^j\sqrt{\binom Nj3^j}$, and let
${\operatorname{Sph}}^{(i)}$ be ${\operatorname{Mat}_i}$ before the
congruence, that is, \eqref{eq:binary-lifted-matrix} with
${\operatorname{Sph}_\ell}$ in place of ${\operatorname{Gram}_\ell}$.
Fix $u_0>0$ and put $a\coloneqq a(u_0)$, $b\coloneqq b(u_0)$. For every column $k\le m$,
$1\le(a/\tau)^{k-m}$, since $a<\tau$. Multiplying the summands of a
weighted finite row of ${\operatorname{Sph}}^{(i)}$ by these factors
and then extending the nonnegative sum to all columns $k\ge0$ gives
$(a/\tau)^{j-m}$ times the extended row for the weights
$a^k\sqrt{\binom Nk3^k}$. Evaluating the lifted kernel at $w=a$, as in
Step~1 of Appendix~\ref{app:binary-signed-rows}, gives the upper bound
\[
 \frac{(a/\tau)^{j-m}}{\binom Ni}[u^i]
 (1+au)^{N-j}(1+bu)^j(1+u)^r,
\]
in which $a$ and $b$ are fixed numbers. This polynomial in $u$ has
nonnegative coefficients, so the coefficient of $u^i$ is at most its
value at $u=u_0$ divided by $u_0^i$. Moreover
\[
 (a/\tau)^{j-m}(1+au_0)^{N-j}(1+bu_0)^j
 =(a/\tau)^{-m}(1+au_0)^N\left(\frac{(a/\tau)(1+bu_0)}{1+au_0}\right)^j,
\]
and the base equals $\tau/a>1$, because
$3a^2v(1+bu_0)=1+au_0$ by~\eqref{eq:binary-adaptive-root}. Its maximum
is at $j=m$, where the bound is $\Phi(u_0)/(\binom Niu_0^i)$; the
infimum over $u_0$ gives~\eqref{eq:binary-adaptive-B}. Thus every
weighted row of ${\operatorname{Sph}}^{(i)}$ is at most $\mathcal B_i$.

\smallskip\noindent\emph{Step 2 (The congruence and the signed margins).}
As in Step~2 of the proof of Lemma~\ref{lem:binary-strip}, now with
$\tau$ in place of $\tau_0$, the weighted row $j$ of
${\operatorname{Mat}_i}$ is at most $1+\tau$ times the row $j$ plus
$\theta_j^2/\tau$ times the row $j-1$ of ${\operatorname{Sph}}^{(i)}$.
Dropping the negative congruence term therefore costs at most the factor
$\kappa_N\coloneqq1+\tau+\theta_m^2/\tau$, and every signed ${\operatorname{Gram}_0}$ row
is at least
$s_N\coloneqq1-m/(N-m+1)-\tau-\theta_m^2/\tau$; these are the analogues of
$91/60$ and $17/60$ in that proof.
Since $\nu\ge11/2$, we have $\tau^2\le2/27$, $\theta_m^2<\tau^2$ and
$m/(N-m+1)<3\tau^2$, so
\[
 s_N>1-2/9-\sqrt{8/27}>1/5,
 \qquad \kappa_N<1+\sqrt{8/27}<8/5.
\]
Every signed margin is at least $s_N{U_i}$ minus the weighted row of
${\operatorname{Mat}_i}$, hence at least $s_N{U_i}-\kappa_N\mathcal B_i$,
which is positive when $\mathcal B_i/{U_i}\le1/8<s_N/\kappa_N$.
Thus~\eqref{eq:binary-four-signs} proves the claim.
\end{proof}

\smallskip\noindent\emph{The saddle point.}
To compare $\mathcal B_i$ with ${U_i}$ we study the minimizing $u$
in~\eqref{eq:binary-adaptive-B}. Define, with $a=a(u)$,
\begin{equation}\label{eq:binary-saddle-functions}
 f_0(u)\coloneqq\frac{2va(1+a)}{1+au},\quad
 f(u)\coloneqq f_0(u)+\frac\rho{1+u},\quad
 \alpha(u)\coloneqq uf(u),\quad {\operatorname{Saddle}}(u)\coloneqq\frac{f(u)}{\nu-uf(u)};
\end{equation}
$f_0$ is the part of $f$ that does not depend on $\rho$.
In the variables $y\coloneqq\log u$ and $\zeta\coloneqq\log a$, regarding $a$ for the
moment as free and putting $b\coloneqq(2+a^{-1})/3$, the normalized
logarithm $m^{-1}\log\Phi$ is
\[
 v\log(1+e^{y+\zeta})+
 \log(1+\tfrac23e^y+\tfrac13e^{y-\zeta})+
 \rho\log(1+e^y).
\]
The second term has positive-definite Hessian: its exponent
vectors $(0,0),(1,0),(1,-1)$ are affinely independent and all
three weights are positive. The other two terms are convex. For fixed
$y$ the derivative in $\zeta$ vanishes exactly when
$3a^2v(1+bu)=1+au$, which is~\eqref{eq:binary-adaptive-root}; so the
unique stationary point in $\zeta$ is $a=a(u)$, the minimizing tilt.
The positive Schur complement of the Hessian shows that the minimized
expression is strictly convex in $y$. By the envelope theorem its
derivative in $y$ is $u[va/(1+au)+b/(1+bu)+\rho/(1+u)]$ at $a=a(u)$,
and the root relation $1+bu=(1+au)/(3a^2v)$ turns this into
$uf(u)=\alpha(u)$. Hence $\alpha$ is strictly increasing. As $u\to0$,
$a\to\tau$ and $\alpha\to0$; as $u\to\infty$, $au\to1/(v-1)$ and
$\alpha\to\rho+2$. Thus every $0<\alpha<\rho+2$ has a unique saddle,
the $u>0$ with $\alpha(u)=\alpha$; for $i=\alpha m$ it is the minimizing
point in~\eqref{eq:binary-adaptive-B}.

\smallskip\noindent\emph{The saddle function decreases.}
We also need that ${\operatorname{Saddle}}$ decreases in $u$; this is
used in two regions of the proof of Lemma~\ref{lem:binary-affine-gap}.
The following rational identity provides this fact with its signs
explicit. Put
\[
 T\coloneqq v-1+2va,\quad {\delta_{\rm aux}}\coloneqq(1-a)(1+3a),\quad
 W\coloneqq v-1+4va+3v(v-1)a^2.
\]
Then $u=(1-3va^2)/(aT)$, $1+au=v{\delta_{\rm aux}}/T$,
$f_0=2a(1+a)T/{\delta_{\rm aux}}$, and, with the prime denoting $d/du$
along $a=a(u)$,
\begin{equation}\label{eq:binary-saddle-derivative}
 -f_0'-f_0^2/4=\frac{a^2T^2{\operatorname{Pos}}}{{\delta_{\rm aux}}^2W}>0,
\end{equation}
where
\[
{\operatorname{Pos}}\coloneqq\{v-1+va(2+a)\}(1-3va^2)
 +(4v-2)a+(20v-9)a^2+9va^3(2-a)>0;
\]
this is a rational-function identity in $v$ and $a$, checked by direct
expansion. All denominators are positive for $v>1$ and
$0<a<1/\sqrt{3v}$. Since $-f_0'>f_0^2/4$ and the derivative of
$\rho/(1+u)$ is $-\rho/(1+u)^2$, the weighted Cauchy--Schwarz
inequality $f_0^2/4+(\rho/(1+u))^2/\rho\ge f^2/(\rho+4)$, in which the
second term is omitted when $\rho=0$, gives $-f'>f^2/(\rho+4)$.
Since $\nu>\rho+4$ in the finite domain,
${\operatorname{Saddle}}'=(\nu f'+f^2)/(\nu-uf)^2<0$.

\smallskip\noindent\emph{The exponent.}
Let ${H_2(x)}\coloneqq-x\log x-(1-x)\log(1-x)$ denote entropy with
natural logarithms. The normalized saddle exponent is
\begin{equation}\label{eq:binary-entropy-exponent}
 E_{\nu,\rho}(\alpha)\coloneqq\inf_{u>0}
 \{(\nu-1)\log(1+au)+\log(1+bu)+\rho\log(1+u)
                         -\alpha\log u\}
 -\nu {H_2(\alpha/\nu)},
\end{equation}
with $a=a(u)$ and $b=b(u)$. By
Lemma~\ref{lem:binary-binomial-prefactor},
$\mathcal B_i\le3\sqrt i\exp(mE_{\nu,\rho}(i/m))$; see the proof of
Proposition~\ref{prop:binary-high-band}. Since $a(u)$ minimizes over the
tilt, the infimum is a joint infimum over $u$ and $a$. By the envelope
theorem and $H_2'(x)=\log((1-x)/x)$, the derivative of $E_{\nu,\rho}$ is
$-\log u+\log(\alpha/(\nu-\alpha))$ at the saddle $u$, which is
$\log {\operatorname{Saddle}}(u)$ since $\alpha=uf(u)$; and
$E_{\nu,\rho}(0)=0$.

\subsection{A strengthened continuous comparison}
\label{app:binary-continuous}

The ratio recurrence~\eqref{eq:binary-ratio-recurrence} has a
continuous model. For real $\nu\ge6+2\rho$ and $\rho\ge0$, which we call
the continuous domain, and for $0\le s\le\rho+2$, let
$R_\nu(s)$ be the larger root of
\begin{equation}\label{eq:binary-continuous-root}
 3(\nu-s)R^2-(3\nu-2s-4)R+s=0.
\end{equation}
This is the fixed-point form of~\eqref{eq:binary-ratio-recurrence},
divided by $m$, with $N=\nu m$, $x=sm$ and $R_x=R_{x-1}=R$, and with
the terms of order $1/m$ omitted. Thus $R_\nu(s)$ models $R_x$ at $x=sm$,
and $L_\nu(\alpha)\coloneqq2\int_0^\alpha\log R_\nu(s)\,ds$ models
$m^{-1}\log{U_i}$ at $i=\alpha m$ (Lemma~\ref{lem:binary-midpoint}).
The root lies in $(0,1)$ and decreases with $s$. Indeed, the quadratic
equals $s\ge0$ at $R=0$ and $4$ at $R=1$, and its vertex lies in
$(0,1)$, since $0<3\nu-2s-4$ and $4s<3\nu+4$; the discriminant
$\Delta\coloneqq(3\nu-2s-4)^2-12s(\nu-s)$ decreases in $s$
through the stated range and is positive at its right endpoint.
Implicit differentiation gives
\begin{equation}\label{eq:binary-R-derivatives}
 \partial_\nu R=\frac{3R(1-R)}{\sqrt\Delta}>0,
 \qquad
 \partial_sR=-\frac{(1-R)(3R+1)}{\sqrt\Delta}<0.
\end{equation}
At $s=(\nu-2)/2$ the larger root is $(\nu-2)/(\nu+2)$;
the other is $1/3$. Since $R_\nu$ decreases in $s$, this supplies the
lower bound $R_\nu(s)\ge(\nu-2)/(\nu+2)\ge1/2$ throughout
$s\le\rho+2\le(\nu-2)/2$; it reappears as $R_{\min}$ below and in the
proof of Lemma~\ref{lem:binary-midpoint}. The next lemma compares the
derivatives of $L_\nu$ and $E_{\nu,\rho}$, namely $\log R_\nu(\alpha)^2$
and $\log{\operatorname{Saddle}}$, with an affine margin that integrates
to the gap $13\alpha/100$ used in
Proposition~\ref{prop:binary-high-band}.

\begin{lemma}\label{lem:binary-affine-gap}
On this continuous domain, with $\alpha=\alpha(u)$,
\begin{equation}\label{eq:binary-affine-gap}
 \frac{R_\nu(\alpha)^2}{{\operatorname{Saddle}}(u)}>
 \frac{10}{9}\left(1+\frac{2\alpha}{3\nu}\right).
\end{equation}
If $\alpha/\nu\ge1/13$ and
$L_\nu(\alpha)=2\int_0^\alpha\log R_\nu(s)\,ds$, then
\begin{equation}\label{eq:binary-integrated-gap}
 L_\nu(\alpha)-E_{\nu,\rho}(\alpha)>13\alpha/100.
\end{equation}
\end{lemma}
\begin{proof}
\smallskip\noindent\emph{Step 1 (The extremal $\rho$).}
The tilt $a=a(u)$ depends only on $v$ and $u$, not on $\rho$. Hence, at
fixed $\nu,u$, increasing $\rho$ increases $f$, $\alpha$,
${\operatorname{Saddle}}$ and the affine factor, and decreases
$R_\nu(\alpha)$. It suffices to take $\rho=(\nu-6)/2$, or $v=\nu-1\ge5$,
$\rho=(v-5)/2$. We call the set of pairs $(v,a)$ with $v\ge5$ and
$0<a<1/\sqrt{3v}$ the physical domain: these are exactly the pairs
with $a=a(u)$ for some $u>0$.

\smallskip\noindent\emph{Step 2 (A rational parametrization).}
Define the following polynomials in $v,a$; they are local to this
proof and unrelated to the ratios ${U_i}$:
\begin{gather*}
 U\coloneqq1-3va^2,\quad V\coloneqq a(v-1+2va),\quad S\coloneqq U+V,\quad {\delta_{\rm aux}}\coloneqq(1-a)(1+3a),\\
 {\operatorname{Aux}}\coloneqq4(1+a)S+(v-5){\delta_{\rm aux}},\quad {\operatorname{Num}}\coloneqq U{\operatorname{Aux}},\quad {\operatorname{Den}}\coloneqq2{\delta_{\rm aux}}S,\\
 {\operatorname{Disc}_0}\coloneqq(v+1){\operatorname{Den}}-{\operatorname{Num}},\qquad {\operatorname{Disc}_1}\coloneqq(3v-1){\operatorname{Den}}-2{\operatorname{Num}}.
\end{gather*}
Solving~\eqref{eq:binary-adaptive-root} for $u$ gives $u=U/V$. Then
$uf_0=2(1+a)U/{\delta_{\rm aux}}$ and $\rho u/(1+u)=\rho U/S$, which give
$\alpha=uf={\operatorname{Num}}/{\operatorname{Den}}$; and
${\operatorname{Saddle}}=\alpha/(u(\nu-\alpha))=V{\operatorname{Aux}}/{\operatorname{Disc}_0}$.
Moreover $R_\nu(\alpha)\ge(v-1)/(v+3)\eqqcolon R_{\min}$, the lower bound above.
All denominator factors are positive on
$0<a<1/\sqrt{3v}$: in particular
${\operatorname{Disc}_0}={\operatorname{Den}}(\nu-\alpha)>0$ and
${\operatorname{Disc}_1}={\operatorname{Den}}(3\nu-2\alpha-4)>0$.

\smallskip\noindent\emph{Step 3 (A sufficient polynomial condition).}
Put
\[
 \begin{gathered}
 F_*\coloneqq10[3(v+1){\operatorname{Den}}+2{\operatorname{Num}}],\quad G_*\coloneqq27(v+1){\operatorname{Den}},
 \\
 Z\coloneqq F_*G_*V{\operatorname{Disc}_1}^2-{\operatorname{Disc}_0}{\operatorname{Aux}}(3F_*V+G_*U)^2,
 \end{gathered}
\]
so that $F_*/G_*=\tfrac{10}9(1+2\alpha/(3\nu))$ is the affine factor.
For $Y\coloneqq((F_*/G_*){\operatorname{Saddle}})^{1/2}>0$, exact multiplication
gives
\[
 {\operatorname{Disc}_1}^2Y^2-(3{\operatorname{Disc}_0}Y^2+{\operatorname{Num}})^2={\operatorname{Aux}}Z/(G_*^2{\operatorname{Disc}_0}).
\]
If $Z>0$, then, since $Y$, ${\operatorname{Disc}_1}$, ${\operatorname{Aux}}$,
${\operatorname{Disc}_0}$ and $3{\operatorname{Disc}_0}Y^2+{\operatorname{Num}}$
are positive, taking square roots gives
$3{\operatorname{Disc}_0}Y^2-{\operatorname{Disc}_1}Y+{\operatorname{Num}}<0$.
Division by ${\operatorname{Den}}$ turns this into
$3(\nu-\alpha)Y^2-(3\nu-2\alpha-4)Y+\alpha<0$.
Thus $Y$ lies between the two roots of
\eqref{eq:binary-continuous-root} at $s=\alpha$, so $Y<R_\nu(\alpha)$,
proving~\eqref{eq:binary-affine-gap}. Since $R_\nu(\alpha)\ge R_{\min}$,
the inequality $R_{\min}^2>(F_*/G_*){\operatorname{Saddle}}$, or any
inequality implying it, is also sufficient. Step~4 proves $Z>0$ for
$a\ge1/20$; Steps~5 and~6 use the second sufficient condition.

\smallskip\noindent\emph{Step 4 (The region $a\ge1/20$).}
First suppose $a\ge1/20$. The image of $t,w\ge0$ under
\begin{equation}\label{eq:binary-affine-substitution}
 v=\frac{15a^2+(1-15a^2)z}{3a^2},\qquad
 a=\frac{6+31t}{120(1+t)},\qquad z=\frac{w}{1+w},
 \qquad t,w\ge0,
\end{equation}
contains the physical domain with $a\ge1/20$.
Indeed $z=3a^2(v-5)/(1-15a^2)\in[0,1)$ there, and
$1/\sqrt{15}<31/120$, so the range $[1/20,31/120)$ of $a$ contains every
physical $a\ge1/20$.
Multiply $Z$ by $3^9a^{18}$ after the first substitution; this clears
the denominators $3a^2$, since $Z$ has degree $9$ in $v$, and
the result has degrees $9$ in $z$ and $40$ in $a$.
After the other substitutions, multiplication by the positive
$120^{40}(1+t)^{40}(1+w)^9$, which clears the denominators of $a$ and
$z$, gives an integer polynomial
with exactly $410$ nonzero coefficients, all positive.
Its minimum coefficient is
\[
54844392567923673464097498788936967716760071626997807185920000000000000.
\]
The defining factored polynomial and substitutions specify the
array uniquely; the certificate package~\cite{ComputationalCertificates}
contains its complete exact expansion.
Positivity is read off the coefficients directly: no Bernstein degree
elevation and no sampled parameter values are used.

\smallskip\noindent\emph{Step 5 (The region $a\le1/20$, $5\le v\le105$).}
For $a\le1/20$ and $5\le v\le105$, the point $a=1/20$
is physical, since $3v/400<1$. Since ${\operatorname{Saddle}}$ decreases
in $u$, by~\eqref{eq:binary-saddle-derivative} and the Cauchy--Schwarz
step after it, and $u=U/V$ decreases in $a$, ${\operatorname{Saddle}}$
increases with $a$, so it is at most its value at $a=1/20$.
Also $F_*/G_*\le20(2v+1)/[27(v+1)]$ because
$\alpha\le\rho+2=(v-1)/2$. With these two bounds, the sufficient
comparison $R_{\min}^2>(F_*/G_*){\operatorname{Saddle}}$ has, at
$a=1/20$, the numerator
\[
 27(v+1){\operatorname{Disc}_0}(v-1)^2-20(2v+1)V{\operatorname{Aux}}(v+3)^2.
\]
Multiplying by $20^5$ and setting $v=5+s$ gives, in increasing powers
of $s$, the coefficients
\[
(38749968000,28932616320,7804473896,909543604,43157284,708776).
\]
They are positive, so $R_{\min}^2>(F_*/G_*){\operatorname{Saddle}}$ on this
region. The cut at $v=105$, that is, at $\rho=50$, is where Step~6
begins.

\smallskip\noindent\emph{Step 6 (The region $v\ge105$).}
For $v\ge105$, write $\rho=(v-5)/2\ge50$.
Since ${\operatorname{Saddle}}(u)$ decreases, it is bounded by its limit
${\operatorname{Saddle}}_0$ as $u\to0$; moreover $R_\nu(\alpha)\ge R_{\min}$,
and $F_*/G_*\le k_{\max}$, its value at $\alpha=\rho+2$. Here
\[
 {\operatorname{Saddle}}_0=\frac{\rho+2/3+2\sqrt{(2\rho+5)/3}}{2\rho+6},\quad
 R_{\min}=\frac{\rho+2}{\rho+4},\quad
 k_{\max}=\frac{10(8\rho+22)}{27(2\rho+6)}.
\]
To prove $R_{\min}^2>k_{\max}{\operatorname{Saddle}}_0$, put
\[
{\operatorname{Pos}_{\rm tail}}\coloneqq81(2\rho+6)^2(\rho+2)^2
       -10(8\rho+22)(3\rho+2)(\rho+4)^2.
\]
After clearing denominators, the comparison reads
\[
 {\operatorname{Pos}_{\rm tail}}>60(8\rho+22)(\rho+4)^2\sqrt{(2\rho+5)/3}.
\]
After $\rho=50+s$, the coefficients of ${\operatorname{Pos}_{\rm tail}}$,
in increasing powers of $s$, are
\[
 (590514624,45867600,1336148,17300,84),
\]
so both sides are positive.
Squaring, with $60^2/3=1200$, the comparison is equivalent to the
positivity of
\[
 {\operatorname{Pos}_{\rm tail}}^2-1200(8\rho+22)^2(2\rho+5)(\rho+4)^4,
\]
whose coefficients, in increasing powers of $s$, are
\begin{gather*}
(157910984845157376,29169688734259200,2277869452418304,\\
99202381776000,2651644326736,44728877600,466316464,2752800,7056).
\end{gather*}
They are positive. The three regions of Steps~4--6 cover the physical
domain and prove~\eqref{eq:binary-affine-gap}.

\smallskip\noindent\emph{Step 7 (Integration).}
Both exponents vanish at zero, and their derivatives are
$\log(R_\nu(\alpha)^2)$ and $\log {\operatorname{Saddle}}(u)$, the latter
at the saddle $u$ of $\alpha$. Integrate the affine comparison over
$[0,\alpha]$, and put $x\coloneqq\alpha/\nu$, which is less than $1/2$ since
$\alpha<\rho+2\le(\nu-2)/2$. The inequality
$\log(1+2s/(3\nu))\ge2s/(3\nu)-2s^2/(9\nu^2)$ yields
\[
 \frac{L_\nu(\alpha)-E_{\nu,\rho}(\alpha)}{\alpha}>
 \log(10/9)+x/3-2x^2/27.
\]
The polynomial part increases on $[0,1/2]$.
If $x\ge1/13$, use $\log(10/9)\ge17/162$ to obtain
\[
 \frac{L_\nu(\alpha)-E_{\nu,\rho}(\alpha)}{\alpha}>
 \frac{17}{162}+\frac1{39}-\frac2{4563}
 =\frac{13}{100}+\frac{193}{1368900}>\frac{13}{100}.
\]
This proves~\eqref{eq:binary-integrated-gap}.
\end{proof}

\subsection{Finite normalization and the midpoint correction}

The next bound is~\eqref{eq:binomial-three} of
Appendix~\ref{app:qutrit-circle-factors}, together with the case $i=N$;
we include its short proof.

\begin{lemma}\label{lem:binary-binomial-prefactor}
For integers $N\ge1$ and $1\le i\le N$,
\[
 \binom Ni\ge\frac{\exp(N{H_2(i/N)})}{3\sqrt i}.
\]
\end{lemma}
\begin{proof}
For $1\le i<N$, use the finite Stirling inequalities
\cite[equations~(1)--(2)]{Robbins}
\[
 \sqrt{2\pi}\,k^{k+1/2}e^{-k}<k!
 <\sqrt{2\pi}\,k^{k+1/2}e^{-k+1/(12k)}.
\]
Applying the lower bound to $N!$ and upper bounds to the two
denominator factorials gives
$\binom Ni e^{-N{H_2(i/N)}}>e^{-1/6}/\sqrt{2\pi i}>1/(3\sqrt i)$.
For the last strict inequality, $\pi<22/7$ follows from
$\int_0^1x^4(1-x)^4/(1+x^2)\,dx=22/7-\pi>0$.
The exponential series gives $e<87/32<(7/5)^3$;
thus $2\pi e^{1/3}<(44/7)(7/5)<9$.
At $i=N$ the conclusion is immediate.
\end{proof}

In the next lemma $\alpha=i/m$; in Lemma~\ref{lem:binary-affine-gap} it is
attained as $\alpha(u)$ at the saddle $u$. The length $\bar\nu=(N+1)/m$
replaces $\nu$ because of the midpoint rule in Step~4 of the proof.

\begin{lemma}\label{lem:binary-midpoint}
Set $\bar\nu\coloneqq(N+1)/m\ge6+2\rho$ and $\alpha\coloneqq i/m$.
For every interior index $1\le i\le D-1$,
\begin{align}
 \log {U_i}&\ge mL_{\bar\nu}(\alpha)-2\log(14/9),
 \label{eq:binary-midpoint-H}\\
 mE_{\nu,\rho}(\alpha)&\le
 mE_{\bar\nu,\rho}(\alpha)+\log2.
 \label{eq:binary-midpoint-E}
\end{align}
\end{lemma}
\begin{proof}
\smallskip\noindent\emph{Step 1 (Discrete comparison roots).}
Let $\hat R_x$ be the larger root of
\[
 3(N-x)R^2-(3N-1-2x-4m)R+(x-1)=0.
\]
This is $m$ times~\eqref{eq:binary-continuous-root} with $\eta=(N-1)/m$
in place of $\nu$ and $s=(x-1)/m$; as $\eta$ may lie below $6+2\rho$,
its properties are checked directly.
The discriminant decreases with $x\le D$ and increases with
$N$; at $N=6m+2r-1,x=D$ it is $4(m+r-1)^2>0$.
The middle coefficient $3N-1-2x-4m$ is positive.
The vertex is below one, because
$3N-4x+4m+1\ge14m+2r-2>0$, and the quadratic at one is
$4m>0$. Thus its larger root lies in $(0,1)$.
For the normalized discriminant
$\Delta(\eta,s)\coloneqq(3\eta-2s-4)^2-12s(\eta-s)$ the derivatives are
$\partial_s\Delta=32s-24\eta+16<0$ and
$\partial_\eta\Delta=18\eta-24s-24>0$ on
$\eta\ge(N-1)/m$, $0\le s\le(D-1)/m$;
their signs follow already at the smallest length and largest $s$,
where $m\partial_s\Delta=-64m-16r+16$ and
$m\partial_\eta\Delta=36m+12r-12$.
Equation~\eqref{eq:binary-R-derivatives} consequently proves
that the larger roots decrease with $x$ on this domain.

\smallskip\noindent\emph{Step 2 (The discrete bound for $\log{U_i}$).}
Since $R_1=\hat R_1$ and
$3(N-x)\hat R_x=3N-1-2x-4m-(x-1)/\hat R_x$, the
recurrence~\eqref{eq:binary-ratio-recurrence}, which increases in its
preceding ratio, proves $R_x\ge\hat R_x$ by induction: if
$R_{x-1}\ge\hat R_{x-1}\ge\hat R_x$, its right side is at least
$3(N-x)\hat R_x$.
Also $R_0\ge R_i$: each factor $1-1/(\xi_j-x)$ of $R_x$ decreases with
$x$. Thus
\begin{equation}\label{eq:binary-midpoint-discrete}
 \log {U_i}=2\sum_{x=1}^i\log R_x+\log(R_0/R_i)
          \ge2\sum_{x=1}^i\log\hat R_x.
\end{equation}

\smallskip\noindent\emph{Step 3 (Passage to $\bar\nu$ and to midpoints).}
For fixed $x\le i\le D-1$, follow the segment
\[
 \eta(t)\coloneqq\bar\nu-2/m+2t/m,\qquad
 s(t)\coloneqq(x-1)/m+3t/(2m),\qquad0\le t\le1,
\]
from $((N-1)/m,(x-1)/m)$, where the larger root is $\hat R_x$, to
$(\bar\nu,(x+1/2)/m)$. The slopes are chosen so that the two changes
below have a definite sign. The discriminant derivative on this
segment is exactly $-24/m$. At its endpoint, $\bar\nu\ge6+2\rho$ and
$s(1)=(x+1/2)/m<\rho+2$. On the continuous domain $\Delta$ increases in
$\eta$ and decreases in $s$, so there $\Delta$ is at least its value
$4(\rho+1)^2\ge4$ at $(6+2\rho,\rho+2)$. The discriminant is
consequently positive throughout the segment.
Equation~\eqref{eq:binary-R-derivatives} gives
\[
 \frac{dR}{dt}=-\frac{3(1-R)^2}{2m\sqrt\Delta}\le0,
 \qquad \hat R_x\ge R_{\bar\nu}((x+1/2)/m).
\]

\smallskip\noindent\emph{Step 4 (Concavity, the midpoint rule and the endpoint loss).}
For fixed $\eta=\bar\nu$, set $\phi(s)\coloneqq\log R_\eta(s)$.
It is decreasing and concave on $0\le s\le\rho+2$.
Indeed solving for $s$ gives
\[
 s=\eta-\frac{\eta-1}{1+3R}-\frac1{1-R},\qquad
 \frac{d^2s}{d(\log R)^2}
 =\frac{3(\eta-1)R(1-3R)}{(1+3R)^3}
   -\frac{R(1+R)}{(1-R)^3}<0.
\]
Here $R\ge1/2$, by the lower bound
after~\eqref{eq:binary-R-derivatives}, and $ds/d\log R<0$, so the
inverse second-derivative formula
$\phi''=-(d^2s/d\phi^2)/(ds/d\phi)^3$ gives $\phi''<0$.
By concavity, $\phi((x+1/2)/m)\ge m\int_{x/m}^{(x+1)/m}\phi(s)\,ds$
(the midpoint inequality), and~\eqref{eq:binary-midpoint-discrete}
yields
\[
 \log {U_i}\ge2m\int_{1/m}^{(i+1)/m}\phi(s)\,ds
 \ge mL_{\bar\nu}(\alpha)+2[\phi((i+1)/m)-\phi(0)];
\]
the second inequality moves the interval of integration to
$[0,\alpha]$ and uses that $\phi$ decreases. Since
$R_\eta(0)=(3\eta-4)/(3\eta)$ and $R_\eta(s)\ge(\eta-2)/(\eta+2)$,
\[
 \frac{R_\eta(0)}{R_\eta(s)}
 \le\frac{(3\eta-4)(\eta+2)}{3\eta(\eta-2)}
 =1+\frac{8(\eta-1)}{3\eta(\eta-2)}\le14/9,
\]
and equation~\eqref{eq:binary-midpoint-H} follows.
The endpoint $(i+1)/m\le D/m$ explains the restriction to
interior indices.

\smallskip\noindent\emph{Step 5 (The length shift in the exponent).}
At fixed $\rho,\alpha$, the infimum in~\eqref{eq:binary-entropy-exponent}
is a joint infimum over $u$ and the tilt $a$
(Appendix~\ref{app:binary-adaptive}), so the envelope theorem gives
\[
 \partial_\nu E=\log(1+au)-\log\frac\nu{\nu-\alpha};
\]
the stationarity in both $a$ and $u$ removes their derivatives, and
$\partial_\nu[-\nu H_2(\alpha/\nu)]=\log(1-\alpha/\nu)$.
Along $\nu\le\eta\le\bar\nu$, one has
$\eta\ge5.5+2\rho$ and $\alpha\le\rho+2<\eta/2$.
Thus this derivative is greater than $-\log2$.
The length displacement is $1/m$, which proves
\eqref{eq:binary-midpoint-E}.
\end{proof}

\begin{proposition}\label{prop:binary-high-band}
For every $55\le i\le D-1$, one has
${\operatorname{Mat}_i}\prec {U_i}{\operatorname{Gram}_0}$.
\end{proposition}
\begin{proof}
Lemma~\ref{lem:binary-strip} covers $i\le X$.
Suppose $i>X$, and put $\varepsilon\coloneqq1/m$, $\nu\coloneqq N/m$,
$\bar\nu\coloneqq\nu+\varepsilon$ and $\alpha\coloneqq i/m$.
By~\eqref{eq:binary-strip-endpoint},
$X/m=\nu(2\nu-3+6\varepsilon)/(3(8\nu-9))$.
Since $\nu\ge6+2\rho-\varepsilon$,
\[
 13\nu(2\nu-3+6\varepsilon)
 -3(\nu+\varepsilon)(8\nu-9)
 =2\nu(\nu-6)+\varepsilon(54\nu+27)>0.
\]
The last expression is at least $\varepsilon(52\nu+27)$.
All denominators are positive, so
$\alpha>X/m>\bar\nu/13$.

Lemma~\ref{lem:binary-binomial-prefactor}, with
$NH_2(i/N)=m\nu H_2(\alpha/\nu)$, and~\eqref{eq:binary-adaptive-B}
give $\mathcal B_i\le3\sqrt i\exp(mE_{\nu,\rho}(\alpha))$.
Lemma~\ref{lem:binary-affine-gap} applies at $\bar\nu$, since
$\bar\nu\ge6+2\rho$, $0<\alpha<\rho+2$ and $\alpha/\bar\nu>1/13$; it
gives $L_{\bar\nu}(\alpha)-E_{\bar\nu,\rho}(\alpha)>13\alpha/100$. With
Lemma~\ref{lem:binary-midpoint}, which bounds
$mE_{\nu,\rho}(\alpha)\le mE_{\bar\nu,\rho}(\alpha)+\log2$ and
$\log{U_i}\ge mL_{\bar\nu}(\alpha)-2\log(14/9)$, this gives
\[
 \frac{\mathcal B_i}{{U_i}}
 <3\sqrt i\,2(14/9)^2e^{-13i/100}
 =\frac{392}{27}\sqrt i\,e^{-13i/100}<\frac18.
\]
At $i=55$ the last comparison follows from
\[
 (3136/27)\sqrt{55}<7840/9<(8/3)^7<e^7<e^{143/20}.
\]
The function $13i/100-\tfrac12\log i$ increases for $i\ge55$,
so the comparison persists. Lemma~\ref{lem:binary-adaptive-row}
closes the entire complement of the strip.
\end{proof}

\section{The binary cases through seven errors}\label{app:binary-small}

This appendix proves Proposition~\ref{prop:binary-small-radii}: for
an original binary parameter, that is, a tuple $(t,e,\sigma,n)$ as
in~\eqref{eq:original-parameters} with $Q=4$, satisfying
\eqref{eq:binary-length}, with $2\le t\le7$ and $m\ge2$, the rounded
polynomial $g=\binom{n-x}{r}h$ satisfies the coefficient
criterion~\eqref{eq:coefficient-criterion}. These are the cases with
$D=2t+\sigma\le15$, below the range $D\ge16$ of
Proposition~\ref{prop:binary-index-cover}, and the proof does not use
the matrix comparisons of Appendices~\ref{app:binary-matrices}
and~\ref{app:binary-high}. Below, a length is a residual length
$N=n-r$.

For a polynomial $\binom{n-x}{r}{\widetilde h}(N-x)$,
Lemma~\ref{lem:binary-geometric} bounds the normalized Krawtchouk
coefficients $|c_i|/c_0$ by a geometric sequence in $i$.
Its main hypothesis places the roots of ${\widetilde h}$ in an
interval $[0,M]$, and the bound is useful when $M$ is well below $N$.
Lemma~\ref{lem:binary-small-uniform} bounds the
reflected Lloyd zeros $N-\xi_j$, applies
Lemma~\ref{lem:binary-geometric}, and so proves the criterion in three
rows of parameters, from $N\ge100$ (used for $t=3,4,5$), $N\ge128$
($t=6$) and $N\ge176$ ($t=7$).
Lemma~\ref{lem:binary-two-error-coarse} treats $t=2$, where $m\ge2$
forces $e=0$, from $N\ge82$ by a separate quartic estimate, and
Lemma~\ref{lem:binary-small-finite} is an exact finite calculation for
the shorter lengths. Only $Q=4$ is used; the first two lemmas are
stated for more general $Q$.

In Lemma~\ref{lem:binary-geometric}, ${\widetilde h}$ is a general
polynomial. In its application it is the reflected product
${\widetilde h}(y)=h(N-y)$ of the rounded polynomial; the terminal
factor $\binom{n-x}{r}$ is not reflected. The hypothesis that all
roots of ${\widetilde h}$ lie in $[0,M]$ then says that the roots of
$h$ lie in $[N-M,N]$, far from the origin.

\begin{lemma}\label{lem:binary-geometric}
Let $Q\ge2$, $q\coloneqq1/Q$, and let $N,M,r,\ell$ be nonnegative
integers with $n\coloneqq N+r\ge1$ and $L\coloneqq M+\ell\le N$.
Suppose ${\widetilde h}$ has positive leading coefficient, degree $\ell$,
all its roots in $[0,M]$, nonnegative values on $0,\ldots,N$,
and positive values at integers at least $L-1$.
For $g(x)\coloneqq\binom{n-x}{r}{\widetilde h}(N-x)=\sum c_iP_i^{(n,Q)}(x)$,
\begin{equation}\label{eq:binary-geometric}
 c_0>0,\qquad
 \frac{|c_i|}{c_0}\le
 \left(\frac{r+L+2q(N-L)}{N+r}\right)^i
 \quad(0\le i\le n).
\end{equation}
\end{lemma}
\begin{proof}
\smallskip\noindent\emph{Step 1 (A stochastic comparison).}
Let $W$ have the ${\widetilde h}$-weighted $\Bin(N,q)$ law,
$\Pr(W=w)\propto{\widetilde h}(w)\binom Nwq^w(1-q)^{N-w}$; it is well
defined because ${\widetilde h}\ge0$ on $0,\ldots,N$ and
${\widetilde h}(N)>0$. Let $\zeta_1,\ldots,\zeta_\ell\le M$ be the roots of
${\widetilde h}$. For an integer $y>L$,
\[
 \frac{{\widetilde h}(y-1)}{{\widetilde h}(y)}
 =\prod_{k=1}^\ell\left(1-\frac1{y-\zeta_k}\right)
 \ge1-\frac{\ell}{y-M}\ge\frac{y-L}{y}.
\]
All factors are in $[0,1]$, since $y-\zeta_k\ge y-M\ge\ell+1$, so the
product is at least one minus the sum of the subtracted terms; the last
inequality reduces to $M(y-L)\ge0$. The empty product gives the same
conclusion if $\ell=0$. Put $W^*\coloneqq L+\Bin(N-L,q)$. For $L<y\le N$,
\[
 \frac{\Pr(W=y)}{\Pr(W=y-1)}\Big/\frac{\Pr(W^*=y)}{\Pr(W^*=y-1)}
 =\frac{{\widetilde h}(y)}{{\widetilde h}(y-1)}\cdot\frac{y-L}{y}\le1,
\]
so the likelihood ratio of $W$, conditional on $W\ge L$, to $W^*$ is
nonincreasing. The normalized mass functions therefore cross at most
once, from positive to negative after subtraction, and their upper
tails give $W\mid\{W\ge L\}\le_{\rm st}W^*$, where $\le_{\rm st}$ is the
usual stochastic order (every upper tail probability is smaller).
Unconditioning decreases these tails; below $L$ the comparison
is automatic. Hence $W\le_{\rm st}W^*$. Positivity at $N$ makes the law
and the conditioning well-defined, including $L=N$.

\smallskip\noindent\emph{Step 2 (Reflection and a monotone majorant).}
Let $X\sim\mu_n$ and $Y\coloneqq n-X\sim\Bin(n,q)$. Then
$g(X)=\binom Yr{\widetilde h}(Y-r)$, and by the terminal binomial
identity~\eqref{eq:terminal-binomial}, weighting $Y$ by $g(X)$ gives the
law $r+W$, with normalizer
$c_0=\E_{\mu_n}g=\binom nrq^r\E {\widetilde h}(\Bin(N,q))>0$.
Set
\[
 {\operatorname{Maj}_i}(y)\coloneqq[z^i](1+z)^y\Bigl(1+\frac z{Q-1}\Bigr)^{n-y}.
\]
The defining sum~\eqref{eq:kraw} bounds $|P_i^{(n,Q)}(n-y)|/(Q-1)^i$ by
this majorant ${\operatorname{Maj}_i}(y)$, and
\[
 {\operatorname{Maj}_i}(y+1)-{\operatorname{Maj}_i}(y)
 =\Bigl(1-\frac1{Q-1}\Bigr)[z^{i-1}](1+z)^y\Bigl(1+\frac z{Q-1}\Bigr)^{n-y-1}\ge0.
\]

\smallskip\noindent\emph{Step 3 (The coefficient bound).}
By orthogonality, $c_i=\E_{\mu_n}[gP_i^{(n,Q)}]/v_i(n)$ with
$v_i(n)=\binom ni(Q-1)^i$. Since $g\ge0$ on the grid,
$|c_i|\le\E_{\mu_n}[g\,|P_i^{(n,Q)}(n-Y)|]/v_i(n)
\le\E_{\mu_n}[g\,{\operatorname{Maj}_i}(Y)]/\binom ni$, so
$|c_i|/c_0\le\E\,{\operatorname{Maj}_i}(r+W)/\binom ni$. As
${\operatorname{Maj}_i}$ is nondecreasing and $W\le_{\rm st}W^*$, this is
at most $\E\,{\operatorname{Maj}_i}(r+L+\Bin(N-L,q))/\binom ni$. Each of
the $N-L$ binomial coordinates contributes the factor
$q(1+z)+(1-q)(1+z/(Q-1))=1+2qz$, so
\[
 \frac{|c_i|}{c_0}
 \le\frac{[z^i](1+z)^{r+L}(1+2qz)^{N-L}}{\binom ni}.
\]

\smallskip\noindent\emph{Step 4 (The symmetric-mean inequality).}
The numerator is the $i$th elementary symmetric function $e_i$ of
$r+L$ ones and $N-L$ copies of $2q$. We use Maclaurin's inequality,
$e_i\le\binom ni\bar x^i$ for $n$ nonnegative entries with mean $\bar x$,
and recall a proof. Since $e_i$ is affine in the product of any two
entries at fixed sum, with a nonnegative coefficient, averaging two
nonnegative entries at fixed sum increases their product and hence
this symmetric function. Repeated averaging of the largest and
smallest entries converges to equal entries: the sum of squared
deviations decreases by half the squared spread. The resulting
arithmetic-mean bound, with mean $(r+L+2q(N-L))/(N+r)$,
proves~\eqref{eq:binary-geometric}. The cases $i=0,1$ are immediate.
This step, Maclaurin's inequality, is also used in the proof of
Lemma~\ref{lem:binary-strip} in Appendix~\ref{app:binary-high}.
\end{proof}

In the next lemma each row of the table is the conjunction of its
conditions; the length condition $N\ge6m+2r-1$ holds automatically in
each row, and the three length thresholds are the cutoffs
${N^{\rm cut}_t}$ of Lemma~\ref{lem:binary-small-finite}. Only $Q=4$ is
used.

\begin{lemma}\label{lem:binary-small-uniform}
For real $Q\ge4$, the rounded polynomial satisfies
$|c_i|/c_0\le u^i<v^i\le g(i)/g(0)$ for $1\le i\le D$
in each row of the following table.
\begin{center}
\begin{tabular}{cccccc}\toprule
$N\ge$&$1\le m\le$&$0\le r\le$&$D\le$&$u$&$v$\\\midrule
100&5&9&11&$86/109$&$45/56$\\
128&6&11&13&$551/695$&$287/352$\\
176&7&13&15&$83/105$&$409/484$\\\bottomrule
\end{tabular}
\end{center}
Here $D=2m+r$, and $g(0),c_0>0$ throughout.
\end{lemma}
\begin{proof}
\smallskip\noindent\emph{Step 1 (The root envelope).}
Put ${N_0}\coloneqq N-1$, $q\coloneqq1/Q$, $v_0\coloneqq Nq(1-q)$, the variance of $\Bin(N,q)$,
and $R_j(y)\coloneqq j!q^jP_j^{({N_0},Q)}({N_0}-y)$. Its exponential generating
series is $(1+(1-q)z)^y(1-qz)^{{N_0}-y}$.
Formal differentiation gives
\[
 R_{j+1}=[y-{N_0}q-j(1-2q)]R_j-j(N-j)q(1-q)R_{j-1}.
\]
Thus $R_j(y)=\det(yI-J^{(j)})$, where $J^{(j)}$ is the leading $j\times j$
block of the symmetric Jacobi matrix $J$ with diagonal ${N_0}q+j(1-2q)$
and off-diagonal squares $j(N-j)q(1-q)$; at $Q=4$,
$R_j=4^{-j}{\operatorname{Sturm}_j}$ of
Lemma~\ref{lem:binary-small-finite}. The zeros of $R_m$ are the
reflected Lloyd zeros $y_j\coloneqq N-\xi_j$. The leading principal minors of $J$
are $(-1)^jR_j(0)=({N_0})_jq^j>0$, so $J$ is positive definite and every
reflected zero is positive.
Ordering the off-diagonal part of $J$ by parity gives blocks
$\left(\begin{smallmatrix}0&{J_{\rm odd}}\\{J_{\rm odd}}^\mathsf T&0\end{smallmatrix}\right)$.
Therefore its operator norm equals that of ${J_{\rm odd}}$, and its square
is at most ${\|{J_{\rm odd}}\|_{\mathrm F}}^2\le m(m-1)v_0/2$, since
$j(N-j)\le jN$. The largest diagonal entry is strictly less than
$Nq+m-1$. By Weyl's inequality, every reflected root $y_j=N-\xi_j$
consequently satisfies
\begin{equation}\label{eq:binary-small-root-envelope}
 0<y_j<Nq+m-1+\sqrt{m(m-1)v_0/2}.
\end{equation}

\smallskip\noindent\emph{Step 2 (The rounded roots and the row constants).}
Write $k_j\coloneqq\lceil y_j\rceil=N-\lfloor\xi_j\rfloor$,
which remains exact at integral roots, and
${\widetilde h}(y)\coloneqq\prod_j(y-k_j)(y-k_j+1)$. Then
${\widetilde h}(N-x)=h(x)$ for the rounded $h$
of~\eqref{eq:rounded-polynomial}, so
$g(x)=\binom{n-x}r{\widetilde h}(N-x)$ is the rounded $g$.
Since $q\le1/4$ and $v_0\le3N/16$,
\eqref{eq:binary-small-root-envelope} and $\lceil y\rceil<y+1$ give
$k_j<N/4+m+\sqrt{3m(m-1)N/32}$. With the bound $m\le5,6,7$ of the
corresponding row of the table in the statement, this gives the
following bounds:
\begin{center}
\begin{tabular}{c|c|c|c}
$N$ range&$k_j<M_0$&$M\coloneqq\lfloor M_0\rfloor$
 &$L\coloneqq M+2m$ at most\\\hline
$N\ge100$&$11N/25$&$\lfloor11N/25\rfloor$&$27N/50$\\
$N\ge128$&$9N/20$&$\lfloor9N/20\rfloor$&$11N/20$\\
$N\ge176$&$9N/20$&$\lfloor9N/20\rfloor$&$11N/20$.
\end{tabular}
\end{center}
For the first row use
$\sqrt{15N/8}\le7N/50$ and $5\le N/20$, so that
$k_j<N/4+N/20+7N/50=11N/25$;
the square-root inequality needs only $N\ge9375/98$.
For the second use $\sqrt{45N/16}\le3N/20$ and
$6\le N/20$, valid for $N\ge125$ and $N\ge120$.
For the third use $\sqrt{63N/16}\le3N/20$ and
$7\le N/20$, valid for $N\ge175$ and $N\ge140$.
The bounds on $L$ also use $10\le N/10$, $12\le N/10$
and $14\le N/10$, respectively. Strictness follows from
\eqref{eq:binary-small-root-envelope} and $\lceil y\rceil<y+1$.

\smallskip\noindent\emph{Step 3 (The coefficient bound).}
All roots $k_j,k_j-1$ of ${\widetilde h}$ lie in $[0,M]$: $k_j\ge1$
since $y_j>0$, and $k_j\le M$ since $k_j$ is an integer below $M_0$.
The paired factors are nonnegative at integers, and ${\widetilde h}$ is
positive above $M$. Since $L-1=M+2m-1\ge M+1$ and $L<N$,
Lemma~\ref{lem:binary-geometric} applies with $\ell=2m$.
In the base of the power in~\eqref{eq:binary-geometric}, replace $2q$ by
$1/2$ and $L$ by its bound in the table. The result,
$(r+N/2+L/2)/(N+r)$, increases with $r$ and decreases with $N$, so it is
largest at the smallest $N$ and the largest $r$ of each row. This gives
\[
 \frac{77N/100+r}{N+r}\le\frac{86}{109},\qquad
 \frac{31N/40+r}{N+r}\le\frac{551}{695},\qquad
 \frac{31N/40+r}{N+r}\le\frac{83}{105},
\]
respectively. The maxima occur at $(N,r)=(100,9),(128,11)$
and $(176,13)$.

\smallskip\noindent\emph{Step 4 (The value bound and the comparison).}
By~\eqref{eq:normalized-value} and ${\widetilde h}(N-x)=h(x)$,
$g(i)/g(0)$ is the product of the $2m$ factors
$(1-i/(N-k_j))(1-i/(N-k_j+1))$ and the $r$ terminal-binomial factors
$1-i/(n-t')$, $0\le t'<r$. Every one of these $D$ factors has the form
$1-i/B_h$ with $B_h\ge B\coloneqq N-M$, since $k_j\le M$ and $n-t'\ge N+1$.
As $M\le M_0$, the smallest $N$ of each row gives the respective bounds
$B\ge56$, $B\ge352/5$ and $B\ge484/5$. Hence all bases are positive,
and Bernoulli's inequality $(1-x)^p\ge1-px$, for $0\le x\le1$ and
$p=D/i\ge1$, gives
\[
 \frac{g(i)}{g(0)}\ge(1-i/B)^D\ge(1-D/B)^i.
\]
With the largest $D$ of each row, $1-D/B$ is at least $45/56$,
$287/352$ and $409/484$, the three lower constants $v$ in the table.
The exact cross-product differences are $89$, $5513$ and
$2773$, all positive. The lemma follows.
\end{proof}

\begin{lemma}\label{lem:binary-two-error-coarse}
For $t=2,e=0$, the coefficient criterion holds at $Q=4,N\ge82$.
\end{lemma}
\begin{proof}
Here $m=2$, $r=\sigma\in\{0,1\}$ and $D=4+\sigma$. We prove the
estimate for real $Q\ge4$ with $v\coloneqq N(Q-1)/Q^2\ge3$; only $Q=4$ is used.

\smallskip\noindent\emph{Step 1 (The Lloyd zeros and their floors).}
Put $\alpha\coloneqq(Q-1)/Q$, $u\coloneqq\alpha(1-\alpha)$,
$\kappa\coloneqq1-2\alpha$, $\beta\coloneqq-1/2+2/Q$ and
$R^2\coloneqq v+\beta^2-2/Q^2$; thus $v=Nu$ is the variance of $\Bin(N,\alpha)$.
The two Lloyd zeros of Section~\ref{sec:framework}, the zeros of
$P_2^{(N-1,Q)}(x-1)$, are $N\alpha+\beta\pm R$.
For their floors $a_1\le a_2$, set
\[
 {z_1}\coloneqq a_1+\tfrac12-N\alpha=\beta-R+\varepsilon_1,
 \quad {z_2}\coloneqq a_2+\tfrac12-N\alpha=\beta+R+\varepsilon_2,
 \quad |\varepsilon_j|\le\tfrac12,
\]
and $s\coloneqq{z_1}+{z_2}$, ${\Pi}\coloneqq{z_1}{z_2}$, $E\coloneqq{\Pi}+v$.
Then $|s|\le2$, since $|\beta|\le1/2$. Since
$v+\beta^2-R^2=2/Q^2$, we have
$E=2/Q^2+\beta(\varepsilon_1+\varepsilon_2)+R(\varepsilon_1-\varepsilon_2)+\varepsilon_1\varepsilon_2$,
so $|E|\le R+7/8\le\sqrt v+11/8$.

\smallskip\noindent\emph{Step 2 (Centred form and the constant coefficient).}
For $z\coloneqq x-N\alpha$,
\[
 h(x)=z^4-2sz^3+(s^2+2{\Pi}-\tfrac12)z^2
      +s(\tfrac12-2{\Pi})z+{\Pi}^2-\tfrac14s^2+\tfrac12{\Pi}+\tfrac1{16}.
\]
The central moments $v$, $\kappa v$ and $3v^2+v(1-6u)$ of
$\Bin(N,\alpha)$, of orders two to four, give
\[
 b_0=2v^2+v(s^2-2\kappa s-6u)-\tfrac14s^2
                   +(E+\tfrac14)^2
 \ge2v^2-\tfrac{17}8v-1\ge v^2;
\]
the first bound uses $s^2-2\kappa s\ge-\kappa^2\ge-1$, $u\le3/16$ and
$s^2\le4$, and the second uses $v\ge3$.

\smallskip\noindent\emph{Step 3 (The remaining coefficients).}
Using Lemma~\ref{lem:finite-difference}, put
${b^{\rm sc}_i}\coloneqq(-1)^iQ^ib_i$, which is $\E_{\mu_{N-i}}\Delta^ih$
by~\eqref{eq:finite-difference}. Direct expansion gives
\begin{align*}
{b^{\rm sc}_1}={}&v(8\kappa-4s)+2E(\kappa-s)+s^2\kappa
 -\tfrac32s\kappa^2+\tfrac12\kappa^3-u(10\kappa-6s),\\
{b^{\rm sc}_2}={}&8v+4E+12\kappa^2-12s\kappa+2s^2+1-24u,\\
{b^{\rm sc}_3}={}&36\kappa-12s,\qquad {b^{\rm sc}_4}=24.
\end{align*}
Since $|\kappa|\le1$, $u\le3/16$ and $v\ge3$,
\[
 |{b^{\rm sc}_1}|\le16v+6\sqrt v+159/8\le27v,\quad
 |{b^{\rm sc}_2}|\le8v+4\sqrt v+55\le29v,\quad |{b^{\rm sc}_3}|\le60.
\]
Division by $Q^ib_0$, with $b_0\ge v^2$, $Qv=N(Q-1)/Q$ and
$Q/(Q-1)\le4/3$, yields
\[
 |b_1|/b_0\le36/N,\quad |b_2|/b_0\le10/N,
 \quad |b_3|/b_0\le27/N^2,\quad |b_4|/b_0\le3/N^2.
\]

\smallskip\noindent\emph{Step 4 (Values, the lift and the conclusion).}
For $N\ge64$, the root formula implies
$a_1,a_2\ge3N/4-\sqrt N/2-2\ge N/2$.
Thus $h(i)/h(0)\ge1-8i/N$ and, by~\eqref{eq:normalized-value}, since
$r=0,1$ and $i\le5$, $g(i)/g(0)\ge1-45/N$.
For $r=0$, $c_i=b_i$. For $r=1$, the exact lift~\eqref{eq:binomial-lift}
is $c_i=Q^{-1}[(n-i)b_i+ib_{i-1}]$, $c_0=nb_0/Q$, where $b_5=0$ since
$h$ has degree four. With the bounds of Step~3 it gives
$c_i/c_0\le37/N$ for $1\le i\le D$; the largest bound occurs at
$i=1$ and $r=1$. Finally $37/N\le1-45/N$ when $N\ge82$.
At $Q=4$ the premise $v\ge3$ is automatic in this length range.
\end{proof}

\begin{lemma}\label{lem:binary-small-finite}
For $2\le t\le7$, $0\le e\le t-2$ and $\sigma=0,1$,
the binary rounded polynomial satisfies the coefficient criterion
on the finite domain
\begin{equation}\label{eq:binary-small-finite-domain}
 \begin{gathered}
 6t-1-2e+4\sigma\le N<{N^{\rm cut}_t},\\
 ({N^{\rm cut}_2},{N^{\rm cut}_3},{N^{\rm cut}_4},{N^{\rm cut}_5},{N^{\rm cut}_6},{N^{\rm cut}_7})\coloneqq(82,100,100,100,128,176).
 \end{gathered}
\end{equation}
\end{lemma}
\begin{proof}
\smallskip\noindent\emph{Step 1 (Domain and summary of the certificate).}
The lower bound in~\eqref{eq:binary-small-finite-domain} is $n-r$ for
$n=6t-1+5\sigma$, the least code length allowed by~\eqref{eq:binary-length};
it exceeds $6m+2r-1$ by $2\sigma$. The cutoffs are the lengths from which
Lemma~\ref{lem:binary-two-error-coarse} ($t=2$) and the rows of
Lemma~\ref{lem:binary-small-uniform} ($t\ge3$) apply.
The calculation uses integer root counts and integer coefficient
deficits, as detailed below. Its complete domain has the following
numbers of tuples, ordered by $e=0,\ldots,t-2$:
\begin{center}
\begin{tabular}{cl r}\toprule
$t$&Counts by $e$&Total\\\midrule
2&138&138\\
3&162,166&328\\
4&150,154,158&462\\
5&138,142,146,150&576\\
6&182,186,190,194,198&950\\
7&266,270,274,278,282,286&1656\\\midrule
Total& &4110\\\bottomrule
\end{tabular}
\end{center}
The finite certificate~\cite{ComputationalCertificates} records every
tuple, its reflected ceilings, scaled coefficients and deficits
$\Delta_i$ (Step~4). For all these tuples the constant and
origin value are positive and every required deficit is nonnegative.
The remaining steps give the exact construction and verification of
those assertions.

\smallskip\noindent\emph{Step 2 (Sturm polynomials and root location).}
Set ${N_0}\coloneqq N-1$, $m\coloneqq t-e$, $r\coloneqq2e+\sigma$, $D\coloneqq2t+\sigma$ and $n\coloneqq N+r$.
Define the integer polynomials
\begin{equation}\label{eq:binary-finite-root-recurrence}
 \begin{gathered}
 {\operatorname{Sturm}_0}(y)\coloneqq1,\quad {\operatorname{Sturm}_1}(y)\coloneqq4y-{N_0},\\
 \begin{aligned}
 {\operatorname{Sturm}_j}(y)&\coloneqq[4y-{N_0}-2(j-1)]{\operatorname{Sturm}_{j-1}}(y)\\
              &\quad-3(j-1)({N_0}-j+2){\operatorname{Sturm}_{j-2}}(y).
 \end{aligned}
 \end{gathered}
\end{equation}
The generating function proves ${\operatorname{Sturm}_j}=j!P_j^{({N_0},4)}({N_0}-y)$.
Thus ${\operatorname{Sturm}_j}=4^jR_j$ with $R_j$ of Step~1 of the proof of
Lemma~\ref{lem:binary-small-uniform} at $Q=4$, and
${\operatorname{Sturm}_j}(y)={\operatorname{Rec}_j}(N-y)$ with
${\operatorname{Rec}_j}$ of Step~5 of the proof of Lemma~\ref{lem:binary-top}.
These are $4^j$ times the principal characteristic determinants
$\det(yI-J^{(j)})$ of the reflected Jacobi matrix $J$ of Step~1 of the
proof of Lemma~\ref{lem:binary-small-uniform}, at $Q=4$. Its positive off-diagonal squares give strict interlacing.
Leading principal minors of both $J$ and ${N_0}I-J$ are positive, since
${\operatorname{Sturm}_j}(0)=(-1)^j({N_0})_j$ and ${\operatorname{Sturm}_j}({N_0})=3^j({N_0})_j$.
Every reflected root therefore lies strictly in $(0,{N_0})$.

\smallskip\noindent\emph{Step 3 (Exact ceilings from sign variations).}
At any integer $y$, omit zeros from the sequence
${\operatorname{Sturm}_0}(y),\ldots,{\operatorname{Sturm}_m}(y)$ and let $V(y)$ be its sign-variation count.
An internal zero has opposite nonzero neighbors; consecutive zeros
are impossible. At a terminal zero, simplicity and interlacing give
${\operatorname{Sturm}_m}'(y)/{\operatorname{Sturm}_{m-1}}(y)>0$, so omitting it gives the count
just to the right of $y$. Jacobi's rule for the inertia of a symmetric
matrix through the signs of its leading principal minors, applied to
$yI-J$, consequently gives
\[
 {\operatorname{Count}(y)}\coloneqq m-V(y)=\#\{\text{reflected roots at most }y\}.
\]
For every $k=1,\ldots,N$, take ${\operatorname{Count}(k)}-{\operatorname{Count}(k-1)}$ copies of $k$
as reflected ceilings. This certifies the half-open intervals
$(k-1,k]$ even when a root is integral.

\smallskip\noindent\emph{Step 4 (Exact coefficients and deficits).}
Multiply the reflected rounded polynomial
${\widetilde h}(y)=\prod_j(y-k_j)(y-k_j+1)$ exactly and expand it
in falling factorials, ${\widetilde h}(y)=\sum_{j=0}^{2m}{\widetilde h}_j(y)_j$.
Multiplying~\eqref{eq:factorial-extraction} by $4^D$ gives the integers
\begin{equation}\label{eq:binary-finite-coefficients}
 {\operatorname{Int}_i}\coloneqq4^Dc_i=\frac1{r!}\sum_{j=0}^{2m}
 {\widetilde h}_j(r+j)!4^{2m-j}\binom{n-i}{r+j-i},\qquad0\le i\le D.
\end{equation}
Every division is exact. Two independent cross-checks of these integers
follow; they are not needed for the argument. First, expand $g$ in
monomials and solve its triangular Krawtchouk expansion, whose leading
coefficient at degree $i$ is $(-4)^i/i!$; the two coefficient lists
agree. Second, orthogonality over the full grid $x=0,\ldots,n$ gives
\[
 4^n\binom ni3^i c_i
   =\sum_{x=0}^n\binom nx3^xg(x)P_i^{(n,4)}(x).
\]
Finally form the deficits $\Delta_i\coloneqq{\operatorname{Int}_0}g(i)-{\operatorname{Int}_i} g(0)$,
which are integers since the values $g(i)$ are, and check their signs:
$\Delta_i\ge0$ is the criterion $c_0g(i)\ge c_ig(0)$ multiplied by $4^D$.
No floating-point root or coefficient comparison enters the calculation.

\smallskip\noindent\emph{Step 5 (Domain completeness).}
The domain loops are exactly those in
\eqref{eq:binary-small-finite-domain}; their tuple sets are compared
with the stored keys of the certificate without duplicates. For fixed
$t,e$ their count is $2{N^{\rm cut}_t}-12t+4e-2$, which gives the
displayed table. Only $m\ge2$ is included: the empty and one-root cases
have separate proofs. The upper boundary $N={N^{\rm cut}_t}$ belongs to
an analytic lemma, Lemma~\ref{lem:binary-two-error-coarse} for $t=2$
and a row of Lemma~\ref{lem:binary-small-uniform} for $t\ge3$, and
every lower boundary comes from~\eqref{eq:binary-length}.
These observations verify that the finite certificate covers precisely
the required remainder.
\end{proof}

In the next proposition, an original binary parameter is a tuple
$(t,e,\sigma,n)$ as in~\eqref{eq:original-parameters} with $Q=4$, and
the criterion is meant for its rounded polynomial.

\begin{proposition}\label{prop:binary-small-radii}
For $2\le t\le7$, every original binary parameter satisfying
\eqref{eq:binary-length} and $m\ge2$ satisfies the coefficient criterion.
\end{proposition}
\begin{proof}
For $t=2$, use Lemma~\ref{lem:binary-two-error-coarse} when
$N\ge82$. For $t=3,4,5$, the first row of
Lemma~\ref{lem:binary-small-uniform} applies at $N\ge100$:
$m\le5$, $r\le2t-4+\sigma\le7$ for the retained $m\ge2$,
and $D\le11$.
Its second and third rows apply to $t=6,7$ at the respective
cutoffs $128,176$: there $m\le t$, $r\le2t-4+\sigma\le9$ and $11$,
and $D\le13$ and $15$. All smaller lengths are exactly
Lemma~\ref{lem:binary-small-finite}.
\end{proof}

\section{Binary grid minima and stronger coefficient estimates}
\label{app:bg-minima}

This appendix proves the root and low-index statements used for a
binary minimizing vertex, and then proves the coefficient estimates
in Proposition~\ref{prop:bg-coefficients} and the small-degree comparison
in Proposition~\ref{prop:bg-small-degree}. The first two coefficient
arguments use the positive quadrature of the minimizing vertex.
The later coefficient estimates hold for every real polynomial of
degree at most \(m\); they are combined with the endpoint forms of
Appendix~\ref{app:bg-endpoints} only after returning to the minimizer.

We keep the manuscript convention
\(D=2m+r\), \(n=N+r\), and \(H=n-D=N-2m\).
For a minimizing vertex, write
\[
 F(x)\coloneqq\frac{\binom{n-x}{r}}{\binom nr}\frac{h(x)}{h(0)},
 \qquad h(x)\coloneqq q(x)q(x-1),\qquad
 q(x)\coloneqq\prod_{j=1}^{m}(x-a_j),
 \qquad \beta\coloneqq\E_{\mu_n}F=c_0.
\]
The terminal block is maximal. Thus the residual roots are distinct
integers, and the lower zeros satisfy \(1\le a_j\le N-2\).
The forms \(\mathcal G,\mathcal C_i,\mathcal E_i\) are those used in
the binary grid-minimum argument: respectively the adjacent-product
expectation, the effective lifted coefficient, and \(q(i)q(i-1)\).
In particular \(\mathcal C_0=\mathcal G\).
The Lloyd ratios \(U_i\) retain the normalization
\eqref{eq:binary-endpoint}.

\subsection{The roots of a minimizing vertex}
\label{sec:bg-roots}

The mean estimate needed here is the \(Q=4\) version of
Lemma~\ref{lem:paired-mean}, whose statement assumes \(Q\ge9\)
but whose proof uses only \(Q>2\). We retain the calculation here
to make its bounds at \(L=4k\) explicit.

\begin{lemma}\label{lem:bg-paired-mean}
Let \(L\ge1\) and \(k\ge0\) be integers with \(L\ge4k\), and let
\(q\ne0\) be a real polynomial of degree at most \(k\).
For \(X\sim\Bin(L,3/4)\) and \(h(x)\coloneqq q(x)q(x-1)\),
\[
 \E h(X)>0,\qquad
 \frac{\E[Xh(X)]}{\E h(X)}
 \ge \frac{(\sqrt{3(L-k)}-\sqrt{k})^2}{4}.
\]
\end{lemma}
\begin{proof}
For \(k=0\) both assertions follow directly and the second is an
equality. Suppose \(k\ge1\), and put
\[
 \rho\coloneqq\sqrt{\frac{k}{3(L-k)}}\le\frac13,\qquad
 a\coloneqq\frac{1-\rho}{1+3\rho}>0.
\]
For \(Y\sim\Bin(L-1,3/4)\), recall the forms of
Section~\ref{sec:paired-forms} at \(Q=4\):
\[
 \begin{gathered}
 \operatorname{Forw}(q)=\E[q(Y+1)q(Y)],\quad
 \operatorname{Back}(q)=\E[q(Y)q(Y-1)],\\
 W(q)=\E[q(Y)^2]>0.
 \end{gathered}
\]
In \eqref{eq:forward-gram}, the correlation parameters are at most
\(3\rho\). For \(3\rho<1\), \eqref{eq:correlation-lower} gives the
lower eigenvalue bound \((1-3\rho)/(1+3\rho)\) for the correlation
matrix, and consequently
\[
 \operatorname{Forw}(q)\ge
 \left\{\frac13+\frac23\frac{1-3\rho}{1+3\rho}\right\}W(q)=aW(q).
\]
If \(3\rho=1\), positive semidefiniteness of the correlation matrix
gives the same lower bound \(W(q)/3=aW(q)\).
For the backward matrix, its correlation parameters are at most
\(\rho<1\). Its geometric row sums are at most
\((1+\rho)/(1-\rho)\), whereas
\eqref{eq:correlation-lower} bounds its least eigenvalue below by
\((1-\rho)/(1+\rho)\). Thus the lower bound for
\(\operatorname{Back}(q)/W(q)\) is
\(2(1-\rho)/(1+\rho)-1=(1-3\rho)/(1+\rho)\ge0\).
Using \eqref{eq:backward-gram} therefore gives
\[
 0\le \operatorname{Back}(q)\le
 \left\{2\frac{1+\rho}{1-\rho}-1\right\}W(q)=a^{-1}W(q).
\]
Conditioning on the last Bernoulli variable and using binomial size
bias now yield
\[
 \begin{gathered}
 \E h=\frac{3\operatorname{Forw}(q)+\operatorname{Back}(q)}{4}>0,\qquad
 \E[Xh(X)]=\frac{3L}{4}\operatorname{Forw}(q),\\
 \frac{\E[Xh]}{\E h}\ge\frac{3L}{3+a^{-2}}.
 \end{gathered}
\]
Finally,
\[
 3(1-\rho)^2+(1+3\rho)^2=4(1+3\rho^2),
 \qquad 1+3\rho^2=\frac{L}{L-k},
\]
reduce the last fraction to
\(3(L-k)(1-\rho)^2/4\), which is the claimed expression.
\end{proof}

\begin{proof}[Proof of Lemma~\ref{lem:bg-root}]
The hypothesis \(n\ge3D-1\) gives
\(N\ge6m+2r-1\). Fix the lower zero \(a\) of a residual pair
and write
\(h(x)=(x-a)(x-a-1)w(x)\). Put
\(M_j\coloneqq\E_{\mu_N}[X^jw(X)]\).
Moving the pair to \(a+1,a+2\) gives a feasible competitor: an
adjacent pair is nonnegative at every integer, collisions do not
change that fact, and the terminal factor vanishes above \(N\).
The residual normalized objective at a lower zero \(a\) is
\[
 J(a)\coloneqq\frac{M_2-(2a+1)M_1+a(a+1)M_0}{a(a+1)w(0)}.
\]
Since the terminal lift multiplies every residual objective by the
same factor \(4^{-r}\), minimality gives \(J(a+1)\ge J(a)\).
Direct subtraction gives
\begin{equation}\label{eq:bg-pair-move}
 J(a+1)-J(a)=
 \frac{2((a+1)M_1-M_2)}{a(a+1)(a+2)w(0)}.
\end{equation}
Here \(w(0)>0\) and \(M_1>0\): all remaining pairs are nonnegative
on the grid and \(w(N)>0\).
Write \(w(x)=q_0(x)q_0(x-1)\), with \(\deg q_0=m-1\).
Size bias gives
\[
 \frac{M_2}{M_1}
 =1+\frac{\E_{\mu_{N-1}}[Yw(Y+1)]}
              {\E_{\mu_{N-1}}w(Y+1)}.
\]
Apply Lemma~\ref{lem:bg-paired-mean} to \(q_0(y+1)\), with
\(L=N-1\) and \(k=m-1\). Its hypothesis holds because
\(N-1\ge6m+2r-2\ge4(m-1)\).
Together with \eqref{eq:bg-pair-move}, this proves
\[
 a\ge A(N,m)\coloneqq\frac{(\sqrt{3(N-m)}-\sqrt{m-1})^2}{4}.
\]
To locate this bound relative to the coefficient indices, put
\(k\coloneqq m-1\). The exact identity
\[
 (H+1+4k)^2-12k(H+1+k)=(H+1-2k)^2>0
\]
and \(H+1>2k\) give
\[
 A(N,m)>\frac{H+1}{2}\ge D.
\]
For \(k=0\) the same strict comparison follows directly from this
identity. Thus every lower zero, and hence every residual root,
exceeds \(D\). The terminal roots exceed \(D\) as well, since
\(N+1>D\). This last conclusion also covers \(m=0\).
\end{proof}

\subsection{The first two coefficients}
\label{sec:bg-low-indices}

\begin{proof}[Proof of Lemma~\ref{lem:bg-low-indices}]
Section~\ref{sec:bg-minimum} proves the first index with a feasible
competitor; we first record its positive-quadrature proof, whose
notation the second index reuses.
All expectations in the
quadrature calculations below are at length \(n\).
Let \(z_1<\cdots<z_D\) be the roots of the normalized minimizing
vertex and put \(u\coloneqq F(1)\). Their integer distinctness and
Lemma~\ref{lem:bg-root} give
\(z_j\ge D+j\), and therefore
\begin{equation}\label{eq:bg-first-product}
 u=\prod_{j=1}^D\frac{z_j-1}{z_j}
 \ge\prod_{j=1}^D\frac{D+j-1}{D+j}=\frac12.
\end{equation}
The polynomial \(R_1(x)\coloneqq(n-x)(F(x+1)-F(x))\) has degree at
most \(D\). At a positive quadrature root \(z<n\), its value is
\((n-z)F(z+1)\ge0\); at \(z=n\) it is zero.
Since the origin weight in Lemma~\ref{lem:positive-quadrature}
is \(\beta\), positive quadrature gives
\(\E R_1\ge\beta n(u-1)\).
For \(s\coloneqq3/4\), the binomial mass ratio gives
\[
 s\E[(n-X)F(X+1)]=(1-s)\E[XF(X)],
 \qquad
 \E[(X-sn)F(X)]=s\E R_1(X).
\]
As \(P_1^{(n,4)}(x)=3n-4x\) and its squared norm is \(3n\),
\[
 \frac{c_1}{\beta}
 =1-\frac{\E[XF(X)]}{sn\beta}\le1-u\le u.
\]
This proves the first-index conclusion, including a root at \(n\);
no value of \(F(n+1)\) is used.

For the second index, the backward difference polynomial
\(x(F(x-1)-F(x))\) also has degree at most \(D\), is zero at the
origin, and is nonnegative at every positive quadrature root.
Its expectation is \(3nc_1\), so \(c_1\ge0\); this will let us
discard a negative multiple of \(c_1\) in the second-index estimate.
Now set
\[
 R_2(x)\coloneqq(n-x)^2(F(x+1)-F(x))-DxF(x).
\]
The terms of degree \(D+1\) cancel, so \(\deg R_2\le D\).
At the positive quadrature roots \(R_2\ge0\), including the zero
value at \(n\), and \(R_2(0)=n^2(u-1)\). Hence
\(\E R_2\ge\beta n^2(u-1)\).
Shifting the binomial sum expresses this expectation as
\(\E[F(X)S(X)]\), where
\[
 S(x)\coloneqq\frac{x(n-x+1)}3-(n-x)^2-Dx.
\]
Using \(P_2=(P_1^2-2P_1-3n)/2\), the expansion at length \(n\) is
\[
 S=-\frac34DnP_0+\frac{3D-n-3}{12}P_1-\frac16P_2.
\]
Orthogonality, with squared norms \(1,3n,9\binom n2\), gives
\begin{equation}\label{eq:bg-second-operator}
 \E R_2=-\frac34Dn\beta
 +\frac{n(3D-n-3)}4c_1-\frac34n(n-1)c_2.
\end{equation}
Since \(n\ge3D-1\), the coefficient of \(c_1\) is negative.
Consequently
\begin{equation}\label{eq:bg-second-upper}
 \frac{c_2}{\beta}
 \le\frac{(4n/3)(1-u)-D}{n-1}.
\end{equation}
The roots exceed \(D\), so Bernoulli's inequality yields
\[
 \frac{F(2)}{u^2}
 =\prod_{j=1}^D\left(1-\frac1{(z_j-1)^2}\right)
 \ge(1-D^{-2})^D\ge1-D^{-1}.
\]
It is therefore enough to prove
\begin{equation}\label{eq:bg-second-scalar}
 (1-D^{-1})(1-n^{-1})u^2+\frac43u+\frac Dn-\frac43>0.
\end{equation}
We first establish this inequality for \(m\ge20\) and for
\(N\ge15m\), then use a sharper version on their finite complement
\(2\le m\le19,\ N<15m\). These sufficient thresholds make the
explicit scalar lower bounds below positive.

\paragraph{A logarithmic budget from minimality.}
The normalized residual \(h/h(0)\) attains \(\beta_{N,2m,4}\),
the minimum expectation on \(\{0,\ldots,N\}\):
a better grid-nonnegative residual polynomial of degree at most
\(2m\) could be lifted by the same terminal factor to improve \(F\).
As in the competitor comparison of Section~\ref{sec:bg-minimum},
the normalized rounded Lloyd competitor \(h_L/h_L(0)\) of
Lemma~\ref{lem:radau-rounded},
applicable since \(2m\le N\), thus gives
\(\mathcal G(q)/h(0)\le1/V_m(N)\).
For monic \(q\), Lemma~\ref{lem:binary-monic-moment}, applied at
auxiliary length \(N+1\) to \(q(x-1)\), gives the first bound below:
the lemma averages a forward product at binomial length
\(L_{\rm aux}-1\), and these choices turn that product into
\(q(X)q(X-1)\) with \(X\sim\mu_N\). Combining it with the
competitor comparison gives the second bound:
\[
 \mathcal G(q)\ge\frac23m!(N)_m(3/16)^m,\qquad
 h(0)\ge\frac23(N)_m^2(3/4)^{2m},
\]
where we used \(V_m(N)\ge\binom Nm3^m\).
If \(t_1,\ldots,t_{2m}\) are the residual roots, then
\(h(0)=\prod_jt_j\). Taking logarithms of the preceding lower
bound and using \((N)_m=\prod_{j=0}^{m-1}(N-j)\) gives
\begin{equation}\label{eq:bg-second-budget}
 \sum_{j=1}^{2m}\log\frac N{t_j}\le B,\qquad
 B\coloneqq\log\frac32+2m\log\frac43+
                 2\sum_{j=0}^{m-1}\log\frac N{N-j}.
\end{equation}
For \(0\le\ell\le\log(N/A(N,m))\), the function
\(g(\ell)\coloneqq\log(1-e^\ell/N)\) is concave, and \(-g'\) is
positive and convex. Concavity places \(g\) above its secant
between \(0\) and \(\log(N/A(N,m))\). The magnitude of that
secant slope is the average of \(-g'\), at most the average of its
endpoint values \(1/(N-1)\) and \(1/(A(N,m)-1)\). Hence
\[
 g(\ell)\ge g(0)
 -\frac{\ell}{2}\left(\frac1{N-1}+\frac1{A(N,m)-1}\right).
\]
Summing and using \eqref{eq:bg-second-budget} gives
\begin{equation}\label{eq:bg-second-log-product}
 \log\frac{h(1)}{h(0)}
 \ge-2m\log\frac N{N-1}
 -\frac B2\left(\frac1{N-1}+\frac1{A(N,m)-1}\right).
\end{equation}
All denominators are positive. A convenient weaker bound, using
\(A(N,m)>(H+1)/2\), is
\[
 \log\frac{h(1)}{h(0)}
 \ge-\frac{2m}{N-1}
 -\frac B2\left(\frac1{N-1}+\frac2{N-2m-1}\right),
 \quad
 B\le\log\frac32+2m\log\frac43+
                         \frac{m(m-1)}{N-m+1}.
\]

\paragraph{The region \(m\ge20\).}
Put
\[
 C\coloneqq\frac{2301}{2380}=\frac{39}{40}\frac{118}{119},
 \quad \epsilon\coloneqq\frac1{20},\quad x\coloneqq\frac rm,\quad y\coloneqq\frac Nm,
 \quad c\coloneqq\frac{5963}{10000},
\]
and define
\[
 E(y)\coloneqq\frac2{y-\epsilon}
 +\frac12\left(c+\frac1{y-1}\right)
       \left(\frac1{y-\epsilon}+\frac2{y-2-\epsilon}\right).
\]
Here \(D\ge40,n\ge119\), so
\((1-D^{-1})(1-n^{-1})\ge C\).
The constant \(\epsilon\) bounds \(1/m\), while \(c\) bounds
\(2\log(4/3)+\log(3/2)/m\), as verified just below. Thus
\[
 \frac Bm\le c+\frac{m-1}{N-m+1}\le c+\frac1{y-1},\qquad
 \frac m{N-1}\le\frac1{y-\epsilon},\qquad
 \frac m{N-2m-1}\le\frac1{y-2-\epsilon}.
\]
Also \(y\ge6+2x-\epsilon\). Substituting these bounds in the
preceding logarithmic estimate, and using
\(u=(N/n)h(1)/h(0)\), gives
\[
 u\ge\frac{y}{y+x}e^{-E(y)}.
\]
For the constant in \(E\), the series
\[
 \log\frac{1+z}{1-z}
 =2\sum_{k\ge0}\frac{z^{2k+1}}{2k+1}
 \le2z+\frac{2z^3}{3(1-z^2)}
\]
at \(z=1/7,1/5\) gives
\(\log(4/3)<36/125\), \(\log(3/2)<203/500\);
hence \(2\log(4/3)+\log(3/2)/20<c\).

For fixed \(y\), write \(v\coloneqq e^{-E(y)}\) and \(w\coloneqq(y+x)^{-1}\).
Inserting the lower bound \(u\ge yvw\) into
\(Cu^2+4u/3+D/n-4/3\) gives
\[
 Cy^2v^2w^2+\left(\frac43yv-y+2\right)w-\frac13.
\]
Its leading coefficient is positive and its constant negative.
It has exactly one positive root, so positivity at the smallest
allowed positive \(w\) implies positivity for all larger \(w\).
It suffices to set \(x=(y-6+\epsilon)/2\), where
\[
 \frac Dn=\frac{y-2+\epsilon}{3y-6+\epsilon}>\frac13,
 \qquad
 u\ge U_*(y)\coloneqq\frac{ye^{-E(y)}}{3y/2-3+\epsilon/2}.
\]
We claim that \(U_*(y)>27/50\) for \(y\ge6-\epsilon\).
Differentiation, discarding positive terms, gives
\[
 -E'(y)\ge
 \frac{2+\frac32(c+1/(y-1))}{y^2},
 \qquad
 \frac{d}{dy}\log U_*(y)
 =-E'(y)-\frac{3-\epsilon/2}
                   {y(3y/2-3+\epsilon/2)}.
\]
On \(6-\epsilon\le y\le7\), the first displayed lower bound
exceeds \(87/(28y^2)\), using \(c>4/7\), whereas the term
subtracted is below \(3/y^2\).
On \(y\ge7\), the respective bounds are \(20/(7y^2)\)
and \(14/(5y^2)\). Thus \(U_*\) increases.
At \(y_0\coloneqq119/20\) its rational prefactor is one, and
\[
 E(y_0)=\frac{278522909}{455598000}<\frac{123}{200},
 \qquad
 \log\frac{50}{27}>
 2\frac{23}{77}+\frac23\left(\frac{23}{77}\right)^3
 =\frac{842536}{1369599}>\frac{123}{200}.
\]
This proves the claim. At the largest allowed \(x\), the desired
scalar expression is therefore greater than
\[
 C(27/50)^2+\frac43(27/50)-1
 =\frac{11429}{5950000}>0.
\]
The quadratic argument extends this to every allowed \(x\).
Thus \eqref{eq:bg-second-scalar} holds for \(m\ge20\).

\paragraph{The region \(N\ge15m\).}
This region needs only Lemma~\ref{lem:bg-root} and integer distinctness.
If \(a\) is the smallest residual root, ordered roots and telescoping
give
\begin{equation}\label{eq:bg-second-packed}
 \frac{h(1)}{h(0)}
 \ge\frac{a-1}{a+2m-1}>
       \frac{N-2m-1}{N+2m-1}.
\end{equation}
Since \(m\ge2\), we have \(1/m\le1/2\); put \(x\coloneqq r/m\), \(y\coloneqq N/m\ge15\).
The last fraction decreases with \(1/m\), so
\[
 u>\frac{y}{y+x}\frac{y-5/2}{y+3/2},
 \qquad 0\le x\le\frac{y-6+1/m}{2}\le\frac{y-11/2}{2}.
\]
Now \(D\ge4,n\ge30\), whence
\((1-D^{-1})(1-n^{-1})\ge29/40\).
The same quadratic argument, with \(C\coloneqq29/40\), reduces the scalar
inequality to the largest allowed \(x\). There \(D/n>1/3\) and
\[
 u>\frac{4y(2y-5)}{(6y-11)(2y+3)}>\frac{23}{40}.
\]
The second comparison is equivalent to
\(44y^2-708y+759>0\). This polynomial is \(39\) at \(y=15\)
and its derivative \(88y-708\) is positive thereafter.
The scalar expression is consequently greater than
\[
 \frac{29}{40}(23/40)^2+\frac43(23/40)-1
 =\frac{1223}{192000}>0.
\]

\paragraph{The finite complement.}
The remaining parameters satisfy
\[
 2\le m\le19,\quad 0\le r\le\lfloor9m/2\rfloor,\quad
 6m+2r-1+2(r\bmod2)\le N<15m.
\]
The parity correction is exactly that of the length bound at the
start of Section~\ref{sec:bg-minimum}.
For a sharper scalar test set \(a\coloneqq\lceil A(N,m)\rceil\).
The packed residual roots give
\[
 \frac{h(2)h(0)}{h(1)^2}
 \ge\frac{(a-2)(a+2m-1)}{(a-1)(a+2m-2)}.
\]
The terminal factor has first value \(N/n\) and
second-to-first-square ratio \(n(N-1)/(N(n-1))\). Consequently,
with \(B\) from \eqref{eq:bg-second-budget},
\begin{align}
 \rho_*&\coloneqq\frac{(a-2)(a+2m-1)}{(a-1)(a+2m-2)}
                  \frac{n(N-1)}{N(n-1)},\nonumber\\
 E&\coloneqq2m\log\frac N{N-1}
       +\frac B2\left(\frac1{N-1}+\frac1{A(N,m)-1}\right),\nonumber\\
 u&\ge u_*\coloneqq\frac Nn
      \max\left(e^{-E},\frac{a-1}{a+2m-1}\right),
 \qquad F(2)\ge\rho_*u^2.
 \label{eq:bg-second-finite-bound}
\end{align}
It suffices that
\((1-n^{-1})\rho_*u_*^2+4u_*/3+D/n-4/3>0\).
Lemma~\ref{lem:bg-second-finite} proves this strict inequality
throughout the displayed finite domain except for
\[
 \begin{aligned}
 (m,r,N)={}&(2,0,11),(2,0,12),(2,0,13),(3,0,17),(3,0,18),\\
           &(4,0,23),(4,0,24),(5,0,29),(6,0,35).
 \end{aligned}
\]
At those nine parameters, the table in the proof of
Lemma~\ref{lem:bg-second-finite} specifies the lower zeros of the candidate's pairs.
For each listed candidate the lemma provides strictly positive
cardinal quadrature weights and a direct strict second-coefficient
comparison. The weights prove uniqueness: every normalized minimizer
must vanish at every positive quadrature node, so its \(D\) roots
and origin normalization determine it. Thus these cases cover the
selected vertex as well. The two infinite regions and this finite
complement prove \(c_2<\beta F(2)\) for \(m\ge2\), and complete
Lemma~\ref{lem:bg-low-indices}.
\end{proof}

\subsection{Stronger comparisons for the coefficient forms}
\label{sec:bg-coefficients}

\begin{proof}[Proof of Proposition~\ref{prop:bg-coefficients}]
Throughout this subsection \(m\ge2\), \(r\ge0\),
\(D=2m+r\), and \(N\ge6m+2r-1\) are integers.
In the congruence of \eqref{eq:binary-congruence},
\(\mathcal G\) and \(\mathcal C_i\) have matrices
\(\operatorname{Gram}_0\) and \(\operatorname{Mat}_i\).
Thus a positive definite difference
\(aU_i\operatorname{Gram}_0-\operatorname{Mat}_i\) proves
\(\mathcal C_i(q)<aU_i\mathcal G(q)\) for every nonzero \(q\).
Here \(\operatorname{Gram}_\ell\) represents the residual
Krawtchouk coefficient \(b_\ell\), and the lift is
\[
 \operatorname{Mat}_i=\frac1{\binom Ni}
   \sum_{\ell=0}^{2m}\binom N\ell\binom r{i-\ell}
                     \operatorname{Gram}_\ell.
\]
The kernels and lift are exactly
\eqref{eq:binary-C-kernel}--\eqref{eq:binary-lifted-kernel};
in particular the lift denominator is \(\binom Ni\).
Their relevant entries are nonnegative, as proved after
\eqref{eq:binary-matrix-target}, and the four signed-row
criterion is \eqref{eq:binary-four-signs}: because
\(\operatorname{Gram}_0\) is tridiagonal, only its two nearest
off-diagonal positions require choices of signs in each row.
These facts allow smaller factors multiplying \(U_i\) without changing the
coefficient matrices.

For positive weights \(w_j\), we write
\[
 \operatorname{row}_j(M)\coloneqq\sum_kM_{jk}\frac{w_k}{w_j},\qquad
 \operatorname{signedrow}_j(M)\coloneqq M_{jj}
       -\sum_{k\ne j}|M_{jk}|\frac{w_k}{w_j}.
\]
The second quantity is the least of the signed margins for that row.

For \(D\ge16\) we prove the linear factor \(1-2i/(3D)\) at every
interior index \(2\le i<D\), and the quarter factor at
\(100\le i<D\). The top-index argument holds for all \(D=2m+r\)
under the standing assumptions, with the single exception
\((m,r,N)=(2,0,11)\), which has \(D=4\) and so does not occur in
Proposition~\ref{prop:bg-coefficients}. The proof is divided as follows.
\begin{center}\small
\begin{tabular}{lll}
\toprule
Factor&Indices&Argument below\\
\midrule
Linear&\(2\le i\le\min(54,D-1)\)&Cleared polynomial arrays\\
Linear&\(55\le i<D\), inside the strip&Recurrence rows\\
Linear&\(55\le i\le61\), outside the strip&Finite row tests\\
Linear&\(62\le i<D\), outside the strip&Exponential estimate\\
Quarter&\(100\le i<D\)&Universal quarter bound\\
Quarter&\(i=D\)&Top-index calculation\\
\bottomrule
\end{tabular}
\end{center}
For the minimizing vertex, the first two indices already follow from
Lemma~\ref{lem:bg-low-indices}, and the quarter factor serves
\(i\ge100\). The additional linear ranges here hold for arbitrary
\(q\) and are recorded alongside the comparisons used for the minimizer.

\paragraph{Indices two through fifty-four.}
In each of the existing cleared signed-row expressions write the
numerator as \(P\coloneqq A_{\rm end}-A_{\rm row}\), where \(A_{\rm end}\)
is the contribution from the \(U_i\operatorname{Gram}_0\) term,
using its rational lower bound. It is unrelated to the evaluation
forms \(\mathcal E_i\) in the endpoint inequalities.
The replacement of \(U_i\) by
\((1-2i/(3D))U_i\), after an additional positive multiplication
by \(3D\), changes this numerator to
\begin{equation}\label{eq:bg-strengthened-arrays}
 P^{\rm new}=3DP-2iA_{\rm end}.
\end{equation}
More explicitly, the contributions from that term in
\eqref{eq:binary-second-numerator},
\eqref{eq:binary-four-sign-polynomial} and
\eqref{eq:binary-scalar-polynomial} are respectively
\[
 N(N-1)a_{\rm row}F,\qquad
 12S_i s_{\rm raw}E_i,\qquad 4^iSE_i.
\]
The \(E_i\) here is the clearing polynomial defined in
\eqref{eq:binary-endpoint-clearing}; it is distinct from the
evaluation form \(\mathcal E_i\).
All the original clearing multipliers are positive.
The signed \(\operatorname{Gram}_0\) factors are positive, so using
the lower bound \(v_i^{\rm end}\le U_i\) defined in
\eqref{eq:binary-rational-endpoint} still gives a sufficient
margin. The scalar reduction to \(j=m-1,m\) in
\eqref{eq:binary-two-scalars} is unchanged, since the new factor is
positive and independent of the row index.

Use the complete orthant partitions from
Appendices~\ref{app:binary-second},
\ref{app:binary-signed-rows} and~\ref{app:binary-scalar}.
An array is the list of coefficients of a cleared numerator after
the parameters are expressed as fixed lower bounds plus nonnegative
variables. Nonnegative coefficients and a positive constant prove
positivity on that entire orthant; shifted orthants cover the
unbounded ranges, and exact matrix tests cover their finite complements.
Lemma~\ref{lem:bg-low-coeff-finite} proves coefficient positivity
of the strengthened arrays, with the following shifted orthants
and their exact finite complements.
For the last row with \(m=8+a,r=c,N=6m+2r-1+s\), the arrays at
\(i=11,12\) are positive if \(a\ge1\), \(c\ge1\), or \(s\ge1\).
At \(i=13\), the sufficient union is \(a\ge1\), \(c\ge2\), or
\(s\ge1\). At \(m=7,r=2+c,i=13\), it is \(c\ge1\) or \(s\ge1\).
These leave five integer parameter cases.
The second-index last-row array with lower sign \(-1\) is positive
after a unit shift in any of \(a,r,s\), where \(m=2+a\);
its complement is \((N,m,r,i)=(11,2,0,2)\).
Finally, each of the forty-two affected scalar arrays has fixed
\(m,r\) and is positive after its listed shift \(s\mapsto s+t\),
\(t\in\{1,2\}\); their integer complements give forty-five further
cases. The finite lemma lists those shifts and proves the
corresponding fifty-one distinct matrix differences positive definite.
It also specifies the complete coefficient tests for the
\(35\,476\) low-band arrays and eight second-index arrays.
Thus \eqref{eq:bg-strengthened-arrays}, the exhaustive orthant cover,
and strict weighted diagonal dominance prove
\begin{equation}\label{eq:bg-low-linear-coefficient}
 \mathcal C_i(q)<
 \left(1-\frac{2i}{3D}\right)U_i\mathcal G(q),
 \qquad 2\le i\le\min(54,D-1),\quad D\ge16.
\end{equation}
The second-index array argument itself does not need \(D\ge16\);
it supplies the same comparison for arbitrary \(q\), although the
proof for the minimizer uses Lemma~\ref{lem:bg-low-indices} at this index.

\paragraph{Rows in the recurrence strip.}
Put
\[
 X\coloneqq\frac{N(2N-3m+6)}{3(8N-9m)},\qquad
 \operatorname{gap}_{\rm bin}\coloneqq1-\frac{3m}{2N},\qquad
 \lambda\coloneqq\frac14+\frac{7m}{4N}.
\]
Here \(\lambda\), and \(\tau\) below, retain their meanings from
Appendix~\ref{app:binary-high}; they are distinct from the multipliers
in Lemma~\ref{lem:bg-cap-comparison}.
The recurrence and row calculations
\eqref{eq:binary-strip-H} and~\eqref{eq:binary-strip-row}
apply before their final numerical cutoff: for \(1\le i\le
\min(D,\lfloor X\rfloor)\), they give
\[
 U_i\ge {\operatorname{gap}_{\rm bin}}^{2i},\qquad
 \operatorname{row}_j(\operatorname{Mat}_i)
 \le\frac{91}{60}\left(\lambda+\frac r{N-i+1}\right)^i,
 \qquad
 \operatorname{signedrow}_j(\operatorname{Gram}_0)\ge\frac{17}{60}.
\]
Here the weights are \(\sqrt{v_j(N)}4^{-j}\), as in
Steps~1, 2 and~4 of the proof of Lemma~\ref{lem:binary-strip}.
For these weights the row multiplier and the signed Gram-row lower
bound come respectively from
\[
 1+\frac14+\frac{\theta_j^2}{1/4},\qquad
 1-\frac{j}{N-j+1}-\frac14-\frac{\theta_j^2}{1/4},
 \qquad \theta_j^2=\frac{j}{3(N-j+1)}.
\]
Taking the largest first expression and the smallest second expression
over \(j\le m\) gives the bounds \(91/60\) and \(17/60\).
The lower index \(55\) in Lemma~\ref{lem:binary-strip}
enters only its scalar rate comparison and final threshold.

We first prove a quarter comparison for \(N\ge17m\).
Then \(j/(N-j+1)\le1/16\) and
\(\theta_j^2=j/[3(N-j+1)]\le1/48\), so the same row
calculations improve the row multiplier \(91/60\) and the
signed Gram-row lower bound \(17/60\), respectively, to
\[
 1+\frac14+\frac{1/48}{1/4}=\frac43,\qquad
 1-\frac1{16}-\frac14-\frac{1/48}{1/4}=\frac{29}{48}.
\]
To compare the base \(\lambda+r/(N-i+1)\) of the row bound
with \({\operatorname{gap}_{\rm bin}}^2\), set
\[
 \begin{aligned}
 Z_{\rm strip}&\coloneqq22N^2-24Nm+18N-27m,\\
 F^{\rm strip}_{a,b}&\coloneqq a(2N-3m)^2Z_{\rm strip}
 -b\{N(N+7m)Z_{\rm strip}+6N^2(N-6m+1)(8N-9m)\}.
 \end{aligned}
\]
Thus \(F^{\rm strip}_{32,33}=F_{\rm strip}\) of
Appendix~\ref{app:binary-high}.
Since \(N-X+1=Z_{\rm strip}/[3(8N-9m)]>0\), substitution of
\(r\le(N-6m+1)/2\), \(i\le X\), gives
\begin{equation}\label{eq:bg-rate-polynomial}
 \frac ab {\operatorname{gap}_{\rm bin}}^2-
 \left\{\lambda+\frac{N-6m+1}{2(N-X+1)}\right\}
 =\frac{F^{\rm strip}_{a,b}}{4bN^2Z_{\rm strip}}.
\end{equation}
At \(m=2+A,N=17m+s\), the polynomial \(F^{\rm strip}_{7,8}\) is
\begin{align*}
 13246120&+29674508A+24642582A^2+9009761A^3+1225666A^4\\
 &+(3241640+5072952A+2641722A^2+457828A^3)s\\
 &+(221232+225084A+57234A^2)s^2
 +(5944+2984A)s^3+56s^4.
\end{align*}
All its coefficients are positive. Thus the row base is less than
\((7/8){\operatorname{gap}_{\rm bin}}^2\), and every signed row of
\((U_i/4)\operatorname{Gram}_0-\operatorname{Mat}_i\) is at least
\[
 {\operatorname{gap}_{\rm bin}}^{2i}
 \left\{\frac{29}{192}-\frac43(7/8)^i\right\}>0
 \quad(i\ge17),
\]
because \(29\cdot8^{17}-256\cdot7^{17}=5748783016147200>0\).
Consequently
\begin{equation}\label{eq:bg-quarter-long-strip}
 \mathcal C_i(q)<\frac{U_i}{4}\mathcal G(q),
 \qquad N\ge17m,\quad17\le i\le\min(D-1,\lfloor X\rfloor).
\end{equation}

For the remaining lengths we sharpen the rate differently.
The two regions used below suffice because \(55\le i\le X\)
will force \(m\ge87\) when \(N<8m\), and \(m\ge20\) when
\(8m\le N<17m\), as checked after the coefficient table.
The complete coefficients of \(A^js^k\) in \(F^{\rm strip}_{28,29}\)
at \(m=80+A,N=6m-1+s\), and in \(F^{\rm strip}_{19,20}\)
at \(m=3+A,N=8m+s\), are displayed below; all other
coefficients vanish.
\begin{center}\small
\begin{tabular}{crr}
\toprule
\((j,k)\)&\(F^{\rm strip}_{28,29}\)&\(F^{\rm strip}_{19,20}\)\\
\midrule
\((0,0)\)&7950612732&2608605\\
\((0,1)\)&26267734090&3276558\\
\((0,2)\)&202895658&415746\\
\((0,3)\)&517086&18360\\
\((0,4)\)&434&272\\
\((1,0)\)&2278716335&6425757\\
\((1,1)\)&997648854&3548556\\
\((1,2)\)&5100327&283068\\
\((1,3)\)&6484&6104\\
\((2,0)\)&78124071&4686687\\
\((2,1)\)&12628677&1273518\\
\((2,2)\)&32052&48162\\
\((3,0)\)&947070&1368999\\
\((3,1)\)&53280&151576\\
\((4,0)\)&3888&141376\\
\bottomrule
\end{tabular}
\end{center}
Thus \eqref{eq:bg-rate-polynomial} proves the respective rates
\(28/29\) and \(19/20\) on these entire nonnegative orthants.
For either rate \(a/b\), the row comparison gives
\begin{equation}\label{eq:bg-strip-strengthened-rate}
 \mathcal C_i(q)<\frac{91}{17}\left(\frac ab\right)^iU_i\mathcal G(q).
\end{equation}

To check that the cones cover the required strip, \(X\) increases
with \(N\): its derivative has the sign of
\(48N^2-108mN+81m^2-162m\).
At \(m=2+A,N=6m-1+s\), this is
\[
 3432+4014A+1161A^2+(840+468A)s+48s^2>0.
\]
If \(N<8m\) and \(55\le i\le X\), then
\[
 55<X(8m)=\frac{104m+48}{165},
\]
so \(m\ge87\); the first cone applies. Moreover
\[
 1-\frac{2i}{3D}\ge1-\frac{i}{3m}>
 \frac{391}{495}-\frac{16}{165m}\ge\frac{1952}{2475},
 \qquad
 \frac{91}{17}(28/29)^{55}<\frac{1952}{2475}.
\]
The last comparison is the exact integer inequality
\(1952\cdot17\cdot29^{55}>2475\cdot91\cdot28^{55}\).
If \(8m\le N<17m\), then \(55\le X\le N/6\) gives \(m\ge20\).
The second cone applies, and
\[
 X\le X(17m-1)<\frac{7m}{5},
 \qquad 1-\frac{2i}{3D}>\frac8{15},
 \qquad \frac{91}{17}(19/20)^{55}<\frac8{15}.
\]
Here the first strict inequality is equivalent to
\(32m^2-353m+20>0\), whose substitution \(m=12+A\)
is \(392+415A+32A^2>0\).
The last is \(8\cdot17\cdot20^{55}>15\cdot91\cdot19^{55}\).
Both rates decrease with \(i\).
For \(N\ge17m\), \eqref{eq:bg-quarter-long-strip} suffices since
\(1-2i/(3D)>1/3\) at an interior index.
These three regions and \eqref{eq:bg-strip-strengthened-rate}
prove the linear coefficient bound for every
\(55\le i<D\) with \(i\le X\).

\paragraph{Indices outside the strip.}
In the notation of \eqref{eq:binary-adaptive-B},
Lemma~\ref{lem:binary-adaptive-row} bounds the coefficient row
by \(\kappa_N\mathcal B_i\) and the signed Gram row below by \(s_N\),
where
\[
 \begin{gathered}
 \tau\coloneqq\sqrt{\frac{m}{3(N-m)}},\qquad
 \theta_m^2=\frac{m}{3(N-m+1)},\\
 \kappa_N\coloneqq1+\tau+\frac{\theta_m^2}{\tau}<\frac85,\qquad
 s_N\coloneqq1-\frac{m}{N-m+1}-\tau-\frac{\theta_m^2}{\tau}>\frac15.
 \end{gathered}
\]
For \(55\le i<D\) and \(i>X\), the combination of
Lemmas~\ref{lem:binary-binomial-prefactor},
\ref{lem:binary-affine-gap} and~\ref{lem:binary-midpoint}
in the proof of Proposition~\ref{prop:binary-high-band} gives
\begin{equation}\label{eq:bg-adaptive-entropy}
 \frac{\mathcal B_i}{U_i}<
       \frac{392}{27}\sqrt{i}\,e^{-13i/100}.
\end{equation}
This use is restricted to interior indices, as required by the
midpoint estimate.

At \(i=62\), the right side of \eqref{eq:bg-adaptive-entropy}
is less than \(1/24\): indeed \(\sqrt{62}<8\) and
\[
 e^{806/100}>e^8>(163/60)^8>\frac{25088}{9},
 \qquad
 9\cdot163^8-25088\cdot60^8=270982108063090089>0.
\]
Here \(e>163/60\) is its positive series through degree five.
Since the logarithmic derivative of
\(\sqrt{i}e^{-13i/100}\) is \(1/(2i)-13/100<0\) for \(i\ge62\),
the comparison persists. Every signed row of
\((U_i/3)\operatorname{Gram}_0-\operatorname{Mat}_i\)
is then strictly greater than
\(U_i/15-(8/5)(U_i/24)=0\). Thus
\[
 \mathcal C_i(q)<\frac{U_i}{3}\mathcal G(q)
 <\left(1-\frac{2i}{3D}\right)U_i\mathcal G(q),
 \qquad i>X,\quad62\le i<D.
\]
We also record the stronger quarter comparison from \(i=64\).
At that index,
\[
 \begin{gathered}
 e^{208/25}> (163/60)^8(33/25)>\frac{100352}{27},\\
 27\cdot33\cdot163^8-100352\cdot25\cdot60^8
       =22613408077445918811>0.
 \end{gathered}
\]
This gives \(\mathcal B_i/U_i<1/32\) for all \(i\ge64\),
and the quarter signed margin is greater than
\(U_i/20-(8/5)(U_i/32)=0\).

It remains to treat \(55\le i\le61\) outside the strip.
Since \(X>N/13\), every such parameter belongs to the finite
projection
\begin{equation}\label{eq:bg-outside-projection}
 \begin{gathered}
 55\le i\le61,\quad m\ge2,\quad
 r_{\min}\coloneqq\max(0,i+1-2m),\qquad D_{\min}\coloneqq2m+r_{\min},\\
 N_{\min}\coloneqq6m+2r_{\min}-1\le N<13i,\qquad
 r_{\min}\le r\le r_{\max}\coloneqq
           \left\lfloor\frac{N-6m+1}{2}\right\rfloor.
 \end{gathered}
\end{equation}
There are \(870\) possible pairs \((m,i)\) with \(N_{\min}<13i\).
For fixed \(m\), \(\kappa_N\) decreases and \(s_N\) increases
with \(N\). The only non-immediate term is
\(\theta_m^2/\tau=\sqrt{m/3}\sqrt z/(z+1)\), \(z\coloneqq N-m>1\),
whose derivative has the sign of \(1-z\).
Hence \(\kappa_N/s_N\) decreases.
Since \(D_{\min}\le D\), the factor
\(1-2i/(3D_{\min})\) is at most \(1-2i/(3D)\).
Lemma~\ref{lem:bg-outside-finite} supplies, for each pair, a
certified boundary \(N_*\) at which
\[
 \frac{\kappa_{N_*}}{s_{N_*}}\,
       \frac{392}{27}\sqrt{i}e^{-13i/100}<1-\frac{2i}{3D_{\min}}.
\]
Monotonicity proves the coefficient comparison at all target
lengths \(N\ge N_*\). The remaining lengths, after imposing
\(i>X\), give \(190\) triples: \(147\) at \(i=55\),
\(40\) at \(i=56\), and \(3\) at \(i=57\).

For completeness, the finite row test at those triples covers all
lifts \(r\), not just their endpoints. Let \(K_\ell\) be the raw
kernel matrix of \eqref{eq:binary-C-kernel}, \(G\coloneqq K_0\), and
\[
 A(r)\coloneqq\frac1{\binom Ni}
       \sum_{\ell=0}^{2m}\binom N\ell\binom r{i-\ell}K_\ell.
\]
All relevant \(K_\ell\) entries are nonnegative, and
\(\binom r{i-\ell}\) increases with the nonnegative integer \(r\).
Thus \(0\le A(r)\le A(r_{\max})\) entrywise.
With weights \(w_j\coloneqq4^{-j}\), the finite lemma proves, for every row,
\begin{equation}\label{eq:bg-outside-rows}
 \gamma_j\coloneqq G_{jj}-\sum_{k\ne j}|G_{jk}|\frac{w_k}{w_j}>0,
 \qquad
 \sum_k A(r_{\max})_{jk}\frac{w_k}{w_j}
                 <\left(1-\frac{2i}{3D_{\min}}\right)U_i\gamma_j.
\end{equation}
The triangle inequality gives each signed row of
\((1-2i/(3D))U_iG-A(r)\) at least
\(\bigl(1-2i/(3D_{\min})\bigr)U_i\gamma_j-
\sum_kA(r_{\max})_{jk}w_k/w_j>0\).
This is an entrywise reduction followed by weighted diagonal
dominance; no matrix-order monotonicity in \(r\) is needed.
It proves the missing linear comparisons. Together with
\eqref{eq:bg-low-linear-coefficient} and the strip argument,
the linear bound is established for every interior \(2\le i<D\)
when \(D\ge16\).

\paragraph{The universal quarter bound from index one hundred.}
For \(i\le X\) and \(i\ge100\), the rate \(32/33\) proved in
Step~3 of Lemma~\ref{lem:binary-strip} applies: its regions
\(m\ge29\) and \(N\ge10m\) cover the strip, since
\(55\le i\le X\le N/6\) forces \(N\ge330\), and \(m\le28\)
then gives \(N>10m\). Combining this rate with the row constants
\(91/60\) and \(17/60\) recalled above gives every signed margin of
\((U_i/4)\operatorname{Gram}_0-\operatorname{Mat}_i\) at least
\[
 \frac{{\operatorname{gap}_{\rm bin}}^{2i}}{240}\{17-364(32/33)^i\}>0,
\]
since \(17\cdot33^{100}>364\cdot32^{100}\) and the power
decreases thereafter. For \(i>X\), the quarter comparison just
proved applies already from \(i=64\). Thus all interior
\(100\le i<D\) have the asserted quarter coefficient bound.

\paragraph{The top index.}
The calculation in Lemma~\ref{lem:binary-top} applies to arbitrary
\(q\), and gives
\[
 \frac{\mathcal C_D(q)}{\mathcal G(q)}
 \le\operatorname{Top}(m,r,N)
 \coloneqq\frac{3D!}{2m!(N)_m3^m(N-2m)_r}.
\]
For degree below \(m\) the numerator is zero.
For degree \(m\), scale to monic \(q\) and use
Lemma~\ref{lem:binary-monic-moment}; only the coefficient of
degree \(2m\) survives the terminal lift.
We use the two monotonicities proved in Steps~3 and~6 of
Lemma~\ref{lem:binary-top}. They concern only the functions
\(\operatorname{Top}(m,r,N)\) and \(U_D\) of these parameters,
so they apply equally to the present parameters of the minimizing vertex:
\(\operatorname{Top}/U_D\) decreases with \(N\) at fixed \(m,r\),
and on \(N=6m+2r-1\) it decreases with \(r\).
Thus it remains to strengthen its initial comparisons.

For \(m\ge15\), that proof gives \(U_{2m}\ge(5/48)^m\) at
\(r=0,N=6m-1\), and a decreasing sequence
\(R_m=\operatorname{Top}(m,0,6m-1)/(5/48)^m\), with
\[
 R_{15}=\frac{79551583817872441344}
                    {345374486968994140625}<\frac14.
\]
For \(m=3,\ldots,14\), every exact fraction in the initial table
in Step~5 of that proof satisfies \(U_{2m}/\operatorname{Top}>4\);
the smallest is \(606050/103089>4\).
For \(m=2\), the additional starting values are
\[
 \frac{\operatorname{Top}(2,0,12)}{U_4}
       =\frac{3155}{13542}<\frac14,\qquad
 \frac{\operatorname{Top}(2,1,13)}{U_5}
       =\frac{20405}{98982}<\frac14.
\]
They follow directly by substituting
\[
 Q_N(x)\coloneqq(3N-1-4x)(3N+1-4x)-3(N-1),\qquad
 U_i=\frac{Q_N(i)Q_N(i+1)}{Q_N(0)Q_N(1)},
\]
where \(Q_N=2!P_2^{(N-1,4)}(x-1)\).
The first starting value covers \(r=0,N\ge12\);
the second covers every \(r\ge1,N\ge6m+2r-1\) by the two
monotonicities. This proves the quarter bound at the top except
for \((m,r,N)=(2,0,11)\), where the coarse ratio is \(207/473\).
That parameter is handled by a matching half bound below.
The case \(m=2,r=0\) has \(D=4\); its values above complete the
top-index calculation on the standing domain, while the small-degree
proof below verifies it again as part of its matched comparisons.
The proof of Proposition~\ref{prop:bg-coefficients} is complete.
\end{proof}

\subsection{Small total degree and the infinite length tail}
\label{sec:bg-small-degree}

\begin{proof}[Proof of Proposition~\ref{prop:bg-small-degree}]
It remains to
cover \(m\ge2,D\le15\). The first two indices are already covered
by Lemma~\ref{lem:bg-low-indices}. The remaining coverage is
\[
 \begin{array}{c|c|l}
 \text{indices}&\text{length}&\text{argument}\\\hline
 1,2&\text{all allowed lengths}&\text{Lemma~\ref{lem:bg-low-indices}}\\
 3,\ldots,D&N\ge13D&\text{analytic tail below}\\
 3,\ldots,D&N<13D&\text{Lemma~\ref{lem:bg-small-finite}}.
 \end{array}
\]
The tail estimate uses only \(N\ge13D\); we apply it here to
\(D\le15\).

Suppose \(N\ge13D\). Then \(m/N\le1/26\), and \(X>N/13\ge D\),
so every required index lies in the recurrence strip.
For its fixed weights,
\[
 \frac{j}{N-j+1}<\frac1{25},\qquad
 \theta_j^2<\frac1{75}.
\]
The row multiplier and signed Gram-row bound, from the generic
formulas in Section~\ref{sec:bg-coefficients}, are therefore
\[
 1+\frac14+\frac4{75}=\frac{391}{300},\qquad
 1-\frac1{25}-\frac14-\frac4{75}=\frac{197}{300}.
\]
Furthermore
\[
 \operatorname{gap}_{\rm bin}\ge\frac{49}{52},\qquad
 \frac14+\frac{7m}{4N}\le\frac{33}{104},\qquad
 \frac r{N-i+1}<\frac1{12}
 \quad(1\le i\le D),
\]
using \(r\le D\) and \(N\ge13D\).
It follows that
\[
 \frac{1/4+7m/(4N)+r/(N-i+1)}{{\operatorname{gap}_{\rm bin}}^2}
 <\frac{3250}{7203}.
\]
The recurrence and row calculation in Steps~1 and~2 of
Lemma~\ref{lem:binary-strip} require only
\(1\le i\le\min(D,\lfloor X\rfloor)\), so they include \(i=D\)
here. The restriction to interior indices in
\eqref{eq:bg-adaptive-entropy} belongs to the separate midpoint
estimate outside the strip. The row calculation now gives
\begin{equation}\label{eq:bg-small-tail}
 \mathcal C_i(q)<
 \frac{391}{197}\left(\frac{3250}{7203}\right)^i
 U_i\mathcal G(q)<\frac{U_i}{4}\mathcal G(q),
 \qquad 3\le i\le D.
\end{equation}
Indeed it suffices to check \(i=3\), and
\[
 197\cdot7203^3-4\cdot391\cdot3250^3
       =19932619122119>0.
\]
The quarter endpoint in Proposition~\ref{prop:bg-endpoints}, proved
analytically in Lemma~\ref{lem:bg-quarter-uniform}, applies since
\(N\ge13D\ge26m>6m-1\) and \(m\ge2\). Also \(n\ge3D-1\) gives
\(H+1\ge2D\), so every \(i\le D\) lies in its required range
\(i\le\lfloor(H+1)/2\rfloor\).
For the minimizer, \(\mathcal E_0\ge V_m(N)\mathcal G>0\).
Lemma~\ref{lem:bg-cap-comparison}, with
\(\lambda=U_i/4,\tau=U_i\), combines this endpoint inequality
with \eqref{eq:bg-small-tail} to prove
\(\mathcal C_i/\mathcal G<\mathcal E_i/\mathcal E_0\)
for every \(3\le i\le D\).

When \(N<13D\) and \(D\le15\), every remaining parameter belongs
to the exact finite domain
\[
 2\le m\le7,\quad 0\le r\le15-2m,\quad
 6m+2r-1\le N<13D.
\]
This is a superset of the code parameters: dropping the nonnegative
parity term \(2(r\bmod2)\) only enlarges the domain to be checked.
Lemma~\ref{lem:bg-small-finite} supplies, for \(3\le i<D\),
both the coefficient and endpoint forms
\[
 \mathcal C_i<
 \left(1-\frac{2i}{3D}\right)U_i\mathcal G,\qquad
 \frac{\mathcal E_i}{V_m(N)U_i}
 -\frac{\mathcal E_0}{V_m(N)}
 +\frac{2i}{3D}\mathcal G>0.
\]
At \(i=D\) it supplies the quarter pair except at
\((m,r,N,i)=(2,0,11,4)\), where it supplies instead
\[
 \mathcal C_i<\frac{U_i}{2}\mathcal G,\qquad
 \frac{\mathcal E_i}{V_m(N)U_i}
 -\frac{3\mathcal E_0}{2V_m(N)}+\mathcal G>0.
\]
These are matched pairs in Lemma~\ref{lem:bg-cap-comparison}:
their \((\lambda,\tau)\) are respectively
\[
 \left(\left(1-\frac{2i}{3D}\right)U_i,
                    \frac{2i}{3D}U_i\right),\qquad
 (U_i/4,U_i),\qquad (U_i/2,U_i).
\]
As noted before Lemma~\ref{lem:bg-small-finite}, the ordinary endpoint
forms also follow from Proposition~\ref{prop:bg-endpoints}:
\(H+1\ge2D\) gives \(4i/[3(H+1)]\le2i/(3D)\), and
\(i<D\le15\) is within its linear range. The ordinary top
coefficient bound was proved above. The finite certificates check
both members of each pair together, including the exceptional half
comparison, and thereby verify their matching constants directly.
Thus every finite pair gives the coefficient comparison for the minimizer
with the exact normalized lift from the main text.
The finite domain ends at \(N=13D-1\); equality \(N=13D\) lies
in \eqref{eq:bg-small-tail}. Together with the separate first two
indices, these ranges prove Proposition~\ref{prop:bg-small-degree}.
\end{proof}

\section{Endpoint inequalities for adjacent products}\label{app:bg-endpoints}

Throughout this appendix, $m\ge2$, $N\ge6m-1$, and $H\coloneqq N-2m$.
For a real polynomial $q$ of degree at most $m$, recall
\[
 \mathcal G(q)=\E_{\mu_N}q(X)q(X-1),\qquad
 \mathcal E_i(q)=q(i)q(i-1),\qquad V=V_m(N).
\]
We prove the two assertions of Proposition~\ref{prop:bg-endpoints}:
for every nonzero $q$,
\begin{align}
 \frac{\mathcal E_i(q)}{VU_i}-\frac54\frac{\mathcal E_0(q)}V
                 +\mathcal G(q)&>0,
 &1\le i\le\left\lfloor\frac{H+1}2\right\rfloor,
 \label{eq:bg-quarter-target}\\
 \frac{\mathcal E_i(q)}{VU_i}-\frac{\mathcal E_0(q)}V
          +\frac{4i}{3(H+1)}\mathcal G(q)&>0,
 &1\le i\le\min\left(99,\left\lfloor\frac{H+1}2\right\rfloor\right).
 \label{eq:bg-linear-target}
\end{align}
These statements concern every polynomial in the indicated space.
The minimizing property used in the main article enters only when these
form inequalities are combined with the coefficient bounds.
We first identify their matrices in the length-$(N-1)$ Krawtchouk
basis. The quarter bound is proved analytically for all parameters
(Lemma~\ref{lem:bg-quarter-uniform}). Differentiation gives the linear
bound near the origin. The following ranges, within the standing
hypotheses, cover both targets; the quarter row uses the full index range
$1\le i\le\lfloor(H+1)/2\rfloor$.
\[
\begin{array}{c|l|l}
 \text{target}&\text{range}&\text{result}\\\hline
 \text{quarter}&\text{all}&\text{Lemma~\ref{lem:bg-quarter-uniform}}\\
 \text{linear}&12i\le H+1&\text{Lemma~\ref{lem:bg-linear-near}}\\
 \text{linear}&2i\le H+1<12i,\ i\le99
    &\text{Lemma~\ref{lem:bg-finite-linear}}
\end{array}
\]
The finite lemma in Appendix~\ref{app:bg-finite} is certified on a
larger domain of $71{,}940{,}002$ triples by outward-rounded dyadic
intervals, with independent re-evaluation at a second precision; only
its $5{,}671{,}852$ triples with $2i\le H+1<12i$ are needed here. We
derive below the sufficient scalar inequality that it verifies.

\subsection{Normalized evaluation forms and their inverses}

Use the local notation
\[
 p_j(x)\coloneqq P_j^{(N-1,4)}(x),\qquad
 w_j\coloneqq\binom{N-1}j3^j,\qquad
 \rho_j^2\coloneqq\frac{w_{j-1}}{w_j}=\frac{j}{3(N-j)}.
\]
The generating function~\eqref{eq:kraw-generating}, interpreted as a
formal power series also at $x=-1$, gives
\[
 p_m(0)=w_m,\qquad
 p_m(-1)=[z^m]\frac{(1+3z)^N}{1-z}
            =\sum_{j=0}^m\binom Nj3^j=V.
\]
Consequently $U_i$ in~\eqref{eq:binary-endpoint} is exactly
$p_m(i)p_m(i-1)/(w_mV)$, since $L(x)=(m!/4^m)p_m(x-1)$.
Write $\operatorname{sym}(ab^{\mathsf T})\coloneqq(ab^{\mathsf T}+ba^{\mathsf T})/2$.
We use the correlation matrices of Lemma~\ref{lem:paired-grams}:
their diagonal entries are one and their $(j,k)$ entry for $j<k$ is
the product of the parameters from $j+1$ through $k$.
The same definition allows signed parameters; in particular
$\operatorname{Cor}_{-3\rho}=\operatorname{diag}((-1)^j)
\operatorname{Cor}_{3\rho}\operatorname{diag}((-1)^j)$.

\begin{lemma}\label{lem:bg-endpoint-normalization}
In the basis $p_j/\sqrt{w_j}$, $0\le j\le m$, the matrix of
$\mathcal G$ is
\begin{equation}\label{eq:bg-lloyd-gram}
 B=\frac{\operatorname{Cor}_\rho+
                  \operatorname{Cor}_{-3\rho}}2,
 \qquad
 B_{jk}=\frac{1+(-3)^{k-j}}2\sqrt{\frac{w_j}{w_k}}\quad(j<k),
 \qquad B_{jj}=1.
\end{equation}
It satisfies $B>\frac12I$. Define
\begin{equation}\label{eq:bg-evaluation-vector}
 u_j(x)\coloneqq\sqrt{\frac{w_m}{w_j}}\frac{p_j(x)}{p_m(x)}.
\end{equation}
Whenever its denominators are positive, the matrices of
$\mathcal E_i/(VU_i)$ and $\mathcal E_0/V$ are respectively
$\operatorname{sym}(u(i)u(i-1)^{\mathsf T})$ and
$\operatorname{sym}(u(0)u(-1)^{\mathsf T})$.
\end{lemma}
\begin{proof}
Using formal power series also when $X=N$ or $X-1=-1$, the
two-variable generating function for the unsymmetrized product is
\[
 \E\left[\sum_{j\ge0}p_j(X)z^j
                  \sum_{k\ge0}p_k(X-1)y^k\right]
       =\frac{(1+3zy)^N}{(1+3z)(1-y)}.
\]
For $j\le k\le m\le N-1$, coefficient extraction gives
\[
 \E[p_j(X)p_k(X-1)]=w_j,\qquad
 \E[p_k(X)p_j(X-1)]=(-3)^{k-j}w_j.
\]
Indeed $\sum_{a=0}^j\binom Na3^a(-3)^{j-a}=w_j$, by
$\sum_{a=0}^j(-1)^{j-a}\binom Na=\binom{N-1}j$.
Symmetrizing proves~\eqref{eq:bg-lloyd-gram} for every row.

For the lower bound, put $r\coloneqq\sqrt{2/27}$ and temporarily allow any
parameters $0\le t_k\le r$, writing
$B_t\coloneqq(\operatorname{Cor}_t+\operatorname{Cor}_{-3t})/2$.
The choice $t=\rho$ gives $B$, since for $m\ge2$
\[
 \rho_j^2\le\frac m{3(N-m)}\le\frac m{3(5m-1)}\le\frac2{27}.
\]
For constant parameters $t_k=r$, the entries are
\[
 (B_r)_{jk}=\frac{r^{|j-k|}+(-3r)^{|j-k|}}2.
\]
Since $3r<1$, the two geometric Fourier series converge absolutely.
Their mean symbol, with $z=\cos\theta$, is
\[
 \Phi(z)\coloneqq\frac12\left(
 \frac{1-r^2}{1+r^2-2rz}+\frac{1-9r^2}{1+9r^2+6rz}\right).
\]
For every finite complex vector $v=(v_0,\ldots,v_m)$ these
expansions give
\[
 v^*B_rv=\frac1{2\pi}\int_0^{2\pi}\Phi(\cos\theta)
           \Bigl|\sum_{j=0}^m v_je^{\mathrm ij\theta}\Bigr|^2\,d\theta.
\]
Both denominators are positive, and direct algebra gives
\[
 2\Phi(z)-1=
 \frac{12r^2(z+r)^2+1-10r^2-39r^4}
 {(1+r^2-2rz)(1+9r^2+6rz)},
\]
where $1-10r^2-39r^4=11/243>0$. The two denominator factors are at
most $(1+r)^2$ and $(1+3r)^2$, so
\[
 \Phi(z)\ge\frac12+\varepsilon,\qquad
 \varepsilon\coloneqq\frac{11/243}{2(1+r)^2(1+3r)^2}>0.
\]
Parseval's identity therefore gives $B_r\succeq(1/2+\varepsilon)I$
in every finite dimension.

For general parameters put $\delta_k\coloneqq t_k/r\in[0,1]$ and
$\Gamma\coloneqq\operatorname{Cor}_\delta$. By the autoregressive construction
in the proof of Lemma~\ref{lem:paired-grams}, which allows
$\delta_k=1$, $\Gamma$ is positive semidefinite with unit diagonal.
Entrywise, $B_t=B_r\circ\Gamma$, where $\circ$ is the entrywise
product. If $M\succeq0$ and $\Gamma=\sum_\ell\zeta_\ell\zeta_\ell^{\mathsf T}$,
then $M\circ\Gamma=\sum_\ell\operatorname{diag}(\zeta_\ell)M
\operatorname{diag}(\zeta_\ell)\succeq0$. Applying this to
$M=B_r-(1/2+\varepsilon)I$ and using $I\circ\Gamma=I$ gives
\[
 B_t\succeq\Bigl(\frac12+\varepsilon\Bigr)I>\frac12I,
\]
in particular for $B=B_\rho$. If $c$ is the coordinate vector
of $q$ in the basis $p_j/\sqrt{w_j}$, then
\[
 q(x)=\frac{p_m(x)}{\sqrt{w_m}}\langle u(x),c\rangle.
\]
Multiplication at $i,i-1$ and at $0,-1$ proves the two evaluation
formulas, using $p_m(i)p_m(i-1)=w_mVU_i$ and $p_m(0)p_m(-1)=w_mV$.
\end{proof}

We will use the inverse inner product
$\langle a,b\rangle_{B^{-1}}\coloneqq a^{\mathsf T}B^{-1}b$.
Reverse the coordinates, writing $k\coloneqq m-j$, and set
\[
 \eta_k\coloneqq\frac{m-k+1}{3(N-m+k-1)},\quad
 \beta_k\coloneqq\sqrt{\eta_k},\quad
 g_0\coloneqq1,\quad g_k\coloneqq\prod_{\ell=1}^k\beta_\ell,
 \qquad \eta_0\coloneqq0.
\]
Thus $\eta_k=\rho_{m-k+1}^2$ is the parameter joining coordinates
$k-1,k$, and $g_k=\sqrt{w_{m-k}/w_m}$. In these reversed
coordinates, $u(0)=g$ and
$B=(\operatorname{Cor}_\beta+\operatorname{Cor}_{-3\beta})/2$.
Let $S_{k,k-1}\coloneqq\beta_k$ be the strictly lower bidiagonal shift and set
$\mathsf D\coloneqq(I-S)(I+3S)=I+2S-3S^2$.
Direct multiplication gives the tridiagonal matrix
\begin{equation}\label{eq:bg-filter}
 \begin{gathered}
 \operatorname{Tri}\coloneqq\mathsf D B\mathsf D^{\mathsf T},\qquad
 \operatorname{Tri}_{00}=1,\\
 \operatorname{Tri}_{kk}=1-9\eta_k\eta_{k-1},\qquad
 \operatorname{Tri}_{k,k-1}=\beta_k(1+3\eta_{k-1}).
 \end{gathered}
\end{equation}
To obtain this reduction, write
$\Lambda_c\coloneqq\operatorname{diag}(1,1-c\eta_1,\ldots,1-c\eta_m)$.
The two correlation factorizations are
\[
 \operatorname{Cor}_\beta=(I-S)^{-1}\Lambda_1(I-S^{\mathsf T})^{-1},
 \qquad
 \operatorname{Cor}_{-3\beta}=(I+3S)^{-1}\Lambda_9(I+3S^{\mathsf T})^{-1}.
\]
Multiplication by $\mathsf D$ and its transpose gives
\[
 \operatorname{Tri}=\tfrac12\bigl[(I+3S)\Lambda_1(I+3S)^{\mathsf T}
                    +(I-S)\Lambda_9(I-S)^{\mathsf T}\bigr],
\]
whose entries are those in~\eqref{eq:bg-filter}.
The formula includes both finite boundaries; entries outside $0,\ldots,m$
are absent.
Its forward $LDL^{\mathsf T}$ pivots and a useful multiplier are
\begin{equation}\label{eq:bg-pivots}
 d_0\coloneqq1,\qquad
 d_k\coloneqq1-9\eta_k\eta_{k-1}
       -\frac{\eta_k(1+3\eta_{k-1})^2}{d_{k-1}},\qquad
 \kappa_k\coloneqq\frac{1+3\eta_k}{d_k}.
\end{equation}
They are positive because $\operatorname{Tri}$ is positive definite.
In the factorization $\operatorname{Tri}=\mathsf L\operatorname{diag}(d_0,\ldots,d_m)
\mathsf L^{\mathsf T}$ with unit lower triangular $\mathsf L$,
the only nonzero subdiagonal multiplier is
$\mathsf L_{k+1,k}=\beta_{k+1}\kappa_k$.
For an input $\mathsf Dv$, its forward innovations mean the entries
of $\mathsf L^{-1}\mathsf Dv$, obtained by forward substitution.
Thus inverse-Gram products are sums of products of innovations
divided by the corresponding pivots.

The finite proofs in Appendix~\ref{app:bg-finite} use the following
rational evaluation formulas. The analytic arguments resume with
Lemma~\ref{lem:bg-ratio-hull}. The three-term recurrence
$jp_j=(3N-2j-1-4x)p_{j-1}-3(N-j+1)p_{j-2}$ gives
\begin{equation}\label{eq:bg-degree-ratio}
 T_j(x)\coloneqq\frac{p_{j-1}(x)}{p_j(x)}
       =\frac{j}{\Delta_j(x)},\qquad
 \Delta_j(x)\coloneqq3N-2j-1-4x-3(N-j+1)T_{j-1}(x),\quad T_0=0.
\end{equation}
Put $h_0(x)\coloneqq1$, $h_{-1}(x)\coloneqq0$, and, for $j=m-k+1$, define
\begin{equation}\label{eq:bg-rational-innovation}
 \begin{split}
 h_k(x)&\coloneqq h_{k-1}(x)\frac{N-j}{\Delta_j(x)},\qquad z_0(x)\coloneqq1,\\
 z_k(x)&\coloneqq h_k(x)+\frac23h_{k-1}(x)-\frac13h_{k-2}(x)
                  -\frac{\kappa_{k-1}}3z_{k-1}(x),\\
 W_0&\coloneqq1,\qquad W_k\coloneqq9\eta_kW_{k-1}.
 \end{split}
\end{equation}
At $k=1$, $\kappa_0=1$ and the term $h_{-1}$ is zero.
Since $u_{m-k}(x)=3^kg_kh_k(x)$, $z_k(x)$ is the forward
innovation of $\mathsf D u(x)$ divided by $3^kg_k$.
The weights satisfy $W_k=(3^kg_k)^2$.
Thus
\begin{equation}\label{eq:bg-inverse-product}
 \langle u(x),u(y)\rangle_{B^{-1}}
           =\sum_{k=0}^m\frac{W_kz_k(x)z_k(y)}{d_k}.
\end{equation}
The same identity applies to linear combinations of the vectors by
taking the corresponding linear combinations of their innovations.
No tail or coordinate is omitted.

\subsection{Positive degree ratios and a convex hull}

\begin{lemma}\label{lem:bg-ratio-hull}
All the denominators in~\eqref{eq:bg-degree-ratio} are positive for
$-1\le x\le(H+1)/2$. On $0\le x\le(H+1)/2$, put
$K_j(x)\coloneqq T_j(x)/\rho_j^2$. Then
\[
 1\le K_1(x)\le\cdots\le K_m(x)\le3.
\]
Every vector $u(x)$ on this interval is a convex combination of
\begin{equation}\label{eq:bg-steps}
 a^{(t)}_{m-k}\coloneqq3^{\min(k,t)}g_k,\qquad 0\le t,k\le m.
\end{equation}
Each coordinate of $u(x)$ is positive, nondecreasing, and convex on
$(-\infty,(H+1)/2]$.
\end{lemma}
\begin{proof}
The positivity and convexity assertions hold on the larger half-line
in the statement; the normalized-ratio bounds and convex hull use
only its nonnegative part. The degree-$m$ Jacobi matrix for
$\Bin(N-1,3/4)$ has diagonal
$(3(N-1)-2j)/4$ and off-diagonal absolute values
$\sqrt{3j(N-j)}/4$, with row indices $0,\ldots,m-1$.
Its smallest diagonal is at $j=m-1$, and every row has at most two
off-diagonal entries, each at most
$\sqrt{3(m-1)(N-m+1)}/4$ in absolute value.
Gershgorin therefore bounds the least zero $a$
of $p_m$ below by
\[
 a\ge\frac{3N-2m-1}4
       -\frac12\sqrt{3(m-1)(N-m+1)}.
\]
The difference between this expression and $(H+1)/2$ increases with
$N\ge6m-1$: its derivative is
$1/4-3(m-1)/(4\sqrt{3(m-1)(N-m+1)})>0$.
At $N=6m-1$ the difference is
$2m-1-(\sqrt{15}/2)\sqrt{m(m-1)}>(m-1)/16$,
using $\sqrt{m(m-1)}<m-1/2$ and $\sqrt{15}<31/8$.
Hence $a>(H+1)/2$.
All lower-degree zeros lie above $a$ by the variational principle
for principal Jacobi blocks. If $\mathcal J_j$ is the degree-$j$
block, its last diagonal resolvent entry gives
\[
 T_j(x)=\frac j4[(\mathcal J_j-xI)^{-1}]_{j-1,j-1}
       =\frac j4\sum_{\nu=1}^j\frac{\mu_\nu}{\xi_\nu-x},
 \qquad \mu_\nu>0,\quad\sum_\nu\mu_\nu=1,\quad\xi_\nu\ge a.
\]
Here $\mu_\nu$ is the square of the last coordinate of a normalized
eigenvector; the nonzero Jacobi off-diagonals make it positive.
For $x<a$ this proves positivity of the ratios, their denominators,
and every derivative.
Products preserve positivity, monotonicity and convexity, giving the
last assertion for $u_j=\prod_{\ell=j+1}^m(T_\ell/\rho_\ell)$.

For $x\le(H+1)/2$, substitution of
$T_{j-1}\le(j-1)/(N-j+1)$ in~\eqref{eq:bg-degree-ratio} gives
\[
 \Delta_j(x)\ge N-5j+4m\ge N-j>0.
\]
Starting with $j=1$, induction gives $T_j\le j/(N-j)$, or $K_j\le3$.
The normalized recurrence is
\[
 K_j(x)=\left(1-
 \frac{4x+(j-1)(K_{j-1}(x)-1)}{3(N-j)}\right)^{-1},
 \qquad K_0=1.
\]
For $x\ge0$, denote the right side, as a function of the previous
value, by $\phi_j(K)$. It increases with $K\ge1$ and with $j$
wherever its denominator is positive. Inductively,
\[
 K_j=\phi_j(K_{j-1})\ge\phi_j(K_{j-2})
       \ge\phi_{j-1}(K_{j-2})=K_{j-1}\ge1.
\]
The denominator at $K_{j-2}$ is at least the positive one at
$K_{j-1}$, which justifies each comparison.

Set $b_k\coloneqq u_{m-k}(x)/g_k$. Then $b_0=1$ and
$b_k/b_{k-1}=K_{m-k+1}$ decreases with $k$ and lies in $[1,3]$.
Define
\[
 h_k^*\coloneqq\frac{b_k-b_{k-1}}{2\,3^{k-1}},\quad 1\le k\le m,
 \qquad h_{m+1}^*\coloneqq0.
\]
The identity
$h_k^*=(K_{m-k+1}-1)\prod_{\ell<k}(K_{m-\ell+1}/3)/2$
shows $1\ge h_1^*\ge\cdots\ge h_m^*\ge0$.
The weights $\lambda_0\coloneqq1-h_1^*$ and
$\lambda_t\coloneqq h_t^*-h_{t+1}^*$ sum to one. Their mixture of
$3^{\min(k,t)}$ has increment
$2\,3^{k-1}\sum_{t\ge k}\lambda_t=b_k-b_{k-1}$ and initial value
one. It is therefore $b_k$, proving the asserted convex combination.
\end{proof}

\subsection{The quarter endpoint inequality}

\begin{lemma}\label{lem:bg-quarter-uniform}
For integers $m\ge2$, $N\ge6m-1$, $1\le i\le\lfloor(H+1)/2\rfloor$, and every nonzero
real polynomial $q$ of degree at most $m$, inequality~\eqref{eq:bg-quarter-target} holds.
\end{lemma}

The proof reduces the form to fixed weights, bounds the covariance
from below by a three-dimensional block, treats zero and positive
profiles separately, and is completed at the end of this subsection.

\paragraph{Reduction to fixed weights.}
Lemma~\ref{lem:bg-endpoint-normalization}, applicable since its
denominators are positive by Lemma~\ref{lem:bg-ratio-hull}, identifies
this form with
\[
 B+\operatorname{sym}(u(i)u(i-1)^{\mathsf T})
 -\frac54\operatorname{sym}(u(0)u(-1)^{\mathsf T}).
\]
In the reversed coordinates $k=m-j$ introduced above, $u(0)=g$ and
$B=\frac12(\operatorname{Cor}_\beta+\operatorname{Cor}_{-3\beta})$.
Define the dominating weights
\[
 \bar\eta_1\coloneqq\frac2{27},\qquad \bar\eta_k\coloneqq\frac1{15}\ (k\ge2),
 \qquad \bar g_k\coloneqq\prod_{j=1}^k\sqrt{\bar\eta_j}.
\]
Indeed $\eta_1\le2/27$, while, for $k\ge2$,
\[
 \eta_k\le\frac{m-k+1}{3(5m+k-2)}\le\frac1{15};
\]
the last comparison is equivalent to $21\le18k$.

\begin{lemma}[Covariance comparison]\label{lem:bg-quarter-covariance-comparison}
Let $D_g\coloneqq\operatorname{diag}(g_0,\ldots,g_m)$ and define $\bar B$ from the
dominating weights. Then
\[
 D_g^{-1}BD_g^{-1}\succeq D_{\bar g}^{-1}\bar B D_{\bar g}^{-1}.
\]
\end{lemma}
\begin{proof}
Put $v_j\coloneqq g_j^{-2}$, $(s_j)_k\coloneqq\mathbf1_{k\ge j}$, and
$(t_j)_k\coloneqq(-3)^{k-j}\mathbf1_{k\ge j}$. The correlation innovation
factorizations give exactly
\[
 D_g^{-1}BD_g^{-1}=\frac12(s_0s_0^{\mathsf T}+t_0t_0^{\mathsf T})
 +\frac12\sum_{j=1}^m\bigl[(v_j-v_{j-1})s_js_j^{\mathsf T}
                          +(v_j-9v_{j-1})t_jt_j^{\mathsf T}\bigr].
\]
For $c=1,9$, all the coefficients are nonnegative, and
\[
 v_j-cv_{j-1}=
 \frac{1/\eta_j-c}{g_{j-1}^2}
 \ge\frac{1/\bar\eta_j-c}{\bar g_{j-1}^2}.
\]
The result follows term by term.
\end{proof}

Lemma~\ref{lem:bg-ratio-hull} says
that $D_g^{-1}u(x)$ is a convex combination of the fixed profiles
$k\mapsto3^{\min(k,t)}$, $0\le t\le m$, for $0\le x\le(H+1)/2$.
This applies to both $i$ and $i-1$.
For the origin, let $R_j\coloneqq p_j(-1)/w_j$.
The generating function gives $p_j(-1)=V_j(N)$, as for $p_m(-1)=V$ above,
and $\binom Nj3^j=w_j+3w_{j-1}$; hence
\[
 R_0=1,\qquad R_j=1+\rho_j^2(3+R_{j-1}).
\]
Since $\rho_j^2$ increases with $j$, subtracting consecutive recurrences
shows $R_j\ge R_{j-1}$. Also $\rho_j^2\le2/27$, and $33/25$ is the
fixed point of $R\mapsto1+(2/27)(3+R)$; hence $R_j\le33/25$ by induction.
Thus $\omega_k\coloneqq u_{m-k}(-1)/g_k$ satisfies
\[
 \omega_k=\frac{R_{m-k}}{R_m},\qquad
 1=\omega_0\ge \omega_1\ge\cdots\ge \omega_m\ge\frac{25}{33}.
\]
Every such sequence is a convex combination of
$k\mapsto25/33+(8/33)\mathbf1_{k\le h}$, $0\le h\le m$:
use the successive differences divided by $8/33$, assigning the
remaining mass to $h=m$.

By Lemma~\ref{lem:bg-quarter-covariance-comparison} and these three independent convex
representations, it suffices to prove the following fixed-weight inequality
for all $0\le s,t,h\le m$:
\begin{equation}\label{eq:bg-quarter-universal-model}
 \bar B+\operatorname{sym}(\bar a^{(s)}\bar a^{(t)\mathsf T})
 -\frac54\operatorname{sym}(\bar g b^{(h)\mathsf T})\succ0,
\end{equation}
where
\[
 \bar a^{(s)}_k\coloneqq3^{\min(k,s)}\bar g_k,
 \qquad b^{(h)}_k\coloneqq\bar g_k\left(\frac{25}{33}
                                  +\frac8{33}\mathbf1_{k\le h}\right).
\]
The reduction is bilinear in each pair of evaluation vectors; a convex
combination of the positive matrices in~\eqref{eq:bg-quarter-universal-model} remains positive.

\paragraph{A three-dimensional covariance lower bound.}
For a test vector $x$, set $c_k\coloneqq\bar g_kx_k$,
$Y_j\coloneqq\sum_{k=j}^m c_k$ and $Y_{m+1}\coloneqq0$. The positive correlation energy is
\[
 Y_0^2+\frac{25}{2}Y_1^2+\sum_{j=2}^m189\,15^{j-2}Y_j^2.
\]
Normalize it by $z_0\coloneqq Y_0$, $z_1\coloneqq(5/\sqrt2)Y_1$, and
$z_j\coloneqq\sqrt{189}\,15^{(j-2)/2}Y_j$ for $j\ge2$, and write
$e_0,e_1,\ldots$ for the coordinate basis of $z$.
The profile functionals become
\begin{equation}\label{eq:bg-quarter-sum-profiles}
 a_0=e_0,\qquad
 a_s=e_0+\frac{2\sqrt2}{5}e_1+
       \frac2{\sqrt{21}}\sum_{j=2}^s r^{j-2}e_j\quad(s\ge1),
 \qquad r\coloneqq\sqrt{\frac35}.
\end{equation}
The mixed origin is $b_h=e_0-\theta_he_{h+1}$ for $h<m$, while
$b_m=e_0$, where
\begin{equation}\label{eq:bg-quarter-origin-prefix}
 \theta_0\coloneqq\frac{8\sqrt2}{165},\qquad
 \theta_h\coloneqq\frac8{99\sqrt{21}\,15^{(h-1)/2}}\quad(h\ge1).
\end{equation}

\begin{lemma}[Exact tail compression]\label{lem:bg-quarter-tail-compression}
In these normalized sum coordinates, $\bar B$ dominates the compression
of an operator $\mathcal L$ on $\ell^2(\mathbb N_0)$ that equals $I/2$ off the span
of $e_0,e_1,\widehat\varphi$, where
\[
 \varphi\coloneqq\sum_{j\ge2}(-r)^{j-2}e_j,\qquad\widehat\varphi\coloneqq\sqrt{\frac25}\varphi.
\]
On this three-dimensional span its matrix is
\begin{equation}\label{eq:bg-quarter-fixed-block}
 \mathcal L=\begin{pmatrix}
 1&-2\sqrt2/5&\sqrt{210}/21\\
 -2\sqrt2/5&33/25&-11\sqrt{105}/105\\
 \sqrt{210}/21&-11\sqrt{105}/105&17/7
 \end{pmatrix},
\end{equation}
with inverse
\begin{equation}\label{eq:bg-quarter-fixed-inverse}
 \mathcal L^{-1}=\begin{pmatrix}
 539/390&47\sqrt2/156&-11\sqrt{210}/780\\
 47\sqrt2/156&205/156&7\sqrt{105}/156\\
 -11\sqrt{210}/780&7\sqrt{105}/156&35/52
 \end{pmatrix}.
\end{equation}
In particular $\mathcal L\succeq I/2$.
\end{lemma}
\begin{proof}
Put $\tilde Y_j\coloneqq\sum_{k=j}^m(-3)^{k-j}c_k$. Summation by parts gives
\[
 \tilde Y_j=Y_j-4\sum_{\ell=j+1}^m(-3)^{\ell-j-1}Y_\ell.
\]
Writing $v_j\coloneqq\bar g_j^{-2}$, the negative correlation energy is
\[
 \tilde Y_0^2+\frac92\tilde Y_1^2+
 \sum_{j=2}^m(v_j-9v_{j-1})\tilde Y_j^2.
\]
On the normalized tail $(z_2,\ldots,z_m)$, the last sum is exactly
$(3/7)\|\mathsf Kz\|^2$, where
\[
 \mathsf K\coloneqq(I-\alpha\mathsf U)(I+3\alpha\mathsf U)^{-1},\qquad \alpha\coloneqq1/\sqrt{15},
\]
and $\mathsf U$ is the finite upper shift. For the finite geometric vector
$\varphi=(1,-3\alpha,(-3\alpha)^2,\ldots)$, triangular multiplication gives
$\mathsf K^{-\mathsf T}\varphi=(1,\alpha,\alpha^2,\ldots)$. For example, the $j$th entry of
$(I-\alpha\mathsf U^{\mathsf T})^{-1}\varphi$ is $\alpha^j[1-(-3)^{j+1}]/4$; multiplying by
$I+3\alpha\mathsf U^{\mathsf T}$ leaves $\alpha^j$. Therefore
\[
 (\varphi^{\mathsf T}z)^2\le\left(\sum_{j\ge0}\alpha^{2j}\right)\|\mathsf Kz\|^2
 =\frac{15}{14}\|\mathsf Kz\|^2,
\]
so the negative tail energy is at least $(2/5)(\varphi^{\mathsf T}z)^2$.
This inequality holds in every finite dimension and uses all tail innovations.

The two retained initial innovations and this tail bound give
$\mathcal L=(I+\psi_0\psi_0^{\mathsf T}+\psi_1\psi_1^{\mathsf T}+\psi_2\psi_2^{\mathsf T})/2$, where the active coordinates are
\[
 \psi_0\coloneqq(1,-4\sqrt2/5,4\sqrt{5/42}),\quad
 \psi_1\coloneqq(0,3/5,-2\sqrt{5/21}),\quad \psi_2\coloneqq(0,0,1).
\]
This proves~\eqref{eq:bg-quarter-fixed-block} and $\mathcal L\succeq I/2$; multiplication verifies
\eqref{eq:bg-quarter-fixed-inverse}. Extending any finite $z$ by zero produces exactly
the same finite quadratic sums. Thus the finite lower bound is a
compression of this single infinite operator, with no interchange of limits.
\end{proof}

It remains to prove, for the profiles in~\eqref{eq:bg-quarter-sum-profiles},
\begin{equation}\label{eq:bg-quarter-reduced-target}
 \mathcal L+\operatorname{sym}(a_sa_t^{\mathsf T})-\frac54e_0e_0^{\mathsf T}
       +\frac54\theta_h\operatorname{sym}(e_0e_{h+1}^{\mathsf T})\succ0.
\end{equation}
The finite-boundary case means $\theta=0$ instead. Every profile $a_s$
has finite support in these coordinates.

\paragraph{The zero-profile cases.}

\begin{lemma}\label{lem:bg-quarter-zero-profile}
Inequality~\eqref{eq:bg-quarter-reduced-target} holds whenever $\min(s,t)=0$.
\end{lemma}
\begin{proof}
Assume $s=0$ and first set $\theta=0$.
For $t=0$, the operator is $\mathcal L-(1/4)e_0e_0^{\mathsf T}\succeq I/4$.
For $t\ge1$, put $k\coloneqq(3/5)^{t-1}$, $x\coloneqq(-3/5)^{t-1}$.
In the inverse inner product induced by $\mathcal L$, the matrix in
\eqref{eq:bg-quarter-fixed-inverse} and geometric sums give
\begin{align*}
 \chi_0&\coloneqq\langle e_0,e_0\rangle=\frac{539}{390},\\
 \tau_0&\coloneqq\langle e_0,a_t\rangle=\frac{2477}{1560}+\frac{11x}{312},\\
 A_t&\coloneqq\langle a_t,a_t\rangle=
 \frac{140507}{43680}+\frac{87x}{1456}
 -\frac{115x^2}{2912}-\frac{20k}{21}.
\end{align*}
Here the projection of $a_t$ onto $\widehat\varphi$ is
$\sqrt{210}(1-x)/84$; the residual squared tail norm is
$10(1-k)/21$ minus the square of that projection, and $\mathcal L^{-1}$ acts
there by $2$.
Since $|x|=k\le1$, $\tau_0\ge1211/780$ and $A_t\le140507/43680$.
For $c\coloneqq a_t-(5/4)e_0$, these give
\[
 F_c\coloneqq\langle e_0,c\rangle\ge-\frac7{40}>-\frac15,
 \quad C_c\coloneqq\langle c,c\rangle\le\frac{5441}{3640}<\frac32,
 \quad \chi_0<\frac75.
\]
The two possibly nonzero eigenvalues of $\operatorname{sym}(vw^{\mathsf T})$ are
$(v^{\mathsf T}w\pm\|v\|\|w\|)/2$. After congruence by $\mathcal L^{-1/2}$,
the least eigenvalue is therefore at least
\[
 1+\frac{F_c-\sqrt{\chi_0C_c}}2>
 1+\frac{-1/5-3/2}{2}=\frac3{20};
\]
indeed $\chi_0C_c<(7/5)(3/2)<9/4$.
Since $\mathcal L\succeq I/2$, the unwhitened operator exceeds $(3/40)I$.
The correction in~\eqref{eq:bg-quarter-reduced-target} has norm
$(5/8)\theta_h\le\sqrt2/33<3/40$, proving the result for every origin prefix.
\end{proof}

\paragraph{Positive profiles.}

\begin{lemma}\label{lem:bg-quarter-positive-profiles}
Let $\mathcal L$, $a_s$ and $\theta_h$ be as in~\eqref{eq:bg-quarter-fixed-block},
\eqref{eq:bg-quarter-sum-profiles} and~\eqref{eq:bg-quarter-origin-prefix}.
For every $s,t\ge1$, the operator
\[
 \mathcal P-\frac54e_0e_0^{\mathsf T}
       +\frac54\theta_h\operatorname{sym}(e_0e_{h+1}^{\mathsf T}),
 \qquad \mathcal P\coloneqq\mathcal L+\operatorname{sym}(a_sa_t^{\mathsf T}),
\]
is strictly positive.
The assertion also holds with the last term omitted.
\end{lemma}
\begin{proof}
By symmetry assume $s\le t$. The representation of $\mathcal L$ as $I/2$ plus
three positive rank-one terms gives $\mathcal L\succeq I/2$. If
$a_+\coloneqq(a_s+a_t)/2$ and $a_-\coloneqq(a_t-a_s)/2$, then
\begin{equation}\label{eq:bg-quarter-resolvent-floor}
 \|a_-\|^2\le\frac14\frac4{21}\sum_{j\ge0}(3/5)^j=\frac5{42},
 \qquad \mathcal P=\mathcal L+a_+a_+^{\mathsf T}-a_-a_-^{\mathsf T}\succeq\frac8{21}I.
\end{equation}
We shall prove
\begin{equation}\label{eq:bg-quarter-resolvent-bounds}
 \gamma_0\coloneqq e_0^{\mathsf T}\mathcal P^{-1}e_0\le\frac{251}{320},\qquad
 C_1\coloneqq e_0^{\mathsf T}\mathcal P^{-1}e_1\ge-\frac{\sqrt2}{8},\qquad
 C_2\coloneqq e_0^{\mathsf T}\mathcal P^{-1}e_2\ge-\frac{\sqrt{21}}9.
\end{equation}
Also, $\gamma_j\coloneqq e_j^{\mathsf T}\mathcal P^{-1}e_j\le21/8$ by~\eqref{eq:bg-quarter-resolvent-floor}.

The inverse of the displayed block of $\mathcal L$ is~\eqref{eq:bg-quarter-fixed-inverse},
while $\mathcal L^{-1}=2I$ on its orthogonal complement. Put
\[
 k\coloneqq(3/5)^{s-1},\quad l\coloneqq(3/5)^{t-1},\quad
 x\coloneqq(-3/5)^{s-1},\quad y\coloneqq(-3/5)^{t-1}.
\]
Thus $0\le l\le k\le1$, $|x|\le k$, $|y|\le l$, and $k\le3/5$
when $s\ge2$. The profiles have $\widehat\varphi$ coordinates
$\sqrt{210}(1-x)/84$ and $\sqrt{210}(1-y)/84$; their tail squared
norms are $10(1-k)/21$ and $10(1-l)/21$, and their tail inner product
is $10(1-k)/21$. Consequently~\eqref{eq:bg-quarter-fixed-inverse} gives the following inner products,
all taken with respect to $\mathcal L^{-1}$:
\begin{align*}
 A_s&\coloneqq\langle a_s,a_s\rangle
   =\frac{140507+2610x-1725x^2-41600k}{43680},\\
 A_t&\coloneqq\langle a_t,a_t\rangle
   =\frac{140507+2610y-1725y^2-41600l}{43680},\\
 A_{st}&\coloneqq\langle a_s,a_t\rangle
   =\frac{140507+1305(x+y)-1725xy-41600k}{43680},\\
 \sigma_0&\coloneqq\langle e_0,a_s\rangle=\frac{2477+55x}{1560},\qquad
 \tau_0\coloneqq\langle e_0,a_t\rangle=\frac{2477+55y}{1560},\qquad
 \chi_0\coloneqq\langle e_0,e_0\rangle=\frac{539}{390}.
\end{align*}
Set $\Omega\coloneqq(2+A_{st})^2-A_sA_t>0$. Positivity follows from~\eqref{eq:bg-quarter-resolvent-floor}, since the smaller
nonzero eigenvalue of the whitened profile perturbation is
$(A_{st}-\sqrt{A_sA_t})/2>-1$. The rank-two inverse formula gives
\begin{equation}\label{eq:bg-quarter-rank-two-diagonal}
 \Omega\gamma_0=\chi_0\Omega-2(2+A_{st})\sigma_0\tau_0+A_t\sigma_0^2+A_s\tau_0^2.
\end{equation}
For $\sigma_j\coloneqq\langle a_s,e_j\rangle$, $\tau_j\coloneqq\langle a_t,e_j\rangle$,
and $\chi_j\coloneqq\langle e_0,e_j\rangle$, the same formula gives
\begin{equation}\label{eq:bg-quarter-rank-two-mixed}
 \Omega C_j=\chi_j\Omega-(2+A_{st})(\sigma_0\tau_j+\tau_0\sigma_j)+A_t\sigma_0\sigma_j+A_s\tau_0\tau_j.
\end{equation}
Here
\[
 \chi_1=\frac{47\sqrt2}{156},\quad
 \sigma_1=\frac{\sqrt2(551-35x)}{624},\quad
 \tau_1=\frac{\sqrt2(551-35y)}{624},\quad
 \chi_2=-\frac{11\sqrt{21}}{390}.
\]
When $s\ge2$,
$\sigma_2=\sqrt{21}(345x+1819)/10920$; when $s=1$, replace $1819$ by
$-261$, because the $e_2$ coordinate of $a_1$ is zero. The analogous
rule gives $\tau_2$.

First,~\eqref{eq:bg-quarter-rank-two-diagonal} yields
\[
 \Omega(251/320-\gamma_0)=\mathcal N_\gamma/880588800,
\]
where
\begin{align*}
 \mathcal N_\gamma\coloneqq{}&3280284+1502953380k+502025820l\\
 &-477376000k(k-l)-41675000kxy-63946600x(k-l)\\
 &+20837500(ky^2+lx^2)+19771985(x^2+y^2)\\
 &+47973530xy+134287860(x+y).
\end{align*}
The inequalities $k(k-l)\le k$, $|kxy|\le k$, $|x(k-l)|\le k$,
$|xy|\le l$, and $x+y\ge-k-l$ imply
\[
 \mathcal N_\gamma\ge3280284+785667920k+319764430l>0.
\]
This proves the first assertion of~\eqref{eq:bg-quarter-resolvent-bounds}.

For $C_1$,~\eqref{eq:bg-quarter-rank-two-mixed} gives $\Omega(C_1/\sqrt2+1/8)=\mathcal N_1/3144960$, where
\begin{align*}
 \mathcal N_1\coloneqq{}&4974732+1216000k(k-l)+(89000k-152510)xy\\
 &+(524580-249800(k-l))x+524580y\\
 &-44500(ky^2+lx^2)-5014860k-92340l-17195(x^2+y^2).
\end{align*}
If $s\ge2$, set $c\coloneqq3/5$. The coefficient of $x$ is positive, so
that term is at least $-524580k$; the $xy$ term is at least
$-152510kl$. All variables have absolute value at most $c$, whence
\[
 \mathcal N_1\ge4974732-5014860c-92340c-1049160c
              -89000c^3-34390c^2-152510c^2=1194408>0.
\]
If $s=1<t$, direct substitution in~\eqref{eq:bg-quarter-rank-two-mixed} instead gives
\[
 \Omega(C_1/\sqrt2+1/8)
 =\frac{159273-122560l-6855y^2+51230y}{349440}>0,
\]
since its numerator is at least $262656/5$ for $l,|y|\le3/5$.
Finally, $s=t=1$ gives $C_1=-275\sqrt2/2562>-\sqrt2/8$.

For $s\ge2$, equation~\eqref{eq:bg-quarter-rank-two-mixed} similarly gives
$\Omega(C_2/\sqrt{21}+1/9)=\mathcal N_2/8255520$, where
\begin{align*}
 \mathcal N_2\coloneqq{}&2810220+620800k(k-l)+(69000k-179070)xy\\
 &+(401640(k-l)-843444)x-843444y\\
 &-34500(ky^2+lx^2)-2624620k+17260l+17085(x^2+y^2).
\end{align*}
The coefficient of $x$ is negative, so its term is at least
$-843444k$; the $xy$ term is at least $-179070kl$. Thus
\[
 \mathcal N_2\ge2810220-2624620c-1686888c-69000c^3-179070c^2=143946>0.
\]
For $s=1<t$, the alternative $\sigma_2$ gives
\[
 \Omega(C_2/\sqrt{21}+1/9)
 =\frac{22080l+1315y^2-14158y+45931}{131040}>0.
\]
For $s=t=1$, one has $C_2=-41\sqrt{21}/1281>-\sqrt{21}/9$.
This proves~\eqref{eq:bg-quarter-resolvent-bounds} in every case.

It remains to add the origin correction. On its at-most-two-dimensional
active subspace after congruence by $\mathcal P^{-1/2}$, the trace and determinant
are positive if
\begin{align*}
 1-\frac58(\gamma_0-\theta_hC_{h+1})&>0,\\
 \Delta\coloneqq1-\frac54\gamma_0+\frac54\theta_hC_{h+1}
       -\frac{25}{64}\theta_h^2(\gamma_0\gamma_{h+1}-C_{h+1}^2)&>0.
\end{align*}
Here $C_j\coloneqq e_0^{\mathsf T}\mathcal P^{-1}e_j$ for every $j\ge1$.
By Cauchy--Schwarz, $|C_j|\le\sqrt{\gamma_0\gamma_j}<3/2$. Since $\gamma_0<4/5$ and
$\theta_h\le\theta_0<1/14$, the trace condition follows from
$\gamma_0-\theta_hC_{h+1}<4/5+3/28<1$. For the determinant,~\eqref{eq:bg-quarter-resolvent-bounds} gives
\[
 \Delta\ge\frac{139}{232320}>0\quad(h=0),\qquad
 \Delta\ge\frac{40435}{5018112}>0\quad(h=1),
\]
after discarding the nonnegative term $25\theta_h^2C_{h+1}^2/64$.
For $h\ge2$, use $\theta_h<1/200$, $|C_j|<3/2$, and $\gamma_0\gamma_j<12/5$:
\[
 \Delta>\frac5{256}-\frac3{320}-\frac3{128000}
       =\frac{1297}{128000}>0.
\]
If the origin correction is absent, $\Delta\ge5/256>0$ directly.
This proves the result.
\end{proof}

\begin{proof}[Proof of Lemma~\ref{lem:bg-quarter-uniform} and of the quarter assertion of Proposition~\ref{prop:bg-endpoints}]
Lemmas~\ref{lem:bg-quarter-positive-profiles} and~\ref{lem:bg-quarter-zero-profile} establish
\eqref{eq:bg-quarter-reduced-target} for every finite pair of profiles and every origin
prefix, including the boundary case $b_m=e_0$.
Lemma~\ref{lem:bg-quarter-tail-compression} consequently proves~\eqref{eq:bg-quarter-universal-model}.
The covariance comparison and three convex representations above
then prove Lemma~\ref{lem:bg-quarter-uniform}, that is,
\eqref{eq:bg-quarter-target}, throughout its stated domain.
\end{proof}

\subsection{The low-index linear endpoint inequality}

\begin{lemma}\label{lem:bg-linear-near}
For $m\ge2$, $N\ge6m-1$ and $1\le i$ with $12i\le H+1$,
inequality~\eqref{eq:bg-linear-target} holds, without a restriction
$i\le99$.
\end{lemma}
\begin{proof}
Let $a>(H+1)/2$ be the least zero of $p_m$, as in
Lemma~\ref{lem:bg-ratio-hull}, and put $R_*\coloneqq6/5$.
The zeros of each $p_j$, hence all poles of $T_j$, are at least
$a$ by Jacobi interlacing. The positive resolvent representation implies
\[
 0<T_j(x)\le R_*T_j(0)=R_*\rho_j^2\quad(-1\le x\le i).
\]
Indeed for $x\ge0$ each pole term is enlarged by at most
$a/(a-i)<6/5$, since $i/a<2i/(H+1)\le1/6$; for negative $x$ use
monotonicity. Since $i<a$, all relevant polynomial values and
denominators are positive.
Logarithmic differentiation of~\eqref{eq:bg-degree-ratio} now gives
\[
 L_j\coloneqq T_j'/T_j\le
 \frac{R_*[4+(j-1)R_*L_{j-1}]}{3(N-j)}.
\]
The initial preceding term is absent. Replacing $j-1$ by $m$ and
$N-j$ by $N-m$ enlarges the right side. Its positive fixed point
solves $L=R_*[4+mR_*L]/[3(N-m)]$, so induction gives
\begin{equation}\label{eq:bg-near-derivative}
 0<L_j\le L_*\coloneqq\frac{4R_*}{3(N-m)-mR_*^2}.
\end{equation}
The denominator is positive since $(N-m)/m\ge9/2>R_*^2/3$.
Set $\eta\coloneqq m/[3(N-m)]$ and $\xi\coloneqq R_*^2\eta<1$. In reverse degree
$k=m-j$, the product formula gives
\[
 0<u_j(x)\le\eta^{k/2}R_*^k,\qquad
 0\le u_j'(x)\le kL_*\eta^{k/2}R_*^k.
\]
Consequently, throughout $[-1,i]$,
\[
 \|u(x)\|\le(1-\xi)^{-1/2},\qquad
 \|u'(x)\|\le L_*\sqrt{\frac{\xi(1+\xi)}{(1-\xi)^3}}.
\]
Let $A(x)\coloneqq\operatorname{sym}(u(x)u(x-1)^{\mathsf T})$.
Both terms of its derivative are symmetrized products of nonnegative
vectors. For vectors $v,w$ with $\langle v,w\rangle\ge0$, for example
entrywise nonnegative ones, the possibly negative eigenvalue of
$\operatorname{sym}(vw^{\mathsf T})$ is
$(\langle v,w\rangle-\|v\|\|w\|)/2\ge-\|v\|\|w\|/2$;
applying this to both terms and adding gives
\[
 A'(x)\succeq-L_*S_*I,\qquad
 S_*\coloneqq\frac{\sqrt{\xi(1+\xi)}}{(1-\xi)^2}.
\]
Integrating from zero to $i$ gives $A(i)-A(0)\succeq-iL_*S_*I$.
Since $B>\frac12I$, that is, $B^{-1}<2I$,
\[
 B^{-1/2}[A(i)-A(0)]B^{-1/2}\succeq-iL_*S_*B^{-1}\succ-2iL_*S_*I.
\]
By the exact endpoint normalization, the target matrix is
$A(i)-A(0)+[4i/(3(H+1))]B$, so it suffices to prove
\begin{equation}\label{eq:bg-near-half-scalar}
 2L_*S_*<\frac4{3(H+1)}.
\end{equation}

Put $y\coloneqq(N-m)/m$, $\delta_*\coloneqq1/m$, and
$c_*\coloneqq R_*^2/3=12/25$, so $\xi=c_*/y$.
Then
\[
 H+1=m(y-1+\delta_*),\qquad
 L_*=\frac{4R_*}{3m(y-c_*)},\qquad
 S_*=\frac{y\sqrt{c_*(y+c_*)}}{(y-c_*)^2}.
\]
Substituting these into~\eqref{eq:bg-near-half-scalar} and
cancelling positive factors gives
\[
 2R_*y(y-1+\delta_*)\sqrt{c_*(y+c_*)}<(y-c_*)^3.
\]
Both sides are positive. Squaring reduces it to the polynomial
inequality
\begin{equation}\label{eq:bg-near-polynomial}
 P_{\delta_*}(y)\coloneqq(y-c_*)^6
       -K_*y^2(y+c_*)(y-1+\delta_*)^2>0,\qquad
 K_*\coloneqq4R_*^2c_*=\frac{1728}{625}.
\end{equation}
If $m\ge3$, then $\delta_*\le1/3$ and $y\ge5-1/m\ge14/3$;
the negative term increases with $\delta_*$, so it suffices to use
$\delta_*=1/3$. For $m=2$, nonemptiness of the integer range
$1\le i\le(H+1)/12$ forces $N\ge15$; thus
$y\ge13/2>6$ and $\delta_*=1/2$.
For fixed $0<\delta<1$ put
\[
 J_\delta(y)\coloneqq\frac{(y-c_*)^6}{y^2(y+c_*)(y-1+\delta)^2},
 \qquad\text{so that}\qquad
 P_\delta(y)=y^2(y+c_*)(y-1+\delta)^2\{J_\delta(y)-K_*\}.
\]
For $y>\max(c_*,3(1-\delta))$ all factors are positive and
\[
 \frac{J_\delta'(y)}{J_\delta(y)}
 =\frac6{y-c_*}-\frac2y-\frac1{y+c_*}-\frac2{y-1+\delta}
 >\frac3y-\frac2{y-1+\delta}
 =\frac{y-3(1-\delta)}{y(y-1+\delta)}>0.
\]
Both cases lie in this range ($14/3>2$ and $6>3/2$), so $J_\delta$
increases and it suffices to prove $P_\delta>0$ at the two left
endpoints. The exact values are
\[
 P_{1/3}(14/3)=\frac{76017013212736}{177978515625}>0,\qquad
 P_{1/2}(6)=\frac{2143476437184}{244140625}>0.
\]
This proves both cases and the lemma.
\end{proof}

We derive the scalar finite test for the complementary range here;
Appendix~\ref{app:bg-finite} applies it with outward-rounded intervals.
The following form avoids subtracting nearly equal quantities.
Put
$e\coloneqq\tfrac12B^{-1/2}(u(0)+u(-1))$, $o\coloneqq\tfrac12B^{-1/2}(u(0)-u(-1))$,
$f\coloneqq\tfrac12B^{-1/2}(u(i)+u(i-1))$ and
$d\coloneqq\tfrac12B^{-1/2}(u(i)-u(i-1))$. Since
$\operatorname{sym}(ab^{\mathsf T})=\frac{a+b}2\frac{(a+b)^{\mathsf T}}2
-\frac{a-b}2\frac{(a-b)^{\mathsf T}}2$,
Lemma~\ref{lem:bg-endpoint-normalization} represents $\mathcal E_i/(VU_i)$
and $\mathcal E_0/V$ by $ff^{\mathsf T}-dd^{\mathsf T}$ and
$ee^{\mathsf T}-oo^{\mathsf T}$. Let
$E\coloneqq\|e\|^2$, $F\coloneqq\|f\|^2$, $C\coloneqq\langle e,f\rangle$,
$G_d\coloneqq\|d\|^2$. Put $s_i\coloneqq4i/[3(H+1)]>0$.
The target matrix is
$s_iI+ff^{\mathsf T}-dd^{\mathsf T}-ee^{\mathsf T}+oo^{\mathsf T}$.
Drop the last positive term and conjugate by
$(s_iI+ff^{\mathsf T})^{-1/2}$. The subtracted positive semidefinite
matrix has trace at most
$[E-C^2/(s_i+F)+G_d]/s_i$. Therefore it suffices that
\begin{equation}\label{eq:bg-linear-gap}
 s_i-E+\frac{C^2}{s_i+F}-G_d
   =\frac{s_i^2+2s_iJ+(s_i-E)W+J^2}{s_i+E+2J+W}-G_d>0,
\end{equation}
where $J\coloneqq\langle e,f-e\rangle$ and $W\coloneqq\|f-e\|^2$.
The equality follows from $F=E+2J+W$ and $C=E+J$.
The denominator is $s_i+F>0$.

These scalars can be evaluated by~\eqref{eq:bg-inverse-product}
without forming a difference of nearly equal endpoint products.
Define
\[
 \epsilon_0(x)\coloneqq0,\qquad
 \epsilon_j(x)\coloneqq
 \frac{4x+(j-1)\epsilon_{j-1}(x)}
 {3(N-j)-4x-(j-1)\epsilon_{j-1}(x)}.
\]
This is exactly $T_j(x)/\rho_j^2-1$ by~\eqref{eq:bg-degree-ratio}.
Let $h_k^0\coloneqq3^{-k}$ and
$h_k^\Delta(x)\coloneqq[u_{m-k}(x)/g_k-1]/3^k$.
Then $h_0^\Delta=0$ and, with $j=m-k+1$,
\[
 h_k^\Delta(x)
   =\frac{(1+\epsilon_j(x))h_{k-1}^\Delta(x)
                 +\epsilon_j(x)h_{k-1}^0}{3}.
\]
Apply the linear innovation recurrence in
\eqref{eq:bg-rational-innovation} to $h^0$ and $h^\Delta$ to obtain
$z^0$ and $z^\Delta$, with initial values $1$ and $0$, respectively.
The innovations of $e$, $f-e$ and $d$ are exactly
\[
 z^e=z^0+z^\Delta(-1)/2,\quad
 z^{f-e}=[z^\Delta(i)+z^\Delta(i-1)-z^\Delta(-1)]/2,\quad
 z^d=[z^\Delta(i)-z^\Delta(i-1)]/2.
\]
Inserting their products into~\eqref{eq:bg-inverse-product} gives
$E,J,W,G_d$ with rational arithmetic throughout.

\begin{proof}[Completion of Proposition~\ref{prop:bg-endpoints}]
The quarter assertion is Lemma~\ref{lem:bg-quarter-uniform}.
For the linear assertion with
$i\le99$, Lemma~\ref{lem:bg-linear-near} includes the boundary
$12i=H+1$. The complementary range is
\[
 N\ge6m-1,\quad1\le i\le99,\quad2i\le H+1<12i.
\]
Since $H+1\ge4m$ and $12i\le1188$, it forces $m\le296$.
For a fixed $m$, $N-2m+1=H+1<12i\le1188$ also forces
$6m-1\le N\le2m+1186$ by integrality.
Lemma~\ref{lem:bg-finite-linear} verifies~\eqref{eq:bg-linear-gap}
throughout a domain containing this finite wedge. The near-origin and finite
ranges exhaust every index in~\eqref{eq:bg-linear-target}.
\end{proof}

\section{Finite certificates for the binary grid minimum}
\label{app:bg-finite}

This appendix gives the finite parts of the binary grid argument.
Each certificate verifies an explicit sufficient inequality from
Appendices~\ref{app:bg-minima} and~\ref{app:bg-endpoints}.
The finite parameter sets and acceptance predicates are specified below;
in particular, a failure of a sufficient scalar test is never counted
as a counterexample to the desired matrix inequality.
The certificate package for the binary grid minimum contains the exact
arrays and the rational interval rectangles.
It is included in version~2.0.0 of the supporting
package~\cite{ComputationalCertificatesV2}.
All sign decisions use integers or outward rational intervals.
The saved rectangles and their bounds can be verified without repeating
the search that found them.

The first three lemmas complete the coefficient arguments of
Appendix~\ref{app:bg-minima}: Lemma~\ref{lem:bg-second-finite}
treats the second index left by $m\ge20$ and $N\ge15m$;
Lemma~\ref{lem:bg-low-coeff-finite} strengthens the polynomial
certificates at indices $2$ through $54$; and
Lemma~\ref{lem:bg-outside-finite} covers indices $55$ through $61$
outside the recurrence strip. The last two comparisons serve
Proposition~\ref{prop:bg-coefficients}.
Lemma~\ref{lem:bg-finite-linear} completes the analytic range of
Appendix~\ref{app:bg-endpoints} for the linear assertion of
Proposition~\ref{prop:bg-endpoints}. Finally,
Lemma~\ref{lem:bg-small-finite} supplies the matched inequalities
for Proposition~\ref{prop:bg-small-degree} when $D\le15$ and $N<13D$.
The complete domains, counts and arithmetic are specified below,
together with the independent verifications where supplied.
The polynomial expansions and finite
matrix tests are exact; the scalar and endpoint tests certify signs
by directed rounding.

\subsection{The finite second-index remainder}

Here $r$ is the size of the maximal terminal block and $m$ is the
number of residual adjacent pairs of the minimizing vertex in
Section~\ref{sec:bg-minimum}. Thus $m\ge2$ already implies $D\ge4$.
The two analytic regions just mentioned leave $2\le m\le19$ and
$N<15m$; the length bound $N\ge6m+2r-1+2(r\bmod2)$ then gives
$r\le\lfloor9m/2\rfloor$. The parity term comes from the code
length bound, so this finite domain is the one needed for the code
argument.

\begin{lemma}\label{lem:bg-second-finite}
Let $2\le m\le 19$, $0\le r\le\lfloor 9 m/2\rfloor$, and
\[
 6 m+2 r-1+2(r\bmod 2)\le N<15 m.
\]
For a normalized minimizing vertex of length $n\coloneqq N+r$ and degree
$D\coloneqq2 m+r$, the second coefficient satisfies $c_2\le c_0F(2)$.
\end{lemma}
\begin{proof}
\smallskip\noindent\emph{Step 1 (Scalar reduction).}
Use Lemma~\ref{lem:bg-root} and the scalar reduction
\eqref{eq:bg-second-upper}--\eqref{eq:bg-second-scalar}.
The quantities in~\eqref{eq:bg-second-finite-bound} are recalled
here with the same notation. In particular, $A=A(N,m)$ of
Appendix~\ref{app:bg-minima}. Put
\[
 A\coloneqq\frac{\bigl(\sqrt{3(N-m)}-\sqrt{m-1}\bigr)^2}{4},\quad
 a\coloneqq\lceil A\rceil,\quad
 \rho_*\coloneqq\frac{(a-2)(a+2 m-1)}{(a-1)(a+2 m-2)}
             \frac{n(N-1)}{N(n-1)}.
\]
The packed-root estimate~\eqref{eq:bg-second-packed} and the
objective estimates~\eqref{eq:bg-second-budget}
and~\eqref{eq:bg-second-log-product} give
\[
 F(2)\ge\rho_* F(1)^2,\qquad
 F(1)\ge\frac Nn\max\left\{e^{-E},
                              \frac{a-1}{a+2 m-1}\right\},
\]
where
\begin{align*}
 B&\coloneqq\log(3/2)+2 m\log(4/3)
                 +2\sum_{j=0}^{m-1}\log\frac N{N-j},\\
 E&\coloneqq2 m\log\frac N{N-1}
       +\frac{B}{2}\left(\frac 1{N-1}+\frac 1{A-1}\right).
\end{align*}
Indeed, \eqref{eq:bg-second-upper} gives
$c_2/c_0\le[\frac{4n}{3}(1-F(1))-D]/(n-1)$.
Combining this with $F(2)\ge\rho_* F(1)^2$, it suffices to prove,
at any certified lower bound $0\le u\le F(1)$ (the local $u$
here is a lower bound, not the exact value $F(1)$),
\[
 (1-1/n)\rho_* u^2+\frac 43 u+\frac Dn-\frac 43>0.
\]
The left side increases for $u\ge 0$.

\smallskip\noindent\emph{Step 2 (Rational evaluation).}
Here is a fully rational way to evaluate this test. Use denominator
$S\coloneqq10^{30}$ and directed rounding. Here $S$ is only the arithmetic
scale. Integer square roots give a lower bound for $A$;
with $T_*\coloneqq3N-2 m-1$ and $R_*\coloneqq3(N-m)(m-1)$, the exact predicate
\[
 A>k\quad\Longleftrightarrow\quad
 T_*-4 k>0\ \text{ and }\ (T_*-4 k)^2>4R_*
\]
determines $a$ exactly. Round the logarithms and $B$ upwards;
the lower bound for $A$ then gives an upper bound for $E$.
For a rational $y\ge1$ and $z\coloneqq(y-1)/(y+1)$, use the first
24 terms of
\[
 \log y=2\sum_{k\ge 0}\frac{z^{2 k+1}}{2 k+1},\qquad
 \text{tail}\le\frac{2 z^{49}}{49(1-z^2)}.
\]
The positive exponential series through degree 40 and the tail bound
$t_{41}/(1-E/42)$, where $t_{41}\coloneqq E^{41}/41!$, give an upper bound for
$e^E$; the test checks
$E<42$ before using it. If the resulting upper bound for $e^E$
is $M/S$, form the integers
\[
 L\coloneqq\max\left\{\left\lfloor\frac{S^2}{M}\right\rfloor,
          \left\lfloor\frac{S(a-1)}{a+2m-1}\right\rfloor\right\},
 \qquad U\coloneqq\left\lfloor\frac{NL}{n}\right\rfloor.
\]
Then $L/S$ bounds the residual factor from below and $U/S\le F(1)$,
including both the maximum and the terminal factor $N/n$.
Now write $\rho_*=p/q$ with positive integers $p,q$; these are the
numerator and denominator of $\rho_*$, independent of the logarithm
arguments above. Multiplying the scalar test by $3nqS^2$ gives the
integer acceptance predicate
\[
 3(n-1)pU^2+4 nqUS+(3D-4 n)qS^2>0.
\]
Every denominator used above is checked positive.

\smallskip\noindent\emph{Step 3 (Counts and remaining cases).}
There are 50,863 parameter triples in the displayed domain.
The scalar predicate succeeds at 50,854 of them. The remaining nine
all have $r=0$ and are given in the table. The listed lower zeros of the pairs
specify candidate polynomials; the quadrature below certifies both
their minimality and their uniqueness, so no assumption about their
discovery is needed.
\begin{center}\small
\begin{tabular}{rrl}\toprule
$m$&$N$&Lower zeros\\\midrule
2&11&6,9\\
2&12&7,10\\
2&13&8,11\\
3&17&9,12,15\\
3&18&9,13,16\\
4&23&11,15,18,21\\
4&24&12,16,19,22\\
5&29&14,18,21,24,27\\
6&35&16,20,24,27,30,32\\\bottomrule
\end{tabular}
\end{center}
\smallskip\noindent\emph{Step 4 (Exact quadrature and replay).}
These nine cases use exact rational quadrature, not the failed scalar
test. For the nodes $z_0=0,z_1,\ldots,z_D$ given by zero and the
paired roots, define their cardinal polynomials and the weights
$\lambda_j$ of Lemma~\ref{lem:positive-quadrature}, indexed here by
the node $z_j$, by
\[
 \ell_j(x)\coloneqq\prod_{k\ne j}\frac{x-z_k}{z_j-z_k},\qquad
 \lambda_j\coloneqq4^{-N}\sum_{x=0}^{N}\binom Nx 3^x\ell_j(x).
\]
Every weight is strictly positive. For
$F(x)=\prod_{j=1}^{D}(1-x/z_j)$, one has
$\lambda_0=\E_{\mu_N}F$. Every node is a grid point, so for any
normalized feasible polynomial $f$ of degree at most $D$, exactness gives
\[
 \E_{\mu_N}f=\lambda_0+\sum_{j=1}^{D}\lambda_jf(z_j)\ge\lambda_0.
\]
Equality forces $f(z_j)=0$ for all $j\ge1$. These $D$ distinct
zeros and $f(0)=1$ determine $f=F$. Thus this is the unique
normalized minimizing polynomial, including the selected vertex of
Section~\ref{sec:bg-minimum}. Directly evaluate
\[
 c_2=\frac{4^{-N}}{9\binom N2}
       \sum_{x=0}^{N}\binom Nx 3^x F(x)P_2^{(N,4)}(x).
\]
In each row the rational difference $\lambda_0F(2)-c_2$ is strictly
positive. The smallest, at $(m,N)=(6,35)$, is
\[
 \frac{976843423961}{6719750886081429504000}>0.
\]
All nine cardinal-weight and coefficient calculations
are included in the certificate. An independent evaluation of the scalar
part with denominator $10^{24}$, 16 logarithm terms and exponential
degree 30 gives the same 50,854 successes and nine remaining cases.
Its tail bounds are $2z^{33}/[33(1-z^2)]$ and
$(E^{31}/31!)/(1-E/32)$, with $E<32$ checked before the latter is used.
Together these computations cover every triple in the lemma.
\end{proof}

\subsection{Strengthened polynomial row certificates}

The signed-row formulas and the scalar row formulas in
Appendix~\ref{app:binary-matrices} are polynomial after their displayed
positive denominators are cleared. Appendix~\ref{app:bg-minima}
multiplies each endpoint contribution by the factor $1-2 i/(3D)$.
Equivalently, if an old numerator is the difference of its endpoint
and coefficient contributions, the new numerator is $3D$ times that
old numerator minus $2 i$ times its endpoint contribution.
This gives the exact strengthened arrays in
\eqref{eq:bg-strengthened-arrays}.
As usual, $A\prec B$ means that $B-A$ is positive definite.
The second-index comparison is retained as an additional matrix
certificate; the grid-minimum proof itself uses
Lemma~\ref{lem:bg-low-indices} at that index.

To distinguish the matrix coordinates used below, put
$v_j\coloneqq\binom Nj3^j$ and $b_j\coloneqq P_j^{(N,4)}-P_{j-1}^{(N,4)}$,
where $P_{-1}\coloneqq0$. The matrices $\operatorname{Gram}_0$ and
$\operatorname{Mat}_i$ represent $\mathcal G$ and $\mathcal C_i$
in the basis $b_j/\sqrt{v_j}$ of~\eqref{eq:binary-congruence}.
If $D_v\coloneqq\operatorname{diag}(\sqrt{v_0},\ldots,\sqrt{v_m})$,
their matrices in the raw basis $b_j$ are respectively
$D_v\operatorname{Gram}_0D_v$ and
$D_v\operatorname{Mat}_iD_v$. They are rational, although the
normalized matrices can contain square roots.

\begin{lemma}\label{lem:bg-low-coeff-finite}
For $m\ge 2$, $r\ge 0$, $N\ge 6 m+2 r-1$ and $D\coloneqq2 m+r\ge 16$,
the strengthened row certificates prove
\[
 \operatorname{Mat}_i\prec
 \left(1-\frac{2 i}{3D}\right)U_i\operatorname{Gram}_0,
 \qquad 2\le i\le\min(54,D-1).
\]
At $i=2$ the same conclusion also holds without $D\ge 16$.
\end{lemma}
\begin{proof}
Use the exact row partition, signs, and nonnegative-orthant
substitutions of Appendix~\ref{app:binary-matrices}. For $3\le i\le 13$
there are 1694 signed-row arrays. For $14\le i\le 54$ there are
33,782 scalar arrays: 82 in the $m\ge28$ family and 33,700 in the
$2\le m\le27$ families (fixed-$m$ tails and fixed lifts) of
Appendix~\ref{app:binary-matrices}. All 35,476 arrays are regenerated from
the defining formulas, and there are eight additional arrays for
the second index. A polynomial with nonnegative coefficients and a
positive constant is positive on its entire nonnegative orthant;
this is why a finite expansion proves an unbounded-parameter claim.

The strengthened 35,476 arrays have 16,610,492 nonzero coefficients.
There are 47 arrays whose original orthant does not give a positive
coefficient certificate: four signed-row arrays, 42 scalar arrays,
and one second-index array. Their finite repairs are as follows.
Write $N=6 m+2 r-1+s$, $s\ge 0$.

All four failed signed-row arrays have $j=m$ and lower-neighbor
sign $\varepsilon_-=+1$ in~\eqref{eq:binary-four-signs}; the upper
neighbor is absent. For the Last row family $m=8+a,r=c$,
indices 11 and 12 have positive
coefficients after any of the shifts $a\ge 1$, $c\ge 1$ or $s\ge 1$.
Index 13 has positive coefficients after $a\ge 1$, $c\ge 2$ or
$s\ge 1$. For the Small $m$ family at $m=7,r=2+c$, index 13 is repaired
by $c\ge 1$ or $s\ge 1$. The unshifted remainder consists of five
matrix cases, in the order $(m,r,N,i)$:
\[
 (8,0,47,11),\ (8,0,47,12),\ (8,0,47,13),
 (8,1,49,13),\ (7,2,45,13).
\]
For the second-index last row, the negative lower
neighbor sign is repaired after any of the unit shifts in
$m-2,r,s$; its only remaining matrix is $(m,r,N,i)=(2,0,11,2)$.

Each failed scalar array has fixed $m,r$ and a single free length
slack $s$. The following list specifies all 42 failed arrays and
the necessary shift $s\ge s_0$. An entry $(r,i;s_0)$ refers to the
last row $j=m$ unless marked ``interior'', meaning the scalar row
$j=m-1$ in~\eqref{eq:binary-two-scalars}.
\begin{center}\small
\begin{tabular}{r p{118 mm}}\toprule
$m$&Failed scalar arrays and shifts\\\midrule
5&$(6,15;1)$\\
6&$(4,14;1)$, $(4,15;1)$, $(5,16;1)$\\
7&$(2,14;1)$, $(2,15;1)$, $(3,14;1)$, $(3,15;1)$, $(3,16;1)$, $(4,17;1)$\\
8&$(0,14;2)$, $(0,15;1)$ interior, $(0,15;2)$,
 $(1,14;1)$, $(1,15;1)$, $(1,16;2)$, $(2,15;1)$, $(2,16;1)$, $(2,17;1)$,
 $(3,17;1)$, $(3,18;1)$, $(4,19;1)$\\
9&$(0,14;1)$, $(0,15;1)$, $(0,16;1)$, $(0,17;2)$,
 $(1,16;1)$, $(1,17;1)$, $(1,18;1)$, $(2,18;1)$, $(2,19;1)$, $(3,20;1)$\\
10&$(0,17;1)$, $(0,18;1)$, $(0,19;1)$, $(1,19;1)$, $(1,20;1)$, $(2,21;1)$\\
11&$(0,20;1)$, $(0,21;1)$, $(1,22;1)$\\
12&$(0,23;1)$\\\bottomrule
\end{tabular}
\end{center}
Every shifted array has nonnegative coefficients and a positive
constant. The 38 scalar arrays with $s_0=1$ and four with $s_0=2$
give $38+2\cdot4=46$ cases at $0\le s<s_0$.
The case $(8,0,47,15)$ occurs in both scalar rows, leaving 45
distinct cases. Adding the five signed-row cases and the one
second-index case gives $45+5+1=51$ finite matrix cases.

At these points the test uses the exact rational value
\[
 U_i=\frac{P_m^{(N-1,4)}(i)P_m^{(N-1,4)}(i-1)}
           {P_m^{(N-1,4)}(0)P_m^{(N-1,4)}(-1)},
\]
whereas the polynomial row expansions use the lower bound
\eqref{eq:binary-rational-endpoint}. The finite matrix represents
$(1-2i/(3D))U_i\mathcal G-\mathcal C_i$ in the raw basis
$P_j^{(N,4)}$, so its entries are rational. The change from this
basis to $b_j$ is the integer triangular matrix whose $j$th column
has entries $1$ in position $j$ and $-1$ in position $j-1$.
Congruence by this matrix, followed by $D_v^{-1}$, gives
$(1-2i/(3D))U_i\operatorname{Gram}_0-\operatorname{Mat}_i$.
Positive rational elimination pivots therefore prove the required
positive definiteness. An independent calculation uses the basis
$\binom{x}{j}$ and direct binomial sums; its change of basis is also
invertible over the rationals, and all 51 matrices again have
positive pivots.

There are $11+3+42=56$ shifted expansions: 11 for the four signed
rows, three for the second index and 42 for the scalar rows.
All 56 expansions and 51 finite matrices are included in the
certificate. For each original orthant, the shifted orthants
together with its listed finite complement cover every integer
point. Comparing this cover with the row partition verifies that
every required row and sign is included. The signed-row criterion
then proves the assertion.
\end{proof}

\subsection{The outside-strip transition at indices 55 through 61}

The recurrence strip in Appendix~\ref{app:bg-minima} is
$i\le X$, where $X\coloneqq N(2N-3m+6)/[3(8N-9m)]$.
The following lemma treats its complement at the seven indices
below the analytic entropy bound's range $i\ge62$.

\begin{lemma}\label{lem:bg-outside-finite}
Let $m\ge 2$, $r\ge 0$, $D\coloneqq2 m+r$, $N\ge 6 m+2 r-1$ and
\[
 55\le i\le\min(61,D-1),\qquad
 i>\frac{N(2N-3 m+6)}{3(8N-9 m)}.
\]
Then $\operatorname{Mat}_i\prec
(1-2 i/(3D))U_i\operatorname{Gram}_0$.
\end{lemma}
\begin{proof}
\smallskip\noindent\emph{Step 1 (Boundary certificates).}
The inequality $X>N/13$ gives $N<13i$ outside the strip, exactly
the finite projection~\eqref{eq:bg-outside-projection}.
For fixed $m,i$ put
\[
 \begin{gathered}
 r_{\min}\coloneqq\max(0,i+1-2 m),\qquad
 r_{\max}\coloneqq\left\lfloor\frac{N-6 m+1}{2}\right\rfloor,\\
 N_{\min}\coloneqq6 m+2 r_{\min}-1,\qquad D_{\min}\coloneqq2m+r_{\min}.
 \end{gathered}
\]
The smallest degree is $D_{\min}$, as in that projection, and
$1-2i/(3D_{\min})\le1-2i/(3D)$ for every admissible $r$.
There are 870 pairs $(m,i)$ with $55\le i\le 61$ and
$N_{\min}<13 i$. For each pair the adaptive row bound is certified
at a length $N_*$ by a rational upper bound strictly smaller than
$1-2i/(3D_{\min})$ for
\[
 \frac{\kappa_N}{s_N}\frac{392}{27}\sqrt{i}\,e^{-13 i/100},
\]
which bounds the ratio of a weighted coefficient row to $U_i$
times its corresponding signed Gram row. Indeed,
Lemma~\ref{lem:binary-adaptive-row} bounds these rows above by
$\kappa_N\mathcal B_i$ and below by $s_N$, respectively, and
\eqref{eq:bg-adaptive-entropy} bounds $\mathcal B_i/U_i$.
Thus the displayed strict comparison gives every required signed
margin a positive lower bound. The quantities are
\[
 \begin{gathered}
 \tau\coloneqq\sqrt{\frac{m}{3(N-m)}},\qquad
 \theta_m^2\coloneqq\frac{m}{3(N-m+1)},\\
 \kappa_N\coloneqq1+\tau+\theta_m^2/\tau,\qquad
 s_N\coloneqq1-\frac{m}{N-m+1}-\tau-\theta_m^2/\tau.
 \end{gathered}
\]
Here $\tau$ and $\theta_m^2$ are the same quantities as in the
adaptive estimates of Appendices~\ref{app:binary-high}
and~\ref{app:bg-minima}. The positive quantity $s_N$ increases
with $N$, and $\kappa_N$ decreases, as proved there. Thus one
accepted boundary covers every larger length still satisfying
$i>X$, where~\eqref{eq:bg-adaptive-entropy} applies; lengths with
$i\le X$ are already covered by the strip argument. Square roots are
enclosed by rationals of denominator $10^{12}$ using integer
square roots. The positive Taylor sum through degree 80 is a lower
bound for $e^{13 i/100}$, providing the required upper bound for
the whole expression. Each denominator and each strict inequality
is checked rationally at the saved boundary; the choice of boundary
by search supplies no mathematical premise. All 870 boundary values
$N_*$ and their rational bounds are listed in the certificate.

\smallskip\noindent\emph{Step 2 (Remaining triples).}
Among the smaller lengths, imposing the outside-strip condition leaves
190 triples $(m,N,i)$:
\begin{center}\small
\begin{tabular}{rrrr}\toprule
$i$&Triples&$m$ range&$N$ range\\\midrule
55&147&23--33&157--198\\
56&40&26--31&165--185\\
57&3&29&173--175\\
58--61&0&---&---\\\bottomrule
\end{tabular}
\end{center}
\smallskip\noindent\emph{Step 3 (The row test).}
These are checked in the unnormalized basis
$P_j^{(N,4)}-P_{j-1}^{(N,4)}$, using the raw kernel matrices
\eqref{eq:binary-C-kernel}. Denote them by $K_\ell$, and put
\[
 G_*\coloneqq K_0,\qquad
 A_i(r)\coloneqq\frac 1{\binom Ni}\sum_\ell
                 \binom N\ell\binom r{i-\ell}K_\ell.
\]
The congruences recalled above are
$G_*=D_v\operatorname{Gram}_0D_v$ and
$A_i(r)=D_v\operatorname{Mat}_iD_v$.
Every entry with $0\le j,k\le m$ and $0\le\ell\le\min(i,2m)$
is nonnegative by Step~6 of Appendix~\ref{app:binary-matrices}:
$N-\ell\ge N-D\ge4m+r-1\ge2m-1$ is its sufficient condition.
The integer generating function also checks these signs directly
in the finite calculation. Therefore
$0\le A_i(r)\le A_i(r_{\max})$ entrywise for
$r_{\min}\le r\le r_{\max}$.
The local weights $w_j\coloneqq4^{-j}$ correspond under this congruence to
the normalized weights $\sqrt{v_j}4^{-j}$ of
Appendix~\ref{app:binary-matrices}: each raw weighted row is $v_j$
times the normalized row. With these weights the certificate verifies, for
every row,
\[
 \gamma_j\coloneqq(G_*)_{jj}-\sum_{k\ne j}|(G_*)_{jk}|\frac{w_k}{w_j}>0,
 \qquad
 \frac{\sum_k A_i(r_{\max})_{jk}w_k/w_j}{U_i\gamma_j}
       <1-2i/(3D_{\min}).
\]
Here $U_i$ is again the exact rational ratio printed in the preceding
proof. All these are exact rational comparisons. They imply strict weighted
diagonal dominance of
$(1-2 i/(3D))U_iG_*-A_i(r)$ for every admissible $r$.
No positive-semidefinite monotonicity in $r$ is needed.
The 870 boundaries and 190 row certificates exhaust the finite
projection, completing the proof.
\end{proof}

\subsection{Rational interval certificates for the endpoint forms}

We use the normalized evaluation vectors and inverse-Gram products of
Appendix~\ref{app:bg-endpoints}. Its basis is
$p_j/\sqrt{w_j}$, where $p_j=P_j^{(N-1,4)}$ and
$w_j=\binom{N-1}{j}3^j$. The identity
$b_j=p_j+2p_{j-1}-3p_{j-2}$, with negative subscripts interpreted
as zero, gives an invertible triangular change from the raw
difference basis above. Thus both coordinate systems represent
the same forms. We recall the sufficient gap and
its derivation so that the finite tests can be read together with
the quantities they enclose. After whitening by the positive
adjacent-product Gram matrix, put
\[
 e\coloneqq\frac{u(0)+u(-1)}2,\quad o\coloneqq\frac{u(0)-u(-1)}2,\quad
 f\coloneqq\frac{u(i)+u(i-1)}2,\quad d\coloneqq\frac{u(i)-u(i-1)}2,
\]
and write $E\coloneqq\|e\|^2$, $F\coloneqq\|f\|^2$, $C\coloneqq\langle e,f\rangle$,
$G_d\coloneqq\|d\|^2$.

For the linear endpoint put $s_i\coloneqq4 i/[3(H+1)]$, with $H=N-2 m$.
The matrix is
$s_iI+ff^{\mathsf T}-dd^{\mathsf T}-ee^{\mathsf T}+oo^{\mathsf T}$.
Dropping its last positive-semidefinite term and applying the
rank-two trace bound after congruence by
$(s_iI+ff^{\mathsf T})^{-1/2}$ gives the sufficient gap
\[
 s_i-E+\frac{C^2}{s_i+F}-G_d>0.
\]
For stable interval evaluation, write
$J\coloneqq\langle e,f-e\rangle$, $W\coloneqq\|f-e\|^2$.
The identical gap, without cancellation of two large terms, is
\[
 \frac{s_i^2+2 s_iJ+(s_i-E)W+J^2}{s_i+E+2J+W}-G_d>0.
\]
Appendix~\ref{app:bg-endpoints} derives this identity and the
square-root-free evaluation formulas
\eqref{eq:bg-inverse-product} and~\eqref{eq:bg-rational-innovation}.
The tridiagonal pivots $d_k$ in~\eqref{eq:bg-pivots} depend only
on $(m,N)$; the ratio denominators $\Delta_j(x)$ in
\eqref{eq:bg-degree-ratio} also depend on $x$. Both must have
positive lower bounds at every accepted rectangle, as must the
denominator of the sufficient gap.
All coordinates through degree $m$ are retained.

Here are the interval conventions used for the certificate.
Represent an interval by $[a,b]/2^{b_*}$, where $a,b$ are integers
and the local precision exponent $b_*$ is fixed for a run.
Addition and
negation are exact. Products use the minimum and maximum of the
four endpoint products, rounded down and up to this denominator.
Division uses the four endpoint quotients with outward rounding
and rejects any denominator interval containing zero. A square
has lower bound zero when its argument interval crosses zero;
this rule applies to the innovation squares in the norms and to
$C^2$, $s_i^2$ and $J^2$ in the gap.
These rules enclose the exact rational expressions at every point
of the input rectangle. Positivity means a strictly positive lower
integer endpoint; no floating tolerance is involved.

For each fixed $m$ a certificate consists of closed integer rectangles
$[N_-,N_+]\times[i_-,i_+]$. Enter $N$ and $i$ as these two
intervals, and derive $i-1$ by subtracting the exact integer one
from the same $i$ interval. All four arguments $0,-1,i,i-1$
use the same $N$ interval, as do the pivots and weights. The
ordinary interval operations enclose these shared-variable
expressions; possible overestimation from repeated occurrences
does not affect the sign implication.

The inverse-Gram sums of~\eqref{eq:bg-inverse-product}, with weights
$W_k/d_k$, are applied to the innovations formed in the proof of
Lemma~\ref{lem:bg-finite-linear}.

The entire rectangle is accepted only
when every denominator test succeeds and the required gap has a
strictly positive lower bound. This includes all real points of
the rectangle, even when some lie outside the required integer
domain, such as above $i=\lfloor(H+1)/2\rfloor$ in a shorter
length column. Such points must pass the same positivity tests;
they are simply not needed for coverage. Splitting a rectangle is a search
device for finding certificates; the final verification simply
re-evaluates each listed rectangle and checks coverage independently.

\begin{lemma}\label{lem:bg-finite-linear}
The linear endpoint inequality in Proposition~\ref{prop:bg-endpoints}
holds throughout
\[
 2\le m\le 1039,\quad N\ge 6 m-1,\quad
 1\le i\le 99,\quad 2 i\le N-2 m+1<42 i.
\]
\end{lemma}
\begin{proof}
This domain is finite: $N\le 2 m+4156$, and
$4 m\le N-2 m+1<4158$ gives $m\le 1039$.
The certificate has 16,699 rectangles, covering 2,154,888 pairs
$(m,N)$ and 71,940,002 integer triples $(m,N,i)$.
There are $4158-4m$ lengths for each $m$, and every resulting
index interval is nonempty; thus the pair count is
$\sum_{m=2}^{1039}(4158-4m)=2,154,888$.
The triple count is the sum of the index-interval sizes printed below.
It evaluates the stable linear gap at 96-bit dyadic precision.
An independent 80-bit replay checks every rectangle, all positivity
conditions on intermediate denominators, and a positive lower gap.
The smallest lower gap in that replay is
\[
 \frac{146595612227992968}{2^{80}}>0.
\]
It belongs to the rectangle $m=326$, $2312\le N\le2401$,
$40\le i\le99$, again as an interval lower bound.

The stable evaluation computes normalized ratio differences directly.
If $T_j(x)$ is the Krawtchouk ratio of Appendix~\ref{app:bg-endpoints}
and $\rho_j^2=j/[3(N-j)]$ is as there, set
$\epsilon_j(x)\coloneqq T_j(x)/\rho_j^2-1$.
The exact recurrence is
\[
 \epsilon_0(x)=0,\qquad
 \epsilon_j(x)=
 \frac{4 x+(j-1)\epsilon_{j-1}(x)}
      {3(N-j)-4 x-(j-1)\epsilon_{j-1}(x)}.
\]
The denominator here equals $\Delta_j(x)$ after substituting
$T_{j-1}(x)=\rho_{j-1}^2(1+\epsilon_{j-1}(x))$; its lower bound
is checked positive in this evaluation as well.
At $x=0$ one has $\epsilon_j(0)=0$. In reversed degree coordinates,
the scaled difference from the origin profile is accumulated by
\[
 h^{\Delta}_0(x)=0,\qquad
 h^{\Delta}_k(x)=
 \frac{(1+\epsilon_j(x))h^{\Delta}_{k-1}(x)
          +\epsilon_j(x)3^{-(k-1)}}{3},\quad j=m-k+1.
\]
Applying the linear innovation filter from
\eqref{eq:bg-rational-innovation} to these differences gives
$z^\Delta_k(x)=z_k(x)-z_k(0)$ directly. In particular, the
innovation of $f-e$ is
\[
 \frac{z^\Delta_k(i)+z^\Delta_k(i-1)-z^\Delta_k(-1)}2,
\]
and that of $d$ is
$[z^\Delta_k(i)-z^\Delta_k(i-1)]/2$.
Together with $z_k(0)+z^\Delta_k(-1)/2$ for $e$, these give
interval enclosures of $E,J,W,G_d$ through the inverse-Gram sum,
without subtracting nearly equal endpoint vectors. The same input
intervals for $N,i$ are used throughout. This recurrence changes
the numerical conditioning, not the mathematical endpoint form.

For every fixed $m$ and every integer $N$ in its range, the accepted rectangles
are checked to cover all integers in
\[
 \left\lfloor\frac{N-2 m+1}{42}\right\rfloor+1
 \ \le i\le\min\left(99,
                   \left\lfloor\frac{N-2 m+1}{2}\right\rfloor\right).
\]
An independent union check and a second sweep over all rectangle
boundaries both verify that every integer in this interval is covered,
with the same pair and triple counts.
Every required point is therefore in a rectangle with a positive
sufficient gap, proving the lemma. The boundary $42 i=N-2 m+1$, like
every index with $12i\le N-2m+1$, is covered by
Lemma~\ref{lem:bg-linear-near}.
\end{proof}

The proof of Proposition~\ref{prop:bg-endpoints} needs only the subset
$2\le m\le 296$, $2 i\le N-2 m+1<12 i$ of this certified domain. It
has $\sum_{m=2}^{296}(1188-4m)=174{,}640$ pairs $(m,N)$ and 5,671,852
triples $(m,N,i)$; the counts and the smallest gap quoted in the proof
refer to the full certified domain.
This finite domain contains the complement, within the index range
of~\eqref{eq:bg-linear-target}, of the analytic linear range
$12 i\le H+1$ of Appendix~\ref{app:bg-endpoints}; the quarter
assertion is proved analytically there for every parameter.
Thus no unbounded parameter search is left in either assertion of
Proposition~\ref{prop:bg-endpoints}.

\subsection{Matched matrix certificates at small total degree}

The following domain contains the one required by
Proposition~\ref{prop:bg-small-degree}: it drops the parity term
$2(r\bmod2)$ from the length lower bound. Its endpoint matrices
also give independent finite checks of inequalities already implied
by Proposition~\ref{prop:bg-endpoints}, except for the exceptional
$3/2$ comparison stated below. Indeed, throughout this domain
$H+1\ge2D$, so $4i/[3(H+1)]\le2i/(3D)$ and
$D\le\lfloor(H+1)/2\rfloor$. The coefficient and endpoint
certificates are kept together to verify their matching constants.

\begin{lemma}\label{lem:bg-small-finite}
Let $2\le m\le 7$, $0\le r\le 15-2 m$, $D\coloneqq2 m+r$, and
\[
 6 m+2 r-1\le N<13D.
\]
For every nonzero real polynomial $q$ of degree at most $m$, the
following matched coefficient and endpoint inequalities hold.
For $3\le i\le D-1$, they are
\[
 \mathcal C_i(q)<\left(1-\frac{2i}{3D}\right)U_i\mathcal G(q),\qquad
 \frac{\mathcal E_i(q)}{U_i}-\mathcal E_0(q)
                      +\frac{2i}{3D}V\mathcal G(q)>0.
\]
At $i=D$ they are
\[
 \mathcal C_D(q)<\frac{U_D}{4}\mathcal G(q),\qquad
 \frac{\mathcal E_D(q)}{U_D}-\frac 54\mathcal E_0(q)
                                  +V\mathcal G(q)>0,
\]
except at $(m,r,N)=(2,0,11)$, where the respective constants
$1/4$ and $5/4$ are replaced by $1/2$ and $3/2$.
\end{lemma}
\begin{proof}
Use the raw basis $b_j(x)=P_j^{(N,4)}(x)-P_{j-1}^{(N,4)}(x)$,
$0\le j\le m$, where $P_{-1}=0$.
These are the raw difference coordinates used above; the local
matrices below represent the indicated quadratic forms in this basis.
Define the integer symmetric matrices
\[
 (K_\ell)_{jk}\coloneqq[z^jw^k]
 (1+z+w-3 zw)(z+w+2 zw)^\ell(1+3 zw)^{N-\ell},
\]
\[
 G_*\coloneqq K_0,\quad d_*\coloneqq\binom Ni,\quad
 A_* \coloneqq\sum_\ell\binom N\ell\binom r{i-\ell}K_\ell.
\]
Thus $A_*/d_*$ represents $\mathcal C_i$ and $G_*$ represents
$\mathcal G$. For $e_x\coloneqq(b_0(x),\ldots,b_m(x))^{\mathsf T}$, put
\[
 F_x\coloneqq e_xe_{x-1}^{\mathsf T}+e_{x-1}e_x^{\mathsf T}.
\]
Then $F_x/2$ represents $\mathcal E_x$. Compute
\[
 U_i=\frac uv=
 \frac{P_m^{(N-1,4)}(i-1)P_m^{(N-1,4)}(i)}
      {P_m^{(N-1,4)}(-1)P_m^{(N-1,4)}(0)},\qquad u,v>0.
\]
The finite sum defining $P_m$ gives integer numerator and denominator;
their positivity is checked before forming the matrices. Here $u,v$
are integers representing this ratio, and $F_x$ is an evaluation
matrix, distinct from the minimizing polynomial $F$ and the norm
$F=\|f\|^2$ used in the preceding subsection.

For an interior index the two integer matrices tested are
\[
 (3D-2 i)u d_*G_*-3DvA_*,\qquad
 3DvF_i-3DuF_0+4 iuVG_*.
\]
They are positive multiples of the coefficient and endpoint matrices
in the lemma, with multipliers $3Dvd_*$ and $6Du$, respectively.
At the top index the corresponding matrices are
\[
 u d_*G_*-4 vA_*,\qquad 4 vF_i-5 uF_0+8 uVG_*.
\]
Their multipliers are $4vd_*$ and $8u$.
For the one exceptional top case they are instead
\[
 u d_*G_*-2 vA_*,\qquad 2 vF_i-3 uF_0+4 uVG_*.
\]
Their multipliers are $2vd_*$ and $4u$.
All scaling factors are positive, so no sign inference is changed by
denominator clearing.

Enumerating the displayed ranges gives 43,995 interior quadruples
and 4,893 top quadruples. Two matrices are tested at each quadruple,
for 97,776 positive-definiteness decisions. For each integer symmetric
matrix, Bareiss fraction-free elimination gives its leading principal minors:
at stage $k$, the update is the determinant cross-product divided by
the preceding pivot (initially one). Every division is checked exact
and every pivot strictly positive. Sylvester's criterion then proves
positive definiteness. A separate implementation constructs the kernel
by polynomial convolution, evaluates the Krawtchouk values by their
recurrence, and uses rational Schur complements. It obtains the same
four complete streams of leading minors: interior coefficient,
interior endpoint, top coefficient and top endpoint. It also obtains
the same counts for each $D$.

The count itself is independent of the matrices: for each $m,r$ the
number of lengths is
$13D-(6 m+2 r-1)=20 m+11 r+1$; summing this over
$2\le m\le 7$, $0\le r\le 15-2 m$ gives 4,893.
Multiplication by $D-3$ before summation gives 43,995 interior cases.
These identities check the complete enumeration. The asserted
quadratic-form inequalities now follow from the positive matrices.
\end{proof}

For a minimizer, $\mathcal E_0\ge V\mathcal G$ turns each matched
pair into the coefficient criterion by
Lemma~\ref{lem:bg-cap-comparison}. In the exceptional case the
coefficient and endpoint multipliers are $U_i/2$ and $U_i$,
respectively, so it is covered by the same implication. The boundary
$N=13D$ belongs to the analytic tail
in Appendix~\ref{app:bg-minima}. The finite lemma, that tail, and
the separate first two indices prove
Proposition~\ref{prop:bg-small-degree}.

\end{document}